\documentclass[a4paper,12pt]{article}
\usepackage{amsfonts,slashed}
\usepackage[greek,english]{babel}
\usepackage{cancel}
\usepackage{upgreek}
\usepackage{float}
\usepackage{url}
\usepackage[sort&compress,numbers]{natbib}
\usepackage{latexsym}
\usepackage{amsfonts}
\usepackage{amsmath}
\usepackage{epsfig}
\usepackage{isomath}
\usepackage{latexsym,amssymb}
  \usepackage{amsmath,amssymb,amsthm}
\usepackage{ifpdf}
\ifx\pdfoutput\undefined
   \pdffalse
   \usepackage{cite}
 \else
   \pdfoutput=1
   \pdftrue
\fi
\DeclareFontFamily{OMS}{rsfs}{\skewchar\font'60}
\DeclareFontShape{OMS}{rsfs}{m}{n}{<-5>rsfs5 <5-7>rsfs7 <7->rsfs10 }{}
\DeclareSymbolFont{rsfs}{OMS}{rsfs}{m}{n}
\DeclareSymbolFontAlphabet{\Scr}{rsfs}

\numberwithin{equation}{section}
\def\be{\begin{equation}}
\def\ee{\end{equation}}
\def\ba{\begin{array}}
\def\ea{\end{array}}

\newcommand{\bea}{\begin{eqnarray}}
\newcommand{\eea}{\end{eqnarray}}

\newcommand{\ft}[2]{{\textstyle\frac{#1}{#2}}}

\def\Id{{\bf 1}}
\def\Zero{{\bf 0}}
\def\N{\mathcal{N}}
\def\Ms{{\Scr M}}

\def\I{\mathcal{I}}
\def\R{\mathcal{R}}
\def\={~=~}
\def\*{{}^*}
\def\nn{\nonumber}
\def\nne{\nonumber\\}
\def\eps{\epsilon}

\def\GN{\mathrm{G}_{\textsc{n}}}
\def\L{{\Scr L}}
\def\LB{{\Scr L}_{\textsc{b}}}
\def\Lscal{{\Scr L}_{\rm scal}}
\def\Lgaug{{\Scr L}_{{\rm gauged}}}
\def\CC{\mathbb{C}}

\def\LL{\mathbb{L}}
\def\P{\mathcal{P}}
\def\Ps{{\Scr P}}
\def\Es{{\Scr E}}

\def\Solv{{\Scr S}}
\def\galg{\mathfrak{g}}
\def\halg{\mathfrak{H}}
\def\kalg{\mathfrak{K}}
\def\Tr{\mathrm{Tr}}
\def\M{\mathcal{M}}

\def\cD{\mathcal{D}}
\def\Gd{\mathcal{G}}

\def\Or{\mathcal{O}}
\def\Mscal{{\Scr M}_{\rm scal}}

\def\MsSK{\Ms_{\textsc{sk}}}
\def\MsQK{\Ms_{\textsc{qk}}}
\def\Rs{{\Scr R}}
\def\Tb{\mathbb{T}}
\def\S{{\bf S}}

\def\dim{\mathrm{dim}}
\def\rank{\mathrm{rank}}
\def\Adj{\mathrm{adj}}
\def\Sp{\mathrm{Sp}}

\def\SU{\mathrm{SU}}
\def\SO{\mathrm{SO}}

\def\N{\mathcal{N}}
\def\Ms{{\Scr M}}

\def\I{\mathcal{I}}
\def\R{\mathcal{R}}
\def\={~=~}
\def\*{{}^*}
\def\nn{\nonumber}
\def\nne{\nonumber\\}
\def\eps{\epsilon}

\def\GN{\mathrm{G}_{{\rm N}}}
\def\Gm{{\Scr G}}
\def\L{{\Scr L}}
\def\Lscal{{\Scr L}_{\rm scal}}
\def\Lgaug{{\Scr L}_{{\rm gauged}}}
\def\CC{\mathbb{C}}
\def\LL{\mathbb{L}}
\def\P{\mathcal{P}}
\def\Ps{{\Scr P}}
\def\Es{{\Scr E}}

\def\Solv{{\Scr S}}
\def\galg{\mathfrak{g}}
\def\halg{\mathfrak{H}}
\def\kalg{\mathfrak{K}}
\def\Tr{\mathrm{Tr}}
\def\M{\mathcal{M}}

\def\cD{\mathcal{D}}
\def\Gd{\mathcal{G}}

\def\Mscal{{\Scr M}_{\rm scal}}

\def\MsSK{\Ms_{\textsc{sk}}}
\def\MsQK{\Ms_{\textsc{qk}}}
\def\Rs{{\Scr R}}
\def\Tb{\mathbb{T}}
\def\S{{\bf S}}

\def\dim{\mathrm{dim}}
\def\rank{\mathrm{rank}}
\def\Adj{\mathrm{adj}}
\def\Sp{\mathrm{Sp}}

\def\SU{\mathrm{SU}}
\def\SO{\mathrm{SO}}

\begin{document}
\begin{center}\phantom{}\vskip 3cm
{\LARGE \bf  Gauged Supergravities}\\
\vspace{60pt}
{\bf {\large Mario Trigiante}$^1$}\\

\vspace{15pt}
{\small
$^1$
Dipartimento di Fisica,
  Politecnico di Torino, Corso Duca degli Abruzzi 24, \\
$~~~$ I-10129
  Turin, Italy and INFN, Sezione di Torino, Italy.\\
\footnotesize{    \texttt{email: mario.trigiante@polito.it}}.
 }
\end{center}
\vskip 1.5cm
\begin{center}
{\bf Abstract}
\end{center}
We give a general review of extended supergravities and their gauging using the duality-covariant embedding tensor formalism. Although the focus is on four-dimensional theories, an overview of the gauging procedure and the related tensor hierarchy in the higher-dimensional models is given. The relation of gauged supergravities to flux compactifications is discussed and examples are worked out in detail.

 \vfill\eject
 \tableofcontents
 \section{General Overview}
 A long-standing problem in high-energy theoretical physics is to formulate a quantum theory unifying all four fundamental interactions. The known elementary particles and their electromagnetic, weak and strong interactions have been given a consistent quantum description within the Standard Model of particle interactions (SM). This picture, however, does not accommodate gravity, which is described, at the classical level, by general relativity. The consistent construction, from Einstein's theory, of a perturbative quantum gravity is hampered by the fact that the coupling constant is $\kappa=\sqrt{8\pi G_N}$, $G_N$ being Newton's constant, which has the dimension of a length (in the natural units $c=\hbar=1$). This makes the theory non-renormalizable, as opposed to the SM: UV divergences occurring in the Feynmann diagrams cannot be disposed of by the introduction of a finite number of counterterms in the Lagrangian. Another feature which distinguishes the three gauge interactions from gravity is the huge hierarchy between the corresponding characteristic energy scales: The SM describing the former is consistently defined at energies of the order of the electro-weak scale ($M_W\sim 100\,{\rm GeV}$) while quantum gravity effects are expected to be important at scales of the order of the Planck mass $M_P=\sqrt{\hbar c/(8\pi\,G_N)}\approx 2.4\times \,10^{18}\,{\rm GeV}/c^2$.
 \paragraph{Why supersymmetry and supergravity?}
 There are various reasons to expect supersymmetry to play an important role in the construction of a unified quantum theory. Supersymmetry is a symmetry between bosons and fermions (see for instance \cite{Martin:1997ns} for an in-depth review of  supersymmetry phenomenology). As such it implies relations between quantum amplitudes whose effect is to soften the UV divergences of the theory. \footnote{For example it is known that a Yang-Mills theory with the maximal amount of supersymmetry compatible with its renormalizablity ($\mathcal{N}=4$), is finite to all orders of perturbation \cite{Mandelstam:1987jr}.}
 This solves for instance the \emph{hierarchy problem} which occurs when trying to extend the SM to energies higher than its characteristic scale (for instance up to $M_P$ at which new physics, associated with gravity, is expected to occur): By virtue of supersymmetry, the bosonic and fermionic contributions to the dangerous quadratic divergences in the radiative corrections to the Higgs mass cancel, leaving just the logarithmic divergences in the energy cut-off. This requires considering at least the minimal supersymmetric extension of the SM (MSSM), whose spectrum comprises, together with the known particles, the corresponding \emph{super-partners}. Besides stabilizing the ratio of the  Higgs mass, now measured to be $m_H\approx 125 \, {\rm GeV}/c^2$, to the Planck mass $M_P$ against quantum corrections, the presence of an underlying supersymmetry also has the beneficial effect of unifying the coupling constants at some higher energy scale: The coupling constants  of the weak, electromagnetic and strong interactions, if extrapolated to high energies through their renormalization-group evolution, meet at an energy scale of about $2\times 10^{16} \, {\rm GeV}$, thus hinting towards a Grand Unified Theory (GUT) of the fundamental interactions. Finally MSSM also contains a natural candidate for the Light Supersymmetric Particle (LSP) to account for the dark matter in our universe.\par
 Supersymmetry was introduced in the $70$'s \cite{Golfand:1971iw,Wess:1974tw,Haag:1974qh} independently of its application to particle physics (for a general introduction to supersymmetry see for instance \cite{Sohnius:1985qm}). It is described as an extension of the Poincar\'{e} algebra, named \emph{super-Poincar\'{e}} algebra,  by \emph{fermionic} (supersymmetry) generators $Q_\alpha$ intertwining between space-time symmetry (Poincar\'{e}) and internal symmetries. Such generators are fermionic in that they transform under Lorentz transformations as spinors ($\alpha$ is a 4-spinor index) and obey \emph{anti-commutation} relations. More specifically the anti-commutator of two supersymmetry generators yields the generators $\hat{\mathcal{P}}_\mu$ of space-time translations:
 \begin{equation}
\{Q_\alpha,\,Q_\beta\}=2\,i\,(\gamma^\mu C)_{\alpha\beta}\,\hat{\mathcal{P}}_\mu\,.\label{QQP}
\end{equation}
This means that a combination of successive supersymmetry transformations results into a space-time translation.
The generators $Q_\alpha$ moreover change the spin-statistics of the field they act on: Their action on a bosonic field yields a fermionic one and vice-versa.\par
 Introducing supersymmetry in a theory of gravity, which is based on \emph{local} Poincar\'{e} invariance, forces us to require invariance also under \emph{local} supersymmetry transformations \cite{VanNieuwenhuizen:1981ae}, since a suitable combination of a local Lorentz transformation on a global supersymmetry one would yield a resulting local supersymmetry transformation. Vice versa, invariance under local supersymmetry implies gravity, since, by virtue of (\ref{QQP}), it implies invariance under local space-time reparametrization, which is the symmetry principle on which General Relativity is based. A theory that is invariant under local supersymmetry transformations (and thus under the whole super-Poincar\'{e} group) is called \emph{supergravity} \cite{Freedman:1976xh,Deser:1976eh}. There are several very good reviews on supergravity, see for instance \cite{VanNieuwenhuizen:1981ae,Nilles:1983ge,Wess:1992cp,Castellani:1991et,deWit:2002vz,Freedman:2012zz,Tanii:1998px,Samtleben:2008pe}. \footnote{We consider for the time being theories in four dimensions} \footnote{For a geometric formulation of supergravity as the ``gauge theory'' of the super-Poincar\`{e} group see \cite{Chamseddine:1976bf},\cite{MacDowell:1977jt},\cite{Neeman:1978nn},\cite{DAdda:1980tc}, the \emph{rheonomic} approach introduced in the last two references is exhaustively reviewed in \cite{Castellani:1991et}.}Supersymmetric theories are characterized by the amount of supersymmetry, namely by the number $\mathcal{N}$ of the fermionic generators $Q_\alpha^A$, $A=1,\dots, \mathcal{N}$ occurring in the supersymmetric extension of the Poincar\'{e} algebra. Theories with $\mathcal{N}=1$ are called \emph{minimal}, in contrast to the \emph{extended} theories which have $\mathcal{N}>1$. \par
 In minimal supergravity the superpartner of the graviton $g_{\mu\nu}$ is a spin-$3/2$ particle $\psi_\mu$ (we suppress the spinor index $\alpha$) called the \emph{gravitino}, which has the role of the \emph{gauge field} associated with local supersymmetry. In an extended supergravity with $\mathcal{N}$ supersymmetries, there are  $\mathcal{N}$ gravitino fields $\psi^A_\mu$, each associated with a supersymmetry generator $Q^A$. Clearly the larger $\mathcal{N}$ the more constrained the theory is by supersymmetry, the larger the irreducible representations of the super-Poincar\'{e} group (super-multiplets). Since the supersymmetry generators carry a spin-$1/2$ representation, their successive action on the lower spin state in a super-multiplet will produce states with higher spins, up to a maximum value which increases with $\mathcal{N}$.
 The limit on the amount $\mathcal{N}$ of supersymmetry in supergravity comes from
the possibility of a consistent coupling to gravity, which restricts the maximum spin of the fields to be 2, and this in turn requires $\mathcal{N}\le 8$ (in four space-time dimensions).  Supersymmetry improves the UV properties of the theory, making pure supergravity finite up to two loops (pure Einstein's gravity is only one-loop finite \cite{'tHooft:1974bx,Goroff:1985sz} and this property is spoiled by the presence of matter).
The maximal (ungauged) $\mathcal{N}= 8$ supergravity, just as the rigid $\mathcal{N}= 4$ super-YM theory, is unique (supersymmetry fixes its field content to be that of the supermultiplet containing the graviton as the maximum spin state). Though its perturbative finiteness  has been tested, so far, up to four loops \cite{Bern:2011qn}, some believe the maximal theory to be perturbatively finite just as its rigid  $\mathcal{N}= 4$ counterpart.
In spite of the beneficial effects of supersymmetry in making the UV divergences less severe, a quantum theory of supergravity in general suffers from the same non-renormalizability problems as general relativity.
\par
There is also another argument in favor of supergravity as a general framework in which to formulate a consistent supersymmetric model for particle interactions, like the MSSM. In the MSSM, supersymmetry is a global (or rigid) symmetry. If it were realized in our universe, it would require a mass degeneracy between the observed particles and their superpartners, which is clearly not observed (none of the  superpartners have been observed so far). It is natural to assume that supersymmetry is spontaneously broken in our universe, namely that we live in a vacuum which is not supersymmetric. Phenomenological constraints imply the masses of the superpartners not to be too high (about $1\,{\rm TeV}$).  In the absence of supergravity, and thus of the gravitino-field, spontaneous global supersymmetry breaking would imply a stringent relation (sum-rule deriving from the vanishing of the \emph{super-trace} of the mass-square matrix) between the (tree-level) masses of the known particles and those of their super-partners which is manifestly at odds with what we observe. The presence of a ``hidden sector'' besides the ``visible'' one of the MSSM,  containing the supergravity multiplet (and thus, in the broken phase, a massive gravitino) allows to relax the phenomenological implications of the super-trace sum rule, allowing for the existence of a desirable mass gap between the observed particles and their superpartners of the order of the gravitino mass.  In this picture supersymmetry breaking occurs in the hidden sector and propagates to the visible one through a gravitational strength interaction, see \cite{Martin:1997ns} and references therein.
\paragraph{Supergravity from Superstring Theory}
In superstring theory \cite{Polc} the fundamental objects are (open or closed) strings of finite length $\ell_s=\sqrt{\alpha'}$ and tension $T\sim 1/\ell_s$. In this picture the known particles and interactions are expected to arise from oscillation modes of these one-dimensional objects. The spectrum of the oscillating strings contains the graviton (closed strings) as well as gauge-vectors (open strings). Moreover the string length provides a natural UV cut-off and there is strong evidence that this theory be finite. Also for this reason superstring theory is generally considered to be a promising candidate to a finite quantum theory unifying the known interactions.\par
Supersymmetry is an essential ingredient for a consistent definition of the theory.  Superstring theory is consistently defined  as a conformal 2-dimensional sigma-model (world-sheet theory) on a suitable ten-dimensional space-time. It is defined perturbatively in its coupling constant $g_s$, which is not a parameter of the theory but a dynamical quantity expressed in terms of the v.e.v. of its scalar excitation (the dilaton field) on the chosen background. The very definition of superstring theory is therefore background-dependent. In the low-energy limit (effected by sending the string tension to infinity or, equivalently, $\ell_s\rightarrow 0$) strings act as point-like particles and their dynamics is captured by an effective field theory. This theory is a ten-dimensional supergravity, to which the superstring background is a solution. The classical supergravity description of superstring dynamics holds to the lowest order in the string length (so that higher-order curvature terms in the effective action can be neglected) and at tree-level in $g_s$. The existence of more than one consistent formulation of this theory (heterotic, Type I/II) seems to be at odds with the concept of a unique unifying quantum theory. In the 90's, however, (non-)perturbative correspondences (named \emph{dualities}) were found between the various superstring theories realized on different backgrounds which allow to think of them as effective descriptions of the same microscopic degrees of freedom and hint towards the existence of an \emph{M-theory} in eleven dimensions as describing the non-perturbative regime of one of the string theories (Type IIA). Although the fundamental degrees of freedom of M-theory are not known, its low-energy effective theory is the well known (and unique) eleven-dimensional supergravity \cite{Cremmer:1978km}. The discovery of dualities opened a window on the non-perturbative regime of string theory and unveiled the existence in its spectrum, besides the fundamental string, of solitonic extended-objects, named D-branes, which minimally couple to the RR fields \cite{Polchinski:1995mt}.\par
If superstring/M-theory is the correct theoretical setting where to search for the unifying quantum theory of the fundamental interactions, it should be possible to derive from it a phenomenologically viable description of our universe. Such an effective description should include the MSSM at low energy scales. These candidate-``microscopic'' theories should also satisfy other phenomenological requirements which are becoming more and more stringent: They should, for instance, account for the observed ``small'' cosmological constant and  provide an inflationary scenario for the early universe in line with the strong experimental constraints provided by the recent high-precision cosmological measurements \cite{Planck:2013jfk,Ade:2013zuv}. Obtaining phenomenological predictions from superstring/M-theory is rather problematic for a number of reasons: First of all these theories are defined in higher dimensions; Their non-perturbative properties are far from being completely understood; Related to this is the problem of finding a (non-perturbative) mechanism which could select the vacuum of our universe from the large number of solutions of these theories;  Moreover superstring (or M-) theories have far more symmetries than we observe in our universe: Type II superstring theories, as well as M-theory, for instance, exhibit the maximal amount of supersymmetry compatible with a consistent theory of gravity, while our universe is not supersymmetric.\par
The most common procedure for deriving an effective four-dimensional model from a higher-dimensional theory, is the Kaluza-Klein compactification\footnote{By Kaluza-Klein compactification we actually mean a generalization of the original mechanism, devised by T. Kaluza and O. Klein in \cite{Kaluza,Klein}, of dimensional reduction on a circle of the five-dimensional Einstein theory.} of the latter on a suitable compact internal manifold $M_{int}$ (see \cite{Duff:1986hr} for an early review on the subject). This amounts to  realizing
 superstring (or eleven-dimensional supergravity), on a space-time solution of the form
\begin{equation}
M=M_{D=4}\times M_{int}\,,
\end{equation}
where $M_{D=4}$ is a non-compact, four-dimensional space-time in which we live and $M_{int}$ is a six-dimensional compact manifold (or seven-dimensional if we consider eleven-dimensional supergravity). The massless modes of superstring theory, or the fields of eleven-dimensional supergravity, are expanded in normal modes of the internal manifold (i.e. eigenfunctions of the Laplace operator on $M_{int}$). The coefficients of this expansion only depend on the four non-compact space-time coordinates $x^\mu$, $\mu=0,\dots, D-1$, and describe fields propagating on $M_{D=4}$. They comprise a finite number of massless fields corresponding to harmonics on the internal manifold, together with infinitely many massive states (the Kaluza-Klein states), with masses of the order of $1/R$, $R$ being the ``size'' of the internal manifold. In some cases a consistent truncation of the low lying modes, together with their mutual interactions, is described by an effective $D=4$ supergravity. In general, by the same token, realizing superstring/M-theory on a space-time of the form $M_{D}\times M_{int}$, we may obtain an effective $D$-dimensional supergravity theory as a consistent truncation of the higher dimensional parent-theory.\footnote{By consistent truncation we mean that all solutions to the lower-dimensional theory precisely correspond to solutions of the
higher-dimensional one.} This effective theory in $D$-dimensions has $M_{D}$ as vacuum solution.  The simplest example of Kaluza-Kelin compactification is the dimensional reduction on an internal torus $T^n$, $n=6$ or $n=7$ if we reduce superstring or M-theory, respectively. In this case the string massless excitations, or the eleven-dimensional fields, are expanded in Fourier-series with respect to the torus-coordinates $x^\upalpha$, $\upalpha=D,\dots, D+n-1$, and the zero-modes, which do not depend on $x^\upalpha$ and are massless $D$-dimensional fields, are described by a supergravity in $D$-dimensions.\par
The general features of the effective $D$--dimensional supergravity emerging from a Kaluza-Klein reduction depend on the original higher-dimensional theory as well as the $M_{D}\times M_{int}$ background. In fact the geometry of the internal manifold $M_{int}$ affects the amount of supersymmetry of the lower-dimensional theory,
as well as its internal symmetries and its field-content, which is arranged in (super-) multiplets with respect to the (super-)symmetries of the vacuum solution $M_{D}$. As far as the internal local symmetries are concerned, this is a general feature of Kaluza-Klein compactifications: Continuous isometries of $M_{int}$ induce, in the low-energy $D$--dimensional supergravity, local symmetries gauged by the vectors which originate from the metric (Kaluza-Klein vectors).
 Toroidal compactifications preserve all the supersymmetries of the parent theory, so that, starting from Type II or eleven-dimensional supergravity, the massless lower-dimensional modes are described by a \emph{maximally-extended} supergravity. In \cite{Cremmer:1978ds} the simplest version  of four-dimensional $\mathcal{N}=8$ supergravity was constructed from dimensional reduction of the eleven-dimensional theory on a seven-dimensional torus $T^7$ and exhibits a maximally-supersymmetric Minkowski vacuum. Compactifications on generic Calabi-Yau six-dimensional  manifolds ($CY_3$) preserve  one-fourth the original amount of supersymmetries and were originally used in \cite{Candelas:1985en} to derive, for the first time, minimal four-dimensional chiral, anomaly-free theories from the heterotic superstring. In this case the lower-dimensional theory only depends, aside from the higher-dimensional one, on the topology of the Calabi-Yau manifold, since it is only defined by the harmonic-forms on the internal space. Moreover, for topological reasons, no Kaluza-Klein vector originates from the reduction. This is consistent with the fact that the compactification induces no extra local internal symmetry (which would be gauged by the Kaluza-Klein vectors) since the internal manifold has no continuous symmetries.  \par Compactifications of Type II-supertring/M-theory on Ricci-flat internal manifolds, like a torus, Calabi-Yau spaces, products of the two (like $K3\times T^2$), toroidal-orbifolds etc.,  typically yield effective four-dimensional supergravities (like the maximal one of  \cite{Cremmer:1978ds}) which have the general feature of being \emph{ungauged}: The $n_v$ vector fields are not minimally coupled to any other field, so that the theory only features a ${\rm U}(1)^{n_v}$ gauge symmetry with respect to which all fields are neutral.\footnote{ The same kind of compactifications of heterotic or Type I superstring will produce four-dimensional theories whose local symmetries gauged by the vector fields only originate from the ten-dimensional gauge group.}\par
 \paragraph{Ungauged Supergravities and Duality} Ungauged supergravities in $D$-dimensions have serious drawbacks from a phenomenological point of view. First of all they are plagued, at the classical level, by the presence of  massless scalar fields, some of which are related to the moduli of the internal manifold, which describe fluctuations in its shape and size. Massless scalars coupled to gravity would produce effects which are not observed in our universe. Moreover such fields enter interaction terms in the Lagrangian, thus spoiling the predictiveness of the theory, being their v.e.v. not fixed by any dynamics. The field content of the ungauged maximal four-dimensional supergravity  constructed in \cite{Cremmer:1978ds},
consists of the only gravitational super-Poincar\'{e} multiplet and contains 70 scalar fields, part of which can be interpreted as the moduli of the internal seven-torus and part as originating from the three-form tensor field in the $D=11$ supergravity. The classical theory features no scalar potential which could fix the values of these scalars on the vacuum. Aside from this serious shortcoming, ungauged supergravities are interesting for a number of reasons. First of all they provide the general framework from which to construct more realistic models through the \emph{gauging procedure} which is the main subject of the present report.
 Secondly these theories feature, at the classical level, a rich structure of global symmetries which were conjectured \cite{Hull:1994ys} to encode the known string/M-theory dualities (see \cite{Vafa:1997pm} for a review on this subject). As mentioned above, the idea behind such dualities is that different compactifications of M- or superstring theories provide distinct descriptions of the same quantum degrees of freedom. These duality-related pictures are characterized by the same spectrum and interactions and thus by the same low-energy effective theory.
In fact the coincidence of the resulting lower-dimensional supergravities is a hint towards the equivalence of the corresponding microscopic constructions within superstring or M-theory and dualities, as correspondences between backgrounds which are solutions to a same effective supergravity, manifest themselves as discrete global symmetries of its field equations.
The simplest such equivalences is T-duality relating a string theory compactified on a circle of radius $R$ to one compactified on a circle of radius $\alpha'/R$ as being described by the same two-dimensional conformal field theory. In a toroidal compactification we can perform T-dualities along any combinations of the $n$ directions of the internal torus $T^n$. Type IIA and IIB compactified down to $D=10-n$ on $T^n$ are described by equivalent $D$-dimensional supergravities which are mapped into one another by the effect of T-dualities along an odd number of directions, while T-dualities along an even number of directions are discrete global symmetries of each theory. The effect of T-duality on the $D$-dimensional fields is implemented by the group
${\rm O}(n,n; \mathbb{Z})$, where the restriction to the integer
numbers is required by the boundary conditions on the coordinates of the torus.
%In particular
%T-duality transformations along an odd number of directions are represented by elements of
%${\rm O}(n,n; \mathbb{Z})$ with negative determinant (in the fundamental representation).
\par
 Another example is S-duality which relates two string theories on backgrounds with coupling constants  $g_s$ and $1/g_s$, respectively. As opposed to T-duality which holds to any order in string perturbation theory, S-duality is non-perturbative. It is conjectured to be a symmetry of  Type IIB superstring theory, connecting its weak and strong-coupling regimes. Evidence for it, aside from the  coincidence of the supergravity descriptions of the massless modes, was gained by comparing part of the spectrum of the corresponding theories, consisting of dyonic states (the BPS states) which satisfy a Bogomolnyi bound between their electric-magnetic charges and their masses. As a consequence of this feature, they preserve an amount of supersymmetry, which protects their masses from perturbative string corrections. Another example of S-duality is the one connecting Type IIA superstring to M-theory. The supergravity description of the former is  obtained by dimensionally reducing eleven-dimensional supergravity  on a circle $S^1$ and the Type IIA coupling constant $g_s$ is related to the radius of the eleventh dimension, so that in the non-perturbative limit $g_s\rightarrow \infty$, the internal circle  decompactifies and the theory becomes effectively eleven-dimensional. Moreover the  massless modes of eleven-dimensional supergravity (low-energy effective description of M-theory) and Type II superstring theories compactified to $D$-dimensions on $T^{11-D}$ and $T^{10-D}$, respectively, are described by the same $D$-dimensional ungauged, maximal supergravity which features S-duality, as well as the T-dualities along an even number of internal directions of $T^{10-D}$, as global symmetries.  S and T-dualities combine in the so-called \emph{U-dualities} which were conjectured in \cite{Hull:1994ys} to unify all dualities and thus to be exact symmetries of the, yet unknown,  background-independent quantum theory underlying superstring and M theories. U-dualities close a discrete  \emph{U-duality group} which is identified in \cite{Hull:1994ys} with the subgroup $G(\mathbb{Z})$ of the continuous global symmetry group $G$ of the classical $D$-dimensional (ungauged) supergravity, which survives quantum corrections. For this reason the study of global symmetries of supergravity models and of their action on the corresponding solutions, plays an important role in understanding the non-perturbative aspects of superstring theories.\par
Ungauged supergravities feature solitonic solutions, namely configurations of the neutral fields which exhibit electric and magnetic charges \cite{Witten:1978mh,Gibbons:1982fy}. These are typically  asymptotically-flat black holes and black branes\footnote{See for instance \cite{Stelle:1996tz}, \cite{Andrianopoli:2006ub} and references therein.}, some of which are realized in terms of D-branes or systems of D-branes in ten dimensions, and arrange in orbits with respect to the action of $G(\mathbb{Z})$. These solutions, which comprise the BPS states mentioned above, have been extensively studied in the literature, for instance in connection to  the microscopic description of the black hole entropy. It is reasonable to think that the latter, being related to the number of black hole microstates, should not depend on their description, and therefore ought to be a duality-invariant quantity. Consistently with the conjecture of \cite{Hull:1994ys}, the entropy of extremal (i.e. with zero Hawking-temperature) black holes was found (at least in the small curvature limit in which it is proportional to the horizon area by  the Bekenstein-Hawking formula) to be described by a quantity which is indeed invariant with respect to the global symmetry group $G(\mathbb{Z})$, see \cite{Andrianopoli:2006ub} and references therein.
One of the major successes of superstring theory is the derivation, in certain limits, of the black hole entropy from a microstate counting, first achieved  by Strominger and Vafa in \cite{Strominger:1996sh} and eventually extended to a large number of supersymmetric solutions (see for instance \cite{Sen:2014aja} for a concise up-to-date review and for the relevant references).
\paragraph{Flux Compactifications and Gauged Supergravities} The progress witnessed since the late nineties in our understanding of the non-perturbative side of superstring theory opened the possibility of a wide variety of new compactification scenarios in which the D-branes became an essential ingredient. In the ``brane-world'' constructions the ``visible sector'', described by a Standard Model-like model, is realized on the world volume of a D-brane or in the intersection of D-branes, where the gauge fields are constrained to propagate \cite{Lust:2004ks}. In this picture gravity and the rest of the  ``hidden sector'', as opposed to the gauge interactions, propagate in the whole space-time (the bulk) and this substantial difference between the to sectors, in the presence of a suitable space-time geometry, could possibly explain the origin of the hierarchy of scales between the gravitational and the Standard Model interactions. Another ingredient in superstring (or M-theory) compactifications is the presence of internal \emph{form-fluxes}, namely of non-vanishing v.e.v. of field strengths $F^{(p+1)}$ of higher-order forms which are present in the massless string/M-theory spectrum, across cycles $\Sigma_{p+1}$ of the internal manifold $M_{int}$:
\begin{equation}
\langle \int_{\Sigma_{p+1}} F^{(p+1)}\rangle\neq 0\,.
\end{equation}
Consistently with the requirement of Lorentz symmetry of $M_D$ we can also allow for a flux of a $D$-form field strength which is proportional to the $\epsilon$-symbol along the corresponding non-compact directions: $\langle F^{(D)}_{\mu_1\dots \mu_D}\rangle\propto \epsilon_{\mu_1\dots \mu_D}.$\par
Some of these fluxes (the RR ones) are sourced by D-branes.
In order to evade no-go theorems \cite{deWit:1986xg,Chamseddine:1997nm,Chamseddine:1997mc,Maldacena:2000mw,Ivanov:2000fg},
consistent compactifications of string theory to D-dimensional space-times with non-negative
cosmological constant, in the presence of fluxes and possibly of branes, may require in superstring theory the introduction
of negative-tension objects which are present in its spectrum: the orientifold planes (or O-planes)\cite{Angelantonj:2002ct}.\par
It has been realized in the last twenty years or so that fluxes and extended objects, which were not present in \emph{traditional} compactifications on Ricci-flat manifolds, are essential ingredients for the construction of realistic models for our universe.
They contribute through their energy-momentum tensors, to the Einstein equation and thus may affect in a non-trivial way the geometry of the background, including the internal manifold $M_{int}$.  \footnote{For a tentative list of early references on the subject of flux-compactifications in the presence of D-branes and O-planes see \cite{Polchinski:1995sm,Dasgupta:1999ss,Gukov:1999ya,Giddings:2001yu,Dall'Agata:2001zh,Frey:2002hf,Kachru:2002he,Kachru:2002sk,Tripathy:2002qw,D'Auria:2002tc,D'Auria:2002th,Kachru:2003aw,D'Auria:2003jk,Berg:2003ri,Angelantonj:2003rq,Angelantonj:2003up,Andrianopoli:2003jf,deWit:2003hq,Angelantonj:2003zx,Lust:2004fi,Lust:2005dy,Lust:2006zg} while for good general reviews see for instance \cite{Grana:2005jc,Douglas:2006es,Blumenhagen:2006ci}.} Flux backgrounds of string theories or $D=11$ supergravity were already known and subject of study in the eighties. This is the case, for instance, of the Freund-Rubin solution \cite{Freund:1980xh} describing the compactification of $D=11$ supergravity on a round seven-dimensional sphere. Variants of the seven-sphere compactification were also studied \cite{Englert:1982vs,Pope:1984bd} (see \cite{Duff:1986hr} for a review), together with compactifying solutions  with general internal homogeneous manifolds \cite{Castellani:1983yg}. The first flux compactification of heterotic string was studied in \cite{Strominger:1986uh}. However, due to the limited knowledge of the geometric properties of $M_{int}$ and of the supergravity model construction techniques, very few of such backgrounds could be described in terms of an effective lower-dimensional model. These include the Freund-Rubin compactification, see below. For this reason, in the eighties and early nineties the attention was mainly focussed on flux-less compactifications on Ricci-flat geometries, whose relation to the lower-dimensional effective supergravity description could be better understood.\par
 In the presence of fluxes and branes one is then led to consider spontaneous compactifications of superstring/M-theory on more general backgrounds which have the form of a warped product $M_{D}\times M_{int}$ of a non-compact $D$--dimensional space-time time, not necessarily Minkowski, and an internal compact manifold, which, in general, is no-longer Ricci-flat. The problem of finding  backgrounds of this kind which preserve a minimal amount of supersymmetries requires considering more general geometries for $M_{int}$, characterized by a restricted structure group (also called $\tt{G}$-structure manifolds) instead of a restricted holonomy, as it is the case for Calabi-Yau manifolds (see \cite{Grana:2005jc} and references therein)\footnote{The existence of residual sypersymmetries amounts to requiring the background to admit a number of Killing spinors, which are spinors along which the supersymmetry transformation of the fermions on the chosen background vanishes. In the absence of fluxes and branes this condition implies the existence of a covariantly constant spinor on $M_{int}$ (with respect to the Levi-Civita connection) which restricts the holonomy of the manifold, for superstring compactifications, to ${\rm SU}(3)$ or less  (Calabi-Yau manifolds). For M-theory compactifications the same condition constrains  the holonomy group to be contained in the ${\rm G}_{2}$ subgroup of ${\rm SO}(7)$. In the presence of fluxes, supersymmetry requires the global existence of a spinor on $M_{int}$ which is covariantly constant with respect to a torsionful connection, the torsion being related to the internal fluxes. The global existence of the Killing spinor only requires a restriction of the structure group of the space rather than its holonomy group. In the case of superstring  compactifications this leaves the possibility of a broad class of manifolds with ${\rm SU}(3)-$ structure  or less which, in general, are no longer Calabi-Yau, K\"ahler or even complex. The same condition on M-theory compactifications leads to consider internal manifolds with  ${\rm G}_{2}$-structure.}. This broader class of internal manifolds comprises geometries characterized by background quantities,  called \emph{geometric fluxes}, which define  topology-deformation of traditional Ricci-flat compactification manifolds like tori or Calabi-Yau spaces. When compactifying on these manifolds, the fields of the lower-dimensional supergravity are obtained by expanding the higher-dimensional ones along the same basis of forms which defined the cohomology of the  undeformed (Ricci-flat) counterpart.
 These forms however, are now no longer closed. An example is the so-called \emph{twisted torus} or Iwasawa manifold \cite{Scherk:1979zr,Kaloper:1999yr,LopesCardoso:2002vpf,Grana:2004bg,Villadoro:2004ci,Andrianopoli:2005jv,Derendinger:2004jn,Grana:2005ny,Hull:2005hk,Dall'Agata:2005ff,dft1,dft2,D'Auria:2005rv,DallAgata:2005zlf,Neupane:2005nb,Villadoro:2005cu,Hull:2006tp,Fre':2006ut,Caviezel:2008ik,Andriot:2010ju,Andriot:2015sia,Andriot:2016rdd}, which is locally a group manifold described by a set of left-invariant one-forms $\sigma^\upalpha$, $\upalpha=D,\dots, D+n-1$, satisfying the Maurer-Cartan equation:
 \begin{equation}
 d\sigma^\upalpha=-\frac{1}{2}\,T_{\upbeta\upgamma}{}^\upalpha\,\sigma^\upbeta\wedge \sigma^\upgamma\,,
 \end{equation}
 where $T_{\upbeta\upgamma}{}^\upalpha$ are the structure constants satisfying the Jacobi identity $T_{[\upbeta\upgamma}{}^\upalpha\,T_{\upsigma]\upalpha}{}^\updelta=0$. This manifold can be viewed as a topological deformation, by means of the constant tensor $T_{\upbeta\upgamma}{}^\upalpha$, of an $n$-torus $T^n$, whose cohomology is described in terms of the closed one-forms $\sigma^\upalpha=dx^\upalpha,\,d\sigma^\upalpha=0$.
  The quantity $T_{\upbeta\upgamma}{}^\upalpha$ can also be described as a \emph{torsion} on the original torus and is an example of geometric flux. The study of the effect of T-dualities on flux-backgrounds unveiled connections between different compactifications.  In particular it was found that T-duality can map a NS-NS three-form flux $H_{\upalpha\upbeta\upgamma}$ into a geometric-flux $T_{\upalpha\upbeta}{}^\upgamma$. The picture of T-dual backgrounds needs to be completed with the introduction of the so-called \emph{non-geometric} fluxes (denoted by $Q_{\upalpha}{}^{\upbeta\upgamma},\,R^{\upalpha\upbeta\upgamma}$ in the literature) \cite{Mathai:2004qq,Hull:2004in,Dabholkar:2002sy,Dabholkar:2005ve,Shelton:2005cf,Wecht:2007wu,Grana:2006hr,Cassani:2008rb,D'Auria:2007ay,Lust:2009mb,Andriot:2011uh,Andriot:2012an,Andriot:2013xca,Andriot:2014uda}
 in the presence of which the geometry of the internal manifold cannot be globally (or even locally) defined. All these descriptions are naturally unified by considering more general internal geometries which feature the whole T-duality group ${\rm O}(n,n)$ as structure group. This is effected by either extending (doubling) the tangent space of $M_{int}$ (\emph{generalized geometry} \cite{Hitchin:2004ut},\cite{Gualtieri:2007ng},\cite{Grana:2005jc}) or extending the space-time manifold itself (\emph{Double Geometry} \cite{Dabholkar:2005ve,Hull:2006va,Hull:2007jy,DallAgata:2007sr} or \emph{Double Field Theory} \cite{Hull:2009mi,Hull:2009zb,Hohm:2010jy,Hohm:2010pp,Hull:2014mxa,Aldazabal:2013sca}, which build on earlier pioneering works \cite{Duff:1989tf,Duff:1990hn,Tseytlin:1990nb,Tseytlin:1990va,Siegel:1993xq,Siegel:1993th}). In both cases the extended tangent space is acted on by a larger structure group which contains ${\rm O}(n,n)$. Along the same lines, new field theoretical constructions have been recently put forward in which the internal structure group is enlarged to contain the whole global symmetry group $G$ of the lower-dimensional theory. This is the case of the \emph{extended generalized geometry} \cite{Hull:2007zu,Pacheco:2008ps} and \emph{exceptional field theory} \cite{Hohm:2013pua,Hohm:2013uia,Hohm:2014qga}
 \par
Let us get back to the discussion of the effective lower-dimensional supergravity description originating from flux-backgrounds.
When considering these more general compactifications,  in many cases the full non-linear dynamics of  the low lying modes, or of a consistent truncation thereof, is described by an effective $D$-dimensional supergravity. In contrast to the traditional compactifications on Ricci-flat manifolds discussed earlier, these effective lower-dimensional theories may be \emph{gauged}: They feature minimal couplings between the vector fields and other fields
which, by consistency, are associated with an internal local symmetry group $G_g$. This is the case of \emph{extended supergravities}, namely theories preserving at least 8 supercharges ($\mathcal{N}>1$ in four-dimensions). $\mathcal{N}=1$ four-dimensional supergravities (preserving only four supercharges) originating from flux-compactifications can still be ungauged since, for instance, the background fluxes may manifest themselves only in the presence of a  \emph{superpotential} and thus an $F$-term scalar potential in the effective theory \cite{Gukov:1999ya}. The gauge symmetry $G_g$ depends on
\begin{itemize}
\item The fluxes on the compactifying background;
\item The structure of the internal manifold.
\end{itemize}
As mentioned earlier, one of the first examples of flux-compactifications yielding a gauged supergravity as a consistent truncation, is that of eleven-dimensional supergravity on a seven-dimensional sphere \cite{Duff:1983gq}, that is on the Freund-Rubin solution \cite{Freund:1980xh} which has the form: $$AdS_4\times S^7\,,$$
where $AdS_4$ denotes a four-dimensional anti-de Sitter space-time. The massless modes on $AdS_4$, together with their full non-linear mutual interactions, are described \cite{deWit:1983vq,deWit:1986oxb} by the gauged $\mathcal{N}=8$ supergravity with gauge group $G_g={\rm SO}(8)$, constructed in \cite{deWit:1981sst,deWit:1982bul}. Consistently with the Kaluza-Klein mechanism, the ${\rm SO}(8)$ local symmetry originates from the isometry group of $S^7$ and is gauged by the 28 vector fields of the $\mathcal{N}=8$ graviton supermultiplet. The Freund-Rubin solution is characterized by the v.e.v.  along the four non-compact space-time directions, $\langle F_{\mu\nu\rho\sigma}\rangle =m\,\epsilon_{\mu\nu\rho\sigma}$, of the four-form field strength $F^{(4)}=dA^{(3)}$ of the eleven-dimensional theory and describes the full back-reaction of this flux on the space-time geometry. The radii of the anti-de Sitter space and the seven-sphere have the same order of magnitude and are both expressed in terms of the flux-parameter $m$, which also fixes the gauge coupling constant and the masses in the four-dimensional theory.\par
The fluxes and the structure of the internal manifold also determine, in the low-energy supergravity, \emph{mass terms}, and a \emph{scalar potential} $V(\phi)$. The former in general  produce spontaneous supersymmetry breaking at tree level, at a scale which is fixed by the fluxes themselves while the latter has another desirable effect: It stabilizes the scalar fields already at the level of classical supergravity, thus removing, completely or in part, the moduli degeneracy characterizing the traditional compactifications. The presence of a potential in the classical low-energy theory
may also induce an effective cosmological constant or determine, under certain conditions, the dynamics of an \emph{inflaton} field whose evolution could trigger the early expansion of our universe.
In certain models, as the ${\rm SO}(8)$-gauged maximal supergravity mentioned above, the scalar potential may define vacua with negative cosmological constant, corresponding to a $D$-dimensional space-time with anti-de Sitter geometry. Such vacua are interesting in light of the AdS/CFT holography conjecture \cite{Maldacena:1997re,Gubser:1998bc,Witten:1998qj}, according to which stable AdS solutions describe conformal critical points of a suitable gauge theory defined on the boundary of the space. In this picture the maximally supersymmetric $AdS_5\times S^5$ solution of Type IIB theory is dual to $\mathcal{N}=4$ supersymmetric Yang-Mills theory on the four-dimensional boundary of $AdS_5$ \cite{Maldacena:1997re}, while the conjectured dual theory to the  maximally supersymmetric Freund-Rubin solution $AdS_4\times S^7$ of $D=11$ supergravity, is the superconformal ABJM model \cite{Aharony:2008ug}.
In this perspective domain wall solutions to the gauged supergravity interpolating between AdS critical points of the potential describe renormalization group (RG) flow \cite{Skenderis:2002wp}, from an ultra-violet to an infra-red fixed point, of the dual gauge theory and gives important insights into its non-perturbative properties. The spatial evolution of such holographic flows is determined by the scalar potential $V(\phi)$ of the theory.
\par
Special care is needed in defining a limit in which a gauged $D$-dimensional supergravity is reliable as a
low-energy description of a flux-compactification, besides  being a consistent truncation of the compactified ten or eleven-dimensional theory. In many cases it suffices for the back-reaction of the fluxes on the background geometry to be negligible. This regime, in the
case of superstring compactifications in the presence of form-fluxes, can be attained if the size of the internal manifold is much larger than
the string scale, so that the following hierarchy of scales is realized:
$$\mbox{(flux-induced masses)} \ll \mbox{Kaluza-Klein masses} \ll \mbox{mass of string excitations}\,,$$
which guarantees the decoupling of the Kaluza-Klein and of the massive string excitations from the supergravity fields.
 We also need the superstring coupling constant $g_s$ to be small on the background since the supergravity description is defined at order zero in $g_s$.
For compactifications of eleven-dimensional supergravity we can only require the decoupling of the Kaluza-Klein modes. This does not occur in the maximally supersymmetric compactification on the seven-sphere where there is no hierarchy between the size of the flux-induced masses in the four-dimensional supergravity and the Kaluza-Klein masses.\par
Supergravities in $D$-dimensions, at the classical level, are consistently defined independently of their string or M-theory origin, and are characterized by
\begin{itemize}
\item{The amount of supersymmetry;}
\item{Their field content;\footnote{Here, in the definition of the field content, we include the non-linear interactions among the scalar fields which define their kinetic terms and which is encoded in the geometry of the scalar manifold. The latter, as we shall see,  defines in the classical theory the on-shell  global symmetry group $G$.}}
\item{The local internal symmetry group $G_g$ gauged by the vector fields,}
\end{itemize}
The latter being a feature of gauged supergravities. When originating from superstring/M-theory compactifications, lower dimensional supergravities offer a unique window into their non-perturbative dynamics. Indeed they provide a field theoretical (and thus well established and controlled) description of the full non-linear interaction among their light modes (and consequently of phenomena such as spontaneous supersymmetry breaking or moduli-fixing), which is totally fixed by general features of the compactification and symmetry requirements (see below).\par
At the classical level, gauged supergravities are obtained from ungauged ones, with the same amount of supersymmetry and field content, through the well defined \emph{gauging} procedure which consists in promoting a suitable subgroup $G_g$ of
the (classical) global symmetry group $G$ to local symmetry of the theory, and to modify the action, by the introduction of fermion mass-terms and of a scalar potential $V(\phi)$, in order to preserve the same number of supersymmetries as the original theory. Provided the chosen gauge group $G_g$ satisfies some consistency constraints, which are the more stringent the larger the amount of supersymmetry, this procedure  uniquely defines the gauged supergravity. In fact the gauging procedure is the only known way for introducing fermion mass-terms or a scalar potential in an extended supergravity without explicitly breaking supersymmetry. When the theory originates from string compactifications, these additional terms in the action, as well as the gauge group $G_g$, are determined by the background fluxes and the structure of $M_{int}$.\par
The way in which the notion of gauged supergravities has been introduced in our discussion should not lead the reader to the incorrect conclusion that these models were first derived from flux-compactifications. Historically the very first instances of gauged extended supergravities were constructed soon after the discovery of supergravity itself, see for instance \cite{Freedman:1976aw},\cite{Freedman:1978ra},\cite{Zachos:1978iw}. These early examples already exhibited some of the generic features of this type of locally supersymmetric models such as fermion masses, spontaneous supersymmetry breaking and a scalar potential (or a cosmological constant when scalars were not present). The relation of gauged supergravities to spontaneous compactifications, however, was object of a later study, which started in the eighties (see for instance \cite{deWit:1983vq} and our earlier discussion).

\paragraph{The Embedding Tensor and Dualities}
As pointed out above, background fluxes typically induce, in the lower dimensional effective theory, minimal couplings as well as mass terms and a scalar potential. The former, involving only the electric vector fields, manifestly break the electric-magnetic duality symmetry of the original ungauged theory.
In the formulation of the gauging construction introduced in \cite{Cordaro:1998tx,Nicolai:2000sc,deWit:2002vt}, and further developed in \cite{deWit:2004nw,deWit:2005hv,deWit:2005ub,deWit:2007mt,deWit:2008ta}, all the deformations of the original ungauged model implied by this procedure, are expressed in terms of a single tensor, covariant with respect to the global symmetries of the theory, called the \emph{embedding tensor} $\Theta$, which defines the embedding of the gauge algebra into the global symmetry one.
%\footnote{In \cite{cgftt} the embedding tensor was first introduced in maximal supergravity as a covariant object only with respect to the symmetry group $G_e$ of the action (off-shell symmetry) which, in four-dimensions, as consequence of electric-magnetic duality, does not coincide with the on-shell symmetry $G$, but is contained in it: $G_e\subset G$. The issue of the different electric-magnetic duality frames of the four-dimensional original ungauged theory was addressed in \cite{dwst1} where a  frame-independent description of the  gauging  in terms of a $G$-covariant embedding tensor was achieved.}
In fact the embedding tensor contains all the information about the choice of $G_g$ inside $G$. This choice is constrained by consistency conditions which depend on the ungauged model we start from. They include the intuitive requirement that the dimension of $G_g$ cannot exceed the number $n_v$ of available vector fields to gauge it, and can be cast in the form of a set of linear and quadratic $G$-covariant (purely group-theoretical) constraints on $\Theta$.

 Given an ungauged model defined by the amount $\mathcal{N}$ of supersymmetry, its field content $\Phi$, and its global symmetry group $G$, a gauged extended model $GT[\mathcal{N},\,\Phi,\,\Theta]$ is uniquely defined once we choose the embedding tensor $\Theta$ associated with the gauge group $G_g$.\par
 The main advantage of this description of the gauging is that the embedding tensor enters the gauged  field equations and Bianchi identities in a manifestly $G$-covariant way, so that, \emph{the on-shell $G$-invariance of the original ungauged theory is formally restored provided $\Theta$ is transformed together with all the fields.} Being, however, $\Theta$ a spurionic object, namely a non-dynamical quantity, this formal invariance should  not be regarded
as a symmetry of the gauged theory, but rather as an equivalence, or proper duality, between different theories:
\begin{align}
GT[\mathcal{N},\,\Phi,\,\Theta]\,\,\equiv\,\,\, GT[\mathcal{N},\,G\star\Phi,\,G\star\Theta]\,,\label{GTequiv}
\end{align}
where $G\star$ denotes the general action of $G$ on the fields and on the embedding tensor, according to the respective representations.

The tensor $\Theta$ contains all the coupling constants and mass parameters characterizing the gauged theory, which derive from the background (form, geometric and non-geometric) fluxes when the theory originates from a flux compactification. This allows us to make a precise statement about the correspondence between the background fluxes and the local internal symmetry of the low-energy effective theory: All the background quantities characterizing the compactification enter the lower-dimensional supergravity as components of the embedding tensor \cite{Angelantonj:2003rq,D'Auria:2003jk,deWit:2003hq}
\begin{equation}
\mbox{Fluxes (form, geometric, non-geometric)}\,\,\subset\,\,\,\Theta\,.\label{fluxtheta}
\end{equation}
In several instances of flux compactifications, in which the back-reaction of the fluxes on the space-time geometry can be made  ``small'' (see discussion above), from general properties of the background and the original higher-dimensional theory, one infers the
amount of supersymmetry and the field content of the lower-dimensional effective supergravity (which also characterize the
corresponding ungauged theory obtained in the zero-flux limit). By general symmetry arguments then, the fluxes can be identified with components of the embedding tensor and the gauging procedure does the rest: The low-energy gauged supergravity is completely determined.
When this back-reaction is not negligible, the flux-induced deformation of the geometry of the internal manifold would amount to additional geometric fluxes which need to be taken into account and identified with other components of the embedding tensor (as it is the case, for instance, of the $S^7$-compactification of $D=11$ supergravity). In all known examples of gauged supergravities originating from superstring or M-theory, the constraints on the embedding tensor required by their consistent definition, reflect corresponding consistency restrictions on the fluxes, like the tadpole cancelation condition.
\par
The description of gauged supergravities based on the embedding tensor provides an ideal field-theoretical framework for studying the effects of $S$, $T$ and in general $U$-dualities on flux-backgrounds and thus to systematically unveil the web of dualities underlying the vast landscape of flux-compactifications. This follows from the identification (\ref{fluxtheta}) of the general fluxes with components of the embedding tensor, which allows to naturally associate them with representations of $G$, and from the interpretation, mentioned above, of a suitable discrete subgroup $G(\mathbb{Z})$ (or of an extension thereof) of $G$ with the $U$-duality group unifying all known string dualities. For example, in toroidal compactifications with ``small'' fluxes, the fields and the background quantities are naturally associated with representations of the ${\rm GL}(n,\mathbb{R})$ group acting transitively on the metric moduli of the internal torus $T^n$. This group describes a global symmetry of the $D$-dimensional ungauged action that one  would obtain in the absence of fluxes. Being the embedding tensor $\Theta$  $G$-covariant, by simply decomposing $G$-representations with respect to ${\rm GL}(n,\mathbb{R})$, the background fluxes are readily identified with components of $\Theta$ and thus associated with representations of $G$. From this purely group-theoretical analysis it follows that the NS-NS, geometric and non-geometric fluxes, $H_{\alpha\beta\gamma},\,T_{\alpha\beta}{}^{\gamma},\,Q_{\alpha}{}^{\beta\gamma},\,R^{\alpha\beta\gamma}$ belong to a same representation of the (continuous version of) the T-duality group ${\rm O}(n,n)$, while the RR fluxes of Type IIA and Type IIB theories complete two chiral representations of ${\rm Spin}(n,n)$. The latter are mapped into one another by \emph{improper} ${\rm O}(n,n)$ transformations which involve T-dualities along an odd number of internal directions, and which interchange the Type IIA and Type IIB descriptions. \par
While, in general, flux compactifications can be given, in certain limits, an effective lower-dimensional supergravity description, there are several (classically) consistent $D$-dimensional gauged supergravities whose superstring/M-theory uplift, and thus their ultra-violet completion, is not known or not understood yet \cite{Vafa:2005ui},\cite{Taylor:2011wt}:
\begin{align}
\mbox{Known string/M-theory compactifications}&\longrightarrow\,\,\,\mbox{Gauged $D$-dimensional supergravity}\,,\nonumber\\
\mbox{Known string/M-theory compactifications}&\stackrel{?}{\longleftarrow}\,\,\,\mbox{Gauged $D$-dimensional supergravity}\,,\nonumber
\end{align}
It is useful, in this respect, to group gauged supergravities into \emph{equivalence classes} (or \emph{orbits}) with respect to the action (\ref{GTequiv}) of $G$. Some of these theories, though having a problematic interpretation as originating from superstring or M-theory compactifications (as it is the case in the presence of the non-geometric $Q$ and $R$-fluxes), can nevertheless be characterized as \emph{dual} to theories whose higher-dimensional origin is well defined.
On the other hand we can define as \emph{intrinsically non-geometric} those models which cannot be related by dualities to the low-energy description of  consistent string or M-theory compactifications. There has been important progress in the recent years in the characterization of intrinsically non-geometric theories and of their duality orbits. Giving these theories, some of which have a rich vacuum structure and interesting physical properties, a microscopic interpretation
amounts to clearing up the ``swampland'' of \cite{Vafa:2005ui} and is one of the major recent challenges in theoretical high-energy physics.

 The aim of the present report is to give a pedagogical review of gauged supergravities, highlighting their
 applications to the description of superstring or M-theory compactifications. At the same time the relevant topics will be dealt with in sufficient detail so as to allow the reader to perform explicit computations. Due to the vastness of the subject and to its many recent developments, a choice among the main issues concerning gauged supergravities will be made, though trying to keep the dissertation as self-consistent as possible. Our focus will mainly be on four-dimensional theories, although general references to higher-dimensional ones will often be made. The reader is required to have a general knowledge of supersymmetry and superstring theory, besides a solid background in general relativity. A basic knowledge of the theory of Lie groups and Lie algebras is also recommended.\par
 The present report can be viewed as divided into three parts:
In the first one, which includes Sections \ref{sec:2}, \ref{sec:3}, \ref{startexa}, \ref{N2sugras} and \ref{glance}, the general structure of four-dimensional supergravities and the gauging procedure are discussed mainly in a way which does not depend on the higher-dimensional origin of the models. Nevertheless references to the string or M-theory interpretation of fields and fluxes will be made, for instance, in Sect. \ref{symfram}, \ref{N8MAB} and \ref{n8fluxc} when referring to toroidal reductions with fluxes yielding maximal models, and in Sect. \ref{rtcy}, where reference is made to the Type II origin of $\mathcal{N}=2$ models and Calabi-Yau reductions. Many of these notions will be clarified in the second part, consisting of Sect. \ref{gsfsmfc} and its subsections, where toroidal reductions and instances of flux-compactifications are discussed in detail; Sect. \ref{vohd} represents the third, final part, where supergravities in $D$-dimensions and their gauging are briefly reviewed.\par
 Let us now illustrate in more detail the organization of this review:\\
 In Section \ref{sec:2} we shall discuss, in a unified framework, the general features of ungauged extended supergravities in four dimensions and their mathematical structure.\footnote{We only consider the formulation of these theories in which all antisymmetric tensor fields are dualized to lower-rank ones, through the Hodge-duality relating the corresponding field strengths.} These include their on-shell global symmetry group (or duality group) $G$ and their non-unique off-shell description in terms of inequivalent Lagrangians, which depend on the definition of the electric vector fields, namely on the choice of the symplectic frame.  The reader who is familiar with this subject can move directly to Section \ref{sec:3}.\\ In Section \ref{sec:3} the gauging procedure is discussed. We choose, for pedagogical reasons, a piecemeal approach to the subject, by initially reviewing the gauging of extended supergravities in the electric frame. In this first analysis only the choice of the gauge group inside $G$ will be characterized in a way which does not depend on the symplectic frame, through a $G$-covariant embedding tensor $\Theta$, subject to a set of $G$-covariant linear and quadratic constraints.\\
 Later in Sect. \ref{sec:4}, we shall deal with a more general formulation of the gauging procedure which no longer depends on the original symplectic-frame and is manifestly $G$-covariant at the level of the field equations and Bianchi identities. The price we have to pay for this is the introduction of fields which are dual to the existing ones in the ungauged model (with the exception of the metric) and which include higher-order antisymmetric tensor fields. These extra fields are associated with new gauge symmetries which ensure the correct number of propagating degrees of freedom. The consistent distribution of the latter among the various fields is defined by the embedding tensor.  This general mechanism is called \emph{hierarchy} and allows a duality covariant formulation of all supergravities in any dimensions. The precise knowledge of how the degrees of freedom are assigned to the fields requires ``looking inside'' $\Theta$ though its \emph{rank-factorization}, discussed, for four-dimensional theories, in Sect. \ref{backelectric}. \par
 Using this formalism, a gauged supergravity can be completely characterized in terms of its amount of supersymmetry, field content and the embedding tensor defining the internal local symmetry gauged by the vector fields. As pointed out earlier in this Introduction, the action of $G$ on $\Theta$ defines an equivalence, or duality, between seemingly different theories: Dual gauged supergravities share the same physics (vacua, mass spectra and interactions). A general discussion of the vacua of gauged supergravities and of the mass matrices, will be given in Sect. \ref{vad}.\\
 Starting from Sect. \ref{startexa}, we discuss applications of the general procedure illustrated in Sect. \ref{sec:4} to the specific four-dimensional extended models, extensively discussing only the maximal and the $\mathcal{N}=2$ models. The former is discussed in Sect. \ref{startexa} while latter will be dealt with in Sect. \ref{N2sugras}. Special emphasis will be put on the new ``dyonic'' gaugings, such as the recently defined ``dyonic'' ${\rm CSO}(p,q,r)$-gaugings in the maximal theory, see Sect. \ref{dyonicg}. A general discussion of the mathematical structures underlying  $\mathcal{N}=2$ supergravities is given in Sect. \ref{N2sugras}: Special and quaternionic K\"ahler geometries are reviewed in Sects. \ref{SK} and \ref{QMans}, respectively. The main facts about the relation between ungauged $\mathcal{N}=2$ supergravities and Calabi-Yau compactifications are recalled in Sect. \ref{rtcy}. The general, symplectic covariant gauging of the $\mathcal{N}=2$  models is discussed in Sects. \ref{gaugspecialg}-\ref{FIsec}. In Sect. \ref{STUsolvq} we consider, as an instructive example, one of the few models in which the constraints on the embedding tensor can be explicitly solved and all possible gaugings classified in orbits of the global symmetry group $G$: the STU model. Some of these orbits feature interesting physics, such as stable de Sitter and anti de Sitter vacua.
  \\
 The relationship between flux-compactifications of string/M-theory and the gauged supergravity describing the dynamics of the low-lying modes (or of a consistent truncation thereof), is dealt with in Sect. \ref{toroidalc}. We start from toroidal reductions from ten and eleven dimensional maximal supergravities and illustrate, in various instances, how the background quantities enter the lower-dimensional effective description as components of the embedding tensor defining the coresponding gauging. The action of dualities on fluxes using the embedding tensor formalism and the issue of non-geoemtric fluxes are discussed in Sect. \ref{Tdualcomp}, where the subjects of (extended) generalized geometry, DFT and exceptional field theory are touched upon.\\
 In Sect. \ref{K3T2Z2} the gauged $\mathcal{N}=2$ supergravity description of the Type IIB theory compactified on a
$K3\times T_2/\mathbb{Z}_2$-orientifold  in the presence of fluxes and space-filling $D3$ and $D7$-branes is reviewed.\\
We end the discussion of the gauged supergravity approach to the study of flux-compactifications with Sect. \ref{mcga} where we deal with a
class of $\mathcal{N}=2$ gauged models which describe flux-compactifications of Type II theories on manifolds with ${\rm SU}(3)\times {\rm SU}(3)$-structure.\\
In Sect. \ref{glance}, we briefly review the main facts (field content, global symmetry and embedding tensor representation) about the four-dimensional $\mathcal{N}=6,5,4,3$ supergravities.\\
We end the report with Sect. \ref{vohd}, in which a general overview of gauged supergravities in higher dimensions is given, with special emphasis on the hierarchy mechanism.\\
The Appendices are devoted to the definition of the conventions used throughout the report as well as to a summary of the useful identities and properties related to the various subjects dealt with in the main text.

\section{Review of Ungauged Supergravities}\label{sec:2}
Let us recall some basic aspects of extended ungauged $D=4$ supergravities. The main references for this section are \cite{Cremmer:1978ds},\cite{deWit:1982bul},\cite{Gaillard:1981rj},\cite{Andrianopoli:1996cm},\cite{Andrianopoli:1996ve}, \cite{D'Auria:2001kv} and \cite{deWit:2007mt}. \par

\subsection{Field Content and Bosonic Action.}
The bosonic sector consists in the graviton $g_{\mu\nu}(x)$, $n_v$ vector fields $A^\Lambda_\mu(x)$, $n_s$ scalar fields $\phi^s(x)$ and is described by a bosonic Lagrangian of the following general form
\footnote{
Using the ``mostly minus'' convention and
\;$8\pi\GN=c=\hbar=1$. Moreover $\epsilon_{0123}=-\epsilon^{0123}=1$. See Appendix \ref{nacv} for the definition of the main notations and conventions.
}
\begin{equation}
\frac{1}{e}\LB=
-\frac{R}{2}
+\frac{1}{2}\,\Gm_{st}(\phi)\,\partial_\mu\phi^s\,\partial^\mu\phi^t
+\frac{1}{4}\,\I_{\Lambda\Sigma}(\phi)\,F^\Lambda_{\mu\nu}\,F^{\Sigma\,\mu\nu}
+\frac{1}{8\,e}\,\R_{\Lambda\Sigma}(\phi)\,\eps^{\mu\nu\rho\sigma}\,F^\Lambda_{\mu\nu} \,F^{\Sigma}_{\rho\sigma}\,,
\label{boslagr}
\end{equation}
where $e=\sqrt{|{\rm det}(g_{\mu\nu})|}$ and the $n_v$ vector field strengths are defined as usual:
\begin{align}
F^\Lambda_{\mu\nu}=\partial_\mu A^\Lambda_\nu-\partial_\nu A^\Lambda_\mu\;.\\ \nn
\end{align}
\bigskip
Let us comment on the general characteristics of the above action.
\begin{itemize}
\item{The scalar fields $\phi^s$ are described by a non-linear $\sigma$-model, that is they are coordinates of a non-compact, \emph{Riemannian} $n_s$-dimensional differentiable manifold (target space), named \emph{scalar manifold} and to be denoted by $\Mscal$. The positive definite metric on the manifold is $\Gm_{st}(\phi)$, where we have used the short-hand notation $\phi\equiv (\phi^s)$. The corresponding kinetic part of the Lagrangian density reads:
   \begin{equation}
     \Lscal=\frac{e}{2}\,\Gm_{st}(\phi)\,\partial_\mu
          \phi^s\partial^\mu \phi^t\,.
    \end{equation}
 The $\sigma$-model Lagrangian is clearly invariant under the action of global (i.e.\ space-time independent) isometries of the scalar manifold:
 \begin{equation}
\phi^s\rightarrow \phi^{\prime s}(\phi)\,:\,\,\,\Gm_{s't'}(\phi'(\phi))\frac{\partial \phi^{\prime s^\prime}}{\partial \phi^s}\frac{\phi^{\prime t^\prime}}{\partial \phi^t}=\Gm_{st}(\phi)\,.\label{Gscalisom}
 \end{equation}
 As we shall discuss below, the isometry group can be promoted to a global symmetry group of the field equations and Bianchi identities (i.e.\ \emph{on-shell global symmetry group}) provided its (non-linear) action on the scalar fields is combined with an electric-magnetic duality transformation on the vector field strengths and their magnetic duals \cite{Gaillard:1981rj}.\par
 In $\mathcal{N}=1$ models \cite{Cremmer:1982en} the $n_s=2n$ real scalar fields are naturally grouped by supersymmetry in $n$ complex fields $z^i$, $i=1,\dots, n$, belonging to $n$ chiral multiplets and which span a complex \emph{K\"ahler} manifold \cite{Zumino:1979et} (in fact a \emph{Hodge-K\"ahler} manifold, see for instance \cite{Freedman:2012zz} for a general review of the subject). The sigma model Lagrangian density becomes:
    \begin{equation}
     \Lscal=e\,g_{i\bar{\jmath}}(z,\bar{z})\,\partial_\mu
          z^i\partial^\mu \bar{z}^{\bar{\jmath}}\,,
    \end{equation}
 where $g_{i\bar{\jmath}}(z,\bar{z})$ is the hermitian target space metric which can be expressed in terms of a real K\"ahler potential $\mathcal{K}(z,\bar{z})$ as follows: $g_{i\bar{\jmath}}=\frac{\partial^2}{\partial z^i\partial \bar{z}^{\bar{\jmath}}}\mathcal{K}$.
 }
 \item{
 The fermionic sector of a supergravity theory consists of $\mathcal{N}$ gravitino fields $\psi_{A\mu}$, $A=1,\dots,\mathcal{N}$, and a number of spin-1/2 fields. These comprise the \emph{gaugini} $\lambda^{IA}$, $I=1,\dots, n$, belonging to the $n$ vector multiplets. In $\mathcal{N}=2$ theories we also have the \emph{hyperinos} $\lambda^\alpha$ in the hypermultiplets (to be denoted later also by $\zeta^\alpha$) and, only for $\mathcal{N}\ge 3$, the \emph{dilatinos} $\chi_{ABC}$ in the supergravity multiplet.\footnote{\label{fotextraferm}For $\mathcal{N}=3$ each vector multiplet (labeled by the index $I$) contains, besides the triplet  $\lambda_{IA}$, also the singlet spin-$1/2$ field $\lambda_{I\,ABC}=\lambda_I\epsilon_{ABC}$ of opposite chirality with respect to the former. In the $\mathcal{N}=5$ model, besides the ten spin-$1/2$ fields $\chi_{ABC}$, the supergravity multiplet also contains a singlet $\chi$ while for $\mathcal{N}=6$, supersymmetry requires, together with the $\chi_{ABC}$, also the presence in the gravity multiplet of extra six fermions $\chi^A$. These additional dilatinos all have opposite chirality with respect to $\chi_{ABC}$. See Sect. \ref{glance} for the general details of the various extended models.}\par
We shall  use the chiral (or Weyl) basis for the fermion fields, in which the full R-symmetry group $H_R$, i.e. the automorphism group of the supersymmetry algebra, is manifest.\par
 The fermionic fields transform non-trivially under the holonomy group $H$ of $\Mscal$, which contains the R-symmetry group $H_R$, in contrast to the bosonic fields which are inert under $H$. Since the action of $H$ is by definition local on the scalar manifold, namely it acts by means of scalar-field-dependent transformations, invariance of the theory under it requires the covariant derivatives of the fermion fields
 to contain a corresponding gauge connection $\mathcal{Q}_\mu$. This is a \emph{composite connection}, namely it is not an independent vector field, but it is defined in terms of the scalar fields $\phi^s$ and their derivatives $\partial_\mu \phi^s$. In $\mathcal{N}=1$ theories, for instance, the fermionic fields are charged under the $H_R={\rm U}(1)$ R-symmetry group of the theory, whose action is associated with the K\"ahler transformations of the scalar manifold. Consistency of this transformation property of the fermionic fields implies the existence of an additional structure and conditions on the scalar manifold which further constrain it to be of \emph{Hodge-K\"ahler} type.}
 \item{The two terms containing the vector field strengths will be called vector kinetic terms. A general feature of supergravity theories is that the scalar fields are non-minimally coupled to the vectors as they enter these terms through the symmetric matrices $\I_{\Lambda\Sigma}(\phi),\,\R_{\Lambda\Sigma}(\phi)$ which contract the vector field strengths. The former $\I_{\Lambda\Sigma}(\phi)$ is negative definite and generalizes the $-1/g^2$ factor in the Yang-Mills kinetic term. The latter $\R_{\Lambda\Sigma}(\phi)$ generalizes the $\theta$-term.\par
     In $\mathcal{N}=1$ supergravity, these two matrices are defined by the real and imaginary parts of a holomorphic matrix $F_{\Lambda\Sigma}(z^i)$. In extended theories ($\mathcal{N}>1$) vector multiplets contain scalar fields and this feature implies an important difference with respect to the minimal case: The $\I_{\Lambda\Sigma}$ and $\R_{\Lambda\Sigma}$ only depend on the scalar fields sitting in the vector multiplets and their form is fixed by supersymmetry (aside from an initial choice of the symplectic frame), so that they are no longer an independent feature of the model.
     }
 \item{There is a ${\rm U}(1)^{n_v}$ gauge invariance associated with the vector fields:
     \begin{equation}
      A_\mu^\Lambda\rightarrow A_\mu^\Lambda+\partial_\mu\zeta^\Lambda\,.
     \end{equation}
     All the fields are neutral with respect to this symmetry group.}
\item{There is no scalar potential. In an ungauged supergravity a scalar potential is allowed only for $\N=1$ (\emph{F-term potential}). In this case it is determined in terms of the holomorphic \emph{superpotential} $W(z^i)$ and its covariant  derivatives
     \begin{equation}
 V(z,\bar{z})=e^{\mathcal{K}}\left(g^{i\bar{\jmath}}\,{\Scr D}_i W\overline{{\Scr D}}_{\bar{\jmath}} \overline{W} -3\,|W|^2 \right)\,,
 \end{equation}
 where the first (positive) term is usually referred to as the ``F''-term contribution to the potential.\par
    In extended supergravities a non-trivial scalar potential can be introduced without explicitly breaking supersymmetry only through the \emph{gauging procedure}, which implies the introduction of a local symmetry group to be gauged by the vector fields of the theory and which will be extensively dealt with in the following.}
\end{itemize}
The fermion part of the action is totally determined by supersymmetry once the bosonic one is given. Let us start discussing in some detail the mathematical description of the different sectors of the theory, starting from the scalar one.

\subsubsection{Scalar Sector and Coset Geometry}\label{ghsect}

As mentioned above the scalar fields $\phi^s$ are coordinates on a Riemannian target space $\Mscal$, with metric $\Gm_{st}(\phi)$.
Their sigma model action therefore features non-linear interactions which are encoded in this metric tensor. The consistent couplings of the scalar fields to the vectors and fermions (gravitino and spin- $1/2$ fields), for a given amount of supersymmetry, requires restrictions on the scalar manifold and additional structures to be defined on it. It is known that fermionic fields on a curved space-time are characterized by the property of transforming in a certain representation (the spinorial one) of the local Lorentz group, which is the structure group of space-time and defines the holonomy of the Levi-Civita connection on it for a generic metric. Similarly, as we mentioned above, they belong to a definite representation of the holonomy group of the scalar manifold, and this determines their couplings to the scalar fields. Physics should not depend on the local holonomy frame either on space-time or on the scalar manifold. For this reason the covariant derivatives of the fermions will contain both the space-time (spin) connection and the the scalar manifold one in the appropriate representation. We will deal with this in Section \ref{fsector}.\par Consistency of the scalar-vector coupling with extended supersymmetry requires, as we shall discuss in Section \ref{gsg}, the definition of a flat symplectic bundle on the scalar manifold which associates with each isometry on $\Mscal$ a constant, symplectic electric-magnetic duality transformation. The same isometry  also implies a corresponding compensating holonomy transformation on the fermions.  These combined actions promote the isometry group to a global symmetry group of the field equations and Bianchi identities.\par
Scalar manifolds can be homogeneous symmetric, namely of the form $G/H$ with $G$ semisimple and $H$ maximal compact in $G$, as it is the case of theories with large enough supersymmetry ($\mathcal{N}>2$). A simple example of a homogeneous symmetric space is the upper- (or lower-) half plane ${\rm SL}(2,\mathbb{R})/{\rm SO}(2)$, see example below. A more complicated example is the scalar manifold ${\rm E}_{7(7)}/({\rm SU}(8)/\mathbb{Z}_2)$ of the $\mathcal{N}=8$ model, see Section \ref{startexa}. This is spanned by the 70 real scalars of the maximal theory. \par In the $\mathcal{N}=2$ models the geometry of the scalar manifold is less constrained and we can have models in which $\Mscal$ is not even homogeneous, namely we cannot move from a point to any other one of the space through a symmetry transformation of the space itself (an isometry). We shall discuss these models in Sect. \ref{N2sugras}.\par
In order to deal with the scalar fields and their interactions we need therefore some mathematical tools, related to the general description of non-compact Riemannian manifolds. Below we shall recall the main facts about homogeneous scalar manifolds in supergravities and the definition of the corresponding sigma-model action. Although particular emphasis will be given to the symmetric case, most of the properties we shall derive also hold for the more general spaces which occur in $\mathcal{N}=2$ models. Some basic notions of differential geometry of Riemannian manifolds are needed at this point, although the relevant facts are recalled in Appendix \ref{nacv} where the notations we shall use are also defined.\par
 The holonomy group $H$ of $\Mscal$ has the general form
\begin{equation}
H\=H_{R}\times H_{\rm matt}\,,\label{Hgroup}
\end{equation}
where $H_{\rm R}$ is the automorphism group of the supersymmetry algebra (R--symmetry group), which is ${\rm U}(\mathcal{N})$ for $\mathcal{N}<8$ and and ${\rm SU}(8)$ for $\mathcal{N}=8$, $H_{\rm matt}$ is a compact group acting on the matter fields. As previously emphasized, the gravitino and spin-$\frac12$ fields transform in representations of the $H$ group. The theories with  $\mathcal{N}\ge 5$ describe the gravitational multiplet only and thus $H=H_{\rm R}$.
The isometry group $G$ of $\Mscal$ clearly defines the global symmetries of the scalar action. If the $G$ has a transitive action on $\Mscal$, the manifold is said to be \emph{homogeneous} and can be written in the form $G/H'$, where $H'$ is the isotropy group, namely the subgroup of $G$ leaving a generic point on the manifold invariant.\par
 If the manifold, besides being homogeneous, is also \emph{symmetric}, then the isotropy and the holonomy groups (locally) coincide:\footnote{ When dealing with the isotropy and holonomy groups, by an abuse of notation, we shall always refer to the corresponding connected components, since we are interested in the local properties of these groups, encoded in their respective Lie algebras.} $H=H'$. In this case it has the general form
\begin{equation}
\Mscal\=\frac{G}{H}\,,
\end{equation}
where $G$ is the semisimple non-compact Lie group of isometries and the isotropy group $H$ is its maximal compact subgroup. In $\N>2$ theories the scalar manifold is constrained by supersymmetry to be of this kind, see Table 1.
In generic homogeneous spaces $\Mscal=G/H'$, $G$ need not be semisimple.
\begin{table}\label{table1}
\begin{center}
{\scriptsize
  \renewcommand{\arraystretch}{1.7}
  \begin{tabular}{ | c|| c | c| c|c|} %last argument to fix a bug!
  \hline
  $\N$ & $\dfrac{G}{H}$ & $n_s$ &$n_v$& ${\Scr R}_v$\\[2ex]
  \hline\hline
  & & & & \\
  8 & $\frac{{\rm E}_{7(7)}}{{\rm SU}(8)}$ & 70 & 28 &{\bf 56} \\
  & & & & \\
  \hline
  & & & & \\
  6 & $\frac{\SO^*(12)}{{\rm U}(6)}$   & 30  &16& ${\bf 32}_c$\\
  & & & & \\
  \hline
  & & & & \\
  5 & $\frac{{\rm SU}(5,1)}{{\rm U}(5)}$  & 10 &10&{\bf 20}\\
  & & & & \\
  \hline
  & & & & \\
  4 & $\frac{{\rm SL}(2,\mathbb{R})}{{\rm SO}(2)}\times \frac{{\rm SO}(6,n)}{{\rm SO}(6)\times {\rm SO}(n)}$  & 6n+2 &n+6 & ${\bf (2,6+n)}$\\
  & & & & \\
  \hline
  & & & & \\
  3 & $\frac{{\rm SU}(3,n)}{{\rm S}[{\rm U}(3)\times{\rm U}(n)]}$  & 6n &3+n & $({\bf 3+n})+({\bf 3+n})'$\\
  & & & & \\
  \hline
\end{tabular}
\caption{Homogeneous symmetric scalar manifolds in $\mathcal{N}>2$ supergravities, their real dimensions $n_s$ and the number $n_v$ of vector fields.}}
\end{center}
\end{table}
\begin{table}
\begin{center}
{\scriptsize
  \renewcommand{\arraystretch}{1.7}
  \begin{tabular}{ | c|| c | c| c|c|} %last argument to fix a bug!
  \hline
  $\N$ & $\dfrac{G}{H}$ & $n_s$ &$n_v$&${\Scr R}_v$ \\[2ex]
  \hline\hline
  & & & & \\
  & $\frac{{\rm SU}(1,n)}{{\rm U}(n)}$  & 2n &n+1&$({\bf 1+n})+({\bf 1+n})'$\\
  & & & & \\
  & $\frac{{\rm SL}(2,\mathbb{R})}{{\rm SO}(2)}\times\frac{{\rm SO}(2,n-1)}{{\rm SO}(2)\times {\rm SO}(n-1)}$ & $2n$ &n+1 & ${\bf (2,n+1)}$\\
  & & & & \\
  & $\frac{{\rm SU}(1,1)}{{\rm U}(1)}$  & 2 &2&${\bf 4}$\\
  & & & & \\
   2, SK & $\frac{{\rm Sp}(6)}{{\rm U}(3)}$  & 12 & 7 &${\bf 14}'$\\
  & & & & \\
 & $\frac{{\rm SU}(3,3)}{{\rm S}[{\rm U}(3)\times{\rm U}(3)]}$ & 18  &10& ${\bf 20}$\\
  & & & & \\
  & $\frac{{\rm SO}^*(12)}{{\rm U}(6)}$ & 30 &16& ${\bf 32}_c$\\
  & & & & \\
  & $\frac{{\rm E}_{7(-25)}}{{\rm U}(1)\times {\rm E}_6}$  & 54 &28& ${\bf 56}$\\
  & & & & \\
  \hline
  & & & & \\
  & $\frac{{\rm SU}(2,n_H)}{{\rm S}[{\rm U}(2)\times {\rm U}(n_H)]}$  & $4n_H$ & &  \\
  & & & & \\
  & $\frac{{\rm SO}(4,n_H)}{{\rm SO }(4)\times {\rm SO}(n_H)}$  & $4n_H$ &  & \\
    & & & & \\
  & $\frac{{\rm G}_{2(2)}}{{\rm SU }(2)\times {\rm SU }(2)}$  & $8$ &  & \\
  & & & & \\
 & $\frac{{\rm F}_{4(+4)}}{{\rm SU}(2)\times {\rm USp}(6)}$  & $28$ &  & \\
    & & & & \\
   2, QK& $\frac{{\rm E}_{6(+2)}}{{\rm SU}(2)\times {\rm SU}(6)}$  & $40$ &  & \\
    & & & & \\
  & $\frac{{\rm E}_{7(-5)}}{{\rm SU}(2)\times {\rm SO}(12)}$  & $64$ &  & \\
    & & & & \\
  & $\frac{{\rm E}_{8(-24)}}{{\rm SU}(2)\times {\rm E}_7}$  & $112$ &  & \\
   & & & & \\
  & $\frac{{\rm USp}(2,2n_H)}{{\rm USp}(2)\times {\rm USp}(2n_H)}$  & $4n_H$ & &  \\
  \hline
\end{tabular}
\caption{Homogeneous symmetric special K\"ahler (SK) and quaternionic K\"ahler (QK) scalar manifolds in $\mathcal{N}=2$ supergravities, their real dimensions $n_s$ and the number $n_v$ of vector fields.}}\label{table2}
\end{center}
\end{table}
Here we recall the main facts about the description of a non-linear sigma-model with a homogeneous target space $G/H'$ \cite{Gaillard:1981rj}. We refer to standard textbooks \cite{Helgason,Nomizu} for an in-depth discussion of the subject and to the Appendix \ref{nacv} for the relevant notations used here.\par The scalars can be described by a space-time dependent element $L(x)$ of $G$ and the sigma-model Lagrangian (and in fact the whole supergravity action) is constructed so as to be invariant under local $H'$-transformations, defined  on $L(x)$ by the right-action of a space-time dependent $H'$-element $h(x)$. The symmetries of the supergravity equations of motion and Bianchi identities also comprise global $G$-transformations ${\bf g}$ acting on $L(x)$ to the left. Therefore the local and global symmetry groups  $H'$ and $G$, respectively, act on $L(x)$ as follows:
\begin{equation}
L(x)\rightarrow L(x)h(x) \,\,;\,\,\,\,L(x)\rightarrow {\bf g}\,L(x)\,\,,\,\,\,{\bf g}\in G\,\,,\,\,\,h(x)\in H'\,.\label{lxtras}\end{equation}
 Invariance under the former transformations allows to group $L(x)$, for each $x^\mu$, into equivalence classes or \emph{left-cosets}, each defining a point on the scalar manifold, whose representative $L(\phi^s(x))$ is obtained by fixing the local right action of $H'$ on $L(x)$ and thus depends on ${\rm dim}(G)-{\rm dim}(H')$ independent local parameters $\phi^s(x)$ which are the scalar fields.
The action of an isometry
transformation ${\bf g}\in G$ on the scalar fields $\phi^r$ parametrizing
$\Mscal$ is defined by means of the coset
representative $L(\phi)\in G/H'$ as follows:
\begin{equation}
{\bf g}\cdot L(\phi^r)=L({\bf g}\star\phi^r)\cdot
h(\phi^r,{\bf g})\,,\label{gLh}
\end{equation}
where ${\bf g}\star\phi^r$ denote the transformed scalar fields,
non-linear functions of the original ones $\phi^r$, and
$h(\phi^r,{\bf g})$ is a \emph{compensator} in the isotropy group $H'$. The coset representative is thus defined modulo the right-action of $H'$ and is fixed by the chosen
parametrization of the manifold.
\par
The Lie algebra $\mathfrak{g}$ of $G$ can be decomposed into two orthogonal subspaces: the Lie algebra $\mathfrak{H}$ generating $H'$, and a coset space $\mathfrak{K}$
\begin{equation}
\mathfrak{g}=\mathfrak{H}\oplus \mathfrak{K}\,,\label{ghkdec}
\end{equation}
and the following general commutation relations hold:
\begin{equation}
[\mathfrak{H},\,\mathfrak{H}]\subset \mathfrak{H}\;;\qquad
[\mathfrak{H},\,\mathfrak{K}]\subset \mathfrak{K}\;;\qquad
[\mathfrak{K},\,\mathfrak{K}]\subset \mathfrak{H}\oplus\mathfrak{K}\;,\label{hkh}
\end{equation}
that is the space $\mathfrak{K}$ supports a representation ${\Scr K}$ of $H'$ with respect to its adjoint action. The space $\mathfrak{K}$ is isomorphic to the tangent space to $G/H'$ at any point and thus supports a representation of the whole holonomy group $H$. \par
 Of particular relevance in supergravity is the so-called \emph{solvable parametrization} \cite{Cremmer:1978ds,Andrianopoli:1996bq,Andrianopoli:1996zg,Cremmer:1997ct}, which corresponds to fixing the action of $H'$ so that $L$ belongs to a solvable Lie group%
\footnote{
 A solvable Lie group $G_S$ can be described (locally) as a the Lie group generated by \emph{solvable Lie algebra} $\Solv$: $G_S=\exp(\Solv) $. A Lie algebra $\Solv$ is solvable iff, for some $k>0$, ${\bf D}^k \Solv=0$, where the \emph{derivative} ${\bf D}$ of a Lie algebra $\mathfrak{g}$ is defined as follows: \,${\bf D}\mathfrak{g}\equiv [\mathfrak{g},\mathfrak{g}]$, \;${\bf D}^n\mathfrak{g}\equiv [{\bf D}^{n-1}\mathfrak{g},{\bf D}^{n-1}\mathfrak{g}]$. In a suitable basis of a given representation, elements of a solvable Lie group or a solvable Lie algebra are all described by upper (or lower) triangular matrices \cite{Helgason}.
}
$G_S=\exp(\Solv)$, generated by a solvable Lie algebra $\Solv$ and defined, if ${\Scr M}_{{\rm scal}}$ is also symmetric, by the Iwasawa decomposition of $G$ with respect to $H'=H$.
The scalar fields are then parameters of the solvable Lie algebra $\Solv$:
\begin{align}
L(\phi^r)&= e^{\phi^r T_r}\in \exp(\Solv)\,,\label{solpar}
\end{align}
where $\{T_r\}$ is a basis of $\Solv$ ($r=1,\dots,\,n_s$). As opposed to $\mathfrak{K}$, the space $\Solv$ is \emph{not} orthogonal to the isotropy algebra $\mathfrak{H}$.
All homogeneous scalar manifolds occurring in supergravity theories admit this (global) parametrization, which is useful when the four-dimensional supergravity originates from the Kaluza-Klein reduction of a higher-dimensional one on some internal compact manifold. The
solvable coordinates directly describe dimensionally reduced fields and moreover this parametrization makes the shift symmetries of the $\sigma$-model metric manifest. We refer the reader to the final paragraph of the present Section for a more detailed discussion on this issue.\par
 An alternative choice of parametrization corresponds to defining the coset representative as an element of $\exp(\mathfrak{K})$:
\begin{align}
L(\phi^r)&= e^{\phi^r K_r} \;\in\, \exp(\mathfrak{K})\,,\label{cartpar}
\end{align}
where $\{K_r\}$ is a basis of $\mathfrak{K}$. As opposed to the solvable parametrization, the coset representative is no-longer a group element, since $\mathfrak{K}$ does not close an algebra, see last of eqs.\ (\ref{hkh}). The main advantage of this parametrization is that the action of $H'$ on the scalar fields is \emph{linear}:
\begin{align}
\forall h\in H'\;:\quad
h\,L(\phi^r)=h\,e^{\phi^r K_r}\,h^{-1}\,h=e^{\phi^r\,h\,K_r\,h^{-1}}\,h=L(\phi^{\prime r})\,h\,,\label{hphiK}
\end{align}
where $\phi^{\prime r}=(h^{-1})_s{}^r\,\phi^s$, and $h_s{}^r$ describes $h$ in the representation ${\Scr K}$. This is not the case for the solvable parametrization since $[\mathfrak{H},\,\Solv]\nsubseteq \Solv$.\par
In all parametrizations, the origin $\Or$ is defined as the point in which the coset representative equals the identity element of $G$ and thus the $H'$-invariance of $\Or$ is manifest: $L(\Or)=\Id$ (corresponding to the coset $H'$).\par
If the manifold, besides being homogeneous, is also \emph{symmetric}, the space $\mathfrak{K}$ can be defined so that:
\begin{equation}
[\mathfrak{K},\,\mathfrak{K}]\subset \mathfrak{H}\,.\label{KKH}
\end{equation}
In this case the Eq.\ (\ref{ghkdec}) defines the Cartan decomposition of $\mathfrak{g}$ into \emph{compact} and \emph{non-compact} generators, in $\mathfrak{H}$ and $\mathfrak{K}$, respectively (recall that ${\Scr M}_{{\rm scal}}$ is a  non-compact manifold). This means that, in a given matrix representation of $\mathfrak{g}$, a basis of the representation space can be chosen so that the elements of $\mathfrak{H}$ and of $\mathfrak{K}$ are represented by anti-hermitian and hermitian matrices, respectively. Unless otherwise stated, we shall restrict our discussion of coset geometry, from now on, to symmetric manifolds for which no distinction is made between $H$ and $H'$. Several relations given in this Section, although derived for the sake of simplicity in this special case, hold in fact for more general manifolds.\par
The geometry of $\Mscal$ is described by vielbein and an $H'$-connection constructed out of the left-invariant one-form
\begin{equation}
\Omega\=L^{-1}\,dL\,\in\,\galg\,,\label{omegapro}
\end{equation}
satisfying the Maurer-Cartan equation:
\begin{equation}
d\Omega+\Omega\wedge \Omega=0\,.\label{MCeq}
\end{equation}
The vielbein and $H$-connection are defined by decomposing $\Omega$ according to (\ref{ghkdec})
\begin{equation}
\Omega(\phi)=\P(\phi)+\mathcal{Q}(\phi)\,; \quad\quad \mathcal{Q}\in\halg\,,\quad \P\in\kalg\,.\label{Vom}
\end{equation}
Let us see how these quantities transform under the action of $G$. For any ${\bf g}\in G$, using Eq.\ (\ref{gLh}), we can write $L({\bf g}\star \phi)={\bf g}\,L(\phi)\,h^{-1}$, so that:
\begin{align}
\Omega({\bf g}\star \phi)&=h\,L(\phi)^{-1}\,{\bf g}^{-1} d({\bf g}\,L(\phi)\,h^{-1})=
h\,L(\phi)^{-1}\,dL(\phi)\;h^{-1}+h\;dh^{-1}\,,
\end{align}
where we have used the fact that ${\bf g}$ is a global transformation: $d{\bf g}=0$.
From (\ref{Vom}) we find:
\begin{align}
&\P({\bf g}\star \phi)+\mathcal{Q}({\bf g}\star \phi)=h\,\P(\phi)\,h^{-1}+h\,\mathcal{Q}(\phi)h^{-1}+h\;dh^{-1}\,.
\end{align}
Since $h\;dh^{-1}$ is the left-invariant 1-form on $\mathfrak{H}$, it has value in this algebra. Projecting the above equation over $\mathfrak{K}$ and $\mathfrak{H}$, we find:
\begin{align}
\P({\bf g}\star \phi)&=h\,\P(\phi)\,h^{-1}\,,\label{Ptra}\\
\mathcal{Q}({\bf g}\star \phi)&=h\,\mathcal{Q}(\phi)\,h^{-1}+h\;dh^{-1}\,.\label{omtra}
\end{align}
We see that $\mathcal{Q}$ transforms as a connection while the matrix-valued one-form $\P$ transforms linearly under the $H$-compensator. These quantities will be used to construct the supergravity action with manifest local $H$-invariance, see Sect. \ref{fsector}.\footnote{Prior to fixing the local isotropy group action to the right of $L(x)$, as in the first of Eqs. (\ref{lxtras}), and defining the scalar fields, we would have written the left invariant one-form as:
\begin{equation}
\Omega(x)=\Omega_\mu(x)\,dx^\mu=L(x)^{-1}\partial_\mu L(x)\,dx^\mu=(\mathcal{P}_\mu(x)+\mathcal{Q}_\mu(x))\,dx^\mu\,.
\end{equation}
Under the action of a generic $h(x)\in H'$ on $L(x)$ to the right $\mathcal{Q}\rightarrow h\,\mathcal{Q}\,h^{-1}+h\;dh^{-1}$, so that $\mathcal{Q}_\mu(x)$ is the connection associated with this local invariance. If we gauge-fix the local action of the isotropy group, $L$ depends on $x$ only through $\phi^s(x)$ and $\mathcal{P}=\mathcal{P}_s(\phi)\,d\phi^s,\,\mathcal{Q}=\mathcal{Q}_s(\phi)\,d\phi^s$, so that the invariance of the theory under the local transformation in (\ref{lxtras}) is clearly no-longer manifest. What is manifest is the invariance under the compensating transformation $h(\phi,\,{\bf g})$ associated with a global isometry ${\bf g}$ \cite{Gaillard:1981rj}.
}
In homogeneous non-symmetric manifolds, $\mathcal{Q}$ defined above does not represent the whole $H$-connection but only its component in the isotropy algebra $\mathfrak{H}$. In this case we would denote it by $\mathcal{Q}'$, see discussion at the end of Appendix \ref{nacv} and in particular Eq. (\ref{QQpDQ}).
\par  The vielbein of the scalar manifold are defined by expanding $\P$ in a basis $\{K_{\underline{s}}\}$ of $\mathfrak{K}$ (underlined indices $\underline{s},\underline{r},\underline{t},\dots$ are rigid tangent-space indices, as opposed to the curved coordinate indices $s,r,t,\dots$):
\begin{equation}
\P(\phi)=\mathcal{P}^{\underline{s}}( \phi)\,K_{\underline{s}}\,.
\end{equation}
From (\ref{Ptra}) it follows that the vielbein 1-forms $\mathcal{P}^{\underline{s}}( \phi)=\mathcal{P}_s{}^{\underline{s}}( \phi)d\phi^s$ transform under the action of $G$ as follows:
\begin{equation}
\mathcal{P}^{\underline{s}}({\bf g}\star \phi)\=\mathcal{P}^{\underline{t}}( \phi)\,(h^{-1})_{\underline{t}}{}^{\underline{s}}\=h^{\underline{s}}{}_{\underline{t}}\mathcal{P}^{\underline{t}}( \phi)\,.\label{Vtra}
\end{equation}
We can also define an $H$-covariant derivative on $L(\phi^s)$ using the connection $Q$, so that, from (\ref{Vom}) we have:
\begin{equation}
{\Scr D}L\equiv dL-L\,\mathcal{Q}=L\,\mathcal{P}\,.\label{DLP}
\end{equation}
From Eq. (\ref{MCeq}) and Eq. (\ref{KKH}) for symmetric spaces, it follows that $\mathcal{Q}$ and $\P$ satisfy the following conditions
\begin{align}
{\Scr D}\P&~\equiv~ d\P+\mathcal{Q}\wedge P+\P\wedge \mathcal{Q}\=0\,,\label{DP}\\
R(\mathcal{Q})&~\equiv~ d\mathcal{Q}+\mathcal{Q}\wedge \mathcal{Q}\=-\P\wedge \P\,,\label{RW}
\end{align}
where we have defined the $H$-covariant derivative ${\Scr D}\P$ of $\P$ and the $\mathfrak{H}$-valued curvature $R(\mathcal{Q})$ of the manifold.
The latter can be written in components:
\begin{equation}
R(\mathcal{Q})=\frac{1}{2}\,R_{rs}\,d\phi^r\wedge d\phi^s \quad\Rightarrow\quad
R_{rs}=-[\P_r,\,\P_s]\in \mathfrak{H}\,.\label{Rcompo}
\end{equation}
Equation (\ref{DP}) for a general Riemannian manifold follows from the first vielbein postulate defining the Levi-Civita connection,  see Eq. (\ref{1vp2}) of Appendix \ref{nacv}.\par
If $\Phi$ is a field on $\Mscal$, only transforming, through its internal indices, in some representation of the holonomy group $H$, the definition of the $H$-covariant derivative generalizes as follows (see also Appendix \ref{nacv}):
\begin{equation}
{\Scr D}_r\Phi=\partial_r\Phi+\mathcal{Q}_r\star \Phi\,,
\end{equation}
where $\star$ indicates the action of $\mathcal{Q}$ on $\Phi$ in the corresponding representation. The reader can easily verify that:
\begin{equation}
{\Scr D}^2\Phi=R(\mathcal{Q})\star\Phi\,\,\Leftrightarrow\,\,\,\,\,\,[{\Scr D}_r,\,{\Scr D}_s]\Phi=R_{rs}\star\Phi\,.
\end{equation}
For homogeneous non-symmetric manifolds, the covariant derivative is defined in terms of the full Levi-Civita $H$-connection $\mathcal{Q}$, which does not coincide with the component in the isotropy algebra $\mathfrak{H}$ of the left-invariant one-form but which is defined by the condition ${\Scr D}\mathcal{P}^{\underline{s}}=0$.\par
Let the metric computed on the basis $\{K_{\underline{s}}\}$ be described by the constant, $H$-invariant matrix $\eta_{\underline{s}\underline{t}}$ which, for symmetric spaces, can be written in the following form: \footnote{For symmetric spaces, $\{K_{\underline{s}}\}$ is a basis of the non-compact generators of $\mathfrak{g}$ and the trace below is proportional to the restriction of the Cartan-Killing metric to these generators, which is positive definite. For homogeneous non-symmetric manifolds this trace-formula does not apply since a basis of $\mathfrak{K}$ may include nilpotent matrices, whose trace would yield a singular matrix. In this case the metric $\eta_{\underline{s}\underline{t}}$ is assigned as a $H'$-invariant, positive definite matrix. See discussion at the end of paragraph ``The geometry of the scalar manifold'' in Appendix \ref{nacv}.}
\begin{equation}
\eta_{\underline{s}\underline{t}}\equiv k\,{\rm Tr}(K_{\underline{s}}\,K_{\underline{t}})>0\,,
\end{equation}
where $k$ is a positive number depending on the matrix representation used, so that the metric in a generic point reads:
\begin{equation}
 ds^2(\phi)\equiv\Gm_{st}(\phi)d\phi^s\,d\phi^t\equiv \mathcal{P}_s{}^{\underline{s}}( \phi)\mathcal{P}_t{}^{\underline{t}}( \phi)\eta_{\underline{s}\underline{t}}\,d\phi^s\,d\phi^t=k\,{\rm Tr}(\P(\phi)\,\P(\phi))\,.
\end{equation}
As it follows from eqs.\ (\ref{Ptra}), (\ref{Vtra}), the above metric is manifestly invariant under global $G$-transformations:
\begin{equation}
ds^2({\bf g}\star \phi)=ds^2(\phi)\;.
\end{equation}
The $\sigma$-model Lagrangian can be written in the form:
\begin{equation}
\Lscal=\frac{e}{2}\, \Gm(\phi)_{st}\partial_\mu\phi^s\,\partial^\mu\phi^t
=\frac{e}{2}\,k\,\Tr\big(\P_\mu(\phi)\,\P^\mu(\phi)\big)\,,
\qquad \P_\mu=\P_s\frac{\partial\phi^s}{\partial x^\mu}\,,\;\quad\label{lagrscal}
\end{equation}
and, just as the metric $ds^2$, it is manifestly invariant under global $G$-transformations acting on $L$ as in Eqs. (\ref{lxtras}), (\ref{gLh}).\par
The bosonic part of the equations of motion for the scalar fields can be derived from the Lagrangian (\ref{boslagr}) and read:
\begin{align}
{\Scr D}_\mu (\partial^\mu
\phi^s)&=\frac{1}{4}\,\Gm^{st}\,\left[F_{\mu\nu}^\Lambda\,
\partial_t\,\I_{\Lambda\Sigma}\,F^{\Sigma\, \mu\nu}+F_{\mu\nu}^\Lambda
\partial_t\, \R_{\Lambda\Sigma}\,{}^*F^{\Sigma\,
\mu\nu}\right]+\dots\,,\label{scaleqs}
\end{align}
where $\partial_s\equiv \frac{\partial}{\partial \phi^s}$ and ${}^*F_{\mu\nu}\equiv \frac{e}{2}\epsilon_{\mu\nu\rho\sigma}\,F^{\rho\sigma}$. The ellipses refer to terms containing fermionic fields. These will be dealt with later in Sections \ref{fsector} and \ref{dcsl}.  The covariant derivative ${\Scr D}_\mu$ also contains the Levi-Civita connection $\tilde{\Gamma}$ on the scalar manifold:
\begin{equation}
{\Scr D}_\mu (\partial_\nu
\phi^s)\equiv \nabla_\mu(\partial_\nu
\phi^s)+\tilde{\Gamma}^s_{t_1 t_2}\partial_\mu \phi^{t_1}\,\partial_\nu\phi^{t_2}\,,
\end{equation}
$\nabla_\mu$ being the covariant derivative only containing the space-time connection, see Appendix \ref{nacv} for a summary of the relevant conventions.\par
$\N=2$ theories deserve a separate discussion and we shall deal with them in detail in Sect \ref{N2sugras}. In this case, the scalar fields may sit either in vector multiplets or in hypermultiplets. The former span a \emph{special K\"ahler manifold} \cite{Strominger:1990pd}, the latter, named hyper-scalars, parametrize a \emph{quaternionic K\"ahler} one \cite{Bagger:1983tt}. The scalar manifold is always factorized in the product of the two:
\begin{equation}
\Mscal^{\scalebox{0.5}{$\,(\N=2)$}}\=\MsSK\times \MsQK\,.\label{SKQK}
\end{equation}
As opposed to $\mathcal{N}>2$ models in which scalar manifolds are homogeneous symmetric, each of these two spaces may not be symmetric or even homogeneous (i.e. the isometry group $G$, if it exists, does not act transitively on them). The holonomy group of $\MsSK$ is $H^{(SK)}={\rm U}(1)\times H^{(SK)}_{{\rm matt}}$ while that of $\MsQK$ is   $H^{(QK)}={\rm SU}(2)\times H^{(QK)}_{{\rm matt}}$ where, by definition of quaternionic K\"ahler spaces,  $H^{(QK)}_{{\rm matt}}$ is contained in ${\rm USp}(2n_H)$, $n_H$ being the number of hypermultiplets. The group ${\rm SU}(2)\times {\rm USp}(2n_H)$ has a linear action on the tangent space to $\MsQK$ according to the representation ${\bf (2,2n_H)}$. This means that we can choose a basis $\{K_{A,\alpha}\}$, $A=1,2$, $\alpha=1,\dots, 2n_H$, of the tangent space at any point and expand the vielbein matrix of $\MsQK$ in components with respect to it:\footnote{Because of a shortage of indices, only when discussing quaternionic K\"ahler manifolds in $\mathcal{N}=2$ theories the indices $\alpha,\,\beta,\,\dots$ will label the fundamental representation of ${\rm USp}(2n_H)$. Otherwise  the same indices will label generic isometry generators.}
\begin{equation}
\P=\P^{A\alpha}\,K_{A,\alpha}=dq^u\,\P_u{}^{A\alpha}\,K_{A,\alpha}\,,\label{Paalpha}
\end{equation}
where we have denoted by $q^u$, $u=1,\dots,\,4n_H$, the (real) hyper-scalars parametrizing $\MsQK$. The vielbein tensor $\P_u{}^{A\alpha}$ is subject to the reality condition:
\begin{equation}
\P_{u\,A\alpha}\equiv (\P_u{}^{A\alpha})^*=\epsilon_{AB}\mathbb{C}_{\alpha\beta}\,\P_u{}^{B\beta}\,.
\end{equation}
The metric ${\Scr G}(q)_{uv}$ of $\MsQK$, in terms of $\P_u{}^{A\alpha}$ has the following form: \footnote{Later, when dealing specifically with the $\mathcal{N}=2$ theories, following the notations of \cite{Andrianopoli:1996cm}, we shall rescale the vielbein matrix and the metric, defining $\mathcal{U}_u{}^{A\alpha}$ and $h(q)_{uv}$ related to $\mathcal{P}_u{}^{A\alpha}$ and ${\Scr G}(q)_{uv}$ as follows: $\mathcal{U}_u{}^{A\alpha}\equiv \frac{1}{\sqrt{2}}\,\mathcal{P}_u{}^{A\alpha}$
and $h_{uv}=\frac{1}{2}\,{\Scr G}_{uv}$.}
\begin{equation}
{\Scr G}(q)_{uv}=\P_u{}^{A\alpha}\P_v{}^{B\beta}\epsilon_{AB}\mathbb{C}_{\alpha\beta}\,.
\end{equation}
Clearly for non-homogeneous spaces, the elements $K_{A,\alpha}$ in (\ref{Paalpha}) are not isometry generators. The connection 1-form of $\MsQK$ is represented by the matrices $\mathcal{Q}^A{}_B$ and $\mathcal{Q}^\alpha{}_\beta$ in the Lie algebras of ${\rm SU}(2)$ and ${\rm USp}(2n_H)$, respectively.
Both $\MsSK$ and $\MsQK$, just as any other scalar manifold in supergravity, are non-compact manifolds with negative curvature. We shall be dealing with their geometry, and in general with $\mathcal{N}=2$ theories, in Sect. \ref{N2sugras}. The homogeneous symmetric special K\"ahler and quaternionic K\"aher manifolds are listed in Table 2. Homogeneous special and quaternionic K\"ahler manifolds were classified in \cite{alek,deWit:1992wf}.

Let us end this paragraph by introducing, in the coset geometry, the Killing vectors describing the infinitesimal action of isometries on the scalar fields. Let us denote by $t_\alpha$ the infinitesimal generators of $G$, defining a basis of its Lie algebra $ \mathfrak{g}$ and satisfying the corresponding commutation relations
\begin{equation}
[t_\alpha,\,t_\beta]={\rm f}_{\alpha\beta}{}^\gamma\,t_\gamma\,,\label{talg}
\end{equation}
${\rm f}_{\alpha\beta}{}^\gamma$ being the structure constants of $\mathfrak{g}$. Under an infinitesimal $G$-transformation generated by $\epsilon^\alpha\,t_\alpha$ ($\epsilon^\alpha\ll 1$):
\begin{equation}
{\bf g}\approx \Id+\epsilon^\alpha\,t_\alpha\,,
\end{equation}
the scalars transform as:
\begin{equation}
\phi^s\rightarrow \phi^s+\epsilon^\alpha\,k^s_\alpha(\phi)\,,
\end{equation}
$k^s_\alpha(\phi)$ being the Killing vector associated with $t_\alpha$.
They satisfy the condition:
\begin{equation}
{\Scr D}_{r}k_\alpha^t\,{\Scr G}_{st}+{\Scr D}_{s}k_\alpha^t\,{\Scr G}_{rt}=0\,.\label{Killcond}
\end{equation}
The action of ${\bf g}$ on the scalars is defined by Eq.\ (\ref{gLh}), neglecting terms of order $O(\epsilon^2)$:
\begin{equation}
(\Id+\epsilon^\alpha\,t_\alpha)\,L(\phi)=L(\phi+\epsilon^\alpha\,k_\alpha)\left(\Id-\frac{1}{2}\,\epsilon^\alpha W_\alpha ^I\,J_I\right)\,,\label{tLkW}
\end{equation}
where $(\Id-\frac{1}{2}\,\epsilon^\alpha W_\alpha ^I\,J_I)$ denotes, expanded to linear order in $\epsilon$, the compensating transformation $h(\phi,{\bf g})$, $\{J_I/2\}$ being a basis of $\mathfrak{H}$. For homogeneous non-symmetric manifolds, we define $\{J_I/2\}$ as a basis of the holonomy algebra, which contains the isotropy algebra $\mathfrak{H}$. Equating the terms linear in $\epsilon^\alpha$, multiplying to the left by $L^{-1}$ and using the expansion (\ref{Vom}) of the left-invariant 1-form, we end up with the following equation:
\begin{equation}
L^{-1}t_\alpha L\=k_\alpha^s\,(\mathcal{P}_s+\mathcal{Q}_s)-\frac{1}{2}\,W_\alpha ^I\,J_I\=k_\alpha^s\,\mathcal{P}_s{}^{\underline{s}}\,K_{\underline{s}}+\frac{1}{2}\,(k_\alpha^s\mathcal{Q}_s^I-W_\alpha ^I)\,J_I\,,\label{kespans}
\end{equation}
where we have expanded the $H$-connection along $J_I/2$ as follows:\footnote{For homogeneous non-symmetric spaces this quantity would be the component $\mathcal{Q}'$ of the $H$-connection $\mathcal{Q}$ on the isotropy algebra $\mathfrak{H}$, see Eq. (\ref{QQpDQ}) in Appendix \ref{nacv}. The derivation below would follow the same lines.}
\begin{equation}
 \mathcal{Q}_s=\frac{1}{2}\,\mathcal{Q}^I_s\,J_I\,.
\end{equation}
Eq.\ (\ref{kespans}) allows to compute $k_\alpha$ for homogeneous scalar manifolds by projecting $L^{-1}t_\alpha L$ along the directions of the coset space $\mathfrak{K}$.
These Killing vectors satisfy the following algebraic relations (note the minus sign on the right hand side with respect to (\ref{talg})):
 \begin{equation}
[k_\alpha,\,k_\beta]=-{\rm f}_{\alpha\beta}{}^\gamma\,k_\gamma\,,\label{kalg}
\end{equation}
We can split, according to the general structure (\ref{Hgroup}), the $H$-generators $J_I$ into $H_{\rm R}$-generators $J_{{\bf a}}$ (${{\bf a}}=1,\dots,{\rm dim}(H_{\rm R})$) and $H_{\rm matt}$-generators $J_{{\bf m}}$ (${\bf m}=1,\dots,{\rm dim}(H_{\rm matt})$), and rewrite (\ref{kespans}) in the form:
\begin{equation}
L^{-1}t_\alpha L\=
k_\alpha^s\,\mathcal{P}_s{}^{\underline{s}}\,K_{\underline{s}}-\frac{1}{2}\,\Ps_\alpha^{{\bf a}}\,J_{{\bf a}}-\frac{1}{2}\,\Ps_\alpha^{{\bf m}}\,J_{{\bf m}}\,.\label{kespans2}
\end{equation}
The quantities
\begin{equation}
\Ps_\alpha^{{\bf a}}=- (k_\alpha^s\mathcal{Q}_s^{{\bf a}}-W_\alpha ^{{\bf a}})\,,\label{PkQW}
\end{equation}
generalize the so-called \emph{momentum maps} in $\N=2$ theories, which provide a Poissonian realization of the isometries $t_\alpha$, see Sect. \ref{N2sugras}.
One can verify the general property:
\begin{equation}
 k_\alpha^s\,R^{{\bf a}}_{st}={\Scr D}_t \Ps_\alpha^{{\bf a}}\,,\label{KRP}
\end{equation}
where ${\Scr D}_s$ denotes the $H$-covariant derivative and we have expanded the curvature $R[\mathcal{Q}]$ defined in (\ref{RW}) along $J_I/2$:
\begin{equation}
 R[\mathcal{Q}]=\frac{1}{2}\,R^I_{st}\,d\phi^s\wedge d\phi^t\,\left(\frac{J_I}{2}\right)\,.
\end{equation}
These objects are important in the gauging procedure since they enter the definition of the gauged connections for the fermion fields as well as gravitino-shift matrix $\mathbb{S}_{AB}$ (see Sect.\ \ref{sec:3}). For all those isometries which do not produce compensating transformations in $H_{\rm R}$, $W_\alpha^{{\bf a}}=0$ and $ \Ps_\alpha^{{\bf a}}$ are easily computed to be $$ \Ps_\alpha^{{\bf a}}=- k_\alpha^s\mathcal{Q}_s^{{\bf a}}\,.$$ This is the case, in the solvable parametrization, for all the isometries in $\Solv$, which include translations in the axionic fields.\par
In $\N=2$ models with non-homogeneous scalar geometries, though we cannot apply the above construction of $k_\alpha,\,\Ps_\alpha^{{\bf a}}$, the momentum maps are constructed from the Killing vectors as solutions to the differential equations (\ref{KRP}).
In general, in these theories, with each isometry $t_\alpha$ of the scalar manifold, we can associate the quantities $\Ps_\alpha^{{\bf a}},\,\Ps_\alpha^{{\bf m}}$ which are related to the corresponding Killing vectors $k_\alpha$ through general
relations (see \cite{Andrianopoli:1996cm} for a comprehensive account of $\N=2$ theories). We shall be dealing with this issue in detail in Sections \ref{isomsk} and \ref{isomqk}.
\paragraph{A worked out example 1.}
Let us discuss the simple example of the lower-half complex plane
\begin{equation}
G/H={\rm SL}(2,\mathbb{R})/{\rm SO}(2)\,.
\end{equation}
This manifold is parametrized by a complex coordinate $z$, with ${\rm Im}(z)<0$.
We can work in the fundamental representation of $G={\rm SL}(2,\mathbb{R})$ and choose the following basis of generators of $\mathfrak{g}=\mathfrak{sl}(2,\mathbb{R})$:
\begin{align}
\mathfrak{sl}(2,\mathbb{R})=\{\sigma^1,\,i\,\sigma^2,\sigma^3\}=\left\{\left(\begin{matrix}0 & 1 \cr 1 & 0\end{matrix}\right),\,\left(\begin{matrix}0 & 1 \cr -1 & 0\end{matrix}\right),\,\left(\begin{matrix}1 & 0 \cr 0 & -1\end{matrix}\right)\right\}\,.\label{sl2basis1}
\end{align}
The isotropy algebra is:
\begin{equation}
\mathfrak{H}=\{J\}=\{-i\,\sigma^2\}\,,
\end{equation}
while its orthogonal complement, isomorphic to the tangent space, is spanned by symmetric traceless matrices:
\begin{equation}
\mathfrak{K}=\{K_{\underline{1}},\,K_{\underline{2}}\}=\{\sigma^3,\,\sigma^1\}\,.
\end{equation}
In this basis the metric reads: $\eta_{\underline{s}\underline{t}}={\rm Tr}(K_{\underline{s}}K_{\underline{t}})={\rm diag}(2,2)$.
Let us illustrate in this simple example the two parametrizations discussed earlier.
The $H={\rm SO}(2)$-covariant parametrization consists in defining the coset-representative as follows:
\begin{equation}
L=e^{\xi^{{s}}\,K_{{s}}}=\exp\left(\frac{\xi}{2}\cos(\alpha)\,\sigma^3+\frac{\xi}{2}\sin(\alpha)\,\sigma^1\right)\,,
\end{equation}
where $K_s=K_{\underline{s}}$ and we have denoted the coordinates by $\xi^1,\,\xi^2$ to distinguish them from those in the solvable parametrization, to be introduced below. We have also set: $\xi^1=\xi \cos(\alpha)/2,\,\xi^2=\xi \sin(\alpha)/2$. The coordinates $\xi^s$ transform linearly under ${\rm SO}(2)$-transformations, which have the form $h(\beta)=\exp(\beta\,J)$, see Eq. (\ref{hphiK}). Indeed one can verify that:
\begin{equation}
h(\beta)L(\xi,\alpha)h(\beta)^{-1}=L(\xi, \alpha+2 \,\beta)\,,
\end{equation}
which amounts to an ${\rm SO}(2)$-rotation on the 2-component vector $\xi^s$. The metric in these coordinates reads:
\begin{equation}
ds^2=\frac{1}{2} \left({d\alpha }^2 \sinh ^2(\xi )+{d\xi }^2\right)\,.
\end{equation}
Let us now define the solvable parametrization by writing the Iwasawa decomposition of $\mathfrak{g}=\mathfrak{sl}(2,\mathbb{R})$ with respect to $\mathfrak{H}=\mathfrak{so}(2)$:
\begin{equation}
\mathfrak{sl}(2,\mathbb{R})=\mathfrak{so}(2)\oplus\,{\Scr S}\,.
\end{equation}
The solvable subalgebra ${\Scr S}$ consists of the follwing upper-triangular generators
\begin{align}
{\Scr S}=\{\sigma^3,\,\sigma^+\}\,\,,\,\,\,\,\sigma^+\equiv \left(\begin{matrix}0 & 1 \cr 0 & 0\end{matrix}\right)\,.
\end{align}
and defines the solvable parametrization $\phi^s=(\varphi,\,\chi)$, in which the coset representative, to be denoted only here by $L^{(S)}$, has the following form:
\begin{align}
L^{(S)}(\varphi,\,\chi)\equiv e^{\chi \sigma^+}\,e^{\frac{\varphi}{2}\sigma^3}=\left(
\begin{array}{ll}
 1 & \chi  \\
 0 & 1
\end{array}
\right)\left(
\begin{array}{ll}
 e^{\varphi /2} & 0 \\
 0 & e^{-\varphi /2}
\end{array}
\right)\,\in\,\, e^{\Solv}\,.\label{LLSol}
\end{align}
The relation between the solvable coordinates and $z$ is
\begin{equation}
z\=z_1+i\,z_2\=\chi-i\,e^{\varphi}\,.
\end{equation}
The metric reads:
\begin{equation}
ds^2=\frac{{d\varphi }^2}{2}+\frac{1}{2}{d\chi }^2 e^{-2 \varphi }=\frac{1}{2z_2^2}\,dz d\bar{z}=2\,g_{z\bar{z}}\,dz d\bar{z}\;,
\end{equation}
and the relation between $\phi^s$ and $\xi^s$ is the following:
\begin{equation}
e^{-\varphi}=\cosh(\xi)-\cos(\alpha)\,\sinh(\xi)\,\,;\,\,\,\,\chi=\frac{\sin(\alpha)\sinh(\xi)}{\cosh(\xi)-\cos(\alpha)\sinh(\xi)}\,,
\end{equation}
as it can be easily ascertained by solving the matrix equation $L^{(S)}(\varphi,\,\chi)\,h(\beta)=L(\xi,\alpha)$ in $\varphi,\,\chi$ and the compensator parameter $\beta$.\par
The vielbein and connection one-forms in the solvable coordinates read:
\begin{align}
L^{(S)\,-1}dL^{(S)}&=\mathcal{P}^{\underline{s}}\,K_{\underline{s}}+\mathcal{Q}\,J\,,\nonumber\\
\mathcal{P}^{\underline{1}}&=\frac{1}{2}\,d\varphi\,,\,\,
\mathcal{P}^{\underline{2}}=\frac{e^{-\varphi}}{2}\,d\chi\,,\,\,\,\mathcal{Q}=
\frac{e^{-\varphi}}{2}\,d\chi\,,
\end{align}
The curvature 2-form is also computed to be:
\begin{equation}
R(\mathcal{Q})=d \mathcal{Q}=\frac{e^{-\varphi}}{2}\,d\varphi\wedge d\chi=i\,g_{z\bar{z}}\,dz\wedge d\bar{z}\,.
\end{equation}
The reader can evaluate the Ricci tensor and scalar to be $\mathcal{R}_{\underline{s}\underline{t}}=-2\eta_{\underline{s}\underline{t}}$ and $\mathcal{R}=-4$, respectively.
 This manifold is an instance of special K\"ahler space, see Sect. \ref{SK},  and $\mathcal{Q}$ is the ${\rm U}(1)$-connection associated with K\"ahler transformations.\par
 The reader can also compute the Killing vectors associated with the isometry generators in the basis (\ref{sl2basis1}) to be:
 \begin{align}
 k_\alpha=k_\alpha^z(z)\,\frac{\partial}{\partial z}+k_\alpha^{\bar{z}}(\bar{z})\,\frac{\partial}{\partial \bar{z}}\,,
 \end{align}
 where $k_\alpha^{\bar{z}}(\bar{z})=(k_\alpha^z(z))^*$ and
 \begin{equation}
 k_1^z=2\,z\,,\,\,k_2^z=1+z^2\,,\,\,\,k_3^z=1-z^2\,.
 \end{equation}
\paragraph{Supersymmetry transformations.}
Let us note that, in coset geometry, any transformation of the coset representative $L$ can be effected by means of the action of a suitable, possibly scalar-field-dependent, $G$-transformation acting on $L$ from the right. This is due to the fact that the right action of $G$ on itself is transitive. Therefore, if $\phi^s\rightarrow \phi^{\prime s}(\phi^t)$ is a generic transformation of the scalar fields, and  $L'=L(\phi')$ is the transformed coset representative ($\phi'\equiv (\phi^{\prime s})$), depending on the transformed scalar fields $\phi^{\prime s}$, we can always write:
\begin{equation}
L(\phi'(\phi))=L(\phi)\,{\bf g}_{{\tt R}}(\phi)\,,\label{generaltransphi}
\end{equation}
being ${\bf g}_{{\tt R}}(\phi)\equiv L(\phi)^{-1}\,L(\phi'(\phi))$. For instance a global isometry transformation resulting from the action of an element ${\bf g}\in G$ on $L$ from the left, using (\ref{gLh}), can be described as the effect of the action of a $\phi$-dependent element ${\bf g}_{{\tt R}}(\phi)$ of $G$ from the right:
 \begin{equation}
 L(\phi')={\bf g}\,L(\phi)\,{ h}^{-1}=L(\phi)\,{\bf g}_{{\tt R}}(\phi)\,,
 \end{equation}
 where ${\bf g}_{{\tt R}}(\phi)\equiv L(\phi)^{-1}\,{\bf g}\,L(\phi)\,{ h}^{-1}$.\par
 Clearly a generic variation of the coset representative could also be effected by a field-dependent element of $G$ acting on $L$ from the left, but this would  not be useful to our discussion.
 Note that property  (\ref{generaltransphi}) also holds for transformations on the scalar fields which are local with respect to space-time, namely which explicitly depend on $x^\mu$. In this case the $G$-transformation ${\bf g}_{{\tt R}}$ will be also $x^\mu$-dependent: ${\bf g}_{{\tt R}}={\bf g}_{{\tt R}}(\phi,x)$.\par
 Let us now write ${\bf g}_{{\tt R}}$  as the product of an element in the coset and an element in $H'$:
 $${\bf g}_{{\tt R}}(\phi,x)={\bf g}_{G/H'}(\phi,x)\,{\bf g}_{H'}(\phi,x)\,,$$
 where ${\bf g}_{H'}(\phi,x)\in H'$ and ${\bf g}_{G/H'}(\phi,x)=\exp(\Sigma(\phi,x))$, with $\Sigma(\phi,x)\in \mathfrak{K}$. General invariance of the theory under local $H'$-transformations on $L$ from the right allows us to ignore the factor ${\bf g}_{H'}(\phi,x)$, so that a generic (local) transformation of the scalar fields can be effected by acting on $L$ to the right by a suitable element ${\bf g}_{G/H'}(\phi,x)=\exp(\Sigma(\phi,x))$. This, for instance, is the case of the (local) supersymmetry transformations of the scalar fields.\par
Therefore the effect of an infinitesimal supersymmetry transformation on the scalar fields can be written in terms of $L$, modulo local $H'$-transformations to the right, as follows:
\begin{equation}
\delta L(\phi)=\delta \phi^s\partial_sL(\phi)=L(\phi)\Sigma(x)\,,\label{delLsig}
\end{equation}
where $\Sigma(x)$ is an element of $\mathfrak{K}$ and is proportional to $(\bar{\epsilon}\cdot {\rm fermions})(x)$, $\epsilon$ being the local supersymmetry parameter. Using Eqs.  (\ref{omegapro}) and (\ref{Vom}) we can rewrite (\ref{delLsig}) as follows:
 \begin{equation}
 \delta\phi^s\,\mathcal{P}_s=\Sigma(x)\,.
 \end{equation}
 By means of the same equations we can also deduce, for symmetric geometries, the variation of the left-invariant 1-form and thus of $\mathcal{P}$:
\begin{equation}
\delta\mathcal{P}=d\Sigma +\mathcal{Q}\,\Sigma-\Sigma\,\mathcal{Q}={\Scr D}\Sigma\,.
\end{equation}
The property that, modulo local rotations in the holonomy group, $\delta \mathcal{P}^{\underline{s}}={\Scr D}\Sigma^{\underline{s}}$, although derived in the homogeneous symmetric case, applies to all scalar manifolds.\footnote{Consider a generic Riemannian manifold with vielbein one-forms $\mathcal{P}^{\underline{s}}$, metric ${\Scr G}_{st}=\mathcal{P}_s{}^{\underline{s}}\mathcal{P}_t{}^{\underline{t}}\,\eta_{\underline{s}\underline{t}}$, and let $\{K_{\underline{s}}\}$ be the basis of the tangent space, dual to $\mathcal{P}^{\underline{s}}$, see Appendix \ref{nacv}. Let $\mathcal{Q}^{\underline{t}}{}_{\underline{s}}$ the metric-compatible, torsionless connection  on the manifold satisfying ${\Scr D}\mathcal{P}^{\underline{t}}\equiv d\mathcal{P}^{\underline{t}}+\mathcal{Q}^{\underline{t}}{}_{\underline{s}}\wedge \mathcal{P}^{\underline{s}}=0$ and $\mathcal{Q}^{\underline{t}}{}_{\underline{r}}\,\eta^{\underline{r}\underline{s}}=-\mathcal{Q}^{\underline{s}}{}_{\underline{r}}\,
\eta^{\underline{r}\underline{t}}$.\par Consider now an infinitesimal diffeomorphism (e.g. a supersymmetry transformation) generated by the vector
$\Sigma=\Sigma^{\underline{s}}\,K_{\underline{s}}$. This transformation induces a variation of $\mathcal{P}^{\underline{t}}$ given by its Lie derivative along $\Sigma$:
\begin{equation}
\delta \mathcal{P}^{\underline{t}}= d\left(\iota_{\Sigma}\mathcal{P}^{\underline{t}}\right)+ \iota_{\Sigma}d\mathcal{P}^{\underline{t}}\,,
\end{equation}
where $\iota_{K_{\underline{s}}}$ denotes the contraction of a form along the vector $K_{\underline{s}}$.
The right hand side can be rewritten as follows:
\begin{align}
\delta \mathcal{P}^{\underline{t}}&= d\left(\Sigma^{\underline{t}}\right)+\mathcal{Q}^{\underline{t}}{}_{\underline{s}}\,\Sigma^{\underline{s}}+
\iota_{\Sigma} {\Scr D}\mathcal{P}^{\underline{t}}-(\iota_{\Sigma}\mathcal{Q})^{\underline{t}}{}_{\underline{s}}\,\mathcal{P}^{\underline{s}}={\Scr D}\Sigma^{\underline{t}}-(\iota_{\Sigma}\mathcal{Q})^{\underline{t}}{}_{\underline{s}}\,\mathcal{P}^{\underline{s}}\,,\label{derdelPsig}
\end{align} where we have used ${\Scr D}\mathcal{P}^{\underline{s}}=0$. The last term on the right hand side of (\ref{derdelPsig}) is a rotation of the vielbein in the holonomy gorup, which drops off in the variation of the metric.
}
Thus the supersymmetry variation of the scalar Lagrangian (\ref{lagrscal}) reads:
\begin{equation}
\delta {\Scr L}_{scal}=k\,{\rm Tr}\left(\mathcal{P}_\mu {\Scr D}^\mu \Sigma \right)\,.
\end{equation}
In Section \ref{dcsl} we shall give for $\Sigma$ an explicit form in a suitable representation.
\paragraph{The solvable parametrization in detail.}
Let us add in this paragraph some more details about the solvable parametrization introduced above.
The precise statement of the correspondence between a homogeneous $\Mscal$ and $G_S$ is the following \cite{alek}. $G_S$ is a solvable group of isometries whose action on  $\Mscal$ is free (i.e. it has no fixed points)  and transitive (homogeneous manifolds admitting such kind of solvable group of isometries are called \emph{normal} \cite{alek}, and all homogeneous scalar manifolds occurring in supergravity are of this kind). If we fix a reference point $\mathcal{O}$ (the \emph{origin}) on $\Mscal$, to any point $P\in \Mscal$ there corresponds an element $g_P$ of $G_S$: $$\forall P\in \Mscal\,\,\longrightarrow\,\,\,\,\,g_P\in G_S\,\,:\,\,\,\,\,\,P=g_P\cdot \mathcal{O}\,.$$
Being the action of $G_S$ on $\Mscal$ free, $g_P$ is \emph{uniquely associated with} $P$.
The origin of $\Mscal$ is then mapped into the unit element ${\bf 1}\in G_S$. This diffeomorphism makes $\Mscal$ and $G_S$ \emph{isometric}, meaning by this that, if we choose the metrics on the tangent spaces of $\Mscal$ and $G_S$ (seen as a metric group manifold) in $\mathcal{O}$ and ${\bf 1}$, respectively, to coincide, they will coincide on every other couple of corresponding points $P$ and $g_P$. This allows to compute metric, affine connection and curvature of $\Mscal$ directly on the group manifold $G_S$, see end of Appendix \ref{nacv}.\par
The algebra $\Solv$, parametrized by the scalar fields of the theory,  has the following general structure:
\begin{equation}
\Solv={\tt C}\oplus {\tt N}\,,\label{SCN}
\end{equation}
where ${\tt C}$ is the Cartan subspace of the coset space $\mathfrak{K}$ and is defined as the maximal set of commuting semisimple generators, and ${\tt N}$ is a nilpotent  subalgebra consisting, in a given matrix representation, of nilpotent generators. When the theory originates from dimensional reduction of a higher dimensional one, the former space is parametrized by the \emph{dilatonic moduli}, typically associated with the size of the internal cycles or the dilaton of ten-dimensional string theory. The latter space ${\tt N}$, on the other hand, is parametrized by \emph{axionic fields}, originating either from components of the internal metric or from higher dimensional antisymmetric tensor fields. Some of these are the so-called Peccei-Quinn fields associated with translational isometries (shift-symmetries) of the manifold. More specifically these are the scalar fields  parametrizing the maximal Abelian ideal of ${\Scr S}$ \cite{Andrianopoli:1996bq}, or, equivalently, the maximal Abelian subalgebra of maximal dimension of $\mathfrak{g}$ \cite{Cremmer:1997ct}. \footnote{The two characterizations are equivalent (see for instance \cite{Papi}, we thank B. Julia for pointing out this reference).}
Being ${\tt N}$ nilpotent, the scalars parametrizing it occur in the sigma-model action polynomially. This is not the case
of the dilatonic scalars parametrizing the  Cartan subalgebra ${\tt C}$.\par
In certain supergravity theories, like the maximally supersymmetric one, the algebra $\mathfrak{g}$ is a \emph{split} Lie algebra. This means that the maximal Abelian subalgebra of diagonalizable elements with real eigenvalues is a Cartan subalgebra $\mathfrak{h}$ of $\mathfrak{g}$ (\emph{non-compact Cartan subalgebra}). Said differently there exists in $\mathfrak{g}$ a Cartan subalgebra consisting only of non-compact generators. In this case ${\Scr S}$ is a \emph{Borel subalgebra} of $\mathfrak{g}$, which is a maximal solvable subalgebra of $\mathfrak{g}$ containing $\mathfrak{h}$. This Borel subalgebra has the general structure (\ref{SCN}), where ${\tt C}$ coincides with the non-compact Cartan subalgebra $\mathfrak{h}$ of $\mathfrak{g}$ and ${\tt N}$ is spanned by the shift-generators corresponding to the positive roots associated with $\mathfrak{h}$. In the general case of a non-split Lie algebra, the maximal space ${\tt C}$ of commuting non-compact generators is only \emph{strictly} contained in a Cartan subalgebra $\mathfrak{h}$ of $\mathfrak{g}$. In other words any Cartan subalgebra of $\mathfrak{g}$ contains a maximal subspace of non-compact generators whose dimension $r$ is strictly less than the rank of $\mathfrak{g}$. The space ${\tt C}$ in ${\Scr S}$ coincides with one of these maximal non-compact subspaces of maximal dimension, its dimension $r$ defining the  rank of the homogeneous manifold \cite{Gilmore}. For instance if $G={\rm SU}(1,n)$, the generators of its Cartan subalgebras can contain at most one non-compact matrix, so that $r=1$, while the rank of $\mathfrak{su}(1,n)$ is $n$. For $n>1$ we have $r<n$ and the manifold ${\rm SU}(1,n)/{\rm U}(n)$ is non-split. In this case ${\tt C}$ is one dimensional.
As prescribed by the standard Iwasawa decomposition \cite{Helgason}, ${\tt N}$ is spanned by the combinations in $\mathfrak{g}$ of the shift generators corresponding to positive roots $\alpha$ of  $\mathfrak{g}^c$, relative to $\mathfrak{h}^c$, which are non-vanishing on ${\tt C}$: $\alpha({\tt C})\neq 0$.\footnote{By $\mathfrak{g}^c$ and $\mathfrak{h}^c$ we denote the complexifications of $\mathfrak{g}$ and $\mathfrak{h}$, respectively: $\mathfrak{g}^c\equiv \mathfrak{g}+i\,\mathfrak{g},\,\mathfrak{h}^c\equiv \mathfrak{h}+i\,\mathfrak{h}$.}
\subsubsection{Vector Sector}
We can associate with the electric field strengths $F_{\mu\nu}^{\Lambda}$ their magnetic duals $ G_{\Lambda\,\mu\nu}$ defined as:\footnote{When computing the derivative of the Lagrangian with respect to $F^\Lambda_{\mu\nu}$ we treat $F^\Lambda_{\mu\nu}$ and $F^\Lambda_{\nu\mu}$ as independent so that $\frac{\partial F^\Lambda_{\mu\nu}}{\partial F^\Sigma_{\rho\sigma}}=\delta^\Lambda_\Sigma\,\delta_\mu^\rho \delta_\nu^\sigma$.}
\begin{align}
 G_{\Lambda\,\mu\nu}&\equiv -\epsilon_{\mu\nu\rho\sigma}
\frac{\partial \L}{\partial
F^\Lambda_{\rho\sigma}}=\R_{\Lambda\Sigma}\,F^\Sigma_{\mu\nu}-\I_{\Lambda\Sigma}\,{}^*F^\Sigma_{\mu\nu}\,,
\label{GF}
\end{align}
where we have omitted fermion currents in the expression of $ G_\Lambda$ since we are only focussing for the time being on the bosonic sector of the theory. We shall include them in Sect. \ref{dcsl}.
The \emph{Hodge-dual} of ${}^*F^\Sigma_{\mu\nu}$ of $F^\Sigma_{\mu\nu}$ is defined as:
$${}^*F^\Sigma_{\mu\nu}\equiv \frac{e}{2}\,\epsilon_{\mu\nu\rho\sigma}\,F^{\Sigma\,\rho\sigma}\,,$$
 and satisfies the property: ${}^{**}=-1$.
In ordinary Maxwell theory (no scalar fields), $\I_{\Lambda\Sigma}=-\delta_{\Lambda\Sigma}$ and $\R_{\Lambda\Sigma}=0$, so that $ G_{\Lambda\,\mu\nu}$ coincides with the Hodge-dual of $F^\Lambda_{\mu\nu}$: $ G_{\Lambda}={}^* F^{\Lambda}$.\par
In terms of $F^\Lambda$ and $ G_{\Lambda}$ the Maxwell equations read
\begin{equation}
\nabla^{\mu}({}^*F^\Lambda_{\mu\nu}) = 0\,;
\qquad \nabla^{\mu }({{}^*G}_{\Lambda\,\mu\nu}) = 0\,,
\label{biafieq}
\end{equation}
In order to set the stage for the discussion of global symmetries, it is useful to rewrite the scalar and vector field equations in a different form.
Using (\ref{GF}) and the property that ${}^*{}^*
F^\Lambda=-F^\Lambda$, we can express ${}^* F^\Lambda$ and ${}^*
 G_\Lambda$ as linear functions of $F^\Lambda$ and $ G_\Lambda$:
\begin{align}
{}^* F^\Lambda&=
\I^{-1\,\Lambda\Sigma}\,(\R_{\Sigma\Gamma}\,F^\Gamma- G_\Sigma)\;;\nonumber\\
{}^* G_\Lambda&=
(\R\I^{-1}\R+\I)_{\Lambda\Sigma}\,F^\Sigma-(\R\I^{-1})_\Lambda{}^\Sigma\, G_\Sigma\,,\label{GF20}
\end{align}
where, for the sake of simplicity, we have omitted the space-time
indices. It is useful to arrange $F^\Lambda$ and $ G_\Lambda$ in a
single $2n_v$-dimensional vector $\mathcal{G}\equiv (\mathcal{G}^M)$
of two-forms:
\begin{equation}
\mathcal{G}= \left(\frac{1}{2}\,\mathcal{G}^M_{\mu\nu}\,dx^\mu\wedge dx^\nu\right) \equiv \left(\begin{matrix}F^\Lambda_{\mu\nu}\cr
 G_{\Lambda\mu\nu}\end{matrix}\right)\,\frac{dx^\mu\wedge dx^\nu}{2}\,,\label{bbF}
\end{equation} in terms of which the Maxwell equations (\ref{biafieq}) read:
\begin{equation}
d\mathcal{G}=0\,,\label{Max}
\end{equation}
and eqs.\ (\ref{GF20}) are easily rewritten in the
following compact form:
\begin{eqnarray}
{}^*\mathcal{G}=-\mathbb{C}\mathcal{M}(\phi^s)\,\mathcal{G}\,,\label{FCMF}
\end{eqnarray}
where
\begin{equation}
\mathbb{C}=(\mathbb{C}^{MN})\equiv\left(\begin{matrix} \Zero & \Id
\cr -\Id & \Zero \end{matrix}\right)\,,\label{C}
\end{equation}
$\Id$, $\Zero$ being the $n_v\times n_v$ identity and zero-matrices, respectively, and
\begin{equation}
\mathcal{M}(\phi)= (\mathcal{M}(\phi)_{MN})\equiv
\left(\begin{matrix}(\R\I^{-1}\R+\I)_{\Lambda\Sigma} &
-(\R\I^{-1})_\Lambda{}^\Gamma\cr -(\I^{-1}\R)^\Delta{}_\Sigma & \I^{-1\,
\Delta \Gamma}\end{matrix}\right)\,,\label{M}
\end{equation}
is a symmetric, negative-definite matrix, function of the scalar
fields. The reader can easily verify that this matrix is also symplectic, namely that:
\begin{equation}
\mathcal{M}(\phi)\mathbb{C}\mathcal{M}(\phi)=\mathbb{C}\,.\label{sympM}
\end{equation}
This matrix contains $\I_{\Lambda\Sigma}$ and $\R_{\Lambda\Sigma}$ as components, and therefore defines the non-minimal couplings of the scalars to the vector fields.\par
Equation (\ref{FCMF}) expressing, in a symplectic covariant form, the dependence of $ G_{\Lambda\,\mu\nu}$ on $F^\Lambda_{\mu\nu},\,{}^*F^\Lambda_{\mu\nu}$ and the scalar fields is called \emph{twisted self-duality} condition \cite{Cremmer:1978ds}. Its complete expression, which includes the fermion bilinears coming from the Pauli terms of the Lagrangian, will be given in Sect. \ref{dcsl}, see Eq. (\ref{FCMF2}).
After some algebra, we can also rewrite eqs.\ (\ref{scaleqs}) in a compact form as follows
\begin{align}
{\Scr D}_\mu (\partial^\mu\phi^s)&=
\frac{1}{8}\,\Gm^{st}\,\mathcal{G}^T_{\mu\nu}\partial_t\mathcal{M}(\phi)\,\mathcal{G}^{\mu\nu}+\dots\,,\label{scaleqs2}
\end{align}

\subsubsection{Coupling to Gravity}
We can now compute the Einstein equations:
\begin{equation}
R_{\mu\nu}-\frac{1}{2}\,g_{\mu\nu}\,R=T^{(S)}_{\mu\nu}+T^{(V)}_{\mu\nu}+T^{(F)}_{\mu\nu}\,,\label{EEQ1}
\end{equation}
where the three terms on the right hand side are the energy-momentum tensors of the scalars, vectors and fermionic fields, respectively. The first two can be cast in the following general form
\begin{align}
T^{(S)}_{\mu\nu}&= \Gm_{rs}(\phi)\,\partial_\mu \phi^r\partial_\nu
\phi^s-\frac{1}{2}\,g_{\mu\nu}\,\Gm_{rs}(\phi)\,\partial_\rho
\phi^r\partial^\rho \phi^s\,,\\
T^{(V)}_{\mu\nu}&={F}^T_{\mu\rho}\,\I\,F_{\nu}{}^\rho-\frac{1}{4}\,g_{\mu\nu}\,(F^T_{\rho\sigma} \I F^{\rho\sigma})\,,\label{tv}
\end{align}
where in the last equation the vector indices $\Lambda,\Sigma$ have been suppressed for the sake of notational simplicity.
It is convenient for our next discussion, to rewrite, after some algebra, the right hand side of (\ref{tv}) as follows
\begin{equation}
T^{(V)}_{\mu\nu}=\frac{1}{2}\,\mathcal{G}^T_{\mu\rho}\,\mathcal{M}(\phi)\,\mathcal{G}_{\nu}{}^\rho\,,
\end{equation}
so that Eq.\ (\ref{EEQ1}) can be finally recast in the following form:
\begin{equation}
R_{\mu\nu}=\Gm_{rs}(\phi)\,\partial_\mu \phi^r\partial_\nu
\phi^s+\frac{1}{2}\,\mathcal{G}^T_{\mu\rho}\,\mathcal{M}(\phi)\,\mathcal{G}_{\nu}{}^\rho+\dots\,,\label{EEQ2}
\end{equation}
where the ellipses refer to fermionic terms.\par
The scalars enter the kinetic terms of the vector fields through the matrices $\I(\phi)$ and $\R(\phi)$. As a consequence of this, a symmetry transformation of the scalar part of the Lagrangian will not in general leave the vector field part invariant.

\subsubsection{Global Symmetry Group}\label{gsg}
In extended supergravity models ($\N>1$) the (identity sector of the) global symmetry group $G$ of the scalar action can be promoted to global invariance \cite{Gaillard:1981rj} of the field equations and the Bianchi identities, provided its (non-linear) action on the scalar fields is associated with a linear transformation on the vector field strengths $F^\Lambda_{\mu\nu}$ and their magnetic duals $ G_{\Lambda\,\mu\nu}$:
\begin{align}
{\bf g}\in G\,:\;
\begin{cases}
\,\,\,\,\,\,\phi^r &\rightarrow \quad {\bf g}\star\phi^r\;\; \qquad\qquad\qquad\qquad\qquad\qquad\qquad\text{(non--linear)},\\[\jot]
\left(\begin{matrix}
F^\Lambda \cr  G_\Lambda
\end{matrix}\right)
&\rightarrow\quad\Rs_v[{\bf g}]\cdot
\left(\begin{matrix}
F^\Lambda\cr  G_\Lambda
\end{matrix}\right)=
\left(\begin{matrix}
A[{\bf g}]^\Lambda{}_\Sigma & B[{\bf g}]^{\Lambda\Sigma}\cr
C[{\bf g}]_{\Lambda\Sigma} & D[{\bf g}]_\Lambda{}^\Sigma
\end{matrix}\right)
\,\left(\begin{matrix}
F^\Sigma\cr  G_\Sigma
\end{matrix}\right)
\,\quad\text{(linear)}.
\end{cases}\label{dual}\nne
\end{align}
The transformations (\ref{dual}) would clearly be a symmetry of the scalar action and of the Maxwell equations ($d\mathcal{G}=0$) if $F^\Lambda$ and $ G_\Lambda$ were independent, since the latter are invariant with respect to any linear transformation on $\mathcal{G}^M$. The definition $ G_\Lambda$ in (\ref{GF}) as a function of $F^\Lambda,\,{}^*F^\Lambda$ and the scalar fields, which is equivalently expressed by the twisted self-duality condition (\ref{FCMF}), however poses constraints on the $2n_v\times 2n_v$ matrix $\Rs_v[{\bf g}]=(\Rs_v[{\bf g}]^M{}_N)$. In order for (\ref{dual}) to be an invariance of the vector equations of motion (\ref{Max}) and (\ref{FCMF}) the following conditions have to be met:
\begin{itemize}
\item[i)]{for each ${\bf g}\in G$ (more precisely in the identity sector of $G$), the matrix $\Rs_v[{\bf g}]$ should be \emph{symplectic}, namely
\begin{equation}
\Rs_v[{\bf g}]^{T}\mathbb{C}\,\Rs_v[{\bf g}]=\mathbb{C}\,;\label{SSym}
\end{equation}
}
\item[ii)]{the symplectic, scalar dependent, matrix $\mathcal{M}(\phi)$ should transform as follows:
\begin{equation}
\mathcal{M}({\bf g}\star \phi)=\Rs_v[{\bf g}]^{-T}\mathcal{M}(\phi)\,\Rs_v[{\bf g}]^{-1}\,,\label{traM}
\end{equation}
where we have used the short-hand notation $\Rs_v[{\bf g}]^{-T}\equiv (\Rs_v[{\bf g}]^{-1})^T$.
}
\end{itemize}
The reader is referred to Appendix \ref{spapp} for the main properties of the symplectic group and its Lie algebra. It is indeed straightforward to verify that conditions i) and ii) are sufficient to guarantee invariance of (\ref{FCMF}) under (\ref{dual}).
The symplectic transformation $\Rs_v[{\bf g}]$, associated with each element ${\bf g}$ of $G$, mixes electric and magnetic field strengths, and therefore acts as a generalized electric--magnetic duality and defines a \emph{symplectic representation} $\Rs_v$ of $G$:
\begin{equation}
\forall {\bf g}\in G\,\,\,\stackrel{\Rs_v}{\longrightarrow}\,\,\,\,\,\Rs_v[{\bf g}]\in {\rm Sp}(2n_v,\,\mathbb{R})\,.
\end{equation}
The field strengths and their magnetic duals thus transform, under the duality action (\ref{dual}) of $G$ in the $2n_v$-dimensional symplectic representation ${\Scr R}_v$.\par
We denote by $\Rs_{v*}=\Rs_v^{-T}$ the representation dual to $\Rs_v$, acting on covariant symplectic vectors, so that, for any ${\bf g}\in G$:
\begin{align}
\Rs_{v*}[{\bf g}]&=(\Rs_{v*}[{\bf g}]_M{}^N)=\Rs_v[{\bf g}]^{-T}=-\mathbb{C}\Rs_v[{\bf g}]\mathbb{C}\,\,\,\Rightarrow \nonumber\\&\Rightarrow\,\,\, \Rs_{v*}[{\bf g}]_M{}^N=\mathbb{C}_{MP}\,\Rs_v[{\bf g}]^P{}_Q\,\mathbb{C}^{NQ}\,,\end{align}
where we have used the property that $\Rs_v$ is a symplectic representation%
\footnote{
The symplectic indices {\small $M,\,N,\dots$} are raised (and lowered) with the symplectic matrix $\mathbb{C}^{MN}$ ($\mathbb{C}_{MN}$) using north-west south-east conventions: $X^{M}=\mathbb{C}^{MN}\,X_{N}$ (and $X_M=\mathbb{C}_{NM}\,X^{N}$)
}.\par
From (\ref{SSym}) and (\ref{traM}), it is straightforward to verify the manifest $G$-invariance of the scalar field equations and the Einstein equations written in the forms (\ref{scaleqs2}) and (\ref{EEQ2}).\par
Conditions i) and ii) are verified in extended supergravities as a consequence of supersymmetry. In these theories indeed supersymmetry is large enough as to connect certain scalar fields to vector fields and, as a consequence of this, symmetry transformations on the former imply transformations on the latter
(more precisely transformations on the vector field strengths $F^\Lambda$ and their duals $ G_\Lambda$). The existence of a symplectic representation $\Rs_v$ of $G$, together with the definition of the matrix $\mathcal{M}$ and its transformation property (\ref{traM}), are built-in in the mathematical structure of the scalar manifold. More precisely they follow from the definition on $\Mscal$ of a \emph{flat symplectic structure}. Supersymmetry totally fixes $\mathcal{M}(\phi)$ and thus the coupling of the scalar fields to the vectors, aside from a freedom in the choice of the basis of the symplectic representation (\emph{symplectic frame}). Different choices of the symplectic frame amount to a change in the definition of $\mathcal{M}(\phi)$ by a constant symplectic transformation $E$:
\begin{equation}
\mathcal{M}(\phi)\rightarrow \mathcal{M}'(\phi)=E\mathcal{M}(\phi)E^T\,.\label{MEtra}
\end{equation}
Clearly if $E\in {\Scr R}_{v*}[G]\subset {\rm Sp}(2n_v,\mathbb{R})$, its effect on $\mathcal{M}(\phi)$ can be offset be a redefinition of the scalar fields, by virtue of Eq.\ (\ref{traM}). On the other hand if $E$ were a block-diagonal matrix, namely an element of ${\rm GL}(n_v,\mathbb{R})\subset {\rm Sp}(2n_v,\mathbb{R})$, it could be reabsorbed by a local redefinition of the field strengths. Inequivalent symplectic frames are then connected by symplectic matrices $E$ defined modulo redefinitions of the scalar and vector fields, namely by matrices in the coset \cite{deWit:2002vt}:
\begin{equation}
E\,\in \,{\rm GL}(n_v,\mathbb{R})\backslash {\rm Sp}(2n_v,\mathbb{R})/ {\Scr R}_{v*}[G]\,,\label{generalE}
\end{equation}
where the quotient is defined with respect to the left-action of ${\rm GL}(n_v,\mathbb{R})$ (local vector redefinitions) and to the right-action of $ {\Scr R}_{v*}[G]$ (isometry action on the scalar fields).\par
A change in the symplectic frame amounts to choosing a different embedding $\Rs_v$ of $G$ inside ${\rm Sp}(2n_v,\,\mathbb{R})$, which in general is not unique.
This affects the form of the action and in particular the coupling of the scalar fields to the vectors. However, at the ungauged level,
it only amounts to a redefinition of the vector field strengths and their duals which has no physical implication.
In the presence of a gauging, namely if vectors are minimally coupled to the other fields, the symplectic frame becomes physically relevant and may lead to different vacuum-structures of the scalar potential.\par
We
emphasize here that the existence of this symplectic structure on the
scalar manifold is a general feature of all extended supergravites in four dimensions,
including those $\N=2$ models in which the scalar manifold is not
even homogeneous. In the $\N=2$ case, as mentioned above, only the scalar fields belonging to the vector multiplets are non-minimally
coupled to the vector fields, namely enter the matrices
$\I(\phi),\,\R(\phi)$, and they span a \emph{special K\"ahler}
manifold. On this manifold a flat symplectic bundle is defined, see Sect. \ref{SK},
which fixes the scalar dependence of the matrices $\I(\phi),\,\R(\phi)$, aside from an initial choice of the symplectic frame, and the matrix
$ \mathcal{M}(\phi)$ defined in (\ref{M}) satisfies the property
(\ref{traM}).\par If the scalar manifold is homogeneous symmetric, we can consider at any point the coset representative $L(\phi)\in G$ in the symplectic, $2n_v$-dimensional representation $\Rs_v $:
\begin{equation}
L(\phi)\,\,\,\stackrel{\Rs_v}{\longrightarrow}\,\,\,\,\,\Rs_v[L(\phi)]\in {\rm Sp}(2n_v,\,\mathbb{R})\,.
\end{equation}
The maximal compact subgroup $H$ of $G$ is mapped through $\Rs_v$ into the maximal compact subgroup $ {\rm U}(n_v)$ of ${\rm Sp}(2n_v,\mathbb{R})$. In general, however, the representation $\Rs_v[H]$ of the isotropy group $H$ may not be orthogonal\footnote{That is unitary, being the representation real.}, that is $\Rs_v[H]\nsubseteq  {\rm SO}(2n_v)$.
In this case we can always change the basis of the representation%
\footnote{
We label the new basis by underlined indices
}
by means of a matrix $\mathcal{S}$%
\begin{equation}
\mathcal{S}=(\mathcal{S}^N{}_{\underline{M}})
\,\in {\rm Sp}(2n_v,\,\mathbb{R})/{\rm U}(n)
\end{equation}
such that, in the equivalent representation $\underline{{\Scr R}}_v\equiv \mathcal{S}^{-1}\Rs_v\,\mathcal{S}$,
\begin{equation}
\underline{{\Scr R}}_v[H]\equiv \mathcal{S}^{-1}\Rs_v[H]\,\mathcal{S}\subset {\rm SO}(2n_v)
\quad\Leftrightarrow\quad
\underline{{\Scr R}}_v[h]^T\underline{{\Scr R}}_v[h]=\Id\;,\quad
\forall h\in H\,.\label{hort}
\end{equation}
Note that $\underline{{\Scr R}}_v$, as a symplectic representation of $H$, \emph{is reducible}.

For any point $\phi$ on the scalar manifold it is useful to define a \emph{hybrid coset-representative matrix} $\LL(\phi)=(\LL(\phi)^M{}_{\underline{N}})$ as follows:
\begin{equation}
\LL(\phi)\equiv \Rs_v[L(\phi)]\mathcal{S}
\quad\Leftrightarrow\quad
\LL(\phi)^M{}_{\underline{N}}\equiv \Rs_v[L(\phi)]^M{}_N\mathcal{S}^N{}_{\underline{N}}\,.\label{hybrid}
\end{equation}
We also define the matrix
\begin{equation}
\LL(\phi)_M{}^{\underline{N}}~\equiv~ \mathbb{C}_{MP}\,\mathbb{C}^{\underline{NQ}}\;\LL(\phi)^P{}_{\underline{Q}}\;.\label{otherL}
\end{equation}
Notice that, as a consequence of the fact that the two indices of $\LL$ refer to two different symplectic bases, $\LL$ itself is not a matrix representation of the coset representative $L$.
From (\ref{gLh}), the property of ${\Scr R}_v$ of being a representation and the definition (\ref{hybrid}) we have:
\begin{equation}
\forall {\bf g}\in G \;:\quad {\Scr R}_v[{\bf g}]\,\LL(\phi)=\LL({\bf g}\star\phi)\,\underline{{\Scr R}}_v[h]\,,\label{gLh2}
\end{equation}
where $h\equiv h(\phi,{\bf g})$ is the compensating transformation. The hybrid index structure of $\LL$ poses no consistency problem since, by (\ref{gLh2}), the coset representative is acted on to the left and to the right by two different groups: $G$ and $H$, respectively. Therefore, in our notations, underlined symplectic indices {\footnotesize $\underline{M},\,\underline{N},\dots$} are acted on by $H$ while non-underlined ones by $G$.\par
The
$ \mathcal{M}(\phi)$ is then expressed in terms of the coset representative as follows:
\begin{equation}
\mathcal{M}(\phi)_{MN}=\mathbb{C}_{MP}\mathbb{L}(\phi)^P{}_{\underline{L}}\mathbb{L}(\phi)^R{}_{\underline{L}}\,\mathbb{C}_{RN}
\;\;\Leftrightarrow\;\;
\mathcal{M}(\phi)=\mathbb{C}\mathbb{L}(\phi)\,\mathbb{L}(\phi)^T\,\mathbb{C}\,,\label{Mcos}
\end{equation}
where summation over the index {\footnotesize $\underline{L}$} is understood. The reader can easily verify that the definition given above of the matrix $\mathcal{M}(\phi)$, as a function of the point $\phi$ on ${\Scr M}_{{\rm scal}}$, is consistent. Indeed it is $H$-invariant, and thus does not depend on the choice of the coset representative, and transforms according to (\ref{traM}):
\begin{align}
\forall {\bf g}\in G \;:\quad \mathcal{M}({\bf g}\star \phi)&=\mathbb{C}\LL({\bf g}\star\phi)\,\LL({\bf g}\star\phi)^T\mathbb{C}=\nonumber\\&=
\mathbb{C}\Rs_v[{\bf g}]\,\LL(\phi)(\underline{\Rs}_v[h]^{-1}\,\underline{\Rs}_v[h]^{-T})\LL(\phi)^T\Rs_v[{\bf g}]^T\mathbb{C}=\nonumber\\&
=\Rs_v[{\bf g}]^{-T}\mathbb{C}\LL(\phi)\,\LL(\phi)^T
\mathbb{C}\Rs_v[{\bf g}]^{-1}=\nonumber\\
&=\Rs_v[{\bf g}]^{-T}\mathcal{M}(\phi)\Rs_v[{\bf g}]^{-1}\,,
\end{align}
where we have used Eq.\ (\ref{gLh2}), the orthogonality property (\ref{hort}) of $\underline{\Rs}_v[h]$ and the symplectic property of $\Rs_v[{\bf g}]$.
From the definition (\ref{Mcos}) of $\mathcal{M}$ in terms of the coset representative, it follows that for symmetric scalar manifolds the scalar Lagrangian (\ref{lagrscal}) can also be written in the equivalent form:
\begin{equation}
\Lscal=\frac{e}{2}\, \Gm_{st}(\phi)\partial_\mu\phi^s\,\partial^\mu\phi^t
=\frac{e}{8}\,k\,\Tr\big(\mathcal{M}^{-1}\partial_\mu\mathcal{M}\,\mathcal{M}^{-1}\partial^\mu\mathcal{M}\big)\,,\label{lagrscalM}
\end{equation}
where $k$ depends on the representation $\Rs_v$ of $G$.
\par
The transformation properties of the matrices $\I_{\Lambda\Sigma}$ and $\R_{\Lambda\Sigma}$ under $G$ can be inferred from (\ref{traM}) and can be conveniently described by defining the complex symmetric matrix \begin{equation}
{\Scr N}_{\Lambda\Sigma}\equiv \R_{\Lambda\Sigma}+i\,\I_{\Lambda\Sigma}\,.\label{Nmatrix}
\end{equation}
Under the action of a generic element ${\bf g}\in G$, \,${\Scr N}$ transforms as follows:
\begin{equation}
{\Scr N}({\bf g}\star\phi)=(C[{\bf g}]+D[{\bf g}]\,{\Scr N}(\phi))(A[{\bf g}]+B[{\bf g}]\,{\Scr N}(\phi))^{-1}\,,\label{Ntra}
\end{equation}
where $A[{\bf g}],\,B[{\bf g}],\,C[{\bf g}]\,,D[{\bf g}]$ are the $n_v\times n_v$ blocks of the matrix $\Rs_v[{\bf g}]$ defined in (\ref{dual}).\par

\paragraph{Parity.} We have specified above that only the elements of $G$ which belong to the identity sector, namely which are continuously connected to the identity, are associated with symplectic transformations. There may exist isometries ${\bf g}\in G$ which do not belong to the identity sector and are associated with \emph{anti-symplectic} matrices ${\bf A}[{\bf g}]$:
\begin{equation}
\mathcal{M}({\bf g}\star  \phi)={\bf A}[{\bf g}]^{-T}\,\mathcal{M}(\phi)\,{\bf A}[{\bf g}]^{-1}\;;
\quad\;
{\bf A}[{\bf g}]^T\mathbb{C}{\bf A}[{\bf g}]=-\mathbb{C}\,.
\end{equation}
Anti-symplectic matrices do not close a group but can be expressed as the product of a symplectic matrix ${\bf S}$ times a fixed anti-symplectic one ${\bf P}$, that is ${\bf A}=\S\,{\bf P}$. In a suitable symplectic frame, the matrix ${\bf P}$ can be written in the following form:
\begin{equation}
{\bf P}=\left(\begin{matrix}
\Id & \Zero \cr \Zero & -\Id \end{matrix}\right)\,.\label{Pmatrix}
\end{equation}
Due to their being implemented by anti-symplectic duality transformations (\ref{dual}), these isometries leave Eq.\ (\ref{FCMF}) invariant up to a sign which can be offset by a \emph{parity transformation}, since under parity one has $\,*\,\rightarrow\,-*\,$\,.\, Indeed one can show that these transformations are a symmetry of the theory provided they are combined with parity.
Notice that this poses no problem with the generalized theta-term since, as parity reverses the sign of $\epsilon^{\mu\nu\rho\sigma}F^\Lambda_{\mu\nu} F^\Sigma_{\rho\sigma}$, under ${\bf P}$ we have:
\begin{equation}
\I_{\Lambda\Sigma}\rightarrow\I_{\Lambda\Sigma}\;;\qquad
\R_{\Lambda\Sigma}\rightarrow -\R_{\Lambda\Sigma}\,,
\end{equation}
see equation (\ref{Ntra}), so that the corresponding term $\epsilon^{\mu\nu\rho\sigma}F^\Lambda_{\mu\nu} F^\Sigma_{\rho\sigma}\R_{\Lambda\Sigma}$ in the Lagrangian is invariant.
The global symmetries of the theory are therefore described by the group
\begin{equation}
G=G_0\times \mathbb{Z}_2=\{G_0,\,G_0\cdot p\}\,,
\end{equation}
where $G_0$ is the \emph{proper duality} group defined by the identity sector of $G$ and $p$ is the element of $G$ which corresponds, in a suitable symplectic frame, to the anti-symplectic matrix ${\bf P}$\,:\; ${\bf P}={\bf A}[p]$.

\paragraph{Example 2.}
Let us consider the simple example of the lower-half complex plane discussed in Example 1.
\begin{equation}
G/H={\rm SL}(2,\mathbb{R})/{\rm SO}(2)\,.
\end{equation}
As a symplectic representation of $G$ we take the fundamental one. Moreover we use the solvable parametrization so that:
\begin{equation}
(\mathbb{L}(\phi)^P{}_{\underline{L}})={\Scr R}_v[L^{(S)}(\varphi,\,\chi)]=L^{(S)}(\varphi,\,\chi)\,,
\end{equation}
where the coset representative $L^{(S)}(\varphi,\,\chi)$ was defined in (\ref{LLSol}).
The matrix $\mathcal{M}(\phi)_{MN}$ has the form:
\begin{align}
\mathcal{M}(z,\,\bar{z})_{MN}=\mathbb{C}_{MP}\,\mathbb{L}(\phi)^P{}_{\underline{L}}\,\mathbb{L}(\phi)^R{}_{\underline{L}}\,
\mathbb{C}_{RN}=
\frac{1}{z_2}\left(
\begin{array}{cc}
 1 & -{z_1} \\
 -{z_1} & |z|^2
\end{array}
\right)\,.
\end{align}
The generic isometry which is continuously connected to the identity is a holomorphic transformation
of the form
\begin{equation}
z\rightarrow z'=\frac{a z +b}{ c z +d}\,,\qquad ad-bc=1\,,
\end{equation}
corresponding to the ${\rm SL}(2,\mathbb{R})$ transformation $\S=({\bf S}^M{}_N)=\left(\begin{matrix}a & b\cr c & d\end{matrix}\right)$ with ${\rm det}(\S)=1$.
The reader can easily verify that:
\begin{equation}
\mathcal{M}(z',\,\bar{z}')=\S^{-T}\mathcal{M}(z,\,\bar{z})\S^{-1}\,.
\end{equation}
We also have the following isometry:
\begin{equation}
z\rightarrow -\bar{z}\,,\label{pisom}
\end{equation}
which is not in the identity sector of the isometry group, and corresponds to the anti-symplectic transformation ${\bf P}={\rm diag}(1,-1)$ in that:
 \begin{equation}
\mathcal{M}(-\bar{z},\,-z)={\bf P}^{-T}\mathcal{M}(z,\,\bar{z}){\bf P}^{-1}\,.
\end{equation}
This corresponds to a parity transformation whose effect is to change the sign of the pseudo-scalar $\chi$ while leaving the scalar $\varphi$ inert:
\begin{equation}
\mbox{parity}:\;\;\chi\rightarrow -\chi\,\,,\,\,\,\,\varphi\rightarrow \varphi\,.
\end{equation}
Notice that the correspondence between the linear transformation ${\bf P}$ and the isometry (\ref{pisom}) exists since ${\bf P}$ is an \emph{outer-automorphism} of the isometry algebra $\mathfrak{g}=\mathfrak{sl}(2,\mathbb{R})$, namely:
\begin{equation}
{\bf P}^{-1}\mathfrak{sl}(2,\mathbb{R}){\bf P}\=\mathfrak{sl}(2,\mathbb{R})\,,
\end{equation}
while ${\bf P}$ is \emph{not} in ${\rm SL}(2,\mathbb{R})$ and the above transformation cannot be offset by any conjugation by ${\rm SL}(2,\mathbb{R})$ elements. Analogous outer-automorphisms implementing parity can be found in other extended supergravities, including the maximal one in which $G={\rm E}_{7(7)}\times \mathbb{Z}_2$\, \cite{Ferrara:2013zga}. We shall come back to this discrete symmetry at the end of Sect. \ref{dcsl} where we illustrate its action on the fermion fields.
 The parity matrix ${\bf P}$ can have a more general form than (\ref{Pmatrix}), allowing an intrinsic parity $\eta_p^\Lambda$ ($(\eta_p^\Lambda)^2=1$) for each vector field. A detailed discussion of this issue in extended supergravities will be given in \cite{AT}.

\paragraph{Solitonic solutions, electric-magnetic charges and duality.}
Ungauged supergravities only contain fields which are neutral with respect to the ${\rm U}(1)^{n_v}$ gauge-symmetry of the vector fields.
These theories however feature \emph{solitonic solutions}, namely configurations of neutral fields which carry ${\rm U}(1)^{n_v}$ electric-magnetic charges \cite{Witten:1978mh,Gibbons:1982fy}. These solutions are typically black holes in four dimensions or black branes in higher and have been extensively studied in the literature.
On a charged dyonic solution of this kind, we define the electric and magnetic charges as the integrals%
\footnote{
The electric and magnetic charges $(e,m)$ are expressed in the rationalized-Heaviside-Lorentz (RHL) system of units
}:
\begin{align}
e_\Lambda\equiv\int_{S^2}  G_{\Lambda}=\frac{1}{2}\,\int_{S^2}  G_{\Lambda\,\mu\nu}\,dx^\mu\wedge dx^\nu \,\,,\nonumber\\
m^\Lambda\equiv\int_{S^2} F^{\Lambda}=\frac{1}{2}\,\int_{S^2} F^{\Lambda}{}_{\mu\nu}\,dx^\mu\wedge dx^\nu \,,
\end{align}
where $S^2$ is a spatial two-sphere. They define a symplectic vector $\Gamma^M$:
\begin{align}
\Gamma=(\Gamma^M)=\left(\begin{matrix}m^\Lambda\cr e_\Lambda\end{matrix}\right)=\int_{S^2} \mathcal{G}^M\,.
\end{align}
These are the \emph{quantized charges}, namely they satisfy the Dirac-Schwinger-Zwanziger quantization condition for dyonic particles \cite{Dirac:1931kp,Schwinger:1966nj,Zwanziger:1968rs}:
\begin{equation}
\Gamma_2^T\mathbb{C}\Gamma_1=m_2^{\Lambda}\,e_{1\Lambda}-m_1^{\Lambda}\,e_{2\Lambda}= 2\pi\,\hbar\,c\,n\,\,\,;\,\,\,\,n\in \mathbb{Z}\,.\label{DZS}
\end{equation}
At the quantum level, the dyonic charges therefore belong to a symplectic lattice and this breaks the duality group $G$ to a suitable discrete subgroup $G(\mathbb{Z})$ which leaves this symplectic lattice invariant:
\begin{equation}
G(\mathbb{Z})\equiv G\cap {\rm Sp}(2n_v,\mathbb{Z})\,.
\end{equation}
This discrete symmetry group (or a suitable extension thereof) surviving quantum corrections was conjectured in \cite{Hull:1994ys} to encode all known string/M-theory dualities.

\subsubsection{Symplectic Frames and Lagrangians}\label{sframes}
As pointed out earlier, the duality action $\Rs_v[G]$ of $G$ depends on which elements, in a basis of the  representation space, are chosen to be the $n_v$ electric vector fields (appearing in the Lagrangian) and which their magnetic duals, namely on the choice of the \emph{symplectic frame} which determines the embedding of the group $G$ inside $\Sp(2n_v,\,\mathbb{R})$. Different choices of the symplectic frame may yield inequivalent Lagrangians (that is Lagrangians that are not related by local field redefinitions) with different global symmetries.
A generic duality transformation of the form (\ref{dual}) can only be a global symmetry of the field equations and Bianchi identities (on-shell global symmetry)
but not of the Lagrangian (off-shell global symmetry). Indeed if $B[{\bf g}]^{\Lambda\Sigma}\neq  0$ the Bianchi identity of the transformed electric field strength $F^{\prime\Lambda}$ is not implied by the Bianchi identities $dF^{\Lambda}=0$ of the original ones since
\begin{equation}
dF^{\prime\Lambda}=A[{\bf g}]^{\Lambda}{}_\Sigma\,dF^\Sigma+B[{\bf g}]^{\Lambda\Sigma}\,d G_\Sigma=B[{\bf g}]^{\Lambda\Sigma}\,d G_\Sigma\,,
\end{equation}
 but also requires the field equation $d G_\Sigma=0$ \cite{Tanii:1998px}.
The global symmetry group of the Lagrangian
\footnote{
Here we only consider \emph{local} transformations on the fields
}
is therefore defined as the subgroup $G_{el}\subset G$, whose duality action is linear on the electric field strengths, namely:
\begin{equation}
{\bf g}\in G_{el}\;:
\quad \Rs_v[{\bf g}]=
\left(\begin{matrix}
A^\Lambda{}_\Sigma & \Zero \cr
C_{\Lambda\Sigma} & D_\Lambda{}^\Sigma
\end{matrix}\right)\,,\label{ge}
\end{equation}
where $D=A^{-T}$ by the symplectic condition (see Eq. (\ref{scfine}) of Appendix  \ref{spapp}), so that
\begin{align}
{\bf g}\in G_{el}\;:\quad
&F^\Lambda \rightarrow\,F^{\prime\Lambda}=A^\Lambda{}_\Sigma\,F^\Sigma\;,\nne
& G_\Lambda \rightarrow\,G'_{\Lambda}=C_{\Lambda\Sigma}\,F^\Sigma+
D_\Lambda{}^\Sigma\, G_\Sigma
\,.\label{Gel}
\end{align}
Indeed, as the reader can verify using Eq.\ (\ref{Ntra}), under the above transformation the matrices $\I,\,\R$ transform as follows:
\begin{equation}
\I_{\Lambda\Sigma}\rightarrow D_\Lambda{}^\Pi D_\Sigma{}^\Delta\,\I_{\Pi\Delta}
\,;\quad\;
\R_{\Lambda\Sigma}\rightarrow D_\Lambda{}^\Pi D_\Sigma{}^\Delta\,\R_{\Pi\Delta}+C_{\Lambda\Pi}\,D_{\Sigma}{}^\Pi\,,
\end{equation}
and the consequent variation of the Lagrangian reads
\begin{equation}
\delta\LB=
\frac{1}{8}\,C_{\Lambda\Pi}\,A^\Pi{}_{\Sigma}\epsilon^{\mu\nu\rho\sigma}\,F^\Lambda_{\mu\nu}F^\Sigma_{\rho\sigma}\,,\label{deltaLC}
\end{equation}
which is a \emph{total derivative} since $C_{\Lambda\Pi}\,A^\Pi{}_{\Sigma}$ is constant. Transformations characterized by a non-vanishing $C_{\Lambda\Sigma}$ block are called of \emph{Peccei-Quinn type} and are associated with shifts in certain axionic scalar fields. They are a symmetry of the classical action, while invariance of the perturbative path-integral requires the variation (\ref{deltaLC}), integrated over space-time, to be proportional through an integer to $2\pi \hbar$. This constrains the symmetries to belong to a discrete subgroup $G(\mathbb{Z})$ of $G$ whose duality action is implemented by integer-valued matrices ${\Rs_v}[{\bf g}]$. Such restriction of $G$ to $G(\mathbb{Z})$ in the quantum theory was discussed earlier as a consequence of the Dirac-Schwinger-Zwanziger quantization condition for dyonic particles (\ref{DZS}).\par
From (\ref{Gel}) we see that, while the vector field strengths $F^\Lambda_{\mu\nu}$ and their duals $ G_{\Lambda\,\mu\nu}$ transform together under $G$ in the ($2n_v$--dimensional) symplectic representation $\Rs_v$, the vector field strengths alone transform linearly under the action of $G_{el}$ in a smaller representation ${\bf n_v}$, defined by the $A$-block in (\ref{ge}).\par\medskip
Different symplectic frames of a same ungauged theory may originate from different compactifications. Examples will be discussed in Sect. \ref{gsfsmfc}.
A distinction here is in order. In $\N\geq 3$ theories, scalar fields always enter the same multiplets as the vector fields. Supersymmetry then implies their non-minimal coupling to the latter and requires the scalar manifold to be endowed with a symplectic structure which associates with each isometry a constant symplectic matrix. \par
 In $\mathcal{N}=2$ models, scalar fields in the hypermultiplets (hyper-scalars) are not connected to vector fields through supersymmetry and thus they do not enter the matrices $\I(\phi)$ and $\R(\phi)$. As a consequence of this the isometries of the quaternionic-K\"ahler manifolds spanned by these scalars are associated with trivial duality transformations
\begin{equation}
{\bf g}\,\in\,\text{isom. of}\;\MsQK
\;\;\;\,\,\,\Rightarrow\quad
\Rs_v[{\bf g}]=\Id\;.\label{qisom}
\end{equation}
Only $\MsSK$ features a flat symplectic structure which defines the embedding of its isometry group inside ${\rm Sp}(2n_v,\mathbb{R})$ and the vector-scalar non-minimal couplings through the matrix $\mathcal{M}(\phi)$. It is important to remark that such structure on a special K\"ahler manifold exists even if the manifold itself is not homogeneous \cite{Andrianopoli:1996cm}. This means that one can still define a matrix $\LL(\phi)$, which generalizes the hybrid coset-representative, in terms of which the matrix $\mathcal{M}(\phi)$, and thus its components $\I_{\Lambda\Sigma}$ and $\R_{\Lambda\Sigma}$, is defined as in (\ref{Mcos}). The matrix $\LL(\phi)$, however, is no longer derived from a coset representative for non-homogeneous manifolds. We shall discuss this issue in detail in Sect. \ref{SK}.\par
It is convenient for later purposes to rewrite the transformation properties of the bosonic fields under the duality group $G$, discussed in this section, in the following infinitesimal form:
\begin{eqnarray}
G\;:\quad
\begin{cases}
\delta\,\phi^s= \Lambda^\alpha\,k_\alpha^s(\phi)\;,\\
\delta\mathcal{M}(\phi)=\Lambda^\alpha\,k_\alpha^s\,\partial_s\mathcal{M}(\phi)=\Lambda^\alpha\,\left({\Scr R}_{v*}[t_\alpha]\,\mathcal{M}(\phi)+\mathcal{M}(\phi){\Scr R}_{v*}[t_\alpha]^T\right)\;,\\
\delta \mathcal{G}^{{M}}_{\mu\nu} =
-\Lambda^{\alpha}\,(t_{\alpha})_{{{N}}}{}^{{{M}}}\;\mathcal{G}_{\mu\nu}^{{{N}}}\,,
\end{cases}
\label{infM}
\end{eqnarray}
in terms of the infinitesimal generators $t_\alpha$ of $G$ introduced earlier, satisfying the relation (\ref{talg}). The second of Eqs. (\ref{infM}) is derived from (\ref{traM}) for infinitesimal transformations ${\bf g}$.
The matrices $(t_{\alpha})_{M}{}^{N}$ define the infinitesimal duality action of $G$ and are symplectic generators (see  Appendix  \ref{spapp})
\begin{equation}
(t_{\alpha})_{M}{}^{N}\,\CC_{NP} = (t_{\alpha})_{P}{}^{N}\,\CC_{NM}\, \qquad\;  M,N,\dotsc=1,\dotsc,\,2n_v\;.\label{sympcondgen}
\end{equation}
This is equivalently stated as the property of the tensor $t_{\alpha\,MN}\equiv (t_{\alpha})_{M}{}^{P}\,\mathbb{C}_{PN}$ of being symmetric in {\footnotesize $M\,N$}:
\begin{equation}
(t_\alpha)_{MN}=(t_\alpha)_{NM}\,.\label{tddsym}
\end{equation}
The above quantity $t_{\alpha\,MN}=(t_\alpha)_{MN}$ can be viewed as a $G$-invariant tensor in the product ${\rm Adj}(G)\times {\Scr R}_{v*}\times_s {\Scr R}_{v*}$, in terms of which we can construct a characteristic rank-4, totally symmetric, $G$-invariant tensor:
\begin{equation}
K_{MNPQ}\equiv t_{\alpha\,(MN}t^\alpha{}_{PQ)}\,,\label{Ktensor}
\end{equation}
where the index $\alpha$ has been raised by means of the invariant metric $\eta_{\alpha\beta}\equiv {\rm Tr}(t_\alpha t_\beta)$: $t^\alpha\equiv \eta^{\alpha\beta}\,t_\beta$,  $\eta^{\alpha\beta}\eta_{\beta\gamma}=\delta^\alpha_\gamma$.
In the description of static, extremal black hole solutions, this tensor plays an important role since the $AdS_2\times S^2$ near-horizon geometry, by virtue of the \emph{attractor} mechanism, see \cite{Andrianopoli:2006ub} and reference therein, is totally described in terms of a  \emph{quartic G-invariant} quantity constructed out of the quantized charges $(\Gamma^M)=(m^\Lambda,\,e_\Lambda)$ as follows:
\begin{equation}
I_4(e,m)\,\propto\,K_{MNPQ}\Gamma^M\Gamma^N\Gamma^P\Gamma^Q\,,\label{qinvar}
\end{equation}
and does not depend on the values of the scalar fields at spatial infinity.

\subsection{The Fermionic Sector} \label{fsector}
As mentioned earlier, we use the Weyl-basis for the fermion fields. In particular we denote by  $\psi_{\mu\,A},\,\chi_{ABC},\,\lambda_{IA}$ and $\lambda_\alpha$ the Weyl components of the fermion fields with positive chirality (see Appendix \ref{nacv} for the conventions about the $D=4$ Clifford algebra), while $\psi_{\mu}^A,\,\chi^{ABC},\,\lambda^{IA}$ and $\lambda^\alpha$ denote the corresponding charge conjugate fields with opposite chirality:\footnote{Later we shall also describe the hyperinos by the fields $\zeta_\alpha\equiv \lambda_\alpha/\sqrt{2}$, to make contact with the notations of \cite{Andrianopoli:1996cm}.}
\begin{equation}
\gamma^5\,\left(\begin{matrix}\psi_{\mu\,A}\cr\chi_{ABC}\cr \lambda_{IA} \cr \lambda_\alpha \end{matrix}\right)=\left(\begin{matrix}\psi_{\mu\,A}\cr\chi_{ABC}\cr \lambda_{IA} \cr \lambda_\alpha \end{matrix}\right)\,\,,\,\,\,A,B,C=1,\dots, \mathcal{N}\,.
\end{equation}
The same convention will be used for the supersymmetry parameter: $\epsilon_A,\,\epsilon^A$.\par
As for the extra gauginos $\lambda_{IABC}=\lambda_I\,\epsilon_{ABC}$ and dilatinos $\chi,\,\chi^A$ in $\mathcal{N}=3,5$ and 6 theories, respectively, see footnote \ref{fotextraferm}, the corresponding chirality is:
\begin{equation}
\gamma^5\lambda_I=-\lambda_I\,\,;\,\,\,\,\gamma^5\chi=-\chi\,\,;\,\,\,\gamma^5\chi^A=-\chi^A\,.
\end{equation}
Occasionally we shall denote the various fermionic fields  by a collective symbol $\lambda_{\mathcal{I}}$ where the calligraphic index $\mathcal{I}$ runs  over all the spin-$1/2$ fields:
\begin{equation}
\lambda_{\mathcal{I}}=\{\chi_{ABC},\,\lambda_{IA},\,\lambda_\alpha,\dots\}\,\,,\,\,\,
\gamma^5\lambda_{\mathcal{I}}=
\lambda_{\mathcal{I}}\,\,,\,\,\,\lambda^{\mathcal{I}}=(\lambda_{\mathcal{I}})_c\,.
\end{equation}
Fermions in supergravity transform covariantly with respect to the holonomy group $H$ of the scalar manifold, which has the general form (\ref{Hgroup}), while they do not transform under $G$, as opposed to the bosonic fields.
 Bosons and fermions have therefore definite transformation properties with respect to different groups of internal symmetry. The matrix $\mathbb{L}$, defined in terms of the coset representative for homogeneous scalar manifolds, transforms under the action of $G$ to the left and of $H$ to the right, according to (\ref{gLh})\footnote{Here we are considering symmetric manifolds of the form $G/H$. In general $\LL$ is acted on to the right by the isotropy group $H'$ which is locally contained in the holonomy group $H$.}
\begin{equation}
G\,\,\rightarrow \,\,\,\LL\,\,\,\leftarrow\,\, H\,,\label{intertwine}
\end{equation}
and thus has the right index structure to ``mediate'' in the Lagrangian between bosons and fermions. This means that we can construct $G$-invariant terms by contracting $\LL$ to the left by bosons (scalars, vectors and their derivatives), and to the right by fermion bilinears
\begin{equation}
(\mbox{Bosons}) \star \mathbb{L}(\phi)\star (\mbox{Fermion bilinears})\,,\label{BLF}
\end{equation}
the two $\star$ symbols denote some contraction of indices: $G$-invariant to the left and $H$-invariant to the right.
The ``Boson'' part of (\ref{BLF}) may also contain $\LL$ and its derivatives.
These are the kind of terms occurring in the action. If under a transformation ${\bf g}\in G$, symbolically:
\begin{equation}
\mbox{Bosons}\;\rightarrow\;\mbox{Bosons}'=\mbox{Bosons}\star {\bf g}^{-1}\,,
\end{equation}
and the \emph{fermions are made to transform under the compensating transformation} $h(\phi,{\bf g})$ in (\ref{gLh}):
\begin{equation}
\mbox{Fermions}\;\rightarrow\;\mbox{Fermions}'=h(\phi,{\bf g})\star \mbox{Fermion bilinears}\,,\label{Hfermi}
\end{equation}
using (\ref{gLh}) we see that (\ref{BLF}) remains invariant:
\begin{equation}
(\mbox{Bosons})' \star \mathbb{L}({\bf g}\star \phi)\star (\mbox{Fermion bilinears})'=(\mbox{Bosons}) \star \mathbb{L}(\phi)\star (\mbox{Fermion bilinears})\,.
\end{equation}
As mentioned earlier, the Lagrangian is manifestly invariant under local $H$-transformations since the covariant derivatives on the fermion fields contain the $H$-connection
\footnote{
We define $\mathcal{Q}_\mu\equiv \mathcal{Q}_s\,\partial_\mu\phi^s$
}
$\mathcal{Q}_\mu$:
\begin{equation}
{\Scr D}_\mu\xi=\nabla_\mu\xi+\mathcal{Q}_\mu\star \xi\,,\label{Dxi}
\end{equation}
where, as usual, the $\star$ symbol denotes the action of the $\mathfrak{H}$-valued connection $\mathcal{Q}_\mu$ on $\xi$ in the corresponding $H$-representation and $\nabla_\mu$ only contains the Levi-Civita connection on space-time, see Appendix \ref{nacv}. The reader can verify that (\ref{Dxi}) is indeed covariant under local $H$-transformations (\ref{Hfermi}), provided $\mathcal{Q}$ is transformed according to (\ref{omtra}). As opposed to the gauge groups we are going to introduce by the gauging procedure, which involve minimal couplings to the vector fields of the theory, the local $H$-symmetry group of the ungauged theory is not gauged by the vector fields, but by the \emph{composite connection} $\mathcal{Q}_\mu$, which is a function of the scalar fields and their derivatives.\par
General $H$-covariance of supergravity allows to describe the couplings between bosons and fermions in terms of $H$-covariant \emph{composite fields}:
\begin{equation}
{\tt f}(\phi,\partial\mbox{Bosons}) \equiv (\partial\mbox{Bosons}) \star \mathbb{L}(\phi)\,,\label{BLF2}
\end{equation}
obtained by \emph{dressing} the derivatives of the bosonic fields with the coset-representative so as to obtain an $H$-covariant quantity with the correct $H$-index structure to contract with fermionic currents. Indeed under a $G$-transformation
\begin{equation}
{\tt f}({\bf g}\star\phi,\partial\mbox{Bosons}')\equiv {\tt f}(\phi,\partial\mbox{Bosons})\star h(\phi,{\bf g})^{-1}\,,
\end{equation}
The manifest $H$-invariance of the supergravity theory requires the supersymmetry transformation properties of the femionic fields to be $H$-covariant. Indeed such transformation rules, which in rigid supersymmetric theories (i.e.\ theories which are invariant only under global supersymmetry) can be schematically described in the form%
\footnote{
This is a schematic representation in which we have suppressed the Lorentz indices and gamma-matrices
}:
\begin{equation}
\delta\mbox{Fermion}=\sum_{ \mbox{{\tiny Bosons}}}\partial\mbox{Boson}\cdot\epsilon\,,
\end{equation}
and in supergravity theories have the following general $H$-covariant form%
\footnote{
The gravitino field has an additional term ${\Scr D}\epsilon$ which is its variation as the gauge field of local supersymmetry
}
\begin{equation}
\delta\mbox{Fermion}=\sum_{ \mbox{{\tiny Bosons}}}{\tt f}(\phi,\partial\mbox{Bosons})\cdot\epsilon\,,
\end{equation}
where the space-time derivatives of the bosonic fields are dressed with the scalars in the definition of ${\tt f}(\phi,\partial\mbox{Bosons})$.
Examples of composite fields ${\tt f}(\phi,\partial\mbox{Bosons})$ are the vielbein of the scalar manifold (pulled back on space-time) $\mathcal{P}_\mu\equiv \mathcal{P}_s\,\partial_\mu\phi^s$, the $H$-connection $\mathcal{Q}_\mu$ in (\ref{Dxi}), the ``dressed'' vector field-strengths $\mathbb{F}_{\mu\nu}(\phi,\,\partial A)$, to be defined in the next Section,
or the \emph{$\Tb$-tensor}, to be introduced later, in which the bosonic field ``dressed'' by the scalars is the \emph{embedding tensor} $\Theta$, which is a non-dynamical quantity defining the choice of the gauge algebra.
\subsection{The Unaguged Lagrangian: A Detailed Discussion}\label{dcsl}
To be more explicit, since the fermion fields transform in complex representations of $H$, it is convenient  to describe the right action of $H$ on the hybrid coset-representative matrix $\mathbb{L}$ in a complex basis, in which this action is block-diagonal. In Sect. \ref{gsg} we introduced a symplectic representation $\underline{{\Scr R}}_v$ defined by the property that $\underline{{\Scr R}}_v[H]\subset {\rm SO}(2n_v)$. Let us change the basis of this representation into a complex one by means of the Cayley matrix $\mathcal{A}$ which maps a real symplectic vector $V^{\underline{M}}=(V^{\underline{\Lambda}},\,V_{\underline{\Lambda}})$ into a complex one ${\bf V}^{\underline{M}}$ as follows:
\begin{align}
{\bf V}^{\underline{M}}&=\left(\begin{matrix} {\bf V}^{\underline{\Lambda}}\cr {\bf V}_{\underline{\Lambda}}\end{matrix}\right)=\frac{1}{\sqrt{2}}\,
\left(\begin{matrix} V^{\underline{\Lambda}}+i\,V_{\underline{\Lambda}}\cr V^{\underline{\Lambda}}-i\,V_{\underline{\Lambda}}\end{matrix}\right)=\mathcal{A}^{\underline{M}}{}_{\underline{N}}\,V^{\underline{N}}\,\,,\,\,\,\,
\mathcal{A}^{\underline{M}}{}_{\underline{N}}&\equiv \frac{1}{\sqrt{2}}\,\left(\begin{matrix}{\bf 1} &
i\,{\bf 1}\cr {\bf 1} & -i\,{\bf 1} \end{matrix}\right)\,.\label{complexV}
\end{align}
As shown in Appendix \ref{spapp}, in this complex basis the maximal compact subgroup ${\rm U}(u_v)$ of ${\rm Sp}(2n_v,\mathbb{R})$ has a block-diagonal action, and the upper and the lower components ( ${\bf V}^{\underline{\Lambda}}$, ${\bf V}_{\underline{\Lambda}}$) of ${\bf V}$ transform in the ${\rm U}(u_v)$- representations $({\bf n_v})_{+1}$ and $(\overline{{\bf n_v}})_{-1}$, respectively, according the branching (\ref{2nvnvnv}).

Let us denote now by $\underline{{\Scr R}}_v^c$ the matrix representation of $G$ in this new basis: $\underline{{\Scr R}}_v^c\equiv \mathcal{A}\underline{{\Scr R}}_v\mathcal{A}^\dagger$. Although this representation is no-longer symplectic, it is convenient for the description of the couplings to the fermion fields since $H$ is represented by block-diagonal matrices, the diagonal blocks defining the representations of $H$ in the decomposition of $\underline{{\Scr R}}_v$ (or, equivalently, of ${\Scr R}_v$).
Recall now the general form of $H$ in Eq. (\ref{Hgroup}) and the fact that the spin-1 states in the gravitational multiplet belong to the 2-times antisymmetric representation ${\bf {\bigwedge}^2\mathcal{N}}\equiv{\bf \mathcal{N}\wedge \mathcal{N}}$ of ${\rm (S)U}(\mathcal{N})= H_R$. On the other hand the spin-1 states in the vector multiplets (being top-spin states) are singlets of $H_R$ and in general transform in a representation ${\bf n}$ of $H_{{\rm matt}}$. The following decompositions then hold:
\begin{align}
\underline{{\Scr R}}_v\,&\stackrel{H}{\longrightarrow}\,\,\, {\bf R}_v+\overline{{\bf R}}_v\,,\nonumber\\
{\bf R}_v\,&\stackrel{H}{\longrightarrow}\,\,\, {\bf ({\bigwedge}^2\mathcal{N},1)}+{\bf (1,n)}\,\,\,\Leftrightarrow\,\,\,\,
(V_{\underline{\Lambda}})=(V_{AB},\,V_I)\,\,;\,\,\,V^{\underline{\Lambda}}=(V_{\underline{\Lambda}})^*=(V^{AB},\,V^I)\,,
\end{align}
where $V^{AB}=-V^{BA},\,\,V_{AB}=-V_{BA}$ and $I=1,\dots, n$. There is an exception to this rule in the  $\mathcal{N}=6$ theory: the gravitational supermultiplet contains, besides  the 15 spin-1 states in the ${\bigwedge}^2{\bf 6}$ of ${\rm SU}(6)$, also an ${\rm SU}(6)$-singlet $A^\bullet_\mu\equiv \epsilon^{A_1\dots A_6} A_{\mu\,A_1\dots A_6}/6!$, that is a spin-1 state with six antisymmetric ${\rm SU}(6)$-indices. This will be taken into account in our discussion by allowing, in this theory, for the presence of a single-valued index $I=\bullet$, keeping in mind that it does not label a vector multiplet. The bosonic sector of the ungauged $\mathcal{N}=6$ theory coincides  with that of the ungauged $\mathcal{N}=2$ model characterized by a special K\"ahler manifold ${\rm SO}^*(12)/{\rm U}(6)$, fifth line of Table 2, and no hypermultiplets. In this case $A^\bullet_\mu$ plays the role of the graviphoton in the gravitational multiplet while the remaining 15 spin-1 fields belong to 15 vector multiplets. The two theories clearly differ in the fermionic sector.
\footnote{In our conventions, we shall use a factor $1/2$ whenever a contraction over an antisymmetric couple is performed, so that:
$${\bf V}^{\underline{\Lambda}} {\bf W}_{\underline{\Lambda}}=\frac{1}{2}\,{\bf V}^{AB}\,{\bf W}_{AB}+{\bf V}^{I}\,{\bf W}_{I}\,.$$
This implies that the identity matrix in the space parametrized by the antisymmetric couple reads: $({\bf 1})^{AB}{}_{CD}=2\,\delta^{AB}_{CD}$.}\par
Written in the complex basis, a generator of $H$ is, as we have seen, block-diagonal, while a non-compact generator in the space $\mathfrak{K}$ is block-off-diagonal of the form
\begin{align}
{\bf k}\in \underline{{\Scr R}}_v[\mathfrak{K}]\,\,,\,\,\,{\bf k}^c &=\mathcal{A}\,{\bf k}\,\mathcal{A}^\dagger =({\bf k}^{\underline{M}}{}_{\underline{N}})=\nonumber\\&=\left(\begin{matrix} {\bf 0} & K^{\underline{\Lambda}\underline{\Sigma}}\cr K_{\underline{\Lambda}\underline{\Sigma}} & {\bf 0}\end{matrix}\right)=\left(\begin{matrix} {\bf 0} & \begin{matrix}K^{AB,CD} & K^{AB,J}\cr K^{I,CD} & K^{IJ}\end{matrix}\cr \begin{matrix}K_{AB,CD} & K_{AB,J}\cr K_{I,CD} & K_{IJ}\end{matrix} & {\bf 0}\end{matrix}\right)\in \underline{{\Scr R}}^c_v[\mathfrak{K}]\,,\label{kmatrixf}
\end{align}
where $K_{\underline{\Lambda}\underline{\Sigma}}=(K^{\underline{\Lambda}\underline{\Sigma}})^*=K_{\underline{\Sigma}\underline{\Lambda}}$. By group theory, the following properties hold: $K_{AB,CD}=K_{[AB,CD]},\,K^{AB,CD}=K^{[AB,CD]}$, so that we shall avoid the comma between the two antisymmetric couples when labeling these components.\par
 Correspondingly the $\mathfrak{K}$-valued vielbein one-form $\mathcal{P}$ in the representation $\underline{{\Scr R}}^c_v$, in the complex basis, reads
\begin{align}
\mathcal{P}^c=\mathcal{A}\,\underline{{\Scr R}}_v[\mathcal{P}]\,\mathcal{A}^\dagger=(\mathcal{P}^{\underline{M}}{}_{\underline{N}})=\left(\begin{matrix} {\bf 0} & \mathcal{P}^{\underline{\Lambda}\underline{\Sigma}}\cr \mathcal{P}_{\underline{\Lambda}\underline{\Sigma}} & {\bf 0}\end{matrix}\right)=\left(\begin{matrix} {\bf 0} & \begin{matrix}\mathcal{P}^{ABCD} & \mathcal{P}^{AB,J}\cr \mathcal{P}^{I,CD} & \mathcal{P}^{IJ}\end{matrix}\cr \begin{matrix}\mathcal{P}_{ABCD} & \mathcal{P}_{AB,J}\cr \mathcal{P}_{I,CD} & \mathcal{P}_{IJ}\end{matrix} & {\bf 0}\end{matrix}\right)\,,\label{Pc}
\end{align}
where
\begin{equation}
\mathcal{P}_{ABCD}=\mathcal{P}_{[ABCD]}\,\,,\,\,\,\,\mathcal{P}^{ABCD}=\mathcal{P}^{[ABCD]}\,.
\end{equation}
Recall, however, that the symplectic  representation ${\Scr R}_v$ of $G$ (and thus $\underline{{\Scr R}}^c_v$), in $\mathcal{N}=2$ theories, is not faithful, since, see (\ref{qisom}), the isometries of the quaternionic K\"ahler part of the scalar manifold in (\ref{SKQK}), are represented by the identity. In this case (\ref{Pc}) only represent the vielbein of the special  K\"ahler space, spanned by the complex scalar fields in the vector multiplets.\par
The composite fields $\mathcal{P}^{ABCD}$ and $\mathcal{P}^{AB,I}$ describe, in the supersymmetry transformation laws, the spin-0 states in the gravitational and vector multiplets, respectively. The former indeed are in the  ${\bf {\bigwedge}^4\mathcal{N}}$ and the latter in the ${\bf {\bigwedge}^2\mathcal{N}}$ of $H_R$.
Depending on the amount of supersymmetry, some of the blocks in (\ref{Pc}) may be zero. We shall discuss below the relevant cases.  The gravitational multiplet features scalar fields only for $\mathcal{N}\ge 4$. They span a submanifold of ${\Scr M}_{scal}$ whose vielbein matrix is $\mathcal{P}^{ABCD}$. The block $\mathcal{P}^{I\,AB}$, on the other hand, defines the vielbein matrix associated with the  scalars in the vector multiplets.
\par
We also write $\mathcal{Q}$ as a matrix in the complex basis:
\begin{align}
\mathcal{Q}^c=\mathcal{A}\,\underline{{\Scr R}}_v[\mathcal{Q}]\,\mathcal{A}^\dagger=(\mathcal{Q}^{\underline{M}}{}_{\underline{N}})=\left(\begin{matrix} \mathcal{Q}^{\underline{\Lambda}}{}_{\underline{\Sigma}} & {\bf 0}\cr {\bf 0} &\mathcal{Q}_{\underline{\Lambda}}{}^{\underline{\Sigma}}\end{matrix}\right)=\left(\begin{matrix} \begin{matrix}\mathcal{Q}^{AB}{}_{CD} & 0\cr 0& \mathcal{Q}^{I}{}_{J}\end{matrix}& {\bf 0}\cr {\bf 0} & \begin{matrix}\mathcal{Q}_{AB}{}^{CD} & 0\cr 0& \mathcal{Q}_{I}{}^{J}\end{matrix}\end{matrix}\right)\,,\label{Qc}
\end{align}
where $\mathcal{Q}^{AB}{}_{CD}$ is the $H_R$-connection while $\mathcal{Q}^{I}{}_{J}$ is the  $H_{{\rm matt}}$ one. For $\mathcal{N}=2$, the matrix $\mathcal{Q}^c$ only belongs to the Lie algebra of the holonomy group $H^{(SK)}$ of the special K\"ahler manifold, and in particular $\mathcal{Q}^{AB}{}_{CD}$  only realizes the ${\rm U}(1)$-factor of $H_R$ while $\mathcal{Q}^{I}{}_{J}$ the part $H^{(SK)}_{{\rm matt}}$ of $H_{{\rm matt}}$.\par Being $H_R={\rm U}(\mathcal{N})$ (${\rm SU}(8)$ for $\mathcal{N}=8$), we can write:
\begin{align}
\mathcal{Q}^{AB}{}_{CD}=-\mathcal{Q}_{CD}{}^{AB}=4\,\delta^{[A}_{[C}\,\mathcal{Q}^{B]}{}_{D]}\,\,\Rightarrow\,\,\,\,
\mathcal{Q}^{AC}{}_{BC}=(\mathcal{N}-2)\mathcal{Q}^A{}_B+\delta^A_B\,\mathcal{Q}^C{}_C\,,
\end{align}
where $\mathcal{Q}^A{}_B$ is the $H_R$-connection in the fundamental representation. Notice that for $\mathcal{N}=2$ only the ${\rm U}(1)$ factor of the ${\rm U}(2)$ R-symmetry group, generated by $\mathcal{Q}^C{}_C$, is represented in $\mathcal{Q}^{AB}{}_{CD}$, while for $\mathcal{N}=8$ $\mathcal{Q}^C{}_C=0$ being $H_R={\rm SU}(8)$.\par
In terms of the connection matrices $\mathcal{Q}^A{}_B,\,\mathcal{Q}^I{}_J$ we can write the $H$-covariant derivatives (\ref{Dxi}) of the gravitinos, the dilatinos and gauginos:
\begin{align}
{\Scr D}_\mu\psi_{A\,\nu}=\partial_\mu\psi_{A\,\nu}-\Gamma_{\mu\nu}^\rho\,\psi_{A\,\rho}+\frac{1}{4}\,\omega_{\mu,\,ab}\gamma^{ab}\psi_{A\,\nu}+
\mathcal{Q}_{\mu\,A}{}^B\,\psi_{B\,\nu}\,,\nonumber\\
{\Scr D}_\mu\chi_{ABC}=\partial_\mu\chi_{ABC}+\frac{1}{4}\,\omega_{\mu,\,ab}\gamma^{ab}\chi_{ABC}+
3\,\mathcal{Q}_{\mu\,[A}{}^D\,\chi_{D|BC]}\,,\nonumber\\
{\Scr D}_\mu\lambda_{AI}=\partial_\mu\lambda_{AI}+\frac{1}{4}\,\omega_{\mu,\,ab}\gamma^{ab}\lambda_{AI}+
\mathcal{Q}_{\mu\,A}{}^D\,\lambda_{DI}+
\mathcal{Q}_{\mu\,I}{}^J\,\lambda_{AJ}\,.\label{genDdef}
\end{align}
Let us discuss the general features of the matrices $\mathcal{P}^c$ and $\mathcal{Q}^c$ in the different extended theories, see also Sect. \ref{glance}.
\begin{itemize}
\item{{\bf $\mathcal{N}>4$ models} only describe the gravitational multiplet, so $n=0$ and, with the exception of the $\mathcal{N}=6$ theory, the index $I$ disappears.
In the maximal (i.e. $\mathcal{N}=8$) theory, $\mathcal{P}^{ABCD}$ transforms in the ${\bf 70}={\bigwedge}^4 {\bf 8}$ and thus satisfies the reality condition:
\begin{equation}
\mathcal{P}^{ABCD}=(\mathcal{P}_{ABCD})^*=\frac{1}{24}\epsilon^{ABCDEFGH}\,\mathcal{P}_{EFGH}\,,\label{realityn8}
\end{equation}
where $A,B,\dots=1,\dots,8$.\par
In the $\mathcal{N}=6$ theory, $\mathcal{P}^{ABCD}=(\mathcal{P}_{ABCD})^*$ and $\mathcal{P}^{AB\bullet}=(\mathcal{P}_{AB\bullet})^*$ are not independent  but are related by the condition:
\begin{equation}
\mathcal{P}^{ABCD}=\frac{1}{2}\,\epsilon^{ABCDEF}\,\mathcal{P}_{EF\bullet}\,,\label{realityn6}
\end{equation}
where $A,B,\dots=1,\dots,6$ and $V^\bullet\equiv \epsilon^{A_1\dots A_6} V_{A_1\dots A_6}/6!$.\par
The $\mathcal{N}=6$ theory can be characterized as a consistent truncation of the maximal one. Indeed let us split the  R-symmetry index of the latter into an index $A=1,\dots, 6$ and the remaining values $7,8$, according to the branching of ${\rm SU}(8)$ into ${\rm U}(6)\times {\rm SU}(2)$, the ${\rm SU}(2)$ factor acting on the components labeled by $7,8$. \emph{The $\mathcal{N}=6$ theory is the consistent truncation of the maximal one obtained by retaining only the  ${\rm SU}(2)$-singlets.} In this perspective we can identify $A^\bullet_\mu$ with $A^{78}_\mu$ and  $\chi_A=\chi_{A78}$. Condition (\ref{realityn6}) then trivially follows from (\ref{realityn8}), being $\mathcal{P}_{EF\bullet}\equiv \mathcal{P}_{EF78}$.\par
Similarly the $\mathcal{N}=5$ theory can be obtained as the truncation of the maximal one defined by branching the ${\rm SU}(8)$-representations in the latter with respect to the subgroup ${\rm U}(5)\times {\rm SU}(3)$ and retaining only the ${\rm SU}(3)$-singlets. We can thus identify $\chi=-\chi_{678}$.
}
\item{{\bf The $\mathcal{N}=4$ case}. This is the only model in which the scalar fields in the gravitational and vector multiplets coexist. In this case we have \cite{Chamseddine:1980cp}:
\begin{equation}
{\Scr M}_{scal}=\frac{{\rm SL}(2,\mathbb{R})}{{\rm SO}(2)}\times \frac{{\rm SO}(6,\,n)}{{\rm SO}(6)\times {\rm SO}(n)}\,,
\end{equation}
where $n$ is the number of vector multiplets and $n_v=6+n$ the number of vector fields, which also comprise the six vectors in the gravitational supermultiplet. The latter contains a complex scalar $S=a-i\,e^{\varphi}$ spanning the $\frac{{\rm SL}(2,\mathbb{R})}{{\rm SO}(2)}$-factor, with vielbein matrix $\mathcal{P}^{ABCD}=\epsilon^{ABCD}\,\mathcal{P}$ while the $6n$ scalar fields in the vector multiplets span the second factor $\frac{{\rm SO}(6,\,n)}{{\rm SO}(6)\times {\rm SO}(n)}$, with vielbein matrix $\mathcal{P}^{I\,AB}$ satisfying a reality condition similar to that of $\mathcal{P}^{ABCD}$ for $\mathcal{N}=8$:
\begin{equation}
\mathcal{P}^{I\,AB}=(\mathcal{P}_{I\,AB})^*=\frac{1}{2}\,\epsilon^{ABCD}\,\mathcal{P}_{I\,CD}\,,\label{realityn4}
\end{equation}
 the index $I=1,\dots, n$ labeling the fundamental representation of ${\rm SO}(n)$.\par
We also have $\mathcal{P}_{IJ}=\bar{\mathcal{P}}\,\delta_{IJ}$. }
\item{{\bf  $\mathcal{N}\le 3$ models.} The gravitational multiplet has no scalar fields and the block $\mathcal{P}^{ABCD}$ is zero. For $\mathcal{N}=3$ the block $\mathcal{P}^{IJ}$ vanishes. As far as  $\mathcal{N}=2$ theories are concerned, we can write the K\"ahler metric $g_{i\bar{\jmath}}$ of the special K\"ahler manifold in terms of complex vielbein $e_{i\,I},\,\bar{e}_{\bar{\imath}}{}^J$: $g_{i\bar{\jmath}}=e_{i\,I}\,\bar{e}_{\bar{\imath}}{}^I$. The non-vanishing blocks in $\mathcal{P}^c$ read:
\begin{equation}
\mathcal{P}^{AB\,I}=\bar{e}_{\bar{\imath}}{}^I\,d\bar{z}^{\bar{\imath}}\,\epsilon^{AB}\,\,,\,\,\,
\mathcal{P}^{IJ}=i\,e^{-1\,Ii}\,e^{-1\,Jj}\,C_{ijk}\,dz^k\,,\label{PeIi}
\end{equation}
where $C_{ijk}$ appear in the Lagrangian as the coefficients of the anomalous magnetic moments of the gauginos.\par
In $\mathcal{N}=2$ theories $\mathcal{Q}^{AB}{}_{CD}$ only contains the K\"ahler ${\rm U}(1)$-connection $\mathcal{Q}$ while $\mathcal{Q}^{I}{}_{J}$ also contains the connection $\hat{\mathcal{Q}}^I{}_J$ associated with the part of $H_{{\rm matt}}$ in the holonomy group of the special K\"ahler manifold:
\begin{align}
\mathcal{Q}^{AB}{}_{CD}&=-2i\,\mathcal{Q}\,\delta^{AB}_{CD}\,\,,\,\,\,\mathcal{Q}^{I}{}_{J}=i\,\mathcal{Q}\,
\delta^{I}_{J}+\hat{\mathcal{Q}}^I{}_J\,,\label{QN2}\\
\mathcal{Q}&\equiv -\frac{i}{2}\,\left(\frac{\partial}{\partial z^i}\mathcal{K}\,dz^i-\frac{\partial}{\partial \bar{z}^{\bar{\imath}}}\mathcal{K}d \bar{z}^{\bar{\imath}}\right)\,,\label{Qconn}
\end{align}
where $\mathcal{K}(z,\bar{z})$ is the K\"ahler potential of the special K\"ahler manifold.\par
The ${\rm SU}(2)$-connection, on the other hand, only pertains to the quaternionic K\"ahler manifold.  In our conventions the ${\rm U}(2)$-connection reads:
\begin{equation}
\mathcal{Q}_A{}^B=\frac{1}{2}\,\omega^x\,(J_x)_A{}^B+\frac{1}{2}\,\mathcal{Q}\,(J_0)_A{}^B=
-\frac{i}{2}\,\omega^x\,(\sigma^x)_A{}^B+\frac{i}{2}\,\mathcal{Q}\delta_A^B\,,\label{Qournot}
\end{equation}
where $\sigma^x$, $x=1,2,3$, are the three Pauli matrices (\ref{Paulim}) and we have renamed $\mathcal{Q}^x$ by $\omega^x$ to comply with the $\mathcal{N}=2$ literature.\par
The covariant derivative of the hyperini in the $\mathcal{N}=2$ models reads:
\begin{equation}
{\Scr D}_\mu \lambda^\alpha=\nabla_\mu \lambda^\alpha+\partial_\mu q^u\mathcal{Q}_u{}^\alpha{}_\beta\,\lambda^\beta+\frac{i}{2}\,\mathcal{Q}_\mu\,\lambda^\alpha\,,
\end{equation}
where $\mathcal{Q}^\alpha{}_\beta$ is the $H^{(QK)}_{{\rm matt}}$-connection 1-form, having value in the Lie algebra of ${\rm USp}(2n_H)$: $\mathbb{C}_{\alpha\gamma}\mathcal{Q}_u{}^\gamma{}_\beta=\mathbb{C}_{\beta\gamma}\mathcal{Q}_u{}^\gamma{}_\alpha$.
 }
\end{itemize}
Since each block in $\mathcal{P}^c$, due to the factorized form of $H$, is an $H$-covariant tensor, when writing the kinetic Lagrangian for the scalar fields, according to the general form (\ref{lagrscal}),
$G$-invariance of the scalar metric allows for different normalizations of each term, which is however fixed by supersymmetry as follows:
\begin{align}
{\Scr L}_{scal}&=\frac{1}{48}\,\mathcal{P}^{ABCD}\mathcal{P}_{ABCD}\,\,\,,\,\,\,\,\mathcal{N}=8\,,\nonumber\\
{\Scr L}_{scal}&=\frac{1}{24}\,\mathcal{P}^{ABCD}\mathcal{P}_{ABCD}\,\,\,,\,\,\,\,\mathcal{N}=5,6\,,\nonumber\\
{\Scr L}_{scal}&=\frac{1}{24}\,\mathcal{P}^{ABCD}\mathcal{P}_{ABCD}+\frac{1}{4}\,\mathcal{P}^{I\,AB}\mathcal{P}_{I\,AB}\,\,\,,\,\,\,\,\mathcal{N}=4\,,\nonumber\\
{\Scr L}_{scal}&=\frac{1}{2}\,\mathcal{P}^{I\,AB}\mathcal{P}_{I\,AB}
\,\,\,,\,\,\,\,\mathcal{N}=3\,,\\
{\Scr L}_{scal}&=\frac{1}{2}\,\mathcal{P}^{I\,AB}\mathcal{P}_{I\,AB}+\frac{1}{2}\,\mathcal{P}^{A\,\alpha}\mathcal{P}_{A\,\alpha}\,\,\,,\,\,\,\,\mathcal{N}=2\,,\label{scalagdk}
\end{align}
where the extra factors $1/2$ in the $\mathcal{N}=4,\,8$ cases are due to the reality conditions (\ref{realityn4}), (\ref{realityn8}) on the vielbein tensors. The vielbein $\mathcal{P}^{A\,\alpha}$ describe the quaternionic K\"ahler manifold in the $\mathcal{N}=2$ models and were defined in Eq. (\ref{Paalpha}). As emphasized earlier, they are not part of the complex matrix $\mathcal{P}^c$.\par
Since the hybrid coset-representative matrix $\mathbb{L}(\phi)$ contracts to the right against fermion fields (see (\ref{intertwine})), which belong to complex representations, and to the left against bosonic fields, which can be real (as the vector fields are), it is useful to express the corresponding matrix by changing only the right index to a complex one and thus defining the following \emph{hybrid complex matrix} (for the components of $\mathbb{L}_c$ we use the notation of \cite{Andrianopoli:1996ve}):
\begin{equation}
\mathbb{L}_c(\phi)=(\mathbb{L}_c^{N}{}_{\underline{M}})\equiv \mathbb{L}(\phi)\mathcal{A}^\dagger=(\mathbb{L}_c^M{}_{AB},\,\mathbb{L}_c^M{}_{I},\,\mathbb{L}_c^{M\,AB},\,\mathbb{L}_c^{M\,I})=\left(\begin{matrix}
{f}^\Lambda{}_{AB} & {f}^\Lambda{}_{I} &  \bar{f}^{\Lambda\,AB} & \bar{f}^{\Lambda\,I}\cr
{h}_{\Lambda \, AB} & {h}_{\Lambda \, I} & \bar{h}_{\Lambda}{}^{AB} & \bar{h}_{\Lambda}{}^{I}
\end{matrix}\right)\,,\label{hybridcomplex}
\end{equation}
where the real matrix $\mathbb{L}(\phi)$ was defined in (\ref{hybrid}).
Equation (\ref{gLh2}) now becomes:
\begin{equation}
\forall {\bf g}\in G \;:\quad {\Scr R}_v[{\bf g}]\,\LL_c(\phi)=\LL_c({\bf g}\star\phi)\,\underline{{\Scr R}}^c_v[h]\,,\label{gL3}
\end{equation}
where now $\underline{{\Scr R}}^c_v[h]$ is a unitary, block-diagonal complex matrix.\par
The reader can verify that this matrix satisfies the following relations (which derive from the symplectic property of $\mathbb{L}(\phi)$ and $\mathbb{L}(\phi)^T$):
\begin{align}
\mathbb{L}_c(\phi)^\dagger \mathbb{C}\,\mathbb{L}_c(\phi)&=\varpi\,\,,\,\,\,\,\mathbb{L}_c(\phi)\, \varpi\,\mathbb{L}_c(\phi)^\dagger =\mathbb{C}\,,\label{propsLc}\\
\varpi&\equiv \mathcal{A}\mathbb{C}\mathcal{A}^\dagger= -i\,\left(\begin{matrix}{\bf 1} & {\bf 0}\cr {\bf 0} & -{\bf 1}\end{matrix}\right)\label{varpi}
\end{align}
In components the first relation reads:
\begin{align}
\mathbb{L}_c^{M\,AB}\,\mathbb{L}_c^{N}{}_{CD}\mathbb{C}_{MN}&=\bar{f}^{\Lambda\,AB}\,{h}_{\Lambda\,CD}-\bar{h}_\Lambda{}^{\,AB}\,{ f}^{\Lambda}_{CD}=-2i\,\delta^{AB}_{CD}\,,\label{iduno}\\
\mathbb{L}_c^{M\,I}\,\mathbb{L}_c^{N}{}_{J}\mathbb{C}_{MN}&=\bar{f}^{\Lambda\,I}\,{h}_{\Lambda\,J}-\bar{h}_\Lambda{}^{\,I}\,{ f}^{\Lambda}_{J}=-i\,\delta^{I}_{J}\,,\label{idue}\\
\mathbb{L}_c^{M\,AB}\,\mathbb{L}_c^{N}{}_{I}\mathbb{C}_{MN}&=\bar{f}^{\Lambda\,AB}\,{h}_{\Lambda\,I}-\bar{h}_\Lambda{}^{\,AB}\,{ f}^{\Lambda}_{I}=0\,,\label{itre}\\
\mathbb{L}_c^{M\,AB}\,\mathbb{L}_c^{N\,CD}\mathbb{C}_{MN}&=\bar{f}^{\Lambda\,AB}\,\bar{h}_{\Lambda}{}^{CD}-\bar{h}_\Lambda{}^{\,AB}\,\bar{ f}^{\Lambda\,CD}=0\,,\label{iquattro}\\
\mathbb{L}_c^{M\,AB}\,\mathbb{L}_c^{N\,I}\mathbb{C}_{MN}&=\bar{f}^{\Lambda\,AB}\,\bar{h}_{\Lambda}{}^{I}-\bar{h}_\Lambda{}^{\,AB}\,\bar{ f}^{\Lambda\,I}=0\label{icinque}\,.
\end{align}
Similarly, the second of (\ref{propsLc}), in components read:
\begin{equation}
(\mathbb{C}\mathbb{L}_c)_M{}^{\underline{\Lambda}}(\mathbb{C}\mathbb{L}_c)_{N\,\underline{\Lambda}}-
(\mathbb{C}\mathbb{L}_c)_{M\,\underline{\Lambda}}(\mathbb{C}\mathbb{L}_c)_N{}^{\underline{\Lambda}}=-i\,\mathbb{C}_{MN}\,,\label{CLCL}
\end{equation}
or, equivalently,
\begin{align}
&\frac{1}{2}\,f^{[\Lambda}{}_{AB}\bar{f}^{\Sigma]\,AB}+f^{[\Lambda}{}_{I}\bar{f}^{\Sigma]\,I}=0=\frac{1}{2}\,h_{[\Lambda\,|AB}\bar{h}_{\Sigma]}{}^{AB}+
h_{[\Lambda\,|I}\bar{h}_{\Sigma]}{}^{I}\,,\label{idsei}\\
&\frac{1}{2}\,f^{\Lambda}{}_{AB}\,\bar{h}_{\Sigma}{}^{AB}+f^{\Lambda}{}_{I}\,\bar{h}_{\Sigma}{}^{I}-\frac{1}{2}\,\bar{f}^{\Lambda\,AB}\,
{h}_{\Sigma\,AB}-\bar{f}^{\Lambda\,I}\,{h}_{\Sigma\,I}=i\,\delta^\Lambda_\Sigma\,.\label{idsette}
\end{align}
It is useful to define the block-matrices ${\bf f}\equiv (f^\Lambda{}_{AB},\,f^\Lambda{}_{I})$, ${\bf h}\equiv (h_{\Lambda\,AB},\,h_{\Lambda\,I})$, and by $\bar{{\bf f}}$ and $\bar{{\bf h}}$ the corresponding complex conjugate matrices.\par
From the abstract definitions (\ref{omegapro}) and (\ref{Vom}) we can derive the left-invariant 1-form $\Omega^c=\mathcal{A}\underline{{\Scr R}}_v[\Omega]\mathcal{A}^\dagger$ in the complex basis in terms of $\mathbb{L}_c$, and from it compute the components of $\mathcal{P}^c$ and $\mathcal{Q}^c$:
\begin{align}
\Omega^c=(\Omega^{\underline{M}}{}_{\underline{N}})=\mathbb{L}_c^{-1}d\mathbb{L}_c=
-\varpi\,\mathbb{L}_c^\dagger\mathbb{C}d\mathbb{L}_c=\mathcal{P}^c+\mathcal{Q}^c\,,\label{omegac}
\end{align}
where we have used (\ref{propsLc}). The components of $\mathcal{P}^c$ and of the connection $\mathcal{Q}^c$,  in (\ref{Pc}) and (\ref{Qc}) respectively, can be read  off the off-diagonal and diagonal blocks of $\Omega^{\underline{M}}{}_{\underline{N}}$:
\begin{align}
\mathcal{P}^{ABCD}&=i\,\left(\bar{f}^{\Lambda\,AB}d \bar{h}_\Lambda{}^{CD}-\bar{h}^{\Lambda\,AB}d \bar{f}_\Lambda{}^{CD}\right)=\mathcal{P}^{[ABCD]}\,\nonumber\\\mathcal{P}^{AB\,I}&=i\,\left(\bar{f}^{\Lambda\,AB}d \bar{h}_\Lambda{}^{I}-\bar{h}^{\Lambda\,AB}d \bar{f}_\Lambda{}^{I}\right)=\mathcal{P}^{I\,AB}\,\,;\,\,\,
\mathcal{P}^{I\,J}=i\,\left(\bar{f}^{\Lambda\,I}d \bar{h}_\Lambda{}^{J}-\bar{h}^{\Lambda\,I}d \bar{f}_\Lambda{}^{J}\right)=\mathcal{P}^{J\,I}\,,\nonumber\\
\mathcal{Q}^{AB}{}_{CD}&=i\,\left(\bar{f}^{\Lambda\,AB}d {h}_{\Lambda\,CD}-\bar{h}_\Lambda{}^{AB}d {f}^{\Lambda\,CD}\right)\,;\,\,\mathcal{Q}^{I}{}_{J}=i\,\left(\bar{f}^{\Lambda\,I}d {h}_{\Lambda\,J}-\bar{h}_\Lambda{}^{I}d {f}^{\Lambda\,J}\right)\,.
\end{align}
From (\ref{DLP}) we can express the $H$-covariant derivative of $\mathbb{L}_c$ in terms of the vielbein matrix \cite{Andrianopoli:1996ve}:
\begin{align}
{\Scr D}\mathbb{L}_c^M{}_{AB}\equiv d\mathbb{L}_c^M{}_{AB}-\frac{1}{2}\,\mathbb{L}_c^M{}_{CD}\,\mathcal{Q}^{CD}{}_{AB}=\frac{1}{2}\,\mathbb{L}_c^{M\,CD}\,\mathcal{P}_{ABCD}+
\mathbb{L}_c^{M\,I}\,\mathcal{P}_{I\,AB}\,.\label{dLP1}\\
{\Scr D}\mathbb{L}_c^M{}_{I}\equiv d\mathbb{L}_c^M{}_{I}-\mathbb{L}_c^M{}_{J}\,\mathcal{Q}^{J}{}_{I}=\frac{1}{2}\,\mathbb{L}_c^{M\,CD}\,\mathcal{P}_{I\,CD}+
\mathbb{L}_c^{M\,J}\,\mathcal{P}_{I\,J}\,.\label{dLP2}
\end{align}
It is useful to write in this basis the components of the $H_R$ and $H_{{\rm matt}}$-curvatures, from Eq. (\ref{RW}):
\begin{align}
R(\mathcal{Q})_{rs}{}^{AB}{}_{CD}&=
-\mathcal{P}_{[r}^{ABEF}\mathcal{P}{}_{s]\,CDEF}-2\,\mathcal{P}_{[r}{}^{AB\,I}\mathcal{P}_{s]\,CD\,I}\,,\label{RHaut}\\
R(\mathcal{Q})_{rs}{}^I{}_{J}&=\partial_r\mathcal{Q}_s{}^I{}_J-\partial_s\mathcal{Q}_r{}^I{}_J+2\,
\mathcal{Q}_{[r}{}^I{}_K\,\mathcal{Q}_{s]}{}^K{}_J=-
\mathcal{P}_{[r}{}^{IEF}\mathcal{P}_{s]\,JEF}-2\,\mathcal{P}_{[r}{}^{I\,K}\mathcal{P}_{s]\,JK}\,,\label{RHmatt}
\end{align}
where
\begin{equation}
R(\mathcal{Q})_{rs}{}^{AB}{}_{CD}=4\,\delta^{[A}_{[C}\,R(\mathcal{Q})_{rs}{}^{B]}{}_{D]}\,\,;\,\,\,\,
R(\mathcal{Q})_{rs}{}^{A}{}_{B}\equiv\partial_r\mathcal{Q}_s^A{}_B-\partial_s\mathcal{Q}_r{}^A{}_B+2\,\mathcal{Q}_{[r}{}^A{}_C\,\mathcal{Q}_{s]}{}^C{}_B\,.
\end{equation}
Finally Eq. (\ref{DP}) can be rewritten (in form notation) as follows \cite{Andrianopoli:1996ve}:
\begin{align}
{\Scr D}\mathcal{P}_{ABCD}=d\mathcal{P}_{ABCD}+4\,\mathcal{Q}_{[A}{}^E\wedge\mathcal{P}_{E|BCD]}=0\,,\\
{\Scr D}\mathcal{P}_{AB\,I}=d\mathcal{P}_{AB\,I}+2\,\mathcal{Q}_{[A}{}^E\wedge\mathcal{P}_{E|B]\,I}+\mathcal{Q}_{I}{}^J\wedge\mathcal{P}_{AB\,J}=0\,,\\
{\Scr D}\mathcal{P}_{IJ}=d\mathcal{P}_{IJ}+2\,\mathcal{Q}_{(I}{}^K\wedge\mathcal{P}_{J)K}=0\,.
\end{align}
From the definition (\ref{Mcos}) of $\mathcal{M}$ we can express this matrix in terms of $\mathbb{L}_c$:
\begin{equation}
\mathcal{M}(\phi)=\mathbb{C}\,\mathbb{L}_c(\phi)\mathbb{L}_c(\phi)^\dagger\,\mathbb{C}\,,\label{MLL2}
\end{equation}
which can also be written as follows:
\begin{align}
\mathcal{M}_{MN}&=
-(\mathbb{C}\mathbb{L}_c)_M{}^{\underline{\Lambda}}(\mathbb{C}\mathbb{L}_c)_{N\,\underline{\Lambda}}-
(\mathbb{C}\mathbb{L}_c)_{M\,\underline{\Lambda}}(\mathbb{C}\mathbb{L}_c)_N{}^{\underline{\Lambda}}=-2 \,(\mathbb{C}\mathbb{L}_c)_{M\,\underline{\Lambda}}(\mathbb{C}\mathbb{L}_c)_N{}^{\underline{\Lambda}}+
i\,\mathbb{C}_{MN}\,,\label{MLc}
\end{align}
where we have used Eq. (\ref{CLCL}).
In terms of the $f$ and $h$-blocks the matrix $\mathcal{M}$ reads:
\begin{equation}
\mathcal{M}(\phi)=\left(\begin{matrix}-2{\bf h}{{\bf h}^\dagger} & 2\,{\bf h}{{\bf f}^\dagger}+i\,{\bf 1}\cr 2\,{\bf f}{{\bf h}^\dagger}-i\,{\bf 1} & -2\,{\bf f}{{\bf f}^\dagger}\end{matrix}\right)\,.
\end{equation}
Comparing the above formula with (\ref{M}), we can derive the expression of the matrices $\mathcal{I},\,\mathcal{R}$ in terms of $\mathbb{L}_c$, namely of
 ${\bf f}$ and  ${\bf h}$:
\begin{equation}
\mathcal{I}=-\frac{1}{2}\,{\bf f}^{-\dagger}\,{\bf f}^{-1}\,\,;\,\,\,\,\mathcal{R}=({\bf h}+\frac{i}{2}\,{\bf f}^{-\dagger}){\bf f}^{-1}\,,\label{Iff}
\end{equation}
from which we derive the simple expression for the complex matrix ${\Scr N}$ defined in Eq (\ref{Nmatrix})
\begin{equation}
{\Scr N}_{\Lambda\Sigma}=\mathcal{R}_{\Lambda\Sigma}+i\,\mathcal{I}_{\Lambda\Sigma}=({\bf h}{\bf f}^{-1})_{\Lambda\Sigma}=\frac{1}{2}\,h_{\Lambda\, AB}\,f^{-1\,AB}{}_{\Sigma}+h_{\Lambda\, I}\,f^{-1\,I}{}_{\Sigma}={\Scr N}_{\Sigma\Lambda}\,,\label{Nfh}
\end{equation}
the symmetry of ${\Scr N}_{\Lambda\Sigma}$ follows from (the complex conjugate of) Eqs. (\ref{iquattro}) and (\ref{icinque}).\par
As mentioned above, in $\mathcal{N}=2$ models, the matrix $\mathbb{L}_c$, and thus the matrix-valued 1-forms $\mathcal{P}^c,\,\mathcal{Q}^c$, only describe the geometry of the special K\"ahler manifold spanned by the scalar fields in the vector multiplets. It is important to emphasize that, as shown in \cite{Andrianopoli:1996ve}, these quantities can also be defined for non-homogeneous manifolds and the geometry of these manifolds expressed in terms of them
as discussed above, although, in the non-homogeneous case, the definition of $\mathbb{L}_c$ cannot be derived from a coset representative. Its components are instead expressed in terms of the \emph{covariantly holomorphic section} $V^M(z,\bar{z})$ of the  special K\"ahler manifold and its covariant derivatives $U_i^{M}\equiv {\Scr D}_iV^M$, to be defined later in Sect. \ref{SK}, as follows:
\begin{equation}
\mathbb{L}_c^M{}_{AB}=V^M\,\epsilon_{AB}\,\,,\,\,\,\mathbb{L}_c^M{}_{I}=\bar{e}^{-1}{}_I{}^{\bar{\imath}}\overline{U}_{\bar{\imath}}^M\,\,;\,\,\,
\mathbb{L}_c^{M\,AB}=\overline{V}^M\,\epsilon^{AB}
\,\,,\,\,\,\mathbb{L}_c^{M\,I}=\bar{e}^{-1\,I i}\,{U}_{i}^M,.
\end{equation}
We shall discuss separately the geometry of special K\"ahler  and quaternionic K\"ahler manifolds in Sections \ref{SK} and \ref{QMans}, respectively.\par
Now we have the mathematical tools we need to complete the duality condition (\ref{FCMF}) by introducing a $\underline{{\Scr R}}^c_v$-vector $\mathcal{O}^{\underline{M}}_{\mu\nu}=-\mathcal{O}^{\underline{M}}_{\nu\mu}$ consisting of fermion bilinears \cite{Gaillard:1981rj,deWit:1981sst}:
\begin{equation}
{}^*\mathcal{G}=-\mathbb{C}\mathcal{M}(\mathcal{G}+\mathbb{L}_c\mathcal{O})=-\mathbb{C}\mathcal{M}\mathcal{G}+
\mathbb{L}_c
\varpi \mathcal{O}\,,\label{FCMF2}
\end{equation}
where the matrix $\varpi$ was defined in (\ref{varpi}).
We denote, as usual, by $\mathcal{O}^{\underline{\Lambda}}_{\mu\nu}$ and $\mathcal{O}_{\underline{\Lambda}\,\mu\nu}=(\mathcal{O}^{\underline{\Lambda}}_{\mu\nu})^*$ the upper and lower components of $\mathcal{O}^{\underline{M}}_{\mu\nu}$, respectively.
These fermion bilinears originate from the Pauli terms in the Lagrangian and the definition of $ G_{\Lambda\mu\nu }$ in (\ref{GF}). It is straightforward to verify that equation (\ref{FCMF2}) is manifestly $G$-covariant provided the symplectic vector $\mathcal{O}^{\underline{M}}$ transforms under the compensating transformation $h=h(\phi,{\bf g})$, see Eq. (\ref{gL3}), as follows:
\begin{equation}
\mathcal{O}_{\mu\nu}\,\stackrel{{\bf g}\in G}{\longrightarrow}\,\mathcal{O}'_{\mu\nu}=\underline{{\Scr R}}^c_v(h)\,\mathcal{O}_{\mu\nu}\,,
\end{equation}
which follows from the fermion transformation rule (\ref{Hfermi}).
Let us define the self-dual and anti-self-dual  components of a rank-2, covariant antisymmetric tensor $F_{\mu\nu}$ as follows:
\begin{equation}
F_{\mu\nu}^{\pm}\equiv \frac{F_{\mu\nu}\pm i\,{}^*F_{\mu\nu}}{2}\,\,\,\,\,\,\Rightarrow\,\,\,\,{}^*F_{\mu\nu}^{\pm}=\mp i\,F_{\mu\nu}^{\pm}\,.
\end{equation}
It is also convenient to introduce the following projectors in the symplectic representation:
\begin{align}
\mathbb{P}^{\pm}&=\frac{1}{2}\left({\bf 1}\mp i\, \mathbb{C}\mathcal{M}(\phi)\right)\,\,;\,\,\,\,
\mathbb{P}^{\pm}\,\mathbb{P}^{\pm}=\mathbb{P}^{\pm}\,\,;\,\,\,\mathbb{P}^{\pm}\,\mathbb{P}^{\mp}=0\,,\label{bbFpm}
\end{align}
where the above properties of the two matrices originate from the symplectic property (\ref{sympM}) of the symmetric matrix $\mathcal{M}$: $(\mathbb{C}\mathcal{M})^2=-{\bf 1}$.
From (\ref{FCMF2}) and the properties (\ref{MLc}), (\ref{propsLc}) of the matrices $\mathcal{M}$ and $\mathbb{L}_c$, we can deduce from (\ref{FCMF2}) the following useful relations
\begin{align}
\mathcal{G}^\pm_{\mu\nu}&=\mathbb{P}^{\pm}\,\mathcal{G}_{\mu\nu}\pm \frac{i}{2}\mathbb{L}_c\,\varpi\,\mathcal{O}_{\mu\nu}\,,\label{GpmO1}\\
\mathcal{G}^\pm_{\mu\nu}&=\mp i\,\mathbb{C}\mathcal{M}\mathcal{G}^\pm_{\mu\nu}\pm i\,\mathbb{L}_c\,\varpi\,\mathcal{O}^\pm_{\mu\nu}\,,\label{GpmO2}
\end{align}
and from the definition of $\mathbb{P}^\pm$ the identities:
\begin{equation}
\mathbb{L}_c^\dagger\mathbb{C}\mathbb{P}^\pm=\frac{1}{2}\,({\bf 1}\pm i\,\varpi)\,\mathbb{L}_c^\dagger\mathbb{C}\,.\label{LCP}
\end{equation}
Next we define the following complex $(2n_V)$-vector of 2-forms by ``dressing'' $\mathcal{G}^M$ in (\ref{bbF}) with scalar fields by using $\mathbb{L}_c$:\footnote{The components $F^{\underline{\Lambda}}$ of $\mathbb{F}$ should not be mistaken for the field strengths $F^\Lambda$, the former being composite fields labeled by the indices $(AB),\,I$.}
\begin{equation}
\mathbb{F}_{\mu\nu}(\phi,\partial A^\Lambda,\,f)=(F^{\underline{M}}_{\mu\nu})\equiv -\mathbb{L}_c(\phi)^\dagger \mathbb{C}\mathcal{G}_{\mu\nu}=-((\mathbb{L}_c(\phi)^{N}{}_{\underline{M}})^*\,\mathbb{C}_{NP}\,\mathcal{G}^P_{\mu\nu})=
\left(\begin{matrix}F^{AB}_{\mu\nu}\cr
F^{I}_{\mu\nu}\cr F_{\mu\nu\,AB}\cr F_{\mu\nu\,I}\end{matrix}\right)\,,\label{Tdef}
\end{equation}
where $f$ generically denotes the spin-$1/2$ fields.
Let us see how the composite field $\mathbb{F}$ transforms under a transformation ${\bf g}$ in $G$ of the elementary fields it depends on:
\begin{equation}
\mathbb{F}(\phi',\partial A^{\prime \Lambda},f')= -\mathbb{L}_c({\bf g}\star \phi)^\dagger \mathbb{C}\mathcal{G}'=-h_c(\phi,{\bf g})\,\mathbb{L}_c( \phi)^\dagger {\Scr R}_v[{\bf g}]^T\mathbb{C}{\Scr R}_v[{\bf g}]\mathcal{G}=h_c(\phi,{\bf g})\,\mathbb{F}(\phi,\partial A^{\Lambda},f)\,,\label{mathbbF}
\end{equation}
where we have used (\ref{gL3}) and denoted by  $h_c(\phi,{\bf g})$ the compensating $H$-transformation $h(\phi,{\bf g})$ in the representation $\underline{{\Scr R}}^c_v$, which is complex and unitary: $(h_c)^{-1}=h_c^\dagger$.
 We see that, just as $\mathcal{Q}(\phi,\partial\phi)$ and $\mathcal{P}(\phi,\partial\phi)$, also this composite field transforms only by the compensating transformation in $H$. By supersymmetry, Weyl fermions, according to their chiralities, transform  either in the self-dual or in anti-self-dual components of the composite field strengths $\mathbb{F}$. To show this let us derive some general properties of these components. \par
 Using Eqs.
 (\ref{GpmO1}) and (\ref{LCP}) we find:
 \begin{align}
 \mathbb{F}^\pm_{\mu\nu}&=-\mathbb{L}_c^\dagger \mathbb{C}\mathcal{G}^\pm_{\mu\nu}=-\mathbb{L}_c^\dagger \mathbb{C}\mathbb{P}^\pm\mathcal{G}_{\mu\nu} \pm\frac{i}{2}\,\mathcal{O}_{\mu\nu}
 =-\frac{1}{2}\,({\bf 1}\pm i\,\varpi)\,\mathbb{L}^\dagger\mathbb{C}\mathcal{G}_{\mu\nu}\pm\frac{i}{2}\,\mathcal{O}_{\mu\nu}=\nonumber\\&=\frac{1}{2}\,({\bf 1}\pm i\,\varpi)\,\mathbb{F}_{\mu\nu}\pm\frac{i}{2}\,\mathcal{O}_{\mu\nu}\,,\label{FpmFO}
 \end{align}
 which also implies
 \begin{equation}
 \frac{1}{2}\,({\bf 1}\mp i\,\varpi)\,\mathbb{F}_{\mu\nu}^\pm=\pm\frac{i}{2}\,\mathcal{O}^\pm_{\mu\nu}\,,
 \end{equation}
that is:
\begin{equation}
\mathcal{O}^-_{\underline{\Lambda}\,\mu\nu}=\mathcal{O}^{+\,\underline{\Lambda}}{}_{\mu\nu}=0\,.\label{Opm0}
\end{equation}
The above equation and (\ref{FpmFO}) imply that the components $F^{+}_{\underline{\Lambda}\mu\nu},\,F^{-\,\underline{\Lambda}}_{\mu\nu}=(F^{+}_{\underline{\Lambda}\mu\nu})^*$ are \emph{purely fermionic}:
\begin{align}
F^{+}_{AB\mu\nu}&=\frac{i}{2}\,\mathcal{O}_{AB\,\mu\nu}^+\,\,;\,\,\,F^{+}_{I\mu\nu}=
\frac{i}{2}\,\mathcal{O}_{I\,\mu\nu}^+\,,\nonumber\\
F^{-\,AB}_{\mu\nu}&=-\frac{i}{2}\,\mathcal{O}^{-\,AB}_{\mu\nu}\,\,;\,\,\,F^{-\,I}_{\mu\nu}=-
\frac{i}{2}\,\mathcal{O}^{-\,I}_{\mu\nu}\,.\label{FpmO}
\end{align}
We can then write the general form of the composite fields $\mathbb{F}^M$:
\begin{align}
\mathbb{F}^{+}_{\mu\nu}&=\left(\begin{matrix}F^{+\,AB}_{\mu\nu}\cr
F^{+\,I}_{\mu\nu}\cr \frac{i}{2}\,\mathcal{O}_{AB\,\mu\nu}\cr \frac{i}{2}\,\mathcal{O}_{I\,\mu\nu}\end{matrix}\right)\,\,\,;\,\,\,\,\,\,
\mathbb{F}^{-}_{\mu\nu}=\left(\begin{matrix}-\frac{i}{2}\,\mathcal{O}^{AB}_{\mu\nu}\cr -\frac{i}{2}\,\mathcal{O}^{I}_{\mu\nu}\cr F^-_{\mu\nu\,AB}\cr F^-_{\mu\nu\,I}\end{matrix}\right)\,.\label{Tpm2}
\end{align}
From (\ref{GpmO2}) and the definition of the matrix ${\Scr N}$ we find:
\begin{align}
 G_{\Lambda}^+&={\Scr N}_{\Lambda\Sigma}\,F^{\Sigma\,+}+i\,\mathcal{I}_{\Lambda\Sigma}{\bar{ f}}^{\Sigma \underline{\Gamma}}\mathcal{O}_{\underline{\Gamma}}={\Scr N}_{\Lambda\Sigma}\,F^{\Sigma\,+}-\frac{i}{2}\,{{ f}}^{-1\,\underline{\Gamma}}{}_\Lambda \mathcal{O}_{\underline{\Gamma}}\,,\label{GpmN1}\\
 G_{\Lambda}^-&=\overline{{\Scr N}}_{\Lambda\Sigma}\,F^{\Sigma\,-}-i\,\mathcal{I}_{\Lambda\Sigma}{{ f}}^{\Sigma}{}_{\underline{\Gamma}}\mathcal{O}^{\underline{\Gamma}}=\overline{{\Scr N}}_{\Lambda\Sigma}\,F^{\Sigma\,-}+\frac{i}{2}\,{\bar{ f}}^{-1}{}_{\underline{\Gamma}\Lambda} \mathcal{O}^{\underline{\Gamma}}\,,\label{GpmN2}
\end{align}
where we have used (\ref{Iff}).
Finally, using Eqs.
 (\ref{GpmO2}) and (\ref{FpmO}) we can derive the following expression for $\mathcal{G}^+_{\mu\nu}$:
 \begin{equation}
 \mathcal{G}^{+\,M}_{\mu\nu}=- i\,\mathbb{L}_c^M{}_{\underline{\Lambda}}\,F^{+\,\underline{\Lambda}}_{\mu\nu}-\frac{1}{2}\,\mathbb{L}_c^{M\,
 \underline{\Lambda}}\,\mathcal{O}_{\underline{\Lambda}\,\mu\nu}^+\,.\label{finqui}
 \end{equation}
 An analogous formula is derived for $\mathcal{G}^-_{\mu\nu}=(\mathcal{G}^+_{\mu\nu})^*$.
 Inverting Eq. (\ref{finqui}) and using (\ref{GpmN1}) we find:
 \begin{equation}
 F^{+\,\underline{\Lambda}}_{\mu\nu}=i\mathbb{L}_c^{-1\,\underline{\Lambda}}{}_M\,\mathcal{G}^{+\,M}_{\mu\nu}=
 -2\,i\,\bar{f}^{\Sigma\underline{\Lambda}}\,\mathcal{I}_{\Sigma\Lambda}\,F^{\Lambda\,+}_{\mu\nu}+{\rm fermi.- bilinears}\label{FGnewuse}
 \end{equation}
 where we have also used Eq. (\ref{Lcm1}). An analogous expression can be derived for $F^{-}_{\underline{\Lambda}\,\mu\nu}$.
 \par
 From Eqs. (\ref{GpmN1}) and (\ref{GpmN2}) we can deduce the  general form of the Pauli terms. This is done by rewriting (\ref{GF}) for the self-dual/anti-self-dual components of the field strengths:
\begin{equation}
{G}^\pm_{\Lambda\,\mu\nu}=\pm \frac{2\,i}{e}\,\frac{\delta {\Scr L}}{\delta F^{\pm\,\Lambda\,\mu\nu}}\,.
\end{equation}
Eqs. (\ref{GpmN1}) and (\ref{GpmN2}) imply for the kinetic Lagrangian of the vector fields and Pauli terms:
\begin{align}
{\Scr L}_{kin.,\,v}&=e\,\frac{i}{4}\left(F^{-\,\Lambda}_{\mu\nu}\overline{{\Scr N}}_{\Lambda\Sigma}F^{-\,\Sigma\,\mu\nu}-F^{+\,\Lambda}_{\mu\nu}{\Scr N}_{\Lambda\Sigma}F^{+\,\Sigma\,\mu\nu}\right)\,,\\{\Scr L}_{Pauli}&=\frac{e}{2}\,F^{+\,\Lambda\,\mu\nu}\mathcal{I}_{\Lambda\Sigma}
{\bar{ f}}^{\Sigma \underline{\Gamma}}\mathcal{O}_{\underline{\Gamma}\,\mu\nu}+h.c.\label{vkinPauli}
\end{align}
where ${\Scr L}_{kin,\,v}$ is an equivalent rewriting of the vector kinetic terms given in (\ref{boslagr}).\par
Denoting by $\lambda_{\mathcal{I}}$ the generic spin-$1/2$ field, the general form of $\mathcal{O}_{\underline{\Lambda}}$ is the following:
\begin{align}
\mathcal{O}_{AB\,\mu\nu}=2\,\bar{\psi}_{A\,\rho}\gamma^{[\rho}\gamma_{\mu\nu}\gamma^{\sigma]}\psi_{B\,\sigma}+
i\,c_{AB;}{}_C{}^{\mathcal{I}}\bar{\psi}_{\rho}^C\gamma_{\mu\nu}\gamma^\rho\lambda_{\mathcal{I}}+
c_{AB;\,\mathcal{I}\mathcal{J}}\bar{\lambda}^{\mathcal{I}}\gamma_{\mu\nu}\lambda^{\mathcal{J}}=
\mathcal{O}^+_{AB\,\mu\nu}\,,\label{o1}\\
\mathcal{O}_{I\,\mu\nu}=
i\,c_{I;}{}_C{}^{\mathcal{I}}\bar{\psi}_{\rho}^C\gamma_{\mu\nu}\gamma^\rho\lambda_{\mathcal{I}}+
c_{I;\,\mathcal{I}\mathcal{J}}\bar{\lambda}^{\mathcal{I}}\gamma_{\mu\nu}\lambda^{\mathcal{J}}=
\mathcal{O}^+_{I\,\mu\nu}\,,\label{o2}
\end{align}
where $c_{AB;}{}_C{}^{\mathcal{I}},\,c_{AB;\,\mathcal{I}\mathcal{J}}$ or $
c_{I;}{}_C{}^{\mathcal{I}},\,c_{I;\,\mathcal{I}\mathcal{J}}$ are model-dependent $H$-covariant tensors fixed by supersymmetry. The last equalities in (\ref{o1}), (\ref{o2}) follow from
the familiar properties (\ref{gamma5mus0}), (\ref{gamprop}) of the gamma matrices. In the maximal theory, for instance, we have \cite{deWit:1982bul,deWit:2007mt}:
\begin{equation}
\mathcal{O}_{AB\,\mu\nu}=2\,\bar{\psi}_{A\,\rho}\gamma^{[\rho}\gamma_{\mu\nu}\gamma^{\sigma]}\psi_{B\,\sigma}+
i\,\bar{\psi}_{\rho}^C\gamma_{\mu\nu}\gamma^\rho\chi_{ABC}+
\frac{1}{72}\epsilon_{ABCDEFGH}\bar{\chi}^{CDE}\gamma_{\mu\nu}\chi^{FGH}\,.
\end{equation}
Just as we did for the quantized charges, we define on a bosonic solution the \emph{central} and \emph{matter} charges as the following integrals over a sphere $S^2_\infty$ at spatial infinity:
\begin{align}
{\Scr Z}_{AB}(\phi,e,m)&\equiv \int_{S^2_\infty}F_{AB}=-\mathbb{L}_c^M{}_{AB}(\phi)\,\,\mathbb{C}_{MN}\,\Gamma^N=h_{\Lambda\,AB}(\phi)\,m^\Lambda-
f^\Lambda{}_{AB}(\phi)\,e_\Lambda\,,\label{Zdeff1}\\
{\Scr Z}_{I}(\phi,e,m)&\equiv \int_{S^2_\infty}F_{I}=-\mathbb{L}_c^M{}_{I}(\phi)\,\,\mathbb{C}_{MN}\,\Gamma^N=h_{\Lambda\,I}(\phi)\,m^\Lambda-
f^\Lambda{}_{I}(\phi)\,\,e_\Lambda\,,\label{Zdeff2}
\end{align}
where we assume that the scalar fields at spatial infinity are constant over $S^2_\infty$. These can be thought of as the physical charges measured on a solution at radial infinity. Together with their complex conjugates, they can be arranged in a vector ${\bf {\Scr Z}}^{\underline{M}}$ in the complex symplectic basis
\begin{equation}
{\bf {\Scr Z}}(\phi,e,m)=({\bf {\Scr Z}}^{\underline{M}}(\phi,e,m))=\left(\begin{matrix}{\Scr Z}^{AB}\cr
{\Scr Z}^{I}\cr {\Scr Z}_{AB}\cr {\Scr Z}_{I}\end{matrix}\right)=-\mathbb{L}_c^\dagger(\phi)\mathbb{C}\Gamma\,.
\end{equation}
Just as $\mathbb{F}_{\mu\nu}$, this vector transforms under $G$ through the compact compensator $h_c(\phi,\,{\bf g})$ in $H$:
\begin{equation}
{\bf {\Scr Z}}({\bf g}\star  \phi;\,{\bf g}\,\Gamma)=h_c(\phi,\,{\bf g})\,{\bf {\Scr Z}}(\phi;\Gamma)\,,\label{Ztransform}
\end{equation}
where we have written $\Gamma$ instead of $(e,m)$ and ${\bf g}\,\Gamma$ instead of ${\Scr  R}[{\bf g}]\,\Gamma$, for the sake of notational simplicity.\par
We can now write the general form (up to higher-order terms in the fermionic fields) of the supersymmetry transformation laws for the bosonic and fermionic fields in the ungauged theory:\footnote{ Supersymmetry closes on-shell on the other local symmetries of the theory. In our conventions, the commutator of two supersymmetry transformations on a generic field $\Phi$ should read:
\begin{equation}
[\delta_1,\,\delta_2]\Phi = i\,(\bar{\epsilon}^A_2\gamma^\mu\epsilon_{1\,A}-\bar{\epsilon}^A_1\gamma^\mu\epsilon_{2\,A})\,{\Scr D}_\mu\Phi+\dots
\end{equation}
where ellipses refer to local symmetry transformations of the theory. In the presence of a gauging, the latter will include gauge transformations.}\footnote{The transformation laws of the fermion fields which are specific to the $\mathcal{N}=2,3,5,6$ theories are:
\begin{align}
\delta \lambda_{I}&=\frac{i}{2} \,\mathcal{P}_{s\,I\,AB}\,\partial_\mu\phi^s\gamma^\mu\,\epsilon_C\,\epsilon^{ABC}+\dots\,\,;\,\,\,
(\mathcal{N}=3)\,,\nonumber\\
\delta \chi&=\frac{i}{24}\,\epsilon^{ABCDE}\,\partial_\mu\phi^s\mathcal{P}_{s\,ABCD}\gamma^\mu\epsilon_E+\dots\,\,;
\,\,\,(\mathcal{N}=5)\,,\nonumber\\
\delta \chi_F&=\frac{i}{24}\,\epsilon_{FABCDE}\,\partial_\mu\phi^s\mathcal{P}_s{}^{ABCD}\gamma^\mu\epsilon^E-\frac{i}{4}\,F_{\mu\nu\,\bullet}^-\gamma^{\mu\nu}
\epsilon_F+\dots\,\,;\,\,\,(\mathcal{N}=6)\,,\nonumber\\
\delta \lambda_\alpha&=i\mathcal{P}^{B\beta}_u\partial_\mu q^u\gamma^\mu \epsilon^A\epsilon_{AB}\mathbb{C}_{\alpha\beta}+\dots  \,\,;\,\,\,(\mathcal{N}=2)
\end{align}
}
\begin{align}
\delta\phi^s\mathcal{P}_s^{ABCD}&=\Sigma^{ABCD}\,\,;\,\,\,\,\delta\phi^s\mathcal{P}_s^{I\,AB}=\Sigma^{I\,AB}\,\,;\,\,\,\,\delta q^u\mathcal{P}_u^{A\alpha}=\Sigma^{A\alpha}\,,\label{traphi}\\
\delta A^\Lambda_\mu&=\mathbb{L}_c^\Lambda{}_{\underline{M}}\,{\Scr O}_\mu^{_{\underline{M}}}=\frac{1}{2}\,f^\Lambda{}_{AB}\,{\Scr O}_\mu^{AB}+f^\Lambda{}_{I}\,{\Scr O}_\mu^{I}+h.c.\,,\label{traA}\\
\delta V_\mu{}^a&=i\,\bar{\epsilon}^A\gamma^a\psi_{\mu\,A}+i\,\bar{\epsilon}_A\gamma^a\psi_{\mu}^A\,,\label{traV}\\
\delta\psi_{A\,\mu}&={\Scr D}_\mu\epsilon_A-\frac{1}{8}\,F^-_{\rho\sigma\,AB}\gamma^{\rho\sigma}\gamma_\mu\epsilon^B+\dots\,,\label{trapsi}\\
\delta \chi_{ABC}&=i\, \partial_\mu\phi^s\,\mathcal{P}_{s\,ABCD}\gamma^\mu\epsilon^D-\frac{3i}{4}\,
F^-_{\mu\nu\,[AB}\gamma^{\mu\nu}\epsilon_{C]}+\dots\,,\label{trachi}\\
\delta \lambda_{I\,A}&=i \,\mathcal{P}_{s\,I\,AB}\,
\partial_\mu\phi^s\gamma^\mu\,\epsilon^B-\frac{i}{4}\,F_{\mu\nu\,I}^-\gamma^{\mu\nu}\epsilon_A+\dots\,,
\label{tralam}
\end{align}
where the ellipses refer to terms which are of higher order in the fermion fields.\footnote{Part of these terms, as usual in supergravity, would contribute, in the supersymmetry transformation laws of the fermionic fields, to the  \emph{super-covariantization} of the vector field strengths and the derivative of the scalars by adding to them fermion bilinears $\bar{\psi}\gamma\psi,\,\bar{\lambda}\gamma\psi$.}
We shall occasionally write the above fermion transformation laws in the following compact form:
\begin{equation}
\delta \lambda_{\mathcal{I}}=i\,\mathcal{P}_{s\,\mathcal{I}A}\partial_\mu\phi^s\,\gamma^\mu\epsilon^A+\dots\,,\label{deltalambdasym}
\end{equation}
where the ellipses now include also the terms depending on the vector fields and $\mathcal{P}_{\mathcal{I}A}$
stands for the different components of the vielbein 1-forms: $\mathcal{P}_{BCDA},\,\mathcal{P}_{I\,A},\,\mathcal{P}_{A\,\alpha}$.
\par The tensors $\Sigma_{ABCD}$  and $\Sigma_{I\,AB}$ in (\ref{trapsi}) are the components of the $\mathfrak{K}$-generator $\Sigma^c=\underline{{\Scr R}}_v^c[\Sigma]$, $\Sigma$ being defined in (\ref{delLsig}), according to the general matrix form (\ref{kmatrixf}). The components of $\Sigma$ entering (\ref{traphi}) read:
\begin{align}
\Sigma^{ABCD}&=-4\,\left(\bar{\epsilon}^{[A}\chi^{BCD]}+\frac{1}{24}\,\epsilon^{ABCDEFGH}\,\bar{\epsilon}_{[A}\chi_{BCD]}\right)\,\,;\,\,\,\,\,\,\,\,(\mathcal{N}=8)\,,\nonumber\\
\Sigma^{ABCD}&=-4\,\bar{\epsilon}^{[A}\chi^{BCD]}-\epsilon^{ABCDEF}\,\bar{\epsilon}_E\chi_F\,\,;\,\,\,\,\,\,\,\,\,(\mathcal{N}=6)\,,\nonumber\\
\Sigma^{ABCD}&=-4\,\bar{\epsilon}^{[A}\chi^{BCD]}+\epsilon^{ABCDE}\,\bar{\epsilon}_E\chi\,\,;\,\,\,\,\,\,\,\,\,(\mathcal{N}=5)\,,\nonumber\\
\Sigma^{ABCD}&=-4\,\bar{\epsilon}^{[A}\chi^{BCD]}\,\,;\,\,\,\,\,\,\,\,\,(\mathcal{N}=4)\,,\nonumber\\
\Sigma^{I\,AB}&=-2  \left(\bar{\epsilon}^{[A}\,\lambda^{I|B]}+\frac{1}{2}\,\epsilon^{ABCD}\,\bar{\epsilon}_{[C}\,\lambda_{I|D]}\right)\,\,;\,\,\,\,\,\,\,\,\,(\mathcal{N}=4)\,,\nonumber\\
\Sigma^{I\,AB}&=-2 \bar{\epsilon}^{[A}\,\lambda^{I|B]}+\bar{\epsilon}_C\lambda_I\,\epsilon^{ABC}\,\,;\,\,\,\,\,\,\,\,\,(\mathcal{N}=3)\,,\nonumber\\
\Sigma^{I\,AB}&=-2  \bar{\epsilon}^{[A}\,\lambda^{I| B]}\,\,;\,\,\,\,\Sigma^{A\alpha}=\bar{\lambda}^\alpha\epsilon^A+\epsilon^{AB}\mathbb{C}^{\alpha\beta}\bar{\lambda}_\beta\epsilon_B\,\,\,\,\,(\mathcal{N}=2)\,.\nonumber\\
\end{align}
The tensors ${\Scr O}_\mu^{AB},\,{\Scr O}_\mu^I$ in (\ref{traA}) are defined as follows:
\begin{align}
{\Scr O}_\mu^{AB}&\equiv i\,\bar{\epsilon}_C\gamma_\mu\chi^{ABC}-4\,\bar{\epsilon}^{[A}\psi^{B]}_\mu\,\,;\,\,\,{\Scr O}_\mu^{I}\equiv i\,\bar{\epsilon}_C\gamma_\mu \lambda^{IC}\,.\\
{\Scr O}_\mu^{\bullet}&\equiv i\,\bar{\epsilon}_C\gamma_\mu\chi^C\,\,\,;\,\,\,\,\,(\mathcal{N}=6)\,,
\end{align}
Notice that the supersymmetry transformation rules are manifestly $H$-covariant, being  written in terms of the fermion fields and the $H$-covariant composite fields $\mathcal{P}^c,\,\mathcal{Q}^c$ and $\mathbb{F}^{\underline{M}}_{\mu\nu}$ so that, when acting by means of $G$ on $\phi^s$ and $\mathcal{G}^M_{\mu\nu}$ and on the fermions by means of the corresponding compensating transformation in $H$, they retain the same form in the transformed quantities.\par
The ungauged supergravity Lagrangian density has the following general form:
\begin{equation}
{\Scr L}={\Scr L}_{scal.,k.}+{\Scr L}_{vect.,k.}+{\Scr L}_{fermi.,k.}+{\Scr L}_{scal.-fermi.}+{\Scr L}_{Pauli}+{\Scr L}_{4f}\,,\label{lagratot}
\end{equation}
where ${\Scr L}_{scal.,k.}+{\Scr L}_{vect.,k.}$ were given in (\ref{boslagr}), see also (\ref{scalagdk}), (\ref{vkinPauli}) and ${\Scr L}_{Pauli}$ was given in (\ref{vkinPauli}) and discussed previously. The kinetic terms of the spin-$3/2$ and  spin-$1/2$ fields in our notations read:\footnote{The kinetic term for each spin-$1/2$ field is normalized with a coefficient $-i/2$, so that the term for $\chi_{ABC}$ reads: $-\frac{ie}{12}\,(\bar{\chi}^{ABC}\gamma^\mu{\Scr D}_{\mu}\chi_{ABC}+\bar{\chi}_{ABC}\gamma^\mu{\Scr D}_{\mu}\chi^{ABC})$. As for the hyperinos, in line with the conventions  of \cite{Andrianopoli:1996cm}, one defines $\lambda^{\mathcal{I}=\alpha}=\sqrt{2}\,\zeta^\alpha$, so that the corresponding kinetic term reads:  $-ie\,(\bar{\zeta}^{\alpha}\gamma^\mu{\Scr D}_{\mu}\zeta_{\alpha}+\bar{\zeta}_{\alpha}\gamma^\mu{\Scr D}_{\mu}\zeta^{\alpha}).$}
\begin{equation}
{\Scr L}_{fermi.,k.}=\epsilon^{\mu\nu\rho\sigma}(\bar{\psi}^A_{\mu}\gamma_\nu{\Scr D}_\rho\psi_{A\sigma}-\bar{\psi}_{A\,\mu}\gamma_\nu{\Scr D}_\rho\psi^A_{\sigma})-\frac{i\,e}{2}\,(\bar{\lambda}^{\mathcal{I}}\gamma^\mu{\Scr D}_{\mu}\lambda_{\mathcal{I}}+\bar{\lambda}_{\mathcal{I}}\gamma^\mu{\Scr D}_{\mu}\lambda^{\mathcal{I}})\,.\label{fermik}
\end{equation}
Finally the scalar-fermion interaction terms ${\Scr L}_{scal.-fermi.}$ read:
\begin{align}
{\Scr L}_{scal.-fermi.}&=
-e\bar{\lambda}^{\mathcal{I}}\gamma^\mu\gamma^\nu\psi^B_\mu\,\partial_\nu\phi^s\mathcal{P}_{s\,\mathcal{I}B}
-e\bar{\lambda}_{\mathcal{I}}\gamma^\mu\gamma^\nu\psi_{B\mu}\,\partial_\nu\phi^s\mathcal{P}_s^{\mathcal{I}B}=\nonumber\\
&=-\frac{e}{6}\,\bar{\chi}^{ABC}\gamma^\mu\gamma^\nu\psi^D_\mu\,\partial_\nu\phi^s\mathcal{P}_{s\,ABCD}-
{e}\,\bar{\lambda}^{IA}\gamma^\mu\gamma^\nu\psi^B_\mu\,\partial_\nu\phi^s\mathcal{P}_{s\,IAB}+\dots\,.\label{scalfermiterms}
\end{align}
The last part ${\Scr L}_{4f}$ of the Lagrangian density (\ref{lagratot}) consists of the quartic terms in the fermion fields, which we shall not consider here. Such terms are given in the maximal and $\mathcal{N}=2$ theories in \cite{deWit:1982bul,deWit:2007mt} and \cite{Andrianopoli:1996cm}, respectively.
\paragraph{Parity.} Let us briefly comment on the action of a parity transformation on the fermion fields in the context of extended supergravities.  We have seen that there may exist isometries on the scalar fields which correspond in a suitable symplectic frame to an anti-symplectic duality transformation ${\bf P}$ of the form (\ref{Pmatrix}). We have shown that these are symmetries of the bosonic sector provided they are combined with a parity transformation on the spatial coordinates: $x^\mu=(x^0,\,\vec{x})\rightarrow x_p^\mu=(x^0,\,-\vec{x})$. The symmetry actually extends to the whole theory. This is apparent if we combine it with a CP-transformation (charge-conjugation+ parity) on generic fermion fields $\xi(x),\,\chi(x)$:
\begin{align}
\xi(x)&\rightarrow\,\xi'(x)= \gamma^0\,\xi_c(x_p)\,,\nonumber\\
i^k\bar{\xi}(x)\gamma^{\mu_1\dots \mu_k} \chi(x)&\rightarrow i^k\bar{\xi}_c(x_p)\gamma^{\mu_1\dots \mu_k} \chi_c(x_p)\eta^{\mu_1\mu_1}\dots\eta^{\mu_k\mu_k}=(i^k\bar{\xi}(x_p)\gamma^{\mu_1\dots \mu_k} \chi(x_p))^*\eta^{\mu_1\mu_1}\dots\eta^{\mu_k\mu_k}\,,\nonumber
\end{align}
where we have used the general properties (\ref{gprops}).
As a consequence of this the $H$-indices of the fermion current invert their (upper or lower) positions. For instance the reader can verify that the currents in the Pauli terms (\ref{vkinPauli}) undergo the following transformation: $$\mathcal{O}_{\underline{\Gamma}\,\mu\nu}(x)\rightarrow \mathcal{O}^{\underline{\Gamma}}_{\,\mu\nu}(x_p)\eta^{\mu\mu}\eta^{\nu\nu}=(\mathcal{O}_{\underline{\Gamma}\,\mu\nu}(x_p))^*\eta^{\mu\mu}\eta^{\nu\nu}\,.$$ This is consistent with the transformation properties of the blocks ${\bf f},\,{\bf h}$ of $\mathbb{L}_c$ under ${\bf P}$. The reader can indeed verify that: ${\bf f}\rightarrow \overline{\bf f}$ and ${\bf h}\rightarrow -\overline{\bf h}$. This amounts to a corresponding change in the position of the $H$-indices in $\mathbb{L}_c$ which contract the fermion currents.
Similarly $\mathcal{P}^{\mathcal{I}A}(x)\rightarrow \,\mathcal{P}_{\mathcal{I}A}(x_p)$. As a result of this transformation
each term in the Lagrangian is mapped into its complex conjugate computed in $x_p^\mu$. Being the Lagrangian real, the total effect on it is: ${\Scr L}(x)\rightarrow {\Scr L}(x_p)$, thus defining a symmetry of the action.
This discrete symmetry \cite{Ferrara:2013zga,AT} is present in a broad class of extended supergravities which include the maximal one, to be discussed in detail later.

\section{Gauging Supergravities} \label{sec:3}
We have reviewed the general features of ungauged supergravities, including their global symmetry group $G$. Now we want to discuss how to construct a gauged theory from an ungauged one.\par
The gauging procedure consists in promoting a suitable global symmetry group $G_g\subset G_{el}$ of the Lagrangian to a local symmetry group gauged by the vector fields of the theory. Besides the introduction of minimal couplings, changes in the Lagrangian and the supersymmetry transformation rules are required in order for the resulting model to feature the same supersymmetries as the original ungauged one. As pointed out in the Introduction, the first examples of gauged extended supergravities, see for instance \cite{Freedman:1976aw},\cite{Freedman:1978ra},\cite{Zachos:1978iw}, date back to the very first years following the discovery of supergravity and feature some of the characteristic properties of this type of models such as fermion masses, spontaneous supersymmetry breaking and a scalar potential (or a cosmological constant when scalars were not present). The first gauged maximal supergravity was constructed in 1982 by de Wit and Nicolai \cite{deWit:1981sst,deWit:1982bul}. It features an ${\rm SO}(8)$ local internal symmetry gauged by the 28 vector fields of the theory and was later put in correspondence with the Freund-Rubin compactification of eleven-dimensional supergravity on a round seven-sphere \cite{deWit:1983vq,deWit:1986oxb}. Generalizations of this model were devised by Hull in the late eighties \cite{Hull:1984yy,Hull:1984vg,Hull:1984rt} and feature gauge groups which are non-compact forms (${\rm SO}(p,q),\,p+q=8$) of ${\rm SO}(8,\mathbb{C})$ and non-semisimple contractions thereof, see Section \ref{startexa}. Meanwhile, still in the eighties, early progress in the understanding of the mathematical structure of $\mathcal{N}=2$ supergravity led to the construction of new gauged models \cite{deWit:1984rvr}. Later developments in the knowledge of the geometry underlying ungauged $\mathcal{N}=2$ theories, in particular of special and quaternionic geometries, allowed to further broaden the class of $\mathcal{N}=2$ gauged models. We shall review this topic in Sect. \ref{N2sugras}.
\par As mentioned earlier, in Sect.\ \ref{sframes}, the choice of the electric-magnetic symplectic frame is not physically relevant in ungauged models: It may affect the Lagrangian description but not the set of field equations and Bianchi identities. The introduction of the minimal couplings, however, explicitly breaks the original electric-magnetic duality invariance and the initial choice of the symplectic frame has physical implications on the resulting gauged model: Different frames correspond to inequivalent ungauged Lagrangians with different global symmetries $G_{el}$ and thus different choices of possible gauge groups; even gauging a same group $G_g$ in different frames may yield inequivalent gauged models. This latter feature was originally exploited in $\mathcal{N}=4$ supergravity \cite{deRoo:1985jh,Wagemans:1990mv}. In the maximal theory the freedom in choosing the original electric-magnetic frame led to the construction of physically interesting gauged models in \cite{Andrianopoli:2002mf,deWit:2002vt,Hull:2002cv,deWit:2007mt} and, more recently, in \cite{Dall'Agata:2011aa,Dall'Agata:2012bb,Dall'Agata:2012sx,Dall'Agata:2014ita}. We shall discuss them in some detail in Section \ref{startexa}.\par
To illustrate the construction of a gauged supergravity we start from an ungauged one in a given symplectic frame, to be referred to as the \emph{electric frame}.
 As we shall see, not all global symmetry groups $G_g\subset G_{el}$ of the corresponding Lagrangian, are admissible gauge groups, that is can be promoted to local symmetries. Consistency conditions on the corresponding Lie algebra have to be imposed, which on the one hand allow to construct a gauged Lagrangian with local symmetry $G_g$, and, on the other hand, guarantee its supersymmetric completion.\par
In the following, we will employ a covariant formalism in which the possible gaugings are encoded into an object called embedding tensor, that can be characterized group-theoretically \cite{Cordaro:1998tx,Nicolai:2000sc,deWit:2002vt}: It defines the embedding of the gauge algebra inside the global symmetry one by associating with each vector field a given combination of the generators of $G_{el}$. It is possible to free its definition
from the original choice of the symplectic frame, i.e. from a specific Lagrangian description, by characterizing it as a formally $G$-covariant tensor. This is done by
writing, through the action of a symplectic transformation $E$, the same embedding tensor in a generic frame, different from the electric one, in which also magnetic vector fields will be involved in the gauging. The embedding tensor then specifies which, among the full set of electric and magnetic vector fields in the new frame, are the gauge vectors associated with which global symmetry generator in $\mathfrak{g}$.
 Interestingly, in terms of this quantity, the consistency conditions on $G_g$ can be formulated as $G$-covariant relations which, as such, do not depend on the initial symplectic frame. By construction any viable gauging of the original ungauged theory solves these constraints. Vice versa, it can be shown that any embedding tensor satisfying the consistency conditions is associated with its own electric frame in which the vectors involved in the gauging are all electric and the group $G_g$ it defines is an off-shell symmetry (i.e. global symmetry of the corresponding ungauged action) and a viable gauge group. \par Since no assumption is made on the original symplectic frame, we can view it as the electric one associated with a generic solution to the $G$-covariant consistency conditions.

\subsection{The Gauging in the Electric Frame}\label{gitef}

For pedagogical reasons we shall adopt a piecemeal approach in reviewing the gauging procedure. We first discuss, in Sect. \ref{gaugingsteps}, the gauging of extended supergravities starting from a theory in a generic symplectic frame and
describing the necessary steps which lead, from the choice of the gauge group $G_g$, to the construction of a locally $G_g$-invariant Lagrangian. In particular we derive the linear and quadratic constraints on the gauge generators which are required by the consistency of this construction. Eventually in Sect. \ref{cgaet2}, these conditions are rewritten in a $G$-covariant form using the embedding tensor description. In this way, as explained above, the choice of the gauge group is completely freed from the symplectic frame.  This is however not the case for the resulting gauged theory, which is formulated in the electric frame of the embedding tensor: Given a solution to the constraints, this has to be rotated, through a suitable symplectic transformation $E$, into its electric frame where the corresponding gauged theory is formulated. Finally, in Sect. \ref{thegaugedlag}, we discuss the supersymmetry completion of the supergravity with local $G_g$-invariance.\par
In Sect. \ref{sec:4}, we shall deal with a more general, frame-independent formulation of the gauging procedure in which the resulting theory is not formulated in the electric frame of the embedding tensor and the symplectic rotation $E$ is not needed. As a consequence of this the minimal couplings involve not only electric vector fields entering the kinetic terms, but also magnetic ones.
\subsubsection{The Gauging Procedure Step-by-Step}\label{gaugingsteps}
Consider an ungauged extended supergravity and let $G_{el}$ describe the global symmetry of its action. We wish to promote a subgroup $G_g$ of it to a local symmetry of the Lagrangian gauged by the vector fields. This implies, as a preliminary requirement, that the number of gauge generators should not exceed the number $n_v$ of vectors
\begin{equation}
\dim(G_g) ~\le~ n_v\;.\label{preliminary}
\end{equation}
The first condition for the global symmetry group $G_g$ to be a viable gauge group, is that there should exist a subset $\{A^{\hat{\Lambda}}\}$ of the vector fields%
\footnote{
We describe by hatted-indices those pertaining to the symplectic frame in which the Lagrangian is defined
}
which transform in the co-adjoint representation of $G_g$ under its global duality action. These fields will become the \emph{gauge vectors} associated with the \emph{generators} $X_{\hat\Lambda}$ of $G_g$. \par
We shall name \emph{electric frame} the symplectic frame defined by the vectors in our ungauged Lagrangian and labeled by hatted indices.\par
Note that, once the gauge group is chosen within $G_{el}$, its action on the various fields is fixed, being it defined by the action of $G_g$ as a global symmetry group of the ungauged theory (duality action on the vector field strengths, non-linear action on the scalar fields and indirect action through $H$-compensators on the fermionic fields): fields are thus automatically associated with representations of $G_g$.\par
After the initial choice of $G_g$ in $G_{el}$, the first part of the procedure is quite standard in the construction of non-Abelian gauge theories: we introduce a gauge-connection, gauge-curvature (i.e.\ non-Abelian field strengths) and covariant derivatives. We will also need to introduce an extra topological term needed for the gauging of the Peccei-Quinn transformations (\ref{deltaLC}). This will lead us to construct a gauged Lagrangian $\Lgaug^{(0)}$ with manifest local $G_g$-invariance. Consistency of the construction will imply constraints on the possible choices of $G_g$ inside $G$. The minimal couplings will however break supersymmetry.\par
The second part of the gauging procedure consists in further deforming the Lagrangian $\Lgaug^{(0)}$ in order to restore the original supersymmetry of the ungauged theory and, at the same time, preserving local $G_g$-invariance.
\smallskip

\paragraph{\textbf{Step 1.} Choice of the gauge algebra.}
We start by introducing the \emph{gauge connection}:
\begin{equation}
\Omega_{g}=\Omega_{g\,\mu}dx^\mu\;;
\quad \Omega_{g\,\mu}\equiv
g\,A^{\hat{\Lambda}}_\mu\,X_{\hat{\Lambda}}\,,\label{gconnection}
\end{equation}
$g$ being the coupling constant.
The gauge-algebra relations can be written in the general form
\begin{equation}
\left[X_{\hat{\Lambda}},\,X_{\hat{\Sigma}}\right]\=f_{{\hat{\Lambda}}{\hat{\Sigma}}}{}^{\hat{\Gamma}}\,X_{\hat{\Gamma}}\,,\label{gaugealg}
\end{equation}
and are characterized by the structure constants
$f_{{\hat{\Lambda}}{\hat{\Sigma}}}{}^{\hat{\Gamma}}$ satisfying the Jacobi identity:
\begin{equation}
f_{[{\hat{\Lambda}}{\hat{\Sigma}}}{}^{\hat{\Gamma}}f_{{\hat{\Delta}}]{\hat{\Gamma}}}{}^{\hat{\Pi}}=0\,.\label{Jacobigeneral0}
\end{equation} This closure condition should be regarded as a constraint on $X_{\hat{\Lambda}}$, since the structure constants are not generic but are fixed in terms of the action of the gauge generators on the vector fields as global symmetry generators of the original ungauged theory.
To understand this, let us recall that $G_g$ is a subgroup of $G_{el}$ and thus its electric-magnetic duality action, as a global symmetry group, will have the form (\ref{ge}) and thus the duality action on the vector field strengths and their duals of the infinitesimal generators $X_{\hat{\Lambda}}$ will be represented by the symplectic matrix in the ${\Scr R}_v$-representation
\footnote{The $n_v\times n_v$-blocks in (\ref{xsymp}) are actually referred to the ${\Scr R}_{v*}$-representation on covariant vectors:
\begin{equation}
\left(X_{\hat{\Lambda}}\right)_{\hat{M}}{}^{\hat{N}}\,=
\left(\begin{matrix}
X_{\hat{\Lambda}\hat{\Sigma}}{}^{\hat{\Gamma}}  &  X_{{\hat{\Lambda}}\,\hat{\Sigma}{\hat{\Gamma}}} \cr
\Zero &  X_{{\hat{\Lambda}}}{}^{\hat{\Sigma}}{}_{\hat{\Gamma}}
\end{matrix}\right)={\Scr R}_{v*}[X_{\hat{\Lambda}}]_{\hat{M}}{}^{\hat{N}}
\,.\label{xsymp2}
\end{equation}
This is the reason for the minus sign in the lower off-diagonal block in (\ref{xsymp}).}
\begin{equation}
\left(X_{\hat{\Lambda}}\right)^{\hat{M}}{}_{\hat{N}}\,=
\left(\begin{matrix}
X_{\hat{\Lambda}}{}^{{\hat{\Sigma}}}{}_{{\hat{\Gamma}}}  &  \Zero \cr
-X_{{\hat{\Lambda}}\,{\hat{\Sigma}}{\hat{\Gamma}}} &  X_{{\hat{\Lambda}}\,{\hat{\Sigma}}}{}^{\hat{\Gamma}}
\end{matrix}\right)={\Scr R}_v[X_{\hat{\Lambda}}]^{\hat{M}}{}_{\hat{N}}
\,,\label{xsymp}
\end{equation}
where $X_{\hat{\Lambda}}{}^{{\hat{\Gamma}}}{}_{{\hat{\Sigma}}}$ and $X_{{\hat{\Lambda}}\,{\hat{\Gamma}}}{}^{\hat{\Delta}}$ are the infinitesimal generators of the $A$ and $D$-blocks in (\ref{ge}) respectively, while $-X_{{\hat{\Lambda}}\,\hat{\Gamma}\hat{\Sigma}}$ describes the infinitesimal $C$-block.
It is worth emphasizing here that we do not identify the generator $X_{{\hat{\Lambda}}}$ with the symplectic matrix defining its electric-magnetic duality action. As pointed our in Sect.\ \ref{sframes}, there are isometries in $\N=2$ models which do not have duality action, see Eq.\ (\ref{qisom}), namely for which the matrix in (\ref{xsymp}) is null.\par
The variation of the field strengths under an infinitesimal transformation $\xi^{{\hat{\Lambda}}}\,X_{{\hat{\Lambda}}}$, whose duality action is described by (\ref{xsymp}), is:
\begin{equation}
\delta \mathcal{G}^{\hat{M}}=\xi^{{\hat{\Lambda}}}\,(X_{{\hat{\Lambda}}}){}^{\hat{M}}{}_{\hat{N}}\,\mathcal{G}^{\hat{N}}
\;\;\Rightarrow\;\;
\begin{cases}\delta F^{\hat{\Lambda}}=
\xi^{{\hat{\Gamma}}}X_{{\hat{\Gamma}}}{}^{\hat{\Lambda}}{}_{\hat{\Sigma}}\,F^{\hat{\Sigma}}\,,\cr \delta G_{\hat{\Lambda}}=-
\xi^{{\hat{\Gamma}}}X_{{\hat{\Gamma}}\,{\hat{\Lambda}}{\hat{\Sigma}}} F^{\hat{\Sigma}}+\xi^{{\hat{\Gamma}}}X_{{\hat{\Gamma}}{\hat{\Lambda}}}{}^{\hat{\Sigma}}\, G_{\hat{\Sigma}}\,.
\end{cases}
\label{deltas}
\end{equation}
The symplectic property of the matrix $\left(X_{\hat{\Lambda}}\right)^{\hat{M}}{}_{\hat{N}}$ implies
the following relations:
\begin{align}
X_{{\hat{\Lambda}} \hat{M}}{}^{\hat{P}}\,\mathbb{C}_{\hat{N} \hat{P}}=X_{{\hat{\Lambda}} \hat{N}}{}^{\hat{P}}\,\mathbb{C}_{\hat{M} \hat{P}}
\;\quad\Leftrightarrow\quad\;
\begin{cases}
X_{\hat{\Lambda}}{}^{{\hat{\Sigma}}}{}_{{\hat{\Gamma}}}\=-X_{{\hat{\Lambda}}{\hat{\Gamma}}}{}^{\hat{\Sigma}}\,,\cr
X_{{\hat{\Lambda}}\,{\hat{\Gamma}}{\hat{\Sigma}}}\=X_{{\hat{\Lambda}}\,{\hat{\Sigma}}{\hat{\Gamma}}}\,.
\end{cases}
\label{sympconde}
\end{align}
The condition that $A^{\hat{\Lambda}}_\mu$ transform in the co-adjoint representation of the gauge group, namely
\begin{equation}
\delta F^{\hat{\Lambda}}=
\xi^{{\hat{\Gamma}}}\,f_{{\hat{\Gamma}}{\hat{\Sigma}}}{}^{\hat{\Lambda}}F^{\hat{\Sigma}}\,,
\end{equation}
together with the transformation properties (\ref{deltas}), lead us to identify the structure constants of the gauge group in (\ref{gaugealg}) with
the diagonal blocks of the symplectic matrices $X_{\hat{\Lambda}}$:
\begin{equation}
f_{{\hat{\Gamma}}{\hat{\Sigma}}}{}^{\hat{\Lambda}}=-X_{{\hat{\Gamma}}{\hat{\Sigma}}}{}^{\hat{\Lambda}}\,,\label{idenfx}
\end{equation}
so that the closure condition reads
\begin{equation}
\left[X_{\hat{\Lambda}},\,X_{\hat{\Sigma}}\right]\=-X_{{\hat{\Lambda}}{\hat{\Sigma}}}{}^{\hat{\Gamma}}\,X_{\hat{\Gamma}}\,,\label{gaugealg2}
\end{equation}
and is a quadratic constraint on the tensor $X_{{\hat{\Lambda}}}{}^{\hat{M}}{}_{\hat{N}}$. The identification (\ref{idenfx}) also implies
\begin{align}
X_{({\hat{\Gamma}}{\hat{\Sigma}})}{}^{\hat{\Lambda}}=0\,.\label{lin1}
\end{align}
\medskip
The closure condition (\ref{gaugealg2}) can thus be interpreted in two equivalent ways:
\begin{itemize}
\item{the vector fields $A^{\hat{\Lambda}}_\mu$ transform in the co-adjoint representation of $G_g$ under its action as global symmetry, namely
     \begin{equation}
     {\bf n_v}=\text{co-adj}(G_g)\,;
     \end{equation}
     }
\item{the gauge generators $X_{{\hat{\Lambda}}}$ are invariant under the action of $G_g$ itself:
\begin{equation}
\delta_{{\hat{\Lambda}}}X_{{\hat{\Sigma}}}\equiv [X_{\hat{\Lambda}},\,X_{{\hat{\Sigma}}}]+X_{{\hat{\Lambda}}{\hat{\Sigma}}}{}^{\hat{\Gamma}}\,X_{\hat{\Gamma}}=0\,.
\end{equation}}
\end{itemize}
\medskip

\paragraph{\textbf{Step 2.} Introducing gauge curvatures and covariant derivatives.}
Having defined the gauge connection (\ref{gconnection}) we also define its transformation property under a local $G_g$-transformation $ {\bf g}(x)\in G_g$:
\begin{equation}
\Omega_g \;\rightarrow\; \Omega_g'={\bf g}\,\Omega_g\,{\bf g}^{-1}+d {\bf g}\,{\bf g}^{-1}=g\,A^{\prime \hat{\Lambda}}\,X_{\hat{\Lambda}}\,.\label{Omtrasg}
\end{equation}
Under an infinitesimal transformation \;${\bf g}(x)\equiv \Id+g\,\zeta^{\hat{\Lambda}}(x)\,X_{\hat{\Lambda}}$,\, Eq.\ (\ref{Omtrasg}) implies the following transformation property of the gauge vectors:
\begin{equation}
\delta A^{\hat{\Lambda}}_\mu\=\cD_\mu\zeta^{\hat{\Lambda}}~\equiv~ \partial_\mu\zeta^{\hat{\Lambda}}+g\,A_\mu^{\hat{\Sigma}}X_{\hat{\Sigma}\hat{\Gamma}}{}^{\hat{\Lambda}}\,\zeta^{\hat{\Gamma}} \,,
\end{equation}
where we have introduced the $G_g$-covariant derivative of the gauge parameter $\cD_\mu\zeta^{\hat{\Lambda}}$.\footnote{In what follows we shall denote by $\mathcal{D}_\mu$ the gauge-covariant derivative containing the Levi-Civita, the $H$ and the gauge connection $\Omega_g$, not to be mistaken for ${\Scr D}_\mu$ which did not contain $\Omega_g$.}\par
As usual in the construction of non-Abelian gauge-theories, we define the gauge curvature%
\footnote{
Here we use the following convention for the definition of the components of a form: $\omega_{(p)}=\frac{1}{p!}\,\omega_{\mu_1\dots\mu_p}\,dx^{\mu_1}\wedge \dots dx^{\mu_p}$
}
\begin{equation}
\mathcal{F}=F^{\hat{\Lambda}}\,X_{\hat{\Lambda}}=\frac{1}{2}\,F^{\hat{\Lambda}}_{\mu\nu}\,dx^\mu\wedge dx^\nu\,X_{\hat{\Lambda}}\equiv \frac{1}{g}\,(d\Omega_g-\Omega_g\wedge \Omega_g)\,,\label{calF}
\end{equation}
which, in components, reads:
 \begin{equation}
F_{\mu\nu}^{{\hat{\Lambda}}} \= \partial_\mu A^{\hat{\Lambda}}_\nu-\partial_\nu A^{\hat{\Lambda}}_\mu -
g\,f_{{\hat{\Gamma}}{\hat{\Sigma}}}{}^{\hat{\Lambda}}\,A^{\hat{\Gamma}}_\mu\,A^{\hat{\Sigma}}_\nu\= \partial_\mu A^{\hat{\Lambda}}_\nu-\partial_\nu A^{\hat{\Lambda}}_\mu +
g\,X_{{\hat{\Gamma}}{\hat{\Sigma}}}{}^{\hat{\Lambda}}\,A^{\hat{\Gamma}}_\mu\,A^{\hat{\Sigma}}_\nu\;.\label{defF}
\end{equation}
The gauge curvature transforms covariantly under a transformation $ {\bf g}(x)\in G_g$:
\begin{equation}
\mathcal{F} \;\rightarrow\; \mathcal{F}'={\bf g}\,\mathcal{F}\,{\bf g}^{-1}\,,\label{gaugecovF}
\end{equation}
and satisfies the Bianchi identity:
\begin{equation}
\cD\mathcal{F}\equiv d\mathcal{F}-\Omega_g\wedge \mathcal{F}+\mathcal{F}\wedge \Omega_g=0
\;\;\Leftrightarrow\;\;
\cD F^{{\hat{\Lambda}}}\equiv dF^{{\hat{\Lambda}}}+g\,X_{{\hat{\Sigma}}{\hat{\Gamma}}}{}^{{\hat{\Lambda}}}A^{{\hat{\Sigma}}}\wedge F^{{\hat{\Lambda}}}=0\,.
\end{equation}
In the original ungauged Lagrangian we then replace the Abelian field strengths by the new $G_g$-covariant ones:
\begin{equation}
\partial_\mu A^{\hat{\Lambda}}_\nu-\partial_\nu A^{\hat{\Lambda}}_\mu\,\,\rightarrow\,\,\,\,\partial_\mu A^{\hat{\Lambda}}_\nu-\partial_\nu A^{\hat{\Lambda}}_\mu +
g\,X_{{\hat{\Gamma}}{\hat{\Sigma}}}{}^{\hat{\Lambda}}\,A^{\hat{\Gamma}}_\mu\,A^{\hat{\Sigma}}_\nu\,.\label{replaceF}
\end{equation}
After having given the gauge fields a $G_g$-covariant description in the Lagrangian through the non-Abelian field strengths, we now move to the other fields.
The next step in order to achieve local invariance of the Lagrangian under $G_g$ consists in replacing ordinary derivatives by gauge-covariant ones
\begin{equation}
\partial_\mu\;\;\longrightarrow\;\;
\partial_\mu -\Omega_{g\mu}=\partial_\mu -g\,A^{\hat{\Lambda}}_\mu\,X_{\hat{\Lambda}} \,,\label{covder}
\end{equation}
thus introducing the \emph{minimal couplings} of the vectors to the other fields.\par
As it can be easily ascertained, the covariant derivatives satisfy the identity which is well known from gauge theories:
\begin{equation}
\cD^2=\dots-g\,\mathcal{F}=\dots-g\,F^{\hat{\Lambda}}\,X_{\hat{\Lambda}}
\quad\Leftrightarrow\quad
[\cD_\mu,\,\cD_\nu]=\dots-g\,F^{\hat{\Lambda}}_{\mu\nu}\,X_{\hat{\Lambda}}\,,\label{D2F}
\end{equation}
where, since $\cD_\mu$ also contains the space-time connection and the composite connection on the scalar manifold, the ellipses refer to terms depending on the space-time curvature and the curvature of the scalar manifold.
Aside from the vectors and the metric, the remaining bosonic fields are the scalars $\phi^s$, whose derivatives are covariantized using the Killing vectors $k_{\hat{\Lambda}}$ associated with
the action of the gauge generator $X_{\hat{\Lambda}}$ as an isometry:
\begin{equation}
\partial_\mu \;\;\longrightarrow\;\;
\cD_\mu\phi^s\=\partial_\mu \phi^s-g\,A_\mu^{\hat{\Lambda}}\,k^s_{\hat{\Lambda}}(\phi)\,,\label{covderphi}
\end{equation}
Care is needed for the fermion fields which, as we have discussed above, do not transform directly under $G$, but under the corresponding compensating transformations in $H$. This was taken into account by writing the $H$-connection $\mathcal{Q}$ in the fermion $H$-covariant derivatives. Now we need to promote such derivatives to $G_g$-covariant ones, by minimally coupling the fermions to the gauge fields. This is effected by modifying the $H$-connection.\par
For homogeneous scalar manifolds redefine the left-invariant 1-form $\Omega$ (pulled-back on space-time), introduced in (\ref{omegapro}), by a \emph{gauged} one obtained by covariantizing the derivative on the coset representative:
\begin{equation}
\Omega_\mu= L^{-1}\partial_\mu L
\;\;\;\longrightarrow\;\;\;
\hat{\Omega}_\mu\equiv L ^{-1}\cD_\mu L =
L^{-1}\left(\partial_\mu-g\,A^{\hat{\Lambda}}_\mu\,X_{\hat{\Lambda}}\right) L =\hat{\mathcal{P}}_\mu+\hat{\mathcal{Q}}_\mu
\label{hatOm}
\end{equation}
where, as usual, the space-time dependence of the coset representative is defined by the scalar fields $\phi^s(x)$:\; $\partial_\mu L\equiv \partial_s L\, \partial_\mu\phi^s$.\par
The \emph{gauged} vielbein and connection are related to the ungauged ones as follows:
\begin{align}
\hat{\mathcal{P}}_\mu&={\mathcal{P}}_\mu-g\,A^{\hat{\Lambda}}_\mu\,{\mathcal{P}}_{\hat{\Lambda}}\;;
\;\;\quad\;\;
\hat{\mathcal{Q}}_\mu=
{\mathcal{Q}}_\mu-g\,
A^{\hat{\Lambda}}_\mu\,{\mathcal{Q}}_{\hat{\Lambda}}\,,\label{gaugedPW}
\end{align}
the matrices ${\mathcal{P}}_{\hat{\Lambda}},\,{\mathcal{Q}}_{\hat{\Lambda}}$ being the projections onto $\mathfrak{K}$ and $\mathfrak{H}$, respectively, of $L^{-1}X_{\hat{\Lambda}}L$:
\begin{align}
{\mathcal{P}}_{\hat{\Lambda}}&\equiv \left.L^{-1}X_{\hat{\Lambda}}L\right\vert_{\mathfrak{K}}\;;
\quad\;
{\mathcal{Q}}_{\hat{\Lambda}}\equiv \left.L^{-1}X_{\hat{\Lambda}}L\right\vert_{\mathfrak{H}}\,.\label{wPproj}
\end{align}
Using Eq.\ (\ref{kespans2}) we can express the above quantities as follows:
\begin{align}
{\mathcal{P}}_{\hat{\Lambda}}&= k_{\hat{\Lambda}}^s\,\mathcal{P}_s{}^{\underline{s}}\,K_{{\underline{s}}}\;;
\quad
{\mathcal{Q}}_{\hat{\Lambda}}= -\frac{1}{2}\,\Ps_{\hat{\Lambda}}^{{\bf a}}\,J_{{\bf a}}-\frac{1}{2}\,\Ps_{\hat{\Lambda}}^{{\bf m}}\,J_{{\bf m}}\,,\label{gaugedPW2}
\end{align}
where $\Ps_{\hat{\Lambda}}^{{\bf a}}$ were defined in Sect.\ \ref{ghsect}.\par\smallskip
For non-homogeneous scalar manifolds we cannot use the construction (\ref{hatOm}) based on the coset representative. Nevertheless
we can still define $\Ps_{\hat{\Lambda}}^{{\bf m}},\,\Ps_{\hat{\Lambda}}^{{\bf a}}$ in terms of the Killing vectors, see discussion below Eq.\ (\ref{KRP}).
From these quantities one then defines gauged vielbein $\hat{P}_\mu$ and $H$-connection $\hat{\mathcal{Q}}_\mu$ using (\ref{gaugedPW}) and (\ref{gaugedPW2}), where now $K_{{\underline{s}}}$ should be intended as a basis of the tangent space of the manifold at the origin (and not as isometry generators) and $\{J_{{\bf a}},\,J_{{\bf m}}\}$ a basis of the holonomy group.\par
Notice that, as a consequence of eqs.\ (\ref{gaugedPW2}) and (\ref{gaugedPW}), the gauged vielbein 1-forms (pulled-back on space-time) can be written as the ungauged ones in which the derivatives on the scalar fields are replaced by the covariant ones (\ref{covderphi}). This is readily seen by applying the general formula (\ref{kespans}) for homogeneous manifolds to the isometry $X_{\hat{\Lambda}}$ in (\ref{hatOm}), and projecting both sides of this equation on the coset space $\mathfrak{K}$:
\begin{equation}
\hat{\mathcal{P}}_\mu=\mathcal{P}_s\,\cD_\mu z^s\,.\label{hatPPDz}
\end{equation}
Consequently the replacement (\ref{covderphi}) is effected by replacing everywhere in the Lagrangian $\P_\mu$ by $\hat{\P}_\mu$.\par
Consider now a local $G_g$-transformation ${\bf g}(x)$ whose effect on the scalars is described by Eq.\ (\ref{gLh}):\; ${\bf g}L(\phi)=L({\bf g}\star\phi)\,h(\phi,{\bf g})$.\; From (\ref{hatOm}) and from the fact that $\cD$ is the $G_g$-covariant derivative, the reader can easily verify that:
\begin{equation}
\hat{\Omega}_\mu({\bf g}\star  \phi)=h\,\hat{\Omega}_\mu(\phi)\,h^{-1}+hdh^{-1}
\;\;\Rightarrow\;\;
\begin{cases}
\hat{\P}({\bf g}\star  \phi)=h\,\hat{\P}(\phi)\,h^{-1}\,,\cr
\hat{\mathcal{Q}}({\bf g}\star  \phi)=h\,\hat{\mathcal{Q}}(\phi)\,h^{-1}+hdh^{-1}\,,
\end{cases}
\label{PWhattra}
\end{equation}
where $ h=h(\phi,{\bf g})$.
By deriving (\ref{hatOm}) we find the \emph{gauged} Maurer-Cartan equations:
\begin{equation}
d\hat{\Omega}+\hat{\Omega}\wedge \hat{\Omega}=-g\,L^{-1}\mathcal{F}L\;,
\end{equation}
where we have used (\ref{D2F}). For symmetric spaces,\footnote{Analogous formulae hold also for non-symmetric spaces.} projecting the above equation onto $\mathfrak{K}$ and $\mathfrak{H}$ we find the gauged version of eqs.\ (\ref{DP}), (\ref{RW}):
\begin{align}
{\Scr D}\hat{\P}&\equiv d\hat{\P}+\hat{\mathcal{Q}}\wedge \hat{P}+\hat{P}\wedge \hat{\mathcal{Q}}=-g\,F^{\hat{\Lambda}}\,{\mathcal{P}}_{\hat{\Lambda}}\,,\label{DP2}\\
\hat{R}(\hat{\mathcal{Q}})&\equiv d\hat{\mathcal{Q}}+\hat{\mathcal{Q}}\wedge \hat{\mathcal{Q}}=-\hat{\P}\wedge \hat{\P}-g\,F^{\hat{\Lambda}}\,{\mathcal{Q}}_{\hat{\Lambda}}\,.\label{RW2}
\end{align}
The above equations are manifestly $G_g$-covariant. Using (\ref{hatPPDz}) one can easily verify that the gauged curvature 2-form (with value in $\mathfrak{H}$) can be written in terms of the curvature components $R_{rs}$ of the manifold, given in Eq.\ (\ref{Rcompo}), as follows:
\begin{equation}
\hat{R}(\hat{\mathcal{Q}})=\frac{1}{2}\,R_{rs}\,\mathcal{ D}\phi^r\wedge \mathcal{ D}\phi^s-g\,F^{\hat{\Lambda}}\,{\mathcal{Q}}_{\hat{\Lambda}}\,.
\end{equation}
The gauge-covariant derivatives, when acting on a generic fermion field $\xi$, is defined using $\hat{\mathcal{Q}}_\mu$, so that (\ref{Dxi}) is replaced by
\begin{equation}
\cD_\mu\xi=\nabla_\mu\xi+\hat{\mathcal{Q}}_\mu\star \xi\,.\label{Dxi2}
\end{equation}
Summarizing, local invariance of the action under $G_g$ requires replacing everywhere in the Lagrangian the Abelian field strengths by the non-Abelian ones, Eq.\ (\ref{replaceF}) and the ungauged vielbein $\mathcal{P}_\mu$ and $H$-connection $\mathcal{Q}_\mu$ by the gauged ones:
\begin{equation}
\mathcal{P}_\mu\;\rightarrow\;\hat{\mathcal{P}}_\mu\;;
\;\quad\;
\mathcal{Q}_\mu\;\rightarrow\;\hat{\mathcal{Q}}_\mu\,.\label{replacePW}
\end{equation}
Clearly supersymmetry of the gauged action requires as a necessary, though not sufficient, condition to perform the above replacements also in the supersymmetry transformation laws of the fields.
\par

\paragraph{\textbf{Step 3.} Introducing topological terms.}
If the symplectic duality action (\ref{xsymp}) of $X_{\hat{\Lambda}}$ has a non-vanishing off-diagonal block $X_{{\hat{\Lambda}}{\hat{\Gamma}}{\hat{\Sigma}}}$, that is if the gauge transformations include Peccei-Quinn shifts, then an infinitesimal (local) gauge transformation $\xi^{\hat{\Lambda}}(x)\,X_{{\hat{\Lambda}}}$ would produce a variation of the Lagrangian of the form (\ref{deltaLC}):
\begin{equation}
\delta\LB=-
\frac{g}{8}\,\xi^{\hat{\Lambda}}(x)X_{{\hat{\Lambda}}{\hat{\Gamma}}{\hat{\Sigma}}}\epsilon^{\mu\nu\rho\sigma}\,
F^{\hat{\Gamma}}_{\mu\nu}F^{\hat{\Sigma}}_{\rho\sigma}\,.\label{deltaLX}
\end{equation}
Being $\xi^{\hat{\Lambda}}(x)$ a local parameter, the above term is no longer a total derivative and thus the transformation is not a symmetry of the action.
In \cite{deWit:1984rvr}, see also discussion in Sect. \ref{sec:4}, it was proven that the variation (\ref{deltaLX}) can be canceled by adding to the Lagrangian a topological term of the form
\begin{equation}
\L_{\rm top.}
=-\frac{1}{3}\,g\,\epsilon^{\mu\nu\rho\sigma}\,X_{{\hat{\Lambda}}{\hat{\Gamma}}{\hat{\Sigma}}}\;A^{\hat{\Lambda}}_\mu\,
A^{\hat{\Sigma}}_\nu\,
\left(\partial_\rho A^{\hat{\Gamma}}_\sigma+\frac{3}{8}\,g\,X_{{\hat{\Delta}}{\hat{\Pi}}}{}^{\hat{\Gamma}}\,A^{\hat{\Delta}}_\rho\,A^{\hat{\Pi}}_\sigma\right)
\,,\label{top}
\end{equation}
provided the following condition holds
\begin{equation}
X_{({\hat{\Lambda}}{\hat{\Gamma}}{\hat{\Sigma}})}\=0\;.\label{xsymmetr}
\end{equation}
We shall see in the following that condition (\ref{xsymmetr}), together with the closure constraint (\ref{gaugealg2}), is part of a set of constraints on the gauge algebra which are also implied by supersymmetry.

\subsubsection{Choice of the Gauge Algebra and the Embedding Tensor}\label{cgaet2}
We have seen that the gauging procedure corresponds to promoting some suitable subgroup $G_g\subset G_{el}$ to local symmetry. This subgroup is defined selecting a subset of generators within the global symmetry algebra $\mathfrak{g}$ of $G$. Now, all the information about the gauge algebra can be encoded in a $G_{el\,}$-covariant object $\Theta$, which expresses the gauge generators as linear combinations of the global symmetry generators $t_\sigma$ of the subgroup $G_{el}\subset G$
\begin{equation}
X_{\hat{\Lambda}}=\Theta_{\hat{\Lambda}}{}^\sigma\,t_\sigma\;; \quad\quad
\Theta_{\hat{\Lambda}}{}^\sigma \in {\bf n_v}\times\Adj(G_{el})\;,
\label{gentheta}
\end{equation}
with \,${\hat{\Lambda}}=1,\,\dotsc,\,n_v$\; and with \,$\sigma=1,\dotsc,\,\dim(G_{el})$. \;The advantage of this description is that the $G_{el\,}$-invariance of the original ungauged Lagrangian $\L$ is restored at the level of the gauged Lagrangian $\L_{\rm gauged}$, to be constructed below, provided $\Theta_{\hat{\Lambda}}{}^\sigma$ is
transformed under $G_{el}$ as well. However, the full global symmetry group $G$ of the field equations and Bianchi identities is still broken, since the parameters $\Theta_{\hat{\Lambda}}{}^\sigma$ can be viewed as electric charges, whose presence manifestly breaks electric-magnetic duality invariance. In other words we are working in a specific symplectic frame defined by the ungauged Lagrangian we started from. \par
\medskip
We shall give later on a definition of the gauging procedure which is completely freed from the choice of the symplectic frame. For the time being, it is useful to give a description of the gauge algebra (and of the consistency constraints on it) which does not depend on the original symplectic frame, namely which is manifestly $G$-covariant. This is done by encoding all information on the initial symplectic frame in a symplectic matrix $E\equiv(E_M{}^N)$ and writing the gauge generators, through this matrix, in terms of new generators
\begin{equation}
X_M=(X_\Lambda,\,X^\Lambda)
\end{equation}
which are at least twice as many as the $X_{\hat{\Lambda}}$:
\begin{eqnarray}
\left(\begin{matrix}
X_{\hat{\Lambda}} \cr 0
\end{matrix}\right)
=E\,\left(\begin{matrix}
X_\Lambda\cr X^\Lambda
\end{matrix}\right)
\;.\label{EXL}
\end{eqnarray}
This description is clearly redundant and this is the price we have to pay in order to have a manifestly symplectic-covariant formalism. We can then rewrite the gauge connection in a symplectic-invariant  fashion
\begin{align}
A_\mu^{\hat{\Lambda}}\,X_{\hat{\Lambda}} = A_\mu^{\hat{\Lambda}}\,E_{\hat{\Lambda}}{}^\Lambda\,X_\Lambda  +A_\mu^{\hat{\Lambda}}\,E_{{\hat{\Lambda}}\,\Lambda}\,X^\Lambda
=A_\mu^\Lambda\,X_\Lambda+A_{\Lambda\,\mu}\,X^\Lambda=A^M_\mu\,X_M\;,\label{syminvmc}
\end{align}
where we have introduced the vector fields $A^\Lambda_\mu$ and the corresponding dual ones $A_{\Lambda\,\mu}$, that can be regarded as components of a symplectic vector
\begin{eqnarray}
A_\mu^M\equiv(A_\mu^\Lambda,\,A_{\Lambda\,\mu})\,.
\end{eqnarray}
 These are not independent, since they are all expressed in terms of the only electric vector fields $A_\mu^{\hat{\Lambda}}$ of our theory (those entering the vector kinetic terms):
\begin{eqnarray}
A_\mu^\Lambda=E_{\hat{\Lambda}}{}^\Lambda\,A^{\hat{\Lambda}}_\mu\;,\qquad
A_{\Lambda\,\mu}=E_{{\hat{\Lambda}}\,\Lambda}\,A^{\hat{\Lambda}}_\mu\;.
\end{eqnarray}
In what follows, it is useful to adopt this symplectic covariant description in terms of $2n_v$ vector fields $A_\mu^M$ and $2n_v$ generators $X_M$, bearing in mind the above definitions through the matrix $E$, which connects our initial symplectic frame to a generic one.\par
\smallskip
The components of the symplectic vector $X_M$ are generators in the isometry algebra $\mathfrak{g}$ and thus can be expanded in a basis $t_\alpha$ of generators of $G$:
\begin{equation}
X_M=\Theta_M{}^\alpha\,t_\alpha\,,\qquad \alpha=1,\dotsc,\,\dim(G)
\,.\label{Thdef}
\end{equation}
The coefficients of this expansion $\Theta_M{}^\alpha$ represent an extension of the definition of $\Theta$ to a $G$-covariant tensor:
\begin{equation}
\Theta_{\hat{\Lambda}}{}^\sigma \,\;\dashrightarrow\;\;
\Theta_M{}^\alpha \equiv (\Theta^{\Lambda\,\alpha},\;\Theta_\Lambda{}^\alpha)\,;
\qquad \Theta_M{}^\alpha\,\in\,\Rs_{v*}\times\Adj(G)
\,,\label{embtens}
\end{equation}
which (locally) describes the explicit embedding of the gauge group $G_g$ into the global symmetry group $G$, and combines the full set of deformation parameters of the original ungauged Lagrangian. The advantage of this description is that it allows to recast all the consistency conditions on the choice of the gauge group into $G$-covariant (and thus independent of the symplectic frame) constraints on $\Theta$.\par
We should however bear in mind that, just as the redundant set of vectors $A^M_\mu$, also the components of $\Theta_M{}^\alpha$ are not independent since, by Eq.\ (\ref{EXL}),
\begin{equation}
\Theta_{\hat{\Lambda}}{}^\alpha=E_{\hat{\Lambda}}{}^M\,\Theta_M{}^\alpha\;,
\;\quad\;
0=\Theta^{\hat{\Lambda}\,\alpha}=E^{\hat{\Lambda}\,M}\,\Theta_M{}^\alpha
\,,\label{elET}
\end{equation}
so that
\begin{equation}
\dim(G_g)\=\rank\left(\Theta_{\hat{\Lambda}}{}^\alpha\right)=\rank(\Theta_M{}^\alpha)\,.
\end{equation}
Being $E$ a symplectic matrix, i.e. $E_{\hat{M}}{}^M\,E_{\hat{N}}{}^N\,\mathbb{C}^{\hat{M}\hat{N}}=\mathbb{C}^{MN}$, the above relations (\ref{elET}) imply for $\Theta_M{}^\alpha$ the following condition:
\begin{equation}
\Theta_\Lambda{}^\alpha\,\Theta^{\Lambda\,\beta}-\Theta_\Lambda{}^\beta\,\Theta^{\Lambda\,\alpha}=0
\quad\Leftrightarrow\quad
\mathbb{C}^{MN}\Theta_M{}^\alpha\Theta_N{}^\beta=0\quad
\,.\label{locality}
\end{equation}
To see this it suffices to note that the above equation is manifestly ${\rm Sp}(2n_v,\mathbb{R})$-invariant. Therefore, being it satisfied by the original embedding tensor $\Theta_{\hat{M}}{}^\alpha=(\Theta_{\hat{\Lambda}}{}^\alpha,\,0)$ in the electric frame, it will be satisfied by the same tensor in a generic frame $\Theta_M{}^\alpha$, being the former related to the latter by the symplectic transformation $E$:
\begin{equation}
0=\Theta_{\hat{\Lambda}}{}^\alpha\,\Theta^{\hat{\Lambda}\,\beta}-\Theta_{\hat{\Lambda}}{}^\beta\,\Theta^{\hat{\Lambda}\,\alpha}=
\mathbb{C}^{\hat{M}\hat{N}}\Theta_{\hat{M}}{}^\alpha\Theta_{\hat{N}}{}^\beta=E_{\hat{M}}{}^M\,E_{\hat{N}}{}^N\,\mathbb{C}^{\hat{M}\hat{N}}\Theta_{{M}}{}^\alpha
\Theta_{{N}}{}^\beta=\mathbb{C}^{MN}\Theta_{{M}}{}^\alpha
\Theta_{{N}}{}^\beta\,,
\end{equation}
where the first equality holds since $\Theta^{\hat{\Lambda}\,\alpha}=0$.\par
Vice versa, one can show that if $\Theta_M{}^\alpha$ satisfies the above conditions, there exists a symplectic matrix $E$ which can rotate it to an electric frame, namely such that eqs.\ (\ref{elET}) are satisfied for some $\theta_{\hat{\Lambda}}{}^\alpha$.\, Equations (\ref{locality}) define the so-called \emph{locality constraint} on the embedding tensor $\Theta_M{}^\alpha$ and they clearly imply:
\begin{equation}
\dim(G_g)=\rank(\Theta)\le n_v\;,
\end{equation}
which is the preliminary consistency condition (\ref{preliminary}). The construction of the matrix $E$ from an embedding tensor satisfying the locality constraint will be discussed in Sect. \ref{backelectric} using the \emph{rank-factorization} of $\Theta_M{}^\alpha$.\par
The electric-magnetic duality action of $X_M$, in the generic symplectic frame defined by the matrix $E$, is described by the tensor:
\begin{equation}
X_{MN}{}^P~\equiv~{\Scr R}_{v*}[X_M]_N{}^P=
\Theta_M{}^\alpha\,t_{\alpha\,N}{}^P\=
E^{-1}{}_M{}^{\hat{M}}E^{-1}{}_N{}^{\hat{N}}\,X_{\hat{M}\hat{N}}{}^{\hat{P}}\,E_{\hat{P}}{}^P\,,\label{XEhatX}
\end{equation}
where we have used Eq. (\ref{Thdef}).
For each value of the index {\footnotesize $M$}, the tensor $X_{MN}{}^P$ should generate symplectic transformations. This implies that:
\begin{equation}
X_{MNP}\equiv X_{MN}{}^Q\mathbb{C}_{QP}=X_{MPN}\,,
\end{equation}
which is equivalent to eqs.\ (\ref{sympconde}). The remaining linear constraints (\ref{lin1}), (\ref{xsymmetr}) on the gauge algebra can be recast in terms of $X_{MN}{}^P$ in the following symplectic-covariant form:
\begin{equation}
X_{(MNP)}=0
\;\quad\Leftrightarrow\quad\;
\begin{cases}
2\,X_{(\Lambda\Sigma)}{}^\Gamma=X^\Gamma{}_{\Lambda\Sigma}\,,\cr
2\,X^{(\Lambda\Sigma)}{}_\Gamma=X_\Gamma{}^{\Lambda\Sigma}\,,\cr X_{(\Lambda\Sigma\Gamma)}=0\,.
\end{cases}
\label{lconstr}
\end{equation}
Notice that the second of equations (\ref{lconstr}) implies that, in the electric frame in which $X^{\hat{\Lambda}}=0$, also the $B$-block (i.e.\ the upper-right one) of the infinitesimal gauge generators ${\Scr R_v}[X_{\hat{\Lambda}}]$ vanishes, being $X_{\hat{\Gamma}}{}^{\hat{\Lambda}\hat{\Sigma}}=0$, so that the gauge transformations are indeed in $G_{el}$. Moreover the antisymmetry of $X_{\hat{\Lambda}\hat{\Sigma}}{}^{\hat{\Gamma}}$ in the first two indices, i.e. Eq. (\ref{lin1}), follows from the first of Eqs. (\ref{lconstr}) in the electric frame. \par
Equations (\ref{lconstr}) can also be written in the following form:
\begin{equation}
X_{(MN)}{}^P=-\frac{1}{2}\,\mathbb{C}^{PL}\,X_{LM}{}^Q\mathbb{C}_{QN}=
-\frac{1}{2}\,\mathbb{C}^{PL}\,\Theta_L{}^\alpha\,t_{\alpha\,MN}\,,\label{lconstrnew}
\end{equation}
where $t_{\alpha\,MN}$ were defined above Eq. (\ref{tddsym}).\par
Finally, also the closure constraints (\ref{gaugealg2}) can be cast in a manifestly symplectic-covariant form:
\begin{equation}
[X_M,\,X_N]=-X_{MN}{}^P\,X_P
\quad\Leftrightarrow\quad
\Theta_M{}^\alpha\Theta_N{}^\beta{\rm f}_{\alpha\beta}{}^\gamma+\Theta_M{}^\alpha\,t_{\alpha\,N}{}^P\Theta_P{}^\gamma=0\,.
\end{equation}
The above condition is equivalently stated as the characteristic property of the \emph{embedding tensor $\Theta_M{}^\alpha$ of being invariant under the action of the gauge group it defines}:
\begin{align}
\delta_M\Theta_N{}^\alpha=0\,.\\
\nn
\end{align}
Summarizing we have found that consistency of the gauging requires the following set of linear and quadratic algebraic, $G$-covariant constraints to be satisfied by the embedding tensor:
\begin{itemize}
\item{\emph{Linear constraint:}
   \begin{align}
   X_{(MNP)}&=0\,,\label{linear2}
   \end{align}
    }
\item{\emph{Quadratic constraints:}
   \begin{align}
   &\mathbb{C}^{MN}\Theta_M{}^\alpha\Theta_N{}^\beta=0\,,\label{quadratic1}\\
   &[X_M,\,X_N]=-X_{MN}{}^P\,X_P\,.\label{quadratic2}
   \end{align}
    }
\end{itemize}
The above conditions on the embedding tensor as well as useful identities deduced from them are summarized in Appendix \ref{apXid}.
The linear constraint (\ref{linear2}) amounts to a projection of the embedding tensor on a specific $G$-representation $\Rs_\Theta$ in the decomposition of the product $\Rs_{v*}\times {\rm Adj}(G)$ with respect to $G$
\begin{equation}
\Rs_{v*}\times {\rm Adj}(G)\;\;\stackrel{G}{\longrightarrow}\;\;
\Rs_\Theta +\, \dots\label{decomvad}
\end{equation}
and thus can be formally written as follows:
\begin{equation}
\mathbb{P}_\Theta\cdot \Theta=\Theta\,,\label{PTheta}
\end{equation}
where $\mathbb{P}_\Theta$ denotes the projection on the representation $\Rs_\Theta$. For this reason (\ref{linear2}) is also named \emph{representation constraint}.\par
Let us now prove that, if $G$ is simple (as in maximal supergravity), (\ref{linear2}) implies that the trace $X_{NM}{}^N$ must vanish:\footnote{The trace $X_{MN}{}^N$ vanishes by construction, being $(X_M)_N{}^P$ a matrix in the algebra of ${\rm Sp}(2n_v,\mathbb{R})$.}
\begin{equation}
X_{(MNP)}=0\,\,\Rightarrow\,\,\,\,\,X_{NM}{}^N=0\,.\label{lconstr1imp2}
\end{equation}
To show this we note that at least one representation $\Rs_{v*}$ is always present in the decomposition (\ref{decomvad}) of the product $\Rs_{v*}\times {\rm Adj}(G)$. If $G$ is simple, this representation occurs just once in that decomposition.
There are two ways of extracting this $\Rs_{v*}$ component out of the $G$-tensor $X_{MN}{}^P$:
\begin{equation}
\Rs_{v*}:\,\,\,\,X_{NM}{}^N\,\,\,,\,\,\,\,\,K_{NPQM} X^{NPQ}\,,\label{thetwoRv}
\end{equation}
where $K_{MNPQ}$ is the characteristic rank-4, totally symmetric, invariant tensor defined in (\ref{Ktensor}). Being $\Rs_{v*}$ unique,
the above two quantities must be proportional to each other and single out a same component $\theta_M$ of $\Theta_M{}^\alpha$.\footnote{One can indeed show by direct computation that, for simple $G$, $K_{MNPQ}\,t_\alpha{}^{PQ}\propto t_{\alpha\,MN}$.}
Note however that only the totally symmetric part of $X^{NPQ}$, which vanishes by virtue of (\ref{linear2}), contributes to the last term in (\ref{thetwoRv}). We conclude that if $G$ is simple, $\theta_M=0$ and thus the implication (\ref{lconstr1imp2}) holds.
One can prove that there is a notable exception in which (\ref{lconstr1imp2}) is still valid while $G$ is not simple. This is the so-called STU model, to be discussed in detail in Sect. \ref{STUsolvq}, in which $G={\rm SL}(2,\mathbb{R})^3$.\footnote{In those models with scalar manifold of the form $\frac{{\rm SL}(2,\mathbb{R})}{{\rm SO}(2)}\times \frac{{\rm SO}(p,q)}{{\rm SO}(p)\times {\rm SO}(q)}$, in which $G$ is not simple and has the form $G={\rm SL}(2,\mathbb{R})\times {\rm SO}(p,q)$, one can prove that (\ref{lconstr1imp2}) holds only if $p+q=4$, which comprises the case of the STU model in which $p=q=2$, since ${\rm SO}(2,2)\sim {\rm SL}(2,\mathbb{R})^2$. } In this case there are three representations $\Rs_{v*}$ occurring in the decomposition of $\Rs_{v*}\times {\rm Adj}(G)$. Nevertheless the only one entering the decomposition of $X_{(MNP)}$, which is thus set to zero, is exactly $X_{NM}{}^N$.\par

The first quadratic constraint (\ref{quadratic1}) guarantees that a symplectic matrix $E$ exists which rotates the embedding tensor $\Theta_M{}^\alpha$ to an electric frame in which the \emph{magnetic components} $\Theta^{\hat{\Lambda}\,\alpha}$ vanish. The second one (\ref{quadratic2}) is the condition that the gauge algebra close within the global symmetry one $\mathfrak{g}$ and implies that $\Theta$ is a singlet with respect to $G_g$.
In the maximal theory we will show, by group theoretical arguments, that once the representation constraint is implemented, the locality condition is in fact equivalent to the other quadratic constraint (\ref{quadratic2}).
This is not the case for $\mathcal{N}<8$ theories where in general the locality and closure conditions (\ref{quadratic1}),(\ref{quadratic2}) should be imposed independently.
\par\smallskip
The second part of the gauging procedure, which we are going to discuss in Sect. \ref{thegaugedlag} below, has to do with restoring supersymmetry after minimal couplings have been introduced and the $G_g$-invariant Lagrangian $\Lgaug^{(0)}$ have been constructed.
As we shall see, the supersymmetric completion of $\Lgaug^{(0)}$ requires no more constraints on $G_g$ (i.e.\ on $\Theta$) than the linear (\ref{linear2}) and quadratic ones (\ref{quadratic1}), (\ref{quadratic2}) discussed above. In particular the representation condition on the embedding tensor will be required by the cancelation of the $O(g)$-terms originating from the supersymmetry variations of the minimal couplings (in particular the ones involving the fermion fields) by means of new terms depending on the so-called fermion shift-tensors to be introduced in Sect. \ref{thegaugedlag} below. In fact, although we have introduced it here from an analysis of the sole bosonic Lagrangian, in all supergravities the linear constraint is deduced directly from supersymmetry.
\par\medskip
  As a final remark let us prove that the locality constraint (\ref{quadratic1}) is independent of the others only in theories featuring scalar isometries with no duality action, namely in which the symplectic duality representation $\Rs_v$ of the isometry algebra $\mathfrak{g}$ is \emph{not faithful}. This is the case of the quaternionic isometries in $\N=2$ theories, see Eq.\ (\ref{qisom}) of Sect.\ \ref{sframes}. Let us split the generators $t_\alpha$ of $G$ into $t_{\ell}$, which have a non-trivial duality action, and $t_{m}$, which do not:
\begin{equation}
(t_{\ell})_M{}^N\neq 0\;; \;\quad\; (t_{m})_M{}^N= 0\,.
\end{equation}
From equation (\ref{quadratic2}) we derive, upon symmetrization of the {\footnotesize $M,\,N$} indices, the following condition:
\begin{equation}
X_{(MN)}{}^P\,X_{P}=X_{(MN)}{}^P\,\Theta_{P}{}^\alpha\,t_\alpha=0\,,\label{quad2n}
\end{equation}
where $t_\alpha$ on the right hand side are \emph{not} evaluated in the $\Rs_v$ representation and thus are all non-vanishing.
Using the linear constraint (\ref{linear2}) we can then rewrite $X_{(MN)}{}^P$ as follows:
\begin{equation}
X_{(MN)}{}^P=-\frac{1}{2}\,\mathbb{C}^{PQ}\,X_{QMN}=-\frac{1}{2}\,\mathbb{C}^{PQ}\,\Theta_Q{}^\ell t_{\ell\,MN}\,,
\end{equation}
so that (\ref{quad2n}) reads
\begin{equation}
\mathbb{C}^{QP}\,\Theta_Q{}^\ell\Theta_{P}{}^\alpha\,t_{\ell\,MN}\,t_\alpha\,=0
\,.\label{quad2nbis}
\end{equation}
Being $t_\alpha$ and $t_{\ell\,MN}$ independent for any $\alpha$ and $\ell$, conditions (\ref{linear2}) and (\ref{quadratic2}) only imply \emph{part of} the locality constraint (\ref{quadratic1}):
\begin{equation}
\mathbb{C}^{QP}\,\Theta_Q{}^\ell\Theta_{P}{}^\alpha=0\,,\label{quad2n2}
\end{equation}
while the remaining constraints (\ref{quadratic1})
\begin{equation}
\mathbb{C}^{QP}\,\Theta_Q{}^m\Theta_{P}{}^{n}=0\,,\label{quad2n3}
\end{equation}
need to be imposed independently. Therefore in theories in which all scalar fields sit in the same supermultiplets as the vector ones, as it is the case of $\N>2$ or $\N=2$ with no hypermultiplets, the locality condition (\ref{quadratic1}) is not independent but follows from the other constraints.\par

The linear constraint (\ref{linear2}) was related in \cite{DeRydt:2008hw} to the absence, in extended supergravities, of chiral anomalies. This implies that, in certain cases, it can be relaxed in $\mathcal{N}=1$ supergravities, which can be chiral.

\subsubsection{The Gauged Lagrangian}\label{thegaugedlag}
The three steps described above allow us to construct a Lagrangian $\Lgaug^{(0)}$ which is locally $G_g$-invariant starting from the ungauged one. Now we have to check if this deformation is compatible with local supersymmetry. As it stands, as emphasized earlier, the Lagrangian $\Lgaug^{(0)}$ is no longer invariant under supersymmetry, due to the extra contributions in its supersymmetry variation that originate from the vector fields in the covariant derivatives.\par
Consider, for instance, the supersymmetry variation of the (gauged) Rarita-Schwinger term in the Lagrangian
\begin{equation}
\L_{\textsc{rs}}=i\,e\,\bar{\psi}^A_\mu\gamma^{\mu\nu\rho}\cD_\nu\psi_{A\,\rho}~+~\text{h.c.}\;,
\end{equation}
where $\cD_\nu$ is the gauged covariant derivative defined in Eq.\ (\ref{Dxi2}). Under a supersymmetry variation of $\psi_\mu$:
\begin{equation}
\delta\psi_\mu=\cD_\mu \epsilon+\dots\,,\label{deltapsigau}
\end{equation}
$\epsilon$ being the local supersymmetry parameter,
\footnote{
The ellipses refer to terms containing the vector field strengths and the fermion fields.
}
the variation of $\L_{\textsc{rs}}$ produces the following term
\begin{align}
\delta\L_{\textsc{rs}}&\=\dots+2i\,e\,\bar{\psi}^A_\mu\gamma^{\mu\nu\rho}\cD_\nu \cD_{\rho}\epsilon_A~+~\text{h.c.}\=\nne
&\=\dots-i\,g\,e\,\bar{\psi}^A_\mu\gamma^{\mu\nu\rho}F_{\nu\rho}^{\hat{\Lambda}}\,(\mathcal{Q}_{\hat{\Lambda}}\epsilon)_A~+~\text{h.c.}
\;,\label{RSvar}
\end{align}
where we have used the property (\ref{D2F}) of the gauge covariant derivative. Similarly we can consider the supersymmetry variation, see Eq. (\ref{deltalambdasym}), of the spin-$1/2$ fields:
\begin{equation}
\delta \lambda^{\mathcal{I}}=i\,\hat{\mathcal{P}}_\mu^{\mathcal{I}\,A}\,\gamma^\mu\epsilon_A+\dots\,,
\end{equation}
where, as in  (\ref{deltapsigau}) the ellipses denote terms containing the vector fields together with higher-order terms in the fermion fields and $\hat{\mathcal{P}}_\mu^{\mathcal{I}\,A}$ is a specific component of the $\mathfrak{K}$-valued matrix $\hat{\mathcal{P}}_\mu$.
The resulting variation of the corresponding kinetic Lagrangian contains terms of the following form:
\begin{align}
\delta\left(-\frac{ie}{2}\,\bar{\lambda}_{\mathcal{I}}\gamma^\mu \cD_\mu\lambda^{\mathcal{I}}~+~\text{h.c.}\right)&=\dots+e\,\bar{\lambda}_{\mathcal{I}}\gamma^{\mu\nu} \cD_\mu\hat{P}_\nu^{\mathcal{I}\,A}\,\epsilon_A~+~\text{h.c.}=\nonumber\\&=\dots-\frac{g}{2}\,e\,\bar{\lambda}_{\mathcal{I}}\gamma^{\mu\nu} F_{\mu\nu}^{\hat{\Lambda}}\,{\mathcal{P}}_{\hat{\Lambda}}^{\mathcal{I}\,A}\,\epsilon_A~+~\text{h.c.}
\label{lambdatra}
\end{align}
where we have used, in going to the second line, Eq. (\ref{DP2}).
We see that the supersymmetry variation of the minimal couplings in the fermion kinetic terms have produced $O(g)$-terms which contain the tensor
\begin{equation}
F_{\mu\nu}^{\hat{\Lambda}}\,L^{-1}X_{\hat{\Lambda}} L=\mathcal{G}_{\mu\nu}^{M}\,L^{-1}X_M L
\label{genvar}
\end{equation}
projected on $\mathfrak{H}$ and contracted with the $\bar{\psi}\epsilon$ current in (\ref{RSvar}), or restricted to $\mathfrak{K}$ and contracted with the $\bar{\lambda}\epsilon$ current in the second case (\ref{lambdatra}). On the right hand side of (\ref{genvar}) the summation over the gauge generators has been written in the symplectic invariant form defined in Eq.\ (\ref{syminvmc}):\; $\mathcal{G}^M\, X_M\equiv F^{\hat{\Lambda}}\,E_{\hat{\Lambda}}{}^M\,X_M$\,.\, These are instances of the various terms occurring in the supersymmetry variation $\delta\Lgaug^{(0)}$.
Just as (\ref{RSvar}) and (\ref{lambdatra}), these terms are proportional to an $H$-tensor defined as follows%
\footnote{
In the formulas below we use the coset representative in which the first index (acted on by $G$) is in the generic symplectic frame defined by the matrix $E$ and which is then related to the same matrix in the electric frame (labeled by hatted indices) as follows:
\begin{equation}
L(\phi)_{\hat{M}}{}^{\underline{N}}~=~E_{\hat{M}}{}^P\,L(\phi)_{P}{}^{\underline{N}}
\quad\Rightarrow\quad
\mathcal{M}(\phi)_{\hat{M}\hat{N}}=E_{\hat{M}}{}^PE_{\hat{N}}{}^Q\mathcal{M}(\phi)_{PQ}\,,
\end{equation}
last equation being (\ref{MEtra})
}:
\begin{align}
\Tb(\Theta,\phi)_{\underline{M}}&~\equiv~ \LL(\phi)^{-1}{}_{\underline{M}}{}^N\,L(\phi)^{-1}X_N\,L(\phi)\=
\LL(\phi)^{-1}{}_{\underline{M}}{}^N\,\Theta_N{}^\beta\,L(\phi)_\beta{}^\alpha\,t_\alpha\=\nonumber\\
&\=\Tb(\Theta,\phi)_{\underline{M}}{}^\alpha\,t_\alpha
\,,\label{TT}
\end{align}
where
\begin{equation}
\Tb(\Theta,\phi)_{\underline{M}}{}^\alpha\equiv \LL(\phi)^{-1}{}_{\underline{M}}{}^N\,\Theta_N{}^\beta L(\phi)_\beta{}^\alpha=
(L^{-1}(\phi)\star \Theta)_{\underline{M}}{}^\alpha\,,
\end{equation}
where $\star$ denotes the action of $L^{-1}$ as an element of $G$ on $\Theta_M{}^\alpha$ in the corresponding $\Rs_\Theta$-representation. The matrix $\LL(\phi)_M{}^{\underline{M}}$ appearing in the above formulas is obtained by complexifying the right index of the matrix defined in (\ref{otherL}) and is related to the $\mathbb{L}_c$-matrix as follows (see Appendix \ref{spapp}) :
\begin{equation}
(\LL(\phi)_M{}^{\underline{N}})=-i\,(\mathbb{C}_{MP}\mathbb{C}^{\underline{NL}}  \mathbb{L}_{c}^{P}{}_{\underline{L}})=i\mathbb{C}\mathbb{L}_{c}\mathbb{C}=\mathbb{L}_{c}^{-T}\,.\label{otherL2}
\end{equation}
The tensor $\Tb(\phi,\,\Theta)=L^{-1}(\phi)\star \Theta$ is called the \emph{$\Tb$-tensor} and was first introduced in \cite{deWit:1981sst}.\par
From its definition and from (\ref{wPproj}) we can also write the $\Tb$-tensor as follows:
 \begin{align}
\Tb(\Theta,\phi)_{\underline{M}}&=\LL(\phi)^{-1}{}_{\underline{M}}{}^M\, \left({\mathcal{P}}_M+ {\mathcal{Q}}_M\right)\,,\label{TTPQ}
\end{align}
where, as usual, we have written $\hat{\mathcal{Q}}_{\hat{\Lambda}}=E_{\hat{\Lambda}}{}^M\,\hat{\mathcal{Q}}_{M},\,
\hat{\mathcal{P}}_{\hat{\Lambda}}=E_{\hat{\Lambda}}{}^M\,\hat{\mathcal{P}}_{M}$. In components, using (\ref{otherL2})
 we can write:
 \begin{align}
 \mathbb{T}_{AB}&=\mathbb{L}_c^M{}_{AB}\,({\mathcal{P}}_M+ {\mathcal{Q}}_M)=f^\Lambda{}_{AB}({\mathcal{P}}_\Lambda+ {\mathcal{Q}}_\Lambda)+h_{\Lambda\,AB}({\mathcal{P}}^\Lambda+ {\mathcal{Q}}^\Lambda)\,,\label{T11}\\
\mathbb{T}_{I}&=\mathbb{L}_c^M{}_{I}\,({\mathcal{P}}_M+ {\mathcal{Q}}_M)=f^\Lambda{}_{I}({\mathcal{P}}_\Lambda+ {\mathcal{Q}}_\Lambda)+h_{\Lambda\,I}({\mathcal{P}}^\Lambda+ {\mathcal{Q}}^\Lambda)\,.
 \end{align}
If $\Theta$ and $\phi$ are simultaneously transformed with $G$, the $\Tb$-tensor transforms under the corresponding $H$-compensator:
\begin{align}
\forall {\bf g}\in G\;:\quad
&\Tb({\bf g}\star\phi,\,{\bf g}\star\Theta)= L^{-1}({\bf g}\star\phi)\star ({\bf g}\star\Theta)=\nonumber\\
&=(h({\bf g},\phi)\,L^{-1}(\phi)\, {\bf g}^{-1})\star ({\bf g}\star\Theta)=h({\bf g},\phi)\star\Tb(\phi,\,\Theta)\,.
\end{align}
Thus the quantity $\Tb$ naturally belongs to a representation of the group $H$ and is an example of \emph{composite field}
discussed at the end of Sect.\ \ref{fsector}.\par
If, on the other hand, we fix $\phi$ and only transform $\Theta$, $\Tb$ transforms in the same $G$-representation $\Rs_\Theta$ as $\Theta$, being $\Tb$ defined by acting on the embedding tensor with the $G$-element $L^{-1}$.
As a consequence of this, $\Tb$ satisfies the same constraints (\ref{linear2}), (\ref{quadratic1}) and (\ref{quadratic2}) as $\Theta$:
\begin{align}
\Tb_{(\underline{MNP})}\=0\,,\label{lcTT}\\
\mathbb{C}^{\underline{MN}}\,\Tb_{\underline{M}}{}^\alpha\,\Tb_{\underline{N}}{}^\beta\=0\,,\nonumber\\
[\Tb_{\underline{M}},\,\Tb_{\underline{N}}]+\Tb_{\underline{MN}}{}^{\underline{P}}\,
\Tb_{\underline{P}}\=0\,,\label{Tids}
\end{align}
where we have defined $\Tb_{\underline{MN}}{}^{\underline{P}}\equiv \Tb_{\underline{M}}{}^\alpha\,t_{\alpha \underline{N}}{}^{\underline{P}}$ and $\Tb_{\underline{MNP}}\equiv \Tb_{\underline{MN}}{}^{\underline{Q}}\mathbb{C}_{\underline{QP}}$. Recall that in $\mathcal{N}=2$ theories the tensor
\begin{equation}
\Tb_{\underline{MN}}{}^{\underline{P}}=\LL^{-1}{}_{\underline{M}}{}^M\,\LL^{-1}{}_{\underline{N}}{}^N\,X_{MN}{}^P\,\LL_P{}^{\underline{P}}\,,
\end{equation}
only represents the part of $\Tb_{\underline{M}}$ associated with the gauged special-K\"ahler isometries, the tensor $\Tb_{\underline{M}}$, appearing in the last two of (\ref{Tids}), also having contributions from gauged quaternionic-K\"ahler isometries. Equations (\ref{Tids}) have been originally derived within maximal supergravity in \cite{deWit:1981sst,deWit:1982bul}, and dubbed \emph{T-identities}%
\footnote{
Recall that in maximal supergravity the locality constraint follows from the linear and the closure ones.
}. Just as for the bare embedding tensor $\Theta$, also in this case the linear constraint (\ref{lcTT}) implies, in the case of $G$ simple (or for the STU model), the tracelessness of $\Tb$: $\Tb_{\underline{MN}}{}^{\underline{M}}=0$ (see Eq. (\ref{lconstr1imp2})).\par
Notice that, using eqs.\ (\ref{wPproj}) and (\ref{gaugedPW2}), we can rewrite the $\Tb$-tensor in the following form \cite{Gallerati:2016oyo}:\footnote{Some of the general formulae given for all extended supergravities were independently derived by the authors of \cite{Bandos:2016smv}.}
\begin{equation}
\Tb_{\underline{M}}\=\LL^{-1}{}_{\underline{M}}{}^N\,\Theta_N{}^\alpha\left(k_{\alpha}^s\,\mathcal{P}_s{}^{\underline{s}}\,K_{{\underline{s}}}-\frac{1}{2}\,\Ps_{\alpha}^{{\bf a}}\,J_{{\bf a}}-\frac{1}{2}\,\Ps_{\alpha}^{{\bf m}}\,J_{{\bf m}}\right)\,, \label{Tgen}
\end{equation}
which can be extended to $\N=2$ theories with non-homogeneous scalar manifolds of the form (\ref{SKQK}). In this case we cannot define a coset representative. However, as discussed in Sect. \ref{dcsl}, one can still define a symplectic matrix $\LL_c^M{}_{\underline{N}}$ (which is no longer related to a coset representative) depending on the complex scalar fields in the vector multiplets. We can then define the $\Tb$-tensor in these theories as in (\ref{Tgen})
where $\{K_{{\underline{s}}}\}$ should be intended as a basis of the tangent space to the origin (and not as isometry generators), while $\{J_I\}=\{J_{{\bf a}},\,J_{{\bf m}}\}$ are holonomy group generators%
\footnote{
The $H_{ R}={\rm U}(2)$-generators $\{J_{{\bf a}}\}$ naturally split into a ${\rm U}(1)$-generator $J_0$ of the K\"ahler transformations on $\MsSK$ and ${\rm SU}(2)$-generators $J_x$ ($x=1,2,3$) in the holonomy group of the quaternionic K\"ahler manifold $\Ms_{\textsc{qk}}$
}. Recall that $\{\Ps_{\alpha}^{{\bf a}},\,\Ps_{\alpha}^{{\bf m}}\}$ enter the definition of the gauged composite connection (\ref{gaugedPW2}) on the scalar manifold and, as mentioned earlier, are related to the Killing vectors by general properties of the spacial K\"ahler and quaternionic K\"ahler geometries \cite{Andrianopoli:1996cm}, see Sect. \ref{N2sugras}.\par\smallskip
 \par\smallskip
\paragraph{Step 4.: The supersymmetric completion of the gauged Lagrangian.} To cancel the supersymmetry variations of $\Lgaug^{(0)}$ and to construct a gauged Lagrangian $\Lgaug$ preserving the original supersymmetries, one can apply the general Noether method (see \cite{VanNieuwenhuizen:1981ae} for a general review) which consists in adding new terms to $\Lgaug^{(0)}$ and to the supersymmetry transformation laws, iteratively in the gauge coupling constant.
In our case the procedure converges by adding terms of order one ($\Delta\Lgaug^{(1)}$) and two ($\Delta\Lgaug^{(2)}$) in $g$, so that
\begin{equation}
\Lgaug=\Lgaug^{(0)}+\Delta\Lgaug^{(1)}+\Delta\Lgaug^{(2)}\,.
\end{equation}
The additional $O(g)$-terms are mass terms (or, more precisely, \emph{Yukawa terms}) and have the general form:
\begin{equation}
e^{-1}\Delta\Lgaug^{(1)}=g\left(2\bar{\psi}^A_\mu\;\gamma^{\mu\nu}\;\psi_\nu^B\;\mathbb{S}_{AB}
~+~i\,\bar{\lambda}^{\mathcal{I}}\;\gamma^\mu\;\psi_{\mu\,A}\;\mathbb{N}_{\mathcal{I}}{}^A
~+~\bar{\lambda}^{\mathcal{I}}\,\lambda^{\mathcal{J}}\;\mathbb{M}_{\mathcal{IJ}}\right)
~+~\text{h.c.}
\;,\label{fmassterms}
\end{equation}
characterized by the scalar-dependent matrices $\mathbb{S}_{AB}=\mathbb{S}_{BA}$ and $\mathbb{N}_{\mathcal{I}}{}^A$ called \emph{fermion-shift matrices}, and a matrix $\mathbb{M}_{\mathcal{IJ}}$ that can be rewritten in terms of the previous mixed-mass tensor $\mathbb{N}_{\mathcal{I}}{}^{ A}$ by supersymmetry, see discussion below Eq. (\ref{SNTid}). For these tensors we shall also use the notation: $\mathbb{N}^{\mathcal{I}}{}_{A}\equiv (\mathbb{N}_{\mathcal{I}}{}^{A})^*$,  $\mathbb{S}^{AB}\equiv (\mathbb{S}_{AB})^*$ and $\mathbb{M}^{\mathcal{IJ}}\equiv (\mathbb{M}_{\mathcal{IJ}})^*$\par
The $O(g^2)$-terms consist of a scalar potential:
\begin{equation}
e^{-1}\Delta\Lgaug^{(2)}=-V(\phi) \,.\label{spot}
\end{equation}
At the same time the fermionic supersymmetry transformations need to be suitably modified. To this end, we shall \emph{add order--$g$ terms to the fermion supersymmetry transformation rules} of the gravitino ($\psi_{\mu A}$) and of the other fermions ($\chi^\mathcal{I}$)
\begin{align}
%&0& &\=&\;
\delta_\epsilon\psi_{\mu A}&\=\cD_\mu\epsilon_A
+i\,g\;\mathbb{S}_{AB}\;\gamma_\mu\,\epsilon^B+\dotsc\,,\nn\\
%&0&&\=&
\delta_\epsilon\lambda_{\mathcal{I}}&\=g\,\mathbb{N}_{\mathcal{I}}{}^{A}\,\epsilon_A+\dotsc
\label{fermshifts}
\end{align}
depending on the same matrices $\mathbb{S}_{AB},\,\mathbb{N}_{\mathcal{I}}{}^{A},\,\mathbb{M}_{\mathcal{IJ}}$ entering the mass terms.
The fermion shift-matrices are composite fields belonging to some appropriate representations $\Rs_S,\,\Rs_N,\,\Rs_M$ of the $H$ group, such that (\ref{fmassterms}) is $H$-invariant. Some properties of these representations are fixed by the way these tensors appear in the Lagrangian and the supersymmetry transformation rules: For instance the gravitino fermion-shift matrix $\mathbb{S}_{AB}$ belongs to the two-times symmetric product of the fundamental representation of $H_R={\rm (S)U}(\mathcal{N})$, the dilatino one $\mathbb{N}_{ABC}{}^D$ belongs to the $(\bigwedge^3{\bf \mathcal{N}})\times \overline{{\bf \mathcal{N}}}$ of $H_R$, $\mathbb{M}_{\mathcal{I},\mathcal{J}}$ is symmetric in its two indices etc. Other properties will be fixed by supersymmetry, namely by the condition that the supersymmetry variations of the minimal coupling terms (in particular those involving the fermion terms), which depend on the $\mathbb{T}$-tensor, be canceled by the new terms depending on the fermion-shift matrices. This condition, as mentioned earlier, poses a restriction on the $\mathbb{T}$-tensor, which amounts, see below, to the representation constraint on embedding tensor. \par
Indeed the additional terms in the Lagrangian and supersymmetry transformation laws are enough to cancel the original $O(g)$ variations in $\delta\Lgaug^{(0)}$ --- like (\ref{RSvar}) and (\ref{lambdatra}), together with new $O(g)$ terms depending on $\mathbb{S}$ and $\mathbb{N}$ in the supersymmetry variation of $\Lgaug^{(0)}$ --- provided the shift-tensors $\mathbb{S}_{AB},\,\mathbb{N}^{\mathcal{I} A}$ are identified with suitable $H$-covariant components of the $\Tb$-tensor, according to the branching:
\begin{equation}
\Rs_\Theta\;\stackrel{H}{\longrightarrow}\;\Rs_N+\Rs_S+\Rs_M+\Rs_{\mbox{{\tiny other}}}\,,
\end{equation}
and that additional $H$-representations $\Rs_{ \mbox{{\tiny other}}}$ in the $\Tb$-tensor do not enter the supersymmetry variations of the Lagrangian. This can be formulated as a $G$-covariant restriction on the representation $\Rs_\Theta$ of the $\Tb$-tensor or, equivalently, of embedding tensor, which can be shown to be, for $\mathcal{N}>2$, no more than the representation constraint (\ref{linear2}) discussed earlier. This condition also defines the properties, mentioned above, of the representations $\Rs_N,\,\Rs_S$ and $\Rs_M$, see example below, in Eq. (\ref{Tlinab}). 
For $\mathcal{N}=2$ models, the linear constraint (\ref{linear2}) only fixes the components of the $\Tb$-tensor associated with the gauging of the special K\"ahler isometries, in terms of the fermion-shift tensors, while the corresponding expressions of the remaining components is fixed by supersymmetry. We shall give the explicit relations below.\par
The identification with components of the $\Tb$-tensor defines the expression of the fermion shift-tensors as $H$-covariant composite fields in terms of the embedding tensor and the scalar fields:
\begin{equation}
\mathbb{S}_{AB}=\mathbb{S}_{AB}(\phi,\Theta)=\left.\Tb(\phi,\Theta)\right\vert_{\Rs_S}\,;
\quad\;
\mathbb{N}_{\mathcal{I}}{}^A=\mathbb{N}_{\mathcal{I}}{}^A(\phi,\Theta)=\left.\Tb(\phi,\Theta)\right\vert_{\Rs_N}
\,.\label{SNTid}
\end{equation}
The precise identifications will be given below, see Eqs. (\ref{identificationsSNT}).\par
A relation between the fermion mass-matrix $\mathbb{M}_{\mathcal{I}\mathcal{J}}$ and $\mathbb{N}_{\mathcal{I}}{}^A$ follows from requiring the cancelation of the terms
(\ref{lambdatra}) against $O(g)$-terms produced by the fermion variations (see Eqs. (\ref{trachi})-(\ref{tralam})) in the mass- terms (\ref{fmassterms}) and in the (gauged) Pauli terms of the Lagrangian. This cancelation also fixes the identification (\ref{SNTid}) as far as the $\mathbb{N}$-tensor is concerned.\par
The first of the (\ref{SNTid}) follows from requiring the cancelation  of the (\ref{RSvar}) terms with $O(g)$-ones of the form $F\mathbb{S}\bar{\psi}\epsilon$, $F$ standing for the vector field strengths, originating from the variation of the (gauged) Pauli terms and of the gravitino mass term in (\ref{fmassterms}).\par
As we shall discuss below, supersymmetry further requires differential relations between the tensors $\mathbb{S},\,\mathbb{N}$ and $\mathbb{M}$, to be dubbed ``gradient flow'' equations \cite{D'Auria:2001kv}. These relations also follow from the identifications (\ref{SNTid}), the definition of the $\Tb$-tensor and, for homogenous scalar manifolds, from the differential equations (\ref{DLP}), (\ref{dLP1}), (\ref{dLP2}) satisfied by $L$ and $\mathbb{L}_c$. For non-homogeneous geometries in $\mathcal{N}=2$ models the same relations originate from properties of special and quaternionic K\"ahler manifolds.

\paragraph{A detailed analysis.} Let us add some more details on how the supersymmetry variations of the deformed Lagrangian vanish, to all orders in $g$, as a consequence of the identifications (\ref{SNTid}). Following \cite{Ferrara:1985gj} and \cite{D'Auria:2001kv}, in the variation of the Lagrangian we can distinguish between two kinds of terms: those with one derivative and those with no derivatives.
 As far as the former are concerned, consider the $O(g)$-terms of the form $\bar{\psi} \partial(\mathbb{S}\epsilon)$ and $\bar{\lambda} \partial(\mathbb{N}\epsilon)$ in the variations of the gravitino and spin-$1/2$ kinetic terms, respectively. The $\bar{\psi}\partial \epsilon  \mathbb{S}$ and $\bar{\lambda}\partial \epsilon \mathbb{N} $ -terms have the form:
 \begin{align}
 &-4e\,g\,\bar{\psi}^A_\mu\gamma^{\mu\nu}\mathcal{D}_\nu\epsilon^B\,\mathbb{S}_{AB}+h.c.\label{psideS}\\
 &-i\,e\,g\,\bar{\lambda}^{\mathcal{I}}\,\gamma^\mu\mathcal{D}_\mu\epsilon_A\,\mathbb{N}_{\mathcal{I}}{}^A+h.c.\label{lamdeN}
 \end{align}
and are canceled by the gravitino variation of the Yukawa-terms (\ref{fmassterms}). Next consider the cancelation of the
$\bar{\psi}\epsilon  \partial \mathbb{S}$ terms
 \begin{align}
 &-4e\,g\,\bar{\psi}^A_\mu\gamma^{\mu\nu}\epsilon^B\,\mathcal{D}_\nu\mathbb{S}_{AB}+h.c.\label{psiepsilondS}
 \end{align}
 which occurs by effect of $O(g)$ contributions from the variation of the scalar-fermion terms and the Yukawa-terms (\ref{fmassterms}):
 \begin{align}
 \mbox{from scalar-fermion terms:}&\,\,\,\,-e\,g\,\mathbb{N}^{\mathcal{I}}{}_A\bar{\psi}^B_\mu\gamma^\nu\gamma^\mu\epsilon^A\,
 \hat{\mathcal{P}}_{\nu\,\mathcal{I}B}+h.c.\label{cont1}\\
 \mbox{from Yukawa terms:}&\,\,\,\,e\,g\,\mathbb{N}^{\mathcal{I}}{}_B\bar{\psi}^B_\mu
 \gamma^\mu\gamma^\nu\epsilon^A\,\hat{\mathcal{P}}_{\nu\,\mathcal{I}A}+h.c.\label{cont2}
 \end{align}
Writing $\gamma^\mu\gamma^\nu=\gamma^{\mu\nu}+\eta^{\mu\nu}$ in the above terms, the reader can easily verify that the $\gamma^{\mu\nu}$-contributions sum up to
\begin{equation}
2\,e\,g\,\bar{\psi}^A_\mu\gamma^{\mu\nu}\epsilon^B\,\hat{\mathcal{P}}_{\nu\,\mathcal{I}(A}\mathbb{N}^{\mathcal{I}}{}_{B)}+h.c.\,,
\end{equation}
which cancel, to order $O(g)$, against (\ref{psiepsilondS}) provided the following differential relation between the fermion-shift tensors hold:
\begin{equation}
{\Scr D}_s\mathbb{S}_{AB}=\frac{1}{2}\,{\mathcal{P}}_{s\,\mathcal{I}(A}\mathbb{N}^{\mathcal{I}}{}_{B)}\,,\label{GF1}
\end{equation}
where ${\Scr D}_s$ is the $H$-covariant derivative on the scalar manifold defined in Sect. \ref{ghsect}. The $\eta^{\mu\nu}$-contributions in (\ref{cont1}) and (\ref{cont2}) sum up to:
\begin{equation}
-2\,e\,g\,\mathbb{N}^{\mathcal{I}}{}_B\,\bar{\psi}_\mu^{[A}\epsilon^{B]} \hat{\mathcal{P}}^\mu_{\mathcal{I} A}+h.c.\,,
\end{equation}
which is canceled by a corresponding term coming from the variation of the vector fields in the scalar kinetic Lagrangian:
\begin{equation}
-2e\,g\,{\Scr G}_{rs}\mathcal{D}^\mu\phi^r \mathbb{L}_c{}^M{}_{AB} k_M^s\bar{\psi}_\mu^A\epsilon^B+h.c.
\end{equation}
provided the following relation holds:
\begin{equation}
 \mathbb{L}_c{}^M{}_{AB} k_M^s=-{\Scr G}^{sr}\,\mathcal{P}_{r\,\mathcal{I}[A}\mathbb{N}^{\mathcal{I}}{}_{B]}\,.
\end{equation}
Next consider the cancelation of the $\bar{\lambda}\epsilon \partial \mathbb{N} $ -terms from the $\lambda^{\mathcal{I}}$-kinetic action. It occurs, in part, by effect of $O(g)$-contributions which originate from the variation of the vector fields in the scalar kinetic Lagrangian, of the form:
\begin{align}
&i\,e\,g\,{\Scr G}_{rs}\mathcal{D}_\mu\phi^r \mathbb{L}_c{}^M{}_{I} k_M^s\bar{\lambda}^{IA}\gamma^\mu\epsilon_A+\frac{i}{2}\,e\,g\,{\Scr G}_{rs}\mathcal{D}_\mu\phi^r \mathbb{L}_c{}^M{}_{AB} k_M^s\bar{\chi}^{ABC}\gamma^\mu\epsilon_C+h.c=\nonumber\\
&=i\,e\,g\,{\Scr G}_{rs}\mathcal{D}_\mu\phi^r \mathbb{L}_c{}^M{}_{\mathcal{I}}{}^A k_M^s\bar{\lambda}^{\mathcal{I}}\gamma^\mu\epsilon_A+h.c.
\end{align}
where, in the last line, we have used the collective symbol $\lambda^{\mathcal{I}}$ to denote all the dilatinos and gauginos, corresponding to $\mathcal{I}=(IA)$ or $(ABC)$, respectively and have set\footnote{Recall that when the index $\mathcal{I}$ equals the antisymmetric triplet $[ABC]$ labeling the dilatini, summation over it requires a normalization factor $1/3!$.}
$$\mathbb{L}_c{}^M{}_{\mathcal{I}}{}^A =\begin{cases}\mathbb{L}_c{}^M{}_{IB}{}^A\equiv\mathbb{L}_c{}^M{}_{I}\delta_B{}^A\cr \mathbb{L}_c{}^M{}_{ABC}{}^A\equiv 3\,\mathbb{L}_c{}^M{}_{[AB}\delta_{C]}{}^A\,.\end{cases}$$
The cancelation of the $\bar{\lambda}\epsilon \partial \mathbb{N} $ -terms also occurs by effect of the following contributions:
  \begin{align}
 \mbox{from scalar-fermion terms:}&\,\,\,\,2i\,e\,g\,\bar{\lambda}^{\mathcal{I}}_\mu\gamma^\mu\epsilon_A \mathbb{S}^{AB}\,\hat{\mathcal{P}}_{\mu\,\mathcal{I} B}+h.c.\label{cont21}\\
 \mbox{from Yukawa terms:}&\,\,\,\,2i\,e\,g\,\mathbb{M}_{\mathcal{I}\mathcal{J}}\bar{\lambda}^{\mathcal{I}}
 \gamma^\mu\epsilon_B\,\hat{\mathcal{P}}_{\mu}{}^{\mathcal{J}B}+h.c.\label{cont22}
 \end{align}
 provided the following differential relation holds:
 \begin{equation}
 {\Scr D}_r \mathbb{N}_{\mathcal{I}}{}^A=
 {\Scr G}_{rs}\,\mathbb{L}_c{}^M{}_{\mathcal{I}}{}^A\,k_M{}^s+2 \mathcal{P}_{r\,\mathcal{I}B}\,\mathbb{S}^{BA}+2\,
 \mathbb{M}_{\mathcal{I}\mathcal{J}}\mathcal{P}_r{}^{\mathcal{J}\,A}\,.\label{GF2}
 \end{equation}
Finally, in order to cancel the $O(g^2)$-contributions resulting from the variations (\ref{fermshifts}) in (\ref{fmassterms})
we need to add an \emph{order-$g^2$ scalar potential} $V(\phi)$ which is totally determined by supersymmetry as a bilinear function of the shift matrices by the condition
\begin{equation}
\delta_B{}^A\,V(\phi)\;=\; g^2\,\left(\mathbb{N}_{\mathcal{I}}{}^{A}\,\mathbb{N}^{\mathcal{I}}{}_{B}-12\;\mathbb{S}^{AC}\,
\mathbb{S}_{BC}\right)\,,\label{WID}
\end{equation}
To see this consider the $g^2\,\bar{\psi}\epsilon$-contributions resulting from the variations (\ref{fermshifts}) in (\ref{fmassterms}), which have the form:
\begin{equation}
i\,e\,g^2\bar{\psi}^A_\mu\gamma^\mu \epsilon_C\left(12\,\mathbb{S}_{AB}\mathbb{S}^{BC}-\mathbb{N}^{\mathcal{I}}{}_{A}\mathbb{N}_{\mathcal{I}}{}^{C}\right)+h.c.\,
\end{equation}
these are canceled by the variation $\delta(e)\,V$ of the potential term, being
\begin{equation}
\delta(e)=e\,V_a{}^\mu\delta V_\mu{}^a=i\,e\,(\bar{\epsilon}^A\gamma^\mu\psi_{\mu\,A}+\bar{\epsilon}_A\gamma^\mu\psi_{\mu}^A)\,,
\end{equation}
provided Eq. (\ref{WID}) holds.
This condition is called \emph{potential Ward identity} \cite{Ferrara:1985gj,Cecotti:1984wn}. It defines the scalar potential as a quadratic function of the embedding tensor and non-linear function of the scalar fields:
\begin{equation}
V(\phi,\Theta)\;=\;\frac{g^2}{\mathcal{N}} \,\left(\mathbb{N}_{\mathcal{I}}{}^{A}\,\mathbb{N}^{\mathcal{I}}{}_{A}-12\;\mathbb{S}^{AC}\,\mathbb{S}_{AC}\right)
\,.\label{Pot}
\end{equation}
 As a constraint on the fermion shifts, it is part of a set of quadratic conditions required for the cancelation of the supersymmetry variations to order-$g^2$ and can be shown,  once the fermion-shift tensors have been identified with components of the $\Tb$-tensor, to follow from the $\Tb$-identities (\ref{Tids}) or, equivalently, from the quadratic constraints (\ref{quadratic1}), (\ref{quadratic2}) on $\Theta$.
We shall explicitly derive the Ward identity from the quadratic constraints on the embedding tensor in the $\mathcal{N}=8$ and $\mathcal{N}=2$ theories.
\par
 Equations (\ref{GF1}) and (\ref{GF2}) are general differential ``gradient flow'' relations \cite{D'Auria:2001kv} among the fermion-shift tensors which also follow from the identification of the fermion shifts with  components of the $\Tb$-tensor.\par
Let us now make the relations (\ref{SNTid}) more precise. The cancelation of terms like (\ref{RSvar}) and (\ref{lambdatra}) leads to the following identifications:
\begin{align}
(\mathbb{T}^{AB})^{CD}{}_{EF}&=4\,\delta^{[C}_{[E}T_{F]}{}^{D] AB}\,\,\,;\,\,\,\,\,(\mathbb{T}_{AB})^{CD}{}_{EF}=-4\,\delta^{[C}_{[E}T^{D]}{}_{F] AB}\,,\nonumber\\
T_{C}{}^{D AB}&= \mathbb{L}_c^{M\,AB}\,\mathcal{Q}_M{}^D{}_C=-\frac{1}{2}\,\mathbb{L}_c^{M\,AB}\,\Theta_M{}^\alpha {\Scr P}_\alpha^{{\bf a}}\,(J_{{\bf a}})^D{}_C=-\frac{1}{2}\,\mathbb{N}^{D AB}{}_C-2 \mathbb{S}^{D[A}\delta^{B]}_C\,,\nonumber\\
(\mathbb{T}_{AB})^{CDEF}&=-\mathbb{L}_c^{M}{}_{AB}\,\Theta_M{}^\alpha\,k_\alpha^s\,\mathcal{P}_s^{CDEF}=-4\,
\delta_{[A}^{[C}\mathbb{N}^{DEF]}{}_{B]}\,,\nonumber\\
(\mathbb{T}_{AB})^{CDI}&=-\mathbb{L}_c^{M}{}_{AB}\,\Theta_M{}^\alpha\,k_\alpha^s\,\mathcal{P}_s^{CD\,I}=-
2\,\delta_{[A}^{[C}\mathbb{N}^{D]I}{}_{B]}\,,\nonumber\\
(\mathbb{T}_{I})_A{}^{B}&=-\frac{1}{2}\mathbb{L}_c^{M}{}_I\,\Theta_M{}^\alpha\,
{\Scr P}_\alpha^{{\bf a}} (J_{{\bf a}})_A{}^B =
\frac{1}{2}\,\mathbb{N}_{IA}{}^B\,,\nonumber\\
(\mathbb{T}^{AB})_{C\,\alpha}&=\mathbb{L}_c^{M\,AB}\,\Theta_M{}^m\,k_m^s\,\mathcal{P}_{s\,C\alpha} =-
\delta^{[A}_C\mathbb{N}_\alpha{}^{B]}\,\,, \,\,\alpha=1,\dots, 2n_H\,\,,\,\,\,\,\,\,(\mathcal{N}=2)\,,\label{identificationsSNT}
\end{align}
where $\mathbb{N}_\alpha{}^{A}$ is the hyperino-shift tensor: $\delta \lambda_\alpha=\dots +g\,\mathbb{N}_\alpha{}^{A}\,\epsilon_A$. In the last line we have labeled the quaternionic isometries by the index $m$ while the index $\alpha=1,\dots, 2n_H$, in this particular case, has a different definition: it is the index of the symplectic representation of the $H^{(QK)}_{{\rm matt}}$-factor in the holonomy group.\par
Notice that, for $\mathcal{N}=2$ theories, not all the above components of $\Tb_{\underline{M}}$ are part of the tensor
$\Tb_{\underline{M}\underline{N}}{}^{\underline{P}}$. For instance the $\mathbb{S}$-dependent part of $(\mathbb{T}_{AB})_C{}^{D}$ does not contribute to $(\mathbb{T}_{AB})_{CD}{}^{EF}$, while $(\mathbb{T}^{AB})_{C\,\alpha}$ is totally absent from $\Tb_{\underline{M}\underline{N}}{}^{\underline{P}}$, being the components of $\Tb_{\underline{M}}$ along the tangent space to the quaternionic K\"ahler manifold with basis $K_{C\,\alpha}$.
The reader can verify that the above identifications are consistent with the linear constraints on the $\Tb$-tensor as expressed in (\ref{lcTT}). In particular the expression of the $\Tb_{\underline{M}\underline{N}}{}^{\underline{P}}$ components in terms of the fermion-shift tensors is totally fixed by these conditions. For instance it is straightforward to verify that:
\begin{equation}
(\Tb_{AB})_{CD}{}^{EF}+(\Tb_{CD})_{AB}{}^{EF}=(\Tb^{EF})_{ABCD}\,,\label{Tlinab}
\end{equation}
provided $\mathbb{N}^{ABC}{}_D=\mathbb{N}^{[ABC]}{}_D$ and $\mathbb{S}_{AB}=\mathbb{S}_{BA}$.\par
Notice that, in $\mathcal{N}=2$ models, we have
\begin{equation}
(\mathbb{T}^{AB})^{CD}{}_{EF}=4\,\delta^{[C}_{[E}T_{F]}{}^{D] AB}=0\,,
\end{equation}
given the expression of $T_{C}{}^{D AB}$ in terms of the fermion-shift tensors, since $\mathbb{N}^{ABC}{}_D=0$ and the gravitino-shift matrix  gives no contribution. On the other hand, form the definition of the $\Tb$-tensor we have
\begin{equation}
(\mathbb{T}^{AB})^{CD}{}_{EF}=\frac{1}{2}\,\mathbb{L}_c^{M\,AB}\,\Theta_M{}^\alpha\,
{\Scr P}_\alpha^{{\bf a}} (J_{{\bf a}})_{EF}{}^{CD}=2i\,\mathbb{L}_c^{M\,AB}\,\Theta_M{}^\alpha\,
{\Scr P}_\alpha^{0} \delta_{EF}{}^{CD}\,.
\end{equation}
Being $\mathbb{L}_c^{M\,AB}=\overline{V}^M\epsilon^{AB}$ we deduce that in $\mathcal{N}=2$ models the linear constraint implies:
$V^M\,\Theta_M{}^\alpha\,{\Scr P}_\alpha^{0}=0$. We shall derive this property also using special geometry, see Eqs. (\ref{newidentities}) of Sect. \ref{gaugspecialg}.\par
In homogeneous models the gradient-flow equations in the fermion-shift matrices, given the identifications (\ref{identificationsSNT}), follow from the very definition of the $\Tb$-tensor and the general property (\ref{DLP}), or (\ref{dLP1}), (\ref{dLP2}):
\begin{align}
{\Scr D}_s\Tb_{\underline{M}}&=(\mathcal{P}^c_{s})^{\,\,\underline{N}}{}_{\underline{M}}\mathbb{L}_c^M{}_{\underline{N}} L^{-1}X_ML+\mathbb{L}_c^M{}_{\underline{M}} [L^{-1}X_ML,\,\mathcal{P}_s]=\Tb_{\underline{N}}\,(\mathcal{P}^c_{s})^{\,\,\underline{N}}{}_{\underline{M}}+[\Tb_{\underline{M}},\,
\mathcal{P}_s]\,,\label{DT}
\end{align}
where $\mathcal{P}^c$ is the matrix-valued 1-form defined in (\ref{Pc}).\par
By deriving the general expression for the scalar potential $V$ given by (\ref{Pot}), and using the gradient-flow equations, we can write the gradient of $V$ in terms of the fermion shift-tensors and the mass-matrix:
\begin{equation}
\frac{\partial V}{\partial\phi^s}=\frac{g^2}{\mathcal{N}}\,
\left(-4\,\mathcal{P}_{s\,\mathcal{I}C}\mathbb{S}^{CA}\mathbb{N}^{\mathcal{I}}{}_A+2\,
\mathbb{M}_{\mathcal{I}\mathcal{J}}\mathcal{P}_{s}^{\mathcal{J}C}\mathbb{N}^{\mathcal{I}}{}_C\right)+c.c.\label{DWID}
\end{equation}
This condition is required for the cancelation, in the supersymmetry variation of the Lagrangian, of the
$O(g^2)$-terms of the form $f(\phi)\,\bar{\chi}\epsilon$.  We shall use the above identity in Sect. \ref{vmm} for proving the consistent definition of the mass-matrix for the spin-$1/2$ fields. It will indeed be crucial to show that the Goldstino fields, associated with the spontaneous supersymmetry breaking, correspond to the zero-eigenvalues of that matrix (i.e. are massless fermions).\par
To derive equation (\ref{DWID}) we need to use the following conditions:
\begin{equation}
\Theta_M{}^\alpha\mathbb{L}_c^M{}_{\mathcal{I}}{}^A\,\mathbb{N}^{\mathcal{I}}{}_A+c.c.=0\,\,,\,\,
\,\,\,\mathbb{S}_{AB}=\mathbb{S}_{BA}\,.\label{csNS}
\end{equation}
which also follow from the constraints on the embedding tensor once the fermion-shift tensors have been identified with components of the $\mathbb{T}$-tensor. In the maximal theory the first of Eqs. (\ref{csNS}) follow from the property $\mathbb{N}^{ABC}{}_A=0$ which in turn is implied by the linear constraints just as, in a generic supergravity, is the symmetry property of $\mathbb{S}_{AB}$.
\paragraph{Example 1.: The scalar potential of minimal supergravity.}
The most widely known supergravity theory, as well as the first to be constructed, is the minimal one. As mentioned earlier, in this theory, being the scalar fields not connected to the vector ones by supersymmetry, there is no built-in symplectic structure in the definition of the scalar manifold. As a consequence of this, the non-minimal couplings of the scalars to the vectors are not fixed by supersymmetry, as it happens (modulo an initial choice of the symplectic frame) for extended theories. This in turn implies that there is no built-in duality symmetry. The Gaillard-Zumino mechanism of promoting the isometry group of the scalar manifold to on-shell global symmetry, through its electric-magnetic duality action of the vector fields strengths and their duals, requires a specific choice of the holomorphic function $F_{\Lambda\Sigma}(z^i)=\overline{{\Scr N}}_{\Lambda\Sigma}(z^i)=\R_{\Lambda\Sigma}-i\,\I_{\Lambda\Sigma}$, defining the vector kinetic terms. The presence of a scalar potential does not require the introduction of a gauge group. As mentioned earlier, we can have an F-term potential defined in terms of a holomorphic superpotential $W(z)$, or a \emph{covariantly holomorphic} superpotential $L(z,\bar{z})\equiv e^{\frac{\mathcal{K}}{2}}\,W(z)$. The superpotential defines the gravitino shift tensor $\mathbb{S}$ (which is a $1\times 1$ complex matrix) as follows ( we reabsorb $g$ in the definition of $\mathbb{S}$):
  \begin{equation}
  \delta \psi_{\bullet \mu}=\dots+i\,\mathbb{S}\,\gamma_\mu\epsilon\,\,\,,\,\,\,\,\,\,\,\,\mathbb{S}=\frac{L}{2}=e^{\frac{\mathcal{K}}{2}}\,\frac{W}{2}\,,
  \end{equation}
 as well as shift-tensors for  the chiralini (fermions in the $n$ chiral multiplets):\footnote{The tensor $\mathbb{N}^{I}$ are the non-vanishing values of the complex auxiliary fields $F_I$ of the chiral multiplets.}
  \begin{equation}
  \delta\lambda^{\bullet I}=\dots +\mathbb{N}^{I}\,\epsilon^\bullet\,\,\,,\,\,\,\,\,\,\,\,\mathbb{N}^{I}=e^{-1\,Ii}{\Scr D}_{i}{L}\,.
  \end{equation}
  We see that ${\Scr D}_{i}\mathbb{S}=\frac{1}{2}\,e_{i\,I}\,\mathbb{N}^{I}$, which is the gradient flow equation (\ref{GF1}). The gauging of isometries of the K\"ahler scalar manifold implies further deformations of the theory by inducing non-vanishing values of the auxiliary fields $D_{ \boldsymbol{\Lambda}}$ of the $n_v$ vector supermultiplets, thus implying the presence of fermion-shift tensors for the gauginos $\chi_{\boldsymbol{\Lambda}}$ (the coupling constant has been reabsorbed in the embedding tensor):
  \begin{equation}
  \delta \chi_{\bullet \boldsymbol{\Lambda}}=\dots+\mathbb{N}_{ \boldsymbol{\Lambda}}\,\epsilon_\bullet\,\,\,,\,\,\,\,\,\,\mathbb{N}_{ \boldsymbol{\Lambda}}=i\,U_{\boldsymbol{\Lambda}}{}^M\,\Theta_M{}^\alpha {\Scr P}_\alpha\,\,\,,\,\,\,\,\,\boldsymbol{\Lambda}=1,\dots, n_v\,\,;\,\,\,\,M=1,\dots, 2\,n_v\,,
  \end{equation}
  where ${\Scr P}_\alpha$ are the moment maps associated with the isometries of the K\"ahler manifold\footnote{Later we shall give a general definition of moment maps for non-homogeneous geometries.}, $\Theta_M{}^\alpha $ is the embedding tensor defining the gauging and $U_{\boldsymbol{\Lambda}}{}^M$ is a complex, scalar-dependent matrix satisfying the property:
  \begin{equation}
  \mathcal{M}^{MN}(z,\bar{z})=-(U_{\boldsymbol{\Lambda}}{}^MU^{\boldsymbol{\Lambda}N}+U_{\boldsymbol{\Lambda}}{}^NU^{\boldsymbol{\Lambda}M})\,\,;\,\,\,\,\,
  U^{\boldsymbol{\Lambda}M}\equiv (U_{\boldsymbol{\Lambda}}{}^M)^*\,,
  \end{equation}
  and $\mathcal{M}^{MN}$ is the inverse of the symplectic, symmetric matrix, defined in (\ref{M}) in terms of the real and imaginary parts of ${\Scr N}(\bar{z})$.
  From the general form (\ref{WID}) of the scalar potential in terms of the fermion-shift tensors we have:
  \begin{equation}
V=e^\mathcal{K}\left(g^{i\bar{\jmath}}{\Scr D}_i W {\Scr D}_{\bar{\jmath}} \overline{W}-3 |W|^2 \right)-\frac{1}{2}\,\mathcal{M}^{MN}\Theta_M{}^\alpha\Theta_N{}^\beta \,{\Scr P}_\alpha{\Scr P}_\beta\,.
  \end{equation}
  The last term is known as the ``D''-term contribution to the scalar potential and only depends on the gauged isometries through the embedding tensor.
In terms of the embedding tensor in the electric frame, the potential reads
  \begin{equation}
V=e^\mathcal{K}\left(g^{i\bar{\jmath}}{\Scr D}_i W {\Scr D}_{\bar{\jmath}} \overline{W}-3 |W|^2 \right)-\frac{1}{2}\,\I^{\hat{\Lambda}\hat{\Sigma}}\Theta_{\hat{\Lambda}}{}^\alpha\Theta_{\hat{\Sigma}}{}^\beta \,{\Scr P}_\alpha{\Scr P}_\beta\,,
  \end{equation}
  which is the well known general form of the $\mathcal{N}=1$ scalar potential, originally found in \cite{Cremmer:1982en}, see also \cite{Freedman:2012zz} for a review.
\paragraph{Example 2.: The gradient flow equations in maximal supergravity.}
As a second example we can work out the gradient-flow equations for the fermion shifts in maximal supergravity from the coset geometry and the definition of the $\Tb$-tensor. In this case Eq. (\ref{DT}) has the following component:
\begin{align}
{\Scr D}_s(\Tb_{AB})_{CD}{}^{EF}&=\frac{1}{2}\,\mathcal{P}_{s\,ABA'B'}\,(\Tb^{A'B'})_{CD}{}^{EF}+
\frac{1}{2}\,\mathcal{P}_{s\,CDA'B'}\,(\Tb_{AB})^{A'B'EF}-
\frac{1}{2}\,(\Tb_{AB})_{CDA'B'}\,\mathcal{P}_{s}{}^{A'B' EF}\,,\label{DTmax}
\end{align}
from which, using the identifications (\ref{identificationsSNT}), after some algebra, we derive:
\begin{align}
{\Scr D}_s\mathbb{S}_{AB}&=-\frac{1}{12}\,\mathbb{N}^{CDE}{}_{(A}\mathcal{P}_{s|B)CDE}\,,\nonumber\\
{\Scr D}_s\mathbb{N}_{ABD}{}^F&=-2\mathbb{S}^{FE}\,\mathcal{P}_{s\,EABD}-\frac{3}{2}\,\mathbb{N}^{FEG}{}_{[A}\,
\mathcal{P}_{s|BD]EG}+\frac{1}{2}\delta^{F}_{[A}\,\mathbb{N}^{EGH}{}_{B}\mathcal{P}_{s|D]EGH}\,.\label{GFN8}
\end{align}
which correspond to the general equations (\ref{GF1}) and (\ref{GF2}). As far as the latter equation is concerned, to see the correspondence one has to use  the property that  dilatino mass matrix in the maximal theory is expressed in terms of $\mathbb{N}$ as follows:
\begin{equation}
\mathbb{M}_{ABC,DEF}=-\frac{1}{4}\,\epsilon_{ABC,A'B'C'[DE}\,\mathbb{N}^{A'B'C'}{}_{F]}\,.\label{matbbmmax}
\end{equation}
This relation derives, as previously pointed out, from the requirement that the terms (\ref{lambdatra}) be canceled by corresponding
variations of the dilatino mass term and the (gauged) Pauli terms.\par
The  derivation of Eqs. (\ref{GFN8}) requires the following useful identities, which follow from the reality condition on the ${\rm SU}(8)$-four-fold antisymmetric representation:
{\small \begin{align}
\Sigma_{1\,CDA'B'}\Sigma_{2}^{A'B' EF}-\Sigma_{2\,CDA'B'}\Sigma_{1}^{A'B' EF}&=\frac{1}{6}\,\delta^{EF}_{CD}\,\Sigma_{1\,A'B' C'D'}\Sigma_{2}^{A'B' C'D'}-\frac{4}{3}\,\delta^{[E}_{[C}\,\Sigma_{1}^{F]A'B'C'}\,\Sigma_{2\,D]A'B'C'}=\nonumber\\
&=\frac{2}{3}\left(\delta^{[E}_{[C}\,\Sigma_{2}^{F]A'B'C'}\,\Sigma_{1\,D]A'B'C'}-
\delta^{[E}_{[C}\,\Sigma_{1}^{F]A'B'C'}\,\Sigma_{2\,D]A'B'C'}\right)\label{sigmaprops}
\end{align}}
We shall discuss the scalar potential of maximal supergravity and its vacua in a later section devoted to this theory.\par
It is a characteristic of supergravity theories that -- in contrast to globally supersymmetric ones -- by virtue of the negative contribution due to the gravitino shift-matrix, the scalar potential is in general not positive definite, but may, in particular, feature AdS vacua. These are maximally symmetric solutions whose negative cosmological constant is given by the value of the potential at the corresponding extremum:\, $\Lambda=V_0<0$.\; Such vacua are interesting in light of the AdS/CFT holography conjecture \cite{Maldacena:1997re,Gubser:1998bc,Witten:1998qj}, mentioned in the Introduction, according to which stable AdS solutions describe conformal critical points of a suitable gauge theory defined on the boundary of the space.
In this perspective, domain wall solutions to the gauged supergravity interpolating between AdS critical points of the potential describe renormalization group (RG) flow (from an ultra-violet to an infra-red fixed point) of the dual gauge theory and give important insights into its non-perturbative properties. The spatial evolution of such holographic flows is determined by the scalar potential $V(\phi)$ of the gauged theory.\par
In some
cases the effective scalar potential $V(\phi)$, at the classical
level, is non--negative and defines vacua with vanishing
cosmological constant in which supersymmetry is spontaneously
broken and part of the moduli are fixed. Models of this type are
generalizations of the so-called ``no--scale'' models
\cite{Cremmer:1983bf}, \cite{Ellis:1984bm}, \cite{Barbieri:1985wq} which were subject to intense study during the eighties.

\subsection{Duality Covariant Gauging}\label{sec:4}
Let us summarize what we have learned so far.
\begin{itemize}
\item{The most general local internal symmetry group $G_g$ which can be introduced in an extended supergravity is defined by an embedding tensor $\Theta$, covariant with respect to the on-shell global symmetry group $G$ of the ungauged model and (locally) defining the embedding of $G_g$ inside $G$. Since a scalar potential $V(\phi)$ can only be introduced through the gauging procedure, $\Theta$ also defines the most general choice for \,$V=V(\phi,\Theta)$. }
\item{Consistency of the gauging at the level of the bosonic action requires $\Theta$ to satisfy a number of (linear and quadratic) $G$-covariant constraints. The latter, besides completely determining the gauged bosonic action, also allow for its consistent (unique) supersymmetric extension.}
\item{Once we find a solution $\Theta_M{}^\alpha$ to these algebraic constraints, a suitable symplectic matrix $E$, which exists by virtue of (\ref{quadratic1}), will define the corresponding electric frame, in which its magnetic components vanish.}
\end{itemize}
Although we have freed our choice of the gauge group from the original symplectic frame, the resulting gauged theory is still defined in an electric frame and thus depends on the matrix $E$: whatever solution $\Theta$ to the constraints is chosen for the gauging, the kinetic terms of the gauged Lagrangian are always written in terms of the only \emph{electric} vector fields $A^{\hat{\Lambda}}_\mu$, namely of the vectors effectively
involved in the minimal couplings, see Eq.\ (\ref{syminvmc}). We shall discuss in the present section a more general formulation of the gauging procedure in four-dimensions which was developed in \cite{deWit:2005ub,deWit:2007mt} and which no longer depends on the matrix $E$, so that the kinetic terms are not written in terms of the vector fields in the electric frame.
\paragraph{Steps 1, 2 and 3 revisited.}
We start from a symplectic-invariant gauge connection of the form%
\footnote{
Here, for the sake of simplicity, we reabsorb the gauge coupling constant $g$ into $\Theta$: $g\,\Theta\rightarrow \Theta$.
}:
\begin{equation}
 \Omega_{g\mu}\equiv A^M_\mu\,X_M=A^\Lambda_\mu\,X_\Lambda+A_{\Lambda\,\mu}\,X^\Lambda=A^M_\mu\,\Theta_M{}^\alpha\,t_\alpha\,,\label{newcon}
\end{equation}
where $\Theta_M{}^\alpha$ satisfies the constraints (\ref{linear2}), (\ref{quadratic1}), (\ref{quadratic2}). The fields $A^\Lambda_\mu$ and $A_{\Lambda\,\mu}$ are now taken to be independent. This is clearly a redundant choice and, as we shall see, half of them play to role of auxiliary fields. Eq.\ (\ref{quadratic1}) still implies that at most $n_v$ linear combinations $A^{\hat{\Lambda}}_\mu$ of the $2n_v$ vectors $A^\Lambda_\mu,\,A_{\Lambda\,\mu}$ effectively enter the gauge connection (and thus the minimal couplings):
 \begin{equation} A^M_\mu\,X_M=A^{\hat{\Lambda}}_\mu\,X_{\hat{\Lambda}}\,.\label{ATHcomb}\end{equation}
Indeed, as a consequence of Eq.\ (\ref{quadratic1}), the symplectic matrix $E_{\hat{M}}{}^N$ exists which rotates
the embedding tensor to its electric frame, namely for which Eq.\ (\ref{elET}) holds. In the new frame the gauge generators are then given by:
 \begin{equation}
 X_{\hat{\Lambda}}=E_{\hat{\Lambda}}{}^\Sigma\,X_\Sigma+E_{\hat{\Lambda}\Sigma}\,X^\Sigma\,\,,\,\,\,\,0=X^{\hat{\Lambda}}=
 E^{\hat{\Lambda}\Sigma}\,X_\Sigma+E^{\hat{\Lambda}}{}_{\Sigma}\,X^\Sigma\,,
 \end{equation}
where we have used the fact that, in the electric frame, the magnetic components $\Theta^{\hat{\Lambda}\alpha}$ of the embedding tensor vanish, so that $X^{\hat{\Lambda}}=\Theta^{\hat{\Lambda}\alpha}\,t_\alpha=0$. Using the matrix $E$ we can then write
  \begin{equation} A^M_\mu\,X_M= A^M_\mu\,E^{-1}{}_M{}^{\hat{N}}\,X_{\hat{N}}=A^M_\mu\,E^{-1}{}_M{}^{\hat{\Lambda}}\,X_{\hat{\Lambda}}=
  A^{\hat{\Lambda}}_\mu\,X_{\hat{\Lambda}}
  \,,
  \nonumber\end{equation}
  where we have defined: $A^{\hat{\Lambda}}_\mu\equiv E^{-1}{}_M{}^{\hat{\Lambda}}\,A^M_\mu$.\par
 In the new formulation we wish to discuss, however, the vectors $A^\Lambda_\mu$, instead of $A^{\hat{\Lambda}}_\mu$, enter the kinetic terms.
 The covariant derivatives are then defined in terms of (\ref{newcon}), as in Step 2 of the Section \ref{gaugingsteps}, and, as prescribed there, should replace ordinary derivative everywhere in the action.
The infinitesimal gauge variation of $A^M$ reads:
\begin{equation}
\delta A^M_\mu=\mathcal{D}_\mu\zeta^M\equiv \partial_\mu\zeta^M+\,A^N_\mu X_{NP}{}^M\,\zeta^P\,,\label{deltaA}
\end{equation}
where, as usual, $X_{MP}{}^R\equiv {\Scr R}_{v*}[X_{M}]_{P}{}^R$.
 We define for this set of electric-magnetic vector fields a symplectic covariant generalization $\mathcal{G}^M$ of the non-Abelian field strengths $F^{\hat{\Lambda}}$ (\ref{defF}):
 \begin{equation}
 {F}^M_{\mu\nu}\equiv \partial_\mu A^M_\nu-\partial_\nu A^M_\mu+\,X_{[NP]}{}^M\,A^N_\mu A^P_\nu\,\,\Leftrightarrow\,\,\,\,{F}^M\equiv dA^M+\frac{g}{2}\,X_{NP}{}^M\,A^N\wedge A^P\,,\label{FMdef}
 \end{equation}
where in the last equation we have used the form-notation for the fields strengths. The gauge algebra-valued curvature $\mathcal{F}$ is defined as in (\ref{calF}):
 \begin{equation}
 \mathcal{F}\equiv {F}^M\,X_M\,.\label{gcurv}
 \end{equation}
 The first problem one encounters in describing the vectors $A^\Lambda_\mu$ in the kinetic terms is that, in a symplectic frame which is not the electric one, such fields are not well defined since their curvatures fail to satisfy the Bianchi identity.
 This comes with no surprise since the components $\Theta^{\Lambda\,\alpha}$ of the embedding tensor are nothing but \emph{magnetic charges}.
 One can indeed verify that:
\begin{equation}
\mathcal{D}{F}^M\equiv d{F}^M+\,X_{NP}{}^M\,A^N\wedge {F}^P=\,X_{(PQ)}{}^M\,A^P\wedge\left( dA^Q+\frac{g}{3}\,X_{RS}{}^QA^R\wedge A^S\right)\neq 0\,.\label{Bianchifail}
\end{equation}
In particular $\mathcal{D}F^\Lambda\neq 0$ since, by Eq. (\ref{lconstrnew}), $X_{(PQ)}{}^\Lambda=-\frac{1}{2}\,\Theta^{\Lambda\alpha}\,t_{\alpha\,P}{}^N\mathbb{C}_{NQ}\neq 0$, being in the non-electric frame $\Theta^{\Lambda\alpha}\neq 0$. To deduce (\ref{Bianchifail}) we have used the quadratic constraint (\ref{quadratic2}) on the gauge generators $X_M$ in the ${\Scr R}_{v*}$-representation, which reads:
\begin{equation}
X_{MP}{}^R X_{NR}{}^Q-X_{NP}{}^R X_{MR}{}^Q+X_{MN}{}^R X_{RP}{}^Q=0\,.\label{quadrdual}
\end{equation}
 From the above identity, after some algebra (see Appendix \ref{apXid}), one finds:
\begin{equation}
X_{[MP]}{}^R X_{[NR]}{}^Q+X_{[PN]}{}^R X_{[MR]}{}^Q+X_{[NM]}{}^R X_{[PR]}{}^Q=-(X_{NM}{}^R\,X_{(PR)}{}^Q)_{[MNP]}\,,\label{nojacobi}
\end{equation}
that is the \emph{generalized structure constants} $X_{[MP]}{}^R$ entering the definition (\ref{FMdef}) do not satisfy the Jacobi identity, and this feature is at the root of (\ref{Bianchifail}).

Related to this is the non-gauge covariance of ${F}^M$. The reader can indeed verify that (we use the form-notation):
\begin{equation}
\delta F^M=-\,X_{NP}{}^M\,\zeta^N\,F^P+\,\left(2 \,X_{(NP)}{}^M\,\zeta^N\,F^P-X_{(NP)}{}^M\,A^N\wedge \delta A^P\right)\neq -\,X_{NP}{}^M\,\zeta^N\,F^P\,,\label{deltsFnc}
\end{equation}
where $\delta A^M$ is given by (\ref{deltaA}) and we have used the general property:
\begin{equation}
\delta F^M=\mathcal{D}\delta A^M-X_{(PQ)}{}^M\,A^P\wedge \delta A^Q\,,\label{deltaFgen}
\end{equation}
valid for generic $\delta A^M$.\par
We also observe that the obstruction to the Bianchi identity (\ref{Bianchifail}), as well as the non-gauge covariant terms in (\ref{deltsFnc}), are proportional to a same tensor $X_{(MN)}{}^P$. This quantity, as a consequence of Eq. (\ref{quadratic2}), see Eq.\ (\ref{quad2n}), vanishes if contracted with the gauge generators $X_M$, namely with the first index of the embedding tensor: $X_{(MN)}{}^P\,\Theta_P{}^\alpha=0$.
Being the true electric vector fields $A^{\hat{\Lambda}}_\mu$, defined as the combinations of the $A^M_\mu$ singled out by the contraction $A^M_\mu \,\Theta_M{}^\alpha$, see Eq. (\ref{ATHcomb}), they are perfectly well defined, since the non-vanishing terms in $\mathcal{D}F^M$ and the non-covariant terms in $\delta F^M$ vanish upon contraction with $\Theta_M{}^\alpha$. Consequently also the gauge connection, which only depends on $A^{\hat{\Lambda}}_\mu$, is well defined. Indeed, one can easily show, using the matrix $E$, that the gauge curvature (\ref{gcurv}) only contains the field strengths $F^{\hat{\Lambda}}$ associated with $A^{\hat{\Lambda}}$ and defined in (\ref{defF}):
\begin{equation}
 \mathcal{F}\equiv {F}^M\,X_M=F^{\hat{\Lambda}}\,X_{\hat{\Lambda}}\,.
\end{equation}
On the other hand, using (\ref{Bianchifail}) and (\ref{quad2n}) we have:
\begin{equation}
 \mathcal{D}\mathcal{F}= \mathcal{D}{F}^M\,X_M=0\,.\label{BianchiFgauge}
\end{equation}
The gauge covariance (\ref{gaugecovF}) of $\mathcal{F}$, and thus of $F^{\hat{\Lambda}}$, is also easily verified by the same token, together with Eq.\ (\ref{D2F}): $\mathcal{D}^2=-\mathcal{F}$.\par
In order to construct gauge-covariant quantities describing the vector fields, we combine the vector field strengths $F^M_{\mu\nu}$ with a set of  massless antisymmetric tensor fields $B_{\alpha\,\mu\nu}$
\footnote{
These fields will also be described as 2-forms $B_\alpha\equiv \frac{1}{2}\,B_{\alpha\,\mu\nu}\,dx^\mu\wedge dx^\nu$.
}
in the adjoint representation of $G$ through the matrix
\begin{equation}
Z^{M\,\alpha}\equiv \frac{1}{2}\,\mathbb{C}^{MN}\,\Theta_N{}^\alpha\,.\label{defZ}
\end{equation}
and define the following new field strengths:\footnote{Restoring the coupling constant $g$ we would have: $\mathcal{H}^M_{\mu\nu}\equiv F^M_{\mu\nu}+g\,Z^{M\,\alpha}\,B_{\alpha\,\mu\nu}$.}
\begin{align}
\mathcal{H}^M_{\mu\nu}\equiv F^M_{\mu\nu}+Z^{M\,\alpha}\,B_{\alpha\,\mu\nu}\;:\;\;
\begin{cases}
\mathcal{H}^\Lambda=F^\Lambda+\frac{1}{2}\,\Theta^{\Lambda\alpha}\,B_\alpha\,,\cr
\mathcal{H}_\Lambda=F_\Lambda-\frac{1}{2}\,\Theta_{\Lambda}{}^{\alpha}\,B_\alpha\,.
\end{cases}
\label{HZB}
\end{align}
From the definition (\ref{defZ}) and (\ref{quadratic1}) we have:
\begin{equation}
Z^{M\,\alpha}\,\Theta_{M}{}^\beta=0\,\,\,\Leftrightarrow\,\,\,\,Z^{M\,\alpha}\,X_{M}=0\,.\label{Zort}
\end{equation}
In terms of the new tensor $Z^{M\,\alpha}$ the linear constraint (\ref{linear2}), or equivalently (\ref{lconstrnew}), read:
\begin{equation}
X_{(NP)}{}^M=-Z^{M\,\alpha}\,t_{\alpha\,NP}\,.\label{linear22}
\end{equation} The reason for considering the combination (\ref{HZB}) is that the non-covariant terms in the gauge variation of $F^M_{\mu\nu}$, being proportional to $X_{(NP)}{}^M$, that is to $Z^{M\,\alpha}$, can be canceled by a corresponding variation of the tensor fields $\delta B_{\alpha\mu\nu}$:
\begin{align}
\delta \mathcal{H}^M&=\,X_{PN}{}^M\,\zeta^N\,F^P+Z^{M\alpha}\,\left(\delta B_\alpha+t_{\alpha NP}\,A^N\wedge \delta A^P\right)=\nonumber\\
&=X_{PN}{}^M\,\zeta^N\,\mathcal{H}^P+Z^{M\alpha}\,\left(\delta B_\alpha+t_{\alpha NP}\,A^N\wedge \delta A^P\right)=\nonumber\\
&=-X_{NP}{}^M\,\zeta^N\,\mathcal{H}^P+2\,X_{(NP)}{}^M\,\zeta^N\,\mathcal{H}^P+Z^{M\alpha}\,\left(\delta B_\alpha+t_{\alpha NP}\,A^N\wedge \delta A^P\right)=\nonumber\\
&=-X_{NP}{}^M\,\zeta^N\,\mathcal{H}^P+Z^{M\alpha}\,\left[\delta B_\alpha+t_{\alpha NP}\,(A^N\wedge \delta A^P-2\,\zeta^N\,\mathcal{H}^P)\right]\,,\label{deltaHM0}
\end{align}
where in going from the first to the second line we have used (\ref{Zort}), so that: $X_{PN}{}^M\,F^P=X_{PN}{}^M\,\mathcal{H}^P$.
If we define:
\begin{equation}
\delta B_\alpha\equiv t_{\alpha NP}\,(2\,\zeta^N\,\mathcal{H}^P-A^N\wedge \delta A^P)\,,\label{Btra1}
\end{equation}
the term proportional to $Z^{M\,\alpha}$ in (\ref{deltaHM0}) vanishes and $ \mathcal{H}^M$ transforms covariantly.
The kinetic terms in the Lagrangian are then written in terms of $\mathcal{H}^\Lambda_{\mu\nu}$:
\begin{equation}
\frac{1}{e}\L_{v,\,kin}=
\frac{1}{4}\,\I_{\Lambda\Sigma}(\phi)\,\mathcal{H}^\Lambda_{\mu\nu}\,\mathcal{H}^{\Sigma\,\mu\nu}
+\frac{1}{8\,e}\,\R_{\Lambda\Sigma}(\phi)\,\eps^{\mu\nu\rho\sigma}\,\mathcal{H}^\Lambda_{\mu\nu} \,\mathcal{H}^{\Sigma}_{\rho\sigma}\,,
\label{bosoniclagr}
\end{equation}
The above transformation property (\ref{Btra1}) should however be modified since the quantity we want to transform covariantly is not quite $\mathcal{H}^M$ but rather the symplectic vector:
\begin{equation}
\Gd^M\equiv \left(\begin{matrix}\mathcal{H}^\Lambda\cr  G_\Lambda\end{matrix}\right)\,\,;\,\,\,\, G_{\Lambda\,\mu\nu}\equiv -\epsilon_{\mu\nu\rho\sigma}
\frac{\partial \L}{\partial
\mathcal{H}^\Lambda_{\rho\sigma}}\,,\label{GMHdef}
\end{equation}
corresponding, in the ungauged theory, to the field-strength-vector $\mathcal{G}^M$ of Eq.\ (\ref{bbF}), satisfying the twisted self-duality condition given by (\ref{FCMF}), or, in its complete form, by (\ref{FCMF2}).
Consistency of the construction will then imply that the two quantities $\mathcal{H}^M$ and $\Gd^M$, which are off-shell different since the former depends on the ``magnetic'' vector fields $A_\Lambda$ as opposed to the latter, \emph{will be identified on-shell}. As we shall see, the equations of motion for $B_{\alpha\mu\nu}$ have indeed the form: \begin{equation}(\mathcal{H}^M-\Gd^M)\,\Theta_M{}^\alpha=(\mathcal{H}_\Lambda- G_\Lambda)\,\Theta^{\Lambda\,\alpha}=0\,.\label{HG0} \end{equation}
These equations in particular identify the field strengths of the auxiliary fields $A_\Lambda$ in $\mathcal{H}_\Lambda$ with the duals $G_\Lambda$ to $\mathcal{H}^\Lambda$.\par
We the require $\Gd^M$ to be the gauge covariant object, namely to transform covariant under $G_g$, upon use of (\ref{HG0}).
To this end we modify Eq.\ (\ref{Btra1}) as follows:
\begin{equation}
\delta B_\alpha\equiv t_{\alpha NP}\,(2\,\zeta^N\,\Gd^P-A^N\wedge \delta A^P)\,,\label{Btra12}
\end{equation}
so that the variations of the symplectic vectors $\mathcal{H}^M$ and $\Gd^M$ read:
\begin{align}
\delta \mathcal{H}^M_{\mu\nu}&=-X_{NP}{}^M\,\zeta^N\,\mathcal{H}^P_{\mu\nu}+\mbox{non-covariant terms}\,,\nonumber\\
\delta \Gd^M_{\mu\nu}&=-X_{NP}{}^M\,\zeta^N\,\Gd^P_{\mu\nu}+\mbox{non-covariant but on-shell vanishing terms}\,.
\end{align}
More specifically we have:
\begin{align}
\delta \mathcal{H}^M_{\mu\nu}&=-X_{NP}{}^M\,\zeta^N\,\mathcal{H}^P_{\mu\nu}-2\,X_{(NP)}{}^M\,(\mathcal{G}_{\mu\nu}-\mathcal{H}_{\mu\nu})^N\,\zeta^P\,,\label{deltazetaH}\\
\delta G_{\Lambda\,\mu\nu}&=-X_{NP\Lambda}\,\zeta^N\,\Gd^P_{\mu\nu}-\zeta^N\,X_{PN}{}^\Sigma\,\left[\R_{\Sigma\Lambda}(\mathcal{G}-\mathcal{H})^P_{\mu\nu}-
\I_{\Sigma\Lambda}\,({}^*\mathcal{G}-{}^*\mathcal{H})^P_{\mu\nu}\right]\,.\label{deltazetaG}
\end{align}
The last two terms on the right hand side of (\ref{deltazetaG}) vanish upon use of the on-shell condition (\ref{HG0}). The field-strengths $\mathcal{H}_{\mu\nu}^\Lambda$ are the upper component of both $\mathcal{H}^M_{\mu\nu}$ and $\mathcal{G}^M_{\mu\nu}$. However they should transform covariantly only as the upper components of $\mathcal{G}^M_{\mu\nu}$, once (\ref{HG0}) is imposed. Their transformation properties can be deduced from (\ref{deltazetaH}) and read:
\begin{equation}
\delta \mathcal{H}^\Lambda_{\mu\nu}=-X_{NP}{}^\Lambda\,\zeta^N\,\mathcal{G}^P_{\mu\nu}-(\mathcal{G}-\mathcal{H})^P_{\mu\nu}\,X_{PN}{}^\Lambda\,\zeta^N\,.\label{deltazetaHup}
\end{equation}
A consistent definition of $B_\alpha$ requires the theory to be gauge-invariant with respect to transformations parametrized by 1-forms: $\Xi_\alpha=\Xi_{\alpha\mu}\,dx^\mu$.
Such transformations should in turn be $G_g$-covariant and leave $\mathcal{H}^M$ unaltered:
\begin{equation}
A^M\rightarrow A^M+\delta_\Xi A^M\,\,;\,\,\,\,B_\alpha\rightarrow B_\alpha+\delta_\Xi B_\alpha\,\,\,\Rightarrow\,\,\,\,\,\delta_\Xi\mathcal{H}^M=0\,.
\end{equation}
Let us use (\ref{deltaFgen}) then to write
\begin{equation}
\delta_\Xi\mathcal{H}^M=\mathcal{D}\delta_\Xi A^M+Z^{M\,\alpha}\,\left(\delta_\Xi B_{\alpha}+t_{\alpha NP}\,A^N\wedge \delta_\Xi A^P\right)\,.
\end{equation}
If we set
\begin{equation}
\delta_\Xi A^M=-Z^{M\alpha}\,\Xi_\alpha\,,\label{deltaxi1}
\end{equation}
the invariance of $\mathcal{H}^M$ implies:
\begin{equation}
\delta_\Xi B_{\alpha}=\mathcal{D}\Xi_\alpha-t_{\alpha NP}\,A^N\wedge \delta_\Xi A^P\,,\label{deltaxi2}
\end{equation}
where
\begin{equation}
\mathcal{D}\Xi_\alpha\equiv d\Xi_\alpha+\Theta_M{}^\beta\,{\rm f}_{\beta\alpha}{}^\gamma A^M\wedge\Xi_\gamma\,.
\end{equation}
We have thus introduced, together with the new fields $A_{\Lambda\,\mu}$ and $B_{\alpha\,\mu\nu}$, extra gauge symmetries, parametrized by $\zeta_\Lambda$ and $\Xi_{\alpha\,\mu}$. This ensures the correct number of propagating degrees of freedom. We can verify that the commutator between these gauge transformations on the fields to close as follows \cite{deWit:2007mt}:
\begin{align}
[\delta(\zeta_1),\delta(\zeta_2)]&=\delta(\zeta_3)+\delta(\Xi_3)\,\,;\,\,\,\,[\delta(\zeta),\delta(\Xi)]=\delta(\tilde{\Xi})\,,\nonumber\\
\zeta_3^M&=\zeta_1^P\zeta_2^Q\,X_{[PQ]}{}^M\,\,;\,\,\,\,\Xi_{3\,\alpha}=t_{\alpha MN}(\mathcal{D}\zeta_1^M \zeta_2^N-\mathcal{D}\zeta_2^M \zeta_1^N)\,,\nonumber\\
\tilde{\Xi}_{\alpha}&=\zeta^P\,((X_{P})_{\alpha}{}^\beta-2\,t_{\alpha\, PM}\,Z^{M\,\beta})\,\Xi_\beta\,.
\end{align}
Using the quadratic constraints (\ref{quadratic2}) on the embedding tensor one can verify that $\delta(\tilde{\Xi})A^M=0$.\par
Let us now introduce field strengths for the 2-forms:
\begin{equation}
\mathcal{H}^{(3)}_\alpha\equiv \mathcal{D}B_\alpha-t_{\alpha PQ}A^P\wedge\left( dA^Q+\frac{1}{3}\,X_{RS }{}^Q\,A^R\wedge A^S\right)\,.
\end{equation}
Writing the forms in components:
\begin{equation}
\mathcal{H}^{(3)}_\alpha=\frac{1}{3!}\,\mathcal{H}_{\alpha\,\mu\nu\rho}\,dx^\mu\wedge dx^\nu\wedge dx^\rho\,\,;\,\,\,\,
\mathcal{D}B_\alpha=\frac{1}{2}\,\mathcal{D}_{\mu}B_{\alpha\,\nu\rho}\,dx^\mu\wedge dx^\nu\wedge dx^\rho\,,
\end{equation}
we have:
\begin{equation}
\mathcal{H}_{\alpha\,\mu\nu\rho}=3\,\mathcal{D}_{[\mu}B_{\alpha\,\nu\rho]}-6\,t_{\alpha PQ}\left(A^P_{[\mu}\partial_\nu A_{\rho]}^Q+\frac{1}{3}\,X_{RS }{}^Q\,A^P_{[\mu}A^R_\nu A^S_{\rho]}\right)\,.\label{defH3}
\end{equation}
The reader can verify, using the relations summarized in Appendix \ref{apXid}, that the following Bianchi identities hold:\footnote{Due to the fact that $\mathcal{H}^M_{\mu\nu}$ do not transform covariantly under gauge transformations, also $\mathcal{D}\mathcal{H}^M\equiv d\mathcal{H}^M-X_{PQ}{}^M\,A^P\wedge \mathcal{H}^Q$ is not covariant. The Bianchi identities (\ref{Bid1n}) and (\ref{Bid2n}) can be written in a manifestly gauge-covariant way by redefining $\mathcal{D}\mathcal{H}^M$ and $\mathcal{H}^{(3)}_\alpha$ correspondingly, see \cite{deWit:2007mt}:
 \begin{equation}
 \mathcal{D}\mathcal{H}^H\rightarrow  \mathcal{D}\mathcal{H}^H+X_{NP}{}^M\,A^P\wedge (\mathcal{G}-\mathcal{H})^N\,\,;\,\,\,\,\mathcal{H}^{(3)}_\alpha\rightarrow \mathcal{H}^{(3)}_\alpha-t_{\alpha \,MN}\,A^M\wedge (\mathcal{G}-\mathcal{H})^N\,.
 \end{equation}
 We shall not do it here since we content ourselves with showing gauge invariance of the action and thus of the field equations.}
\begin{align}
\mathcal{D}\mathcal{H}^M&= Z^{M\alpha}\,\mathcal{H}^{(3)}_\alpha\,,\label{Bid1n}\\
Z^{M\,\alpha}\,\mathcal{D}\mathcal{H}^{(3)}_\alpha &= X_{NP}{}^M\,\mathcal{H}^N\wedge \mathcal{H}^P\,.\label{Bid2n}
\end{align}
Just as in Step 3. of Section \ref{gaugingsteps}, gauge invariance of the bosonic action requires the introduction of topological terms, so that the final gauged bosonic Lagrangian reads:
\begin{eqnarray}
{\Scr L}_{B} &=& -\frac{e}{2}\,R+\frac{e}{2}\,\Gm_{st}(\phi)\,\mathcal{D}_\mu\phi^s\,\mathcal{D}^\mu\phi^t+
\frac{e}{4} \, {\cal
I}_{\Lambda\Sigma}\,\mathcal{H}_{\mu\nu}{}^{\Lambda}
\mathcal{H}^{\mu\nu\,\Sigma} +\frac{1}{8} {\cal
R}_{\Lambda\Sigma}\;\varepsilon^{\mu\nu\rho\sigma}
\mathcal{H}_{\mu\nu}{}^{\Lambda}
\mathcal{H}_{\rho\sigma}{}^{\Sigma}+
 \nonumber\\
&&+{\Scr L}_{top,\,B}+{\Scr L}_{GCS}\,,\label{boslag2}
\end{eqnarray}
where we have defined:
\begin{align}
{\Scr L}_{top,\,B}&\equiv -\frac{1}{8}\, \varepsilon^{\mu\nu\rho\sigma}\,
\Theta^{\Lambda\alpha}\,B_{\mu\nu\,\alpha} \, \Big(
2\,\partial_{\rho} A_{\sigma\,\Lambda} + X_{MN\,\Lambda}
\,A_\rho{}^M A_\sigma{}^N
-\frac{1}{4}\,\Theta_{\Lambda}{}^{\beta}B_{\rho\sigma\,\beta}
\Big)\,,\label{topB}\\
{\Scr L}_{GCS}&\equiv-\frac{1}{3}\,
\varepsilon^{\mu\nu\rho\sigma}X_{MN\,\Lambda}\, A_{\mu}{}^{M}
A_{\nu}{}^{N} \Big(\partial_{\rho} A_{\sigma}{}^{\Lambda}
+\frac{1}{4}  X_{PQ}{}^{\Lambda}
A_{\rho}{}^{P}A_{\sigma}{}^{Q}\Big)
\nonumber\\[.9ex]
&{} -\frac{1}{6}\,
\varepsilon^{\mu\nu\rho\sigma}X_{MN}{}^{\Lambda}\, A_{\mu}{}^{M}
A_{\nu}{}^{N} \Big(\partial_{\rho} A_{\sigma}{}_{\Lambda}
+\frac{1}{4}\, X_{PQ\Lambda}
A_{\rho}{}^{P}A_{\sigma}{}^{Q}\Big)\,.\label{GCS}
\end{align}
The \emph{Chern-Simons terms} in ${\Scr L}_{GCS}$ generalize those in Eq. (\ref{top}). On top of them, gauge invariance of the action requires the introduction of new topological terms, depending on the $B$-fields, which appear in ${\Scr L}_{top,\,B}$. Notice that if the magnetic charges $\Theta^{\Lambda\,\alpha}$ vanish (i.e.\ we are in the electric frame), $B_\alpha$ disappear from the action, since (\ref{topB}) vanishes as well as the $B$-dependent Stueckelberg term in $\mathcal{H}^\Lambda$. In this limit one can verify that ${\Scr L}_{GCS}$ reduces to (\ref{top}). \par
To understand the role of the various new terms in the bosonic Lagrangian, it is useful to evaluate the variation of ${\Scr L}_{GCS}$ and ${\Scr L}_{top,\,B}$ corresponding to generic variations of the vector and tensor fields:
\begin{equation}
A^M_\mu\rightarrow A^M_\mu+\delta A^M_\mu\,\,;\,\,\,\,\,B_{\alpha\,\mu\nu}\rightarrow B_{\alpha\,\mu\nu}+\delta B_{\alpha\,\mu\nu}\,.
\end{equation}
After some algebra, and using the constraints (\ref{linear2}), (\ref{quadratic1}) and (\ref{quadratic2}) on the embedding tensor, one finds:
\begin{align}
\delta {\Scr L}_{GCS}&=-\frac{1}{2}\epsilon^{\mu\nu\rho\sigma}\left(F^\Lambda_{\mu\nu}\mathcal{D}_\rho \delta A_{\Lambda\sigma}-X_{(PQ)}{}^\Lambda F_{\Lambda\,\mu\nu}A^P_\rho \delta A^Q_\sigma\right)\,,\nonumber\\
\delta {\Scr L}_{top,\,B}&=-\frac{1}{8}\epsilon^{\mu\nu\rho\sigma}\left[2\,\Theta^{\Lambda\,\alpha} B_{\alpha\,\mu\nu}\mathcal{D}_\rho \delta A_{\Lambda\sigma}+Z_{\Lambda}{}^\alpha B_{\alpha\,\mu\nu}\,\Theta^{\Lambda\,\beta}\left(\delta B_{\beta\,\rho\sigma}+2\,t_{\beta PQ}\,A^P_\rho\delta A^Q_\sigma\right)
\right.+\nonumber\\
&+\left.F_{\Lambda\,\mu\nu}\,\Theta^{\Lambda\,\beta}\delta B_{\beta\,\rho\sigma}\right]=\nonumber\\
&=-\frac{1}{2}\epsilon^{\mu\nu\rho\sigma}\left[\mathcal{H}^\Lambda_{\mu\nu} \mathcal{D}_\rho \delta A_{\Lambda\sigma}+\frac{1}{2}\,\mathcal{H}_{\Lambda\,\mu\nu}\,Z^{\Lambda\,\beta}\left(\delta B_{\beta\,\rho\sigma}+2\,t_{\beta PQ}\,A^P_\rho\delta A^Q_\sigma\right)\right]-\delta {\Scr L}_{GCS}\,,
\end{align}
so that, summing the two variations, we have:
\begin{equation}
\delta({\Scr L}_{top,\,B}+{\Scr L}_{GCS})=-\frac{1}{2}\epsilon^{\mu\nu\rho\sigma}\left[\mathcal{H}^\Lambda_{\mu\nu} \mathcal{D}_\rho \delta A_{\Lambda\sigma}+\frac{1}{2}\,\mathcal{H}_{\Lambda\,\mu\nu}\,Z^{\Lambda\,\beta}\left(\delta B_{\beta\,\rho\sigma}+2\,t_{\beta PQ}\,A^P_\rho\delta A^Q_\sigma\right)\right]\,.\label{deltatop}
\end{equation}
On the other hand the variation of the kinetic terms ${\Scr L}_{g,k}$ for the gauge fields $A^M_\mu$ reads:
\begin{equation}
\delta {\Scr L}_{g,k}=\frac{1}{4}\epsilon^{\mu\nu\rho\sigma}G_{\Lambda\,\mu\nu}\delta \mathcal{H}^\Lambda_{\rho\sigma}=
\frac{1}{2}\epsilon^{\mu\nu\rho\sigma}G_{\Lambda\,\mu\nu}\left[\mathcal{D}_\rho \delta A^\Lambda_{\sigma}+\frac{1}{2}\,Z^{\Lambda\,\alpha}\left(\delta B_{\alpha\,\rho\sigma}+2\,t_{\alpha\,PQ}A^P_\rho\delta A^Q_\sigma\right)\right]\,.\label{deltagk}
\end{equation}
Summing up (\ref{deltatop}) and (\ref{deltatop}) we finally find:
\begin{align}
\delta({\Scr L}_{g,k}+{\Scr L}_{top,\,B}+{\Scr L}_{GCS})&=-\frac{1}{2}\epsilon^{\mu\nu\rho\sigma}\left[
\mathcal{G}^M_{\mu\nu}\,\mathcal{D}_\rho \delta A^N_{\sigma}\,\mathbb{C}_{MN}+\right.\nonumber\\&\left.+\frac{1}{2}(\mathcal{H}_{\Lambda\,\mu\nu}-G_{\Lambda\,\mu\nu})\,Z^{\Lambda\,\alpha}\left(\delta B_{\alpha\,\rho\sigma}+2\,t_{\alpha\,PQ}A^P_\rho\delta A^Q_\sigma\right)\right]=\nonumber\\
&=-\frac{1}{2}\epsilon^{\mu\nu\rho\sigma}\left[-\mathcal{D}_\mu\mathcal{G}^M_{\nu\rho}\, \delta A^N_{\sigma}\,\mathbb{C}_{MN}+\right.\nonumber\\&\left.+\frac{1}{2}(\mathcal{H}_{\Lambda\,\mu\nu}-G_{\Lambda\,\mu\nu})\,Z^{\Lambda\,\alpha}\left(\delta B_{\alpha\,\rho\sigma}+2\,t_{\alpha\,PQ}A^P_\rho\delta A^Q_\sigma\right)\right]\,,\label{deltaLtot}
\end{align}
where we have integrated by parts the term $\mathcal{G} \mathcal{D}\delta A$. The above formula will be useful in order to compute the field equations for $A^M_\mu$ and $B_{\alpha\,\mu\nu}$. We notice that, as observed earlier, in the electric frame in which $\Theta^{\Lambda\,\alpha}=0$, ${\Scr L}_B$ does not depend on the tensor fields since the coefficient of $\delta B_\alpha$ in (\ref{deltaLtot}) vanishes. On  the other hand the magnetic vectors $A_{\Lambda\,\mu}$ drop off from the minimal couplings and only remain in ${\Scr L}_{top,\,B}+{\Scr L}_{GCS}$. However $\delta A_{\Lambda\,\mu}$ enters (\ref{deltaLtot}) multiplied by $\mathcal{D}_{[\mu}F^\Lambda_{\nu\rho]}$ which vanishes in the electric frame by the Bianchi identity, being there no magnetic charges. This shows that, in the duality-covariant formulation under discussion, the dependence of the Lagrangian on the extra fields $A_{\Lambda\,\mu},\,B_{\alpha\,\mu\nu}$ needed  for the construction, is related to the \emph{dyonic} nature of the embedding tensor, namely to the presence of the corresponding magnetic components $\Theta^{\Lambda\,\alpha}\neq 0$. If we restrict ourselves to \emph{electric gaugings}, characterized by $\Theta^{\Lambda\,\alpha}= 0$, the extra fields disappear altogether from the Lagrangian and we are back to the construction illustrated in Sect. \ref{gitef}.\par
Before deriving the field equations for the vector and antisymmetric tensor fields, let us comment on the gauge invariance of the action in order to appreciate the interplay between the various terms in the (\ref{boslag2}). The reader who is not interested in these details can directly move to Subsection \ref{ageneraldiscussion}.
\paragraph{Gauge invariance of the action.}
In order to compute the variation of the gauge-kinetic terms ${\Scr L}_{g,k}$ under a vector-gauge transformation parametrized by $\zeta^M(x)$, we need to add to the term in (\ref{deltagk}) the contribution due to the gauge variation of the scalar fields in $\I_{\Lambda\Sigma}$ and $\R_{\Lambda\Sigma}$. We can write these additional terms in a symplectic invariant form as follows:
\begin{align}
\delta_\zeta {\Scr L}_{g,k}&=\frac{1}{4}\epsilon^{\mu\nu\rho\sigma}G_{\Lambda\,\mu\nu}\delta \mathcal{H}^\Lambda_{\rho\sigma}+\frac{e}{4}\,\zeta^M\,\mathcal{G}^T_{\mu\nu} X_M\mathcal{M}\mathcal{G}^{\mu\nu}=\nonumber\\
&=\frac{1}{4}\epsilon^{\mu\nu\rho\sigma}\left(G_{\Lambda\,\mu\nu}\delta \mathcal{H}^\Lambda_{\rho\sigma}+\frac{1}{2}\zeta^M\,\mathcal{G}^T_{\mu\nu} X_M\mathbb{C}\mathcal{G}_{\rho\sigma}\right)=\nonumber\\&=
\frac{1}{4}\epsilon^{\mu\nu\rho\sigma}\zeta^P\left(X^\Sigma{}_P{}^\Lambda\,G_{\Lambda\,\mu\nu}\,\mathcal{H}_{\Sigma\,\rho\sigma}-\frac{1}{2}\,
\,X_{P\,\Lambda\Sigma}\,\mathcal{H}^\Lambda_{\mu\nu}\mathcal{H}^\Sigma_{\rho\sigma}\right)\,,\label{deltazetaLgk}
\end{align}
where we have suppressed the symplectic indices $M,N$ and we have used, in going from the first to the second line, the twisted self-duality condition (\ref{FCMF}). We have also used the transformation properties of $\I_{\Lambda\Sigma}$ and $\R_{\Lambda\Sigma}$ deduced from (\ref{traM}) and (\ref{infM}):
\begin{equation}
\delta_\zeta \mathcal{M}=\zeta^P\,k_P^s\,\partial_s\mathcal{M}=\zeta^P\,(X_P\,\mathcal{M}+\mathcal{M}\,X_P^T)\,,
\end{equation}
and the transformation law of $\mathcal{H}^\Lambda_{\mu\nu}$ given in (\ref{deltazetaHup}). Notice that in the electric frame ($\Theta^{\Lambda\,\alpha}=0$) the above variation reduces to
\begin{align}
\delta_\zeta {\Scr L}_{g,k}&=-\frac{1}{8}\,\zeta^\Gamma\,X_{\Gamma\Lambda\Sigma}\,\epsilon^{\mu\nu\rho\sigma}
\mathcal{H}^\Lambda_{\mu\nu}\mathcal{H}^\Sigma_{\rho\sigma}\,,\label{deltazetaLgkel}
\end{align}
which corresponds to Eq. (\ref{deltaLC}) being $C_{\Lambda\Pi}\,D_{\Sigma}{}^\Pi=-\zeta^\Gamma\,X_{\Gamma\Lambda\Sigma}$ and $\mathcal{H}^\Lambda_{\mu\nu}=F^\Lambda_{\mu\nu}$.\par
The gauge variation of ${\Scr L}_{top,\,B}+{\Scr L}_{GCS}$ can be computed from (\ref{deltatop}), using the identities in  Appendix \ref{apXid}:
 \begin{align}
 \delta_\zeta ({\Scr L}_{top,\,B}+{\Scr L}_{GCS})&=-\frac{1}{2}\epsilon^{\mu\nu\rho\sigma}\left(
 \mathcal{H}^\Lambda_{\mu\nu}\mathcal{D}_{\rho}\mathcal{D}_{\sigma}\zeta_{\Lambda}-\zeta^Q\,X_{(PQ)}{}^\Lambda\mathcal{H}_{\Lambda\,\mu\nu}\mathcal{G}^P_{\rho\sigma}\right)
 \nonumber\\&=-\frac{1}{4}\epsilon^{\mu\nu\rho\sigma}\zeta^Q\,\left(\mathcal{H}^\Lambda_{\mu\nu}\mathcal{H}^P_{\rho\sigma}
 X_{PQ\Lambda}-2\,X_{(PQ)}{}^\Lambda\mathcal{H}_{\Lambda\,\mu\nu}\mathcal{G}^P_{\rho\sigma}\right)\,.
 \end{align}
 Expanding the above expression and using the linear constraints (\ref{lconstr}), the reader can verify that it precisely cancels (\ref{deltazetaLgk}).\par In the electric frame this variation is only due to ${\Scr L}_{GCS}$, being ${\Scr L}_{top,\,B}=0$,  and has the
simple form
\begin{align}
\delta_\zeta {\Scr L}_{GCS}&=
-\frac{1}{4}\epsilon^{\mu\nu\rho\sigma}\,F^\Lambda_{\mu\nu}F^\Gamma_{\rho\sigma}\,X_{ \Gamma\Lambda\Sigma}\zeta^\Sigma=
\frac{1}{8}\epsilon^{\mu\nu\rho\sigma}\,F^\Lambda_{\mu\nu}F^\Gamma_{\rho\sigma}\,X_{\Sigma \Gamma\Lambda}\zeta^\Sigma\,,
\end{align}
where we have used (\ref{lconstr}): $X_{(\Gamma\Lambda)\Sigma}=-\frac{1}{2}\,X_{\Sigma \Gamma\Lambda}$. This variation precisely cancels (\ref{deltazetaLgkel}), as it was first proven in \cite{deWit:1984rvr}.\par
In a similar way, using Eqs. (\ref{deltatop}) and (\ref{deltagk}) and the constraints on the embedding tensor, one can prove the invariance of the bosonic action with respect to the $\Xi$-gauge transformations.

\subsection{A General Discussion}\label{ageneraldiscussion}
Before going on with our analysis of the field equations, let us pause for a moment to highlight the main features of the duality-covariant construction discussed so far. The constraints (\ref{linear2}), (\ref{quadratic1}) and (\ref{quadratic2}) are needed for the consistent definition of the gauged bosonic action, which is uniquely determined. Just as discussed in Sect.\ \ref{gaugingsteps}, they are also enough to guarantee its consistent supersymmetric completion through Step 4, which equally applies to this more general construction, see Subect. \ref{suscompdua}.\par
Some comments are in order:
\begin{itemize}
\item[i)]{The construction we are discussing in this Section requires the introduction of additional fields: $n_v$ magnetic potentials $A_{\Lambda\mu}$ and a set of antisymmetric tensors $B_{\alpha\,\mu\nu}$. These new fields come together with extra gauge-invariances (\ref{deltaA}), (\ref{deltaxi1}), (\ref{deltaxi2}) which guarantee the correct counting of physical degrees of freedom. As we shall discuss below these fields can be disposed of using their equations of motion;}
\item[ii)]{It is known that in $D$-dimensions there is a duality relating $p$-forms to $(D-p-2)$-forms, the corresponding field strengths having complementary order and being related by a Hodge-like duality, see Sect \ref{vohd}. In four dimensions vectors are dual to vectors while scalars are dual to antisymmetric tensor fields. From this point of view we can understand the 2-forms $B_\alpha$ as ``dual'' to the scalars in the same way as $A_{\Lambda}$ are ``dual'' to $A^\Lambda$. This relation can be schematically illustrated as follows:
    $$\partial_{[\mu} B_{\nu\rho]}\propto e\,\epsilon_{\mu\nu\rho\sigma}\partial^\sigma \phi+\dots\,.$$
  More precisely we can write the non-local relation between $B_\alpha$ and $\phi^s$ in a $G$-covariant fashion as a Hodge-like duality between $\mathcal{H}^{(3)}_\alpha$ and the Noether current ${\bf j}_\alpha$ of the sigma model describing the scalar fields, associated with the generator $t_\alpha$:
\begin{equation}
\mathcal{H}_{\alpha\,\mu\nu\rho}\propto e\,\epsilon_{\mu\nu\rho\sigma}\,{\bf j}_\alpha^\sigma+\dots\,\,;\,\,\,\,{\bf j}_\alpha^\mu\equiv \frac{\delta \L_{scal.}}{\delta \partial_\mu\phi^s}\,k_\alpha^s\,,\label{duaBphi}
\end{equation}
  $k_\alpha^s$ being the Killing vector corresponding to $t_\alpha$. This motivated the choice of the 2-forms in the adjoint representation of $G$. In the gauged theory we will find a $G_g$-invariant version of (\ref{duaBphi}), see discussion below.}
\item[iii)]{It was shown, see discussion below Eq. (\ref{deltaLtot}), that the presence of the extra fields $B_\alpha$ and $A_{\Lambda}$ in the action is related to non-vanishing magnetic components $\Theta^{\Lambda\,\alpha}$ of the embedding tensor. In the electric frame in which $\Theta^{\Lambda\,\alpha}=0$, these fields disappear altogether from the Lagrangian and we are back to the gauged action described in Sect.\ \ref{gitef};}
\item[iv)]{The kinetic terms in the Lagrangian only describe fields in the ungauged theory while the extra fields enter topological terms, or Stueckelberg-like couplings, and satisfy first order equations, as we shall show in the next Subsection. This feature is common to the $G$-covariant construction of gauged supergravities in any dimensions \cite{deWit:2004nw,deWit:2005hv,Samtleben:2005bp,deWit:2008ta};}
   \item[v)]{The dyonic\footnote{Here we use the word ``dyonic''  for the embedding tensor in a somewhat improper way, since $\Theta_M{}^\alpha$ can be seen as a collection of electric-magnetic charge-vectors, labeled by the index $\alpha$, which are, by virtue of the quadratic constraint (\ref{quadratic1}), all mutually local. } embedding tensor $\Theta_M{}^\alpha$ determines a splitting of the $2n_v$ vector fields $A^M_\mu$ into the truly electric ones $A^{\hat{\Lambda}}_\mu$, which are singled out by the combination $A^M_\mu \Theta_M{}^\alpha$ and thus define the gauge connection, and the remaining ones $\tilde{A}^M_\mu$, corresponding to non-vanishing components of $Z^{M\,\alpha}$, that is to the components along which the Jacobi identity is not satisfied, see (\ref{nojacobi}). These latter vectors, of which there are at most $n_v$ independent, can be then written as $\tilde{A}^M_\mu=Z^{M\,\alpha} A_{\alpha\,\mu}$ and are ill-defined since the corresponding field strengths do not satisfy the Bianchi identity. Another problem with the vectors $\tilde{A}^M_\mu$ is that they are not part of the gauge connection but in general are charged under the gauge group, that is are minimally coupled to $A^{\hat{\Lambda}}_\mu$. These fields cannot therefore be consistently described as vector fields. However this poses no consistency problem for the theory since $\tilde{A}^M_\mu$ can be gauged away by a transformation (\ref{deltaxi1}), (\ref{deltaxi2}) proportional to $\Xi_\alpha$. In a vacuum they provide the two degrees of freedom needed by some of the tensor fields $B_\alpha$ to become massive according to the \emph{anti-Higgs} mechanism \cite{Townsend:1981nu,Cecotti:1987qr}. In the electric frame these vectors become magnetic $A_{\hat{\Lambda}\,\mu}$ and disappear from the action. This phenomenon also occurs in higher dimensions: 
       The vectors $\tilde{A}^M_\mu$ which do not participate in the gauge connection but are charged with respect to the gauge group, are gauged away by a transformation associated with some of the antisymmetric tensor fields which, in a vacuum, become massive;}
\item[vi)]{An important role in this construction was played by the linear constraint (\ref{linear2}), in particular by the property (\ref{linear22}) implied by it, which allowed to cancel the non-covariant terms in the gauge variation of $F^\Lambda$ by a corresponding variation of the antisymmetric tensor fields. It turns out that a condition analogous to (\ref{linear22}) represents the relevant linear constraint on the embedding tensor needed for the construction of gauged theories in higher dimensions \cite{deWit:2004nw,deWit:2005hv,Samtleben:2005bp,deWit:2008ta} (see Sect. \ref{thi}).}
\end{itemize}
\subsection{The Field Equations for $A^M_\mu$ and $B_{\alpha\,\mu\nu}$ }\label{generalfeqs}
Let us now discuss the bosonic field equations for the antisymmetric tensor fields and the vectors. As it follows from (\ref{deltaLtot}), the variation of the action with respect to $B_{\alpha\,\mu\nu}$ yields equations (\ref{HG0}). By fixing the $\Xi_\alpha$-gauge freedom, we can gauge away the ill-defined vectors $\tilde{A}^M_\mu=Z^{M\,\alpha} A_{\alpha\,\mu}$ and then solve eqs.\ (\ref{HG0}) in $B_\alpha$ as functions of the remaining field strengths, which are a combination of the $F^{\hat{\Lambda}}$ only. Substituting this solution in the action, the latter will only describe the $A^{\hat{\Lambda}}_{\mu}$ vector fields and no longer contain magnetic ones or antisymmetric tensors. In other words by eliminating $B_\alpha$ through equations (\ref{HG0}) we effectively perform the rotation to the electric frame and find the action discussed in Sect.\ \ref{gaugingsteps}.\par
Using Eqs. (\ref{deltaLtot}) and (\ref{HG0}), we find for  $A^M_\mu$ the following field equations:
\begin{equation}
\epsilon^{\mu\nu\rho\sigma}\,\mathcal{D}_{\nu}\Gd^M_{\rho\sigma}=2\, e\, \mathbb{C}^{MN}\,{\bf J}_N^\mu\,,\label{Max2}
\end{equation}
which are the manifestly $G$-covariant form of the Maxwell equations.
We have introduced, on the right-hand-side of (\ref{Max2}), the current:
\begin{equation}
{\bf J}_M^\mu\equiv e^{-1}\,\frac{\delta {\Scr L}_{{\rm matter}}}{\delta A^M_\mu}\,,
\end{equation}
$ {\Scr L}_{{\rm matter}}$ denoting the part of the Lagrangian depending on the matter fields (scalars and fermions). The current ${\bf J}_M^\mu$ originates from the minimal couplings and has the following form:
\begin{align}
{\bf J}_M^\mu&=-\Theta_M{}^\alpha\,{\bf j}_\alpha^\mu- \frac{2}{e}\,\epsilon^{\mu\nu\rho\sigma}\bar{\psi}_{A\,\nu}\gamma_\rho\mathcal{Q}_{M\,A}{}^B\,\psi_{B\sigma}+i\,
\bar{\lambda}^{\mathcal{I}}\gamma^\mu\,\mathcal{Q}_{M\,\mathcal{I}}{}^{\mathcal{J}}\,\lambda_{\mathcal{J}}+\nonumber\\
&+\bar{\lambda}^{\mathcal{I}}\gamma^\nu\gamma^\mu\psi^{B}_{\nu}\,\mathcal{P}_{s\,\mathcal{I}B}\,k_M^s+
\bar{\lambda}_{\mathcal{I}}\gamma^\nu\gamma^\mu\psi_{B\,\nu}\,\mathcal{P}_{s}{}^{\mathcal{I}B}\,k_M^s\,.
\end{align}
The first term on the right-hand side contains the Noether current ${\bf j}_\alpha^\mu$ associated with the $\sigma$-model isometries $k_\alpha^s$:
$${\bf j}_\alpha^\mu\equiv e^{-1}\, \frac{\delta {\Scr L}_{scal}}{\delta \partial_\mu\phi^s}\,k_\alpha^s=\mathcal{D}^\mu \phi^s\Gm_{sr}\,k_\alpha^r\,.$$
The next two terms on the first line originate from the gravitino and spin-$1/2$ kinetic terms while the remaining ones come from the scalar-fermion coupling terms. Note that the whole current ${\bf J}_M^\mu$ is proportional to $\Theta_M{}^\alpha$ since $k_M^s=\Theta_M{}^\alpha k_\alpha^s$ and $\mathcal{Q}_M=\Theta_M{}^\alpha\,\mathcal{Q}_\alpha$. We can then write it as ${\bf J}_M^\mu=\Theta_M{}^\alpha \,{\bf J}_\alpha^\mu$.
If we contract both sides of (\ref{Max2}) with $\Theta_M{}^\alpha$ we are singling out the Bianchi identity for the fields strengths $F^{\hat{\Lambda}}$ of the vectors which actually participate in the minimal couplings. By using the locality condition on $\Theta$, we find:
\begin{equation}
\epsilon^{\mu\nu\rho\sigma}\mathcal{D}_{[\nu}\Gd^M_{\rho\sigma]}\,\Theta_M{}^\alpha=2\,e\,\mathbb{C}^{MN}\,\Theta_M{}^\alpha\,
\Theta_N{}^\beta\,{\bf J}_\beta^\mu=0\,,\label{Bianchigauge}
\end{equation}
which are nothing but the Bianchi identities for $F^{\hat{\Lambda}}$. This is consistent with our earlier discussion, see Eq.\ (\ref{BianchiFgauge}), in which we showed that the locality condition implies that the Bianchi identity for the gauge curvature have no magnetic source term, so that the gauge connection is well defined%
\footnote{
In our earlier discussion we showed that $\mathcal{D}\mathcal{H}^M\,\Theta_M{}^\alpha=\mathcal{D}F^M\,\Theta_M{}^\alpha=0$. This is consistent with Eq.\ (\ref{Bianchigauge}) since on-shell $\mathcal{H}^M\Theta_M{}^\alpha=\Gd^M\Theta_M{}^\alpha$.
}.\par
Now we can use the Bianchi identity (\ref{Bid1n}) to rewrite Eqs.\ (\ref{Bianchigauge}) as \emph{dualization equations} relating the antisymmetric tensor fields to the scalars. To this end we consider only the upper components of (\ref{Bianchigauge}), corresponding to the field equations for $A_{\Lambda\,\mu}$:
\begin{align}
Z^{\Lambda\alpha}\,\epsilon^{\mu\nu\rho\sigma}\mathcal{H}_{\alpha\,\nu\rho\sigma}&=-12\,e\,Z^{\Lambda\alpha}\,
\mathcal{D}^\mu \phi^s\Gm_{sr}\,k_\alpha^r+\dots\,,\label{Hduaga}
\end{align}
where the ellipses refer to the terms containing the fermion fields.
Using the explicit  expression (\ref{defH3}) for $\mathcal{H}_{\alpha\,\nu\rho\sigma}$ we find:
\begin{align}
\epsilon^{\mu\nu\rho\sigma}\left(Z^{\Lambda\alpha}\,\mathcal{D}_{\nu}B_{\alpha\,\rho\sigma}+2\,X_{(PQ)}{}^\Lambda\,(A^P_\nu\partial_\rho A^Q_\sigma+\frac{1}{3}\,X_{RS}{}^Q A^P_\nu A^R_\rho A^S_\sigma)\right)&=-4\,e\,Z^{\Lambda\alpha}\,
\mathcal{D}^\mu \phi^s\Gm_{sr}\,k_\alpha^r+\dots\,.\label{Hduaga2}
\end{align}
As a final remark, expanding Eqs. (\ref{Max2}) about a vacuum $\phi_0$ to first order in the vector-field fluctuations we can derive a symplectic-covariant formula for the vector squared-mass matrix, to be discussed later, see Eq. (\ref{MvectMN}) .\par
The gauged theory we have discussed in this section features a number of non-dynamical extra fields. This is the price we have to pay for a manifest $G$-covariance of the field equations and Bianchi identities.
The embedding tensor then defines how the physical degrees of freedom are distributed within this larger set of fields, by fixing the gauge symmetry associated with the extra fields and solving the corresponding non-dynamical field equations (\ref{HG0}), (\ref{Hduaga}).
\subsection{Back to the Electric Frame: The Rank Factorization of $\Theta$}\label{backelectric}
As mentioned above, we can always go to the electric frame by eliminating the antisymmetric tensor fields through Eqs. (\ref{HG0}), after fixing the $\Xi$-gauge \cite{deWit:2005ub}.
This amounts to changing the symplectic frame to the electric one by means of the symplectic matrix $E_{\hat{M}}{}^N$. Here we shall not explicitly solve  Eqs. (\ref{HG0}) but rather use an equivalent path to the formulation of the theory in the electric frame. There is a simple, intrinsic way of constructing the matrix $E$ out of the embedding tensor and at the same time defining those components of $A^M_\mu$ which contribute to the gauge connection, distinguishing them from those which can be set to zero by a $\Xi$-transformation. Let $r$ be the rank of $\Theta_M{}^\alpha$, which coincides with the dimension of the gauge group $G_g$. We can always write this rectangular matrix in the following way (\emph{rank-factorization}):
\begin{equation}
\Theta_M{}^\alpha=\sum_{I=1}^r\,\xi_M{}^I\,\xi_I{}^\alpha\,,
\end{equation}
where $\xi_M{}^I$ can be viewed as a set of $r$ symplectic vectors encoding our freedom in the choice of the symplectic frame, while the vectors  $\xi_I{}^\alpha$ encode our freedom in choosing the $r$ combinations of isometries to be gauged. Condition (\ref{quadratic1}) implies the following orthogonality relations:
\begin{equation}
\xi_M{}^I\,\mathbb{C}^{MN}\,\xi_N{}^J=0\,\,,\,\,\,\,\,\,\forall I,J=1,\dots,\, r\,,
\end{equation}
as a consequence of which $r\le n_v$. \par
The symplectic-invariant connection (\ref{newcon}) reads:
\begin{equation}
 \Omega_{g\mu}\equiv A^M_\mu\,X_M=A^I_\mu\,X_I\,,\label{newcon2}
\end{equation}
where
\begin{equation}
A^I_\mu\equiv \xi_M{}^I\,A^M_\mu\,\,\,,\,\,\,\,\,\,\,X_I\equiv \xi_I{}^\alpha\,t_\alpha\,.
\end{equation}
We see that $A^I_\mu$ are the only gauge fields, namely the vector fields participating in the minimal couplings and $X_I$ the corresponding gauge generators.\par
If $r<n_v$ we can still define a set of mutually orthogonal $n_v-r$ vectors $\xi_M{}^a$, $a=1,\dots, n_v-r$ such that:
\begin{equation}
\xi_M{}^I\,\mathbb{C}^{MN}\,\xi_N{}^a=0\,\,,\,\,\,\,\,\xi_M{}^a\,\mathbb{C}^{MN}\,\xi_N{}^b=0\,\,\,,\,\,\,\,\forall I,\,a,\,b\,.
\end{equation}
We can also define a dual set of vectors $\eta_{M\,I},\,\eta_{M\,a}$ such that:
\begin{equation}
\xi_M{}^I\,\mathbb{C}^{MN}\,\eta_{N\,J}=\delta^I_J\,\,,\,\,\,\xi_M{}^a\,\mathbb{C}^{MN}\,\eta_{N\,b}=\delta^a_b\,,
\end{equation}
all other symplectic products being zero. Using the vectors $\xi_M{}^I,\,\xi_M{}^a,\,\eta_{M\,I},\,\eta_{M\,a}$ we define the symplectic matrices:
\begin{equation}
E_{\hat{M}}{}^N\equiv \mathbb{C}^{NP}(\eta_{P\,I},\,\eta_{P\,a},\,-\xi_P{}^I,\,-\xi_P{}^a)\,\,;\,\,\,\, E^{-1}{}_M{}^{\hat{M}}=\left(\xi_M{}^I,\,\xi_M{}^a,\,\eta_{M\,I},\,\eta_{M\,a}\right)\,,\label{EEm1}
\end{equation}
where we have defined the index $\hat{M}$ as follows:
\begin{equation}
V^{\hat{M}}=(V^{\hat{\Lambda}},\,V_{\hat{\Lambda}})=(V^I,\,V^a,\,V_I,\,V_a)\,,
\end{equation}
being $\hat{\Lambda}=(I,\,a)$.
The reader can indeed verify that:
\begin{equation}
E_{\hat{M}}{}^N\,E^{-1}{}_N{}^{\hat{N}}=\delta_{\hat{M}}^{\hat{N}}\,\,;\,\,\,\,\mathbb{C}_{MN}\,E_{\hat{M}}{}^M E_{\hat{N}}{}^N=\mathbb{C}_{\hat{M}\hat{N}}\,,
\end{equation}
The matrix $E$ defined above maps $\Theta_M{}^\alpha$ to the electric frame since:
\begin{equation}
\Theta_{\hat{M}}{}^\alpha=E_{\hat{M}}{}^M\,\Theta_{M}{}^\alpha=\left(\begin{matrix}\Theta_{\hat{\Lambda}}^\alpha\cr {\bf 0}_{n_v}\end{matrix}\right)\,\,;\,\,\,\,\Theta_{I}{}^\alpha=\xi_I{}^\alpha\,\,;\,\,\,\,\Theta_{a}{}^\alpha=0\,.\label{Thetaxi}
\end{equation}
The tensor $Z^{M\,\alpha}$ can be written in the following way:
\begin{equation}
Z^{M\,\alpha}=\frac{1}{2}\mathbb{C}^{MN}\xi_N{}^I\,\xi_I{}^\alpha\,.
\end{equation}
we see that the antisymmetric tensor fields only enter the theory through the $r$ independent combinations $B_{I\,\mu\nu}\equiv \xi_{I}{}^\alpha\,B_{\alpha\,\mu\nu}$: $$Z^{M\,\alpha}\,B_{\alpha\,\mu\nu}=\frac{1}{2}\,\mathbb{C}^{MN}\xi_N{}^I\,B_{I\,\mu\nu}\,.$$
Consequently, if we write
\begin{equation}A^M_\mu=E_{\hat{M}}{}^M\,A^{\hat{M}}_\mu=\mathbb{C}^{NP}\left(\eta_{P\,I}\,A^I_\mu+\eta_{P\,a}\,A^a_\mu-\xi_P{}^I\,A_{I\,\mu}-\xi_P{}^a\,A_{a\,\mu}
\right)\,,\end{equation}
the field strengths $\mathcal{H}^M_{\mu\nu}$ read (we suppress the Lorentz indices):
\begin{align}
\mathcal{H}^M &=E_{\hat{M}}{}^M\,F^{\hat{M}}+\frac{1}{2}\,\mathbb{C}^{MN}\Theta_N{}^\alpha\,B_{\alpha}=E_{\hat{M}}{}^M\,\mathcal{H}^{\hat{M}}=\nonumber\\&=
\mathbb{C}^{MP}\left(\eta_{P\,I}\,F^I+\eta_{P\,a}\,F^a-\xi_P{}^I\,\left(F_{I}-\frac{1}{2}\,B_I\right)-\xi_P{}^a\,F_{a}\right)\,,
\end{align}
where we have used (\ref{Thetaxi}), the symplectic property of the matrix $E$ and the definition of $\mathcal{H}^{\hat{M}}_{\mu\nu}$:
\begin{equation}
\mathcal{H}^{\hat{M}}_{\mu\nu}\equiv F^{\hat{M}}_{\mu\nu}+Z^{\hat{M}\,\alpha}\,B_{\alpha\,\mu\nu}=\left(\begin{matrix}F^{I}_{\mu\nu}\cr F^{a}_{\mu\nu}\cr
F_{I\,\mu\nu}-\frac{1}{2}\,B_{I\,\mu\nu}\cr F_{a\,\mu\nu}\end{matrix}\right)\,.
\end{equation}
We see that the only vectors which are Stueckelberg-combined with the antisymmetric tensors are $A_{I\,\mu}$, which become magnetic in the electric frame.
They are the vector fields whose field strengths fail to satisfy the Bianchi identities.\par
Subtler is the redefinition of $\mathcal{G}^M_{\mu\nu}$ which contains the information about the Lagrangian. We can write $\mathcal{G}^M_{\mu\nu}=E_{\hat{M}}{}^M\,\mathcal{G}^{\hat{M}}_{\mu\nu}$, where $\mathcal{G}^{\hat{M}}_{\mu\nu}$ satisfies the twisted self-duality condition
(\ref{FCMF}) (or (\ref{FCMF2})) with the matrix $\mathcal{M}$ written in the electric-frame:
\begin{equation}
\mathcal{M}_{\hat{M}\hat{N}}(\phi)=E_{\hat{M}}{}^M\,E_{\hat{N}}{}^N\,\mathcal{M}_{MN}(\phi)\,.
\end{equation}
This follows from the manifest symplectic -covariance of (\ref{FCMF}), and implies that also the lower-components of $ G_{\hat{\Lambda}\,\mu\nu}$ can be expressed, as in (\ref{GF}), in terms of the derivatives of a new Lagrangian density $\hat{\L}$ with respect to its upper components $G^{\hat{\Lambda}}_{\mu\nu}$. The vector kinetic terms of $\hat{\L}$ are written in terms of the matrices $\I_{\hat{\Lambda}\hat{\Sigma}}(\phi),\,\R_{\hat{\Lambda}\hat{\Sigma}}(\phi)$, related to $\mathcal{M}_{\hat{M}\hat{N}}(\phi)$ by (\ref{M}). Consistency then requires that the upper components $G^{I}_{\mu\nu}$ be the field strengths $F^{I}_{\mu\nu}$ of the gauge fields $A^I_\mu$, which are the upper components $\mathcal{H}^{I}_{\mu\nu}$ of $\mathcal{H}^{\hat{M}}_{\mu\nu}$.
To show this we recall that, by definition, the
upper components $\mathcal{H}_{\mu\nu}^\Lambda$ of $\mathcal{H}_{\mu\nu}^M$ and of $\mathcal{G}_{\mu\nu}^M$, in our construction, coincide. This implies the coincidence of the upper components $G^{I}_{\mu\nu}$ and $\mathcal{H}^{I}_{\mu\nu}=F^{I}_{\mu\nu}$  of the corresponding symplectic vectors in the electric frame, \emph{provided Eqs. (\ref{HG0}) are satisfied}. Indeed
let us write $\mathcal{H}_{\mu\nu}^{I}$ and $G^{I}_{\mu\nu}$ in terms of $\mathcal{H}^M_{\mu\nu}$ and $\mathcal{G}^M_{\mu\nu}$ through the matrix $E$:
\begin{equation}
\mathcal{H}_{\mu\nu}^{I}=E^{-1}{}_M{}^{I}\mathcal{H}^M_{\mu\nu}\,\,;\,\,\,\,G^{I}_{\mu\nu}=
E^{-1}{}_M{}^{I}\mathcal{G}^M_{\mu\nu}\,.
\end{equation}
Given the expression of the matrix $E^{-1}{}_M{}^I$, they coincide provided:
\begin{equation}
\xi_{M}{}^{I}(\mathcal{H}^M_{\mu\nu}-\mathcal{G}^M_{\mu\nu})=0\,.
\end{equation}
which are nothing but Eqs. (\ref{HG0}).
Therefore, by writing in the original field equations and Bianchi identities the sympliectic covariant quantities in the hatted frame, through the matrix $E$, one ends up with the corresponding equations in the electric frame, derived from a Lagrangian $\hat{{\Scr L}}$   of the form discussed in Sect. \ref{gitef}.\par
In the electric frame one can show, using the linear constraints (\ref{lconstr}) and the condition that $X_{\hat{M}\hat{N}}{}^{\hat{P}}$ be a symplectic matrix in the last two indices, that the most general form of this tensor is \cite{deWit:2007mt}:
\begin{align}
X_{\hat{M}\hat{N}}{}^{\hat{P}}=\left(\begin{matrix}-f_{IJ}{}^K & h_{IJ}{}^a & C_{IJK} & C_{IJa}\cr {\bf 0} & {\bf 0} & C_{IbJ} & {\bf 0}\cr
{\bf 0} & {\bf 0} & f_{IK}{}^J & {\bf 0}\cr {\bf 0} & {\bf 0} & -h_{IK}{}^b & {\bf 0}\end{matrix}\right)\,,\label{Xtenselec}
\end{align}
where $C_{IbJ}=C_{IJb}$ and
\begin{equation}
h_{(IJ)}{}^a=f_{(IJ)}{}^K=C_{(IJ)a}=C_{I[JK]}=C_{(IJK)}=f_{IJ}{}^J=0\,.
\end{equation}
The closure relation (\ref{quadratic2}) further implies:
\begin{align}
f_{[I_1 I_2}{}^J \,f_{I_3] J}{}^K&=f_{[I_1 I_2}{}^J \,h_{I_3] J}{}^a=0\,,\nonumber\\
f_{I_1 I_2}{}^J\,C_{J J_1 J_2}-4\,f_{(J_1[I_1}{}^J\,C_{I_2]J_2)J}+4 \,h_{(J_1[I_1}{}^a\,C_{I_2]J_2)a}&=0\,,\nonumber\\
f_{[I_1 I_2}{}^J \,C_{I_3] Ja}&=0\,.
\end{align}
As emphasized in Sect. \ref{gaugingsteps}, the first block $X_{\hat{\Lambda}\hat{\Sigma}}{}^{\hat{\Gamma}}$ of the tensor in (\ref{Xtenselec}) is related to the structure constants of the gauge algebra: $X_{\hat{\Lambda}\hat{\Sigma}}{}^{\hat{\Gamma}}=-f_{\hat{\Lambda}\hat{\Sigma}}{}^{\hat{\Gamma}}$. We have distinguished in these constants two set of components, denoted by $f_{IJ}{}^K $  and $h_{IJ}{}^a$, which characterize the structure of an algebra with generators $X_I,\,X_a$ closing the following commutation relations:
\begin{equation}
[X_I,\,X_J]=f_{IJ}{}^K\,X_K-h_{IJ}{}^a\,X_a\,\,;\,\,\,\,[X_a,\,X_I]=[X_a,\,X_b]=0\,.\label{gaualgcc}
\end{equation}
Notice that the generators $X_a$, which exist whenever $h_{IJ}{}^a\neq 0$, are \emph{central charges}. Moreover, being $\Theta_a{}^\alpha=0$ in the electric frame, $X_a$ are not expressed in terms of isometries. If fact $X_a$ \emph{are abstract generators with no action on the physical fields}. They generate an Abelian ideal $\mathfrak{I}$ of the gauge group $G_g$, so that the gauge transformations of the physical fields are described by the quotient $G_g/\mathfrak{I}$. This is the group which is actually embedded in $G$. More precisely, due to the presence of a non-trivial Abelian ideal in the gauge algebra, when $h_{IJ}{}^a\neq 0$, the adjoint representation of the gauge group is not faithful and we have:
\begin{equation}
{\rm Adj}(G_g)={\rm Adj}(G_g/\mathfrak{I})\hookrightarrow \,{\Scr R}_{v}[G]\,,
\end{equation}
where last embedding is defined by the gauge-invariant embedding tensor and is the statement that the symplectic duality representation ${\Scr R}_{v}$ of $G$, if branched with respect to the gauge group, contains its adjoint representation.
The parameters $\zeta^a(x)$ of $X_a$ only enter the gauge transformations of the vector fields $A^a_\mu$:
\begin{equation}
\delta A^I_\mu=\partial_\mu\zeta^I+\zeta^J\,f_{JK}{}^I\,A^K_\mu\,\,,\,\,\,\,\,\,\delta A^a_\mu=\partial_\mu\zeta^a-\zeta^J\,h_{JK}{}^a\,A^K_\mu\,,
\end{equation}
and thus disappear altogether from the corresponding variations of the physical field strengths:
\begin{equation}
\delta F^I_{\mu\nu}=\zeta^J\,f_{JK}{}^I\,F^K_{\mu\nu}\,\,,\,\,\,\,\,\,\delta  F^a_{\mu\nu}=-\zeta^J\,h_{JK}{}^a\,F^K_{\mu\nu}\,.
\end{equation}
This situation is not uncommon in gaugings originating from flux compactifications \cite{Angelantonj:2003rq,Angelantonj:2003up}, see the examples of compactifications in the presence of an $H$-flux discussed in Sect. \ref{toroidalc}. The structure constants $f_{JK}{}^I$ of $G_g/\mathfrak{I}$ can be expressed, using (\ref{quadratic2}), as follows:
\begin{equation}
f_{IJ}{}^K=-\xi_I{}^\alpha t_{\alpha\,M}{}^N\,\xi_N{}^K\,\mathbb{C}^{MP}\eta_{P\,J}\,.
\end{equation}

\subsection{Solving the Dualization Equations  in a Special Case}\label{solvingspecial}
When the gauging involves translational isometries \cite{deWit:2005ub}, $\phi^I\rightarrow \phi^I+c^I$, the above equations can be solved in the fields $A_\Lambda$ contained in the covariant derivative. This is done by first using the $\zeta$-gauge freedom associated with $A_\Lambda$ to gauge away the scalar fields $\phi^I$ acted on by the translational isometries. Eqs.\ (\ref{Hduaga}) are then solved in the fields $A_\Lambda$ which are expressed in terms of the remaining scalars, the vectors $A^\Lambda$ and the field strengths of the antisymmetric tensors. Substituting this solution in the action we obtain a theory in which no vectors $A_\Lambda$ appear and the scalar fields $\phi^I$ have been effectively dualized to corresponding tensor fields $B_{I\,\mu\nu}$. The latter become dynamical being described by their own kinetic terms.
These theories were first constructed in the framework of $\N=2$ supergravity in \cite{Dall'Agata:2003yr,D'Auria:2004yi}, generalizing previous results \cite{Louis:2002ny}.\par
Let us illustrate this mechanism in the simple case in which the gauged translational isometries have no duality action, as it is the case for Abelian isometries of the quaternionic K\"ahler manifold in an $\mathcal{N}=2$ theory:
\begin{equation}
X_M=\Theta_M{}^\alpha\,t_\alpha=\xi_M{}^I\,\xi_I{}^\alpha\,t_\alpha\,.
\end{equation}
The gauged isometries are $t_I\equiv \xi_I{}^\alpha\,t_\alpha$ and the condition that they be Abelian amounts to requiring:
\begin{equation}
[t_I,\,t_J]=0\,\,\Leftrightarrow\,\,\,\,\,\,\xi_I{}^\alpha\xi_J{}^\beta {\rm f}_{\alpha\beta}{}^\gamma=0\,.
\end{equation}
The closure constraint (\ref{quadratic2}) further requires:
\begin{equation}
[X_M,\,X_N]=0\,\,\Rightarrow\,\,\,\,\,X_{MN}{}^P\,X_P=0\,,
\end{equation}
which is certainly satisfied for gauged quaternionic isometries, being $t_{I\,M}{}^N=0$. \footnote{Note that gauging non-Abelian isometries also requires the gauging of isometries with a non-trivial duality action, being $X_{MN}{}^P\neq 0$.}
From the condition (\ref{quadratic1}) it follows that:
\begin{equation}
\xi_M{}^I\,\mathbb{C}^{MN}\,\xi_N{}^J=0\,.
\end{equation}
We can choose a parametrization $\phi^s$ of the scalar manifold so that the gauged translational isometries $t_I$ only act on scalars $\phi^I$ and not on the remaining ones $\phi^S$:
\begin{equation}
\phi^I\rightarrow \phi^I+c^J\,k_J{}^I=\phi^I+c^I\,\,;\,\,\,\,\,\phi^S\rightarrow \phi^S+c^J\,k_J{}^S=0\,,
\end{equation}
that is the only non-vanishing components of $k_I$ are $k_I{}^J=\delta_I^J$. The presence of these isometries associated with constant shifts in the scalars $\phi^I$ implies that the scalar metric does not depend on $\phi^I$.\par
In order to economize on indices, we use the form-notation for the fields and the Lagrangian and work with the Lagrangian 4-form ${\Scr L}^{(4)}\equiv d^4 x\,{\Scr L}$.
The scalar-tensor Lagrangian 4-form reads:
\begin{equation}
{\Scr L}^{(4)}_{scal.,\,tens.}=\frac{1}{2}\,\mathcal{D}\phi^r\wedge {}^*\mathcal{D}\phi^r\,G_{rs}(\phi)-\frac{1}{2}\,\xi^{\Lambda\,I}\,B_I\wedge dA_\Lambda+\frac{1}{8}\,\xi^{\Lambda\,I}\xi_\Lambda{}^{J}\,B_I\wedge B_J\,,\label{stact}
\end{equation}
where we have used the fact that, in this example, ${\Scr L}_{GCS}=0$. The covariant derivatives of the scalar fields read:
\begin{equation}
\mathcal{D}\phi^I=d\phi^I-A^M\,\xi_M{}^J\,k_J{}^I=d\phi^I-A^M\,\xi_M{}^I\,\,;\,\,\,\,\mathcal{D}\phi^S=d\phi^S\,.
\end{equation}
The dualization equation (\ref{Hduaga2}) has the simple form:
\begin{equation}
{}^*dB_I=-2\,(G_{IJ}\,\mathcal{D}\phi^I+G_{IS}\,d\phi^S)\,\,\Rightarrow\,\,\,\,\,\mathcal{D}\phi^I=-G^{IJ}\,\left(\frac{1}{2}{}^*dB_J+G_{JS}\,d\phi^S\right)\,.\label{solvingdua}
\end{equation}
Note that we can solve this equation in the magnetic vector fields $A_{\Lambda\,\mu}\,\xi^{\Lambda\,I}$, after using the gauge invariance associated with this field to eliminate $\phi^I$:
\begin{equation}
d\phi^I-A_{\Lambda\,\mu}\,\xi^{\Lambda\,I}\rightarrow -A_{\Lambda\,\mu}\,\xi^{\Lambda\,I}\,.
\end{equation}
The second of Eqs. (\ref{solvingdua}) then yields:
 \begin{equation}
 A_{\Lambda\,\mu}\,\xi^{\Lambda\,I}=-A^\Lambda_\mu\,\xi_{\Lambda}{}^I+G^{IJ}\,\left(\frac{1}{2}{}^*dB_J+G_{JS}\,d\phi^S\right)\,.
 \end{equation}
 We substitute this solution in the scalar-tensor action (\ref{stact}) to find, after some algebra:
 \begin{align}
{\Scr L}^{(4)}_{scal.,\,tens.}&=  \frac{1}{2}\,\mathcal{D}\phi^I\wedge {}^*\mathcal{D}\phi^J\,G_{IJ}(\phi)+\mathcal{D}\phi^I\wedge {}^*\mathcal{D}\phi^S\,G_{IS}(\phi)+\frac{1}{2}\,d\phi^S\wedge {}^* d\phi^T\,G_{ST}(\phi)+\nonumber\\
&+\frac{1}{2}\,\xi^{\Lambda\,I}\,dB_I\wedge A_\Lambda+\frac{1}{8}\,\xi^{\Lambda\,I}\xi_\Lambda{}^{J}\,B_I\wedge B_J=\nonumber\\
&=\frac{1}{8}\,dB_I\wedge {}^*dB_{J}\,G^{IJ}-\frac{1}{2}G^{IJ}\,G_{JS}\,d\phi^S\wedge  dB_I+\frac{1}{2}\,d\phi^S\wedge {}^* d\phi^T\,\hat{G}_{ST}(\phi)+\nonumber\\&+\frac{1}{2}\,\xi_\Lambda{}^I\,B_I\wedge \left(dA^\Lambda+\frac{1}{4}\,\xi^{\Lambda\,J}\,B_J\right)\,,
 \end{align}
 where we have defined:
 \begin{equation}
 \hat{G}_{ST}(\phi)\equiv G_{ST}-G_{IS}\,G^{IJ}\,G_{JT}\,.
 \end{equation}
We recover the bosonic action derived in \cite{Dall'Agata:2003yr,D'Auria:2004yi}.\par A gauging of this kind, involving Abelian subalgebras of a characteristic Heisenberg algebra of quaternionic isometries, was the starting point for the construction of the low-energy effective $\mathcal{N}=2$ description of flux-compactifications of Type II theories on manifolds with ${\rm SU}(3)\times {\rm SU}(3)$-structure, see \cite{Grana:2006hr},\cite{D'Auria:2007ay}. We shall discuss this model later on as an example of gauged $\mathcal{N}=2$ models, in Sect. \ref{mcga}.\par
A less trivial example of solution to the dualization equation was discussed in \cite{dft2,D'Auria:2005rv}, where the low-energy effective supergravity description of $D=11$ supergravity  on a twisted torus with fluxes was studied, see end of Sect. \ref{n8fluxc}. In that case the magnetic components of the embedding tensor, in the symplectic frame originating from direct dimensional reduction, coincide with the internal torsion $T_{\alpha\beta}{}^\gamma$.
\subsection{Supersymmetric Completion}\label{suscompdua}
In this section we discuss the supersymmetric completion of the bosonic Lagrangian constructed above, up to quartic terms on the fermionic fields.
In Sect. \ref{thegaugedlag} we discussed in some detail the supersymmetric completion of the gauged Lagrangian ${\Scr L}_{{\rm gauged}}^{(0)}$ in the electric frame. This was effected using the Noether method which consisted in introducing additional terms $\Delta{\Scr L}_{{\rm gauged}}^{(1)}+\Delta{\Scr L}_{{\rm gauged}}^{(2)}$ (fermion mass terms and a scalar potential) in the Lagrangian and in the supersymmetry transformation laws of the gravitino and spin- $1/2$ fields, depending on the fermion-shift tensors. These latter quantities, by supersymmetry, were identified with components of the $H$-covariant $\mathbb{T}$-tensor, depending linearly on the embedding tensor and non-linearly on the scalar fields. These deformations of the original Lagrangian, as well as the fermion-shifts, were written, using the matrix $E$, in a form which is manifestly $G$-invariant, provided the embedding tensor is transformed under $G$ together with all the other fields, in the appropriate representation. For this reason the $\mathbb{T}$-tensor-dependent deformations $\Delta{\Scr L}_{{\rm gauged}}^{(1)}+\Delta{\Scr L}_{{\rm gauged}}^{(2)}$ will be the same for the duality-covariant gauging.\par
We now start from a different ${\Scr L}_{{\rm gauged}}^{(0)}$ which consists in the bosonic Lagrangian ${\Scr L}_{B}$ in (\ref{boslag2}), the kinetic terms of the spin- $3/2$ and the spin- $1/2$ fields (\ref{fermik}), the scalar-fermion interaction terms and the Pauli-terms in which the derivatives and the vector field-strengths are covariantized:
 \begin{align}
 {\Scr L}_{{\rm gauged}}^{(0)}&={\Scr L}_{B}+{\Scr L}_{fermi.,k.}+{\Scr L}_{scal.-fermi.}+{\Scr L}_{Pauli}+{\Scr L}_{4f}\,,\\
 {\Scr L}_{fermi.,k.}&=\epsilon^{\mu\nu\rho\sigma}(\bar{\psi}^A_{\mu}\gamma_\nu\mathcal{D}_\rho\psi_{A\sigma}-\bar{\psi}_{A\,\mu}\gamma_\nu\mathcal{D}_\rho
 \psi^A_{\sigma})-\frac{i\,e}{2}\,(\bar{\lambda}^{\mathcal{I}}\gamma^\mu\mathcal{D}_{\mu}\lambda_{\mathcal{I}}+\bar{\lambda}_{\mathcal{I}}\gamma^\mu\mathcal{D}_{\mu}\lambda^{\mathcal{I}})\,.\nonumber\\
{\Scr L}_{scal.-fermi.}&=-e\bar{\lambda}^{\mathcal{I}}\gamma^\mu\gamma^\nu\psi^B_\mu\,\mathcal{D}_\nu\phi^s
\mathcal{P}_{s\,\mathcal{I}B}
-e\bar{\lambda}_{\mathcal{I}}\gamma^\mu\gamma^\nu\psi_{B\mu}\,\mathcal{D}_\nu\phi^s\,
\mathcal{P}_{s}^{\mathcal{I}B}=\nonumber\\
&=-\frac{e}{6}\,\bar{\chi}^{ABC}\gamma^\mu\gamma^\nu\psi^D_\mu\,\mathcal{D}_\nu\phi^s\mathcal{P}_{s\,ABCD}-
{e}\,\bar{\lambda}^{IA}\gamma^\mu\gamma^\nu\psi^B_\mu\,\mathcal{D}_\nu\phi^s\mathcal{P}_{s\,IAB}+\dots\,,\nonumber\\
{\Scr L}_{Pauli}&=\frac{e}{2}\,\mathcal{H}^{+\,\Lambda\,\mu\nu}\mathcal{I}_{\Lambda\Sigma}
{\bar{ f}}^{\Sigma \underline{\Gamma}}\mathcal{O}_{\underline{\Gamma}\,\mu\nu}+h.c.\,,\label{duagauggen1}
 \end{align}
 where the fermion bilinears $\mathcal{O}_{\underline{\Gamma}\,\mu\nu}$ were defined in (\ref{o1}) and (\ref{o2}). \par
 There is a subtlety here regarding the coupling of the fermionic fields to the vector ones, as well as the supersymmetry transformations connecting the two kinds of fields. In all these terms the vectors enter through $H$-covariant tensors $\mathbb{H}^{\underline{M}}_{\mu\nu}\equiv(H^{\underline{\Lambda}}_{\mu\nu},\,H_{\underline{\Lambda}\,\mu\nu})$, which are defined by dressing the symplectic vector $\mathcal{G}^M_{\mu\nu}$ with scalar fields by means of the coset representative in the ${\Scr R}_v$ representation, as in Eq. (\ref{Tdef}):
 \begin{equation}
\mathbb{H}_{\mu\nu}(\phi,\partial A^\Lambda,\,B_\alpha)\equiv -\mathbb{L}_c(\phi)^\dagger \mathbb{C}\mathcal{G}_{\mu\nu}=
\left(\begin{matrix}H^{AB}_{\mu\nu}\cr
H^{I}_{\mu\nu}\cr H_{\mu\nu\,AB}\cr H_{\mu\nu\,I}\end{matrix}\right)\,.\label{Tdef2}
\end{equation}
All equations (\ref{mathbbF})-(\ref{finqui}) hold, provided we replace the Abelian $\mathcal{G}^M_{\mu\nu}$ by its non-Abelian counterpart and  $\mathbb{F}\rightarrow \mathbb{H},\,F^{\underline{\Lambda}}\rightarrow H^{\underline{\Lambda}},\,F_{\underline{\Lambda}}\rightarrow H_{\underline{\Lambda}}$. Moreover the supersymmetry transformations in of gauged theory are obtained from those of the ungauged one, (\ref{traphi})-(\ref{tralam}), besides covariantizing the derivatives, by adding the fermion-shift terms according to Eqs. (\ref{fermshifts}) and  replacing the composite field strengths $F^\pm_{\mu\nu}$ by the $H^\pm_{\mu\nu}$ defined above. Note that $\mathbb{H}^{\underline{M}}_{\mu\nu}$ is constructed in terms of the symplectic vector $\mathcal{G}^M_{\mu\nu}$, which is the truly gauge covariant quantity, and not of $\mathcal{H}^M_{\mu\nu}$. The latter, in this construction, enters the gauge curvature:
\begin{equation}
\mathcal{F}_{\mu\nu}=F^M_{\mu\nu}\,X_M=\mathcal{H}^M_{\mu\nu}\,X_M\,.\label{ddH}
\end{equation}
As a consequence of this, when computing to first order in coupling constant (i.e. in $\Theta$), the supersymmetry variation of the gauged Lagrangian, the kinetic terms of the spin- $3/2$ and the spin- $1/2$ fields, ${\Scr L}_{RS},\,{\Scr L}_{\lambda,\,k}$ respectively, due to (\ref{ddH}), yield terms of the form (\ref{RSvar}) and (\ref{lambdatra}) in which $$F^{\hat{\Lambda}}_{\mu\nu}\,\mathcal{P}_{\hat{\Lambda}\,\mathcal{I}\,A}\rightarrow\,\,\mathcal{H}^M_{\mu\nu}\,\mathcal{P}_{M\,\mathcal{I}\,A}
\,\,;\,\,\,\,F^{\hat{\Lambda}}_{\mu\nu}\,\mathcal{Q}_{\hat{\Lambda}\,A}{}^B\rightarrow\,\,\mathcal{H}^M_{\mu\nu}\,\mathcal{Q}_{M\,A}{}^B\,,
$$
in other words they produce terms containing the $\mathbb{T}$-tensor contracted with $\mathcal{H}^M_{\mu\nu}$ and not $\mathcal{G}^M_{\mu\nu}$, i.e. of the form
$$
\mathcal{H}_{\mu\nu}^{M}\,L^{-1}X_M L\star (\bar{f}\gamma\epsilon)\,,
$$
where $f$ generically stands either for $\psi$ or for $\lambda$. These terms therefore also depend on the vector fields $A_{\Lambda\mu}$.\par
Explicit computation shows that Eqs. (\ref{RSvar}) and (\ref{lambdatra}) now become:
\begin{align}
\delta{\Scr L}_{RS}&=\dots+2i\,e\,\left(\mathcal{H}^{+M\,\mu\nu}-\mathcal{H}^{-M\,\mu\nu}\right)\left(\bar{\psi}_{A\,\mu}\gamma_{\nu}\epsilon^B+\bar{\psi}^B_{\mu}\gamma_{\nu}
\epsilon_A\right)\,\mathcal{Q}_{M\,B}{}^A\,,\label{tras1RS}\\
\delta{\Scr L}_{\lambda,\,k}&=\dots-\frac{e}{2}\,\left(\bar{\lambda}^{\mathcal{I}}\gamma^{\mu\nu}\epsilon^A\,\mathcal{H}^{+M}_{\mu\nu}\,\mathcal{P}_{M\,\mathcal{I}A}+
\bar{\lambda}_{\mathcal{I}}\gamma^{\mu\nu}\epsilon_A\,\mathcal{H}^{-M}_{\mu\nu}\,\mathcal{P}_M{}^{\mathcal{I}A}\right)\,.\label{tras2lam}
\end{align}
The variation of the Pauli and Yukawa (or mass) terms produce contributions of the same form as (\ref{tras1RS}) and (\ref{tras2lam}) but with $\mathcal{H}^M_{\mu\nu}$ replaced with $\mathcal{G}^M_{\mu\nu}$. Thus, as opposed to the electric gauging discussed in Sect. \ref{thegaugedlag}, the variation of the spin- $3/2$ and the spin- $1/2$ fields in ${\Scr L}_{{\rm gauged}}^{(0)}+\Delta{\Scr L}_{{\rm gauged}}^{(1)}$ now produce the following non-vanishing terms:
\begin{align}
&2i\,e\,\left([(\mathcal{H}-\mathcal{G})^{+M\,\mu\nu}-(\mathcal{H}-\mathcal{G})^{-M\,\mu\nu}\right]\left(\bar{\psi}_{A\,\mu}\gamma_{\nu}\epsilon^B+\bar{\psi}^B_{\mu}\gamma_{\nu}
\epsilon_A\right)\,\mathcal{Q}_{M\,B}{}^A-\nonumber\\
&-\frac{e}{2}\,\left[\bar{\lambda}^{\mathcal{I}}\gamma^{\mu\nu}\epsilon^A\,(\mathcal{H}-\mathcal{G})^{+M}_{\mu\nu}\,\mathcal{P}_{M\,\mathcal{I}A}+
\bar{\lambda}_{\mathcal{I}}\gamma^{\mu\nu}\epsilon_A\,(\mathcal{H}-\mathcal{G})^{-M}_{\mu\nu}\,\mathcal{P}_M{}^{\mathcal{I}A}\right]=\nonumber\\
&2i\,e\,\left[(\mathcal{H}-\mathcal{G})_{\Lambda}{}^{+\,\mu\nu}-(\mathcal{H}-\mathcal{G})_{\Lambda}{}^{-\,\mu\nu}\right]\left(\bar{\psi}_{A\,\mu}\gamma_{\nu}\epsilon^B+\bar{\psi}^B_{\mu}\gamma_{\nu}
\epsilon_A\right)\,\mathcal{Q}^\Lambda{}_{B}{}^A-\nonumber\\
&-\frac{e}{2}\,\left[\bar{\lambda}^{\mathcal{I}}\gamma_{\mu\nu}\epsilon^A\,(\mathcal{H}-\mathcal{G})_{\Lambda}{}^{+\,\mu\nu}\,\mathcal{P}_{\Lambda\,\mathcal{I}A}+
\bar{\lambda}_{\mathcal{I}}\gamma_{\mu\nu}\epsilon_A\,(\mathcal{H}-\mathcal{G})_{\Lambda}{}^{-\,\mu\nu}\,\mathcal{P}^{\Lambda\,\mathcal{I}A}\right]\,,\label{newvar}
\end{align}
where we have used the property that $(\mathcal{H}-\mathcal{G})^{M}_{\mu\nu}\,\Theta_M{}^\alpha=(\mathcal{H}-\mathcal{G})_{\Lambda\,\mu\nu}\,\Theta^{\Lambda\,\alpha}$. In order to cancel (\ref{newvar}), we can use Eq. (\ref{deltaLtot}) and devise a supersymmetry transformation law for the antisymmetric tensor fields. In particular
the part of Eq. (\ref{deltaLtot}) depending on $\delta B_{\alpha\,\mu\nu}$ can be written in the following form:
\begin{equation}
\delta {\Scr L}_{B}=\dots+\frac{ie}{4}\left[(\mathcal{H}-\mathcal{G})_{\Lambda}{}^{+\,\mu\nu}-(\mathcal{H}-\mathcal{G})_{\Lambda}{}^{-\,\mu\nu}\right]\,\Theta^{\Lambda\,\alpha}
\delta B_{\alpha\,\mu\nu}\,.
\end{equation}
In order for the above variation to cancel the terms (\ref{newvar}) we need to define:
\begin{align}
\Theta^{\Lambda{}\alpha}\,\delta B_{\alpha\,\mu\nu}&=2i\left(\bar{\lambda}_{\mathcal{I}}\gamma_{\mu\nu}\epsilon_A\,
\mathcal{P}^{\Lambda\,\mathcal{I}A}-\bar{\lambda}^{\mathcal{I}}\gamma_{\mu\nu}\epsilon^A\,
\mathcal{P}^\Lambda{}_{\mathcal{I}A}\right)-8\,\left(\bar{\psi}_{A\,[\mu}\gamma_{\nu]}\epsilon^B+
\bar{\psi}^B_{[\mu}\gamma_{\nu]}
\epsilon_A\right)\,\mathcal{Q}^\Lambda{}_{B}{}^A-\nonumber\\
&-2X^\Lambda{}_P{}^M\mathbb{C}_{MN}\,A^P_{[\mu}\,\delta A^N_{\nu]}\,,\label{ThetadeltaB}
\end{align}
where last term originates from the fact that $\delta B_{\alpha\,\mu\nu}$ enters Eq. (\ref{deltaLtot}) in the combination: $\delta B_{\alpha\,\rho\sigma}+2\,t_{\alpha\,P}{}^M\mathbb{C}_{MQ}\,A^P_\rho\delta A^Q_\sigma$.\par
The discussion about identification of the fermion-shift tenors $\mathbb{S}_{AB},\,\mathbb{N}^{\mathcal{I}}{}_A $ with components of the $\mathbb{T}$-tensor, done in Sect. \ref{thegaugedlag}, holds equally well in this more general setting, as well as the potential Ward identity (\ref{WID}) relating the scalar potential $V(\phi)$ to $\mathbb{S}_{AB},\,\mathbb{N}^{\mathcal{I}}{}_A$ and their complex conjugates. Thus the $G$-invariant expression of $V(\phi,\,\Theta)$ as a function of the scalars and the embedding tensor is the same as the one found in our previous treatment.\par
We recall here that we have not been dealing with terms in the Lagrangian and in the supersymemtry transformation laws which are of higher order in the fermionic fields.
Below we give the supersymmetry transformation laws of the various fields in  the gauged model (we restore the coupling constant $g$, $\Theta\rightarrow g\,\Theta$):\footnote{The transformation laws for the fermionic fields which are specific to $\mathcal{N}=2,3,5,6$ theories read
\begin{align}
\delta \lambda_{I}&=\frac{i}{2} \,\mathcal{P}_{s\,I\,AB}\,\mathcal{D}_\mu\phi^s\gamma^\mu\,\epsilon_C\,\epsilon^{ABC}++
g\,\mathbb{N}_{I}{}^A\epsilon_A\dots\,\,;\,\,\,
(\mathcal{N}=3)\,,\nonumber\\
\delta \chi&=\frac{i}{24}\,\epsilon^{ABCDE}\,\mathcal{D}_\mu\phi^s\mathcal{P}_{s\,ABCD}\gamma^\mu\epsilon_E+
g\,\mathbb{N}^A\epsilon_A\dots\,\,;
\,\,\,(\mathcal{N}=5)\,,\nonumber\\
\delta \chi_F&=\frac{i}{24}\,\epsilon_{FABCDE}\,\mathcal{D}_\mu\phi^s\mathcal{P}_s{}^{ABCD}\gamma^\mu\epsilon^E-
\frac{i}{4}\,H_{\mu\nu\,\bullet}^-\gamma^{\mu\nu}
\epsilon_F+
g\,\mathbb{N}_F{}^A\epsilon_A\dots\,\,;\,\,\,(\mathcal{N}=6)\,,\nonumber\\
\delta \lambda_\alpha&=i\mathcal{P}^{B\beta}_u\mathcal{D}_\mu q^u\gamma^\mu \epsilon^A\epsilon_{AB}\mathbb{C}_{\alpha\beta}+
g\,\mathbb{N}_\alpha{}^A\epsilon_A\dots  \,\,;\,\,\,(\mathcal{N}=2)\,.\nonumber
\end{align}
}
\begin{align}
\delta\phi^s\mathcal{P}_s^{\mathcal{I}\,A}&=\Sigma^{\mathcal{I}\,A}\,,\label{traphi2}\\
\delta A^M_\mu&=\mathbb{L}_c^M{}_{\underline{M}}\,{\Scr O}_\mu^{_{\underline{M}}}=\frac{1}{2}\,\mathbb{L}_c^M{}_{AB}\,{\Scr O}_\mu^{AB}+\mathbb{L}_c^M{}_{I}\,{\Scr O}_\mu^{I}+h.c.\,,\label{traA2}\\
\delta V_\mu{}^a&=i\,\bar{\epsilon}^A\gamma^a\psi_{\mu\,A}+i\,\bar{\epsilon}_A\gamma^a\psi_{\mu}^A\,,\label{traV2}\\
\delta\psi_{A\,\mu}&=\mathcal{D}_\mu\epsilon_A-\frac{1}{8}\,H^-_{\rho\sigma\,AB}
\gamma^{\rho\sigma}\gamma_\mu\epsilon^B+i\,g\,\mathbb{S}_{AB}\,\gamma_\mu\,\epsilon^B\dots\,,\label{trapsi2}\\
\delta \chi_{ABC}&=i\, \mathcal{D}_\mu\phi^s\,\mathcal{P}_{s\,ABCD}\gamma^\mu\epsilon^D-\frac{3i}{4}\,
H^-_{\mu\nu\,[AB}\gamma^{\mu\nu}\epsilon_{C]}+g\,\mathbb{N}_{ABC}{}^D\epsilon_D\dots\,,\label{trachi22}\\
\delta \lambda_{I\,A}&=i \,\mathcal{P}_{s\,I\,AB}\,
\mathcal{D}_\mu\phi^s\gamma^\mu\,\epsilon^B-\frac{i}{4}\,H_{\mu\nu\,I}^-\gamma^{\mu\nu}\epsilon_A+
g\,\mathbb{N}_{{I}}{}^A\epsilon_A\dots\,,
\label{tralam22}\\
\Theta^{\Lambda{}\alpha}\,\delta B_{\alpha\,\mu\nu}&=2i\,g\,\left(\bar{\lambda}_{\mathcal{I}}\gamma_{\mu\nu}\epsilon_A\,
\mathcal{P}^{\Lambda\,\mathcal{I}A}-\bar{\lambda}^{\mathcal{I}}\gamma_{\mu\nu}\epsilon^A\,
\mathcal{P}^\Lambda{}_{\mathcal{I}A}\right)-8\,g\,\left(\bar{\psi}_{A\,[\mu}\gamma_{\nu]}\epsilon^B+
\bar{\psi}^B_{[\mu}\gamma_{\nu]}
\epsilon_A\right)\,\mathcal{Q}^\Lambda{}_{B}{}^A-\nonumber\\
&-2X^\Lambda{}_P{}^M\mathbb{C}_{MN}\,A^P_{[\mu}\,\delta A^N_{\nu]}\,,\label{traB2}
\end{align}
where, as usual, the ellipses refer to terms which are of higher order in the fermion fields.
\subsection{Dual Gauged Supergravities}
All the deformations of the ungauged model required by the gauging procedure depend on $\Theta$ in a manifestly $G$-covariant way. This means that, if we transform all the fields $\Phi$ (bosons and fermions) of the model under $G$ (the fermions transforming under corresponding compensating transformations in $H$) and at the same time transform $\Theta$, the field equations and Bianchi identities -- which we collectively denote by $\Es(\Phi,\,\Theta)=0$ -- are left invariant:\footnote{In the non-duality-covariant formulation discussed in Sect. \ref{gitef}, the equations of motion (in particular the Maxwell ones) still depended on the symplectic matrix $E$ connecting the electric frame to a generic one and thus the formal on-shell $G$-invariance discussed here would require $G$-transforming $E$ together with the fields and the embedding tensor: $$\forall\,{\bf g}\in G\,\,\,:\,\,\,\,\,\,E\rightarrow E'=E\,\Rs_{v}[{\bf g}]^T.$$ }
\begin{equation}
\forall {\bf g}\in G\;:\;\;
\Es(\Phi,\,\Theta)=0
\;\;\Leftrightarrow\;\;
\Es({\bf g}\star\Phi,\,{\bf g}\star\Theta)=0\,.\label{dualgaug}
\end{equation}
Since the embedding tensor $\Theta$ is a \emph{spurionic}, namely non-dynamical, object, the above on-shell invariance should not be regarded as a symmetry of a single theory, but rather as an equivalence (or proper duality) between two different theories, one defined by $\Theta$ and the other by ${\bf g}\star  \Theta$.\, Gauged supergravities are therefore classified in \emph{orbits} with respect to the action of $G$ (or better $G(\mathbb{Z})$) on $\Theta$.
\subsection{Vacua and Dualities}\label{vad}
A (Lorentz preserving) vacuum of a supergravity theory is a maximally symmetric solution, that is it can, depending on the value of the cosmological constant $\Lambda$, exhibit Minkowski ($\Lambda=0$), de Sitter  ($\Lambda>0$) or anti-de Sitter  ($\Lambda<0$) space-time geometry.  Due to the maximal space-time symmetry, only scalar fields are allowed to have a non-vanishing (uniform) v.e.v.:
  $$\langle \phi^s(x)\rangle\equiv \phi^s_0\,,$$
the vector and the fermionic fields vanishing on the solution. This v.e.v. defines a point $\phi_0\equiv (\phi_0^s)$ in the moduli space which is an extremum of the scalar potential $V(\phi)$:
\begin{equation}
\left.\frac{\partial V}{\partial\phi^s}\right\vert_{\phi_0}=0\,,\label{dV0}
\end{equation}
and the value $V(\phi_0)$ of the scalar potential  on the vacuum provides the effective cosmological constant for the underlying space-time geometry:
\begin{equation}
\Lambda=V(\phi_0)\,.\label{LV0}
\end{equation}
The Riemann tensor reads (see Appendix \ref{nacv} for the relevant conventions):
\begin{equation}
R_{\mu\nu\rho\sigma}=-\frac{\Lambda}{3}\left(g_{\mu\rho} g_{\nu\sigma}-g_{\mu\sigma} g_{\nu\rho}\right)\,,\label{RiemL}
\end{equation}
so that the Ricci tensor is ${R}_{\mu\nu}=R_{\mu\rho\nu}{}^\rho=-\Lambda\,g_{\mu\nu}$.\par
In extended supergravity theories the scalar potential is given by Eq. (\ref{Pot}).
Being expressed as an $H$-invariant combination of composite fields (the fermion shifts),
it is invariant under the simultaneous action of $G$ on $\Theta$ and $\phi^s$:
\begin{equation}
\forall {\bf g}\in G \;:\quad
V({\bf g}\star\phi,{\bf g}\star\Theta)=V(\phi,\Theta)\,.\label{Vinvar}
\end{equation}
This means that, if $V(\phi,\Theta)$ has an extremum in $\phi_0$
\begin{equation}
\left.\frac{\partial}{\partial\phi^s}V(\phi,\Theta)\right\vert_{\phi_0}=0\,,
\end{equation}
$V(\phi,{\bf g}\star\Theta)$ has an extremum at $\phi'_0={\bf g}\star \phi_0$ with the same properties (value of the potential at the extremum and its derivatives):
\begin{equation}
\left.\frac{\partial}{\partial\phi^s}V(\phi,{\bf g}\star\Theta)\right\vert_{{\bf g}\star\phi_0}=0\,,
\qquad \forall {\bf g}\in G\;.
\end{equation}
If the scalar manifold is homogeneous, we can map any point $\phi_0$ to the origin $\Or$, where all scalars vanish, by the inverse of the coset representative $L(\phi_0)^{-1}\in G$. We can then map a generic vacuum $\phi_0$ of a given theory (defined by an embedding tensor $\Theta$) to the origin in a theory defined by \,$\Theta'=L(\phi_0)^{-1}\star \Theta$.\;
As a consequence of this, when looking for vacua with given properties (residual (super)symmetry, cosmological constant, mass spectrum etc.), with no loss of generality we can compute all quantities defining the gauged theory -- fermion shifts and mass matrices -- at the origin: \begin{equation}
\mathbb{N}(\Or,\,\Theta)\,,\;\;\mathbb{S}(\Or,\,\Theta)\,,\;\;\mathbb{M}(\Or,\,\Theta)\,,
\end{equation}
and translate the properties of the vacuum in conditions on $\Theta$. In this way, we can search for the vacua by scanning through all possible gaugings \cite{inverso,Dibitetto:2011gm,Dall'Agata:2011aa}.\par
Let us also note the following useful property. Using the R-symmetry group ${\rm (S)U}(\mathcal{N})$, on a given point $\phi_0$ in the moduli space we can always diagonalize $\mathbb{S}_{AB}(\phi_0)$ \footnote{This can be proven using the \emph{Autonne-Takagi factorization}, see point (c) of Corollary 4.4.4 in \cite{HJ}.} and bring it to the form:
\begin{equation}
\mathbb{S}_{AB}(\phi_0)\stackrel{U\in H_R}{\longrightarrow } \,(U\mathbb{S}U^T)_{AB}=s_A\,\delta_{AB}\,,\label{Sdiag}
\end{equation}
where $s_A$ are complex numbers. If the R-symmetry group is ${\rm U}(\mathcal{N})$, $s_A$ can all be made real, while if the group is ${\rm SU}(\mathcal{N})$, as for the maximal theory, we cannot remove the overall phase
${\rm Arg}(s_1\,s_2\dots\,s_{\mathcal{N}})$.
The quantities $|s_A|$ and, for $\mathcal{N}=8$, ${\rm Arg}(s_1\,s_2\dots\,s_{8})$, seen as functions of the scalar fields and of the embedding tensor, are $G$-invariant.
\subsubsection{Supersymmetric Vacua}
A vacuum $\phi_0$ can be supersymmetric, namely can preserve an amount of supersymmetry. In this case there should exist a local supersymmetry parameter $\epsilon_A(x)$ along which the supersymmetry variation, evaluated on the solution, of the fermionic fields vanish. This follows from the property that, if $\vert 0 \rangle$ is the vacuum state, along the direction of the preserved supersymmetry we have $\bar{\epsilon}\,Q\,\vert 0 \rangle=0$, and thus
\begin{equation}
\delta_\epsilon f(x)=\langle 0\vert [\bar{\epsilon}\,Q,\,\hat{f}(x)]\vert 0 \rangle=0\,,\nonumber
\end{equation}
where $f(x)$ denotes a generic fermionic field and $\hat{f}(x)$ the corresponding field operator. The right-hand-side of the above equation depends on the v.e.v. $\phi_0^s$ of the scalar fields and the space-time geometry of the vacuum solution. The analogous condition on the supersymmetry variations of the bosonic fields would be trivially satisfied since these are expressed in terms of the fermionic fields which vanish on the background. Explicitly the above conditions read:
\begin{align}
\delta\psi_{\mu\,A}&=\nabla_\mu \epsilon_A+ig\,\mathbb{S}_{AB}\gamma_\mu\epsilon^B=0\,,\nonumber\\
\delta \lambda_{\mathcal{I}}&=g\,\mathbb{N}_{\mathcal{I}}{}^A\,\epsilon_A=0\,,\label{KSeqs}
\end{align}
where the tensors $\mathbb{S}_{AB},\,\mathbb{N}_{\mathcal{I}}{}^A$ are evaluated at $\phi_0$.
 These are the Killing spinor equations for the vacuum. Notice that, being the gauge-connection trivial and the scalar fields uniform  on the solution, we have $\hat{\mathcal{Q}}_\mu=0$ and the covariant derivative in the first of (\ref{KSeqs}) is the Levi-Civita one on space-time.
 The background preserves
 a number $\mathcal{N}'\le \mathcal{N}$ of the $\mathcal{N}$ supersymmetries of the theory if the Killing spinor equations admit $\mathcal{N}'$ distinct solutions (Killing spinors). We can work out the integrability condition on the first of (\ref{KSeqs}):
 \begin{align}
 0=\nabla_{[\mu}\delta\psi_{\nu]A}&= \nabla_{[\mu} \nabla_{\nu]}\epsilon_A+i\,g\,\mathbb{S}_{AB}\gamma_{[\nu}
 \nabla_{\mu]} \epsilon^B=\nonumber\\
 &=\frac{1}{8}\,R_{\mu\nu\rho\sigma}\,\gamma^{\rho\sigma}\epsilon_A-g^2\,
 \gamma_{\mu\nu}\mathbb{S}_{AB}\mathbb{S}^{BC}\,\epsilon_C\,,\label{intKS}
 \end{align}
 Without loss of generality we may assume the $\mathcal{N}'$ Killing spinors to correspond to the first $\mathcal{N}'$ directions of the supersymmetry parameter space. This amounts to splitting the index $A$ into $a$ and $a'$ with $a=1,\dots, \mathcal{N}'$ and $a'=\mathcal{N}'+1,\dots, \mathcal{N}$ and to write the  $\mathcal{N}'$  Killing spinors as: $\epsilon_A^{(a)}(x)=h_a(x)\delta^a_A$.
 The second of Eqs. (\ref{KSeqs}) implies
 \begin{equation}
 \mathbb{N}_{\mathcal{I}}{}^a=0\,\,,\,\,\,\,a=1,\dots, \mathcal{N}'\,,\label{KSN}
 \end{equation}
 while from the  integrability condition (\ref{intKS}) it follows that:
 \begin{equation}
\frac{1}{8}\,R_{\mu\nu}{}^{\rho\sigma}\,\delta_a^b=g^2\,
\mathbb{S}_{aB}\mathbb{S}^{Bb}\,\delta_{\mu\nu}^{\rho\sigma}\,\,;\,\,\,\,\mathbb{S}_{a'B}\mathbb{S}^{Bc}=0\,.
 \end{equation}
We see that the Riemann tensor has the form (\ref{RiemL}) with cosmological constant $\Lambda$
given by:
\begin{equation}
{\Lambda}\,\delta_a^c=-12\,g^2\, \mathbb{S}_{aB}\mathbb{S}^{Bc}\le 0\,\,;\,\,\,\,\mathbb{S}_{a'B}\mathbb{S}^{Bc}=0\,,
\end{equation}
In other words we can write
\begin{equation}
g^2\,\mathbb{S}\mathbb{S}^\dagger=\left(\begin{matrix}-\frac{1}{12}\Lambda\, \delta_a^b &{\bf 0}\cr {\bf 0} & A_{a'}{}^{b'} \end{matrix}\right)\,.
\end{equation}
Being $\mathbb{S}\mathbb{S}^\dagger$ a  non-negative matrix, then see that \emph{supersymmetric vacua can only be Minkowski ($\Lambda=0$) or anti-de Sitter ($\Lambda<0$).}\par
Using ${\rm SU}(\mathcal{N}-\mathcal{N}')$  we can further reduce the block $A_{a'}{}^{b'}$ to a real non-negative diagonal matrix. By the proof of the aforementioned  Autonne -- Takagi factorization (see previous footnote),  we can write the first $\mathcal{N}'\times \mathcal{N}'$ block $\mathbb{S}_{ab}$ of $\mathbb{S}_{AB}$, on the background, in the form (\ref{Sdiag}) with
\begin{equation}
g\,\mathbb{S}_{ab}=e^{\varphi_a}\,\sqrt{-\frac{\Lambda}{12}}\,\delta_{ab}\,,\label{Sdiagx}
\end{equation}
where, when $\Lambda< 0$,  $\varphi_a$ can all be set to zero except for the case $\mathcal{N}'=\mathcal{N}=8$.
The Ward identity (\ref{WID}), computed on the solution and  restricted to the indices $a,c$, taking into account (\ref{KSN}), reads:
\begin{equation}
V(\phi_0)\delta_a^c=-12\,g^2\,\mathbb{S}_{ab}\mathbb{S}^{bc}=\Lambda\,\delta_a^c\,,
\end{equation}
which yields the identification  (\ref{LV0}). Finally we notice that  bosonic backgrounds, on which the only non-vanishing fields, aside from the metric, are uniform scalar fields, and which preserve supersymmetry are also vacua of the theory, namely satisfy the field equations. To see this we prove that, as a consequence of the Killing spinor equations, the corresponding point in the moduli space is an extremum for  $V$. This is easily seen by computing the derivative of the Ward identity, restricted to the $a,c$ indices, on $\phi_0$:
\begin{equation}
g^{-2}\,\delta_a^c\left.\partial_s V\right\vert_{\phi_0}=-12\,{\Scr D}_s\mathbb{S}_{ab}\mathbb{S}^{bc}+{\Scr D}_s\mathbb{N}^{\mathcal{I}}{}_a\mathbb{N}_{\mathcal{I}}{}^c+c.c=-6\,\mathcal{P}_{s\,\mathcal{I}(a}\mathbb{N}^{\mathcal{I}}_{c)}\,
\mathbb{S}^{bc}+{\Scr D}_s\mathbb{N}^{\mathcal{I}}{}_a\mathbb{N}_{\mathcal{I}}{}^c+c.c=0\,,
\end{equation}
where we have used the gradient flow equation (\ref{GF1}), as well as (\ref{Sdiagx}) and (\ref{KSN}).\par
\subsubsection{Vacua and Mass Matrices}\label{vmm}
A four-dimensional de Sitter (dS) and anti-de Sitter (AdS) space-time can be described as connected hyperboloids in $\mathbb{R}^5$ defined by the equations:
\begin{align}
{\rm AdS}_4&=\frac{{\rm O}(2,3)}{{\rm O}(1,3)}\,: \,\,\,y_0^2-y_1^2-y_2^2-y_3^2+y_4^2=L^2\,\nonumber\\
{\rm dS}_4&=\frac{{\rm O}(1,4)}{{\rm O}(1,3)}\,: \,\,\,y_0^2-y_1^2-y_2^2-y_3^2-y_4^2=-L^2\,,
\end{align}
where $L$ is the ``\emph{radius}'' of the space-time. The cosmological constant is related to this length as follows: $\Lambda=\frac{3}{L^2}$ for  de Sitter and $\Lambda=-\frac{3}{L^2}$ for anti-de Sitter. The dynamics of the scalar modes on a vacuum $\phi_0$ is determined by expanding  the scalar potential $V$ about the corresponding extremum:
\begin{equation}
\phi^s= \phi_0^s+\delta \phi^s\,\,;\,\,\,\,V=V(\phi_0)+\frac{1}{2}\left.\partial_{r}\partial_sV\right\vert_{\phi_0}\,\delta\phi^r
\delta\phi^s+\dots
\end{equation}
where we have used (\ref{dV0}). The squares of the scalar masses on the solution  are given by the eigenvalues of the matrix
\begin{equation}
M_s{}^r={\Scr G}^{rt}\,\left.\partial_{t}\partial_sV\right\vert_{\phi_0}\,.
\end{equation}
Minkowski backgrounds ($\Lambda=V(\phi_0)=0$), as well as  de Sitter ones, are stable vacua if the squared masses of the fluctuations are non-negative (namely there are no tachyonic modes). This in particular implies that $V(\phi)$ must have a  local minimum at the critical point $\phi_0$.\par Anti-de Sitter (AdS) solutions, on the other hand, can be perturbatively stable even if $\phi_0$ is a saddle point or a local maximum, provided the negative eivenvalues of the matrix $M_s{}^r$, i.e. the negative squared scalar masses $m^2$, be not too large in absolute value. More specifically the stability condition on the scalar spectrum is:
\begin{equation}
m^2\,L^2\ge -\frac{9}{4}\,.\label{BFBound}
\end{equation}
This is the Breitenlohner-Freedman (BF) bound first found in \cite{Breitenlohner:1982jf}. It can be shown that supersymmetric AdS backgrounds are also stable, namely that the BF condition is satisfied. For example, the maximal supergravity with ${\rm SO}(8)$-gauging, constructed  in \cite{deWit:1981sst}, features a maximally supersymmetric (i.e. preserving all the eight supersymmetries of the model) AdS vacuum at the origin of the scalar manifold, which is a local maximum of the scalar potential.
The vacuum is however stable since the squared masses of the scalar fields are not too negative, i.e. they satisfy the BF bound, being
\begin{equation}
m^2=-\frac{2}{L^2}>-\frac{9}{4\,L^2}\,.
\end{equation}
Being maximally supersymmetric, the spin-$1/2$ shift-tensor vanishes, $\mathbb{N}^{\mathcal{I}}{}_A=0$, while $\mathbb{S}_{AB}$ is proportional to the identity.\par
Linearizing the field equations on a given vacuum $\phi_0$, one derives the mass matrices for the various fields, as functions of $\phi_0$ and the embedding tensor $\Theta_M{}^\alpha$. In particular, expanding the duality-covariant Maxwell equations (\ref{Max2}) about the vacuum in which $\langle A_\mu\rangle=0$,  we can derive for the vector fields the following symplectic-frame-independent for of the squared-mass matrix:
\begin{equation}
M^{(v)}{}_M{}^N(\phi_0,\Theta)=-\left.g^2\,\Theta_M{}^\alpha\Theta_P{}^\beta\,k_\alpha^r k_\beta^s\,{\Scr G}_{rs}\,\mathcal{M}^{PN}\right\vert_{\phi_0}\,.\label{MvectMN}
\end{equation}
Note that $M^{(v)}(\phi_0,\Theta)$ is a non-negative $2n_v\times 2n_v$ matrix (recall that $\mathcal{M}_{MN}$ is negative-definite, its expression being given in (\ref{MLL2})). However the locality  constraint (\ref{locality}) on the embedding tensor guarantees that its rank be less than $n_v$, its non-vanishing eigenvalues being the vector squared masses.\par
Let us now discuss the fermion masses. The masses of the gravitini fields signal spontaneous supersymmetry breaking: $k=\mathcal{N}-\mathcal{N}'$ of the $\mathcal{N}$ supersymmetries of the theory are broken on a vacuum if and only if the corresponding $k$ gravitini become massive. This occurs by virtue of the super-Higgs phenomenon: $k$ massless fermion fields (Goldstini) are ``eaten'' by the $k$ gravitini thus providing the spin-$1/2$ component they need to become massive. The Goldstino fields $\eta_A$ are associated with the directions in the supersymmetry parameter space (labeled by $a',b',\dots=1,\dots, k$ earlier) along which the supersymmetry is broken, and can be written in the form (see also the discussion below):
\begin{equation}
\eta_A\propto\mathbb{N}^{\mathcal{I}}{}_A\,\lambda_{\mathcal{I}}\,,\label{Gold}
\end{equation}
the tensor $\mathbb{N}^{\mathcal{I}}{}_A$ being evaluated on the vacuum. From Eq. (\ref{Gold}) and (\ref{KSN}) it indeed follows that $\eta_a=0$, $a=1,\dots, \mathcal{N}'$, namely the Goldstini have non-vanishing components only along the broken supersymmetries.\par
The fermion masses are obtained by linearizing about the vacuum the field equations of the fermion fields (gravitini and spin-$1/2$ fields). As emphasized earlier, in extended models the masses only originate from the gauging and, more specifically, from the $O(g)$-terms (\ref{fmassterms}) depending on the tensors $\mathbb{S},\,\mathbb{N},\,\mathbb{M}$.
The fermion masses can be read off the linearized field equations once they are put in the general forms:
\begin{align}
\epsilon^{\mu\nu\rho\sigma}\,\gamma_\nu \mathcal{D}_\rho\psi_{A\sigma}&=\,M^{\left[\frac{3}{2}\right]}_{AB}\,\gamma^{\mu\nu}\,\psi_\nu^B\,,\nonumber\\
i\gamma^\mu\mathcal{D}_\mu\lambda_{\mathcal{I}}&=\,M^{\left[\frac{1}{2}\right]}_{\mathcal{I}\mathcal{J}}\,\lambda^{\mathcal{J}}\,. \label{feqs}
\end{align}
Due to the interaction term $i\,eg\,\mathbb{N}^{\mathcal{I}}{}_A\,\bar{\lambda}_{\mathcal{I}}\gamma^\mu\psi_{A\mu}+h.c.$, in order to write the fermion equations in the form (\ref{feqs}), we need to redefine the gravitino. Consider for instance a Minkowski vacuum, $V_0=0$, in which all supersymmetries are broken. In this case the matrix $\mathbb{S}$, which we shall consider to be diagonal in the following, is invertible  and the reader can verify that the $\bar{\lambda}\psi$-interaction term can be disposed of by redefining the gravitini as follows:
\begin{equation}
\psi_\mu^A=\psi^{\prime \,A}_{\mu}+\frac{i}{12}\,\sum_C\mathbb{S}^{-1\,AC}\mathbb{N}^{\mathcal{I}}{}_C\,\gamma_\mu\lambda_{\mathcal{I}}\,,\label{psired}
\end{equation}
where, as usual, all the fermion shift-tensors are computed on the vacuum.
If the supersymmetry breaking is only partial, then $\mathbb{S}$ is singular and the sum on the right hand side of (\ref{psired}) should be intended over the index $c'$ of the broken supersymmetries, since the gravitini $\psi_{a\mu}$ do not couple to $\lambda_{\mathcal{I}}$ by virtue of (\ref{KSN}). This restriction is automatically implemented by the contraction with the $\mathbb{N}^{\mathcal{I}}{}_C$ tensor which is indeed non-vanishing only if $C=c'$.
The gravitino mass matrix reads:\footnote{In the case of anti-de Sitter vacua, the eigenvalues of the mass matrix along the preserved supersymmetries are $\sqrt{-\frac{V_0}{12}}$. Nevertheless the AdS-mass of the corresponding gravitini is zero.}
\begin{equation}
M^{\left[\frac{3}{2}\right]}_{AB}=-2\,g\,\mathbb{S}_{AB}\,,\label{32mass}
\end{equation}
where $\mathbb{S}_{AB}$ is taken to be diagonal, see (\ref{Sdiag}): $\mathbb{S}_{AB}=\delta_{AB}\,s_A$.\par
The reader can verify that, upon the redefinition (\ref{psired}), the spin-$1/2$ mass matrix in (\ref{feqs}) reads:
\begin{equation}
M^{\left[\frac{1}{2}\right]}_{\mathcal{I}\mathcal{J}}=2\,g\,\mathbb{M}_{\mathcal{I}\mathcal{J}}-\frac{g}{3}\,{\sum_{AB}}'
\mathbb{S}^{-1}_{AB}\,\mathbb{N}_{\mathcal{I}}{}^A\mathbb{N}_{\mathcal{J}}{}^B\,,\label{12mass}
\end{equation}
where $\sum'$ is the sum over the broken supersymmetries only. As pointed out above, this restriction is already effected through the contraction with the $\mathbb{N}$-tensors and therefore we shall omit in what follows the prime on the sum. \par
If the vacuum is of anti-de Sitter type ($V_0< 0$) the above expression is modified as follows:
\begin{equation}
M^{\left[\frac{1}{2}\right]}_{\mathcal{I}\mathcal{J}}=2\,g\,\mathbb{M}_{\mathcal{I}\mathcal{J}}-\frac{g}{3}\,\sum_{AB}
\left(\frac{\mathbb{S}}{|\mathbb{S}|^2+\frac{V_0}{12\,g^2}\,{\bf 1}}\right)_{AB}\,\mathbb{N}_{\mathcal{I}}{}^A\mathbb{N}_{\mathcal{J}}{}^B\,.\label{12massads}
\end{equation}
The above expression is consistent since the (diagonal) matrix $\mathbb{S}\bar{\mathbb{S}}+\frac{V_0}{12\,g^2}\,{\bf 1}$, if restricted to the indices $a',b'$ of the broken supersymmetries, is invertible. Indeed from our previous discussion only $|\mathbb{S}_{aa}|^2=-\frac{V_0}{12\,g^2}$, while $|\mathbb{S}_{a'a'}|^2\neq -\frac{V_0}{12\,g^2}$. The generalization (\ref{12massads}) of (\ref{12mass}) can be derived by the requiring it to have zero eigenvalues along the Goldstini directions (as it should since the Goldstino fields are massless fermions). To prove this we can contract the mass matrix with the following combinations of Goldstinos: $\lambda^{(G)\,\mathcal{J}}=\mathbb{N}^{\mathcal{J}}{}_A\eta^A$:\footnote{Summation over repeated indices in now understood.}
\begin{align}
g^{-1}M^{\left[\frac{1}{2}\right]}_{\mathcal{I}\mathcal{J}}\mathbb{N}^{\mathcal{J}}{}_A\eta^A&=2
\mathbb{M}_{\mathcal{I}\mathcal{J}}\mathbb{N}^{\mathcal{J}}{}_A\eta^A
-\frac{1}{3}\,\sum_{CD}
\left(\frac{\mathbb{S}}{|\mathbb{S}|^2+\frac{V_0}{12\,g^2}\,{\bf 1}}\right)_{CD}\,\mathbb{N}_{\mathcal{I}}{}^C\mathbb{N}_{\mathcal{J}}{}^D\mathbb{N}^{\mathcal{J}}{}_A\eta^A\,.
\end{align}
Let us apply to the last term on the right hand side the Ward identity (\ref{WID})
\begin{align}
g^{-1}M^{\left[\frac{1}{2}\right]}_{\mathcal{I}\mathcal{J}}\mathbb{N}^{\mathcal{J}}{}_A\eta^A&=
2\mathbb{M}_{\mathcal{I}\mathcal{J}}\mathbb{N}^{\mathcal{J}}{}_A\eta^A-\frac{V_0}{3\,g^2}\,
\left(\frac{\mathbb{S}}{|\mathbb{S}|^2+\frac{V_0}{12\,g^2}\,{\bf 1}}\right)_{CA}\mathbb{N}_{\mathcal{I}}{}^C\eta^A-\nonumber\\
&-4\,
\left(\frac{\mathbb{S}}{|\mathbb{S}|^2+\frac{V_0}{12\,g^2}\,{\bf 1}}\right)_{CD}\,\mathbb{N}_{\mathcal{I}}{}^C\mathbb{S}^{DE}\,\mathbb{S}_{EA}\eta^A=\nonumber\\
&=(2\mathbb{M}_{\mathcal{I}\mathcal{J}}\mathbb{N}^{\mathcal{J}}{}_A-4\,\mathbb{N}_{\mathcal{I}}{}^B\mathbb{S}_{BA})\,\eta^A\propto
g^{-2}\frac{\partial V}{\partial\phi^s}\mathcal{P}^{-1}{}_{\mathcal{I}A}{}^s\,.\label{Mgold}
\end{align}
In the last equality we have used Eq. (\ref{DWID}).\footnote{Equation (\ref{DWID}) can indeed also be written in the equivalent form $$
g^{-2}\,\frac{\partial V}{\partial\phi^s}\mathcal{P}^{-1}{}_{\mathcal{I}A}{}^s\propto
2\mathbb{M}_{\mathcal{I}\mathcal{J}}\mathbb{N}^{\mathcal{J}}{}_A\eta^A-4\,\mathbb{N}_{\mathcal{I}}{}^B\mathbb{S}_{BA}\,,$$
where we have defined the inverse $\mathcal{P}^{-1}$ of the vielbein matrix $\mathcal{P}$.} The expression (\ref{Mgold}) vanishes, being proportional to the  gradient of the scalar potential on the vacuum $\phi_0$.

\section{Examples of Gaugings:  The $\N=8$\,,\; $D=4$ Supergravity}\label{startexa}
In what follows we illustrate how the general procedure discussed above is applied to four-dimensional extended supergravities. We shall give particular emphasis on the maximal and $\mathcal{N}=2$ theories, starting in this Section from the former.
\subsection{The Ungauged Theory}
The four-dimensional maximal supergravity is characterized by
having $\N=8$ supersymmetry (that is $32$ supercharges), which is
the maximal amount of supersymmetry allowed for a consistent theory of gravity.\par\smallskip
We shall start restricting ourselves to the (ungauged) $\N=8$ theory with no antisymmetric tensor field. The theory, first constructed in \cite{Cremmer:1978ds}, describes a single massless graviton supermultiplet in Minkowski space-time, consisting of the graviton $g_{\mu\nu}$, $8$ spin-$3/2$ gravitini $\psi^A_\mu$ ($A=1,\dots, 8$) transforming in the
fundamental representation of the R--symmetry group $\SU(8)$,
$28$ vector fields $A^\Lambda_\mu$ (with $\Lambda=0,\dotsc,\,27$), $56$ spin-$1/2$ dilatini $\chi_{ABC}$ in the ${\bf
56}$ of $\SU(8)$ and $70$ real scalar fields $\phi^r$:
\begin{align}
\big[\;\;
1\;\times\;\underbrace{g_{\mu\nu}}_{{j=2}}\;,\quad
8\;\times\;\underbrace{\psi^A_\mu}_{{j=\frac32}}\;,\quad
28\;\times\;\underbrace{A^\Lambda_\mu}_{{j=1}}\;,\quad
56\;\times\;\underbrace{\chi_{ABC}}_{{j=\frac12}}\;,\quad
70\;\times\;\underbrace{\phi^r}_{{j=0}}
\;\;\;\big]
\;.\label{N8}
\end{align}
The scalar fields are described by a non-linear $\sigma$-model on the Riemannian manifold $\Mscal$ of the form
\begin{eqnarray}
\Mscal\=\frac{G}{H}\=\frac{{\rm E}_{7(7)}}{\SU(8)}\,,
\end{eqnarray}
the isometry group being $G={\rm E}_{7(7)}$, and $H=\SU(8)$
being the R--symmetry group.\footnote{More precisely the isotropy group is ${\rm SU}(8)/\mathbb{Z}_2$. This is because the compact subgroup of ${\rm E}_{7(7)}$ is defined through its ${\bf 28}$ representation, see Eq. (\ref{su8ine7}), namely its action on rank-2 antisymmetric tensors $V_{AB}=-V_{BA}$, in which a $\mathbb{Z}_2$ subgroup is trivially realized. The image of ${\rm SU}(8)$ through such representation is then isomorphic to ${\rm SU}(8)/\mathbb{Z}_2$. Similarly, in the maximal $D=5$ supergravity, the isotropy group is the ${\rm USp}(8)/\mathbb{Z}_2$ subgroup of ${\rm E}_{6(6)}$. We shall be sloppy with this and omit the quotient by $\mathbb{Z}_2$ when writing these isotropy groups.} The bosonic Lagrangian has the
usual form (\ref{boslagr}).
The global on-shell symmetry group is therefore $G={\rm E}_{7(7)}$ and has 133 generators $t_\alpha$. The (Abelian) vector field strengths $F^\Lambda=dA^\Lambda$ and their magnetic duals $ G_\Lambda$ together transform in the ${\Scr R}_v={\bf 56}$ fundamental representation of the ${\rm E}_{7(7)}$ duality group with generators $(t_\alpha)_M{}^N$, so that
\begin{equation}
\delta \mathcal{G}^M_{\mu\nu} ~=
\left(\begin{matrix}
\delta F^\Lambda_{\mu\nu} \cr \delta  G_{\Lambda\,\mu\nu}
\end{matrix}\right)
=~ -\Lambda^\alpha\,(t_\alpha)_N{}^M\,\mathcal{G}^N_{\mu\nu}\;.
\end{equation}
This is a symplectic representation which defines an embedding of ${\rm E}_{7(7)}$ into ${\rm Sp}(56,\,\mathbb{R})$. For a comparison of our notations with those used in the literature on maximal supergravity (for instance in \cite{deWit:2007mt}) see Appendix \ref{apcomparison}.\par
As far as the infinitesimal generators of ${\rm E}_{7(7)}$ are concerned, they close the Lie algebra $\mathfrak{g}=\mathfrak{e}_{7(7)}$ which splits, according to the Cartan decomposition, into the Lie algebra $\mathfrak{H}=\mathfrak{su}(8)$ and a 70-dimensional space $\mathfrak{K}$ supporting the representation ${\Scr K}={\bf 70}$ under the adjoint action of ${\rm SU}(8)$. Generic elements of the two spaces, in the $\underline{{\Scr R}}_v^c$-representation, introduced in Sect. \ref{dcsl}, read
\begin{align}
\Lambda\in \mathfrak{su}(8)\,\longrightarrow\,\,\,\underline{{\Scr R}}_v^c[\Lambda]=\left(\begin{matrix} \Lambda^{AB}{}_{CD} & {\bf 0}\cr {\bf 0} & \Lambda_{AB}{}^{CD}\end{matrix}\right)\,,\label{su8ine7}\\
\Sigma\in \mathfrak{K}\,\longrightarrow\,\,\,\underline{{\Scr R}}_v^c[\Sigma]=\left(\begin{matrix} {\bf 0}& \Sigma^{ABCD}\cr \Sigma_{ABCD} & {\bf 0}\end{matrix}\right)\,,
\end{align}
where $\Lambda^{AB}{}_{CD} =4\,\delta^{[A}_{[C}\Lambda^{B]}{}_{D]}=-\Lambda_{CD}{}^{AB}$ is an ${\rm SU}(8)$ generator in the ${\bf 28}$-representation, and  $\Lambda^A{}_B=-\Lambda_B{}^A$ the corresponding generator in the  fundamental one. The tensors $\Sigma_{ABCD}$ and $\Sigma^{ABCD}=(\Sigma_{ABCD})^*$ transform in the ${\bf 70}$ of ${\rm SU}(8)$ and are related by the reality condition (\ref{realityn8}):
\begin{equation}
\Sigma^{ABCD}=\frac{1}{24}\epsilon^{ABCDEFGH}\,\Sigma_{EFGH}\,.\label{realityn82}
\end{equation}
Two generators $\Sigma_1,\,\Sigma_2$ in $\mathfrak{K}$ close, under commutation, on an $\mathfrak{su}(8)$-element, according to  the property (\ref{KKH}) of symmetric spaces, by virtue of  the relation (\ref{sigmaprops}).
\paragraph{Parametrizations of the scalar manifold.} As pointed out in Sect. \ref{ghsect}, we can use for a homogeneous manifold two kinds of parametrizations. One, which we have called \emph{solvable}, consists in fixing at each point of the manifold the right action of $H$ so that $L(\phi^r)$ belongs to a solvable subgroup $G_S$ of $G$, see  Eq. (\ref{solpar}). We shall discuss examples of this parametrization when dealing with dimensional reductions.\par
The other parametrization consists in defining the coset representative as follows, see Eq. (\ref{cartpar}):
\begin{equation}
L=\exp(\phi^{ABCD}\,K_{ABCD})\,,\label{Hcovpar}
\end{equation}
where $\{K_{ABCD}\}$ is a basis of $\mathfrak{K}$. As pointed out in Sect. \ref{ghsect}, the scalar fields
$\phi^{ABCD}$ transform linearly under the action of $H$ in the representation ${\Scr K}$. In our case they transform in the ${\bf 70}$ of ${\rm SU}(8)$ and satisfy the reality condition (\ref{realityn82}):
\begin{equation}
\phi^{ABCD}=\frac{1}{24}\epsilon^{ABCDEFGH}\,(\phi^{EFGH})^*\,.\label{realityn83}
\end{equation}
\subsection{Symplectic Frames} \label{symfram}
As discussed in Sect. \ref{sframes}, different choices of the symplectic frame lead to ungauged theories which, although on-shell equivalent, may be described by inequivalent Lagrangians. These Lagrangians are characterized by different global symmetry groups $G_{el}$. Inequivalent symplectic frames typically originate from different dimensional reductions.
In general any compactification is characterized by a group $G_{int}$ associated with the internal manifold (in the absence of fluxes).\footnote{Here and throughout this paper, we are not considering quantum corrections, so that the global symmetry groups are all continuous Lie groups.} As it was mentioned earlier, maximal supergravities originate from the dimensional reduction of the maximal eleven and ten -dimensional theories on tori. We shall be dealing with toroidal compactifications in some detail later in Sect. \ref{toroidalc}. Here we just recall some general facts. For a compactification on an $n$-torus, $G_{int}={\rm GL}(n,\mathbb{R})$, acting transitively on the internal metric moduli $g_{\upalpha\upbeta }$, $\upalpha,\,\upbeta=1,\dots,\,n$. These are $n(n+1)/2$ and parametrize the manifold:
\begin{equation}
g_{\upalpha\upbeta  }\in \frac{{\rm GL}(n,\mathbb{R})}{{\rm SO}(n)}\,.\label{GinGL}
\end{equation}
The ${\rm SO}(1,1)$ factor in ${\rm GL}(n,\mathbb{R})$ acts as a rescaling of the internal metric: $g_{\upalpha\upbeta }\,\rightarrow\,e^{2\lambda}\,g_{\upalpha\upbeta }$.
 Starting from a theory in $D$-dimensions featuring a global symmetry group $G_D$ of the action, dimensional reduction on $T^n$ to four dimensions ($D=4+n$) yields a formulation of the four-dimensional supergravity in which the global symmetry group of the Lagrangian contains $G_D\times G_{int}$:
\begin{equation}
G_D\times G_{int}\,\subset\, G_{el}\,.\label{GDGint}
\end{equation}
For instance the global symmetry group of five-dimensional maximal supergravity  (which is a symmetry of the action since there is no electric-magnetic duality in five dimensions) is $G_5={\rm E}_{6(6)}$ and, upon dimensional reduction of the theory on a circle we find a formulation of the $\mathcal{N}=8,\,D=4$ theory with  ${\rm E}_{6(6)}\times{\rm SO}(1,1)\subset G_{el}$ manifest symmetry of the Lagrangian, where the factor ${\rm SO}(1,1)$ acts as a rescaling of the radius of the fifth compact dimension.
Let us summarize some of the main symplectic frames of the maximal four-dimensional theory.
\begin{itemize}
\item{\emph{The ${\rm SL}(8,\mathbb{R})$-frame.} It originates from dimensional reduction of eleven dimensional supergravity on a seven-torus, once all forms are dualized to lower-order ones \cite{Cremmer:1978ds} (see Sect. \ref{vohd} for a discussion on the issue of dualization). The electric group is $G_{el}={\rm SL}(8,\mathbb{R})$ with respect to which the vector fields transform in the ${\bf 28}'$. It contains the ${\rm GL}(7,\mathbb{R})$ group of the internal torus;}
\item{\emph{The $ {\rm E}_{6(6)}$-frame.} The block-diagonal component of the electric duality group $G_{el}$ in this case is ${\rm SO}(1,1)\times {\rm E}_{6(6)}$. As mentioned above, this is the symplectic frame which one obtains upon Kaluza-Klein reduction on a circle of the five-dimensional maximal supergravity;}
\item{\emph{The $ {\rm SU}^*(8)$-frame.}  Another basis considered in
the literature \cite{Hull:2002cv}, is the one in
which ${\rm G_e}={\rm SU}^\star (8)$, the group of $8\times8$ matrices
that are real up to a symplectic matrix. The electric vector fields transform in the ${\bf 28}'$ (real) representation of this group.}
\end{itemize}
These three frames are related by symplectic matrices which are not in ${\rm E}_{7(7)}$ and are given in \cite{deWit:2002vt}.\par
In Sect. \ref{N8MAB} we shall deal in some detail with the frames originating from toroidal reduction of Type IIA, IIB and M-theory.
\paragraph{The ${\rm SL}(8,\mathbb{R})$-frame.} The duality representation ${\Scr R}_v={\bf 56}$ of ${\rm E}_{7(7)}$ and its adjoint representation ${\bf 133}$ branch with respect to ${\rm SL}(8,\mathbb{R})$ as follows:
    \begin{align}
    {\bf 56}&\rightarrow {\bf 28}+{\bf 28}'\,,\nonumber \\
    {\bf 133}&\rightarrow {\bf 63}+{\bf 70}\,.\label{sl8branch}
    \end{align} Although the above branchings are the same as those with respect to ${\rm SU}(8)$, the ${\bf 28}'$ describing the electric vector fields $A^\Lambda_\mu$ and the ${\bf 28}$ describing their magnetic duals $A_{\Lambda\mu}$, are real. Denoting, only in this section,  by ${\tt a},\,{\tt b},\,{\tt c},\,{\tt d},\,\dots =1,\dots,\,8$ the indices labeling the fundamental representation of ${\rm SL}(8,\mathbb{R})$, the index $\Lambda$ of the ${\bf 28}$ is written as the  antisymmetric couple $\Lambda={\tt ab}$ and an ${\rm E}_{7(7)}$-generator in this basis has the general form:
    \begin{equation}
    \Pi\in \mathfrak{e}_{7(7)}\,\longrightarrow\,\,\,{\Scr R}^{({\rm SL}(8))}_v[\Pi]=\left(\begin{matrix} \Lambda^{{\tt a}{\tt b}}{}_{{\tt cd}} & \Sigma^{{\tt ab ef}}\cr \Sigma_{{\tt ghcd}} & \Lambda_{{\tt gh}}{}^{{\tt ef}}\end{matrix}\right)\,,\label{e77sl8}
    \end{equation}
    \begin{align}
\Pi\in \mathfrak{sl}(8)\equiv {\bf 28}\,\longrightarrow\,\,\,{\Scr R}^{({\rm SL}(8))}_v[\Pi]=\left(\begin{matrix} \Lambda^{{\tt a}{\tt c}}{}_{{\tt cd}} & {\bf 0}\cr {\bf 0} & \Lambda_{{\tt gh}}{}^{{\tt ef}}\end{matrix}\right)\,,\nonumber\\
\Pi\in {\bf 70}\,\longrightarrow\,\,\,{\Scr R}_v^{({\rm SL}(8))}[\Pi]=\left(\begin{matrix} {\bf 0}& \Sigma^{{\tt abef}}\cr \Sigma_{{\tt ghcd}} & {\bf 0}\end{matrix}\right)\,,
\end{align}
where $\Lambda^{{\tt a  b}}{}_{{\tt cd}} =4\,\delta^{[{\tt a}}_{[{\tt c}}\Lambda^{{\tt b}]}{}_{{\tt d}]}=-\Lambda_{{\tt cd}}{}^{{\tt ab}} $ is an ${\rm SL}(8,\mathbb{R})$-generator in the ${\bf 28}$-representation, and  $\Lambda^{{\tt a}}{}_{{\tt b}}=-\Lambda_{{\tt b}}{}^{{\tt a}}$ the corresponding generator in the fundamental one. The tensor $\Sigma^{{\tt abcd}}$ transforms in the ${\bf 70}$  of ${\rm SL}(8,\mathbb{R})$ and it is related to $\Sigma_{{\tt abcd}}$ by a duality condition analogous to (\ref{realityn82}):
\begin{equation}
\Sigma^{{\tt abcd}}=\frac{1}{24}\epsilon^{{\tt abcdefgh}}\,\Sigma_{{\tt efgh}}\,.\label{realityn822}
\end{equation}
In order to define the relation between this basis and the ${\rm SU}(8)$ one, associated with  the complex-representation $\underline{{\Scr R}}_v^c$, it is useful to branch the representations of ${\rm SU}(8)$ and of ${\rm SL}(8,\mathbb{R})$ with respect to the common subgroup ${\rm SO}(8)$. The adjoint representation of ${\rm SL}(8,\mathbb{R})$ branches with respect to ${\rm SO}(8)$ as follows:
\begin{equation}
{\bf 63}\rightarrow {\bf 28}+{\bf 35}_v\,,
\end{equation}
where the ${\bf 28}$ is the adjoint representation of ${\rm SO}(8)$, described by antisymmetric $8\times 8$ matrices $(\Lambda^{[{\bf 28}]})^{{\tt a}}{}_{{\tt b}}=-(\Lambda^{[{\bf 28}]})^{{\tt b}}{}_{{\tt a}}$, while ${\bf 35}_v$ is spanned by  symmetric $8\times 8$ traceless matrices $(\Lambda^{[{\bf 35}_v]})^{{\tt a}}{}_{{\tt b}}=(\Lambda^{[{\bf 35}_v]})^{{\tt b}}{}_{{\tt a}}$, so that, the $\Lambda$-block in (\ref{e77sl8}) splits accordingly: $\boldsymbol{\Lambda}=(\Lambda^{{\tt a  b}}{}_{{\tt cd}})=\boldsymbol{\Lambda}^{[{\bf 28}]}+\boldsymbol{\Lambda}^{[{\bf 35}_v]}$. The representation ${\bf 70}$ of $\Sigma^{{\tt abcd}}$ branches as follows:
\begin{equation}
{\bf 70}\rightarrow {\bf 35}_s+{\bf 35}_c\,,
\end{equation}
where ${\bf 35}_s$ and ${\bf 35}_c $ describe the anti-self-dual $\Sigma^{(-)\,{\tt abcd}}$ and the self-dual $\Sigma^{(+)\,{\tt abcd}}$ components of $\Sigma^{{\tt abcd}}$, respectively:
\begin{align}
\Sigma^{(\pm){\tt abcd}}&=\pm\frac{1}{24}\epsilon^{{\tt abcdefgh}}\,\Sigma^{(\pm){\tt efgh}}=\pm \Sigma_{(\pm){\tt abcd}}\,,\nonumber\\
\Sigma^{{\tt abcd}}&=\Sigma^{(+)\,{\tt abcd}}+\Sigma^{(-)\,{\tt abcd}}\,\,\,,\,\,\,\Sigma_{{\tt abcd}}=\Sigma_{(+)\,{\tt abcd}}+\Sigma_{(-)\,{\tt abcd}}=\Sigma^{(+)\,{\tt abcd}}-\Sigma^{(-)\,{\tt abcd}}\,.
\end{align}
The block-representation (\ref{e77sl8}) now reads:
    \begin{equation}
 {\Scr R}_v^{({\rm SL}(8))}[\Pi]=\left(\begin{matrix} \boldsymbol{\Lambda}^{[{\bf 28}]}+\boldsymbol{\Lambda}^{[{\bf 35}_v]} & \boldsymbol{\Sigma}^{(+)}+\boldsymbol{\Sigma}^{(-)}\cr \boldsymbol{\Sigma}^{(+)}-\boldsymbol{\Sigma}^{(-)} & \boldsymbol{\Lambda}^{[{\bf 28}]}-\boldsymbol{\Lambda}^{[{\bf 35}_v]}\end{matrix}\right)\,,\label{e77sl82}
    \end{equation}
    where $\boldsymbol{\Sigma}\equiv (\Sigma^{{\tt abcd}})$.
 Since the ${\rm SU}(8)$-generators are described by antisymmetric symplectic matrices, being the $\boldsymbol{\Sigma}$'s symmetric, the $\mathfrak{su}(8)$ subalgebra of $\mathfrak{e}_{7(7)}$ is defined by the blocks $\boldsymbol{\Lambda}^{[{\bf 28}]}$ and $\boldsymbol{\Sigma}^{(-)}$ only, corresponding to the branching of the ${\rm SU}(8)$-adjoint representation with respect to ${\rm SO}(8)$:
 \begin{equation}
{\bf 63}({\rm SU}(8))\rightarrow {\bf 28}+{\bf 35}_s\,,
\end{equation}
The coset space $\mathfrak{K}$ of symmetric symplectic matrices, on the contrary, is spanned by the remaining blocks  $\boldsymbol{\Lambda}^{[{\bf 35}_v]},\,\boldsymbol{\Sigma}^{(+)}$, consistently with the decomposition:
 \begin{equation}
{\bf 70}({\rm SU}(8))\rightarrow {\bf 35}_v+{\bf 35}_c\,.\label{70su8sl8}
\end{equation}
The relation between the (SUSY) indices $A,\,B,\,\dots$  and ${\tt a},\,{\tt b},\,\dots$ follows from the identification of the former with the indices of the spinorial representation ${\bf 8}_s$ of ${\rm SO}(8)$ and the latter with those of the fundamental one ${\bf 8}_v$. Denoting by $\frac{1}{2}(\Gamma^{{\tt a b}})_{A}{}^B$ the ${\rm SO}(8)$-generators in the spinorial ${\bf 8}_s$ representation, this tensor can be viewed as a $28\times 28 $ orthogonal matrix $\boldsymbol{\Gamma}\equiv (\frac{1}{2}\Gamma^{{\tt a b}}{}_{AB})$ in the two antisymmetric couples of indices $({\tt a b})$ and $(AB)$. The relation between $\underline{{\Scr R}}_v^c$ and ${\Scr R}^{({\rm SL}(8))}_v$ is:
\begin{equation}
\forall \Pi \in {\rm E}_{7(7)}\,\,;\,\,\,\,\,\underline{{\Scr R}}_v^c[\Pi]=\mathcal{S}^\dagger {\Scr R}^{({\rm SL}(8))}_v[\Pi]\mathcal{S}\,,
\end{equation}
where:
\begin{equation}
\mathcal{S}\equiv \left(\begin{matrix} \boldsymbol{\Gamma} & {\bf 0}\cr {\bf 0} & \boldsymbol{\Gamma}\end{matrix}\right)\mathcal{A}^\dagger\,,
\end{equation}
$\mathcal{A}$ being the Cayley matrix defined earlier. The above relation implies that two vectors $V^{\underline{M}}=(V^{AB},\,V_{AB})$ and $V^{M}=(V^{{\tt ab}},\,V_{{\tt ab}})$,  in the bases of the two representations, respectively, are connected as follows:
\begin{equation}
V_{AB}=\frac{1}{4\sqrt{2}}(V^{{\tt ab}}-i\,V_{{\tt ab}})(\Gamma^{{\tt a b}})_{AB}\,\,,\,\,\,\,V^{AB}=(V_{AB})^*\,.
\end{equation}

\paragraph{The $ {\rm E}_{6(6)}$-frame.}
With respect to ${\rm E}_{6(6)}\times {\rm SO}(1,1)$
the ${\bf 56}$ and the adjoint of ${\rm E}_{7(7)}$ decompose as:
\begin{eqnarray}
{\bf 56}&\rightarrow&   {\bf
1}_{-3}+{\bf 27}'_{-1}+{\bf 1}_{+3} +    {\bf 27}_{+1} \,,
\label{56e6}\\
{\bf 133}&\rightarrow&{\bf 27}_{-2}+ {\bf 1}_0+{\bf 78}_0+{\bf 27}'_{+2}\,.\label{133e6}
\end{eqnarray}
where the 28 vector fields $A^\Lambda_\mu$ naturally split into the Kaluza-Klein vector $A_\mu^0$ in the ${\bf 1}_{-3}$, originating from the five-dimensional metric, and the 27 vector fields $A_\mu^\lambda$, $\lambda=1,\dots,\,27$, of maximal five-dimensional theory, in the ${\bf 27}'_{-1}$  of the electric group. The ${\bf 78}_0$ in (\ref{133e6}) defines ${\rm E}_{6(6)}$-generators among the generators of ${\rm E}_{7(7)}$, while the ${\bf 27}'_{+2}$ are commuting isometries acting as translations on the 27 axionic scalars originating from the five-dimensional vector fields through the Kaluza-Klein reduction. These are the remnant in four dimensions of the vector gauge transformations in one dimension higher. Writing a covariant vector in the ${\bf 56}$ as $V_M=(V_0,\,V_\lambda,\,V^0,\,V^\lambda)$, the ${\Scr R}_{v*}$-representation of the generators of ${\rm E}_{7(7)}$ in this basis is:
\begin{align}
D&\in {\bf 1}_0\,\,,\,\,\,\,\,(D_M{}^N)=\left(\begin{matrix}3 & {\bf 0} & 0 & {\bf 0}\cr
{\bf 0}  &{\bf 1} &{\bf 0}  & {\bf 0} \cr 0 &{\bf 0} &-3 & {\bf 0} \cr {\bf 0}  &{\bf 0} &{\bf 0}  & -{\bf 1}\end{matrix}\right)\,,\nonumber\\
\mathcal{E}&\in {\bf 78}_0\,\,,\,\,\,\,\,(\mathcal{E}_M{}^N)=\left(\begin{matrix}0 & {\bf 0} & 0 & {\bf 0}\cr
{\bf 0}  &{\bf E} &{\bf 0}  & {\bf 0} \cr 0 &{\bf 0} &0 & {\bf 0} \cr {\bf 0}  &{\bf 0} &{\bf 0}  & -{\bf E}^T\end{matrix}\right)\,,\nonumber\\
a^\lambda\,t_\lambda&\in {\bf 27}'_{+2}\,\,,\,\,\,\,\,(a^\lambda\,(t_\lambda)_M{}^N)=\left(\begin{matrix}0 & {\bf a} & 0 & {\bf 0}\cr
{\bf 0}  &{\bf 0} &{\bf 0}  & {\bf d} \cr 0 &{\bf 0} &0 & {\bf 0} \cr {\bf 0}  &{\bf 0} &-{\bf a}^T & {\bf 0}\end{matrix}\right)\,,\nonumber\\
a_\lambda\,t^\lambda&\in {\bf 27}_{-2}\,\,,\,\,\,\,\,a_\lambda\,{\Scr R}_{v*}[t^\lambda]=a_\lambda\,{\Scr R}_{v*}[t_\lambda]^T\,,\label{e6frameg}
\end{align}
where ${\bf E}=({\bf E}_\lambda{}^\sigma)$ is the ${\rm E}_{6(6)}$-generator in the ${\bf 27}$ representation, ${\bf a}=(a^\lambda)$ and the dimension of each block is understood as following from the structure of the vector $V_M$ given above. In particular the $27\times 27$ matrix ${\bf d}=(d_{\lambda\sigma})$ is defined as follows: $d_{\lambda\sigma}\equiv d_{\lambda\sigma\gamma}\,a^\gamma$, $d_{\lambda\sigma\gamma}$ being the characteristic  totally symmetric ${\rm E}_{6(6)}$-invariant tensor defining in the $D=5$ parent action the Chern-Simons term: $d_{\lambda\sigma\gamma}F^\lambda\wedge F^\sigma\wedge A^\gamma$. From (\ref{e6frameg}) the reader can easily verify that:
$$[D,\,t^\lambda]=2\,t^\lambda\,\,,\,\,\,\,[t^\lambda,\,t^\sigma]=0\,.$$
The five-dimensional origin of the scalar fields can be made manifest by choosing the parametrization corresponding to the following definition of the coset representative:
\begin{equation}
L(\phi)=e^{a^\lambda\,t_\lambda}\,L_5(\hat{\phi})\,e^{\sigma\,D}=L(a^\Lambda,\hat{\phi},\sigma)\,,\label{Ld4d5}
\end{equation}
where $\hat{\phi}$ collectively denote the five-dimensional scalar fields spanning the symmetric coset space $\frac{{\rm E}_{6(6)}}{{\rm USp}(8)}$, and $L_5(\hat{\phi})$ the corresponding coset representative\footnote{$L_5(\hat{\phi})$, being an element of ${\rm E}_{6(6)}$ is also an element of ${\rm E}_{7(7)}$.}. The scalar $\sigma$ is related to the radius of the internal fifth dimension ($R=e^\sigma$) and $a^\lambda$ are the scalars originating from the components of the 27 five-dimensional vector fields  along the compact internal direction $a^\lambda\equiv A^\lambda_4$. The choice (\ref{Ld4d5}) corresponds to describing the scalar manifold of the theory as isometric to the following manifold (in the sense defined at the end of Sect. \ref{ghsect}):
\begin{equation}
\Mscal \sim \left[{\rm O}(1,1)\times\frac{{\rm E}_{6(6)}}{{\rm USp}(8)} \right]\ltimes \exp({\tt N}^{[{\bf 27}'_{+2}]})\,,
\end{equation}
where $\ltimes$ denotes the semi-direct product and ${\tt N}^{[{\bf 27}'_{+2}]}$ is the 27-dimensional space spanned by $t_\lambda$. If we choose for $L_5(\hat{\phi})$ the solvable parametrization, then (\ref{Ld4d5}) defines the solvable parametrization of $\Mscal$, with solvable Lie algebra ${\Scr S}$. Constructing the vielbein $\mathcal{P}$ out of $L(\phi)$ in (\ref{Ld4d5}), and using the property that the $t_\lambda$ are commuting, one immediately verifies that the scalars $a^\lambda$ only enter the metric ``covered'' by a derivative. As a consequence of this
the constant shifts $a^\lambda\rightarrow a^\lambda+c^\lambda$ are isometries. They are implemented by the ${\rm E}_{7(7)}$-transformation ${\bf g}=e^{c^\lambda\,t_\lambda}$:
\begin{equation}
{\bf g}\,L(a^\Lambda,\hat{\phi},\sigma)=L(a^\Lambda+c^\lambda,\hat{\phi},\sigma)\,.
\end{equation}
These shifts are examples of Peccei-Quinn transformations and the scalars $a^\lambda$ are dubbed Peccei-Quinn scalars. The Abelian nilpotent space ${\tt N}^{[{\bf 27}'_{+2}]}$ is a maximal Abelian subalgebra of $\mathfrak{e}_{7(7)}$ of maximal dimension, and is also a maximal Abelian ideal of the solvable subalgebra ${\Scr S}$, see final paragraph of Sect. \ref{ghsect}.\par As mentioned earlier, the shift symmetries $a^\lambda\rightarrow a^\lambda+c^\lambda$ originate from the gauge invariance of the vector fields in five dimensions: $A^\lambda_{\hat{\mu}}\rightarrow A^\lambda_{\hat{\mu}}+\partial_{\hat{\mu}}\zeta^\lambda$, $\hat{\mu}=0,\dots,4$. It corresponds to choosing the gauge parameter $\zeta^\lambda=c^\lambda\,y$, where $y=x^4$ is the fifth compact coordinate.
\paragraph{The $ {\rm SU}^*(8)$-frame.} Recall that the ${\rm SU}^*(2n)$ group is defined as the group of $2n\times 2n $ complex matrices $U$ satisfying the condition:
\begin{equation}
UJ=JU^*\,,\label{sust}
\end{equation}
where $U^*$ is the complex conjugate of $U$ and $J$ is an antisymmetric matrix such that $J^2=-{\bf 1}_{2n}$. The maximal compact subgroup consists of unitary matrices in ${\rm SU}^*(2n)$. Using the unitarity condition $U^{-1}=U^\dagger$, Eq. (\ref{sust}) becomes $UJU^T=J$, namely a compact matrix $U$ in ${\rm SU}^*(2n)$ is unitary and symplectic, that is the maximal compact subgroup of ${\rm SU}^*(2n)$ is ${\rm USp}(2n)$. The group ${\rm SO}^*(2n)$ on the other hand, is defined as the subgroup of orthogonal matrices in ${\rm SU}^*(2n)$: ${\rm SO}^*(2n)={\rm SU}^*(2n)\bigcap {\rm SO}(2n,\mathbb{C})$. The orthogonality condition $U^{-1}=U^T$ allows to rewrite  Eq. (\ref{sust}) in the form: $UJU^\dagger=J$. The maximal compact subgroup of ${\rm SO}^*(2n)$ is ${\rm U}(n)$. Let ${\tt a}',\,{\tt b}'=1,\dots, 2n$ label the fundamental representation of ${\rm SU}^*(2n)$. Using the matrix $J=(J^{{\tt a}'{\tt b}'})$ one can define a consistent reality condition on rank-2 antisymmetric tensors in the ${\bf 2n}\wedge {\bf 2n}$ representation: $$T^{{\tt a}'{\tt b}'}\equiv (T_{{\tt a}'{\tt b}'})^*=J^{{\tt a}'{\tt c}'}J^{{\tt b}'{\tt d}'}T_{{\tt c}'{\tt d}'}\,.$$
The generators  of $\mathfrak{so}^*(2n)$ can be characterized as the $\mathfrak{so}(2n,\mathbb{C})$ generators $\boldsymbol{\Lambda}_{{{\tt a}'{\tt b}'}}=-\boldsymbol{\Lambda}_{{{\tt b}'{\tt a}'}}$ satisfying the above reality condition.
The representation ${\bf 28}'={\bf 8}'\wedge {\bf 8}'$ and its conjugate ${\bf 28}$ of ${\rm SU}^*(8)$ are therefore real and thus can consistently describe the electric and magnetic vector fields $A^{{\tt a}'{\tt b}'}_\mu,\,A_{{\tt a}'{\tt b}'\,\mu}$.

\subsubsection{Parity}\label{party} In the ${\rm SL}(8,\mathbb{R})$-frame one easily verifies that the $56\times 56$ anti-symplectic matrix $${\bf P}=\left(\begin{matrix}{\bf 1} & {\bf 0}\cr {\bf 0}& -{\bf 1}  \end{matrix}\right)\,,$$
is an automorphism of $\mathfrak{e}_{7(7)}$.
Its effect on a generic  ${\rm E}_{7(7)}$ generator $\Pi$ is:
\begin{equation}
{\bf P}^{-1}{\Scr R}_v^{({\rm SL}(8))}[\Pi]{\bf P}=\left(\begin{matrix} \boldsymbol{\Lambda}^{[{\bf 28}]}+\boldsymbol{\Lambda}^{[{\bf 35}_v]} & -\boldsymbol{\Sigma}^{(+)}-\boldsymbol{\Sigma}^{(-)}\cr -\boldsymbol{\Sigma}^{(+)}+\boldsymbol{\Sigma}^{(-)} & \boldsymbol{\Lambda}^{[{\bf 28}]}-\boldsymbol{\Lambda}^{[{\bf 35}_v]}\end{matrix}\right)\,,\label{parite7}
\end{equation}
namely it switches the sign of the blocks $\boldsymbol{\Sigma}^{(+)},\,\boldsymbol{\Sigma}^{(-)}$, the latter being associated with ${\rm SU}(8)$ generators. We can adopt the ${\rm SU}(8)$-covariant parametrization of the scalar manifold (\ref{Hcovpar}) and write, in light of our previous discussion, and in particular of Eq. (\ref{70su8sl8}),
\begin{equation}
\mathbb{L}(\phi)=\exp(\phi^{{\tt ab}}_v\,K^{(v)}_{{\tt ab}}+\phi^{{\tt abcd}}_c\,K^{(+)}_{{\tt abcd}})\,,
\end{equation}
where $(K^{(v)}_{{\tt ab}})$ is a basis of ${\bf 35}_v$, while $(K^{(+)}_{{\tt abcd}})$ is a  basis of ${\bf 35}_c$, so that $\phi^{{\tt ab}}_v\,K^{(v)}_{{\tt ab}}$ only contains the  $\boldsymbol{\Lambda}^{[{\bf 35}_v]}$ blocks and $\phi^{{\tt abcd}}_c\,K^{(+)}_{{\tt abcd}}$ the $\boldsymbol{\Sigma}^{(+)}$ ones. Eq. (\ref{parite7})  shows that the action of ${\bf P}$ induces the following transformation on the scalar fields:
\begin{equation}
p:\,\,\begin{cases}\phi^{{\tt abcd}}_c\rightarrow\,-\phi^{{\tt abcd}}_c\cr \phi^{{\tt ab}}_v\rightarrow\,\phi^{{\tt ab}}_v\end{cases}\,.\label{phipp}
\end{equation}
This transformation is an isometry of the target-space metric. Being ${\bf P}$ anti-symplectic,  by our discussion at the end of Sect. \ref{gsg}, it describes a \emph{parity transformation}, since it can be promoted to a global symmetry of the theory provided it is combined with a parity transformation on the spatial coordinates. In this respect $\phi^{{\tt ab}}_v$ are proper scalars while $\phi^{{\tt abcd}}_c$ are \emph{pseudo-scalars}.\par
In the $ {\rm E}_{6(6)}$-frame the parity transformation ${\bf P}$ has the following form:
\begin{equation}
{\bf P}=\left(\begin{matrix}-1 & {\bf 0} & 0 & {\bf 0}\cr
{\bf 0}  &{\bf 1} &{\bf 0}  & {\bf 0} \cr 0 &{\bf 0} &1 & {\bf 0} \cr {\bf 0}  &{\bf 0} &{\bf 0}  & -{\bf 1}\end{matrix}\right)\,,
\end{equation}
where, as usual, the splitting of rows and columns is $(1,27,1,27)$. The reader can easily verify that
\begin{equation}
{\bf P}D{\bf P}^{-1}=D\,\,;\,\,\,\,{\bf P}\mathcal{E}{\bf P}^{-1}=\mathcal{E}\,\,;\,\,\,\,{\bf P}t_\lambda{\bf P}^{-1}=-t_\lambda\,\,;\,\,\,\,{\bf P}t^\lambda{\bf P}^{-1}=-t^\lambda\,,
\end{equation}
so that the five-dimensional scalars $\hat{\phi}$ and the modulus $\sigma$ are proper scalars, while $a^\lambda$ are \emph{pseudo-scalars}: $p: a^\lambda\rightarrow -a^\lambda$. Notice that in this frame  the vector $A^0_\mu$ has a different intrinsic parity than the other 27: $\eta_P^0=-1=-\eta_P^\lambda$.\par
In all the cases ${\bf P}$ is an automorphism of $\mathfrak{e}_{7(7)}$:
\begin{equation}
{\bf P}^{-1}\mathfrak{e}_{7(7)}{\bf P}=\mathfrak{e}_{7(7)}\,.
\end{equation}
In fact ${\bf P}$ is an \emph{outer-automorphism} of $\mathfrak{e}_{7(7)}$, since if there existed an element ${\bf g}$ of ${\rm E }_{7(7)}$ whose adjoint action  of $\mathfrak{e}_{7(7)}$ had the same effect as ${\bf P}$, ${\bf P}{\Scr R}_v[{\bf g}]^{-1}$ would commute with all $\mathfrak{e}_{7(7)}$-generators in the ${\Scr R}_v$-representation. By Shur's lemma, being ${\Scr R}_v$ irreducible, ${\Scr R}_v[{\bf g}]$ and ${\bf P}$ would be proportional, which cannot be since ${\bf P}$ is anti-symplectic while  ${\Scr R}_v[{\bf g}]$ is symplectic.
\paragraph{Triality.} Let us mention that there is a $\mathbb{Z}_2$-subgroup of ${\rm E }_{7(7)}$ which is also contained in the  triality symmetry group of the $\mathfrak{so}(8)$-subalgebra of $\mathfrak{e}_{7(7)}$.\footnote{In fact this triality symmetry defines the outer automorphisms of $\mathfrak{so}(8)$ which are inner with respect to $\mathfrak{e}_{7}$ over the complex numbers \cite{Minchenko}. Only a $\mathbb{Z}_2$-subgroup of it however survives within the real form $\mathfrak{e}_{7(7)}$.}

\subsection{M-theory, Type IIA and IIB Descriptions.}\label{N8MAB}
Let us briefly recall the field contents of $D=11$ supergravity and of Type IIA/IIB theories. The former describes a gravitational multiplet whose bosonic content consists of the $D=11$ graviton field $g_{\hat{\mu}\hat{\nu}}$, $\hat{\mu},\,\hat{\nu}=0,\dots, 10$, and an antisymmetric rank-3 field $A_{\hat{\mu}\hat{\nu}\hat{\rho}}$ (128 on-shell bosonic degrees of freedom).\footnote{For the sake of simplicity, we shall refer to antisymmetric rank-p tensors $\omega_{\hat{\mu}_1\dots \hat{\mu}_p}$ also as p-forms $\omega^{(p)}\equiv \frac{1}{p!}\omega_{\hat{\mu}_1\dots \hat{\mu}_p}\,dx^{\hat{\mu}_1}\wedge\dots \wedge dx^{\hat{\mu}_p}$.} The fermionic sector consists of a single gravitino field (128 on-shell fermionic degrees of freedom).\par  The bosonic sector of Type IIA supergravity, low-energy limit of the corresponding ten-dimensional superstring theory, consists of the graviton $g_{\hat{\mu}\hat{\nu}}$, $\hat{\mu},\,\hat{\nu}=0,\dots, 9$, a scalar dilaton field $\phi$ and a rank-2 antisymmetric tensor field $B_{\hat{\mu}\hat{\nu}}$ (the Kalb-Ramond $B$-field) which define the Neveu-Schwarz-Neveu-Schwarz (NS-NS) sector, together with a vector field $C_{\hat{\mu}}$ and an antisymmetric rank-3 tensor field $C_{\hat{\mu}\hat{\nu}\hat{\rho}}$ which make up the Ramond-Ramond (RR) sector. The fermionic sector consists of two Majorana-Weyl gravitinos with opposite chiralities and two \emph{dilatinos} (also having opposite chiralities). As mentioned in the Introduction, the Type IIA supergravity is obtained from the eleven-dimensional one upon compactification on a circle. The radius of the circle being related to the dilaton field $\phi$.\par The bosonic sector of Type IIB supergravity, on the other hand, differs from the Type IIA one only in the RR-fields which consist of an axionic scalar field $\rho$, a rank-2 and a rank-4 antisymmetric tensor fields $C_{\hat{\mu}\hat{\nu}},\,C_{\hat{\mu}\hat{\nu}\hat{\rho}\hat{\sigma}}$. The field strength of the latter is a self-dual five-form $F^{(5)}$: $F^{(5)}={}^*F^{(5)}$. The fermionic sector of Type IIB theory consists of two gravitinos with equal chirality and two dilatinos with the same property. This classical theory features a non-perturbative global ${\rm SL}(2,\mathbb{R})$ symmetry, to be denoted by ${\rm SL}(2,\mathbb{R})_{{\rm IIB}}$. Its restriction to the integers, ${\rm SL}(2,\mathbb{Z})_{{\rm IIB}}$, is a symmetry of the whole string theory and is known as \emph{$S$-duality}. The two scalar fields $\phi,\,\rho$ span the manifold ${\rm SL}(2,\mathbb{R})_{{\rm IIB}}/{\rm SO}(2)$, while the two 2-forms $B_{\hat{\mu}\hat{\nu}},\,C_{\hat{\mu}\hat{\nu}}$ transform under its action as a doublet. The RR 4-form $C_{\hat{\mu}\hat{\nu}\hat{\rho}\hat{\sigma}}$ is an ${\rm SL}(2,\mathbb{R})_{{\rm IIB}}$-singlet. In what follows, in dealing with the dimensional reduction on a torus to four dimensions of the above theories, we shall split the $D=11$ and $D=10$ coordinates as follows:
\begin{align}
D=11&:\,\,\,(x^{\hat{\mu}})=(x^\mu,\,x^{\upalpha})\,\,,\,\,\,\,\,\mu=0,1,2,3\,\,,\,\,\,\,\upalpha=4,\dots, 10\,,\nonumber\\
D=10&:\,\,\,(x^{\hat{\mu}})=(x^\mu,\,x^{{\tt u}})\,\,,\,\,\,\,\,\mu=0,1,2,3\,\,,\,\,\,\,{\tt u}=4,\dots, 9\,,\nonumber
\end{align}
$x^{\upalpha}$ and $x^{{\tt u}}$ being the compact coordinates on $T^7$ and $T^6$, respectively. We now introduce few notions about toroidal reductions just for the sake of making contact with our present discussion of the maximal theory. We refer the reader to Sect. \ref{toroidalc} for a more detailed discussion.
\par
When considering toroidal compactifications of string theory, $G_{int}$ is enlarged to ${\rm O}(n,n)$ which acts transitively on the fields originating from the ten-dimensional metric and the $B$-field and contains the T-duality group ${\rm O}(n,n;\,\mathbb{Z})$ \cite{Buscher:1987sk,Giveon:1994fu}. With an abuse of terminology we shall refer to the continuous group ${\rm O}(n,n)$ as the T-duality group, bearing in mind that the actual T-duality group is defined over the integers. As mentioned in the Introduction, this group ${\rm O}(n,n)$ does not leave the RR-sectors of (dimensionally reduced) Type IIA or Type IIB theories invariant, but only ${\rm SO}(n,n)$ does. Transformations in ${\rm O}(n,n)/{\rm SO}(n,n)$ will map dimensionally reduced Type IIA RR-fields into the Type IIB ones. This is reflected in the fact that in the four-dimensional maximal model, only the ${\rm SO}(6,\,6)$ part of the T-duality group is contained in ${\rm E}_{7(7)}$. In fact the full four-dimensional model is \emph{chiral} with respect to ${\rm SO}(6,\,6)$ in the sense that, depending on wether it originates from Type IIA or Type IIB theories, the relevant ${\rm E}_{7(7)}$ representations branch  with respect to the maximal ${\rm SL}(2,\mathbb{R})\times {\rm SO}(6,\,6)$ subgroup of ${\rm E}_{7(7)}$ as follows:
\begin{align}
\mbox{Type IIA}&:\nonumber\\
&{\bf 133} {\longrightarrow}\,\,\,{\bf (3,1)}+{\bf (1,66)}+{\bf (2,32_c)}\,,
\nonumber\\
&{\bf 56} {\longrightarrow}\,\,\,{\bf (2,12)}+{\bf (1,32_s)}\,,\nonumber\\
\mbox{Type IIB}&:\nonumber\\
&{\bf 133} {\longrightarrow}\,\,\,{\bf (3,1)}+{\bf (1,66)}+{\bf (2,32_s)}\,,
\nonumber\\
&{\bf 56} {\longrightarrow}\,\,\,{\bf (2,12)}+{\bf (1,32_c)}\,,\label{IIABdecs}
\end{align}
where ${\bf 32_s}$ and ${\bf 32_c}$ are the two chiral spinorial representations of ${\rm Spin}(6,6)$, which describe the fields originating from the RR-forms in ten dimensions. They are are  mapped into one another by the outer automorphisms of the $\mathfrak{so}(6,6)$ algebra, in ${\rm O}(6,\,6)/{\rm SO}(6,\,6)$, describing the action of T-duality along an odd number of internal directions \cite{Bertolini:1999uz}. The fields originating from the NS-NS sector, on the other hand, are associated with tensor representations of ${\rm O}(6,\,6)$ and are the same in the Type IIA and IIB pictures.\par  From the branching of the adjoint representation ${\bf 133}$ one can then infer the ten-dimensional origin of the scalars, as deriving from RR or NS-NS fields. This is done by decomposing the solvable Lie algebra ${\Scr S}$ associated with ${\Scr M}_{scal}$ with respect to the  solvable Lie algebra ${\Scr S}'={\Scr S}_{\mbox{{\tiny ${\rm SL}(2)$}}}\oplus {\Scr S}_{{\tiny {\rm SO}(6,6)}}$ associated with the submanifold \cite{Andrianopoli:1996bq}:
\begin{equation}
\frac{{\rm SL}(2,\mathbb{R})}{{\rm SO}(2)}\times \frac{{\rm SO}(6,\,6)}{{\rm SO}(6)\times {\rm SO}(6)}\subset\,\frac{{\rm E}_{7(7)}}{{\rm SU}(8)/\mathbb{Z}_2}\,,\label{subsl2so66}
\end{equation}
where we define the following solvable parametrizations
\begin{equation}
\frac{{\rm SL}(2,\mathbb{R})}{{\rm SO}(2)}\sim \exp({\Scr S}_{\mbox{{\tiny ${\rm SL}(2)$}}})\,\,;\,\,\,\,\,\frac{{\rm SO}(6,\,6)}{{\rm SO}(6)\times {\rm SO}(6)}\sim \exp({\Scr S}_{{\tiny {\rm SO}(6,6)}})\,.
\end{equation}
The former manifold is parametrized by the four-dimensional dilaton $\phi_4$, which is a $T$-duality invariant combination of the ten dimensional dilaton and the volume of the internal manifold in the string frame (see for instance \cite{Polc}):
$$\phi_4\equiv\phi-\frac{1}{2}\,\log({\rm Vol}[M_{int}])\,,$$ and the scalar $\tilde{B}$ dual to the 2-form $B_{\mu\nu}$. The latter is parametrized by the moduli $g_{{\tt uv}},\,B_{{\tt uv}}$, ${\tt u,v}=4,\dots,9$ describing the internal components of metric and of the $B$-field. The solvable Lie algebra decomposition
reads:
\begin{equation}
{\Scr S}=({\Scr S}_{{\tiny {\rm SL}(2)}}\oplus {\Scr S}_{{\tiny {\rm SO}(6,6)}})\oplus_s\,{\tt N}^{[{\bf 32}]}\,,
\end{equation}
where $\oplus_s$ denotes the \emph{semidirect sum} and the 32-dimensional nilpotent space ${\tt N}^{[{\bf 32}]}$ is parametrized by the scalar fields $\phi^s_{RR}$ originating from the RR-sector (${\bf 32}={\bf 32_c}$ and ${\bf 32}={\bf 32_s}$ in the Type IIA and IIB pictures, respectively).
The above decomposition corresponds to writing the following isometric mapping
\begin{equation}
{\Scr M}_{scal}\sim \left[\frac{{\rm SL}(2,\mathbb{R})}{{\rm SO}(2)}\times \frac{{\rm SO}(6,\,6)}{{\rm SO}(6)\times {\rm SO}(6)}\right]\ltimes \exp({\tt N}^{[{\bf 32}]})\,,
\end{equation}
that is choosing the coset representative $L(\phi^s)$ of the following form:
\begin{equation}
L(\phi^s)=L(\phi_{RR}^s)L(\phi_{NS}^s)\,,
\end{equation}
where $L(\phi_{RR}^s)\in \exp({\tt N}^{[{\bf 32}]})$ and $L(\phi_{NS}^s)$ describes the submanifold (\ref{subsl2so66}) spanned by the NS-NS scalars.\par
\begin{figure}[H]
\begin{center}
\centerline{\includegraphics[width=0.8\textwidth]{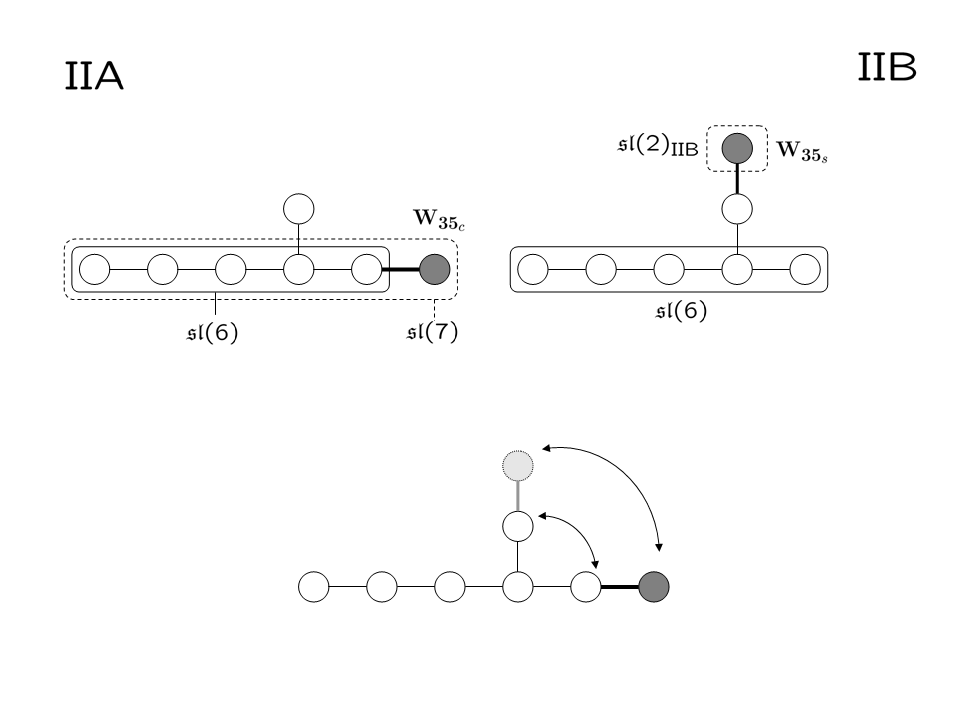}}
 \caption{\small Dynkin diagram of $\mathfrak{e}_{7}$: The diagram in the upper-left corner corresponds to the Type IIA picture, that in the upper-right corner to the Type IIB one. The empty circles define the $D_6$ subalgebra, while the continuous lines enclose the subdiagrams of the two inequivalent $\mathfrak{sl}(6)$ subalgebras.  The two pictures are related by an outer automorphism of $D_6$ subalgebra (lower figure).}\label{fige7}
\end{center}
\end{figure}

A more detailed characterization of the ten-dimensional origin of the four-dimensional fields is effected by further decomposing the ${\rm SO}(6,6)$-representations in (\ref{IIABdecs}) with respect to the ${\rm GL}(6,\mathbb{R})$ group of the six-dimensional internal torus. There are two inequivalent $\mathfrak{ sl}(6,\mathbb{R})$ algebras within $\mathfrak{e}_{7(7)}$. To understand the corresponding embeddings inside the global symmetry algebra, let us recall few general facts about the algebra $\mathfrak{e}_{7}$ over the complex numbers (a basic knowledge of Lie algebras is required at this point, see for instance \cite{Helgason}). Simple complex Lie algebras are totally characterized by a Dynkin diagram.
\begin{figure}[H]
\begin{center}
\centerline{\includegraphics[width=0.6\textwidth]{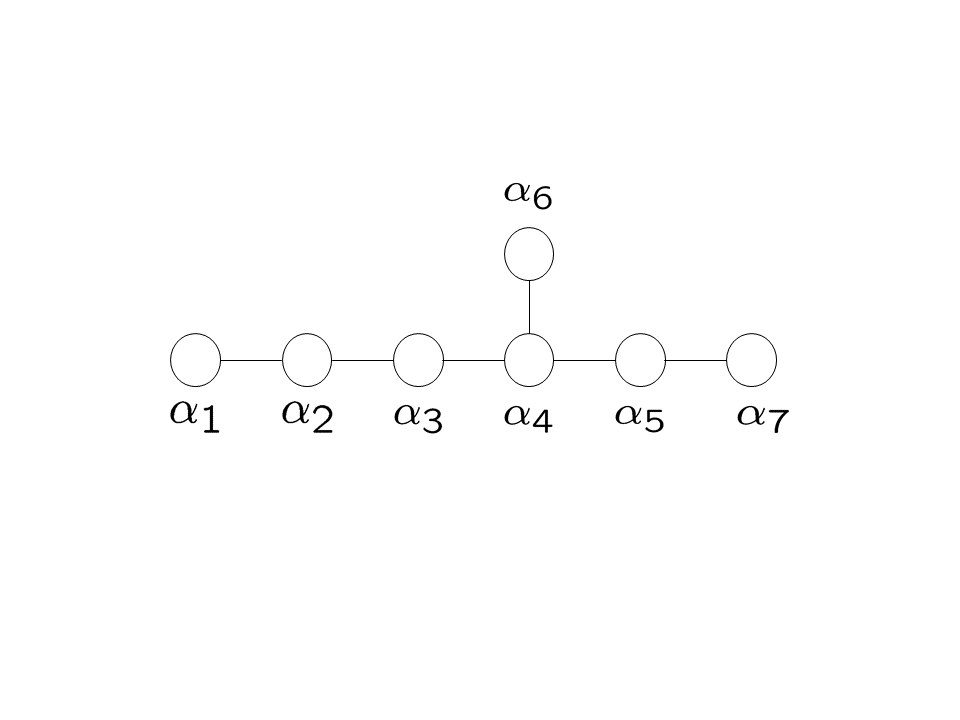}}
 \caption{\small The Dynkin diagram of the complex Lie algebra $\mathfrak{e}_7$.}\label{fige70}
\end{center}
\end{figure}
The diagram of $\mathfrak{e}_{7}$, see Fig. \ref{fige70}, can be obtained from that of $\mathfrak{so}(12,\mathbb{C})$ (also denoted by $D_6$ and describing, once projected to the real numbers, the algebra of the ${\rm SO}(6,6)$-subgroup) by attaching
to one of the two symmetric legs the weight ${\bf W}_{{\bf 32}}$ corresponding to the chiral spinorial representation ${\bf 32}$.
We can either attach ${\bf W}_{{\bf 32}_c}$ to one leg or ${\bf W}_{{\bf 32}_s}$ to the other, the two possibilities corresponding to the Type IIA or Type IIB pictures and being related by the outer automorphism of the $D_6$ subalgebra, see Fig. \ref{fige7}.
The embeddings of the two inequivalent $\mathfrak{ sl}(6)$ subalgebras is illustrated in the same Figure, where they are represented as a same algebra within two different constructions of the $\mathfrak{e}_7$-one.   We see that in the Type IIB picture the  $\mathfrak{ sl}(6)$ algebra, associated with the internal torus, commutes with the $\mathfrak{sl}(2)$ subalgebra defined by the root ${\rm W}_{{\bf 32}_s}$. The latter generates the global symmetry group ${\rm SL}(2,\mathbb{R})_{{\rm IIB}}$ of Type IIB supergravity at the classical level. In fact, in the Type IIB picture, the ${\rm SL}(6,\mathbb{R})$ subgroup of the torus commutes inside ${\rm E}_{7(7)}$ with a larger group: ${\rm SL}(3,\mathbb{R})$. On the other hand, in the type IIA picture, the ${\rm SL}(6,\mathbb{R})$ subgroup of the torus only commutes with a ${\rm GL}(2,\mathbb{R})$ group. Its diagram, however, can be extended to an $\mathfrak{sl}(7)$ one by including the ${\rm W}_{{\bf 32}_c}$-root. The corresponding ${\rm SL}(7,\mathbb{R})$
is associated with the internal seven-torus in the description of the four-dimensional theory as deriving from eleven dimensions. We shall return on this issue later in Sect. \ref{toroidalc} when dealing with toroidal reductions.\par
If we view the four-dimensional theory as originating from the Type IIB theory compactified on a six-torus, the ten-dimensional origin of the fields is then derived by branching the relevant ${\rm E}_{7(7)}$-representations with respect to the characteristic subgroup (\ref{GDGint}) $ {\rm SL}(6,\mathbb{R})\times{\rm SL}(2,\mathbb{R})_{{\rm IIB}}\times{\rm SO}(1,1)$ contained in the maximal subgroup  $ {\rm SL}(6,\mathbb{R})\times{\rm SL}(3,\mathbb{R})$ of ${\rm E}_{7(7)}$:
\begin{align}
 \label{eq:1IIB}
{\bf  56} \rightarrow & ({\bf 6}', {\bf 1})_{-2} + ({\bf 6}, {\bf
  2})_{-1}
+ ({\bf 20}, {\bf 1})_0 + ({\bf 6}', {\bf 2})_{+1} +({\bf 6},{\bf
  1})_{+2} \;.\\
{\bf 133}\rightarrow &({\bf 1},{\bf 2})_{-3}+({\bf 15},{\bf 1})_{-2}+({\bf 15}',{\bf 2})_{-1}+({\bf 1}+{\bf 35},{\bf 1})_{0}+({\bf 1},{\bf 3})_{0}+\nonumber\\
&+({\bf 15},{\bf 2})_{+1}+({\bf 15}',{\bf 1})_{+2}+({\bf 1},{\bf 2})_{+3} \label{eq:2IIB}
\end{align}
The gradings are referred to the ${\rm SO}(1,1)$-factor acting as a rescaling of the internal metric, $g_{{\tt uv}}\rightarrow\,e^{2\lambda}\,g_{{\tt uv}}$. A generic field $\Phi$ is associated with a grading $k$ such that the transformation $\Phi\rightarrow e^{2k\lambda}\,\Phi$, combined with the corresponding transformation of $g_{{\tt uv}}$, is a symmetry of the action.
In (\ref{eq:1IIB}) $({\bf 6}', {\bf 1})_{-2}$ describes the six Kaluza-Klein vectors $G^{{\tt u}}_\mu$, ${\tt u}=4,\dots, 9$, originating from the ten-dimensional metric, $({\bf 6}, {\bf
  2})_{-1}$ the couple of vectors $(C_{{\tt u}\mu},\,B_{{\tt u}\,\mu})$ from the NS-NS $B$-field and the RR 2-form, respectively, and
  $({\bf 20}, {\bf 1})_0$ describes the vectors $C_{\mu\,{\tt uvw}}$ originating from the 4-form RR field. Of the latter, only half are electric vector fields, and are defined by the field strengths which are self-dual with respect to the Hodge-duality operation in the  six internal directions.\par The identification of the scalar fields proceeds through the solvable Lie algebra decomposition:
\begin{align}
{\Scr S}={\Scr S}_{{\tiny {\rm GL}(6)}}\oplus {\Scr S}_{{\tiny {\rm SL}(2)_{{\rm IIB}}}}\oplus {\tt N}^{{\tiny [({\bf 15},{\bf 2})_{+1}]}}\oplus {\tt N}^{{\tiny [({\bf 15}',{\bf 1})_{+2}]}}\oplus {\tt N}^{{\tiny [({\bf 1},{\bf 2})_{+3}]}}\,,
\end{align}
where ${\Scr S}_{{\tiny {\rm GL}(6)}}$ and  ${\Scr S}_{{\tiny {\rm SL}(2)_{{\rm IIB}}}}$ define the solvable parametrization of the submanifolds ${\rm GL}(6,\mathbb{R})/{\rm SO}(6)$ and ${\rm SL}(2,\mathbb{R})_{{\rm IIB}}/{\rm SO}(2)$, respectively. The algebra
${\Scr S}_{{\tiny {\rm SL}(2)_{{\rm IIB}}}}$ is parametrized by the ten-dimensional dilaton $\phi$ and the axion $\rho$,
 ${\Scr S}_{{\tiny {\rm GL}(6)}}$ by the moduli $g_{{\tt uv}}$ of the internal metric, see Eq. (\ref{GinGL}), ${\tt N}^{{\tiny [({\bf 15},{\bf 2})_{+1}]}}$ by the internal components $B_{{\tt uv}},\,C_{{\tt uv}}$ of the $B$-field and the RR 2-form, respectively, ${\tt N}^{{\tiny [({\bf 15}',{\bf 1})_{+2}]}}$ by the internal components $C_{{\tt uvwt}}$ of the 4-form RR field and ${\tt N}^{{\tiny [({\bf 1},{\bf 2})_{+3}]}}$ by the two scalar fields dual to the NS-NS and RR 2-forms $B_{\mu\nu},\,C_{\mu\nu}$. The nilpotent isometries generated by ${\tt N}^{{\tiny [({\bf 15},{\bf 2})_{+1}]}},\,{\tt N}^{{\tiny [({\bf 15}',{\bf 1})_{+2}]}},\,{\tt N}^{{\tiny [({\bf 1},{\bf 2})_{+3}]}}$ are all associated with the scalar fields originating from higher-rank antisymmetric tensor fields in ten dimensions. In fact the corresponding global symmetries in of the four-dimensional theory are the remnant of the gauge symmetries related to these fields.\par
In the Type IIA picture, the branchings are effected with respect to the subgroup (\ref{GDGint}) which has the form ${\rm SO}(1,1)\times {\rm GL}(6,\mathbb{R})$, since in this case the global symmetry group of the ten-dimensional theory is just ${\rm SO}(1,1)$ and not ${\rm SL}(2,\mathbb{R})$ as in the Type IIB case. Let us give these branchings only specifying the grading with respect to the ${\rm SO}(1,1)$-factor acting as a rescaling of the internal metric:
\begin{align}
 \label{eq:1IIA}
{\bf  56} \rightarrow &   {\bf 6}'_{-2}+{\bf 1}_{-\frac{3}{2}}
+ {\bf 6}_{-1}+ {\bf 15}_{-\frac{1}{2}} + {\bf 15}'_{+\frac{1}{2}}  + {\bf 6}'_{+1}+{\bf 1}_{+\frac{3}{2}}+{\bf 6}_{+2} \;.\\
{\bf 133}\rightarrow &{\bf 1}_{-3}+{\bf 6}_{-\frac{5}{2}}+{\bf 20}_{-\frac{3}{2}}+{\bf 15}'_{-1}+{\bf 6}'_{-\frac{1}{2}}+({\bf 35}+{\bf 1}+{\bf 1})_{0}+\nonumber\\
&+{\bf 6}_{+\frac{1}{2}}+{\bf 15}_{+1}+{\bf 20}_{+\frac{3}{2}}+{\bf 6}'_{+\frac{5}{2}}+{\bf 1}_{+3}\,. \label{eq:2IIA}
\end{align}
The representation ${\bf 6}'_{-2}$ in (\ref{eq:1IIA}) describes the Kaluza-Klein vectors $G^{{\tt u}}_\mu$, ${\bf 1}_{-\frac{3}{2}}$ the vector $C_\mu$ originating from the RR 1-form in ten dimensions, $ {\bf 6}_{-1}$ and ${\bf 15}_{-\frac{1}{2}}$ the vectors $B_{{\tt u} \mu }$ and $C_{\mu\,{\tt uv}}$ from the NS-NS $B$-field and the RR 3-form, respectively.
Just as in the previous cases, the identification of the scalar fields proceeds through the solvable Lie algebra decomposition deduced from Eq. (\ref{eq:2IIA}):
\begin{align}
{\Scr S}={\Scr S}_{{\tiny {\rm GL}(6)}}\oplus {\Scr S}_{{\tiny {\rm SO}(1,1)}}\oplus {\tt N}^{{\tiny \left[{\bf 6}_{+1/2}\right]}} \oplus {\tt N}^{{\tiny \left[{\bf 15}_{+1}\right]}}\oplus {\tt N}^{{\tiny \left[{\bf 20}_{+3/2}\right]}}\oplus {\tt N}^{{\tiny \left[{\bf 6}'_{+5/2}\right]}}\oplus {\tt N}^{{\tiny \left[{\bf 1}_{+3}\right]}}\,,
\end{align}
where, as usual, ${\Scr S}_{{\tiny {\rm GL}(6)}}$ is parametrized by the moduli $g_{{\tt uv}}$ of the internal metric, ${\Scr S}_{{\tiny {\rm SO}(1,1)}}$ by the ten-dimensional dilaton field, ${\tt N}^{{\tiny \left[{\bf 6}_{+1/2}\right]}}$ by the six scalar fields originating form the RR 1-form, ${\tt N}^{{\tiny \left[{\bf 20}_{+3/2}\right]}}$ by the scalars $C_{{\tt uvw}}$ from the RR 3-form, ${\tt N}^{{\tiny \left[{\bf 15}_{+1}\right]}}$ by the scalars $B_{{\tt uv}}$ from the $B$-field and ${\tt N}^{{\tiny \left[{\bf 6}'_{+5/2}\right]}},\, {\tt N}^{{\tiny \left[{\bf 1}_{+3}\right]}}$ by the scalars $\tilde{C}^{{\tt u}}$ and $\tilde{B}$ dual to the 2-forms $C_{{\tt u}\mu\nu}$ and $B_{\mu\nu}$, respectively.\par
Let us elaborate on the difference between the branchings (\ref{eq:1IIB}), (\ref{eq:2IIB}) and (\ref{eq:1IIA}), (\ref{eq:2IIA}) in light of our previous discussion about $T$-duality. We have learned that the difference between the Type IIA and the Type IIB descriptions of the maximal four-dimensional supergravity is limited to the fields originating from the RR-sector, which belong to chiral spinorial representations of ${\rm Spin}(6,6)$. This holds non only for the bosonic fields (scalars and vectors), but also, as we shall see, for the different kinds of fluxes, which belong to the ${\bf 912}$ representation of the embedding tensor.
In particular the different branchings in the two pictures originate from the different branchings of the two chiral representations ${\bf 32}_s$ and ${\bf 32}_c$ of ${\rm Spin}(6,6)$ with respect to the ${\rm SL}(6,\mathbb{R})$ subgroup of ${\rm O}(6,6)$:
\begin{align}
{\bf 32}_s&\rightarrow\,{\bf 1}+{\bf 15}+{\bf 15}'+{\bf 1}\,,\label{32s}\\
{\bf 32}_c&\rightarrow\,{\bf 6}+{\bf 20}+{\bf 6}'\,.\label{32c}
\end{align}
We see that ${\bf 32}_c$ decomposes in ${\rm SL}(6,\mathbb{R})$-representations describing rank-k  antisymmetric tensors with $k$ odd (rank-1, rank-3 and rank-5), as it is the case for the scalar fields in the Type IIA description or the vector fields in the Type IIB one. On the other hand the ${\bf 32}_s$ decomposes in antisymmetric tensor-representations with even-rank (rank-0, rank-2, rank-4 and rank-6), as it is the case for the scalar fields in the Type IIB description or the vector fields in the Type IIA one.\par
Finally let us consider the description of the maximal theory as originating from the toroidal compactification of M-theory, that is $D=11$ supergravity. Since the eleven-dimensional action has no global invariance, the group with respect to which to branch the relevant ${\rm E}_{7(7)}$- representations is $G_{int}={\rm GL}(7,\mathbb{R})$. We find:
\begin{align}
\label{eq:1M}
{\bf  56} \rightarrow & \,  {\bf 7}'_{-3}+{\bf 21}_{-1}+ {\bf 7}_{+3}+ {\bf 21}'_{+1}\,,\\
{\bf 133}\rightarrow &\,{\bf 7}_{-4}+{\bf 35}'_{-2}+{\bf 1+48}'_{0}+{\bf 35}_{+2}+{\bf 7}'_{+4}\,. \label{eq:2M}
\end{align}
In Eq. (\ref{eq:1M}) the representations ${\bf 7}'_{-3}+{\bf 21}_{-1}$ describe the Kaluza-Klein vectors $G^{\upalpha}_\mu$, $\upalpha,\upbeta=4,\dots, 10$ and the vectors $A_{\mu\,\upalpha\upbeta}$ from the 3-form field, respectively.
To identify the scalar fields we need to proceed through the solvable Lie algebra decomposition
\begin{align}
{\Scr S}={\Scr S}_{{\tiny {\rm GL}(7)}}\oplus {\tt N}^{{\tiny [{\bf 35}_{+2}]}}\oplus {\tt N}^{{\tiny [{\bf 7}_{+4}]}}\,,\label{solvMth}
\end{align}
${\Scr S}_{{\tiny {\rm GL}(7)}}$ being parametrized by the moduli of the internal metric $g_{{\upalpha\upbeta}}$, ${\tt N}^{{\tiny [{\bf 35}_{+2}]}},\, {\tt N}^{{\tiny [{\bf 7}_{+4}]}}$ by the scalars $A_{{\upalpha\upbeta\upgamma}}$ and the scalars $\tilde{A}^{\upalpha}$ dual to $A_{{\upalpha}\mu\nu}$. Once all forms are dualized to lower-order ones, the on-shell symmetry group ${\rm E}_{7(7)}$ becomes manifest and the off-shell symmetry ${\rm GL}(7,\mathbb{R})$ is enhanced to ${\rm SL}(8,\mathbb{R})$. The resulting symplectic frame is the ${\rm SL}(8,\mathbb{R})$-one discussed above.\par
In the Type II or M-theory descriptions of maximal four-dimensional supergravity, there is a precise characterization of the dimensionally reduced fields and geometric quantities associated with the global symmetry algebra $\mathfrak{g}=\mathfrak{e}_{7(7)}$: The dilatonic scalars, which comprise the ten-dimensional dilaton in the Type II picture and the moduli associated with the radii of the internal torus, are parameters of the Cartan-subalgebra ${\tt C}$, while the axionic scalar fields, deriving from the higher-dimensional antisymmetric tensor fields, parametrize the nilpotent generators of ${\Scr S}$ (in ${\tt N}$) and thus correspond to positive roots of $\mathfrak{e}_{7(7)}$ (see final paragraph of Sect. \ref{ghsect}). The vector fields on the other hand are in one-to-one
correspondence withe the $\mathfrak{e}_{7(7)}$-weights of the ${\bf 56}$-representation. As we shall see, the internal fluxes are associated with weights of the ${\bf 912}$ representation of the embedding tensor. The precise correspondence, together with a detailed discussion of the action of dualities, is given in Appendix \ref{appdualsugras}.
\subsection{The Gauging}\label{N8gaugs}
According to our general discussion of Sect.\ (\ref{gaugingsteps}), the most general gauge group $G_g$ which can be introduced in this theory is defined by an embedding tensor $\Theta_M{}^\alpha$ ({\footnotesize $M$}$=1,\dots, 56$ and $\alpha=1,\dots, 133$), which expresses the gauge generators $X_M$ as linear combinations of the global symmetry group ones $t_\alpha$ (\ref{Thdef}). The embedding tensor encodes all parameters (couplings and mass deformations) of the gauged theory.
This object is solution to the $G$-covariant constraints (\ref{linear2}),(\ref{quadratic1}), (\ref{quadratic2}), which we can be expressed as pure group theoretical constraints on the representation of $\Theta$ and its orbit under the action of ${\rm E}_{7(7)}$, respectively.
\subsubsection{A Group Theoretical Analysis}\label{agtan} The embedding tensor formally belongs to the product
\begin{equation}
\Theta_M{}^\alpha ~\in~ {\Scr R}_{v*}\times\Adj(G)
\={\bf 56}\;\times\;{\bf 133} \=
{\bf 56}\;+\;{\bf 912}\;+\;{\bf 6480}\;.\label{56133}
\end{equation}
The linear constraint (\ref{linear2}) sets to zero all the representations in the above decomposition which are contained in the 3-fold symmetric product of the ${\bf 56}$ representation:
\begin{equation}
X_{(MNP)}~\in~ ({\bf 56}\times{\bf 56}\times{\bf 56})_{{\rm sym.}}
\,\rightarrow\;{\bf 56}+ {\bf 6480}+ {\bf 24320}\,.
\end{equation}
The representation constraint therefore selects the ${\bf 912}$ as the representation ${\Scr R}_\Theta$ of the embedding tensor%
\footnote{
We can relax this constraint by extending the representation to include the ${\bf 56}$ in (\ref{56133}). Consistency would require the gauging
of the scaling symmetry of the theory (which is never an off-shell symmetry), also called \emph{trombone symmetry} \cite{Cremmer:1997xj,LeDiffon:2008sh,LeDiffon:2011wt}. This however leads to gauged theories which do not have an action. We shall not discuss these gaugings here.
}.
The quadratic constraints pose further restrictions on the ${\rm E}_{7(7)}$-orbits of the ${\bf 912}$ representation which $\Theta_M{}^\alpha$ should belong to. As we have seen, since all isometries have a non-trivial symplectic duality action, if the linear constraint holds, the quadratic constraint (\ref{quadratic2}) implies (\ref{quadratic1}). In particular, by virtue of the latter, the embedding tensor can be rotated to an electric frame through a suitable symplectic matrix $E$, see Eq.\ (\ref{elET}) and Sect. \ref{backelectric}. As we shall show below, the two quadratic constraints, provided the linear one holds, are equivalent. This is a feature of maximal supergravity.\par
From a group theoretical point of view, the product of two embedding tensors transforms in the symmetric product of two ${\Scr R}_{\Theta}={\bf 912}$, which decomposes under ${\rm E}_{7(7)}$ as follows:
\begin{equation}
({\bf 912}\times {\bf 912})_{{\rm sym.}}={\bf 133}+{\bf 8645}+{\bf 1463}+{\bf 152152}+{\bf 253935}\,.\label{912sym}
\end{equation}
The locality condition formally transforms in the antisymmetric product of two adjoint representations of ${\rm E}_{7(7)}$, whose decomposition is:  ${\bf 133}\wedge {\bf 133}={\bf 133}+{\bf 8645}$. It thus amounts to setting to zero the representation ${\bf 133}+{\bf 8645}$ on the right hand side of (\ref{912sym}). We can prove that, if the linear constraint holds, the closure condition (\ref{quadratic2}) poses no further restriction on $\Theta$ \cite{deWit:2007mt}.
To show this it is useful to use the explicit form of the projector $\mathbb{P}_\Theta$ in (\ref{PTheta}) implementing the linear constraint as given in Appendix \ref{app-A}. In particular by the second of Eqs. (\ref{projectors}) (in our case ${\bf D}_1={\bf 912}$), with the coefficients $a_1,\,a_2,\,a_3$ given in Table \ref{projs}. Once the ${\bf 56}$ is projected out through the condition $t_{\alpha\,M}{}^N\,\Theta_N{}^\alpha=0$, the action of $\mathbb{P}_\Theta$ simplifies and the representation constraint reads \cite{deWit:2007mt}:
\begin{equation}
(t_\beta t^\alpha)_M{}^N\,\Theta_N{}^\beta=-\frac{1}{2}\,\Theta_M{}^\alpha\,.\label{proj912sim}
\end{equation}
This constraint has strong implications on the form of the tensor $X_{MN}{}^P$ in the electric frame \cite{deWit:2007mt}, discussed at the end of Sect. \ref{backelectric}.\par
 The quadratic condition (\ref{quadratic2}) can be expressed in the following form:
 \begin{equation}
 C_{MN}{}^\alpha=0\,,
 \end{equation}
 where the tensor $C_{MN}{}^\alpha$ is defined as follows:
 \begin{equation}
 C_{MN}{}^\alpha\equiv \Theta_M{}^\beta\,\Theta_N{}^\delta\,{\rm f}_{\beta\delta}{}^\alpha+\Theta_M{}^\beta\,t_{\beta N}{}^P\,\Theta_P{}^\alpha\,.
 \end{equation}
 Using the linear constraint on $\Theta$, namely $t_{\alpha\,M}{}^N\,\Theta_N{}^\alpha=0$, and Eq. (\ref{proj912sim}), one can show after some algebra that \cite{deWit:2007mt}:
 \begin{equation}
 t_{\alpha\,M}{}^N\,C_{PN}{}^\alpha=0\,\,,\,\,\,(t_\beta t^\alpha)_M{}^N\,C_{PN}{}^\beta=-\frac{1}{2}\,C_{PM}{}^\alpha\,,\,\,
 t_{\alpha M}{}^P\,C_{PN}{}^\alpha= t_{\alpha N}{}^P\,C_{PM}{}^\alpha\,.
 \end{equation}
 The above properties imply that $C_{PN}{}^\alpha$, once the linear constraint is implemented on $\Theta$, belongs to the product ${\bf 56}\times {\bf 912}$ which decomposes as follows:
 \begin{equation}
 {\bf 56}\times {\bf 912}\rightarrow {\bf 133} + {\bf 8645} + {\bf 1539} + {\bf 40755}\, ,
 \end{equation}
 The only representations on the right hand side which occur also in the decomposition (\ref{912sym}) are the ${\bf 133} + {\bf 8645}$.\par
 \emph{This proves that the  quadratic constraints (\ref{quadratic1}) and (\ref{quadratic2}), if the representation constraint holds, are equivalent and amount to setting to zero the representation ${\Scr R}_{\Theta\Theta}={\bf 133}+{\bf 8645}$ on the right hand side of (\ref{912sym}).}
\par
We shall denote by ${\Scr R}_{\Theta\Theta}$ this representation.\par
From Eq. (\ref{proj912sim}) we can also deduce that the gauge group cannot have ${\rm U}(1)$ or ${\rm SO}(1,1)$-factors, namely be of the form: $G_g={\rm U}(1)\times \hat{G}_g$, or $G_g={\rm SO}(1,1)\times \hat{G}_g$. Indeed Eq. (\ref{proj912sim}) can be recast in the form:
\begin{equation}
X_{NM}{}^PX_{LP}{}^N=-\frac{1}{2}\,{\rm Tr}(X_M\,X_L)\,.
\end{equation}
Suppose the generator of the ${\rm U}(1)$-factor corresponds, in the above equation, to $L=1$. Since $X_1$ commutes with all the other generators, $[X_1,\,X_M]=0$, from the quadratic constraints we have $X_{1P}{}^N\,X_N=0$ and thus, from (\ref{proj912sim}) we find
\begin{equation}
{\rm Tr}(X_1\,X_M)=0\,.
\end{equation}
We conclude that ${\rm Tr}(X_1\,X_1)=0$, which excludes the presence of a ${\rm U}(1)$ or ${\rm SO}(1,1)$-factor. \par
Steps 1,2 and 3 of Sect. \ref{sec:3} allow to construct the bosonic gauged Lagrangian in the electric frame, while their generalization discussed in  Sect.\ \ref{sec:4} allow a frame-independent formulation of the gauging procedure and thus a manifestly $G$-covariant form of the field equations and Bianchi identities. \par

 \subsubsection{Fermion-Shift Tensors, $\mathbb{T}$-Tensor and the Scalar Potential.} The complete supersymmetric gauged Lagrangian is then obtained by adding fermion mass terms, a scalar potential and additional terms in the fermion supersymmetry transformation rules, according to the prescription given in Step 4. All these deformations depend on the fermion shift matrices $\mathbb{S}_{AB},\,\mathbb{N}_\mathcal{I}{}^A$.
In the maximal theory $\mathcal{I}=[ABC]$ labels the spin-$1/2$ fields $\chi_{ABC}$ and the two fermion shift-matrices are conventionally denoted by the symbols $A_1=(A_{AB}),\,A_2=(A^D{}_{ABC})$. The precise correspondence is:
\begin{equation}
\mathbb{S}_{AB}=-\frac{1}{\sqrt{2}}\,A_{AB}\,;\quad
\mathbb{N}_{ABC}{}^D=-\sqrt{2}\,A^D{}_{ABC}{}\,.
\end{equation}
where
\begin{equation}
A_{AB}=A_{BA}\,;\quad\;
A_{ABC}{}^D=A_{[ABC]}{}^D\,;\quad\;
A_{DBC}{}^D=0\,.\label{propsAA}
\end{equation}
The above properties identify the ${\rm SU}(8)$ representations of the two tensors:
\begin{equation}
A_{AB}\in {\bf 36}\,;\quad\; A_{ABC}{}^D\in {\bf 420}\,.
\end{equation}
The $\Tb$-tensor, defined in (\ref{TT}), as an ${\rm E}_{7(7)}$-object transforms in ${\Scr R}_\Theta={\bf 912}$, while as an ${\rm SU}(8)$-tensor it belongs to the following sum of representations:
\begin{equation}
\mathbb{T}~\in~ {\bf 912}\;
\stackrel{{\rm SU}(8)}{\longrightarrow}\;\,
{\bf 36}\;\oplus\;\overline{{\bf 36}}\;\oplus\;
{\bf 420}\;\oplus\;\overline{{\bf 420}}\;,
\end{equation}
which are precisely the representations of the fermion shift-matrices and their conjugates $A_{AB},\,A^{AB},\,A^A{}_{BCD},\,A_A{}^{BCD}$. We can thus view the properties (\ref{propsAA}) as following from the linear constraint on $\Tb$.
This in particular guarantees that the $O(g)$-terms in the supersymemtry variation of $\L_{{\rm gauge}}^{(0)}$, which depend on the $\Tb$-tensor, only contain ${\rm SU}(8)$-structures which can be canceled by the new terms containing the fermion shift-matrices. This shows that the linear condition $\Theta \in {\Scr R}_\Theta$ is also required by supersymmetry. The same holds for the quadratic constraints (\ref{quadratic2}) which imply the $\Tb$-identities (\ref{Tids}) and, in particular, the Ward identity (\ref{WID}) for the potential \cite{deWit:1982bul,deWit:2007mt}:
\begin{equation}
V(\phi)\,\delta^B_A\=
\frac{g^2}{6}\,\mathbb{N}^{CDE}{}_A\mathbb{N}_{CDE}{}^B-12\,g^2\,\mathbb{S}_{AC}\mathbb{S}^{BC}
\=\frac{g^2}{3}A^B{}_{CDE} A_A{}^{CDE}-6\,g^2\, A_{AC}\,A^{BC}\,,
\end{equation}
from which we derive:
\begin{equation}
V(\phi)=g^2\,\left(
\frac{1}{24}\,|A^B{}_{CDE}|^2-\frac{3}{4}\, |A_{AB}|^2\right)\,,\label{potN8}
\end{equation}
The scalar potential can also be given a manifestly $G$-invariant form \cite{deWit:2007mt}\,:
\begin{equation}
V(\phi)=
-\frac{g^2}{672}\,\Big(X_{MN}{}^{R}\,X_{PQ}{}^{S}\,\M^{MP}\,\M^{NQ}\,\M_{RS}
+ 7\,X_{MN}{}^{Q}\,X_{PQ}{}^{N}\,\M^{MP}\Big)\;,
\label{potentialN8}
\end{equation}
where $\mathcal{M}^{MN}$ is the inverse of the (negative definite) matrix $\mathcal{M}_{MN}$ defined in (\ref{M}) and, as usual, $X_{MN}{}^{R}$ describe the symplectic duality action of the generators $X_M$ in the ${\Scr R}_{v*}$-representation:\, $X_{MN}{}^{R}\equiv {\Scr R}_{v*}[X_M]_N{}^P$.
\paragraph{Properties of the fermion-shift tensors.}
Let us write the fermion-shift tensors as components of the $\mathbb{T}$-tensor, using Eqs. (\ref{identificationsSNT})
\begin{align}
(\mathbb{T}^{AB})^{CD}{}_{EF}&=4\,\delta^{[C}_{[E}T_{F]}{}^{D] AB}\,\,\,;\,\,\,\,\,(\mathbb{T}_{AB})^{CD}{}_{EF}=-4\,\delta^{[C}_{[E}T^{D]}{}_{F] AB}\,,\nonumber\\
T_{C}{}^{D AB}&= \mathbb{L}_c^{M\,AB}\,\mathcal{Q}_M{}^D{}_C=-\frac{1}{2}\,\mathbb{N}^{D AB}{}_C-2 \mathbb{S}^{D[A}\delta^{B]}_C=\frac{1}{\sqrt{2}}\,A_C{}^{D AB}+\sqrt{2} A^{D[A}\delta^{B]}_C\,,\nonumber\\
(\mathbb{T}_{AB})^{CDEF}&=-4\,
\delta_{[A}^{[C}\mathbb{N}^{DEF]}{}_{B]}=4\sqrt{2}\,\delta_{[A}^{[C}A_{B]}{}^{DEF]}\,.\nonumber
\end{align}
The quadratic constraints (\ref{Tids}), expressed in terms of the  $\mathbb{T}$-tensor, imply the following conditions on the fermion-shift tensors:
\begin{eqnarray}
 0&=&  A{}^C{}_{DAB} \,A_{F}{}^{EAB} - A_{D}{}^{CAB} \, A{}^E{}_{FAB}
  -4\,A{}^{(C}{}_{DFA}A^{E)A}-4A_{(F}{}^{ECA}A_{D)A} \nn\\
%%%%%%%%%%%%%
  &&{}-2\,\delta_{D}^{E}\,A_{FA}A{}^{CA}+2\,\delta_{F}^{C}\,A_{DA}A{}^{EA}
 \;,\label{ida1} \\[1ex]
%%%%%%%%%%%%%
0&=&   A{}^A{}_{BC[E} \,A{}^C{}_{FGH]}
  +A_{BC}\delta^{A}_{[E}A{}^{C}{}_{FGH]}
  -A_{B[E}A^{A}{}_{FGH]} \label{ida2}\\
 && {}+\frac1{24}\,\varepsilon_{EFGH A_1 A_2 A_3 A_4}\,
  \left(A_{B}{}^{AC A_1}\, A_{C}{}^{A_2 A_3 A_4}
  +A^{AC}\delta_{B}^{A_1}A_{C}{}^{A_2 A_3 A_4}-A^{A A_1}A_{B}{}^{A_2 A_3 A_4}\right)  \;,
 \nonumber\\[1ex]
0&=& A^{A_1}{}_{ABC} \,A_{A_1}{}^{EFG} - 9\,A^{[E}{}_{A_1[AB}\,
A_{C]}{}^{FG]A_1}
- 9\, \delta_{[A}{}^{[E} \, A^F{}_{|A_1 A_2|B}\,A_{C]}{}^{G]A_1 A_2} \nn\\
&&{}- 9\, \delta_{[AB}{}^{[EF}\,A^{|A_3|}{}_{C]A_1 A_2}\, A_{A_3}{}^{G] A_1 A_2}
+\delta{}_{ABC}{}^{EFG} \,A^{A_3}{}_{A_1 A_2 A_4} \, A_{A_3}{}^{A_1 A_2 A_4}  \;.
\label{ida3}
\end{eqnarray}
By suitably contracting the indices of Eqs. (\ref{ida1}) and (\ref{ida3}), one finds after some algebra the following relation:
\begin{equation}
\frac{1}{24}\,A^A{}_{CDE}\,A_B{}^{CDE}-\frac{3}{4}\,A^{AC}\,A_{BC}=\frac{1}{8}\,\delta_B^A\,
\left(\frac{1}{24}\,A^F{}_{CDE}\,A_F{}^{CDE}-\frac{3}{4}\,A^{FC}\,A_{FC}\right)\,,
\end{equation}
which is the potential Ward identity (\ref{WID}) for the maximal supergravity, the potential being given by (\ref{potN8}).\par
Using the coset geometry and Eq. (\ref{DTmax}) we derive the ``gradient flow'' equations (\ref{GFN8}) which, in terms of $A_1,\,A_2$, read:
\begin{align}
{\Scr D}_s A_{AB}&=-\frac{1}{6}\,A_{(A}{}^{CDE}\mathcal{P}_{s|B)CDE}\,,\nonumber\\
{\Scr D}_s A^F{}_{ABD}&=-A^{FE}\,\mathcal{P}_{s\,EABD}-\frac{3}{2}\,A_{[A}{}^{FEG}\,
\mathcal{P}_{s|BD]EG}+\frac{1}{2}\delta^{F}_{[A}\,A_{B}^{EGH}\mathcal{P}_{s|D]EGH}\,.\label{GFN8A}
\end{align}
Using the above relations and the expression (\ref{potN8}) of the scalar potential, one can write
the gradient of the potential in the following form:
\begin{align}
\frac{\partial}{\partial \phi^s}V&=-\frac{g^2}{12}\mathcal{P}_s^{BEFG}\,\mathcal{C}_{BEFG}+c.c.=\nonumber\\
&=-\frac{g^2}{12}\mathcal{P}_s^{BEFG}\,\left(\mathcal{C}_{BEFG}+\frac{1}{24}\epsilon_{BEFGA_1 A_2 A_3 A_4}\mathcal{C}^{A_1A_2 A_3 A_4}\right)\,\label{DVPC}
\end{align}
where the tensor $\mathcal{C}_{BEFG}$ reads:
\begin{equation}
\mathcal{C}_{BEFG}\equiv A^A{}_{[BEF}\,A_{G]A}+\frac{3}{4}\,A^A{}_{D[BE}\,A^D{}_{FG]A}\,,
\end{equation}
and we have used the reality condition (\ref{realityn8}) on $\mathcal{P}_s^{BEFG}$.
Equation (\ref{DVPC}), which expresses Eq. (\ref{DWID}) in the maximal model,  implies that extremal points $\phi_0$ of the scalar potential correspond to values of the scalar fields for which the tensor $\mathcal{C}_{BEFG}$ is anti-selfdual:
\begin{equation}
\frac{\partial}{\partial \phi^s}V=0\,\,\Leftrightarrow\,\,\,\,\,\mathcal{C}_{BEFG}+\frac{1}{24}\epsilon_{BEFGA_1 A_2 A_3 A_4}\mathcal{C}^{A_1A_2 A_3 A_4}=0\,.\label{Casd}
\end{equation}
Recall that $\mathcal{C}_{BEFG}$ is a quadratic function of $\Theta$ and a non-linear function of the scalar fields. As mentioned in Sect. \ref{vad}, when looking for vacua of gauged maximal supergravities with certain features, like residual symmetry, one can, with no loss of generality, restrict oneself to the origin of the manifold (in which all scalar fields are zero) and solve Eqs. (\ref{Casd}) as quadratic conditions on the  embedding tensor \cite{inverso,Dibitetto:2011gm,Dall'Agata:2011aa}. This allows to search for the vacua through all gauged models at once.\par
General arguments based on the quadratic constraints (\ref{ida1}), (\ref{ida2}) and (\ref{ida3}), exclude the existence of $AdS$ vacua preserving $4<\mathcal{N}'<8$ supersymmetries \cite{Gallerati:2014xra} (and thus $\mathcal{N}'=6$ AdS vacua, consistently with the results of \cite{Andrianopoli:2008ea}).\footnote{The only known instances of $\mathcal{N}'=3$ and $\mathcal{N}'=4$ AdS vacua were recently found in \cite{Gallerati:2014xra}. They occur in models with \emph{dyonic gaugings} to be dealt with in Sect. \ref{dyonicg}.}

\paragraph{Mass matrices.} Let us give the mass matrices for the various fields \cite{LeDiffon:2011wt}.\par
\emph{Scalar masses.}
Using the ${\rm SU}(8)$-covariant parametrization (\ref{Hcovpar}) of the scalar manifold and linearizing the scalar field equations by expanding the scalar fields about a vacuum solution $\phi_0=(\phi^{ABCD}_0)$:
\begin{equation}
\phi^{ABCD}=\phi^{ABCD}_0+\delta\phi^{ABCD}\,,
\end{equation}
one finds:
\begin{equation}
\Box\, \delta\phi_{ABCD} =
{M}_{ABCD}{}^{EFGH}\,\delta\phi_{EFGH} + {\cal O}(\delta\phi^2)
\;,
\end{equation}
where we have used the property $\mathcal{P}_\mu^{ABCD}=\partial_\mu\delta \phi^{ABCD} + {\cal O}(\delta\phi^2)$ and
the scalar squared-mass matrix ${ M}_{ABCD}{}^{EFGH}$ is given by
\bea
{M}_{ABCD}{}^{EFGH}\,\delta\phi^{ABCD}\delta\phi_{EFGH} &=&
6g^2 \left(A_{E}{}^{ABC}A^{D}_{ABF}\!-\!\ft14A_{A}{}^{BCD}A^{A}_{BEF}\right)\delta\phi^{EFGH}\delta\phi_{CDGH}
\nonumber\\
&&{}
+g^2\left(\ft5{24}\,A_{A}{}^{BCD}A^{A}{}_{BCD}-\ft12A_{AB}A^{AB}\right) \delta\phi^{EFGH}\delta\phi_{EFGH}
\nonumber\\
&&{}
-\ft23\,g^2\,A_{A}{}^{BCD} A^{E}{}_{FGH} \, \delta\phi^{AFGH}\delta\phi_{BCDE}\nonumber\\[1ex]
&=& 48\,V^{(2)}(\delta\phi)
\;,
\label{Mscalar_sym}
\eea
where we have denoted by $V^{(2)}(\delta\phi)$ the terms of the scalar potential (\ref{potN8})
which are of second order in $\delta\phi$ upon expansion around the vacuum $\phi_0$:
\begin{equation}
V(\phi)~=~V_0+V^{(2)}(\delta\phi)+{\cal O}(\delta\phi^3)\,.
\end{equation}

\emph{Vector masses.} Besides the general expression (\ref{MvectMN}) for the vector squared-mass matrix we can derive formulas for it which are specific to the maximal model. In the ${\rm SU}(8)$-basis, the matrix reads
\bea
{ M}_{ v} &=&
\left(
\begin{array}{cc}
{ M}^{AB}{}_{CD} & {M}^{ABCD}
\\
 { M}_{ABCD}&
 { M}_{AB}{}^{CD}
\end{array}
\right)
\;,
\label{M_vector}
\eea
with
\bea
{ M}^{AB}{}_{CD}
&=&
-\ft16 g^2\,A^{[A}{}_{EGH} \delta^{B]}_{[C} A_{D]}{}^{EGH}
+\ft12 g^2\, A^{[A}{}_{GH[C}A_{D]}{}^{B]GH}\,,
\nonumber\\[2ex]
{ M}^{ABCD} &=&
\ft1{36}  g^2\,A^{[A}{}_{A_1A_2A_3} \epsilon^{B]A_1A_2A_3 B_1B_2B_3[C} A^{D]}{}_{B_1B_2B_3}\,.
\eea
We can also give this matrix a manifestly symplectic covariant form
\begin{equation}
{ M}_{{ v}\,M}{}^N=-\frac{1}{24}\,g^2\,\left[{\rm Tr}(X_M\,X_P)+{\rm Tr}(\mathcal{M}^{-1}\,X_M\,\mathcal{M}\,(X_P)^T)\right]\,\mathcal{M}^{PN}\,.
\end{equation}
\emph{Fermion masses.} Finally, the gravitino and spin-$1/2$ mass matrices were discussed in a general theory in Sect. \ref{vmm}. They are given, in an AdS vacuum, by Eqs. (\ref{32mass}) and (\ref{12massads}). Specializing these general formulas to the maximal theory we find \cite{Dall'Agata:2012cp,Gallerati:2014xra}
\begin{align}
{ M}^{\left[\frac{3}{2}\right]}{}_{AB} &= \sqrt{2}\,g\,A_{AB}\,,\nonumber\\
{M}^{\left[\frac{1}{2}\right]}{}_{ABC,\,EFG} &=~\frac{1}{3}\left(\mathbb{M}_{ABC,EFG}+\frac{\sqrt{2}}{3}\,\sum_{G,H}{}^\prime  \left(\frac{A}{|A|^2+\frac{V_0}{6\,g^2}\,{\bf 1}}\right)_{GH}\,A^G{}_{ABC} A^H{}_{EFD}\right)=\nonumber\\
&=\frac{\sqrt{2}}{12}\left(\epsilon_{ABC A_1A_2A_3[EF}  A_{G]}{}^{A_1A_2A_3}+\frac{4}{3}\,\sum_{G,H}{}^\prime  \left(\frac{A}{|A|^2+\frac{V_0}{6\,g^2}\,{\bf 1}}\right)_{GH}\,A^G{}_{ABC} A^H{}_{EFD}\right)\,,
\label{Mferm}
\end{align}
where the normalization of the dilatino mass matrix is fixed by writing the corresponding linearized field equation in the form
\begin{equation}
i\gamma^\mu\mathcal{D}_\mu\chi_{ABC}={M}^{\left[\frac{1}{2}\right]}{}_{ABC,\,EFG}\,\chi^{EFG}\,.
\end{equation}
The Minkowski vacua correspond to the case $V_0=0$.

\subsection{The Gauged Action and Supersymmetry Transformations}\label{gaun8action}
We give, for the sake of completeness, the gauged Lagrangian of the
maximal theory, in the duality covariant formalism:\footnote{We refer the reader to \cite{deWit:2007mt} for the complete Lagrangian which includes higher order terms in the fermion fields.}
{\small \begin{align}
{\Scr L}_{B} &= -\frac{e}{2}\,R+\frac{e}{48}\,\hat{\mathcal{P}}_\mu^{ABCD}\hat{\mathcal{P}}^\mu_{ABCD}+
\frac{e}{4} \, {\cal
I}_{\Lambda\Sigma}\,\mathcal{H}_{\mu\nu}{}^{\Lambda}
\mathcal{H}^{\mu\nu\,\Sigma} +\frac{1}{8} {\cal
R}_{\Lambda\Sigma}\;\varepsilon^{\mu\nu\rho\sigma}
\mathcal{H}_{\mu\nu}{}^{\Lambda}
\mathcal{H}_{\rho\sigma}{}^{\Sigma}+
 \nonumber\\
&-\frac{1}{8}\, \varepsilon^{\mu\nu\rho\sigma}\,
\Theta^{\Lambda\alpha}\,B_{\mu\nu\,\alpha} \, \Big(
2\,\partial_{\rho} A_{\sigma\,\Lambda} + X_{MN\,\Lambda}
\,A_\rho{}^M A_\sigma{}^N
-\frac{1}{4}\,\Theta_{\Lambda}{}^{\beta}B_{\rho\sigma\,\beta}
\Big)\nonumber\\
&-\frac{1}{3}\,
\varepsilon^{\mu\nu\rho\sigma}X_{MN\,\Lambda}\, A_{\mu}{}^{M}
A_{\nu}{}^{N} \Big(\partial_{\rho} A_{\sigma}{}^{\Lambda}
+\frac{1}{4}  X_{PQ}{}^{\Lambda}
A_{\rho}{}^{P}A_{\sigma}{}^{Q}\Big)
\nonumber\\[.9ex]
&{} -\frac{1}{6}\,
\varepsilon^{\mu\nu\rho\sigma}X_{MN}{}^{\Lambda}\, A_{\mu}{}^{M}
A_{\nu}{}^{N} \Big(\partial_{\rho} A_{\sigma}{}_{\Lambda}
+\frac{1}{4}\, X_{PQ\Lambda}
A_{\rho}{}^{P}A_{\sigma}{}^{Q}\Big)\nonumber\\
& +\epsilon^{\mu\nu\rho\sigma}(\bar{\psi}^A_{\mu}\gamma_\nu\mathcal{D}_\rho\psi_{A\sigma}-
 \bar{\psi}_{A\,\mu}\gamma_\nu\mathcal{D}_\rho
 \psi^A_{\sigma})-
 \frac{i\,e}{12}\,(\bar{\chi}^{ABC}\gamma^\mu\mathcal{D}_{\mu}\chi_{ABC}+\bar{\chi}_{ABC}
 \gamma^\mu\mathcal{D}_{\mu}\chi^{ABC})\nonumber\\
 &-\frac{e}{6}\,\bar{\chi}^{ABC}\gamma^\mu\gamma^\nu\psi^D_\mu\,\mathcal{D}_\nu\phi^s\mathcal{P}_{s\,ABCD}
 -\frac{e}{6}\,\bar{\chi}_{ABC}\gamma^\mu\gamma^\nu\psi_{D\mu}\,\mathcal{D}_\nu\phi^s\mathcal{P}_{s}^{ABCD}\nonumber\\
&+\frac{e}{4}\,\mathcal{H}^{+\,\Lambda\,\mu\nu}\mathcal{I}_{\Lambda\Sigma}
{\bar{ f}}^{\Sigma AB}\mathcal{O}_{AB\,\mu\nu}+\frac{e}{4}\,\mathcal{H}^{-\,\Lambda\,\mu\nu}\mathcal{I}_{\Lambda\Sigma}
{{ f}}^{\Sigma }{}_{AB}\mathcal{O}^{AB}_{\mu\nu}\nonumber\\
&-g\,e\,\left(\sqrt{2}\bar{\psi}^A_\mu\;\gamma^{\mu\nu}\;\psi_\nu^B\;A_{AB}
~+\frac{\sqrt{2}}{6}~i\,\bar{\chi}^{BCD}\;\gamma^\mu\;\psi_{\mu\,A}\;A^A{}_{BCD}
~-\frac{1}{36}~\bar{\chi}^{ABC}\,\chi^{EFG}\;\mathbb{M}_{ABC,\,EFG}\right)
~+~\text{h.c.}\nonumber\\
&-e\,V(\phi)+{\Scr L}_{4f}\,,
\end{align}}
where:
\begin{equation}
\mathcal{O}_{AB\,\mu\nu}=2\,\bar{\psi}_{A\,\rho}\gamma^{[\rho}\gamma_{\mu\nu}\gamma^{\sigma]}\psi_{B\,\sigma}+
i\,\bar{\psi}_{\rho}^C\gamma_{\mu\nu}\gamma^\rho\chi_{ABC}+
\frac{1}{72}\epsilon_{ABCDEFGH}\bar{\chi}^{CDE}\gamma_{\mu\nu}\chi^{FGH}\,,
\end{equation}
and the tensor $\mathbb{M}_{ABC,\,EFG}$ was given in (\ref{matbbmmax})
\begin{equation}
\mathbb{M}_{ABC,DEF}=\frac{1}{2\sqrt{2}}\,\epsilon_{ABC,A'B'C'[DE}\,A_{F]}{}^{A'B'C'}\,.
\end{equation}
The supersymmetry variations of the fields, see Eqs. (\ref{traphi2})-(\ref{traB2}), read
{\small \begin{align}
\delta\phi^s\mathcal{P}_s^{ABCD}&=\Sigma^{ABCD}\,,\label{traphi28}\\
\delta A^M_\mu&=\mathbb{L}_c^M{}_{\underline{M}}\,{\Scr O}_\mu^{_{\underline{M}}}=\frac{1}{2}\,\mathbb{L}_c^M{}_{AB}\,{\Scr O}_\mu^{AB}+h.c.\,,\label{traA28}\\
\delta V_\mu{}^a&=i\,\bar{\epsilon}^A\gamma^a\psi_{\mu\,A}+i\,\bar{\epsilon}_A\gamma^a\psi_{\mu}^A\,,\label{traV28}\\
\delta\psi_{A\,\mu}&=\mathcal{D}_\mu\epsilon_A-\frac{1}{8}\,H^-_{\rho\sigma\,AB}
\gamma^{\rho\sigma}\gamma_\mu\epsilon^B+i\,g\,\mathbb{S}_{AB}\,\gamma_\mu\,\epsilon^B\dots\,,\label{trapsi28}\\
\delta \chi_{ABC}&=i\, \mathcal{D}_\mu\phi^s\,\mathcal{P}_{s\,ABCD}\gamma^\mu\epsilon^D-\frac{3i}{4}\,
H^-_{\mu\nu\,[AB}\gamma^{\mu\nu}\epsilon_{C]}+g\,\mathbb{N}_{ABC}{}^D\epsilon_D\dots\,,\label{trachi228}\\
\Theta^{\Lambda{}\alpha}\,\delta B_{\alpha\,\mu\nu}&=\frac{i}{3}\left(\bar{\chi}_{ABC}\gamma_{\mu\nu}\epsilon_D\,
\mathcal{P}^{\Lambda\,ABCD}-\bar{\chi}^{ABC}\gamma_{\mu\nu}\epsilon^D\,
\mathcal{P}^\Lambda{}_{ABCD}\right)-\nonumber\\
&-8\,\left(\bar{\psi}_{A\,[\mu}\gamma_{\nu]}\epsilon^B+
\bar{\psi}^B_{[\mu}\gamma_{\nu]}
\epsilon_A\right)\,\mathcal{Q}^\Lambda{}_{B}{}^A-2X^\Lambda{}_P{}^M\mathbb{C}_{MN}\,A^P_{[\mu}\,\delta A^N_{\nu]}\,,
\end{align}
}
where
\begin{align}
\Sigma^{ABCD}&=-4\,\left(\bar{\epsilon}^{[A}\chi^{BCD]}+\frac{1}{24}\,\epsilon^{ABCDEFGH}\,\bar{\epsilon}_{[A}\chi_{BCD]}\right)\,,\nonumber\\
{\Scr O}_\mu^{AB}&\equiv i\,\bar{\epsilon}_C\gamma_\mu\chi^{ABC}-4\,\bar{\epsilon}^{[A}\psi^{B]}_\mu\,,
\end{align}
and we have used Eq. (\ref{ThetadeltaB}).
\subsection{Old and New Gaugings}\label{oandng}
As mentioned in Sect.\ \ref{gaugingsteps}, different symplectic frames (i.e.\ different ungauged Lagrangians) correspond to different choices for the viable gauge groups and may originate from different compactifications.
\subsubsection{The ${\rm SL}(8,\mathbb{R})$-Frame. } In the ${\rm SL}(8,\mathbb{R})$-frame the possible gaugings are conveniently discussed by branching the ${\bf 912}$ with respect to $G_{el}$:
\begin{equation}
{\bf 912}\;
\stackrel{{\tiny {\rm SL}(8,\mathbb{R})}}{\longrightarrow}\;\,
{\bf 36}\;+\;{{\bf 36}}'\;+\;
{\bf 420}\;+\;{{\bf 420}}'\;,\label{912sl8}
\end{equation}
the branchings of the ${\bf 133}$ and the ${\bf 56}$ representations were given in Eqs. (\ref{sl8branch}).
It is useful to arrange the representations in Eq. (\ref{912sl8}) in the following table \cite{deWit:2002vt,deWit:2007mt}
\begin{equation}
\begin{tabular}{c|cc}
&${\bf 28}$ &${\bf 28}'$
\\
\hline
${\bf 63}$
&${\bf 36}+{\bf 420}$
&${\bf 36}'+{\bf 420}'$
\\
${\bf 70}$
&${\bf 420}'$
&${\bf 420}$
\\
\end{tabular}
\label{decblock}
\end{equation}
where the left column describes the ${\rm E}_{7(7)}$ generators (in the ${\bf 133}$), grouped in ${\rm SL}(8,\mathbb{R})$-representations, while the top row contains the same description of the electric and magnetic vector fields (in the ${\bf 56}$). The table illustrates which (electric or magnetic) vector field is associated  by the embedding tensor with which generator, if we choose to switch on any of the representations in the branching of the ${\bf 912}$ separately. If we are interested in the gaugings within the ${\rm SL}(8,\mathbb{R})$-frame, only the electric vectors $A^{\Lambda}_\mu=A^{{\tt ab}}_\mu$ in the ${\bf 28}'$ can be involved. This mounts to considering an embedding tensor in which the only non-vanishing block is $\Theta_{{\tt ab}}{}^\alpha$, namely the index $M$ labels the ${\bf 28}$. Table \ref{decblock} tells us then that the index $\alpha$ can only run in the ${\bf 63}$ (namely the ${\rm SL}(8,\mathbb{R})$-generators) since otherwise the representation ${\bf 420}'$ would be switched on, which also involves the magnetic vector fields. By the same token we single out in the branching of the ${\bf 912}$ only the representation ${\bf 36}$, excluding the ${\bf 420}$. Any element in the
${\bf 36}$ defines a viable gauging, since the locality constraint (\ref{quadratic1}) is satisfied by construction. A closer analysis \cite{Cordaro:1998tx,deWit:2002vt} shows that, modulo an ${\rm
  SL}(8,\mathbb{R})$-conjugation, the general form of $\Theta\in{\bf
  36}$ is given by
\begin{eqnarray}
  \label{eq:cso}
  \Theta_M{}^\alpha=\Theta_{[{\tt ab}]\,{\tt d}}{}^{{\tt c}} &\!=\!&
  \delta^{{\tt c}}_{[{\tt a}}\,\theta^{~}_{{\tt b}]{\tt d}} \;, \qquad \theta_{{\tt ab}} = {\rm diag}\{
  \underbrace{1, \dots,1}_p, \underbrace{-1, \dots, -1}_q,
  \underbrace{0, \dots, 0}_r \} \;,
\end{eqnarray}
with ${\tt a},\,{\tt b} = 1, \dots, 8$, and correspond to gauge groups of the form ${\rm CSO}(p,q,r)$, $p+q+r=8$ \cite{Hull:1984yy,Hull:1984vg,Hull:1984rt,Cordaro:1998tx}. \footnote{In \cite{Cordaro:1998tx}, introducing the embedding tensor representation, it was proven that the ${\rm CSO}(p,q,r)$ groups represent the most general gauging in the ${\rm SL}(8,\mathbb{R})$-frame. In \cite{Hull:1984yy,Hull:1984vg,Hull:1984rt,Cordaro:1998tx} it is also discussed how these groups can be obtained though a singular ${\rm SL}(8,\mathbb{C})$-transformation (implying an In\"on\"u-Wigner contraction) on the $X_{\hat{\Lambda}\hat{\Sigma}}{}^{\hat{\Gamma}}$ tensor of the ${\rm SO}(8)$ gauge group.}
Such groups are in general non-semisimple and are defined as the subgroups of ${\rm SL}(8,\mathbb{R})$ preserving the metric $\theta_{{\tt ab}}$. They have the following structure:
\begin{equation}
{\rm CSO}(p,q,r)={\rm SO}(p,q)\ltimes \exp({\tt N}^{(p+q,r)})={\rm SO}(p,q)\ltimes T^{r(p+q)}\,,
\end{equation}
where ${\tt N}^{(p+q,r)}$ is a $r(p+q)$-dimensional Abelian, nilpotent subspace of $\mathfrak{sl}(8,\mathbb{R})$ transforming under (the adjoint action of) the ${\rm SO}(p,q)$-subgroup in $r$-copies of its fundamental representation ${\bf p+q}$. It generates a subgroup of $r(p+q)$  translations $T^{r(p+q)}$. The group ${\rm CSO}(p,q,1)$, $p+q=7$, is also denoted by ${\rm ISO}(p,q)$. \par
A special case is ${\rm CSO}(8,0,0)$, which coincides with ${\rm SO}(8)$ and corresponds to the first gauging of the maximal theory by de Wit and Nicolai \cite{deWit:1981sst,deWit:1982bul}. The vacua of these models, which have not been completely classified yet, have been the subject to a long investigation which started in the eighties, see for instance \cite{Warner:1983vz,Hull:1984ea,Hull:1984wa,Hull:1988jw,Fischbacher:2010ec}. The ${\rm SO}(8)$-gauged theory, for instance, besides the maximally supersymmetric AdS vacuum at the origin, corresponding to the compactification of the eleven dimensional theory on a seven-sphere, features a number of other vacua which can be systematically studied in terms of their residual symmetry. Only some of them were put in correspondence with variants of the seven-sphere compactification. As an example, the ${\rm G}_{2}$-invariant vacua are illustrated in Fig. \ref{fign8}.\footnote{Under the action of the ${\rm G}_2$ subgroup of ${\rm SO}(8)$, all the representations ${\bf 8}_{s,c,v}$ branch as ${\bf 8}_{s,c,v}\rightarrow {\bf 1}+{\bf 7}$. Similarly we have: ${\rm 35}_{s,c,v}\rightarrow {\bf 1}+{\bf 7}+{\bf 27}$. Therefore, parametrizing the scalar manifold in terms of $\phi^{{\tt ab}}_v,\,\phi^{{\tt abcd}}_c$ in the ${\bf 35}_v$ and ${\bf 35}_c$, respectively, there are only two scalars which are ${\rm G}_2$-singlets: $\phi_1$ in the ${\bf 35}_v$ (proper scalar) and $\phi_2$ in the ${\bf 35}_c$ (pseudo-scalar), see Eq. (\ref{phipp}).\label{fot}} The other ${\rm CSO}(p,q,r)$-models do not feature particularly interesting vacua: There are unstable de Sitter vacua for $G_g={\rm SO}(5,3)$ and ${\rm SO}(4,4)$, and a Minkowski vacuum for $G_g={\rm CSO}(2,0,6)$, which is an example of \emph{flat group}, see below. Just as the ${\rm SO}(8)$-model originates from a reduction on a seven-sphere of eleven-dimensional supergravity,  ${\rm CSO}(p,q,r)$-models can be interpreted as consistent truncations of the same theory on  backgrounds which are warped products of an Einstein space-time and a non-compact internal space of the form $$\mathcal{M}_{int}=\mathcal{H}^{(p,q)}\times \mathbb{R}^r\,,$$
where, for $q>0$, $\mathcal{H}^{(p,q)}$ is a hyperboloid\footnote{The space $\mathcal{H}^{(p,q)}$ is defined as the surface $(X^1)^2+\dots+(X^{p})^2-(X^{p+1})^2-\dots-(X^{p+q})^2=L^2$ in $\mathbb{R}^{p,q}$.} \cite{Hull:1988jw}. This statement, for generic $p,q,r$, has been recently put on rigorous grounds within the framework of \emph{exceptional field theory} \cite{Hohm:2014qga}. Certain ${\rm CSO}(p,q,r)$-models, featuring a runaway potential (and thus no vacua), have been considered in the context of domain-wall solutions \cite{Boonstra:1998mp,Hull:2001yg}.

\subsubsection{The ${\rm E}_{6(6)}$-Frame.}\label{gaugE6}
In order to discuss the gaugings in the ${\rm E}_{6(6)}$-frame we start by decomposing the ${\bf 912}$ with respect to the block-diagonal group
${\rm E}_{6(6)}\times {\rm SO}(1,1)$:
\begin{eqnarray}
{\bf 912} &\rightarrow &
{\bf 78}_{-3}+ {\bf 27}'_{-1}+{{\bf 351}}'_{-1}+ {\bf
  351}_{+1}+ {\bf 27}_{+1}+ {\bf 78}_{+3} \,,
\label{Tdec45}
\end{eqnarray}
the decompositions of the ${\bf 133}$ and ${\bf 56}$ being given in (\ref{133e6}) and (\ref{56e6}).
Just as we did for the ${\rm SL}(8,\mathbb{R})$-frame, we arrange the representations in the above branching in the following table \cite{deWit:2002vt,deWit:2007mt}
%%%%%%%%%%%%%%%%%%%%%%%%%%%%%%%%%%%%%%%%%%
\begin{equation}
\begin{tabular}{c|cccc}
&${\bf 1}_{-3}$ &${\bf 27}'_{-1}$  & ${\bf 27}_{+1}$
&${\bf 1}_{+3}$
\\
\hline
${\bf 27}_{-2}$
&
&${\bf 78} _{-3}$
&${\bf 351}'_{-1}+ {\bf 27}' _{-1}$
&${\bf 27}_{+1}$
\\
%\hline
${\bf 78}_0$
&${\bf 78} _{-3}$
&${\bf 351}'_{-1}+ {\bf 27}'_{-1}$
&$ {{\bf 351}}_{+1}+ {\bf {27}} _{+1}$
&${\bf 78} _{+3}$
\\
%\hline
${\bf 1}_0$
&
& ${\bf 27}'_{-1}$
& ${\bf 27}_{+1}$
&
\\
%\hline
${\bf 27}'_{+2}$
& ${\bf 27}'_{-1}$
&$ {{\bf 351}}_{+1}+ {\bf{27}} _{+1}$
&${\bf 78} _{+3}$
&
\\ %\hline
\end{tabular}
\label{decblock-e6}
\end{equation}
which summarizes the couplings between the vector fields and the generators induced by the various components of the embedding tensor. Recall that the electric vector fields $(A^\Lambda_\mu)=(A^0_\mu,\,A^\lambda_\mu)$ transform in the ${\bf 1}_{-3}+{\bf 27}'_{-1}$. Therefore the most general gauging involving only these vectors must be defined by an embedding tensor of the form $(\Theta_\Lambda{}^\alpha)=(\Theta_0{}^\alpha,\,\Theta_\lambda{}^\alpha)$, whose first index labels the ${\bf 1}_{+3}+{\bf 27}_{+1}$. Inspection of the above table shows that such gauging can only live in the ${\bf 78}_{+3}$ representation. Vice versa, every
such embedding tensor automatically satisfies the quadratic
constraint~(\ref{quadratic1}) and thus defines a viable
gauging.
These are the theories originating from five dimensions by
\emph{Scherk-Schwarz reduction} on a circle \cite{Scherk:1979zr,css,Sezgin:1981ac,Andrianopoli:2002mf}.\footnote{This reduction was first considered in \cite{css} and should be more appropriately named \emph{Cremmer-Scherk-Schwarz} or \emph{generalized Scherk-Schwarz} reduction, to distinguish it from the original Scherk-Schwarz mechanism introduced in \cite{Scherk:1979zr} as a generalization of toroidal dimensional reduction for a generic model describing Einstein's gravity coupled to matter.}
This kind of dimensional reduction was originally devised in \cite{Scherk:1979zr,css}
 as a possible mechanism for producing an effective four-dimensional supergravity featuring spontaneous supersymmetry
 breaking at various scales. It represents a generalized type of compactification
 in which the ansatz for the five-dimensional fields, on a space-time of the form $\mathbb{R}^{1,3}\times
 S^1$,
 contains a dependence on the internal $S^1$ coordinate $y$ through a
 global symmetry transformation of the five-dimensional
 Lagrangian, called Scherk--Schwarz ``twist''
 \begin{eqnarray}
\Phi(x^\mu,y)&=&e^{M\,y}\cdot \Phi(x^\mu)\,,
\end{eqnarray}
where $e^{M\,y}$ is the \emph{twist matrix} and $M$ is a global symmetry
generator of the five-dimensional theory which has a non-trivial
action on the field $\Phi$. This property of $M$
guarantees that the dependence on $y$ ultimately disappears in the
four-dimensional theory. However since $y$ has the dimension of an
inverse-mass, $M$ has the dimension of a mass and will induce mass
deformations in the lower-dimensional theory. They originate from
terms, in the $D=5$ Lagrangian, containing derivatives with
respect to $y$: $\partial^2_y\Phi(x,y)=M^2\cdot \Phi(x,y)$. If we
start from the five-dimensional \emph{maximal} (ungauged)
supergravity, whose Lagrangian has an ${\rm E}_{6(6)}$ global
symmetry group, we can perform a Scherk--Schwarz reduction by taking as $M$
any generator of ${\rm E}_{6(6)}$. The resulting four-dimensional
supergravity is a gauged supergravity, as was first shown in
\cite{Andrianopoli:2002mf}. This model is an instance of a ``no--scale''
supergravity as it features a
 non--negative scalar potential. The only possible vacua are of Minkowski type
 and are defined by the points in the scalar manifold in which the potential vanishes. These points exist
only if $M^T=-M$, namely if $M$ is a generator of the maximal
compact subgroup ${\rm USp}(8)$ of ${\rm E}_{6(6)}$, see footnote \ref{fotss} below, in which case
the gauge group is called ``flat'' group. The embedding tensor description for
this theory, in terms of the ${\bf 78}_{+3}$ component, was first given in \cite{deWit:2002vt}.
The gauge generators have the following form:
\begin{eqnarray}
X_\Lambda &=&\begin{cases}X_0=\theta_{0}{}^{\alpha'}\,t_{\alpha'}\cr
X_\lambda=\theta_\lambda{}^\delta\,t_\delta\end{cases}\,\,\,\,;\,\,\,\,
X_{0\lambda}{}^\delta =(X_0)_\lambda{}^\delta=-\theta_\lambda{}^\delta=-M_\lambda{}^\delta\in {\rm E} _{6(6)}\,.
\end{eqnarray}
where $M_\lambda{}^\delta$ is the twist-matrix depending in general on $78$
parameters, $t_{\alpha'}$, $\alpha'=1,\dots, 78$, are the ${\rm E} _{6(6)}$ generators, and
$t_\lambda$ are ${\rm E} _{7(7)}$ generators in the ${{\bf
27}}^\prime_{+2}$, defined in (\ref{e6frameg}). The non-vanishing components of $X_{MN}{}^P$
 are:
\begin{eqnarray}
X_{0\lambda}{}^\delta &=&-X_{\lambda0}{}^\delta=-X_{0}{}^\delta{}_\lambda=X_\lambda{}^\delta{}_0=-M_\lambda{}^\delta\,\,\,\,\,;\,\,\,\,\,\,
X_{\lambda \delta \gamma}=M_\lambda{}^{\lambda^\prime}\,d_{\lambda^\prime \delta \gamma}\,,\label{Xss}
\end{eqnarray}
where $d_{\lambda \delta \gamma}$ denotes the rank-3, symmetric, invariant tensor
 of ${\rm E} _{6(6)}$, introduced earlier.
 To obtain Eqs. (\ref{Xss})
we have used the property
 $(t_\lambda)_{ \delta \gamma}=d_{\lambda \delta \gamma}$.
The gauge algebra has the following structure:
\begin{eqnarray}
[X_0,\,X_\lambda]&=& M_\lambda{}^\delta\,X_\delta\,,
\end{eqnarray}
all other commutators vanishing. The linear constraint $X_{(MNP)}=0$ is satisfied since $X_{(\lambda\sigma\delta)}=M_{(\lambda}{}^\pi\,d_{\sigma\delta)\pi}=0$, which in turn follows from the fact that $M$ is a generator of ${\rm E}_{6(6)}$ and $d_{\sigma\delta\pi}$ is an ${\rm E}_{6(6)}$-invariant tensor.\par If $M $ is non--compact the
corresponding theory effectively depends only on six parameters
and the potential is of run--away type, namely there is no vacuum
solution. If, on the other hand, $M$ is compact, the theory has
Minkowski vacua and depends effectively on four mass parameters
$m_1,\,m_2,\,m_3,\,m_4$, since $M$ can always be reduced to an
element of the maximal torus of  ${\rm USp}(8)$.
\footnote{\label{fotss} The scalar potential is obtained from Eq. (\ref{potentialN8}) and the explicit from of the $X_{MN}{}^P$ tensor. It is conveniently computed using the parametrization of the scalar manifold defined in (\ref{Ld4d5}). We can easily convince ourselves that the $\mathbb{T}$-tensor does not depend on the Peccei-Quinn scalars $a^\lambda$. Indeed, when dressing the embedding tensor $\Theta$ in the ${\bf 78}_{+3}$ by means of the coset representative $L(\phi)$, the leftmost factor $e^{a^\lambda\,t_\lambda}$ acts as an ${\rm E}_{7(7)}$-transformation on $\Theta$ generated by the shift-generators $t_\lambda$ in the ${\bf 27}'_{+2}$. The result would be $\Theta$ plus terms, containing $a^\lambda$, with ${\rm O}(1,1)$-grading $+5$, which are however absent in the branching (\ref{Tdec45}) of the ${\bf 912}$ and thus do not occur in the $\mathbb{T}$-tensor. Thus the action of $e^{a^\lambda\,t_\lambda}$ on $\Theta$ leaves it unaltered so that the $\mathbb{T}$-tensor only depends on $\sigma$ and the five-dimensional scalar fields $\hat{\phi}$. The scalar potential is computed, using (\ref{potentialN8}) to have the general form:
\begin{equation}
V(\phi)=V(\sigma,\hat{\phi})\propto e^{-6\,\sigma}{\rm Tr}(\hat{M}(\hat{M}+\hat{M}^T))=\frac{ e^{-6\,\sigma}}{2}\,{\rm Tr}\left((\hat{M}+\hat{M}^T)^2\right)\ge 0\,,
\end{equation}
where $\hat{M}=L_5^{-1}M L_5$ depends on the five-dimensional scalar fields, being $L_5=L_5(\hat{\phi})$ the coset representative in five dimensions in the ${\bf 27}$ representation. The derivative of $V$ with respect to $\sigma$ vanishes if and only if $\hat{M}=-\hat{M}^T$, namely $\hat{M}$ is the Lie algebra $\mathfrak{usp}(8)$, maximal compact subalgebra of $\mathfrak{e}_{6(6)}$. The reader can easily verify that this condition also implies the vanishing of the derivatives of $V$ with respect to the five-dimensional scalar fields. With no loss of generality we can therefore fix $M$ to be in $\mathfrak{usp}(8)$ and the condition $\hat{M}=-\hat{M}^T$ fixes all five-dimensional scalar fields except those parametrizing ${\rm E}_{6(6)}$ generators which commute with $M$. The latter are flat directions of the potential. The potential therefore features a locus of extremal points parametrized by the flat directions. At these points the potential vanishes, so that they define Minkowski vacua, and being $V\ge 0$, these vacua are minima of $V$.
} These mass
parameters fix the scale of spontaneous supersymmetry breaking,
which can yield an $\mathcal{N}'=6,4,2$ or $\mathcal{N}'=0$ effective theory.
\subsubsection{The ${\rm SU}^*(8)$-Frame.} These gaugings were first explored in \cite{Hull:2002cv}. The gauge group is contained in ${\rm G_e}={\rm SU}^* (8)$. The branchings of the ${\rm E}_{7(7)}$-representations with respect to this group are formally the same as for the ${\rm SL}(8,\mathbb{R})$-case. In particular we can write a table of the same form as Table \ref{decblock}.
Just as in the ${\rm SL}(8,\mathbb{R})$-case, the gaugings involving only the electric vector fields $A^{{\tt a'b'}}_\mu$, are defined by $\Theta$ in the ${\bf 36}$ representation. This component of the embedding tensor, by means of an ${\rm SU}^*(8)$-transformation, can be cast in the form:
 \begin{equation}
 \theta_{{\tt a'b'}} = {\rm diag}\{
  \underbrace{1, \dots,1}_{2p},\,
  \underbrace{0, \dots, 0}_{2q} \} \;,
 \end{equation}
 The corresponding gauge group ${\rm CSO}^*(2p,\,2q)$ is defined as the subgroup of ${\rm SU}^*(8)$ preserving $\theta_{{\tt a'b'}}$ and contains the group ${\rm SO}^*(2p)$ as well as commuting translations generated by nilpotent generators. The case $p=4$ and $q=0$ corresponds to $G_g={\rm SO}^*(8)\sim {\rm SO}(2,6)$. The theory with gauge group ${\rm CSO}^*(6,2)$ has a Minkowski vacuum preserving $\mathcal{N}'=2$ supersymmetries and originating from a corresponding vacuum of the ${\rm SO}^*(6)$-gauged five-dimensional maximal supergravity, upon dimensional reduction on a circle \cite{Hull:2002cv}. In fact all the ${\rm CSO}^*(2p,\,2q)$ models feature Minkowski vacua preserving $\mathcal{N}'=2q$ supersymmetries \cite{Catino:2013ppa}.

\subsubsection{Gaugings from Flux Compactifications}\label{n8fluxc}
As we shall discuss in more detail in Sect. \ref{toroidalc}, besides the dimensionally reduced fields, also all possible fluxes, namely the background quantities which can be switched on in a toroidal compactification, naturally fit representations of the characteristic group $G_{int}={\rm GL}(n,\mathbb{R})$ of the internal torus $T^n$. These comprise the form-fluxes, namely the fluxes of field-strengths of ten-dimensional antisymmetric tensor fields.  Clearly the back-reaction of such quantities on the background metric will alter the geometry of the internal manifold which in general will be different from a torus. Such interactions are captured, in a manifestly $G_{int}$-covariant way, by the lower-dimensional gauged theory, in which the gauge group and all the couplings, encoded in a single embedding tensor, are induced by the fluxes. As anticipated in the Introduction, the embedding tensor provides the precise relation between the (generalized) background fluxes and the gauging of the lower-dimensional theory: The former are characteristic components of the embedding tensor \cite{Angelantonj:2003rq,D'Auria:2003jk,deWit:2003hq}. This identification can be made precise by associating all possible background quantities with representations of $G_{int}$ and identifying such representations in the branching of the embedding tensor representation ${\bf 912}$ with respect to the same group. The effect of any such generalized fluxes  in the compactification is reproduced by simply switching on the corresponding components of $\Theta$. The gauging procedure does the rest, leading, in a unique way, to the construction of the gauged model.
 Since $G$ encodes the conjectured string-duality group $G(\mathbb{Z})$, by virtue of property (\ref{dualgaug}), the duality covariant formulation of gauged supergravities provides a suitable framework where to systematically study the effect of string dualities on flux compactifications.\par
As discussed in Sect. \ref{N8MAB}, when deriving the maximal four-dimensional theory from Type II superstring theory, the resulting description, with respect to the $T$-duality group (or better ${\rm Spin}(6,6)$), is chiral: The RR fields are grouped in spinorial representations of ${\rm Spin}(6,6)$ whose chiralities depend on wether the higher dimensional theory is Type IIA or Type IIB. This also holds for the background quantities which can be switched on in the compactification (generalized fluxes), and which are encoded in the embedding tensor. The branching of the embedding tensor representation with respect to ${\rm SL}(2,\mathbb{R})\times {\rm SO}(6,6)$ is:
\begin{align}
\mbox{Type IIA}&:\,\,\,{\bf 912}\rightarrow\,{\bf (3,32_s)}+ {\bf (2,220+12)}+{\bf(1,352_c)}\,,\label{so66912a}\\
\mbox{Type IIB}&:\,\,\,{\bf 912}\rightarrow\,{\bf (3,32_c)}+ {\bf (2,220+12)}+{\bf(1,352_s)}\,.\label{so66912b}
\end{align}
Each ${\rm SO}(6,6)$-representation above represents a subset of components of the embedding tensor which is invariant under the action of proper T-duality transformations, while transformations in ${\rm O}(6,6)/{\rm SO}(6,6)$, which comprise T-dualities along an odd-number of internal directions, will alter the chirality of the spinorial representations ${\bf 32}$ and ${\bf 352}$ and map the two pictures into one another. From dimensional reduction one can deduce the $G_{int}$-representation
of the form-fluxes and pinpoint them in the branching of the ${\bf 912}$ representation. For instance the fluxes of the RR field-strength k-forms across $k$-dimensional cycles $\Sigma_{{\tt u_1\dots u_k}}$ of the torus transform in the following ${\rm SL}(6,\mathbb{R})$-representations:
\begin{equation}
F_{{\tt u_1\dots u_k}}\equiv \langle \int_{\Sigma_{{\tt u_1\dots u_k}}}\hat{F}^{(k)}\rangle\in \bigwedge{}^k \,{\bf 6}\,.
\end{equation}
where $\hat{F}^{(k)}$ are field-strengths of RR $(k-1)$-form fields $C_{(k-1)}$, which are present in the two theories in the \emph{democratic formulation} of \cite{Bergshoeff:2001pv}:\footnote{In this formulation the R-R sector consists of $C^{(p)}$-forms with $p=1,\,3,\,5,\,7,\,9$ in the Type IIA theory and $p=0,\,2,\,4,\,6,\,8$ in the Type IIB one. This is a redundant description since $p$ and $(8-p)$-forms are dual to one another, so duality relations have to be imposed to ensure the correct counting of degrees of freedom.} even-k tensors in Type IIA theory and odd-k in the Type IIB one.\par
These occur in the decomposition of one of the three ${\bf 32}$ representations in (\ref{so66912a}) and (\ref{so66912b}). According to (\ref{32s}) and consistently with the interpretation of their ten-dimensional origin, in the Type IIA picture there are only even-rank fluxes, since only the representation ${\bf 32}_s$ is present, while in the Type IIB picture the embedding tensor  only contains odd-rank fluxes:
\begin{align}
\mbox{Type IIA}&:\nonumber\\
&{\bf 32}_s\rightarrow\,\left.\{\hat{F}^{(0)},\,\hat{F}^{(2)},\,\hat{F}^{(4)},\,\hat{F}^{(6)}\}\right\vert_{T^6}\,,\nonumber\\
\mbox{Type IIB}&:\nonumber\\
&{\bf 32}_c\rightarrow\,\left.\{\hat{F}^{(1)},\,\hat{F}^{(3)},\,\hat{F}^{(5)}\}\right\vert_{T^6}\,.\label{32s32cF}
\end{align}
In Type IIA theory the 0-form flux $m=\hat{F}^{(0)}$ is also known as the \emph{Romans' mass}, in the presence of which the ten-dimensional theory is the so-called \emph{massive Type IIA} and was constructed in \cite{Romans:1985tz}.\par
The rank-3 flux-tensor $H_{{\tt uvw}}$ of the field-strength $H^{(3)}=dB^{(2)}$  of the $B$-field, belongs to the ${\rm SL}(6,\mathbb{R})$-representation ${\bf 20}$ which appears in the decomposition of one of the ${\bf 220}$ in both (\ref{so66912a}) and (\ref{so66912b}). The same $T$-duality representation also contains other ${\rm GL}(6,\mathbb{R})$-components describing tensors with different index structures:
\begin{equation}
{\bf 220}\rightarrow {\bf 20}[H_{{\tt uvw}}]+({\bf 84}+{\bf 6})[T_{{\tt uv}}{}^{{\tt w}}]+({\bf 84}'+{\bf 6}')[Q_{{\tt u}}{}^{{\tt vw}}]+{\bf 20}'[R^{{\tt uvw}}]\,,\label{220into}
\end{equation}
where the tensor $T_{{\tt uv}}{}^{{\tt w}}$ can be obtained from $H_{{\tt uvw}}$ through a T-duality $T^{({\tt w})}$ along the direction ${\tt w}=4,\dots, 9$ of the internal torus. As mentioned in the Introduction, it is an instance of \emph{geometric flux} and defines the geometry of a \emph{twisted-torus}, which we shall further deal with in Sect. \ref{toroidalc}. It can also be viewed as an internal \emph{torsion} \cite{Andrianopoli:2005jv}. If we only switch on this component, the quadratic constraint implies
$$T_{[{\tt u_1 u_2}}{}^{{\tt u}}\,T_{{\tt u_3]u}}{}^{{\tt v}}=0\,,$$
which is the Jacobi identity on $T_{{\tt uv}}{}^{{\tt w}}$, seen as the structure constants of the six-dimensional Lie group locally describing the internal manifold.
By further applying T-dualities along the ${\tt v}$ and the ${\tt u}$ directions one obtains the other two structures $Q_{{\tt u}}{}^{{\tt vw}},\,R^{{\tt uvw}}$:
\begin{align}
\begin{matrix}H_{{\tt uvw}} & \stackrel{T^{({\tt w})}}{\longrightarrow}& T_{{\tt uv}}{}^{{\tt w}}& \stackrel{T^{({\tt v})}}{\longrightarrow}&Q_{{\tt u}}{}^{{\tt vw}} &\stackrel{T^{({\tt u})}}{\longrightarrow}&R^{{\tt uvw}}\,,\cr
& & & &&&\cr
{\small ({\rm IIA/B})} & & {\small ({\rm IIB/A})}& &{\small ({\rm IIA/B})}&& {\small ({\rm IIB/A})}\end{matrix}
\end{align}
 which are the instances of the \emph{non-geometric fluxes} mentioned in the Introduction. What the branching (\ref{220into}) implies is that geometric as well as non-geometric fluxes are required to define a minimal set of background quantities, which includes the NS-NS 3-form flux $H$, and is closed under T-duality.\par
 The effect of a $T$-duality transformation along one direction $x^{{\tt u}}$ of the internal manifold changes the rank of the RR fluxes by one unit:
 \begin{equation}
 F_{{\tt u}{\tt u}_1\dots {\tt u}_k}\,\,\,\stackrel{T^{({\tt u})}}{\longleftrightarrow}\,\,\,\,\,F_{{\tt u}_1\dots {\tt u}_k}\,,
 \end{equation}
 thus interchanging the ${\bf 32}_s$ with the ${\bf 32}_c$, according to (\ref{32s32cF}).
Consider the maximal theory in the Type IIB picture discussed in Sect. \ref{N8MAB}. The branching of the ${\bf 56}$ and of the ${\bf 133}$ with respect to the relevant symmetry group ${\rm GL}(6,\mathbb{R})\times {\rm SL(2,\mathbb{R})_{IIB}}$ were given in Eqs. (\ref{eq:1IIB}) and (\ref{eq:2IIB}), while the branching of the ${\bf 912}$ with respect to the same group, capturing the most general gauging in this picture, is summarized in the following table \cite{deWit:2007mt}
%%%%%%%%%%%%%%%%%%%%%%%%%%%%%%%%%%%%%%%%%%%%%%%%%%%%%%%%%%%%%%
\begin{center}
{\scriptsize %%%%%%%%%%%%%%%%%%%%%%%%%%%%%%%%%%%%%%%%%%%%%%%%%%
\begin{tabular}{c|ccccc}
&$\!({\bf 6}',{\bf 1})_{-2}\!$ & $\!({\bf 6},{\bf 2})_{-1}\!$ &
$\!({\bf 20},{\bf 1})_{0}\!$ &$\!({\bf 6}',{\bf 2})_{+1}\!$ &
$\!({\bf 6},{\bf 1})_{+2}\!$
\\[.1mm]
\hline
%%%%%%%%%%%%%%%%%%%%%%%%%%%%%%%%%%%%%%%%%%%%%%%%%%%%%%%%%%%
$({\bf 1},{\bf2})_{-3}$ %%%%%%%%%%%%%%%%%%%%%%%%%%%%%%%%%%%
& {~} & $({\bf 6},{\bf 1})_{-4}$ & $({\bf 20},{\bf 2})_{-3}$ &
$({\bf 6}',{\bf 3}+{\bf 1})_{-2}$ & $({\bf 6},{\bf 2})_{-1}$
\\[1mm]
%%%%%%%%%%%%%%%%%%%%%%%%%%%%%%%%%%%%%%%%%%%%%%%%%%%%%%%%%%%%
$({\bf 15},{\bf 1})_{-2}$ %%%%%%%%%%%%%%%%%%%%%%%%%%%%%%%%%%
& $({\bf 6},{\bf 1})_{-4}$ & $({\bf 20},{\bf 2})_{-3}$ & $({\bf
6}'\!+\!{\bf 84}',{\bf 1})_{-2}$ & $({\bf 6}\!+\!{\bf 84},{\bf
2})_{-1}$ & $({\bf 70}\!+\!{\bf 20},{\bf 1})_{0}$
\\[1mm]
%%%%%%%%%%%%%%%%%%%%%%%%%%%%%%%%%%%%%%%%%%%%%%%%%%%%%%%%%%%
$({\bf 15}',{\bf 2})_{-1}$ %%%%%%%%%%%%%%%%%%%%%%%%
& $({\bf 20},{\bf 2})_{-3}$ & $({\bf 6}'\!+\!{\bf 84}',{\bf
1})_{-2}+ ({\bf 6}',{\bf 3})_{-2}$ & $({\bf 6}\!+\!{\bf 84},{\bf
2})_{-1}$ & $({\bf 20},{\bf 3}\!+\!{\bf 1})_{0}
   +({\bf 70}',{\bf 1})_0$
& $({\bf 6}'\!+\! {\bf 84}',{\bf 2})_{+1}$
\\[1mm]
%%%%%%%%%%%%%%%%%%%%%%%%%%%%%%%%%%%%%%%%%%%%%%%%%%%%%%%%%%%
$({\bf 1},{\bf 1})_{0}$ %%%%%%%%%%%%%%%%%%%%%%%%%%%%%%%%%%%
&$({\bf 6}',{\bf 1})_{-2}$ & $({\bf 6},{\bf 2})_{-1}$ & $({\bf
20},{\bf 1})_{0}$ & $({\bf 6}',{\bf 2})_{+1}$ & $({\bf 6},{\bf
1})_{+2}$
\\[1mm]
%%%%%%%%%%%%%%%%%%%%%%%%%%%%%%%%%%%%%%%%%%%%%%%%%%%%%%%%%%%%
$({\bf 35},{\bf 1})_{0}$ %%%%%%%%%%%%%%%%%%%%%%%%%%%%%%%%%%%
& $({\bf 6}'\!+\! {\bf 84}',{\bf 1})_{-2}$ & $({\bf 6}\!+\! {\bf
84},{\bf 2})_{-1}$ & $({\bf 70}\!+\!{\bf 70}'\!+\!{\bf 20},{\bf
1})_{0}$ & $({\bf 6}'\!+\! {\bf 84}',{\bf 2})_{+1}$ & $({\bf
6}\!+\! {\bf 84},{\bf 1})_{+2}$
\\[1mm]
%%%%%%%%%%%%%%%%%%%%%%%%%%%%%%%%%%%%%%%%%%%%%%%%%%%%%%%%%%%%
$({\bf 1},{\bf 3})_{0}$ %%%%%%%%%%%%%%%%%%%%%%%%%%%%%%%%%%%%
& $({\bf 6}',{\bf 3})_{-2}$ & $({\bf 6},{\bf 2})_{-1}$ & $({\bf
20},{\bf 3})_{0}$ & $({\bf 6}',{\bf 2})_{+1}$ & $({\bf 6},{\bf
3})_{+2}$
\\[1mm]
%%%%%%%%%%%%%%%%%%%%%%%%%%%%%%%%%%%%%%%%%%%%%%%%%%%%%%%%%%%%
$({\bf 15},{\bf 2})_{+1}$ %%%%%%%%%%%%%%%%%%%%%%%%%%%%%%%%%%
& $({\bf 6}\!+\!{\bf 84},{\bf 2})_{-1}$ & $({\bf 20},{\bf
3}\!+\!{\bf 1})_{0}+({\bf 70},{\bf 1})_0$ & $({\bf 6}'\!+\!{\bf
84}',{\bf 2})_{+1}$ & $({\bf 6}\!+\!{\bf 84},{\bf 1})_{+2} + ({\bf
6},{\bf 3})_{+2}$ & $({\bf 20},{\bf 2})_{+3}$
\\[1mm]
%%%%%%%%%%%%%%%%%%%%%%%%%%%%%%%%%%%%%%%%%%%%%%%%%%%%%%%%%%%%%
$({\bf 15}',{\bf 1})_{+2}$ %%%%%%%%%%%%%%%%%%%%%%%%%%
& $({\bf 70}'\!+\!{\bf 20},{\bf 1})_{0}$ & $({\bf 6}'\!+\! {\bf
84}',{\bf 2})_{+1}$ & $({\bf 6}\!+\!{\bf 84} ,{\bf 1})_{+2}$ &
$({\bf 20},{\bf 2})_{+3}$ & $({\bf 6}',{\bf 1})_{+4}$
\\[1mm]
%%%%%%%%%%%%%%%%%%%%%%%%%%%%%%%%%%%%%%%%%%%%%%%%%%%%%%%%%%%%%%
$({\bf 1},{\bf 2})_{+3}$ %%%%%%%%%%%%%%%%%%%%%%%%%%%%%%%%%%%%%
& $({\bf 6}',{\bf 2})_{+1}$ & $({\bf 6},{\bf 3}\!+\!{\bf 1})_{+2}$
& $({\bf 20},{\bf 2})_{+3}$ & $({\bf 6}',{\bf 1})_{+4}$ & {~}
\\
%%%%%%%%%%%%%%%%%%%%%%%%%%%%%%%%%%%%%%%%%%%%%%%%%%%%%%%%%%%%%%%
\end{tabular}\label{IIBtab}
}%%%%%%%%%%%%%%%%%%%%%%%%%%%%%%%%%%%%%
\end{center}
%%%%%%%%%%%%%%%%%%%%%%%%%%%%%%%%%%%%%%%%%%%%%%%%%%%%%%%%%%%%%
Only a restricted number of entries in the above Table can be interpreted in terms of fluxes of field-strengths of  Type IIB form-fields, or of $T$-duals thereof.
Within the ${\bf 912}$ all the components in the above table appear with multiplicity~1
apart from the $({\bf 6},{\bf 2})_{-1}$ and $({\bf 6}',{\bf 2})_{+1}$
which appear with multiplicity~2. It follows from the table that an
embedding tensor in the $({\bf 6}',{\bf 1})_{+4}$ defines a purely
electric gauging which thus automatically satisfies the quadratic
constraint. This corresponds to the theory induced by a five-form
flux. The ${\bf (20,2)}_{+3}$ representation describes the RR and NS-NS 3-form fluxes $(\theta_{{\tt uvw}}{}^\tau)=(F_{{\tt uvw}},\,H_{{\tt uvw}})$, which belong to a doublet, labeled by $\tau=1,2$, with respect to the Type IIB $S$-duality group ${\rm SL(2,\mathbb{R})_{IIB}}$. From the above table we see that switching only this component on, the gauging will involve the electric and magnetic vectors in the ${\bf (20,1)}_0$. The quadratic constraint (\ref{quadratic1}) then implies the following condition \cite{Angelantonj:2003rq,deWit:2003hq}:
\begin{equation}
\epsilon_{\tau\sigma}\epsilon^{{\tt u_1 u_2 u_3 v_1 v_2 v_3}} \theta_{{\tt u_1 u_2 u_3}}{}^\tau\theta_{{\tt v_1 v_2 v_3}}{}^\sigma=0\,\,\Leftrightarrow\,\,\,\,\int_{T^6} \hat{F}^{(3)}\wedge  \hat{H}^{(3)}=0\,,\label{tad}
\end{equation}
which is nothing but the tadpole cancelation condition \cite{Grana:2005jc}, in the absence of localized objects.\footnote{In general the presence of D-branes or O-planes is incompatible with an effective description within four-dimensional maximal supergravity.}
Condition (\ref{tad}) prevents the existence of vacua in the corresponding gauged supergravity \cite{deWit:2003hq}, whose scalar potential is positive (no-scale model) but has no stationary points.\par
 This gauged model has however a consistent $\mathcal{N}=4$ truncation, see end of Sect. \ref{Tdualcomp}, which no-longer contains the vectors  in the ${\bf (20,1)}_0$ originating from the 4-form field. This smaller model describes the flux-compactification of Type IIB theory on a $T^6/\mathbb{Z}_2$-orientifold \cite{Giddings:2001yu,Frey:2002hf,D'Auria:2002tc,D'Auria:2003jk} and the fields $A_{\mu {\tt uvw}}$, together with other fields,  are projected out by the orientifold projection. This feature has the desired implication of relaxing the quadratic constraint (\ref{tad}), thus allowing for the existence of Minkowski vacua with spontaneous supersymmetry breaking to $\mathcal{N}'=0,1,2,3$ at different scales. From the ten-dimensional point of view, the quadratic condition on the internal fluxes is relaxed by the presence of the space-filling O3-planes, which are not directly seen in the effective four-dimensional theory \cite{Grana:2005jc}:
\begin{equation}
N_{O3}=\frac{4}{(2\pi)^4 (\alpha')^2}\,\int_{T^6} \hat{H}^{(3)}\wedge  \hat{F}^{(3)}\,.
\end{equation}
In \cite{D'Auria:2003jk} a duality was found between this $\mathcal{N}=4$ model and a truncation of the maximal supergravity with Scherk-Schwarz gauging discussed above. This truncation is obtained as follows. Recall that the Scherk-Schwarz gauging is defined, in the ${\rm E}_{6(6)}$-frame, by an embedding tensor in the ${\bf 78}_{+3}$. We can interpret the maximal five-dimensional theory as
originating from the compactification of M--theory on a
six--torus, by branching the relevant ${\rm E}_{6(6)}$--representations with respect to the ${\rm
SL}(2,\mathbb{R})\times {\rm SL}(6,\mathbb{R})$ subgroup of ${\rm E} _{6(6)}$. In particular
the representation ${\bf 78}_{+3}$ of $\Theta$ branches as follows
\begin{eqnarray}
{\bf 78} \longrightarrow ({\bf 35},{\bf 1}) + ({\bf 1},{\bf 3}) + ({\bf
20},{\bf 2})\label{sl6decomp}
\end{eqnarray}
In \cite{D'Auria:2003jk} it was
observed that if the embedding tensor is restricted  to the
representation ${\bf (20,2)}$ in (\ref{sl6decomp}), upon an $\mathcal{N}=4$
truncation, the resulting theory
coincides with the $\mathcal{N}=4$ gauged supergravity mentioned above, describing Type IIB
superstring compactified on a $T^6/\mathbb{Z}_2$--orientifold in
the presence of RR and NS-NS 3--form fluxes $\hat{F}^{(3)},\,H^{(3)}$.

One can apply a similar analysis to the study of M-theory flux-compactifications \cite{Dall'Agata:2005ff,dft1,dft2,D'Auria:2005rv}, see also \cite{Fre':2006ut,Hull:2006tp}.
In the toroidal compactification of eleven-dimensional supergravity to four-dimensions the relevant group with respect to which to branch the ${\rm E}_{7(7)}$-representations is $G_{int}={\rm GL}(7,\mathbb{R})$, see end of Sect. \ref{N8MAB}.\par
 The most general gauging is described in this frame by arranging the irreducible ${\rm GL}(7,\mathbb{R})$-components of the embedding tensor in the  following table
 \begin{center}
\begin{tabular}{c|cccc}
    & ${\bf 7}'_{-3}$ & ${{\bf 21}}_{-1}$ & ${\bf 21}'_{+1}$ & $ {{\bf
  7}}_{+3}$ \\\hline
  ${{\bf 7}}_{-4}$ & ${\bf 1}_{-7}$ & ${{\bf 35}}_{-5}$ & $({\bf
  140}'+{\bf 7}')_{-3}$ & $({{\bf 28}}+{{\bf 21}})_{-1}$ \\
  ${\bf 35}'_{-2}$ & ${{\bf 35}}_{-5}$ &${\bf 140}'_{-3}$ & $({{\bf
  21}}+{{\bf 224}})_{-1}$ &$({{\bf 21}}'+{{\bf 224}}')_{+1}$ \\
  ${\bf 48}_0$ & $({\bf 140}'+{\bf 7}')_{-3}$ & $({{\bf 21}}+{{\bf
  28}}+{{\bf 224}})_{-1}$ &
   $({{\bf 21}}'+{{\bf 28}}'+{{\bf 224}}')_{+1}$
  & $({{\bf 140}}+{{\bf 7}})_{+3}$\\
  ${\bf 1}_0$ & ${\bf 7}'_{-3}$ & ${{\bf 21}}_{-1}$ & ${\bf 21}'_{+1}
  $ & $ {{\bf 7}}_{+3}$ \\
  ${{\bf 35}}_{+2}$ & $({{\bf 21}}+{{\bf 224}})_{-1}$ & $({{\bf
  21}}'+{{\bf 224}}')_{+1}$ &
  ${{\bf 140}}_{+3}$ & ${\bf 35}'_{+5}$ \\
  ${\bf 7}'_{+4}$ & $({{\bf 28}}'+{{\bf 21}}')_{+1}$ & $({{\bf 140}}+
  {{\bf 7}})_{+3}$ & ${\bf 35}'_{+5}$ & ${\bf 1}_{+7}$ \\
\end{tabular}
\end{center}
From this table we can infer that purely electric gaugings, which automatically satisfy the quadratic constraint, can only involve an embedding tensor in the ${\bf 1}_{+7}$ and in
the ${{\bf 35}}'_{+5}$. These components reproduce the effect, in the dimensional reduction,
of a seven-form $g_7$ \cite{Aurilia:1980xj} and a four-form flux $g_{\upalpha\upbeta\upgamma\updelta}$,
respectively. The ${\bf 140}_{+3}$ representation, on the other hand, corresponds to a geometric-flux (or internal torsion) $T_{\upalpha\upbeta}{}^\upgamma$, which also involves the magnetic vector fields $A^{\upalpha\upbeta}_\mu$ in the ${\bf 21}'_{+1}$. In the presence of $g_7$, $g_{\upalpha\upbeta\upgamma\updelta}$ and $T_{\upalpha\upbeta}{}^\upgamma$, the gauge algebra closes provided the following conditions are satisfied:
\begin{eqnarray}
T_{[\upalpha_1\upalpha_2}{}^\upbeta\,T_{\upalpha_3] \upbeta}{}^\upgamma=0\,,\\
T_{[\upalpha_1\upalpha_2}{}^\upbeta\,g_{\upalpha_3\upalpha_4\upalpha_5]\upbeta}=0\,.
\end{eqnarray}
The former is the Jacobi identity for the gauge algebra which describes the local geometry of the internal manifold. Both conditions, which in the four-dimensional theory are implied by the quadratic constraints, follow from the Bianchi identity of the eleven-dimensional four-form field-strength on the chosen background \cite{dft2,D'Auria:2005rv}.

\subsubsection{New Dyonic Gaugings}\label{dyonicg}
Exploiting the freedom in the initial choice of the sympectic frame it was recently possible to discover a new class of gaugings generalizing the original ${\rm CSO}(p,q,r)$ ones \cite{Dall'Agata:2011aa,Dall'Agata:2012sx,Dall'Agata:2012bb,Dall'Agata:2014ita}.
In \cite{Dall'Agata:2011aa} gaugings defined by an embedding tensor with non-vanishing components in both the ${\bf 36}$ and of the ${\bf 36}'$ representations of ${\rm SL}(8,\mathbb{R})$ were considered (see the previous paragraph on the gaugings in the  ${\rm SL}(8,\mathbb{R})$-frame). Such embedding tensor features both electric and magnetic components (\emph{dyonic embedding tensor}) and thus the quadratic constraint (\ref{quadratic1}) poses non-trivial restrictions on its entries. Moreover the electric frame is different from the ${\rm SL}(8,\mathbb{R})$-one. The ${\bf 36}$ and ${\bf 36}'$ components are described by two symmetric matrices $\theta_{{\tt ab}},\,\xi^{{\tt ab}}$, respectively, so that the embedding tensor reads:
\begin{eqnarray}
  \label{eq:csod}
 \Theta_M{}^\alpha=\left(\begin{matrix}\Theta_{[{\tt ab}]\,{{\tt d}}}{}^{{\tt c}}\cr \Theta^{[{\tt ab}]}{}_{{\tt d}}{}^{{\tt c}}\end{matrix}\right) &\!=\!&
  \left(\begin{matrix}\delta^{{\tt c}}_{[{\tt a}}\,\theta_{{\tt b}]{\tt d}}\cr  \delta_{{\tt d}}^{[{\tt a}}\,\xi^{{\tt b}]{\tt c}}\end{matrix}\right)\;,
\end{eqnarray}
where the index $\alpha$ only labels the $\mathfrak{sl}(8,\mathbb{R})$-generators, which are involved in the gauging. The quadratic constraint implies \cite{Dall'Agata:2011aa}
\begin{equation}
(\theta\xi)_{{\tt a}}{}^{{\tt c}}\delta_{{\tt b}}^{{\tt d}}-(\theta\xi)_{{\tt b}}{}^{{\tt d}}\delta_{{\tt a}}^{{\tt c}}=0\,,\label{xieta1}
\end{equation}
from which one derives
\begin{equation}(\theta\xi)_{{\tt a}}{}^{{\tt c}}=\frac{1}{8}\,{\rm Tr}(\theta\xi)\,\delta_{{\tt a}}^{{\tt c}}\,. \label{xieta2} \end{equation}
Following the same reference, we distinguish between two cases: $\theta_{{\tt ab}}$ non-singular and $\theta_{{\tt ab}}$ singular matrix. In the former case the quadratic constraint is satisfied by setting $\xi=c\, \theta^{-1}$. For the sake of notational simplicity, we shall denote by $\theta^{{\tt ab}}$ the components of the matrix $\theta^{-1}$, so that $\xi^{{\tt ab}}=c\,\theta^{{\tt ab}}$.
The matrix  $\theta_{{\tt ab}}$, by means of an ${\rm SL}(8,\mathbb{R})$-transformation, can be brought in the form (\ref{eq:cso}), where now the number $r$ of vanishing eigenvalues is zero and $p+q=8$. It therefore defines  ${\rm SO}(p,q)$-gauging and so does $\xi^{{\tt ab}}$. The reader can easily verify that $X^{{\tt ab}}=-\theta^{{\tt ae}}\theta^{{\tt bf}}\,X_{{\tt ef}}$. The difference with respect to the previous case in which $c=0$ is that the same ${\rm SO}(p,q)$-generators $X_{{\tt ab}}$, in the ${\rm SL}(8,\mathbb{R})$-frame, are gauged by a combination of electric and magnetic vector fields:
\begin{equation}
\Omega_{g\,\mu}=g\,A^M_{\mu} X_M=\frac{g}{2}\,(A^{{\tt ab}}_{\mu}\,X_{{\tt ab}}+A_{{\tt ab}\,\mu}\,X^{{\tt ab}})=
\frac{g}{2}\,(A^{{\tt ab}}_{\mu}\,-c\,\theta^{{\tt ae}}\theta^{{\tt bf}}\,A_{{\tt ef}\,\mu})\,X_{{\tt ab}}=g\,A_\mu^{\hat{\Lambda}}\,X_{\hat{\Lambda}}\,,\label{AcA}
\end{equation}
where the true\footnote{Here we denote by $A_\mu^{\hat{\Lambda}}$ the \emph{true electric vectors}, since they refer to the electric frame defined by the matrix $E$, to distinguish them from the electric and magnetic vector fields $A^{{\tt ab}}_{\mu},\,A_{{\tt ab}\,\mu}$ in the ${\rm SL}(8,\mathbb{R})$-frame.} electric vector fields are now $A_\mu^{\hat{\Lambda}}\equiv A^{{\tt ab}}_{\mu}\,-c\,\theta^{{\tt ae}}\theta^{{\tt bf}}\,A_{{\tt ef}\,\mu}$ and correspond to a different symplectic frame. We shall come back to these gaugings below and describe them as resulting from a symplectic transformation of the $X_{MN}{}^P$ tensor.\par
As far as the singular case ${\rm det}(\theta)=0$ is concerned, using ${\rm SL}(8,\mathbb{R})$ we can bring $\theta$ in the form  (\ref{eq:cso}), and Eqs. (\ref{xieta1}), (\ref{xieta2}) imply that $\xi$ has non-vanishing entries only in the  $r$-dimensional kernel of $\theta$. Acting then on $\xi$ by means of ${\rm GL}(r,\mathbb{R})$ (${\rm SL}(r,\mathbb{R})$ being the little group of $\theta$) and suitably rescaling the coupling constant $g$, we can further reduce $\xi$ to the form
\begin{eqnarray}
  \label{eq:csoeta}
\xi^{{\tt ab}} = {\rm diag}\{
  \underbrace{0, \dots,0}_{8-p'-q'}, \underbrace{+1, \dots, +1}_{p'},
  \underbrace{-1, \dots, -1}_{q'} \} \;,
\end{eqnarray}
where now the multiplicities of the  $+1$ and $-1$ diagonal entries are $p',\,q'$, respectively, with $p'+q'\le r$. We see that the electric vector fields $A^{{\tt ab}}_{\mu}$ in the ${\rm SL}(8,\mathbb{R})$-frame gauge a ${\rm CSO}(p,q,r)$ group defined by the $\theta$-matrix, while the magnetic ones $A_{{\tt ab}\,\mu}$ a ${\rm CSO}(p',q',r')$ group.  This is to be contrasted to the non-singular (i.e. $r=0$) case, in which the electric and magnetic vectors in the original frame gauge the same ${\rm SO}(p,q)$-group. Indeed in the singular case the two gauge algebras, ${\rm CSO}(p,q,r)$ and ${\rm CSO}(p',q',r')$, only overlap in a subset of $(p+q)(p'+q')$ commuting nilpotent generators, which are gauged by a linear combination of the electric and magnetic vectors. We have:
\begin{align}
G_g&=[{\rm SO}(p,q)_e\times {\rm SO}(p',q')_m]\ltimes \exp({\tt N})\,,\nonumber\\
{\tt N}&={\tt N}^{(p+q)(r-p'-q')}_e\oplus {\tt N}^{(p+q)(p'+q')}_{e+m}\oplus{\tt N}^{(p'+q')(r-p'-q')}_m\,,
\end{align}
where the subscript $e$ or $m$ means that the corresponding generators are gauged by the electric or magnetic vectors in the original frame while the subscript $e+m$ signifies that the generators are gauged by a linear combination of the two. For instance if $p=7,\,q=0,\,r=1$ and $p'=1,\,q'=0$ the corresponding gauge group is called \emph{dyonic ${\rm CSO}(7,0,1)$}, or simply \emph{dyonic ${\rm ISO}(7)$}: The electric vector fields  $A^{{\tt ab}}_{\mu}$ gauge the ${\rm SO}(7)$ subgroup, while the seven Abelian nilpotent generators are gauged by a linear combination of the  $A^{{\tt ab}}_{\mu}$ and the $A_{{\tt ab}\,\mu}$. The latter thus gauge the ${\rm CSO}(1,0,7)$ subgroup consisting of the seven translations only.
This model has recently attracted considerable interest since it was shown to describe a consistent truncation of the massive Type IIA theory on a background with topology of the form $AdS_4\times S^6$ \cite{Guarino:2015jca,Guarino:2015qaa,Guarino:2015vca}:\footnote{This is not a maximally supersymmetric backgorund, as opposed to the only maximally supersymmetric Freund-Rubin
solutions in $D = 11$ and $D = 10$: $AdS_4\times S^7$, $AdS_7\times S^4$ \cite{Nastase:1999cb}, $AdS_5\times S^5$, which are the near-horizon geometries of the $M2,\,M5$ and $D3$-branes, respectively. } The only non-vanishing component $m=\xi^{88}$ of $\xi^{{\tt ab}}$ is identified with the Romans' mass. The actual value of $m$ has no physical relevance since it can be changed by field-redefinitions and symmetries of the theory. This is consistent with the corresponding property of the dyonic ${\rm ISO}(7)$ gauging in which $\xi^{88}$, if non-vanishing, can always be set to $1$ using an ${\rm SL}(8,\mathbb{R})$-transformation and parity.

As far as the models with non-singular $\theta$ are concerned, as mentioned above, they are obtained by gauging the same ${\rm SO}(p,q)$ group, though in a different frame. Consider then two inequivalent frames admitting $G_g={\rm SO}(p,q)$ as gauge group, namely for each of which ${\rm SO}(p,q)\subset G_{el}$. Let $\hat{{\Scr R}}_v$ and ${{\Scr R}}_v$ be the corresponding symplectic duality representations of $G$. We can safely consider one of them ($\hat{{\Scr R}}_v$) as electric. The duality action of the gauge generators in $\hat{{\Scr R}}_{v*}$ and ${{\Scr R}}_{v*}$ are described by two tensors $X_{\hat{M}\hat{N}}{}^{{\hat{P}}}$ and $X_{MN}{}^P$, respectively, related by a suitable matrix $E$ (\ref{XEhatX}):
\begin{equation}
X_{\hat{M}\hat{N}}{}^{{\hat{P}}}=E_{\hat{M}}{}^M\,E_{\hat{N}}{}^N\,(E^{-1})_P{}^{{\hat{P}}}\,X_{MN}{}^P\,.\label{symdef}
\end{equation}
The matrices $\mathcal{M}(\phi)$ in the two frames are then related by (\ref{MEtra}).
The two embedding tensors describe the same gauge group provided that $\{X_{M}\}$ and $\{E\,X_M\,E^{-1}\}$ define different bases of the same gauge algebra $\mathfrak{g}_g=\mathfrak{so}(p,q)$ in $\mathfrak{e}_{7(7)}$. In other words $E$ should belong to the \emph{normalizer} of $\mathfrak{so}(p,q)$ in ${\rm Sp}(2n_v,\mathbb{R})$. At the same time the effect of $E$ should not be offset by a vector redefinition or a $G$-transformation, see (\ref{generalE}).
The duality action of $G_g$ in both $\hat{{\Scr R}}_{v*}$ and ${{\Scr R}}_{v*}$ is block-diagonal:
\begin{equation}
\hat{{\Scr R}}_{v*}[G_g]={{\Scr R}}_{v*}[G_g]=\left(\begin{matrix}G_g & \Zero \cr\Zero & G_g^{-T}\end{matrix}\right)\,.
\end{equation}
In \cite{Dall'Agata:2014ita} it was shown that the most general $E$ belongs to an ${\rm SL}(2,\mathbb{R})$-subgroup of ${\rm Sp}(56,\mathbb{R})$ and has the general form:\footnote{See \cite{Inverso:2015viq} for a generalization of this analysis to $\mathcal{N}<8$ theories.}
\begin{equation}
E=\left(\begin{matrix}a\,\Id & b\,\kappa\cr c\,\kappa & d\,\Id \end{matrix}\right)\in {\rm Sp}(56,\mathbb{R})\,\,\,;\,\,\,\,ad-bc=1\,,\label{Eimage}
\end{equation}
where $\kappa=(\kappa_{\Lambda\Sigma})$ is the $\mathfrak{so}(p,q)$-Cartan Killing metric, normalized so that $\kappa^2=\Id$. The most general
 ${\rm SL}(2,\mathbb{R})$-matrix can be written, using the Iwasawa decomposition, as follows:
 \begin{equation}
 \left(\begin{matrix}a & b\cr c & d\end{matrix}\right)=\left(\begin{matrix}\lambda & 0\cr 0 & \frac{1}{\lambda}\end{matrix}\right)\left(\begin{matrix}1 & \vartheta \cr 0 & 1\end{matrix}\right)\left(\begin{matrix}\cos(\omega) & \sin(\omega) \cr -\sin(\omega) & \cos(\omega)\end{matrix}\right)\,.
 \end{equation}
The leftmost block corresponds in $E$ to an unphysical rescaling of the vectors (in ${\rm GL}(28,\mathbb{R})$). The middle block realizes, in going from the unhatted frame to the hatted one, a constant shift in the generalized $\theta$-angle matrix $\R$: $\R_{\Lambda\Sigma}\rightarrow \R_{\Lambda\Sigma}+\vartheta\,\kappa_{\Lambda\Sigma}$. This can only have effects at the quantum level but does not affect the field equations \cite{Dall'Agata:2014ita}. The rightmost block has, on the other hand, an important bearing on the physics of the classical theory. Let $E(\omega)$ be the symplectic image (\ref{Eimage}) of this block only and let ${{\Scr R}}_{v}$ be the ${\rm SL}(8,\mathbb{R})$-frame, where the ${\rm SO}(p,q)$ gaugings were originally constructed  \cite{Cremmer:1978ds,deWit:1982bul} and  discussed earlier. For $\omega\neq 0$ this frame is no-longer electric, but is related to the electric one by $E(\omega)$. Using (\ref{elET}) we can write:
\begin{equation}
X_{\hat{\Lambda}}=\cos(\omega) X_\Lambda+ \sin(\omega)\kappa_{\Lambda\Sigma}\,X^\Sigma\,\,;\,\,\,\,0=-\sin(\omega) \kappa^{\Lambda\Sigma}\,X_\Sigma+ \cos(\omega)X^\Lambda\,,
\end{equation}
where $(\kappa^{\Lambda\Sigma})\equiv \kappa^{-1}=\kappa$. The above relation is easily inverted:
\begin{equation}
X_\Lambda=\cos{(\omega)}\,X_{\hat{\Lambda}}\,\,,\,\,\,\,\,\,X^\Lambda=\sin{(\omega)}\,\kappa^{\Lambda\Sigma}X_{\hat{\Sigma}}\,.
\end{equation}
We can then write the symplectic-invariant connection (\ref{syminvmc}) in the following way:
\begin{equation}
\Omega_{g\,\mu}=A^M_\mu\,X_M=A^\Lambda_\mu\,X_\Lambda+A_{\Lambda\,\mu}\,X^\Lambda=(\cos{(\omega)}\,A^\Lambda_\mu+
\sin(\omega)\,\kappa^{\Lambda\Sigma}\,A_{\Sigma\,\mu})
X_{\hat{\Lambda}}=A^{\hat{\Lambda}}_\mu\,X_{\hat{\Lambda}}\,,
\end{equation}
which is Eq. (\ref{AcA}), the parameter $\omega$ being related to $c$.
In other words, as pointed out above, the gauging defined by $X_M$ amounts to gauge, in the ${\rm SL}(8,\mathbb{R})$-frame, the same ${\rm SO}(p,q)$-generators by a linear combination of the electric $A^\Lambda_\mu$ and magnetic $A_{\Lambda\,\mu}$ vector fields. The true electric vectors are all and only those entering the gauge connection, that is $A^{\hat{\Lambda}}_\mu$, and define the electric frame. We shall denote by $\Theta[\omega]$ the corresponding embedding tensor.\par The gauged model can be constructed either directly in the ${\rm SL}(8,\mathbb{R})$-frame, using the covariant formulation discussed in Sect.\ \ref{sec:4}, or in the electric frame, along the lines described in Sect.\ \ref{sec:3}. The range of values of $\omega$ is restricted by the discrete symmetries of the theory. One of these is parity, see Sections\ \ref{gsg} and \ref{party}, whose duality representation ${\bf P}$ in the ${\rm SL}(8,\mathbb{R})$-frame has the form (\ref{Pmatrix}) \cite{Ferrara:2013zga}. The reader can verify that its effect on the $\Tb$-tensor (\ref{TT}) is:
\begin{align}
\Tb(\Theta[\omega],\phi)_{\underline{M}}&={\bf P}\star \mathbb{T}(\Theta[-\omega],\phi_p)\,,\label{PTtras}
\end{align}
by using the properties $${\bf P}_{\hat{M}}{}^{\hat{N}}\,{\bf P}^{-1} X_{\hat{N}}{\bf P}=X_{\hat{M}}\,\,;\,\,\,\,{\bf P}^{-1} E(\omega){\bf P}=E(-\omega)\,\,\,;\,\,\,\,{\bf P}^{-1}\LL(\phi){\bf P}=\LL(\phi_p)\,,$$
where $\phi_p$ denote the parity-transformed scalar fields (\ref{phipp}). Eq.\ (\ref{PTtras}) shows that parity maps $\phi$ into $\phi_p$ and $\omega$ in $-\omega$. In other words $\omega$ is a \emph{parity-odd parameter}. The overall ${\bf P}$ transformation on $\mathbb{T}$ in
(\ref{PTtras}) is ineffective since it will cancel everywhere in the Lagrangian, being ${\bf P}$ an ${\rm O}(2n_v)$-transformation.
Similarly we can use other discrete global symmetries of the ungauged theory, which include the ${\rm SO}(8)$-triality transformations in $ {\rm E}_{7(7)}$ for the ${\rm SO}(8)$-gauging, to further restrict the range of values of $\omega$.
One finds that \cite{Dall'Agata:2012bb,Dall'Agata:2014ita}
\begin{align}
&\omega\in\left[0,\frac{\pi}{8}\right]\,\,\,,\,\,\,\,\,\,\,\mbox{${\rm SO}(8)$ and ${\rm SO}(4,4)$-gaugings}\,,\nonumber\\
&\omega\in\left[0,\frac{\pi}{4}\right]\,\,\,,\,\,\,\,\,\,\,\mbox{all other non-compact ${\rm SO}(p,q)$-gaugings}\,.
\end{align}
These are called ``$\omega$-rotated'' ${\rm SO}(p,q)$-models or simply ${\rm SO}(p,q)_{\omega}$-models. The ${\rm SO}(8)_\omega$ ones in particular came as a surprise since they contradicted the common belief that the original de Wit-Nicolai ${\rm SO}(8)$-gauged model were unique. The $\omega$-parameter can be viewed as a de
Roo-Wagemans' angle \cite{deRoo:1985jh} in maximal supergravity.\par
Even more surprisingly the new class of gauged models discussed in this Section, both in the cases of singular and non-singular $\theta_{{\tt ab}}$-matrix, feature a broader range of vacua than their purely electric counterparts, defined by  $\xi^{{\tt ab}}=0$. Indeed, in the case of the ${\rm SO}(p,q)_{\omega}$-gaugings, the $\omega\rightarrow 0$ limit can be regarded as a singular one in which some of the vacua move to the boundary of the moduli space at infinity and thus disappear. Consider for instance the ${\rm SO}(8)_{\omega}$-models. They all feature an $AdS_4$, $\N=8$ vacuum at the origin with the same cosmological constant and mass spectrum as the original  ${\rm SO}(8)$ theory. The parameter $\omega$ manifests itself in the higher-order interactions of the effective theory.\par
\begin{figure}[H]
\begin{center}
\centerline{\includegraphics[width=0.8\textwidth]{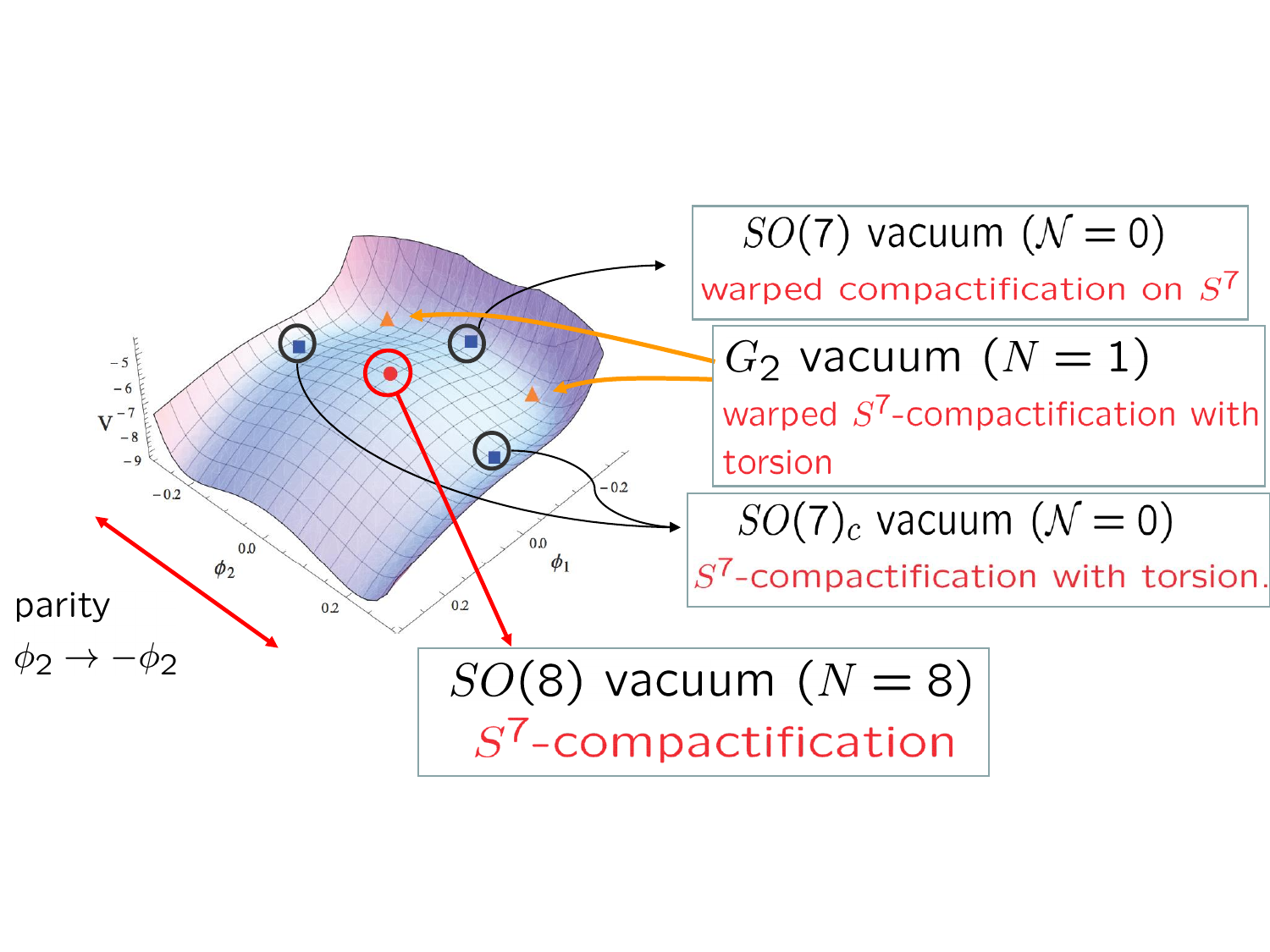}}
 \caption{\small The $G_2$-invariant anti-de Sitter bosonic backgrounds of the de Wit-Nicolai model, with their interpretation in terms of compactifications of the eleven-dimensional theory. The non-supersymmetric ones are unstable.}\label{fign8}
\end{center}
\end{figure}

\begin{figure}[H]
\begin{center}
\centerline{\includegraphics[width=1\textwidth]{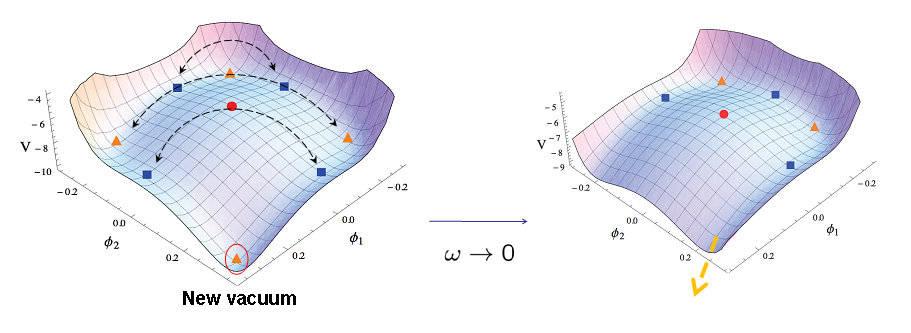}}
 \caption{\small On the left the $G_2$-invariant $AdS$-backgrounds of the ${\rm SO}(8)_\omega$ model, with $\omega=\frac{\pi}{8}$. The dashed black lines
 represent identifications of vacua due to a discrete symmetry of the theory which is a combination of triality and parity: $\omega\rightarrow \frac{\pi}{4}-\omega$, $\phi_1\leftrightarrow \phi_2$. All of them have an $\omega=0$ counterpart, except the lowest one, marked by a circle, which disappears in the $\omega\rightarrow 0$ limit. This vacuum is stable, non-supersymmetric.}\label{fign82}
\end{center}
\end{figure}

 The ${\rm SO}(8)_\omega$ models also feature new vacua which do not have counterparts in the $\omega=0$ model.
Fig.\ \ref{fign8} illustrates some of the vacua of the de Wit-Nicolai model ($\omega=0$), namely those which have a residual symmetry group $G_2\subset{\rm SO}(8)$, see also Appendix \ref{Aso8om}.
Fig.\ \ref{fign82} shows the $G_2$-invariant vacua of a particular ${\rm SO}(8)_\omega$ model and the disappearance of one of the vacua in the $\omega\rightarrow 0$ limit \cite{Dall'Agata:2012bb}. The two scalars $\phi_1,\,\phi_2$ which are $G_2$-singlets were defined in footnote \ref{fot}.
One may wonder if, just as in the ${\rm SO}(p,8-p)$-case, the gaugings with ${\rm det}(\theta)=0$ discussed earlier could be obtained by a symplectic deformation of the form (\ref{symdef}) from some ${\rm CSO}(p,q,r)$-embedding tensor. The answer is positive only for the ${\rm ISO}(p,7-p)\equiv {\rm CSO}(p,7-p,1)$ group and for the real forms of ${\rm SO}(4,\mathbb{C})^2\ltimes T^{16}$. In the former case the deformation matrix only depends on a discrete $\omega$-parameter and, up to ${\rm E}_{7(7)}\times \mathbb{Z}_2$ ($\mathbb{Z}_2$ representing the parity transformations), there are just two inequivalent cases: $\xi=0$ and $\xi\neq 0$. The latter models, on the other hand, admit a continuous deformation parameter $\omega$: Inequivalent theories are found for $\omega\in (0,\pi/4]$ if the gauge group features the same real form of the two ${\rm SO}(4,\mathbb{C})$-factors, while $\omega\in (0,\pi/2)$ if the real forms are different.\par The Cremmer-Scherk-Schwarz gaugings discussed earlier in Sect. \ref{gaugE6} (defined by the embedding tensor in the ${\bf 78}_{+3}$), as opposed to the above cases, do not admit any symplectic deformation \cite{Dall'Agata:2014ita}. \par
The dyonic counterparts of the ${\rm CSO}^*(2p,\,2q)$-gaugings in the ${\rm SU}^*(8)$-frame were also studied in \cite{Catino:2013ppa}, combining a ${\rm CSO}^*(2p,\,2q)$ gauged by electric vector fields with a ${\rm CSO}^*(2p',\,2q')$ gauged by magnetic ones. These models feature new Minkowski vacua with spontaneous supersymmetry breaking.\par
The vacua of the dyonic models discussed in this Section have been extensively studied \cite{Borghese:2012qm,Borghese:2012zs,Borghese:2013dja,Gallerati:2014xra} also in the context of renormalization group flows interpolating between (or simply originating from) $AdS$-vacua \cite{Guarino:2013gsa,Tarrio:2013qga,Pang:2015mra} and $AdS$-black holes \cite{Anabalon:2013eaa,Wu:2015ska}.\par
Determining a string or M-theory origin of the $\omega$-rotated models is, to date, an open problem \cite{deWit:2013ija,Lee:2015xga}. They seem to provide examples of what we named \emph{intrinsically non-geometric} models in the Introduction. The only exceptions so far are the \emph{dyonic non-semisimple gaugings} ${\rm CSO(p,q,r)}$. Indeed, as mentioned earlier, the ${\rm ISO}(p,\,7-p)$ gaugings were shown to be related to compactifications of massive Type IIA theory \cite{Guarino:2015jca,Guarino:2015qaa,Guarino:2015vca,Cassani:2016ncu}. The $p=7$ theory features $\mathcal{N}'=2$ \cite{Guarino:2015jca} and $\mathcal{N}'=3$ \cite{Gallerati:2014xra,Pang:2015vna,Pang:2015rwd} $AdS$-vacua, all corresponding to backgrounds with topology $AdS_4\times S^6$. The uplift of the generic ${\rm CSO(p,q,r)}$-model to Type IIA or Type IIB theory was eventually achieved in \cite{Inverso:2016eet}.

\section{ The $\N=2$\,,\; $D=4$ Supergravities}\label{N2sugras}
$\N=2$ four-dimensional supergravities are characterized by eight conserved supercharges and the field content consists in the following supermultiplets:
\begin{align}
\mbox{gravitational multiplet}&:\,\,\,\big[\;\;
1\;\times\;\underbrace{g_{\mu\nu}}_{{j=2}}\;,\quad
2\;\times\;\underbrace{\psi^A_\mu}_{{j=\frac32}}\;,\quad
1\;\times\;\underbrace{A^0_\mu}_{{j=1}}
\;\;\;\big]\,,\nonumber\\
\mbox{$n$ vector multiplets}&:\,\,\,\big[\;\;
1\;\times\;\underbrace{A^I_{\mu}}_{{j=1}}\;,\quad
2\;\times\;\underbrace{\lambda^{IA}}_{{j=\frac12}}\;,\quad
1\;\times\;\underbrace{z^i}_{{j=0}}
\;\;\;\big]\,,\nonumber\\
\mbox{$n_H$ hyper-multiplets}&:\,\,\,\big[\;\;
1\;\times\;\underbrace{\lambda^\alpha}_{{j=\frac{1}{2}}}\;,\quad
1\;\times\;\underbrace{q^u}_{{j=0}}
\;\;\;\big]\,,\label{N2m}
\end{align}
where $I,i=1,\dots, n$, $\alpha=1,\dots, 2n_H$ and $u=1,\dots,4n_H$. There are $n_v=n+1$ vector fields $A^\Lambda=(A^0_\mu,\,A^I_\mu)$.
As mentioned in Sect. \ref{ghsect} the scalar manifold of an $\mathcal{N}=2$ model has the following factorized form:
\begin{equation}
{\Scr M}_{{\rm scal}}={\Scr M}_{{\rm SK}}\times {\Scr M}_{{\rm QK}}\,,
\end{equation}
 where the special K\"ahler submanifold ${\Scr M}_{{\rm SK}}$ is parametrized by the $n$ complex scalar fields $z^i$ in the vector multiplets (the number of real scalar fields is therefore $n_s=2n$) while quaternionic K\"ahler one by the $4n_H$ real scalars $q^u$, $u=1,\dots, 4n_H$ in the hypermultimplets (\emph{hyper-scalars}).\par
 The holonomy group $H$ of the scalar manifold, as in all extended theories, splits according to (\ref{Hgroup}) into $H_R={\rm U}(2)$ and ${H}_{{\rm matt}}$, the latter acting on the fields in the vector and hyper-multiplets. At the same time $H$ is the product of the holonomy groups $H^{(SK)},\,H^{(QK)}$ of ${\Scr M}_{{\rm SK}},\, {\Scr M}_{{\rm QK}}$, respectively:
 \begin{align}
 H&=H^{(SK)}\times H^{(QK)}\,,
 \end{align}
 where $H_R$ and $H_{{\rm matt}}$ are split between these two groups as follows:
 \begin{align}
 H^{(SK)}&={\rm U}(1)\times H^{(SK)}_{{\rm matt}}\,\,;\,\,\,\,\,H^{(QK)}={\rm SU}(2)\times H^{(QK)}_{{\rm matt}}\,,\label{Hsplits}
 \end{align}
 where ${H}_{{\rm matt}}=H^{(SK)}_{{\rm matt}}\times H^{(QK)}_{{\rm matt}}$.\par
 Consistently with the notations adopted in the literature on $\mathcal{N}=2$ models we redefine the gaugini $\lambda_{IA}$ and the hyperini $\lambda_\alpha$ as follows:
 \begin{equation}
 \lambda^{iA}\equiv e^{-1\,Ii}\epsilon^{BA}\,\lambda_{IB}\,\,;\,\,\,\,\zeta_\alpha\equiv \frac{1}{\sqrt{2}}\,\lambda_\alpha\,.\label{redefsf}
 \end{equation}
 where $ e_{iI}$ is the complex vielbein matrix introduced above Eq. (\ref{PeIi}) of Section \ref{dcsl}. For the same reasons we redefine the metric and vielbein of the quaternionic K\"ahler manifold as follows:
 \begin{equation}
 {\Scr G}_{uv}\rightarrow \,h_{uv}=\frac{1}{2}\,{\Scr G}_{uv}\,\,;\,\,\,\,\,\mathcal{P}^{A\alpha}\rightarrow \,\mathcal{U}^{A\alpha}=\frac{1}{\sqrt{2}}\,\mathcal{P}^{A\alpha}\,.
 \end{equation}
 Below we shall first review the general structure of the ungauged theory, recalling the main facts about the special K\"ahler and quaternionic K\"ahler manifolds. The construction of the duality-covariant gauged theory will proceed along the same lines discussed earlier in Sect. \ref{sec:3}.
 The general electric gauging of an $\mathcal{N}=2$ model is discussed in \cite{Andrianopoli:1996cm} using a coordinate-independent, manifestly symplectic-covariant description of the special-K\"ahler manifold. More general gaugings were constructed in \cite{Andrianopoli:2011zj}. The duality-covariant gaugings have been constructed, in the framework of superconformal calculus, in \cite{deWit:2011gk}. A generic gauged $\mathcal{N}=2$ Poincar\'{e} supergravity can then be obtained from this analysis by suitably fixing the superconformal symmetry. The direct construction of the duality-covariant gauged $\mathcal{N}=2$ Poincar\'{e} supergravities was discussed in \cite{Andrianopoli:2015rpa}.\par
 Let us start recalling the definitions and main properties of special and quaternionic K\"ahler manifolds.

\subsection{Special K\"ahler Manifolds}\label{SK}
A special K\"ahler manifold \cite{deWit:1984rvr,Strominger:1990pd,D'Auria:1990fj,Ceresole:1995jg,Andrianopoli:1996cm} ${\Scr M}_{SK}$ is a Hodge- K\"ahler manifold endowed with  a flat, symplectic, holomorphic bundle  satisfying certain defining properties. Let us recall the main definitions. Consider a complex $n$-dimensional manifold with hermitian metric which, in a local patch parametrized by complex coordinates $z\equiv (z^i)$, has the following form:
\begin{equation}
ds^2_{{\tiny scal}}=2\,g_{i\bar{\jmath}}(z,\bar{z})\,dz^i\,d\bar{z}^{\bar{\jmath}}\,,
\end{equation}
where $g_{i\bar{\jmath}}=g_{\bar{\jmath}i}=(g_{\bar{\imath}j})^*=e_{i\,I}\,\bar{e}_{\bar{\jmath}}{}^I$.\par
 \emph{Definition: The manifold is K\"ahler if the 2-form:
\begin{equation}
K\equiv i\,g_{i\bar{\jmath}}\,dz^i\wedge d\bar{z}^{\bar{\jmath}}\,,\label{Kform}
\end{equation}
is closed}:
\begin{equation}
dK=0\,.\label{defKal}
\end{equation}
This property implies that \emph{locally} the metric can be written in terms of a \emph{K\"ahler potential} $\mathcal{K}(z,\bar{z})$:
\begin{equation}
g_{i\bar{\jmath}}=\partial_i\partial_{\bar{\jmath}}\mathcal{K}\,,\label{kmetric}
\end{equation}
From the above metric one derives the Levi-Civita connection with the following non-vanishing entries:
\begin{equation}
\Gamma_{jk}^i=g^{i\bar{\ell}}\partial_ig_{\bar{\ell}k}\,\,;\,\,\,\,\Gamma_{\bar{\jmath}\bar{k}}^{\bar{\imath}}
=g^{\bar{\imath}\ell}\partial_{\bar{\imath}}g_{{\ell}\bar{k}}\,,
\end{equation}
so that $\Gamma_{ki}^i=\partial_k\log({\rm det}(g_{i\bar{\jmath}}))$.\footnote{
In terms of the Levi-Civita connection one computes the Riemann tensor whose non-vanishing components have the characteristic expression:
$$R_{i\bar{\jmath}k\bar{\ell}}=g_{p\bar{\jmath}}\,\partial_{\bar{\ell}}\Gamma_{ik}^p\,,$$
while the Ricci tensor reads: $\mathcal{R}_{i\bar{\jmath}}\equiv R_{i\bar{\ell}\bar{\jmath}k}g^{k\bar{\ell}}=-\partial_i\partial_{\bar{\jmath}}\log({\rm det}(g_{k\bar{\ell}}))$.}

The K\"ahler potential is defined modulo a \emph{K\"ahler transformation}: if $U_{({\tt m})},\,U_{({\tt n})}$ are two overlapping patches on the manifold in which the K\"ahler potential is described by the functions  $\mathcal{K}_{({\tt m})}$ and $\mathcal{K}_{({\tt n})}$, respectively, in the intersection  $U_{({\tt m})}\cap U_{({\tt n})}$ we have
\begin{equation}
\mathcal{K}_{({\tt m})}=\mathcal{K}_{({\tt n})}-f_{({\tt m,n})}(z)-\bar{f}_{({\tt m,n})}(\bar{z})\,.  \label{fK}
\end{equation}
where $f_{({\tt m,n})}=f_{({\tt m,n})}(z)$ is a holomorphic transition function.
The defining condition (\ref{defKal}) also implies that, in a given patch, we can write $K$ in terms of a 1-form $\mathcal{Q}$:
\begin{equation}\label{connec}
K=d\mathcal{Q}\,\,;\,\,\,\,\,
  \mathcal{Q}= -\frac {i}{2}\left[\partial_i  \mathcal{K} \,dz^i- \partial_{\bar{\imath}}  \mathcal{K} \,d\bar{z}^{\bar{\imath}}\right]\,.
\end{equation}
Under a K\"ahler transformation $ \mathcal{Q}$ transforms  as a connection
\begin{equation}
\mathcal{Q}_{({\tt m})}=\mathcal{Q}_{({\tt n})}+\frac{i}{2}\left(df_{({\tt m,n})}-d\bar{f}_{({\tt m,n})}\right)=\mathcal{Q}_{({\tt n})}-d\left({\rm Im}(f_{({\tt m,n})})\right)\,,  \label{fQ}
\end{equation}
and was introduced earlier in Eq. (\ref{Qconn}).\par
K\"ahler manifolds describe complex scalar fields in $\mathcal{N}=1$ and $2$ rigid supersymmetric models \cite{Zumino:1979et}. The description of scalar fields in $\mathcal{N}=1$ supergravity requires an additional structure: The fermionic fields (gravitino and spin-$1/2$ fields) are minimally coupled to the composite connection $\mathcal{Q}$ and thus transform under K\"ahler transformations by means of corresponding ${\rm U}(1)$ transformations with given weights. The same holds for the gravitini, the gaugini and the hyperini in $\mathcal{N}=2$  supergravities. As $\mathcal{K}\rightarrow \mathcal{K}-f-\bar{f}$, $f=f(z)$, we have:
\begin{align}
\psi_{A\mu}&\rightarrow\,\,e^{\frac{i}{2}\,{\rm Im}(f)}\,\psi_{A\mu}\,,\nonumber\\
\lambda^{iA}&\rightarrow\,\,e^{-\frac{i}{2}\,{\rm Im}(f)}\,\lambda^{iA}\,,\nonumber\\
\zeta_\alpha &\rightarrow\,\,e^{-\frac{i}{2}\,{\rm Im}(f)}\,\zeta_\alpha\,.\label{Kweightsf}
\end{align}
In other words the gravitini, the gaugini and the hyperini transform as sections of a \emph{${\rm U}(1)$-bundle} ${\tt U}$ defined on the K\"ahler manifold, the ${\rm U}(1)$-connection being $\mathcal{Q}$.\par In general a field $\Phi(z,\bar{z})$ on the K\"ahler manifold is a \emph{section of the ${\rm U}(1)$-bundle ${\tt U}$ of weight $p$} if it transforms under a K\"ahler transformation $\mathcal{K}\rightarrow \mathcal{K}-f-\bar{f}$ as follows:
\begin{equation}
\Phi(z,\bar{z})\rightarrow e^{{i}\,p\,{\rm Im}(f)}\,\Phi(z,\bar{z})\,.
\end{equation}
Correspondingly we define a covariant derivative on the bundle as follows:
 \begin{align}
 {\Scr D}^{[{\rm U}(1)]}\Phi&\equiv d\Phi+i\,p\,\mathcal{Q}\,\Phi=\left(dz^i\,{\Scr D}^{({\rm U}(1))}_i+d\bar{z}^{\bar{\imath}}\,{\Scr D}^{[{\rm U}(1)]}_{\bar{\imath}}\right)\Phi\,:\,\,\,\begin{cases}{\Scr D}^{[{\rm U}(1)]}_i\Phi \equiv \left(\partial_i+\frac{p}{2}\,\partial_i\mathcal{K}\right)\Phi\cr {\Scr D}^{[{\rm U}(1)]}_{\bar{\imath}}\Phi \equiv \left(\partial_{\bar{\imath}}-\frac{p}{2}\,\partial_{\bar{\imath}}\mathcal{K}\right)\Phi\end{cases}\,.\label{DKp}
 \end{align}
 From Eqs. (\ref{Kweightsf}) we conclude that $p=1/2$ for $\psi_{A\mu}$ and $-1/2$ for $\lambda^{iA}$. The corresponding covariant derivatives in (\ref{genDdef})  are consistent with the above definition of the ${\rm U}(1)$-connection (\ref{DKp}).
 Together with the ${\rm U}(1)$-bundle, we define a \emph{holomorphic line-bundle} $\texttt{L}$ , whose sections $\tilde{\Phi}(z)$ are holomorphic functions transforming  under K\"ahler transformations as follows:
 \begin{equation}
\tilde{\Phi}(z)\rightarrow e^{p\,f(z)}\, \tilde{\Phi}(z)\,.
 \end{equation}
 We can map sections of the two bundles into one another as follows:
 \begin{equation}
 \Phi(z,\bar{z})=e^{\frac{p}{2}\mathcal{K}}\,\tilde{\Phi}(z)\,.\label{phitphi}
 \end{equation}
 The covariant derivative on $\texttt{L}$ is then easily derived:
 \begin{equation}
 {\Scr D}^{[\texttt{L}]}\tilde{\Phi}\equiv d\tilde{\Phi}+p\,\partial_i\mathcal{K}\,dz^i\,\tilde{\Phi}\,.
 \end{equation}
 An example of a section of $\texttt{L}$ is the holomorphic superpotential $W(z)$ with weight $p=1$. A section $\Phi$ of the ${\rm U}(1)$-bundle, with weight $p$, is said to be \emph{covariantly holomorphic} if
 \begin{equation}
 {\Scr D}^{[{\rm U}(1)]}_{\bar{\imath}}\Phi=\left(\partial_{\bar{\imath}}-\frac{p}{2}\,\partial_{\bar{\imath}}\mathcal{K}\right)\Phi=0\,.
 \end{equation}
The ${\rm U}(1)$-bundle (or, equivalently, the holomorphic line-bundle) is thus an additional structure defined on the K\"ahler manifold which is implied by local supersymmetry and which fixes the new couplings occurring in supergravity between the scalars and the fermion fields. Consistency of the theory then requires the first Chern class $c_1$ of $\texttt{L}$ to coincide, modulo an appropriate factor, with the de Rham cohomology class $[K]$ of the K\"ahler form.\footnote{The precise relation is $c_1=\frac{1}{4\pi}\,[K]\in \mathbb{Z}$. This is the quantization condition on the ``magnetic charge'' associated with the composite connection $\mathcal{Q}_\mu$ analogous to the Dirac quantization condition in electrodynamics.}
 A K\"ahler manifold endowed with a holomorphic line-bundle (and thus a ${\rm U}(1)$-bundle) satisfying this condition is
called \emph{Hodge-K\"ahler}. The scalar manifolds of $\mathcal{N}=1$ and $\mathcal{N}=2$ supergravities are of this kind.\par
In $\mathcal{N}=2$ theories the larger amount of supersymmetry implies an additional structure on the part of the scalar manifold spanned by $z^i$, which fixes their non-minimal couplings to the vector fields strengths through the matrices $\mathcal{I}_{\Lambda\Sigma}(z,\bar{z}),\,\mathcal{R}_{\Lambda\Sigma}(z,\bar{z})$, aside from an initial choice of the symplectic frame. This structure is a \emph{flat symplectic bundle} ${\mathcal{SV}}$, with structure group ${\rm Sp}(2+2n,\mathbb{R})$, which has to satisfy certain properties. There are more than one equivalent definitions of a special K\"ahler manifold ${\Scr M}_{SK}$ of local type, see \cite{deWit:1995tf,Andrianopoli:1996cm,Craps:1997gp}.\footnote{From now on by special K\"ahler manifold we shall always refer to special K\"ahler manifold of local type. Special K\"ahler manifold of rigid type are relevant to globally supersymmetric $\mathcal{N}=2$ theories \cite{Andrianopoli:1996cm} and will not be dealt with here.} We shall give only one characterization of this kind of manifolds, referring to the above works for a detailed discussion on this topic.\par We can consider the tensor product of the ${\rm U}(1)$-bundle and the symplectic one: ${\tt U}\times {\mathcal{SV}}$. Let $V(z,\bar{z})$ be a  section of it, in some local patch of ${\Scr M}_{SK}$, of ${\rm U}(1)$-weight $p=1$. It is described as a $(2n+2)$-component vector of complex functions $V(z,\bar{z})=(V^M(z,\bar{z}))$:
 \begin{equation}
V(z,\bar{z})=\begin{pmatrix}L^\Lambda(z,\bar{z})\cr M_\Lambda(z,\bar{z})\end{pmatrix}\,\,,\,\,\,\Lambda=0,\dots, n\,,
\end{equation}
The transition functions connecting overlapping coordinate patches $U_{({\tt m})},\,U_{({\tt n})}$ on ${\Scr M}_{SK}$, act on $V(z,\bar{z})$ as follows:
\begin{equation}
V_{({\tt m})}=e^{i\,{\rm Im}(f_{({\tt m,n})})}\,{\Scr R}_{v}^{({\tt m,n})}\,V_{({\tt n})}\,,\label{fV}
\end{equation}
where $f_{({\tt m,n})}=f_{({\tt m,n})}(z)$ is the holomorphic function describing a K\"ahler transformation and ${\Scr R}_{v}^{({\tt m,n})}$ is a constant ${\rm Sp}(2(n+1),\mathbb{R})$ matrix. Consistency requires the transition functions $e^{{\rm Im}(f_{({\tt m,n})})}$ and ${\Scr R}_{v}^{({\tt m,n})}$ should satisfy the cocycle condition on a triple overlap.
The flat symplectic structure, in particular, defines a homomorphic correspondence between isometries ${\bf g}:\,z\rightarrow z'$ on ${\Scr M}_{SK}$ and constant symplectic matrices:
\begin{equation}
{\bf g}\in G^{(SK)}\,\,\longrightarrow\,\,\,\, {\Scr R}_{v}[{\bf g}]\in {\rm Sp}(2+2n,\mathbb{R})\,,
\end{equation}
where ${\Scr R}_{v}[{\bf g}]=({\Scr R}_{v}[{\bf g}]^M{}_N)$ is defined in (\ref{fV}) and $G^{(SK)}$ denotes the isometry group of ${\Scr M}_{SK}$. This mapping ${\Scr R}_{v}$ is a symplectic representation of $G^{(SK)}$ on contravariant vectors. As usual we shall denote by ${\Scr R}_{v*}$ the corresponding representation on covariant vectors.\par
Let us denote by $U_i=(U_i^M)$ the ${\rm U}(1)$-covariant derivative of $V$ (we omit the superscript ${\rm U}(1)$):
\begin{equation}
U_i^M=\left(\begin{matrix}f_i^\Lambda\cr h_{\Lambda\,i}\end{matrix}\right)\equiv {\Scr D}_iV^M=\left(\partial_i+\frac{1}{2}\,\partial_i\mathcal{K}\right)V^M\,.
\end{equation}
Since $U_i$ has a coordinate index, its covariant derivative  should also contain the Levi-Civita connection on the base K\"ahler manifold:
\begin{equation}
{\Scr D}_iU_j\equiv \partial_iU_j+\frac{1}{2}\partial_i\mathcal{K}\,U_j-\Gamma_{ij}^k\,U_k\,.
\end{equation}
\emph{Definition.} We define a special K\"ahler manifold of local type to be a Hodge-K\"ahler manifold endowed with a flat symplectic bundle $\mathcal{SV}$ such that a section $V$ of ${\tt U}\otimes \mathcal{SV}$ exists that satisfies the following properties:\footnote{Not all these conditions are independent. For a discussion of the different definitions of special K\"ahler manifolds in terms of minimal sets of conditions, and their equivalence, see for instance \cite{Craps:1997gp}, and  \cite{Lledo:2006nr}.}
\begin{align}
{\Scr D}_{\bar{\imath}}\,V^M&=0\,,\label{sk1}\\
V^T\mathbb{C} \overline{V}&=i\,,\label{sk2}\\
{\Scr D}_i U_j&=i\,C_{ijk}\,g^{k\bar{k}}\,\overline{U}_{\bar{k}}\,,\label{sk3}\\
{\Scr D}_i\overline{U}_{\bar{\jmath}}&=g_{i\bar{\jmath}}\,\overline{V}\,,\label{sk4}\\
V^T\mathbb{C}U_i&=0\,,\label{sk5}
\end{align}
where $\mathbb{C}=(\mathbb{C}_{MN})$ is the usual ${\rm Sp}(2(n+1),\mathbb{R})$-invariant matrix;
\begin{equation}
\mathbb{C}\equiv \begin{pmatrix}{\bf 0} & {\bf 1}\cr -{\bf 1} & {\bf 0}\end{pmatrix}\,,
\end{equation}
The first condition (\ref{sk1}) is the requirement that $V$ be \emph{covariantly holomorphic}.
The quantity $C_{ijk}$ is a totally symmetric rank-3 tensor which is a covariantly-holomorphic section of the ${\rm U}(1)$-bundle with weight $p=2$: ${\Scr D}_{\bar{\ell}}C_{ijk}=0$. It is a characteristic quantity of the special K\"ahler manifold entering its Riemann tensor and the Pauli terms in the Lagrangian involving the gauginos, thus determining their anomalous magnetic moments.
From (\ref{sk1}) and (\ref{sk2}) it follows that:
\begin{equation}
V^T\mathbb{C}\overline{U}_{\bar{k}}=0\,.
\end{equation}
Given the section $V^M$, using (\ref{phitphi}), we can construct a holomorphic section $\Omega(z)=(\Omega^M(z))$ of the bundle $\texttt{L}\otimes \mathcal{SV}$ as follows
\begin{equation}
\Omega(z)=\begin{pmatrix}X^\Lambda(z)\cr F_\Lambda(z)\end{pmatrix}\equiv e^{-\frac{\mathcal{K}}{2}}\,V^M\,.
\end{equation}
\emph{Property 1.} Equation (\ref{sk2}) then allows to express the K\"ahler potential in terms of $\Omega(z)$ as follows:
\begin{equation}
\mathcal{K}(z,\bar{z})=-\log[i\,\overline{\Omega}(\bar{z})^T\mathbb{C}\Omega(z)]=-
\log[i\,(\overline{X}^\Lambda F_\Lambda-{X}^\Lambda \overline{F}_\Lambda)]\,.\label{Komapp}
\end{equation}
\emph{Definition.} Using $V$ and its covariant derivatives, we can construct the following matrix:
\begin{equation}
\mathbb{L}_c(z,\bar{z})^M{}_{\underline{N}}\equiv(V^M\,\epsilon_{AB},\overline{U}_I{}^M,\,\overline{V}^M\,
\epsilon^{AB},\,U^{I\,M})\,,\label{LcN2}
\end{equation}
$\underline{N}$ being the holonomy group index and
\begin{equation}
\overline{U}_I{}^M\equiv \bar{{ e}}^{-1}{}_{{I}}{}^{\bar{\imath}}\overline{U}^M_{\bar{\imath}}\,\,;\,\,\,\,
U^{I\,M}\equiv {{  e}}^{-1\,Ii}U_i^M\,,
\end{equation}
\emph{Property 2.} Eqs. (\ref{sk1})-(\ref{sk5}) imply the first of properties (\ref{propsLc}) of $\mathbb{L}_c$, from which the second follows as well:
\begin{equation}
\mathbb{L}_c^\dagger \mathbb{C}\mathbb{L}_c=\varpi\,\,;\,\,\,\,\mathbb{L}_c\varpi\mathbb{L}_c^\dagger =\mathbb{C}\,,\label{Lsymp}
\end{equation}
where $\varpi$ was defined in (\ref{varpi}). The matrix $\mathbb{L}_c$ provides a description of the geometry of the scalar manifold, as mentioned in Sect. \ref{dcsl}. In particular one defines the left-invariant one-form\footnote{Note that the quantity $\Omega^c$, although containing the same symbol, should not be mistaken for the holomorphic section $\Omega(z)$.} $\Omega^c\equiv \mathbb{L}_c^{-1}d\mathbb{L}_c$, as in (\ref{omegac}), and decomposes it in a vielbein and a connection matrices $\mathcal{P}_c,\,\mathcal{Q}_c$. These latter matrices contain all the characteristic geometric quantities of the special K\"ahler manifold. To evaluate them we first compute the inverse matrix $\mathbb{L}_c^{-1}$:
\begin{equation}
\mathbb{L}_c^{-1}=-\varpi\mathbb{L}_c^\dagger\mathbb{C}=\left(\begin{matrix}i\,\epsilon^{AB}\overline{V}^T\mathbb{C}\cr i\,U^{I}\mathbb{C}\cr -i\,V^T\mathbb{C}\epsilon_{AB}\cr -i\,\overline{U}_I\mathbb{C}\end{matrix}\right)\,.
\end{equation}
Using properties (\ref{sk1}) - (\ref{sk5}) we find for the components of $\mathcal{P}_c,\,\mathcal{Q}_c$ the expressions given in (\ref{PeIi}) and (\ref{QN2}). Vice versa, from the characteristic geometric data (i.e. metric, $C_{ijk}$, connection) encoded in $\mathcal{P}_c,\,\mathcal{Q}_c$ one can derive $\mathbb{L}_c$ as a solution to (\ref{omegac}), that is to the differential equations (\ref{dLP1}), (\ref{dLP2}). When the $\mathcal{N}=2$ theory is interpreted as originating from the compactification of Type II supergravity on a Calabi-Yau three-fold, the matrix $\mathbb{L}_c$ describes the periods of the internal manifold and Eqs. (\ref{dLP1}), (\ref{dLP2}) are the Picard-Fuchs equations satisfied by them \cite{Ceresole:1993qq,Andrianopoli:1996ve}. The integrability condition of these equations is just
\begin{equation}
d\Omega^c+\Omega^c\wedge \Omega^c={\bf 0}\,,\label{MCeq2}
\end{equation}
which, in the case of homogeneous manifolds $G^{(SK)}/H^{(SK)}$, is the Maurer-Cartan equation (\ref{MCeq}) for $G^{(SK)}$, and in general expresses the existence of a flat symplectic structure on the manifold. From (\ref{MCeq2}) one derives the $H_R$ and $H_{{\rm matt}}$-curvatures of the manifold, see Eqs. (\ref{RHaut}) and (\ref{RHmatt}).
In a coordinate basis, the non-vanishing components of the curvature tensor for a special K\"ahler manifold have the characteristic form:
\begin{equation}
R_{i\bar{\jmath}k\bar{\ell}}=g_{i\bar{\jmath}}\,g_{k\bar{\ell}}+g_{k\bar{\jmath}}\,g_{i\bar{\ell}}-C_{ijp}\,\overline{C}_{\bar{\jmath}\bar{\ell}\bar{q}}\,g^{p\bar{q}}\,.
\end{equation}
Using Eq. (\ref{MLL2}), we construct the symplectic, negative-defined, symmetric matrix $\mathcal{M}_{MN}(z,\bar{z})$ which encodes the non-minimal scalar-vector couplings.
This matrix defines the embedding of ${\Scr M}_{SK}$ in the symmetric manifold ${\rm Sp}(2+2n,\mathbb{R})/{\rm U}(n+1)$:
\begin{equation}
{\Scr M}_{SK}\,\hookrightarrow\,\,\frac{{\rm Sp}(2+2n,\mathbb{R})}{{\rm U}(n+1)}\,.
\end{equation}
In light of this, the matrix $\mathbb{L}_c$, although in  general not itself derived from a coset representative, since ${\Scr M}_{SK}$ may not be homogeneous, can be related to the  restriction to ${\Scr M}_{SK}$ of a coset representative in ${\rm Sp}(2+2n,\mathbb{R})/{\rm U}(n+1)$. This feature allows us to describe the geometry of ${\Scr M}_{SK}$ in terms $\mathbb{L}_c$ by means of the same formulas that we derived for homogeneous manifolds using the corresponding coset representative. Similarly the dependence of the $\mathbb{T}$-tensor (and thus of the fermion-shifts) on the special K\"ahler coordinates is obtained by dressing the $X_{MN}{}^P$ tensor by means of $\mathbb{L}_c(z,\bar{z})$ as if it were an ${\rm Sp}(2+2n,\mathbb{R})$-tensor.\par
\emph{Property 3.} Denoting as usual by $\mathcal{M}^{MN}$ the inverse of $\mathcal{M}_{MN}$, equation (\ref{MLc}), which is derived from (\ref{Lsymp}), implies the following identity:
\begin{equation}
U^{MN}\equiv g^{i\bar{\jmath}}\,U_i^M \overline{U}_{\bar{\jmath}}^N=-\frac{1}{2}\mathcal{M}^{MN}-\frac{i}{2}\,\mathbb{C}^{MN}-\overline{V}^MV^N\,,\label{UMN}
\end{equation}
One can find a symplectic frame in which $n$ out of the $n+1$ upper components $X^\Lambda(z)$ of $\Omega(z)$ are functionally independent. The $X^\Lambda$ can then be interpreted as projective coordinates for the manifold since under a K\"ahler transformation $X^\Lambda(z)\rightarrow e^{f(z)}\,X^\Lambda(z)$. In particular, in a patch in which $X^0\neq 0$, one can define \emph{special coordinates} $\tau^I$, $I=1,\dots,\,n$, as follows:
\begin{equation}
\tau^I\equiv \frac{X^I}{X^0}=\tau^I(z)\,,\label{specord}
\end{equation}
provided the Jacobian $\partial_i \tau^J$ is non-singular: ${\rm det}(\partial_i \tau^J)\neq 0$.
In this symplectic frame, named \emph{special coordinate frame}, we can write the lower part of the holomorphic section as a  function of the upper one, $F_\Lambda (z)=F_\Lambda (X^\Sigma(z))$ and, in particular, as the gradient with respect to $X^\Lambda$ of a holomophic function $F(X^\Lambda)$, called \emph{prepotential}\cite{Andrianopoli:1996cm}:
%\footnote{In the symplecic frame in which $X^\Lambda$ can regarded as homogeneous coordinates, $F_\Lambda$ can be described as functions of $X^\Sigma$. Due to the transformation property of $\Omega^M$ under K\"ahler transformations, $F_\Lambda$ must be homogeneous functions of degree one of $X^\Sigma$: $F_\Lambda(\lambda\,X)=\lambda F_\Lambda(X)$. By Euler's theorem we have: $X^\Lambda\frac{\partial}{\partial X^\Lambda}F_\Sigma=F_\Sigma$. Using this property and $V^T\mathbb{C}U_i=0$, which implies $X^\Lambda\partial_i F_\Lambda=F_\Lambda\partial_i X^\Lambda$, it is straightforward to prove that $ \frac{\partial}{\partial X^\Lambda}F_\Sigma=\frac{\partial}{\partial X^\Sigma}F_\Lambda$. This in turn implies that we can write $F_\Lambda=\frac{\partial}{\partial X^\Lambda}F$, $F=F(X)$ being the prepotential.}
\begin{equation}
F_\Lambda (X)= \frac{\partial}{\partial X^\Lambda}F(X)\,.\label{FLambdaF}
\end{equation}
To see this last feature we consider the property $V^T\mathbb{C}U_i=0$ from which it follows that \begin{equation}\Omega^T\mathbb{C}\partial_i\Omega=X^\Sigma\partial_iF_\Sigma-F_\Sigma \partial_i X^\Sigma=0\,.\end{equation}
Expressing the components of  $\Omega$ as functions of $X^\Lambda$, the above equation implies
\begin{equation}
0=X^\Sigma\frac{\partial}{\partial X^\Lambda} F_\Sigma-F_\Sigma \frac{\partial}{\partial X^\Lambda} X^\Sigma=X^\Sigma\frac{\partial}{\partial X^\Lambda} F_\Sigma-F_\Lambda\,\,\leftrightarrow\,\,\,F_\Lambda=\frac{1}{2}\,\frac{\partial}{\partial X^\Lambda}(X^\Sigma F_\Sigma)\,,\label{Fthings}\end{equation}
which is Eq. (\ref{FLambdaF}) if we identify
\begin{equation}
F(X)\equiv \frac{1}{2}\,X^\Sigma F_\Sigma\,.\label{Fdefxf}
\end{equation}
From (\ref{Fthings}) and the above definition we also find:
\begin{equation}
X^\Lambda\frac{\partial}{\partial X^\Lambda} F(X)=2\,F(X)\,,\label{Fdefxf2}
\end{equation}
which, by Euler's theorem, implies that $F(X)$ is a homogeneous function of degree 2 of $X^\Lambda$: $F(\lambda\, X^\Sigma)=\lambda^2\,F(X^\Sigma)$.

This function is the generalization to local $\mathcal{N}=2$ theories of the holomorphic function $F(X)$ which, in rigid $\mathcal{N}=2$ models,  defines the Lagrangian describing $n$ chiral superfields $\Phi^I$: $\int d^4x \int d^4\theta F(\Phi^I)$.
Using the homogeneity property of $F$ and (\ref{specord}), we can define the holomorphic function ${\Scr F}(\tau^I)$ of the special coordinates:
\begin{equation}
F(X)=(X^0)^2\,{\Scr F}(t)\,,
\end{equation}
in terms of which the lower components $F_\Lambda$ of the holomorphic section read
\begin{equation}
F_0=X^0\,\left(2\,{\Scr F}(t)-\tau^I\,\partial_I{\Scr F}(t)\right)\,\,;\,\,\,\,F_I=X^0\,\partial_I{\Scr F}(t)\,,\label{prepotentialF}
\end{equation}
where $\partial_I{\Scr F}\equiv \frac{\partial}{\partial \tau^I}{\Scr F}$. Using Eq. (\ref{Komapp}), modulo a K\"ahler transformation,  we can rewrite the K\"ahler potential in the following form:
\begin{equation}
\mathcal{K}(t,\bar{t})=-{\rm log}\left[2i\,\left({\Scr F}-\overline{{\Scr F}}\right)+i\,(\bar{\tau}^{\bar{I}}-\tau^I)\left(\partial_I{\Scr F}+\partial_{\bar{I}}\overline{{\Scr F}}\right)\right]\,.\label{Kprepot}
\end{equation}
From (\ref{sk3}) we also find $C_{IJK}=e^{\mathcal{K}}\partial_I\partial_J\partial_K\,{\Scr F}$ and the following expression for the matrix ${\Scr N}_{\Lambda\Sigma}$:
\begin{equation}
{\Scr N}_{\Lambda\Sigma}=\overline{F}_{\Lambda\Sigma}+2i\,\frac{{\rm Im}(F_{\Lambda\Gamma}){\rm Im}(F_{\Sigma\Pi})\,L^\Gamma\,{L}^\Pi}{{\rm Im}(F_{\Pi\Gamma})\,{L}^\Pi L^\Gamma}\,,
\end{equation}
where ${F}_{\Lambda\Sigma}\equiv \frac{\partial^2}{\partial X^\Lambda \partial X^\Sigma}\,F$. This matrix, being a homogeneous function of $X^\Lambda $ of weight zero, has also K\"ahler weight zero, and thus so does ${\Scr N}_{\Lambda\Sigma}$. We therefore see that, in the special coordinate frame, the whole $\mathcal{N}=2$ Lagrangian can be written in terms of a single holomorphic prepotential $F(X)$ and its derivatives.\par
It is important to emphasize that there are symplectic frames in which a prepotential $F(X)$ does not exist. These were used to evade no-go theorems \cite{Witten:1982df,Cecotti:1984rk} according to which \emph{spontaneous $\mathcal{N}=2\rightarrow \mathcal{N}=1$ supersymmetry breaking with vanishing cosmological constant cannot occur within gauged models constructed in frames admitting a prepotential function}. The first gauged $\mathcal{N}=2$ supergravity models \cite{Ferrara:1995xi,Tsokur:1996su,Tsokur:1996vk,Fre:1996js} featuring spontaneous partial supersymmetry breaking on Minkowski space-time were indeed devised in frames with no $F(X)$. This requirement on the symplectic frame can be avoided in the context of duality-covariant gaugings. It was indeed shown in \cite{Louis:2009xd} that partial supersymmetry breaking can be achieved in any symplectic frame (and in particular in  one in which the prepotential does exist), using an embedding tensor with both electric and magnetic components. This is not in contradiction with the aforementioned no-go theorems since in the electric frame associated with such embedding tensor, see Sect. \ref{backelectric}, a prepotential does not exist. We shall be dealing with gaugings of $\mathcal{N}=2$ models in some detail later in Sect. \ref{gaugspecialg} and subsequent Sections.

\subsubsection{Isometries}\label{isomsk}
The isometries of a K\"ahler metric, under some general assumptions which hold for the K\"ahler manifolds we consider,
\footnote{This statement applies to K\"ahler  manifolds which are non Ricci-flat and in which the action of the holonomy group on the tangent space is irreducible.} can either be holomorphic (i.e. $z^i\rightarrow f^i(z^j)$) or anti-holomorphic (i.e. $z^i\rightarrow f^i(\bar{z}^{\bar{\jmath}})$, as it is the case for the parity transformation $z\rightarrow-\bar{z}$ in the example discussed in Sect. \ref{gsg}).
The identity sector of the isometry group can only consists of holomorphic isometries and thus its infinitesimal transformations are described by holomorphic Killing vectors $k_a^i(z^j)$:
\begin{equation}
z^i\rightarrow z^i+\epsilon^a k_a^i(z)\,\,.
\end{equation}
Clearly the infinitesimal variation of $\bar{z}^{\bar{\imath}}$ is defined by $k_a^{\bar{\imath}}(\bar{z})=(k_a^i(z))^*$, so that the Killing vector $k_a$ on the tangent space is
$$k_a=k_a^i(z)\partial_i+k_a^{\bar{\imath}}(\bar{z})\partial_{\bar{\imath}}\,.$$
As isometry generators, $k_a$ satisfy conditions (\ref{Killcond})
\begin{equation}
{\Scr D}_ik_{a\,j}+{\Scr D}_jk_{a\,i}=0\,\,;\,\,\,\,{\Scr D}_ik_{a\,\bar{\jmath}}+{\Scr D}_{\bar{\jmath}}k_{a\,i}=0\,,\label{kcondc}
\end{equation}
where $k_{a\,i}=g_{i\bar{\jmath}}\,k_a^{\bar{\jmath}}$, $k_{a\,\bar{\jmath}}=g_{i\bar{\jmath}}\,k_a^i$.\par
Let us focus on the holomorphic isometries, in the identity sector $G^{(SK)}_{0}$ of $G^{(SK)}$.
By Eqs. (\ref{fK}) and (\ref{fV}), associated with each element ${\bf g}\in G^{(SK)}_{0}$ there is a symplectic matrix ${\Scr R}_v[{\bf g}]$ and a holomorphic function $f_{{\bf g}}(z)$ describing the K\"ahler compensating transformation:
\begin{align}
{\bf g}&: z^i\rightarrow z^{\prime i}(z^j)\,\,;\,\,\,\,\,\begin{cases}
\mathcal{K}(z',\bar{z}')=\mathcal{K}(z,\bar{z})-f_{{\bf g}}(z)-\bar{f}_{{\bf g}}(\bar{z})\,,\cr
\Omega^M(z')=e^{f_{{\bf g}}(z)}\,{\Scr R}_v[{\bf g}]^M{}_N\,\Omega^N(z)\,,\cr
V^M(z',\bar{z}')=e^{i\,{\rm Im}(f_{{\bf g}})}\,{\Scr R}_v[{\bf g}]^M{}_N\, V^N(z,\bar{z})\,.
\end{cases}\label{gRf}
\end{align}
In K\"ahler manifolds invariance of the metric also implies invariance of the K\"ahler form $K$ (\ref{Kform}). Denoting by $\ell_a$ the Lie derivative with respect to the Killing vector $k_a$ and by $\iota_a$ the contraction of a form with $k_a$, invariance of $K$ under infinitesimal isometries, $\ell_a K=0$, implies:
\begin{equation}
\ell_aK=d(\iota_a K)=0\,\,\Rightarrow\,\,\,\,\iota_a K=-d {\Scr P}_a\,,\label{defPa}
\end{equation}
 where last equation holds locally.\par
 \emph{Definition.} Equation (\ref{defPa}) defines the momentum maps ${\Scr P}_a(z,\bar{z})$ and can be written in components as follows:
 \begin{equation}
k_a^i=i\,g^{i\bar{\jmath}}\,\partial_{\bar{\jmath}}{\Scr P}_a\,,\,
\,\,k_a^{\bar{\imath}}=-i\,g^{\bar{\imath}i}\,\partial_{i}{\Scr P}_a\,,\label{kpal}
\end{equation}
Note that the momentum maps are defined by the above relations modulo an additive constant.
Let $t_a$ be the abstract generators of the isometry group $G^{(SK)}$ of ${\Scr M}_{SK}$ and $k_a$ the corresponding Killing vectors.\footnote{\label{abcindex} Due to a shortage of indices we use throughout Section \ref{N2sugras} $a,b,c,\dots$ to label special K\"ahler isometries. Elsewhere we also use them as rigid four-dimensional space-time indices. Their meaning should be clear from the context.} The structure of the isometry algebra is described by Eqs. (\ref{talg}) and (\ref{kalg}):
 \begin{equation}
 [t_a,\,t_b]={\rm f}_{ab}{}^c\,t_c\,\,,\,\,\,\,\,[k_a,\,k_b]=-{\rm f}_{ab}{}^c\,k_c\,.\label{tkalg}
 \end{equation}
 According to Eqs. (\ref{gRf}) with each isometry generator $t_a$ we associate, aside from the Killing vector $k_a$, a holomorphic function $f_a(z)$ and a symplectic matrix $(t_{a\,M}{}^N)\equiv {\Scr R}_{v*}[t_a]$: $t_{a\,M}{}^P\mathbb{C}_{NP}=t_{a\,N}{}^P\mathbb{C}_{MP}$
  \begin{equation}
  t_a\in \mathfrak{g}_{SK}\,\,\longrightarrow\,\,\,\,(k_a,\,f_a(z),\,t_{a\,M}{}^N)\,.
  \end{equation}
  Equations (\ref{gRf}), in terms of infinitesimal isometry generators, are equivalent to
 \begin{eqnarray}
\label{Lie1}\ell_a\mathcal{K}&=&k_a^i \partial_i\mathcal{K}+k_a^{\bar{\imath}}\partial_{\bar{\imath}}\mathcal{K}=-(f_a+\bar{f}_a)\\
\label{Lie2}\ell_a\Omega^M&=&k_a^i\partial_i\Omega^M=-t_{aN}{}^M\,\Omega^N+f_a(z)\Omega^M\,,\\
\label{Lie3}\ell_a V^M&=&(k_a^i \partial_i+k_a^{\bar{\imath}}\partial_{\bar{\imath}})V^M=-t_{aN}{}^M\,V^N+\frac{(f_a-\bar{f}_a)}{2}\,V^M\,,\label{kV}
\end{eqnarray}
\emph{Property 4.} The Killing vectors satisfy the Poisson-bracket relation \cite{D'Auria:1990fj}:
\begin{equation}\label{lhsPP}
\{{\Scr P}_a,\,{\Scr P}_b\}\equiv K\left(k_a,k_b\right)=2\,ig_{i\bar{\jmath}}\,k^i_{[a}\,k^{\bar{\jmath}}_{b]}=
2\,ig^{k\bar{\jmath}}\partial_{\bar{\jmath}}{\Scr P}_{[a}
\partial_{k}{\Scr P}_{b]}= -f_{ab}{}^c {\Scr P}_{c}\,.
\end{equation}
\emph{Proof.} This property can be proven by computing the covariant derivative of the left hand side and using (\ref{kcondc}), the second of Eqs. (\ref{tkalg}), and (\ref{kpal}). One finds:
\begin{equation}
{\Scr D}_kK\left(k_a,k_b\right)=-{\rm f}_{ab}{}^c \partial_k{\Scr P}_{c}\,,
\end{equation}
which implies
\begin{equation}
K\left(k_a,k_b\right)=-\left({\rm f}_{ab}{}^c {\Scr P}_{c}+C_{ab}\right)\,,
\end{equation}
where the constants $C_{ab}$ satisfy the cocycle condition ${\rm f}_{[ab}{}^d\,C_{c]d}=0$. Under general assumptions on the $G^{(SK)}$,\footnote{The condition is that the isometry algebra have a trivial second cohomology group $H^{(2)}(\mathfrak{g}_{SK})$ \cite{Galicki:1986ja}. This certainly holds for semi-simple Lie algebras.} $C_{ab}$ can be written as $C_{ab}={\rm f}_{ab}{}^c\,C_c$ and the constant vector $C_a$ be reabsorbed in a redefinition of ${\Scr P}_a$, so that Eq. (\ref{lhsPP}) follows.\\
\emph{Property 5.} Eqs. (\ref{kpal}) imply:
\begin{eqnarray}
{\Scr P}_a&=&-\frac{i}{2}\,\left(k_a^i \partial_i\mathcal{K}-k_a^{\bar{\imath}}\partial_{\bar{\imath}}\mathcal{K}\right)+{\rm Im}(f_a)=i\,k_a^{\bar{\imath}}\partial_{\bar{\imath}}\mathcal{K}+ i\, \bar{f}_a=-i\,k_a^i \partial_i\mathcal{K}- i\, f_a\,,\label{totti}
\end{eqnarray}
\emph{Proof.} To prove this property, let us invert the metric in one of Eq.s (\ref{kpal}):
\begin{equation}
g_{i\bar{\jmath}}\,k_a^i=i\,\partial_{\bar{\jmath}}{\Scr P}_a\,,\label{kpalA}
\end{equation}
and use (\ref{kmetric}). Recalling the general condition on K\"ahler-manifold isometries $\partial_{\bar\jmath}\,k^i_a(z)=0$, we find:
\begin{equation}
\partial_{\bar\jmath}(k_a^i\,\partial_i\mathcal{K})=i\,\partial_{\bar{\jmath}}{\Scr P}_a\,,
\end{equation}
which implies
\begin{equation}
k_a^i\, \partial_i\mathcal{K}=i\,{\Scr P}_a\,-\hat{f}_a(z)\,. \label{pkA1}
\end{equation}
This would reproduce (\ref{totti}) if $\hat{f}_a(z)=f_a(z)$. To fix the holomorphic functions $\hat{f}_a(z)$, it is sufficient to consider the holomorphic derivative of (\ref{Lie1}), which implies:
\begin{equation}
g_{j\bar\jmath}k^{\bar\jmath}_a+\partial_j(k^i_a\partial_i \mathcal{K})= -\partial_j f_a\,,
\end{equation}
that is, using (\ref{kpal}):
\begin{equation}
-i \,\partial_j {\Scr P}_a\,+\partial_j(k^i_a\partial_i \mathcal{K})= -\partial_j f_a\,. \label{pfA}
\end{equation}
By inserting now (\ref{pkA1}) in (\ref{pfA}), one finally finds $\hat{f}_a(z)=f_a(z)$, modulo an additive constant that, as discussed above, can be absorbed in the definition of ${\Scr P}_a$.\par
Let us recall some more properties of the Killing vectors and their description in terms of momentum maps.\\
 \emph{Property 6.} Using (\ref{kV}) and (\ref{totti}) we find:
\begin{equation}
k_a^i\,U_i^M=-t_{aN}{}^M\,V^N+i\,{\Scr P}_a\,V^M\,.\label{kaUi}
\end{equation}
 \emph{Property 7.} Contracting the above equation with $\mathbb{C}\overline{V}$ and using the special geometry relations $V^T\mathbb{C} \overline{V}=i,\,V^T\mathbb{C}U_i=0$, we find:
\begin{equation}
{\Scr P}_a=-V^N\,t_{aNM}\overline{V}^M=-\overline{V}^N\,t_{aNM}\,V^M\,,\label{PVtV}
\end{equation}
where we have defined, as usual, $t_{aNM}\equiv t_{aN}{}^P\mathbb{C}_{PM}=t_{aMN}$. \\
 \emph{Property 8.} Let us now prove the general property \cite{deWit:1984pk,deWit:2011gk}:
\begin{equation}
t_{a\,MN}\Omega^M\Omega^N=0\,\,,\,\,\,\,\forall t_a\,.\label{tOmOm}
\end{equation}
 \emph{Proof.} This property is readily derived by contracting  (\ref{kaUi}) with $\mathbb{C}\Omega$ and using (\ref{sk5}), i.e. $V^T\mathbb{C}U_i=0$, which implies
\begin{equation}
\Omega^T\mathbb{C} \partial_i \Omega=0\,.
\end{equation}

\subsection{Quaternionic K\"ahler Manifolds}\label{QMans} Here we briefly recall the definition of a quaternionic  K\"ahler manifold\footnote{We shall be interested in non-compact quaternionic K\"ahler manifolds with negative curvature as only these are relevant to supergravity.} $\mathcal{M}_{QK}$ \cite{Bagger:1983tt,Hitchin:1986ea,Galicki:1986ja,D'Auria:2001kv} and fix the notations. A quaternionic-K\"ahler manifold $\mathcal{M}_{QK}$ of real dimension $4n_H$ is defined as a Riemannian manifold whose holonomy group is contained inside ${\rm SU}(2)\times {\rm USp}(2n_H)$. The manifolds of this kind which occur in supergravity have holonomy group of the form:
\begin{equation}
H^{(QK)}={\rm SU}(2)\times H^{(QK)}_{{\rm matt}}\,\,,\,\,\,\,H^{(QK)}_{{\rm matt}}\subset  {\rm USp}(2n_H)\,,\label{Hsu2Hp}
\end{equation}
where ${\rm SU}(2)$, together with the group ${\rm U}(1)$ of the K\"ahler transformations in the holonomy group of $\mathcal{M}_{SK}$, define the ${\rm U}(2)$ R-symmetry group of the supersymmetry algebra, see Eq. (\ref{Hsplits}).\par
 The positive definite metric is denoted by $h_{uv}(q)$, where $q^u$ are the coordinates describing the scalar fields of the hypermultiplets. Let us review the main properties of these spaces.
The action of the ${\rm SU}(2)$ generators on the tangent space defines three complex structures $J^x{}^u{}_v$, $x=1,2,3$, satisfying the quaternionic algebra:
\begin{equation}
J^xJ^y=-\delta^{xy}+\epsilon^{xyz}\,J^z\,.\label{Jstruc}
\end{equation}
The metric $h_{uv}$ is required to be hermitian with respect to all the $J^x$:
\begin{equation}
h_{uw}\,J^x{}^w{}_v=-h_{vw}\,J^x{}^w{}_u\,\,,\,\,\,\,\,x=1,2,3\,.
\end{equation}
From the definition it follows that one can define a Levi-Civita connection, which comprises ${\rm SU}(2)$-connection 1-forms $\omega^x=\omega^x_u\,dq^u$, such that
\begin{equation}
{\Scr D}_u J^x{}^w{}_v\equiv \partial_u J^x{}^w{}_v+\tilde{\Gamma}_{uu'}^w\,J^x{}^{u'}{}_v-\tilde{\Gamma}_{uv}^{u'}\,J^x{}^{w}{}_{u'}+\epsilon^{xyz}\,\omega^y_u\,J^z{}^w{}_v=0\,.\label{DJ0parallel}
\end{equation}
where $\tilde{\Gamma}$ is the Christoffel symbol. In terms of this quaternionic structure, a triplet of 2-forms, named hyper-K\"ahler form, are defined:
\begin{equation}
K^x=K^x_{uv}\,dq^u\wedge dq^v\,\,,\,\,\,K^x_{uv}=h_{uw}\,J^x{}^w{}_v\,.\label{Kxdef}
\end{equation}
The above definition and Eq. (\ref{Jstruc}) imply the following relation:
\begin{equation}
K^x_{uw}h^{ws}K_{sv}^y=-\delta^{xy}\,h_{uv}+\epsilon^{xyz}\,K^z_{uv}\,,\label{Kstruc}
\end{equation}
where, as usual, $h^{uv}$ are the components of the inverse metric.\par
From the ${\rm SU}(2)$-connection 1-forms $\omega^x$  we construct the  ${\rm SU}(2)$-curvature 2-form
 \begin{equation}
R^x\equiv d\omega^x+\frac{1}{2}\epsilon^{xyz}\,\omega^y\wedge \omega^z=\frac{1}{2}\,R^x_{uv}\,dq^u\wedge dq^v\,.
\end{equation}\par
Property (\ref{DJ0parallel}) implies that also $K^x_{uv}$ are parallel:
\begin{equation}
{\Scr D}_u K^x_{vw}=0\,,
\end{equation}
and thus that the 2-forms $K^x$ are covariantly closed with respect to the ${\rm SU}(2)$-connection
\begin{equation}
{\Scr D} K^x=dK^x+\epsilon^{xyz}\,\omega^y\wedge  K^z=0\,.\label{DK0}
\end{equation}
A quaternionic K\"ahler manifold ${\Scr M}_{QK}$ is further characterized by the property that the ${\rm SU}(2)$-curvature components $R^x$ are proportional to the triplet of  covariantly constant 2-forms $K^x$:
\begin{equation}
R^x=\lambda \,K^x\,,\label{OmKx}
\end{equation}
where $\lambda$ is a real coefficient depending on the normalization of the metric. This follows from the fact that quaternionic K\"ahler manifolds are Einstein spaces.
Choosing the standard normalization of the kinetic term
for the hyper-scalars $q^u$ amounts to fixing $\lambda=-1$. The above equation is consistent with (\ref{DK0}) in virtue of the Bianchi identity for  $R^x$:
\begin{equation}
{\Scr D} R^x=dR^x+\epsilon^{xyz}\,\omega^y\wedge R^z=0\,.
\end{equation}
Property (\ref{Hsu2Hp}) implies that we can define the vielbein 1-forms as follows:
\begin{equation}
\mathcal{U}^{A\alpha}=\mathcal{U}^{A\alpha}_u\,dq^u\,,
\end{equation}
where $A=1,2$ is the ${\rm SU}(2)$-doublet index labeling the supersymmetries and $\alpha=1,\dots ,2 n_H$ labels the fundamental representation of
${\rm USp}(2n_H)$. In this basis the rigid tangent space index ${\bf u}$ is a composite one ${\bf u}=(A,\alpha)$ and the rigid metric is $\eta_{{\bf u}{\bf v}}=\epsilon_{AB}\mathbb{C}_{\alpha\beta}$, where $\mathbb{C}_{\alpha\beta}$ is the ${\rm USp}(2n_H)$-invariant matrix, so that:
\begin{equation}
\mathcal{U}^{A\alpha}_u\mathcal{U}^{B\beta}_u\epsilon_{AB}\,\mathbb{C}_{\alpha\beta}=h_{uv}\,,\label{UUh}
\end{equation}
where  $\mathcal{U}^{A\alpha}$ satisfy the following
reality condition:
\begin{eqnarray}
\mathcal{U}_{A\alpha}&\equiv &(\mathcal{U}^{A\alpha})^*=\epsilon_{AB}\mathbb{C}_{\alpha\beta}\,\mathcal{U}^{B\beta}\,.\nonumber
\end{eqnarray}
These 1-forms in fact satisfy a condition which is stronger than (\ref{UUh}) \cite{Bagger:1983tt}:
\begin{align}
\left(\mathcal{U}^{A\alpha}_u \mathcal{U}^{B\beta}_v-\mathcal{U}^{A\alpha}_v \mathcal{U}^{B\beta}_u\right)\mathbb{C}_{\alpha\beta}&=h_{uv}\,\epsilon^{AB}\,.
\end{align}
and they are related to the hyper-K\"ahler 2-form as follows:
\begin{equation}
\mathcal{U}_{A\alpha\,u}\,\mathcal{U}^{B\alpha}_{v}=\frac{1}{2}\,h_{uv}\,\delta_A^B-\frac{i}{2}\,K^x_{uv}\,(\sigma^x)_A{}^B\,,\label{UUKs}
\end{equation}
where the relative sign between the two terms on the right hand side is fixed by (\ref{Kstruc}).\par
By definition of the Levi-Civita connection, the vielbein 1-forms are also covariantly constant:
\begin{equation}
{\Scr D} \mathcal{U}^{A\alpha}\equiv d\mathcal{U}^{A\alpha}+\frac{i}{2}\,(\sigma^x)_B{}^A\,\omega^x\wedge \mathcal{U}^{B\alpha}+\Delta^{\alpha\gamma}\wedge \mathcal{U}^{A\beta}\mathbb{C}_{\gamma\beta}=0\,,
\end{equation}
where $\Delta^{\alpha\beta}=\Delta^{\beta\alpha}$ denote the $H^{(QK)}_{{\rm matt}}\subset USp(2n_H)$-connection 1-forms.\par
Consistently with the defining property that the holonomy group is contained inside ${\rm SU}(2)\times {\rm USp}(2n_H)$, the Riemann tensor of a quaternionic manifold has the following general expression:
\begin{equation}
R_{uv}{}^{A\alpha,\,B\beta}\equiv R_{uv,u'v'}\,\mathcal{U}^{A\alpha\,u'}\,\mathcal{U}^{B\beta\,v'}= R_{uv}^{AB}\mathbb{C}^{\alpha\beta}+\mathbb{R}_{uv}^{\alpha\beta}\,\epsilon^{AB}\,,\label{RgenQK}
\end{equation}
where $\mathcal{U}^{A\alpha\,u}$ is the inverse vielbein matrix used to convert the last two indices of the Rimeann tensor into rigid ones,  $R_{uv}^{AB}$ is the ${\rm SU}(2)$-curvature
\begin{equation}
R_{uv}^{AB}=R_{uv}^{BA}=-\frac{i}{2}\,R^x_{uv}\epsilon^{AC}(\sigma^x)_C{}^B=
2\,\lambda\,\mathbb{C}_{\alpha\beta}\mathcal{U}^{A\alpha}_{[u}\,\mathcal{U}^{B\beta}_{v]}\,,\label{RgenQK1}
\end{equation}
and
$\mathbb{R}^{\alpha\beta}$ denotes instead the $H^{(QK)}_{{\rm matt}}\subset USp(2n_H)$-curvature, defined in terms of the connection one-form $\Delta^{\alpha\beta}$ as follows
\begin{equation}
\mathbb{R}^{\alpha\beta}=\mathbb{R}^{\beta\alpha}\equiv d\Delta^{\alpha\beta}+\mathbb{C}_{\gamma\delta}\Delta^{\alpha\gamma}\wedge\Delta^{\delta\beta}\,.
\end{equation}
This tensor has the following general form \cite{Bagger:1983tt}:
\begin{equation}
R_{uv}{}^{\alpha\beta}=\lambda\,\epsilon_{AB}\,\left(\mathcal{U}_u^{A\alpha}\mathcal{U}_v^{B\beta}-
\mathcal{U}_v^{A\alpha}\mathcal{U}_u^{B\beta}\right)+2\,\epsilon_{AB}\,\mathbb{C}^{\alpha\rho}\,
\mathbb{C}^{\beta\sigma}\mathcal{U}_u^{A\gamma}\mathcal{U}_v^{B\delta}\,\Omega_{\rho\sigma\gamma\delta}\,,\label{RgenQK2}
\end{equation}
where the tensor $\Omega_{\rho\sigma\gamma\delta}$ is totally symmetric in its four indices.
All these quaternionic K\"ahler manifolds are Einstein spaces with negative scalar curvature $\mathcal{R}=8\,\lambda\,n_H(n_H+2)$, see derivation below.
\paragraph{The Ricci tensor.}
Let us evaluate the Ricci tensor of the manifold using Eqs. (\ref{RgenQK}),(\ref{RgenQK1}),(\ref{RgenQK2}). The Riemann tensor in curved indices reads:
\begin{equation}
R_{uv,p}{}^q=R_{uv,A\alpha}{}^{B\beta}\,\mathcal{U}_p^{A\alpha}\,\mathcal{U}_{B\beta}^q\,.
\end{equation}
Tracing over the indices $v$ and $q$ we find:
\begin{align}
\mathcal{R}_{up}&=\left(-i\,\lambda\,K^x_{uv}\,(\sigma^x)_A{}^B\delta_\alpha^\beta+R_{uv,\alpha}{}^\beta \delta_A^B\right)\,\mathcal{U}_p^{A\alpha}\,\mathcal{U}_{B\beta}^q=\nonumber\\
&=3\lambda\,h_{up}+\lambda(2n_H+1)\,h_{up}=2\lambda\,(n_H+2)\,h_{up}\,.\label{RicciQK}
\end{align}
The manifold is therefore of Einstein type. The Ricci scalar being
\begin{equation}
\mathcal{R}=\mathcal{R}_u{}^u=8\lambda\,n_H\,(n_H+2)\,.
\end{equation}
\subsubsection{Isometries}\label{isomqk}
Consider now infinitesimal isometries generated by $t_m$, whose action on the scalar fields is described by Killing vectors $k_m=k_m^u\,\partial_u$. They close the isometry algebra:
\begin{equation}
[t_m,\,t_n]=f_{mn}{}^p\,t_p\,\,\,,\,\,\,\,\,[k_m,\,k_n]=-f_{mn}{}^p\,k_p\,,
\end{equation}
and leave the 4-form $\sum_{x=1}^3 K^x\wedge K^x$ invariant \cite{D'Auria:1990fj}. This condition amounts to requiring:
\begin{equation}
\ell_n K^x=\epsilon^{xyz}\,K^y\,W^z_n\,,\label{elKx}
\end{equation}
where $W^z_n$ is an ${\rm SU}(2)$-compensator. Equation (\ref{elKx}) is solved by writing the Killing vectors $k_n$ in terms of \emph{tri-holomorphic momentum maps} ${\Scr P}_n^x$ as follows \cite{D'Auria:1990fj}:
\begin{equation}
\iota_n K^x=-{\Scr D} {\Scr P}^x_n=-(d{\Scr P}^x_n+\epsilon^{xyz}\omega^y\,{\Scr P}_n^z)\,,\label{inKx}
\end{equation}
provided
\begin{equation}
 {\Scr P}^x_n=\lambda^{-1}(\iota_n \omega^x-W^x_n)=W^x_n-\iota_n \omega^x\,,\label{PHW}
\end{equation}
where we have used $\lambda=-1$. The above equation was derived in \cite{Galicki:1986ja}, see also \cite{D'Auria:1990fj}. We shall derive it below in the homogeneous case.
For those isometries with vanishing compensator, $W^x_n=0$, the momentum maps have the simple expression: ${\Scr P}^x_n=-k^u_n\,\omega^x_u$.\par
Just as for the special K\"ahler manifolds, (see equation (\ref{lhsPP})), the momentum maps satisfy the following Poisson-bracket relations:
\begin{equation}
\{{\Scr P}_m\,{\Scr P}_n\}^x\equiv K^x(k_m,\,k_n)-\lambda\,\epsilon^{xyz}\,{\Scr P}_n^y\,{\Scr P}_m^z=-{\rm f}_{mn}{}^p\,{\Scr P}_p^x\,,
\end{equation}
which amount to the \emph{equivariance condition}:
\begin{equation}
2\,K^x_{uv}\,k^u_n\,k^v_m-\lambda\,\epsilon^{xyz}\,{\Scr P}_n^y\,{\Scr P}_m^z=-{\rm f}_{mn}{}^p\,{\Scr P}_p^x\,.\label{equivar2}
\end{equation}
For homogeneous manifolds $k_n$ and ${\Scr P}^x_n$ can be given a simple geometric characterization \cite{Andrianopoli:2015rpa}.
Indeed if $\mathcal{M}_{QK}$ has the general form:
\begin{equation}
{\Scr M}_{QK}=\frac{G^{(QK)}}{H^{(QK)\prime}}\,,
\end{equation}
where  $G^{(QK)}$ is the isometry group and $H^{(QK)\prime}$ the isotropy group, locally contained in the holonomy group $H^{(QK)}$ . Denoting by $\mathfrak{g}_{QK}$ and $\mathfrak{H}_{QK}$ the Lie algebras of $G^{(QK)}$ and  $H^{(QK)\prime}$, respectively,
we can write the general decomposition (\ref{ghkdec}):
\begin{equation}
\mathfrak{g}_{QK}=\mathfrak{H}_{QK}\oplus\mathfrak{K}_{QK}\,,
\end{equation}
where $\mathfrak{H}_{QK}$ and  $ \mathfrak{K}_{QK}$, satisfy the general commutation relations (\ref{hkh}).
Let $\{K_{{\bf u}}\}$ be a basis of generators for the coset space $\mathfrak{K}_{QK}$.
The generators of $H^{(QK)}$ split into those of ${\rm SU}(2)$ ($J^x$) and those of $H^{(QK)}_{{\rm matt}}$  ($J_{\alpha\beta}=J_{\beta\alpha}$), according to the decomposition (\ref{Hsu2Hp}). These comprise the generators in the isotropy algebra $\mathfrak{H}_{QK}$ and are the generators of $\mathfrak{H}_{QK}$ only in the symmetric case.
In the chosen basis, the last of (\ref{hkh}) reads
\begin{equation}
[K_{{\bf u}},\,K_{{\bf v}}]={\rm f}_{{\bf u}{\bf v}}{}^{x}\,J^x+\frac 12 {\rm f}_{{\bf u}{\bf v}}{}^{\alpha\beta}\,J_{\alpha\beta}+{\rm f}_{{\bf u}{\bf v}}{}^{{\bf w}}\,K_{{\bf w}}\,.\label{KKJ}
\end{equation}
For symmetric manifolds ${\rm f}_{{\bf u}{\bf v}}{}^{{\bf w}}=0$.
We can normalize the generators so that the Cartan-Killing form $(\,,\,)$ of $\mathfrak{g}_{QK}$ is
\begin{equation}
(K_{{\bf u}},\,K_{{\bf v}})=\delta_{{\bf u}{\bf v}}\,\,,\,\,\,\,(J^x,\,J^y)=-\delta^{xy}\,, \,\,\,\,(J_{\alpha\beta},\,J_{\gamma\delta})=-2 \,\mathbb{C}_{\alpha(\gamma}\mathbb{C}_{\delta)\beta}\,.
\end{equation}
The vielbein and $H^{(QK)\prime}$-connections are, as usual, defined by decomposing the left-invariant one-form in components along $\mathfrak{K}_{QK}$ and
 $\mathfrak{H}_{QK}$:
 \begin{equation}
\Omega= L^{-1}dL=\mathcal{P}^{{\bf u}}\,K_{{\bf u}}+\frac{1}{2}\omega^x\,J^x+\frac 12 \Delta^{\alpha\beta}\,J_{\alpha\beta}\,,\label{li1f}
 \end{equation}
where $L$ is the coset representative in some representation of $G^{(QK)}$, so that
\begin{equation}
\mathcal{P}^{{\bf u}}=(K_{{\bf u}},\,\Omega)\,\,,\,\,\,\,\omega^x=-2\,(J^x,\,\Omega)\,,\,\,\,\,\Delta^{\alpha\beta}= (J^{\alpha\beta},\,\Omega)\,.
\end{equation}
From the Maurer-Cartan equations $d\Omega+\Omega\wedge \Omega=0$ we can read off the expression for the curvature and the 2-forms $K^x$:
\begin{equation}
R^x=d\omega^x+\frac{1}{2}\epsilon^{xyz}\,\omega^y\wedge \omega^z=-{\rm f}_{{\bf u}{\bf v}}{}^{x}\mathcal{P}^{{\bf u}}\wedge \mathcal{P}^{{\bf v}}=-K^x\,,
\end{equation}
where we have used (\ref{KKJ}) and (\ref{OmKx}) with $\lambda=-1$. From this we derive the holonomic components of $K^x$:
\begin{equation}
K^x_{uv}={\rm f}_{{\bf u}{\bf v}}{}^{x}\mathcal{P}_u{}^{{\bf u}}\, \mathcal{P}_v{}^{{\bf v}}\,.\label{Kxuvsym}
\end{equation}
We can give the following useful characterization \cite{Andrianopoli:2015rpa} of the Killing vector $k_n$ and the momentum map ${\Scr P}_n^x$ associated with the isometry generator $t_n\in \mathfrak{g}_{QK}$, see Eq. (\ref{kespans2}):
\begin{equation}
L^{-1}t_n L=k_n^u\,\mathcal{P}_{u}{}^{{\bf u}}\,K_{{\bf u}}-\frac{1}{2}{\Scr P}_n^x\,J^x-\frac 12 {\Scr P}_n^{\alpha\beta}\,J_{\alpha\beta}\,,\label{Lm1tL}
\end{equation}
which implies:
\begin{equation}
{\Scr P}_n^x=2\,(L^{-1}t_n L,\,J^x)\,.
\end{equation}
We prove below that $k_n$ and  ${\Scr P}_n^x$  defined in (\ref{Lm1tL}) do satisfy (\ref{inKx}).
From (\ref{Kxuvsym}) and (\ref{li1f}) we find:
\begin{equation}
2 k_n^u\,K_{uv}^x=2\,{\rm f}_{{\bf u}{\bf v}}{}^{x}\,k_n^u\,\mathcal{P}_u{}^{{\bf u}}\, \mathcal{P}_v{}^{{\bf v}}=-2\,([L^{-1} t_n L,\,L^{-1}\partial_v L],J^x)+\epsilon^{xyz}\,{\Scr P}^y_n \omega^z_v\,.\label{equa1q}
\end{equation}
Now let us evaluate ${\Scr D} {\Scr P}_n^x$:
\begin{eqnarray}
{\Scr D}_v {\Scr P}_n^x&=&\partial_v {\Scr P}_n^x+\epsilon^{xyz}\omega_v^y\,{\Scr P}_n^z=2\,(\partial_vL^{-1}t_nL+L^{-1}t_n \partial_v L,J^x)+\epsilon^{xyz}\omega_v^y\,{\Scr P}_n^z=\nonumber\\
&=&2\,([L^{-1} t_n L,\,L^{-1}\partial_v L],J^x)+\epsilon^{xyz}\omega_v^y\,{\Scr P}_n^z=-2 k_n^u\,K_{uv}^x\,,
\end{eqnarray}
where in the last equality we have used (\ref{equa1q}).

 Let us now prove (\ref{PHW}). We shall repeat here the derivation of the corresponding general equation (\ref{PkQW}). From basic coset geometry, see Section \ref{ghsect}, we know that the left action of an isometry on the coset representative $L$ yields $L$ computed in the transformed point, multiplied to the right by a compensator in $H$. For an infinitesimal isometry this
is expressed by the property (see Eq. (\ref{tLkW})):
\begin{equation}
t_n\,L=k_n^u\,\partial_uL+L\,W_n\,.\label{tnL}
\end{equation}
where $W_n\in \mathfrak{H}_{QK}$ is the infinitesimal generator of the compensating transformation, which can be expanded as follows
\begin{equation}
W_n=-\frac{1}{2}\,W_n^xJ^x+\frac 12 W_n^{\alpha\beta}\,J_{\alpha\beta}\,.
\end{equation}
Multiplying (\ref{tnL}) to the left by $L^{-1}$ we find:
\begin{equation}
L^{-1}t_nL=k_n^u\,\Omega_u+W_n=k_n^u\,\mathcal{P}_u{}^{{\bf u}}\,K_{{\bf u}}+\frac{1}{2}k_n^u\,\omega_u^x\,J^x+\frac 12 k_n^u\,\omega_u^{\alpha\beta}\,J_{\alpha\beta}
-\frac{1}{2}\,W_n^xJ^x+\frac 12 W_n^{\alpha\beta}\,J_{\alpha\beta}\,.
\end{equation}
Comparing the above expansion with (\ref{Lm1tL}) we find:
\begin{equation}
{\Scr P}_n^x=W_n^x-k_n^u\,\omega_u^x\,,
\end{equation}
which is (\ref{PHW}). Equations (\ref{PHW}) and (\ref{inKx}) then imply (\ref{elKx}).\par
Consider now a \emph{solvable} parametrization of the coset for which we describe the quaternionic K\"ahler manifold as globally isometric
to a solvable Lie group generated by a solvable Lie algebra ${\Scr S}_{QK}$:
\begin{equation}
\mathcal{M}_{QK}\sim \exp{({\Scr S}_{QK})}\,.
\end{equation}
The coset representative is then an element of $\exp{({\Scr S}_{QK})}$:
\begin{equation}
L(q)=e^{q^u\,T_u}\in \exp{({\Scr S}_{QK})}\,,
\end{equation}
where $T_u$ are the generators of ${\Scr S}_{QK}$. Being $L(q)$ an element of a group, the action on it of any other element of the same group has no compensating transformation:
\begin{equation}
\forall g\in \exp{({\Scr S}_{QK})}\,\,:\,\,\,\,\,gL(q)=L(q')\,.
\end{equation}
Therefore for any $t_n\in {\Scr S}_{QK}$ we have $W_n=0$, i.e.
\begin{equation}
{\Scr P}_n^x=-k_n^u\,\omega^x_u\,.
\end{equation}
Transformations in $\exp({\Scr S}_{QK})$ comprise \emph{translational isometries}. Note that the above properties of ${\Scr P}_n^x$ hold for the analogous quantities defined in a general extended supergravity by Eq. (\ref{Lm1tL}), or equivalently (\ref{kespans2}), where $J^x$ are the $H_R$-generators.\par
Homogeneous special and quaternionic K\"ahler manifolds with negative curvature have been classified in  \cite{alek,deWit:1992wf}. They are all \emph{normal}, see final paragraph of Sect. \ref{ghsect}, namely they feature a solvable group of isometries $G_S=\exp({\Scr S})$ whose action is free and transitive.
\paragraph{Some more properties.}
Let us derive some more general relations involving the momentum maps and the Killing vectors (see for instance \cite{D'Auria:2001kv}).
From the general properties of the Levi-Civita connection and the corresponding curvature on the manifold, we find:
\begin{equation}
{\Scr D}_u{\Scr D}_v\,k_n^q-{\Scr D}_v{\Scr D}_u\,k_n^q=-R_{uv,p}{}^q\,k_n^p\,.
\end{equation}
Tracing over $q$ and $v$ and using the expression (\ref{RicciQK}) for the Ricci tensor we find:
\begin{equation}
{\Scr D}_u{\Scr D}_v\,k_n^v-{\Scr D}_v{\Scr D}_u\,k_n^v=-\mathcal{R}_{up}\,k_n^p=-2\lambda\,(n_H+2)\,k_{n\,u}\,.
\end{equation}
Next use the general property (\ref{Killcond}) of the Killing vectors ${\Scr D}_u k_n^v=-{\Scr D}^v k_{n\,u}$ to rewrite the above equation as follows:
\begin{equation}
{\Scr D}_u{\Scr D}^u\,k_{n\,u}=-\mathcal{R}_{up}\,k_n^p=-2\lambda\,(n_H+2)\,k_{n\,u}\,.
\end{equation}
Let us now evaluate the antisymmetrized double covariant derivative on ${\Scr P}_n^x$:
\begin{equation}
({\Scr D}_u{\Scr D}_v-{\Scr D}_v{\Scr D}_u){\Scr P}_n^x=\epsilon^{xyz}\,R^y_{uv}\,{\Scr P}_n^z=2\,\lambda\,\epsilon^{xyz}\,K^y_{uv}\,{\Scr P}_n^z\,.
\end{equation}
Using now the property $K^x_{uv}K^{y\,uv}=4\,n_H\,\delta^{xy}$, one finds:
\begin{equation}
K^{y\,uv}\,{\Scr D}_u{\Scr D}_v{\Scr P}_n^x=4\,n_H\,\lambda\,\epsilon^{xyz}\,{\Scr P}_n^z\,.
\end{equation}
Finally let us apply Eq. (\ref{inKx}) on the left-hand-side of the above equation to write
\begin{equation}
4\,n_H\,\lambda\,\epsilon^{xyz}\,{\Scr P}_n^z=2\,{\Scr D}_u(k_n^p\,K^x_{vp})\,K^{y\,uv}=2\,({\Scr D}_u k_n^p)\,K^x_{vp}\,K^{y\,uv}=2\,\epsilon^{xyz}\,{\Scr D}_u k_n^p\,K^z{}_p{}^u\,,
\end{equation}
where we have used the property that $K^x_{uv}$ are parallel tensors and Eq. (\ref{Kstruc}).
We therefore obtain the following useful formula for deriving the momentum maps from the Killing vectors:
\begin{equation}
{\Scr P}_n^x=\frac{1}{2n_H \lambda}\,{\Scr D}_u k_n^p\,K^x{}_p{}^u\,.
\end{equation}
\subsection{The Gauging}\label{gaugspecialg}
Let us now consider the gauging of a subgroup $G_g$ of the isometry group of the scalar manifold. The gauge generators $X_M$ are expanded in the generators $\{t_a,\,t_m\}$ of the isometry groups of $\mathcal{M}_{SK}$ and $\mathcal{M}_{QK}$ through the embedding tensor:
\begin{equation}
{ X}_M=\Theta_M{}^a\,t_a+\Theta_M{}^m\,t_m\,.\label{N2gaugen}
\end{equation}
The symplectic electric-magnetic duality action of ${ X}_M$ is, as usual, described by the symplectic matrices: ${ X}_{MN}{}^P=\Theta_M{}^a\,t_{a\,N}{}^P $\,.
The linear and quadratic constraints (\ref{lconstr}), (\ref{quadratic1}), (\ref{quadratic2}) on the embedding tensor read:
\begin{eqnarray}
&&\,\,\,\,{X}_{(MNP)}\equiv { X}_{(MN}{}^Q\mathbb{C}_{Q|P)}=0\,,\label{lc}\\
&&\,\,\,\,\Theta_M{}^a\Theta_N{}^b{\rm f}_{ab}{}^c+{ X}_{MN}{}^P\,\Theta_P{}^c=0\,,\label{qc1}\\
&&\,\,\,\,\Theta_M{}^m\Theta_N{}^n {\rm f}_{mn}{}^p+{ X}_{MN}{}^P\,\Theta_P{}^p=0\,,\label{qc2}\\
&&\,\,\,\,\Theta_M{}^a\mathbb{C}^{MN}\Theta_N{}^b=\Theta_M{}^a\mathbb{C}^{MN}\Theta_N{}^n=\Theta_M{}^m\mathbb{C}^{MN}\Theta_N{}^n=0\,.\label{qc3}
\end{eqnarray}
Conditions (\ref{qc1}), (\ref{qc2}) express the closure condition (\ref{quadratic2}).
The first two equalities in (\ref{qc3}) follow from (\ref{lc}) and (\ref{qc1}), (\ref{qc2}) while the last one has to be imposed independently.
We can define gauge Killing vectors and momentum maps as follows:
\begin{equation}
k^i_M\equiv \Theta_M{}^a\,k^i_a\,,\quad k^u_M\equiv \Theta_M{}^m\,k^u_m\,,\quad{\Scr P}_M\equiv\Theta_M{}^a\,{\Scr P}_a\,,\quad{\Scr P}_M^x\equiv\Theta_M{}^m\,{\Scr P}_m^x\,.
\end{equation}
From the quadratic constraints and Eqs. (\ref{lhsPP}) and (\ref{equivar2}) we find the equivariance conditions \footnote{By setting the parameter $\lambda$ of the quaternionic geometry to $\lambda =-1$.}:
\begin{eqnarray}
i g_{i\bar{\jmath}}\,k^i_{[M}\,k^{\bar{\jmath}}_{N]}=\frac{1}{2}\,{ X}_{MN}{}^P\,{\Scr P}_P\,,\label{gequiv1}\\
2\,K^x_{uv}\,k^u_M\,k^v_N+\epsilon^{xyz}\,{\Scr P}_M^y\,{\Scr P}_N^z={ X}_{MN}{}^P\,{\Scr P}_P^x\,,\label{gequiv2}
\end{eqnarray}
Using the linear constraint (\ref{lc}) on the embedding tensor we can prove the following identities:
\begin{equation}
{\Scr P}_M\Omega^M=0\,\,,\,\,\,k_M^i\,\Omega^M=0\,.\label{newidentities}
\end{equation}
To prove the first one we write (\ref{PVtV}) for the gauge-momentum maps:
\begin{equation}
{\Scr P}_M=-e^\mathcal{K}\,{\rm X}_{MNP}\overline{\Omega}^N\Omega^P\,.
\end{equation}
Contracting both sides with $\Omega^M$ we find:
\begin{equation}
\Omega^M{\Scr P}_M=-e^\mathcal{K}\,\Omega^M {\rm X}_{MNP}\overline{\Omega}^N\Omega^P=\frac{e^\mathcal{K}}{2}\, \overline{\Omega}^N{\rm X}_{NMP}\Omega^M\Omega^P=0\,,\label{OmP}
\end{equation}
where we have used the linear constraint (\ref{lc}) and the symplectic property of the matrices ${\rm X}_{MN}{}^P$:
\begin{equation}
2{\rm X}_{(MP)N}=-{\rm X}_{NMP}\,,
\end{equation}
being ${\rm X}_{MNP}\equiv {\rm X}_{MN}{}^Q\mathbb{C}_{QP}$. Last equality in (\ref{OmP}) then follows from (\ref{tOmOm}).\par
The second of (\ref{newidentities}) is derived in the following way
\begin{equation}
\Omega^M\,k_M^i=i\,g^{i\bar{\jmath}}\,\Omega^M\,\partial_{\bar{\jmath}} {\Scr P}_M=i\,g^{i\bar{\jmath}}\,\partial_{\bar{\jmath}}(\Omega^M\, {\Scr P}_M)=0\,,
\end{equation}
where we have used the first of (\ref{newidentities}).\par
 Let us now prove an other important property which follows from the constraints on the embedding tensor.
 Using Eq. (\ref{kaUi}), for a generic gauge generator we can write:
 \begin{equation}
 k_M^i\,U_i{}^N=-X_{MP}{}^N\,V^P+i\,{\Scr P}_M\,V^N\,.
 \end{equation}
 Contracting with $\overline{V}^M$ and using (\ref{newidentities}) we find:
 \begin{equation}
 \overline{V}^M k_M^i\,U_i{}^N=-X_{MP}{}^N\,\overline{V}^M\,V^P\,.
 \end{equation}
 If both sides of this equations are further contracted with $\Theta_N{}^\alpha$, we can use the antisymmetry of $X_{MP}{}^N\,\Theta_N{}^\alpha$ in its first two indices, $X_{MP}{}^N\,\Theta_N{}^\alpha=-X_{PM}{}^N\,\Theta_N{}^\alpha$, and write:
 \begin{equation}
 \overline{V}^M k_M^i\,U_i{}^N\Theta_N{}^\alpha=-X_{MP}{}^N\,\overline{V}^M\,V^P\Theta_N{}^\alpha=
 X_{MP}{}^N\,{V}^M\,\overline{V}^P\Theta_N{}^\alpha=-(\overline{V}^M k_M^i\,U_i{}^N\Theta_N{}^\alpha)^*\,,\label{idgauk}
 \end{equation}
 namely we find that the expression on the left-hand-side is imaginary. This in particular implies the first of conditions (\ref{csNS}) for $\mathcal{N}=2$ theories, using the expressions of the fermion-shift tensors to be given below.
\subsection{Minimal Couplings, Fermion-Shift Tensors and the Scalar Potential}
After the gauge group has been chosen within the global symmetry group, the construction of the gauged theory proceeds along the lines illustrated in Sect. \ref{sec:3}: Ordinary derivatives are replaced by covariant ones, vector field strengths are covariantized, $O(g)$-terms, defined by the fermion-shift tensors, are added to the fermion supersymmetry transformation rules and to the Lagrangian and an $O(g^2)$ scalar potential is introduced. We shall perform here the duality-covariant gauging discussed in Sect. \ref{sec:4} which requires the introduction of extra fields: the magnetic vector fields $A_{\Lambda\mu}$ and antisymmetric tensor fields $B_{\alpha\,\mu\nu}=(B_{a\,\mu\nu},\,B_{n\,\mu\nu})$ in the adjoint representation of the isometry group $G=G^{(SK)}\times G^{(QK)}$. Following the general procedure we define the vector field field strengths $\mathcal{H}_{\mu\nu}^M$ as in (\ref{HZB}), obtained by combining the (non-Abelian) field strengths (\ref{FMdef}) with the antisymmetric tensors $B_{\alpha\,\mu\nu}$ through the embedding tensor. We also define the symplectic covariant vector of field strengths $\mathcal{G}^M$ as in (\ref{GMHdef}) and its composite counterpart $\mathbb{H}^{\underline{M}}_{\mu\nu}$ obtained by dressing it with the scalar fields, see Eq. (\ref{Tdef2}). Consistency of the construction requires the introduction of the topological terms (\ref{topB}), (\ref{GCS}) in the Lagrangian. We refer to Sect. \ref{sec:4} for the details and the definitions which we shall not repeat here.\par
Let $W^{i\,AC},\,N^{\alpha}_{B},\,\mathbb{S}_{AB}$ denote the shift-tensors of the chiral gaugini $\lambda^{iA}$, hyperini $\zeta^{\alpha}$ and gravitini $\psi_A$ respectively, \footnote{Due to the redefinitions (\ref{redefsf}), the shifts $\mathbb{N}_{IA}{}^B$ and $\mathbb{N}_{\alpha}{}^B$ are related to $W^{i\text{ }AB}$ and $N_{\alpha}^A$ as follows:
\begin{equation}
\mathbb{N}_{IA}{}^B=e_{i\,I}\,\epsilon_{AC}\,W^{i\text{ }CB}\,\,;\,\,\,\,\,\mathbb{N}_{\alpha}{}^B=\sqrt{2}\,N_{\alpha}^A\,.
\end{equation}}
\begin{align}
\delta\lambda^{i\text{ }A}  &  =\dots+g\,W^{i\text{ }AB}\epsilon_{B},\label{shiftF1}\\
\delta\psi_{A\text{ }\mu}  &  =\dots+g\,i\mathbb{S}_{AB}\,\gamma_{\mu}\epsilon^{B},\label{shiftF2}\\
\delta\zeta^{\alpha}  &  =\dots+g\,N^{\alpha}_A\,\epsilon^{A},\label{shiftF3}
\end{align}
where\footnote{Note the different sign used here for the gauge connection in the gauge-covariant derivatives with respect to \cite{D'Auria:2001kv} or \cite{Andrianopoli:1996cm}. Since the fermion mass terms have the same form, all formulas here are obtained from those in these references by inverting the sign of $g$ and of the fermion shift-tensors and mass matrices at the same time.}
\begin{eqnarray}
\mathbb{S}_{AB}&=&\frac{i}{2}\,(\sigma^x)_A{}^C\epsilon_{BC}\,{\Scr P}^x_M\,V^M\,,\label{SAB}\\
W^{i\,AB}&=&-\epsilon^{AB}\,k_M^i\,\overline{V}^M+i\,(\sigma^x)_C{}^B\epsilon^{CA}{\Scr P}^x_M\,g^{i\bar{\jmath}}
\overline{U}^M_{\bar{\jmath}}\,,\label{WiAB}\\
N_\alpha{}^A&=&-2\,\mathcal{U}_{u}^A{}_\alpha\,k^u_M\,\overline{V}^M\,\,,\,\,\,\,N^\alpha{}_A\equiv (N_\alpha{}^A)^*=2\,\,\mathcal{U}_{u\,A}{}^\alpha\,k^u_M\,{V}^M\,,\label{NAa}
\end{eqnarray}
having defined $\mathcal{U}_{u\,A}{}^\alpha\equiv \epsilon_{AB}\,\mathcal{U}_{u}{}^{B\alpha}$ and $\mathcal{U}_{u}^A{}_\alpha=\epsilon^{BA}\,\mathcal{U}_{u\,B\alpha}=-(\mathcal{U}_{u\,A}{}^\alpha)^*$.
Using the explicit form (\ref{Qournot}) of the $H_R$-generators $J_0,\,J_x$: $$(J^0)_A{}^B=i\,\delta_A^B\,,\,\,(J^x)_A{}^B=-i\,(\sigma^x)_A{}^B\,,$$
integrating the first term on the right-hand-side of (\ref{WiAB}) by parts and using (\ref{OmP}), $\mathbb{N}_{IA}{}^B$ can be written in the following compact form:
\begin{equation}
\mathbb{N}_{IA}{}^B=e_{i\,I}\,\epsilon_{AC}\,W^{i\text{ }CB}=-e^{-1}{}_I{}^{\bar{\jmath}}\,\overline{U}_{\bar{\jmath}}{}^M\,{\Scr P}_M^{{\bf a}}\,J_{{\bf a}}=2\,e^{-1}{}_I{}^{\bar{\jmath}}\,\overline{U}_{\bar{\jmath}}{}^M\,\mathcal{Q}_{M\,A}{}^B\,,
\end{equation}
where we have defined ${\Scr P}_M^{0}\equiv {\Scr P}_M$ and used the definition of $\mathcal{Q}_{M\,A}{}^B$ in (\ref{Qournot}). This is consistent with the relation between the gaugino shift-tensor and the $\mathbb{T}$-tensor given in the fourth and fifth lines of (\ref{identificationsSNT}). Similarly one can show that the definition of the gravitino and hyperino shift-tensor are consistent with the identifications in (\ref{identificationsSNT}).\\
\emph{The scalar potential.} The scalar potential has the following general form:
  \begin{equation}
{V}(z,\bar{z},q)=g^2\,\left((k_M^ik_N^{\bar{\jmath}}g_{i\bar{\jmath}}+4\,h_{uv}k_M^uk_N^v)\overline{V}^M\,V^N+(U^{MN}-3\,V^M\overline{V}^N){\Scr P}^x_N{\Scr P}^x_M \right)\,, \label{potentialV}
\end{equation}
which can be derived from the potential Ward identity (\ref{WID}) which now reads:
\begin{equation}
g_{i\bar{\jmath}}\,W^{i\,AC}\,{W}_{BC}^{\bar{\jmath}}+2\, N_\alpha{}^A\,N^{\alpha}{}_{B}-12 \, \mathbb{S}^{AC}\mathbb{S}_{BC}=\delta_B^A\,{V}(z,\bar{z},q)\,g^{-2}\,,\label{wardn2}
\end{equation}
where ${W}_{AB}^{\bar{\imath}}\equiv (W^{i\,AB})^*$.\\
\emph{Proof of the potential Ward identity.} Following \cite{Andrianopoli:2015rpa}, we shall evaluate each term in the left-hand-side of (\ref{wardn2}) separately. From the above definitions we find:
\begin{eqnarray}
W^{i\,AC} {W}^{\bar{\jmath}}_{BC}g_{i\bar{\jmath}}&=&\delta_B^A\,k_M^ik_N^{\bar{\jmath}}g_{i\bar{\jmath}}\overline{V}^M\,V^N-i\,(\sigma^x)_B{}^A\,\left(
k_M^{\bar{\jmath}}\,V^M\,\overline{U}_{\bar{\jmath}}^N-k_M^{i}\,\overline{V}^M\,{U}_{i}^N\right)\,{\Scr P}^x_N+\nonumber\\
&&+(\sigma^x\sigma^y)_B{}^A\,{\Scr P}^x_M{\Scr P}^y_N U^{MN}\,,
\end{eqnarray}
where $U^{MN}\equiv {U}_{i}^N\,g^{i\bar{\jmath}}\,\overline{U}_{\bar{\jmath}}^N$, see (\ref{UMN}). On the right hand side of the above expression we split the terms proportional to $\delta_B^A$ from those proportional to $(\sigma^x)_B{}^A$ and use Eq. (\ref{idgauk}) to find:
\begin{eqnarray}
W^{i\,AC} {W}^{\bar{\jmath}}_{BC}g_{i\bar{\jmath}}&=&\delta_B^A\,\left(k_M^ik_N^{\bar{\jmath}}g_{i\bar{\jmath}}\overline{V}^M\,V^N+{\Scr P}^x_N{\Scr P}^x_M U^{MN}\right)+i\,(\sigma^x)_B{}^A\,\left(-2\,
{\rm X}_{MN}{}^P\overline{V}^M\,V^N\,{\Scr P}^x_P+\right.\nonumber\\&&+\left.\epsilon^{xyz}\,{\Scr P}^y_M{\Scr P}^z_N U^{[MN]}\right)\,.
\end{eqnarray}
Now use Eqs. (\ref{UMN}) and the locality constraint (\ref{qc3}) to write:
\begin{equation}
{\Scr P}^y_M{\Scr P}^z_N U^{[MN]}=-\frac{i}{2}\,{\Scr P}^y_M{\Scr P}^z_N \mathbb{C}^{MN}-{\Scr P}^y_M{\Scr P}^z_N \overline{V}^{[M} V^{N]}=-{\Scr P}^y_M{\Scr P}^z_N \overline{V}^{[M} V^{N]}\,,
\end{equation}
so that we finally find:
\begin{eqnarray}
W^{i\,AC} {W}^{\bar{\jmath}}_{BC}g_{i\bar{\jmath}}&=&\delta_B^A\,\left(k_M^ik_N^{\bar{\jmath}}g_{i\bar{\jmath}}\overline{V}^M\,V^N+{\Scr P}^x_N{\Scr P}^x_M U^{MN}\right)+i\,(\sigma^x)_B{}^A\,\left(-2\,
{\rm X}_{MN}{}^P\overline{V}^M\,V^N\,{\Scr P}^x_P\right.\nonumber\\&&-\left.\epsilon^{xyz}\,{\Scr P}^y_M{\Scr P}^z_N \, \overline{V}^{M} V^{N}\right)\,.
\end{eqnarray}
Let us now move to the evaluation of the square of the hyperini shifts:
\begin{eqnarray}
2\,N_\alpha{}^A\, N^\alpha{}_A=8\,\mathcal{U}^{A\alpha}_u\,\mathcal{U}_{v\,B\alpha}\,k^u_M\,k^v_N\,\overline{V}^M{V}^N=4\left(\delta_B^A h_{uv}+i\,(\sigma^x)_B{}^A\,K^x_{uv}\right)k^u_M\,k^v_N\,\overline{V}^M{V}^N\,,\label{NNesp}
\end{eqnarray}
where we have used Eq. (\ref{UUKs}). Finally let us compute the square of the gravitini shifts:
\begin{align}
-12\, \mathbb{S}^{AC}\,\mathbb{S}_{BC}&=-3\,(\sigma^x\sigma^y)_B{}^A\,{\Scr P}^x_M{\Scr P}^y_N \,V^M\overline{V}^N=-3\,{\Scr P}^x_M{\Scr P}^x_N\,V^M\overline{V}^N\delta_B^A+\nonumber\\
&+3i\,\epsilon^{xyz}\,{\Scr P}^y_M{\Scr P}^z_N \, \overline{V}^{M} V^{N}(\sigma^x)_B{}^A\,.\label{SSesp}
\end{align}
Thus the left-hand-side of (\ref{wardn2}) has the general form:
\begin{equation}
g_{i\bar{\jmath}}\,W^{i\,AC}{W}_{BC}^{\bar{\jmath}}+2\,N_\alpha{}^A\,N^{\alpha}{}_{B}-12 \, \mathbb{S}^{AC}\mathbb{S}_{BC}=\delta_B^A\,{V}(z,\bar{z},q)\,g^{-2}+i\,Z^x\,(\sigma^x)_B{}^A\,,
\end{equation}
where $V$ is given in (\ref{potentialV})
and
\begin{equation}
Z^x=(-2\,
{\rm X}_{MN}{}^P\,{\Scr P}^x_P+2\,\epsilon^{xyz}\,{\Scr P}^y_M{\Scr P}^z_N+4\,K^x_{uv}k^u_M\,k^v_N)\overline{V}^{M} V^{N}\,.
\end{equation}
We see that $Z^x$ vanishes by virtue of the equivariance condition (\ref{gequiv2}), so that Eq. (\ref{wardn2}) holds.
\subsection{Action and Supersymmetry Transformations}
For the sake of completeness, we give the Lagrangian and the supersymmetry transformation laws of the fields (as usual omitting terms which are quartic in the fermion fields in the action and cubic in the transformation laws). The formulas presented here are simply obtained by adaptation of the general formulae discussed in Sect. \ref{sec:3}, see for instance Eqs. (\ref{traphi2})-(\ref{traB2}).
\begin{align} \label{lagn2}
{\Scr L} &=
-\frac{e}{2} R +e\,
g_{i\bar{\jmath}}\mathcal{D}^\mu z^i \mathcal{D}_\mu \bar z^{\bar{\jmath}}+e\,
h_{uv}\mathcal{D}_\mu q^u \mathcal{D}^\mu q^v
+
\frac{e}{4} \, {\cal
I}_{\Lambda\Sigma}\,\mathcal{H}_{\mu\nu}{}^{\Lambda}
\mathcal{H}^{\mu\nu\,\Sigma} +\frac{1}{8} {\cal
R}_{\Lambda\Sigma}\;\varepsilon^{\mu\nu\rho\sigma}
\mathcal{H}_{\mu\nu}{}^{\Lambda}
\mathcal{H}_{\rho\sigma}{}^{\Sigma}+
 \nonumber\\
&-\frac{1}{8}\, \varepsilon^{\mu\nu\rho\sigma}\,
\Theta^{\Lambda\alpha}\,B_{\mu\nu\,\alpha} \, \Big(
2\,\partial_{\rho} A_{\sigma\,\Lambda} + X_{MN\,\Lambda}
\,A_\rho{}^M A_\sigma{}^N
-\frac{1}{4}\,\Theta_{\Lambda}{}^{\beta}B_{\rho\sigma\,\beta}
\Big)-\nonumber\\
&-\frac{1}{3}\,
\varepsilon^{\mu\nu\rho\sigma}X_{MN\,\Lambda}\, A_{\mu}{}^{M}
A_{\nu}{}^{N} \Big(\partial_{\rho} A_{\sigma}{}^{\Lambda}
+\frac{1}{4}  X_{PQ}{}^{\Lambda}
A_{\rho}{}^{P}A_{\sigma}{}^{Q}\Big)-
\nonumber\\[.9ex]
&{} -\frac{1}{6}\,
\varepsilon^{\mu\nu\rho\sigma}X_{MN}{}^{\Lambda}\, A_{\mu}{}^{M}
A_{\nu}{}^{N} \Big(\partial_{\rho} A_{\sigma}{}_{\Lambda}
+\frac{1}{4}\, X_{PQ\Lambda}
A_{\rho}{}^{P}A_{\sigma}{}^{Q}\Big)+\nonumber\\
& +\epsilon^{\mu\nu\rho\sigma}(\bar{\psi}^A_{\mu}\gamma_\nu\mathcal{D}_\rho\psi_{A\sigma}-
 \bar{\psi}_{A\,\mu}\gamma_\nu\mathcal{D}_\rho
 \psi^A_{\sigma})- {\,i\,e\,\over2}g_{i\bar{\jmath}} \left(\bar\lambda^{iA}\gamma^\mu
\mathcal{D}_\mu\lambda^{\bar{\jmath}}_A
+\bar\lambda^{\bar{\jmath}}_A \gamma^\mu \mathcal{D}_\mu\lambda^{iA}\right )-\nonumber\\
&-\,i\,e\,\left (\bar\zeta^\alpha\gamma^\mu\mathcal{D}_\mu\zeta_\alpha
+\bar\zeta_\alpha\gamma^\mu \mathcal{D}_\mu \zeta^\alpha \right)+ e\,\Big\{ -g_{i\bar{\jmath}}
 \mathcal{D}_\mu \bar z^{\bar{\jmath}} \bar \psi^\mu_A \lambda^{i A} -2 {\cal U}^{A\alpha}_u \mathcal{D}_\mu q^u
\bar \psi_A^\mu \zeta _\alpha+\nonumber\\&+g_{i\bar{\jmath}}  \mathcal{D} _\mu \bar z^{\bar{\jmath}}
\bar \lambda^{iA} \gamma^{\mu\nu} \psi_{A\nu}
+ 2\,{\cal U}^{A \alpha}_u \mathcal{D}_\mu q^u
\bar \zeta_\alpha \gamma^{\mu\nu}
\psi_{A\nu}+{\rm h.c.}\Big\}+\nonumber\\&+e\,\frac{1}{2}\{
{\cal H}^{-\Lambda}_{\mu\nu}\,
\mathcal{I}_{\Lambda\Sigma}\,
{\lbrack} 4 L^\Sigma  \bar \psi^{A\mu}
\psi^{B\nu} \epsilon_{AB}-4\,i\,
{\bar f}^\Sigma_{i^\star}\bar \lambda^{i^\star}_A \gamma^\nu
\psi_B^\mu \epsilon^{AB} + \nonumber \\
&+ \frac{1}{2}
\mathcal{D}_i f^\Sigma_j
\bar \lambda^{iA} \gamma^{\mu\nu} \lambda^{jB}\epsilon_{AB}-
L^\Sigma \bar\zeta_\alpha\gamma^{\mu\nu} \zeta_\beta
\mathbb{C}^{\alpha\beta}
{\rbrack}+{\rm h.c.}\}+
\nonumber  \\
&+ e\,\bigl[2g\,\mathbb{S}_{AB}\bar\psi^A_\mu\gamma^{\mu\nu}\psi^B_\nu +
\,i\, g\, g_{i\bar{\jmath}} W^{iAB} \bar\lambda^{\bar{\jmath}}_A\gamma_\mu \psi_B^\mu+
 2\,i\,g\, N^A_\alpha\bar\zeta^\alpha\gamma_\mu \psi_A^\mu +\nonumber \\
&+
g\,{\cal M}^{\alpha\beta}{\bar \zeta}_\alpha
\zeta_\beta +g\,{\cal M}^{\alpha}_{\phantom{\alpha}iB}
{\bar\zeta}_\alpha \lambda^{iB} + g\,{\cal M}_{ij\,AB}
{\bar \lambda}^{iA} \lambda^{j
B} + \mbox{h.c.}\bigr]
-e\,{ V}\bigl ( z, {\bar z}, q \bigr )+{\Scr L}_{4f},
\end{align}
where ${\Scr L}_{4f}$ contains the terms which are quartic in the fermion fields.\par
The matrices ${\cal M}^{\alpha\beta},\,{\cal M}^{\alpha}_{\phantom{\alpha}iB}
,\,{\cal M}_{ij\,AB}$ are defined as follows:
\begin{eqnarray}\label{mass2}
{\cal M}^{\alpha\beta}  &=&{\cal U}^{\alpha A}_u \, {\cal
U}^{\beta B}_v \, \varepsilon_{AB}
\, \mathcal{D}^{[u}   k^{v]}_{M}  \, V^M \\
{\cal M}^{\alpha }_{\phantom{\alpha} iB} &=& 4 \, {\cal
U}^{\alpha}_{B  u} \, k^u_{M} \,
 U_i^M \\
{\cal M}_{AB \,\, ik} &=&  -  \epsilon_{AB} \,
g_{\bar{\ell} [i}
 U_{k]}^M  k^{\bar{\ell}}_M \,+\frac {1}{2}
{\rm {i}}  P_ {M}^x\,(\sigma^x)_A{}^C\epsilon_{BC} \,
\mathcal{D}_i U^M _k  \label{pesamatrice}
\end{eqnarray}
The scalar potential $V$ was given in (\ref{potentialV}).\par
The supersymmetry transformation laws for the fermion fields read:
\begin{eqnarray}
\delta\,\psi _{A \mu} &=& {\cal D}_{\mu}\,\epsilon _A+\frac{1}{2}
\epsilon_{AB} H^-_{\mu \nu}\,\gamma^{\nu}\epsilon^B
 + \,i\, \, g \,\mathbb{S}_{AB}\eta _{\mu \nu}\gamma^{\nu}\epsilon^B+\dots
 \label{trasfgrav} \\
\delta \,\lambda^{iA}&=&
 \,i\,\,\mathcal{D} _ {\mu}\, z^i
\gamma^{\mu} \epsilon^A
+\frac{i}{4}\,H^{-i}_{\mu \nu} \gamma^{\mu \nu} \epsilon _B \epsilon^{AB}\,+\,g\,
W^{iAB}\epsilon _B+\dots
\label{gaugintrasfm}\\
 \delta\,\zeta _{\alpha}&=&\,i\,\,
{\cal U}^{B \beta}_{u}\, \mathcal{D} _{\mu}\,q^u
\,\gamma^{\mu} \epsilon^A
\epsilon _{AB}\,\mathbb{C}_{\alpha  \beta}
\,+g\,N_{\alpha}^A\,\epsilon _A+\dots\,, \label{iperintrasf}
\end{eqnarray}
where we have written $H^-_{AB\,\mu\nu}=\epsilon_{AB}\,H^-_{\mu\nu}$ and, using (\ref{FGnewuse}) and neglecting the fermion-bilinears
\begin{align}
H^-_{\mu\nu}&=2i\,L^\Lambda\,\mathcal{I}_{\Lambda\Sigma}\,
\mathcal{H}^{\Sigma\,-}_{\mu\nu}\,,\nonumber\\
H^{-i}_{\mu \nu}\equiv e^{-1\,Ii}\,H^-_{I\,\mu\nu}&=2i\,g^{i\bar{\jmath}}\,\bar{f}_{\bar{\jmath}}^\Lambda\,\mathcal{I}_{\Lambda\Sigma}\,
\mathcal{H}^{\Sigma\,-}_{\mu\nu}.
\end{align}
Let us consider now the supersymmetry variations of the bosonic fields. We have:
\begin{eqnarray}
\delta\,V^a_{\mu}&=& -\,i\,\,\bar {\psi}_{A
\mu}\,\gamma^a\,\epsilon^A -\,i\,\,\bar {\psi}^A _
\mu\,\gamma^a\,\epsilon_A\\
\delta \,A^M _{\mu}&=&
2 \overline{V}^M \bar \psi _{A\mu} \epsilon _B
\epsilon^{AB}\,+\,2V^M\,\bar\psi^A_{\mu}\epsilon^B \epsilon
_{AB}\nonumber\\
&+&\,i\, \,U^M_i \,\bar {\lambda}^{iA}
\gamma _{\mu} \epsilon^B \,\epsilon _{AB} +\,i\, \,
\overline{U}^{M}_{\bar{\imath}} \,\bar\lambda^{\bar{\imath}}_A
\gamma _{\mu} \epsilon_B \,\epsilon^{AB} \label{gaugtrasf}\\
\delta\,z^i &=& \bar{\lambda}^{iA}\epsilon _A \label{ztrasf}\\
\delta\,\bar{z}^{\bar{\imath}}&=& \bar{\lambda}^{\bar{\imath}}_A \epsilon^A
\label{ztrasfb}\\
  \delta\,q^u &=& {\cal U}^u_{\alpha A} \left(\bar {\zeta}^{\alpha}
  \epsilon^A + \mathbb{C}^{\alpha  \beta}\epsilon^{AB}\bar {\zeta}_{\beta}
  \epsilon _B \right)\,.
 \end{eqnarray}
As for the tensor fields, using (\ref{ThetadeltaB}), one finds:
\begin{align}
\Theta^{\Lambda\alpha}\,\delta B_{\alpha\mu\nu}&=2i\,g\,\Theta^{\Lambda\, a}\,\left(\bar{\lambda}^{iA}\gamma_{\mu\nu}\epsilon_A\,k_a^{\bar{\jmath}}-
\bar{\lambda}^{\bar{\jmath}}_A\gamma_{\mu\nu}\epsilon^A\,k_a^i\right)\,\,g_{i\bar{\jmath}}+\nonumber\\
&+4i\,g\,\Theta^{\Lambda\, n}\,\left(\bar{\zeta}_\alpha\gamma_{\mu\nu}\epsilon_A\,k_n^{u}\,\mathcal{U}_u^{A\alpha}-
\bar{\zeta}^\alpha\gamma_{\mu\nu}\epsilon^A\,k_n^{u}\,\mathcal{U}_{u\,A\alpha}\right)-\nonumber\\
&-4i\,g\,\Theta^{\Lambda\, n}\,{\Scr P}_n^x\,(\sigma^x)_B{}^A\,\left(\bar{\psi}_{A\,[\mu}\gamma_{\nu]}\epsilon^B+
\bar{\psi}^B_{[\mu}\gamma_{\nu]}
\epsilon_A\right)+\nonumber\\
&+4i\,g\,\Theta^{\Lambda\, a}\,{\Scr P}_a\,\left(\bar{\psi}_{A\,[\mu}\gamma_{\nu]}\epsilon^A+
\bar{\psi}^A_{[\mu}\gamma_{\nu]}
\epsilon_A\right)-\nonumber\\
&-2X^\Lambda{}_P{}^M\mathbb{C}_{MN}\,A^P_{[\mu}\,\delta A^N_{\nu]}\,.
\end{align}
where, as usual, $a,b=1,\dots, {\rm dim}(G^{(SK)})$, $n,m=1,\dots, {\rm dim}(G^{(QK)})$ and, in adapting the general formula (\ref{ThetadeltaB}), we have used the expressions for $\mathcal{P}_M,\,\mathcal{Q}_M$:
\begin{align}
\mathcal{P}_M^{IAB}&=\Theta_M{}^a\,k_a^{\bar{\imath}}\,e_{\bar{\imath}}{}^I\epsilon^{AB}\,\,,\,\,\,\,
\mathcal{P}_M^{A\alpha}=\Theta_M{}^n\,k_n^u\,\mathcal{P}_u^{A\alpha}=\sqrt{2}\,\Theta_M{}^n\,k_n^u\,
\mathcal{U}_u^{A\alpha}\,,\nonumber\\
(\mathcal{Q}_M)_A{}^B&=-\frac{1}{2}\Theta_M{}^n\,{\Scr P}^x_n\,(J_x)_A{}^B-\frac{1}{2}\Theta_M{}^a\,{\Scr P}_a\,(J_0)_A{}^B=\nonumber\\&=\frac{i}{2}\,\Theta_M{}^n\,{\Scr P}^x_n\,(\sigma_x)_A{}^B-\frac{i}{2}\,\Theta_M{}^a\,{\Scr P}_a\,\delta_A^B\,,
\end{align}
the matrix form of the generators $J_x,\,J_0$ being given in (\ref{Qournot}). \par
If we only gauge Abelian quaternionic isometries, upon dualizing the quaternionic scalars acted on by the gauge  transformations into tensors, as discussed in Sect. \ref{solvingspecial}, we find the model constructed in \cite{Dall'Agata:2003yr,D'Auria:2004yi}.\footnote{Our antisymmetric tensor fields are eight times the corresponding fields of \cite{Dall'Agata:2003yr,D'Auria:2004yi}.}\par
We notice that the scalar potential can also be written in the following more compact form \cite{Andrianopoli:1996cm}:
  \begin{equation}
{V}(z,\bar{z},q)=g^2\,\left({\Scr G}_{rs}\,k_M^rk_N^s\overline{V}^M\,V^N+(U^{MN}-3\,V^M\overline{V}^N)({\Scr P}^x_N{\Scr P}^x_M-{\Scr P}_N{\Scr P}_M) \right)\,, \label{potentialV2}
\end{equation}
where $(k_M^r)=(k_M^i,\,k_M^{\bar{\imath}},\,k^u_M)$ and :
\begin{equation}
{\Scr G}_{rs}\,k_M^rk_N^s=2\,g_{i\bar{\jmath}}\,k^i_M\,k^{\bar{\jmath}}_N+{\Scr G}_{uv}\,k_M^uk_N^v=2\,g_{i\bar{\jmath}}\,k^i_M\,k^{\bar{\jmath}}_N+2\,h_{uv}\,k_M^uk_N^v\,.
\end{equation}
In deriving (\ref{potentialV2}) we have also used the properties (\ref{newidentities}) to rewrite the term of the form $g_{i\bar{\jmath}}\,k^i_M\,k^{\bar{\jmath}}_N\overline{V}^M\,V^N$ as follows:
\begin{equation}
g_{i\bar{\jmath}}\,k^i_M\,k^{\bar{\jmath}}_N\overline{V}^M\,V^N={\Scr D}_{\bar{\jmath}}{\Scr P}_M{\Scr D}_i{\Scr P}_N\overline{V}^M\,V^N\,g^{i\bar{\jmath}}={\Scr P}_M {\Scr P}_N\,U^{MN}\,.
\end{equation}
\subsection{Fayet-Iliopulos Terms}\label{FIsec}
In the absence of hypermultiplets the ${\rm SU}(2)$ part of $H_R$ becomes a global symmetry of the theory which can still be gauged.
The equivariance condition reads:
\begin{equation}
\epsilon^{xyz}\,\mathcal{P}_M^y\,\mathcal{P}_N^z={ X}_{MN}{}^P\,\mathcal{P}_P^x\,,\label{gequiv3}
\end{equation}
where we have used the fact that, for $n_H=0$, $K^x=0$. The momentum maps $\mathcal{P}_P^x$ are now constant quantities coinciding with the embedding tensor which describes the gauging of the ${\rm SU}(2)$-symmetry: $\mathcal{P}_M^x=\Theta_M{}^x$. They are known as Fayet-Iliopoulos (FI) terms. From (\ref{gequiv3}) we see that we have two possibilities: either we gauge a ${\rm U}(1)$ inside ${\rm SU}(2)$, or the whole ${\rm SU}(2)$. In the former case we take $\Theta_M{}^x$ to have only one non-vanishing component, say $\theta_M=\Theta_M{}^{x=1}$, and choose the remaining gauge algebra so that ${ X}_{MN}{}^P\,\theta_P=0$. We can consider an Abelian gauging for which $X_{MN}{}^P=0$. In this case $\theta_M$ are the only gauge parameters and the scalar potential reads:
\begin{equation}
V=g^2\,\left(U^{MN}-3\,{V}^M\,\overline{V}^N\right)\theta_M\,\theta_N=-\frac{g^2}{2}\,\theta_M\mathcal{M}^{MN}\,
\theta_N-4\,g^2\,{V}^M\,\overline{V}^N\theta_M\,\theta_N\,.\label{VFI}
\end{equation}
The FI terms are like electric-magnetic charges: they are background quantities transforming in the representation ${\Scr R}_{v*}$ of $G^{(SK)}$. Being $k_n^u=0$, however, they do not define scalar-vector minimal couplings while they couple the fermion fields to the vectors. Just as we did for the electric and magnetic charges we can introduce the following composite quantities ${\Scr Z},\,{\Scr Z}_i$
\begin{align}
{\Scr Z}(z,\bar{z},\,\theta)&=V^M(z,\bar{z})\,\,\theta_M=L^\Lambda(z,\bar{z})\,\theta_\Lambda+
M_\Lambda(z,\bar{z})\,\theta^\Lambda\,,\label{Zdeff12}\\
{\Scr Z}_i(z,\bar{z},\,\theta)&\equiv {\Scr D}_i{\Scr Z}=U_i^M(z,\bar{z})\,\,\theta_M=f_i^\Lambda(z,\bar{z})\,\theta_\Lambda+
h_{\Lambda\,i}(z,\bar{z})\,\theta^\Lambda\,,\label{Zdeff22}
\end{align}
which are analogous to central and matter charges defined in Eqs. (\ref{Zdeff1}) and (\ref{Zdeff2}) for black hole solutions.
Using the definition (\ref{MLL2}) of $\mathcal{M}$ we can write the first term in (\ref{VFI}) as follows:
\begin{equation}
-\frac{g^2}{2}\,\theta_M\mathcal{M}^{MN}\,\theta_N=g^2\,\left(|{\Scr Z}|^2+g^{i\bar{\jmath}}\,{\Scr Z}_i\,\overline{{\Scr Z}}_{\bar{\jmath}}\right)\,.\label{VBHV}
\end{equation}
This term is a positive definite function of the scalars $z^i,\,\bar{z}^{\bar{\jmath}}$ and the FI terms which is $H$-invariant and has the same form as the black hole effective potential $V_{BH}$ \cite{Andrianopoli:2006ub}. The scalar potential can therefore be conveniently rewritten in terms of ${\Scr Z},\,{\Scr Z}_i$ as follows:
\begin{equation}
V=g^2\,\left(|{\Scr Z}|^2+g^{i\bar{\jmath}}\,{\Scr Z}_i\,\overline{{\Scr Z}}_{\bar{\jmath}}-4\,|{\Scr Z}|^2 \right)=g^2\,\left(g^{i\bar{\jmath}}\,{\Scr Z}_i\,\overline{{\Scr Z}}_{\bar{\jmath}}-3\,|{\Scr Z}|^2 \right)\,.\label{VFI2}
\end{equation}
Let us remark here the difference between $m^\Lambda,\,e_\Lambda$ and $\theta_M$: the former are solitonic charges, while the latter are background quantities actually entering the Lagrangian.\par
In the case in which the whole global ${\rm SU}(2)$ is gauged, namely $\Theta_M{}^x$ has rank 3, the left-hand-side of  (\ref{gequiv3}) is non-vanishing, and so has to be the right-hand-side. This in turn requires the gauging of a corresponding ${\rm SU}(2)$ inside $G^{(SK)}$.\par
The FI terms are an important ingredient in supergravity model building: They were required in the construction of the first extended supergravity models featuring a stable de Sitter vacuum \cite{Fre:2002pd}; Abelian $\mathcal{N}=2$ models with FI terms have provided useful supergravity frameworks where to study black hole solutions in anti-de Sitter spacetime, since, for suitable choices of $\theta$, they feature vacua of this kind, see for instance \cite{adsblackholes1,adsblackholes2,adsblackholes3,adsblackholes4,adsblackholes5}.
\subsection{Relation to Calabi-Yau Compactifications}\label{rtcy}
$\mathcal{N}=2$ ungauged supergravities originate from compactifications of Type II superstring theory on Calabi-Yau three-folds \cite{Ferrara:1989vp,Strominger:1990pd,Candelas:1990pi,Candelas:1990rm,Candelas:1990qd}. Our discussion of this low-energy effective theory is restricted to the string tree-level approximation.\par
A Calabi-Yau manifold $\mathcal{X}$ is a compact K\"ahler manifold with vanishing first Chern class. This condition restricts the holonomy group to ${\rm SU}(n_c)$, $n_c$ being its complex dimension, and is equivalent to the existence of a Ricci-flat K\"ahler metric on the manifold.  We shall consider $\mathcal{X}$ of complex dimension 3, which has ${\rm SU}(3)$ holonomy.
Considering a ten-dimensional background of the form
$$M_{D=4}\times \mathcal{X}\,,$$
the ${\rm SU}(3)$ holonomy of the internal manifold  guarantees the existence of a covariantly constant spinor and thus of a residual amount of supersymmetry preserved by background, which turns out to be one quarter the supersymmetry of the original ten-dimensional theory. $M_{D=4}$ is the four-dimensional Minkowski vacuum of the resulting effective four-dimensional model with eight conserved supercharges. Superstring theory on this background is described by a super-conformal world-sheet theory with $(2,2)$-supersymmetry.\par
It is known that the moduli parametrizing the deformations of a Calabi-Yau metric split into two classes: The K\"ahler moduli $\delta {\tt g_{s\bar{t}}}$, ${\tt s,t}=1,2,3$, describing the deformations in ``size'' of $\mathcal{X}$, which are in one-to-one correspondence with the harmonic $(1,1)$-forms, i.e. with elements of $H^{(1,1)}(\mathcal{X})$, and the complex structure moduli $\delta {\tt g_{\bar{s}\bar{t}}}$, describing the deformations in ``shape'' of $\mathcal{X}$, which are associated with harmonic $(2,1)$-forms, i.e. with elements of $H^{(2,1)}(\mathcal{X})$, ${\tt g}$ being the Calabi-Yau metric. We shall denote a suitable complexification, see below, of the former set of moduli by $w^a$, where, only in this Section, $a,b=1,\,\dots, h_{1,1}$, $ h_{1,1}={\rm dim}(H^{(1,1)})$, and the latter moduli by $z^i$, $i=1,\,\dots, h_{2,1}$, $h_{2,1}$ being the dimension of $H^{(2,1)}(\mathcal{X})$ (and of $H^{(1,2)}(\mathcal{X})$). In the four-dimensional effective supergravity, these deformation parameters enter as massless scalar fields parametrizing two different special K\"ahler manifolds: ${\Scr M}_{SK}^{(1)},\,{\Scr M}_{SK}^{(2)}$, respectively. Let us briefly recall the relation between the geometries of the two moduli spaces and the topology of the corresponding internal manifold.\par
 A feature of Calabi-Yau manifolds is that $H^{(3,0)}(\mathcal{X})$ (as well as $H^{(0,3)}(\mathcal{X})$) is one-dimensional and contains the characteristic holomorphic $(3,0)$-form $\boldsymbol{\Omega}$. The existence on $\mathcal{X}$ of the unique closed $(3,0)$-form $\boldsymbol{\Omega}$ and $(1,1)$-form $J$, i.e. the K\"ahler form, is strictly related to the manifold having ${\rm SU}(3)$-holonomy.  Let $(\boldsymbol{\alpha}_M)\equiv (\alpha_\Lambda,\,-\beta^\Lambda)$, $\Lambda=0,\dots, h_{2,1}$, be a basis of the third cohomology class $H^{(3)}(\mathcal{X},\mathbb{Z})$. The basis $\boldsymbol{\alpha}_M$ is chosen so that:
\begin{equation}
\int_{\mathcal{X}} \boldsymbol{\alpha}_M\wedge \boldsymbol{\alpha}_N=-\mathbb{C}_{MN}\,.\label{alphabeta}
\end{equation}
 The form $\boldsymbol{\Omega}$ is a holomorphic function of the complex structure moduli $z^i$, $\boldsymbol{\Omega}=\boldsymbol{\Omega}(z)$, and the holomorphic section $\Omega_2(z)^M$ of ${\Scr M}_{SK}^{(2)}$  is defined by expanding $\boldsymbol{\Omega}$ in the basis $\alpha_M$:
 \begin{equation}
 \boldsymbol{\Omega}(z)=\Omega_2^M(z)\,\boldsymbol{\alpha}_M=X_2^\Lambda(z)\,\alpha_\Lambda-F_{2\,\Lambda}(z)\,\beta^\Lambda\,.\label{bOmOm}
 \end{equation}
 The three-form $\boldsymbol{\Omega}$ is defined modulo multiplication by a holomorphic function $\boldsymbol{\Omega}(z)\rightarrow e^{f(z)}\,\boldsymbol{\Omega}(z)$, which amounts to a K\"ahler transformation on the line bundle $\texttt{L}$ over the moduli space.
 Similarly the covariant derivatives ${\Scr D}_i\Omega_2^M(z),\,{\Scr D}_{\bar{\imath}}\overline{\Omega}_2^M(\bar{z})$ of $\Omega_2^M(z)$ and $\overline{\Omega}_2^M(\bar{z})$ describe the components along $\boldsymbol{\alpha}_M$ of $(2,1)$ and $(1,2)$ harmonic forms $\boldsymbol{\chi}_i,\,\bar{\boldsymbol{\chi}}_{\bar{\imath}}$, so that the columns $\Omega_2^M,\,{\Scr D}_{\bar{\imath}} \overline{\Omega}_2^M$, $\overline{\Omega}_2^M,\,{\Scr D}_i\Omega_2^M$ of $e^{-\frac{\mathcal{K}_2}{2}}\,\mathbb{L}_c^M{}_{\underline{N}}$, $\mathcal{K}_2$ being the K\"ahler potential, describe the components, or \emph{periods}, in the chosen basis $\boldsymbol{\alpha}_M$, of the 3-forms $\boldsymbol{\Omega},\,\bar{\boldsymbol{\chi}}_{\bar{\imath}},\,\overline{\boldsymbol{\Omega}},\,\boldsymbol{\chi}_i$. These forms define a basis of
 \begin{equation}
 H^{(3)}(\mathcal{X})=H^{(3,0)}(\mathcal{X})\oplus H^{(1,2)}(\mathcal{X})\oplus H^{(0,3)}(\mathcal{X})\oplus H^{(2,1)}(\mathcal{X})\,.
 \end{equation}
 The symplectic frame is therefore determined by the choice of the basis $\boldsymbol{\alpha}_M$. One can also show that the matrix $\mathcal{M}(z,\bar{z})$ has the following geometric interpretation:
 \begin{equation}
 \int_{\mathcal{X}} \boldsymbol{\alpha}_M\wedge {}^*\boldsymbol{\alpha}_N=-\mathcal{M}^{MN}>0\,.
 \end{equation}
 In the symplectic frames originating from Calabi-Yau compactifications the prepotential always exists, i.e. $X^\Lambda$ can be viewed as homogeneous coordinates for the moduli space. The K\"ahler potential of ${\Scr M}_{SK}^{(2)}$ reads:
 \begin{equation}
 \mathcal{K}_2(z,\,\bar{z})=-\log\left(i\,\int_{\mathcal{X}}\boldsymbol{\Omega}\wedge \overline{\boldsymbol{\Omega}}\right)\,,
 \end{equation}
 which yields, using (\ref{bOmOm}) and (\ref{alphabeta}), Eq. (\ref{Komapp}).\par
 As far as the complexified K\"ahler moduli $w^a$ of the Calabi-Yau are concerned, they are defined as the components, along a basis ${\bf e}_a$ of $H^{(1,1)}$, of the complex combination $B+i\,J$, where $B$ is the Kalb-Ramond 2-form of the ten-dimensional superstring theory with the indices along the internal directions ($B_{{\tt s\bar{t}}}$), and $J$ the K\"ahler 2-form on $\mathcal{X}$:
 \begin{equation}
 B+i\,J=w^a\,{\bf e}_a\,.
 \end{equation}
 Only the imaginary parts $v^a$ of $w^a=u^a+i\,v^a$ parametrize the possible choices of the K\"ahler class, the real parts being related to the $B$-form: $J=v^a\,{\bf e}_a,\, B=u^a\,{\bf e}_a$. Let $d_{abc}$ denote the triple intersection numbers:
 \begin{equation}
 d_{abc}\equiv \int_{\mathcal{X}} {\bf e}_a\wedge {\bf e}_b\wedge {\bf e}_c=d_{(abc)}\,.
 \end{equation}
  The volume ${\rm Vol}[\mathcal{X}]$ of $\mathcal{X}$ then reads: ${\rm Vol}[\mathcal{X}]=\frac{1}{3!}\,\int_{\mathcal{X}}J\wedge J\wedge J=\frac{1}{3!}\,d_{abc}\,v^a\, v^b\, v^c$. The moduli $w^a$ are special coordinates on the special K\"ahler manifold ${\Scr M}_{SK}^{(1)}$ on which the holomorphic section is denoted by $(\Omega_1^m(w))=(X_1^A(w),\,F_{1\,A}(w))$, $A=(0,a)$, with $w^a=X_1^a/X^0_1$, and the prepotential, in its ``bare'' or non-quantum corrected form, is defined by the following cubic polynomial:
  \footnote{Our discussion of the manifold ${\Scr M}_{SK}^{(1)}$ does not take into account quantum corrections.}
  \begin{equation}
  F(X_1)=(X_1^0)^2\,{\Scr F}(w)\,\,;\,\,\,\,\,{\Scr F}(w)\equiv \frac{1}{3!}\,d_{abc}\,w^a w^b w^c\,.\label{cubicF}
  \end{equation}
 From Eq. (\ref{Kprepot}) the  K\"ahler potential is computed to be:
   $$\mathcal{K}_1(w,\bar{w})=-\log\left(\frac{1}{3!}\,d_{abc}\,v^a v^b v^c\right)\,,$$
  and the metric reads:
  \begin{equation}
  g_{a\bar{b}}=-\frac{3}{2}\,\left(\frac{d_{ab}}{d}-\frac{3}{2}\,\frac{d_a\,d_b}{d^2}\right)\,,
  \end{equation}
  where $d\equiv d_{abc}\,v^a\, v^b\, v^c$, $d_a\equiv d_{abc}\,v^b\, v^c$ and $d_{ab}=d_{abc}\,v^c$. Note that the above metric features the following Peccei-Quinn translational isometries:
  \begin{equation}
  u^a\rightarrow u^a+c^a\,\,,\,\,\,\,v^a\rightarrow v^a\,,
  \end{equation}
  where $c^a$ are $h_{1,1}$ constant parameters.
 Finally the characteristic rank-3 symmetric tensor $C_{abc}$ has the simple form: $C_{abc}=e^{\mathcal{K}_1}\,d_{abc}$.
  Special K\"ahler manifolds characterized by a prepotential of the form (\ref{cubicF}) are called \emph{cubic special geometries}. The manifold ${\Scr M}_{SK}^{(1)}$ spanned by the complexified K\"ahler moduli receives world-sheet instanton corrections while that of the complex structure deformations does not.\par Aside from Calabi-Yau compactifications, cubic geometries also originate from toroidal reduction to four dimensions of models in $D=5$ with eight supercharges. In this case the Peccei-Quinn axions $u^a$ derive from the internal components of the five-dimensional vector fields $A^a_{\tilde{\mu}}$ and the Peccei-Quinn symmetry is the remnant in four dimensions of the corresponding gauge symmetry. The tensor $d_{abc}$ defines the Chern-Simons term $d_{abc} F^a\wedge F^b\wedge A^c$ in the five-dimensional Lagrangian and is an invariant of its global symmetry group. In this picture the axions $u^a$ are analogous, in the $\mathcal{N}=2$ case, to the scalars $a^\lambda$ in the maximal theory originating from Kaluza-Klein reduction of the five-dimensional one, see Eqs. (\ref{e6frameg}), and $d_{abc}$ corresponds to the ${\rm E}_{5(5)}$-invariant $d_{\lambda\sigma\gamma}$. \par
The structure of ${\Scr M}_{SK}^{(1)}$ parallels that of of ${\Scr M}_{SK}^{(2)}$. In particular we can associate the symplectic frame on the former with a basis of even forms on $\mathcal{X}$ $({\bf e}_{\mathcal{A}})=({\bf e}_A,\,{\bf e}^A)$:
\begin{align}
({\bf e}_A)=({\bf e}_0,\,{\bf e}_a)\,\,;\,\,\,\,{\bf e}_0=1\in\, H^{(0,0)}(\mathcal{X})\,,\,\,\,{\bf e}_a\in \,H^{(1,1)}(\mathcal{X})\,,\nonumber\\
({\bf e}^A)=({\bf e}^0,\,{\bf e}^a)\,\,;\,\,\,\,{\bf e}^0=-\frac{\sqrt{{\rm det}({\tt g})}}{{\rm Vol}[\mathcal{X}]}\,d^6x\in\, H^{(3,3)}(\mathcal{X})\,,\,\,\,{\bf e}^a\in H^{(2,2)}(\mathcal{X})\,.
\end{align}
This basis is chosen so that  $\int_{\mathcal{X}}{\bf e}_a\wedge {\bf e}^b=\delta_a^b$. The symplectic structure follows from the Mukai pairing $\langle\,,\,\rangle$:
\begin{equation}
\langle \sigma,\,\gamma\rangle\equiv ( \sigma\wedge\lambda(\gamma))_{{\rm top}}\,,
\end{equation}
where the subscript ``top'' stands for the projection on the top-form and $\lambda$ acts on $p$-forms $\omega^{(p)}$ as follows: $\lambda(\omega^{(2n)})=(-1)^n\,\omega^{(2n)}$, $\lambda(\omega^{(2n-1)})=(-1)^n\,\omega^{(2n-1)}$.
We then have for the basis of even forms a relation similar to (\ref{alphabeta}):
\begin{equation}
\int_{\mathcal{X}} \langle\boldsymbol{e}_{\mathcal{A}},\,\boldsymbol{e}_{\mathcal{B}}\rangle=\mathbb{C}_{{\mathcal{AB}}}\,.\label{alphabetaeven}
\end{equation}
Clearly also (\ref{alphabeta}) can be written in terms of the Mukai pairing.\par
Let us consider the dimensional reduction of Type II theories on a Calabi-Yau manifold $\mathcal{X}$ and restrict ourselves to the bosonic sector. Using the property that there are no harmonic one-forms on $\mathcal{X}$, the various ten-dimensional fields yield the following four-dimensional ones (all the four-dimensional fields should be intended as fluctuations about their vacuum values):\footnote{The scalars $\zeta^\Lambda$ should not be confused with the hyperini $\zeta^\alpha$.}
\begin{align}
\mbox{Type IIA:}&\nonumber\\
&C^{(3)}= A^a_\mu\,dx^\mu\wedge {\bf e}_a+\zeta^\Lambda\,\alpha_\Lambda-\tilde{\zeta}_\Lambda\,\beta^\Lambda=A^a_\mu\,dx^\mu\wedge {\bf e}_a+\mathcal{Z}^M\,\boldsymbol{\alpha}_M\,,\nonumber\\
&C^{(1)}\,=\,A^0_\mu\,dx^\mu\,,\nonumber\\
&B_{(2)}\,=\,\frac{1}{2}\,B_{\mu\nu}\,dx^\mu\wedge dx^\nu+u^a\,{\bf e}_{a},\,J=v^a\,{\bf e}_{a}\,,\nonumber\\
&\phi\,\rightarrow\,\,\phi_4=\phi-\frac{1}{2}\,\log\left({\rm Vol}[\mathcal{X}]\right)\,,\nonumber\\
&({\tt g}_{{\tt s{t}}},\,{\tt g}_{{\tt \overline{s{t}}}})\,\rightarrow\,\,z^i\,,\nonumber\\
\mbox{Type IIB:}&\nonumber\\
&C^{(4)}=A_{\mu}^\Lambda\,dx^\mu\wedge \alpha_\Lambda+C_a\,{\bf e}^a\,,\nonumber\\
&C^{(2)}\,=\,\frac{1}{2}\,C_{\mu\nu}\,dx^\mu\wedge dx^\nu+C^a\,{\bf e}_{a}\,,\nonumber\\
&C^{(0)}\,=\,\rho\,,\nonumber\\
&B_{(2)}\,=\,\frac{1}{2}\,B_{\mu\nu}\,dx^\mu\wedge dx^\nu+u^a\,{\bf e}_{a},\,J=v^a\,{\bf e}_{a}\,,\nonumber\\
&\phi\,\rightarrow\,\,\phi_4=\phi-\frac{1}{2}\,\log\left({\rm Vol}[\mathcal{X}]\right)\,,\nonumber\\
&({\tt g}_{{\tt s{t}}},\,{\tt g}_{{\tt \overline{s{t}}}})\,\rightarrow\,\,z^i\,,\label{redsiiba}\\
\end{align}
The resulting four-dimensional theory is an ungauged $\mathcal{N}=2$ supergravity which thus features no scalar potential. The corresponding scalar fields are therefore not dynamically fixed and parametrize a continuum of degenerate Minkowski vacua.\par
In the Type IIA compactification the effective $\mathcal{N}=2$ theory describes, aside from the supergravity multiplet which contains the graviphoton $A^0_\mu$, $n=h_{1,1}$ vector multiplets containing the vectors $A^a_\mu$ and the complexified K\"ahler moduli $w^a$, and $n_H=h_{2,1}+1$ hypermultiplets whose scalar content consists in the $(2\,h_{2,1}+2)$-symplectic vector of scalars $(\mathcal{Z}^M)=(\zeta^\Lambda,\,\tilde{\zeta}_\Lambda)$, the scalar $\tilde{B}$ dual to $B_{\mu\nu}$, the four-dimensional dilaton $\phi_4$, and the complex structure moduli $z^i$. The latter are therefore part of the hyper-scalars $q^u$. In fact the special K\"ahler manifold ${\Scr M}_{SK}^{(2)}$ spanned by $z^i$ is contained in the quaternionic K\"ahler manifold ${\Scr M}_{QK}^{{\rm (IIA)}}$ parametrized by $q^u$.\par
The reduction of Type IIB theory, on the other hand, yields a supergravity multiplet coupled to $n=h_{2,1}$ vector multiplets containing the complex structure moduli $z^i$, and $n_H=h_{1,1}+1$ hypermultiplets. Of the vectors $A^\Lambda_\mu$ originating from the RR four-form, one is contained in the supergravity multiplet. The  hypermultiplets contain the $(2\,h_{1,1}+2)$-symplectic vector of scalars $(\mathcal{Z}^{\mathcal{A}})=(\zeta^A,\,\tilde{\zeta}_A)=(\rho,\,C^a,\,\tilde{C},\,C_a)$, and the NS-NS fields $\tilde{B}$, $\phi_4$, $w^a$. The latter moduli $w^a$ span the manifold ${\Scr M}_{SK}^{(1)}$ which is thus contained in the quaternionic K\"ahler one ${\Scr M}_{QK}^{{\rm (IIB)}}$ parametrized by the hyper-scalars $q^u$. Note that we did not write, in the expansion of the RR-four-form, the components $C_{\mu\nu}^a$ and $A_{\mu\,\Lambda}$ since they are related to $C_a$ and $A^\Lambda_\mu$, respectively, by the self-duality condition on the five-form field strength $\hat{F}^{(5)}={}^*\hat{F}^{(5)}$. \par To summarize, in the two pictures, the scalar manifolds have the following forms:
\begin{align}
\mbox{Type IIA:}&\nonumber\\
&{\Scr M}_{{\rm scal}}={\Scr M}_{SK}^{(1)}\times {\Scr M}_{QK}^{{\rm (IIA)}}\,\,,\,\,\,\,\,{\Scr M}_{SK}^{(2)}\subset {\Scr M}_{QK}^{{\rm (IIA)}}\,,\nonumber\\
\mbox{Type IIB:}&\nonumber\\
&{\Scr M}_{{\rm scal}}={\Scr M}_{SK}^{(2)}\times {\Scr M}_{QK}^{{\rm (IIB)}}\,\,,\,\,\,\,\,{\Scr M}_{SK}^{(1)}\subset {\Scr M}_{QK}^{{\rm (IIB)}}\,.
\end{align}
In both cases half of the quaternionic scalars $q^u$ originate from the RR fields and half from the NS-NS ones, including the dilaton field (in the form of the four-dimensional dilaton $\phi_4$). The lower-dimensional theories also feature two symplectic structures, associated with the groups ${\rm Sp}(2h_{1,1}+2,\mathbb{R})$ and ${\rm Sp}(2h_{2,1}+2,\mathbb{R})$, one of which defines the scalar-vector couplings, the other is ``hidden'' inside the quaternionic manifold. The fact that  the dilaton, which is related to the string coupling constant $g_s=e^\phi$, sits in the quaternionic K\"ahler manifold and the fact that $\mathcal{N}=2$ supersymmetry forbids couplings between vector multiplet scalars and hyper-scalars implies that only the geometry of ${\Scr M}_{QK}$ receives perturbative and non-perturbative string corrections (which depend on $g_s$).
\par The embedding of a special K\"ahler manifold inside a quaternionic one, which occurs in the two pictures (at string tree-level), defines a correspondence between the former class of spaces and the latter one which is called \emph{c-map} \cite{Cecotti:1988qn}:
\begin{equation}
{\Scr M}_{SK}\,\stackrel{{\scriptsize {\rm c-map}}}{\longrightarrow}\, {\Scr M}_{QK}\supset {\Scr M}_{SK}\,.
\end{equation}
In particular the special K\"ahler manifold in the vector multiplet sector from Type IIA or IIB theory is mapped into the quaternionic K\"ahler manifold in the Type IIB or IIA picture, respectivley.
Actually the same correspondence occurs when we further compactify the four-dimensional theory on a circle to three dimensions and dualize the vector fields into scalars \cite{Cecotti:1988qn,Ferrara:1989ik}:\footnote{This kind of dimensional reduction and the geometry of the emerging sigma-model, was first studied in \cite{Breitenlohner:1987dg}. } The scalar fields in the vector multiplets, together with the two scalars originating from the four-dimensional metric and those arising from the vector fields, span the quaternionic K\"ahler manifold image through the c-map of the original special  K\"ahler one.\footnote{Not all quaternionic K\"ahler manifolds are in the image of the c-map.} We refer the reader to Appendix \ref{BGM} for a review of this construction. When going to three dimensions in the Type IIA picture, the special K\"ahler manifold ${\Scr M}_{SK}^{(1)}$ associated with the vector multiplet sector becomes part of the quaternionic K\"ahler manifold ${\Scr M}_{QK}^{{\rm (IIB)}}$, while the hyper-scalars are described in the lower-dimensional theory by the same space ${\Scr M}_{QK}^{{\rm (IIA)}}$. Therefore in the three dimensional theory all bosonic degrees of freedom of the four-dimensional one are scalar fields described by a sigma model with target space:
\begin{equation}
{\Scr M}_{{\rm scal}}^{(D=3)}={\Scr M}_{QK}^{{\rm (IIB)}}\times {\Scr M}_{QK}^{{\rm (IIA)}}\,.\label{MD3qq}
\end{equation}
It is remarkable that performing the reduction in the Type IIB picture one ends up with precisely the same sigma-model. This can be understood from the fact that a T-duality along the internal fourth dimension, which is a symmetry of the three dimensional theory, maps the Type IIA description into Type IIB one. This amounts to an irrelevant exchange of the two factors in (\ref{MD3qq}).\par
In Table 2 the symmetric special and quaternionic K\"ahler manifolds are listed according to the correspondence defined by the c-map: from top to bottom the first special K\"ahler space is mapped into the first quaternionic one with $n_H=n+1$; the second special K\"ahler space into the second quaternionic one with $n_H=n+1$, and so on. The last manifold in the table $\frac{{\rm USp}(2,2n_H)}{{\rm USp}(2)\times {\rm USp}(2n_H)}$ is the only homogeneous quaternionic K\"ahler space not in the image of the c-map.
\par
With each Calabi-Yau manifold $\mathcal{X}$ we can associate a \emph{mirror} manifold $\hat{\mathcal{X}}$ of the same type (see for instance \cite{Greene:1998yu}), such that $h_{1,1}(\mathcal{X})=h_{2,1}(\hat{\mathcal{X}})$ and $h_{2,1}(\mathcal{X})=h_{1,1}(\hat{\mathcal{X}})$ and
\begin{equation}
{\Scr M}_{SK}^{(1)}[\mathcal{X}]={\Scr M}_{SK}^{(2)}[\hat{\mathcal{X}}]\,\,;\,\,\,\,{\Scr M}_{SK}^{(2)}[\mathcal{X}]={\Scr M}_{SK}^{(1)}[\hat{\mathcal{X}}]\,,
\end{equation}
once all the world-sheet instanton corrections are taken into account. The following duality, named \emph{mirror symmetry}, holds: \emph{Type IIA on $\mathcal{X}$ is equivalent to Type IIB on $\hat{\mathcal{X}}$}. In particular the corresponding low-energy effective theories should coincide. Mirror symmetry is thus analogous to T-duality along an odd number of internal directions in toroidal compactifications, see Section \ref{N8MAB}.\footnote{ See \cite{Strominger:1996it} for an interpretation of mirror symmetry as a T-duality.}
A stronger statement of mirror symmetry also involves the quantum corrected quaternionic geometries of the two effective models (recall that the geometry of ${\Scr M}_{QK}$ receives $g_s$-corrections).
\paragraph{Fluxes and ${\rm SU}(3)$-structure manifolds.}
As previously pointed out, the presence of fluxes in a Calabi-Yau compactification, through their back-reaction on the space-time metric, will alter the geometry of the internal manifold, spoiling, in general, its defining property of having ${\rm SU}(3)$-holonomy and possibly inducing a warp-factor, depending on the internal coordinates, multiplying the metric of $M_{D=4}$. We may still require the low-energy dynamics on the backgorund to be captured by an effective $\mathcal{N}=2$ supergravity. This amounts to the condition that the back-reacted internal manifold ${\Scr X}$ admit a \emph{globally defined spinor} and restricts its structure group to ${\rm SU}(3)\subset {\rm SO}(6)$ or less, ${\rm SO}(6)$ being the structure group of a generic 6-dimensional compact Riemannian manifold. The condition on ${\Scr X}$ to have an ${\rm SU}(3)$-structure is less stringent than that on the Levi-Civita connection to have ${\rm SU}(3)$-holonomy, which defines Calabi-Yau manifolds. Manifolds with ${\rm SU}(3)$-structure (or in general ${\tt G}$-structure) received considerable attention
in relation to the study of supersymmetric flux-compactifications, see \cite{Grana:2005jc} and references therein. The effective lower-dimensional theory is a gauged supergravity, the gauging being induced by the internal fluxes. This model describes, as a solution, the original background $M_D$, but may well feature other solutions, possibly associated with different compactifications.

\subsection{Solving the Constraints in a Simple Model}\label{STUsolvq}
In this Section we discuss a special example \cite{Fre:2013tya} in which the constraints on the embedding tensor can be explicitly solved and all possible gaugings classified in orbits of the global symmetry group $G$. We consider the so-called STU model which describes $\mathcal{N}=2$ supergravity  coupled to three vector multiplets and no hypermultiplets. Although simple, this example is instructive since it provides one of the few examples in which the quadratic constraints on the embedding tensor can be explicitly solved and, at the same time, it features gaugings with stable anti-de Sitter and de Sitter vacua.
\subsubsection{The Geometry of the Scalar Manifold}\label{STUgeometry0} The scalar manifold belongs to the infinite series in the second line of Table 2, with $n=3$, and reads:
\begin{equation}
{\Scr M}^{(STU)}_{{\rm scal}}=\left(\frac{{\rm SL}(2,\mathbb{R})}{{\rm SO}(2)}\right)^3\,.
\end{equation}
The global symmetry group of the theory is $ G \, = \, {\rm SL}(2,\mathbb{R})^3\times {\rm SO}(3)$ the latter
factor being the form of ${G}^{(QK)}$ in the absence of hypermultiplets. The duality representation is ${\Scr R}_v={\bf (2,2,2)}$ of ${\rm SL}(2,\mathbb{R})^3$.\par
In the special coordinate frame the coordinates are $(z^i)=(z^1,\,z^2,\,z^3)$, also denoted in the literature by $s,t$ and $u$, and the prepotential is cubic of the form:
\begin{equation}
F(X)=\frac{X^1 X^2 X^3}{X^0}=(X^0)^2\,{\Scr F}(z)\,\,,\,\,\,\,{\Scr F}(z)=z^1\,z^2\,z^3\,.
\end{equation}
We write the complex scalars in terms of real ones $\{\phi^s\}=\{a_i,\,\varphi_i\}$ as follows:
$z^i=a_i-i\,e^{\varphi_i}$. The K\"ahler potential is readily computed to be $e^{-\mathcal{K}}=8\,e^{\varphi_1+\varphi_2+\varphi_3}$, so that the K\"ahler metric in the complex basis reads:
\begin{equation}
g_{i\bar{\jmath}}=-\frac{1}{(z^i-\bar{z}^{\bar{\imath}})^2}\,\delta_{ij}=\frac{e^{-2\varphi_i}}{4}\,\,\delta_{ij}\,.
\end{equation}
Setting $X^0=1$, the holomorphic section $\Omega^M(z)$ has the following form:
\begin{equation}
\Omega^M(z)=\{1,z^1,z^2,z^3,-z^1 z^2 z^3,z^2 z^3,z^1 z^3,z^1
   z^2\}\,.\label{omegastu}
\end{equation}
The special coordinates correspond to a solvable parametrization of the manifold in which the real coordinates $\phi^s$ are parameters of a solvable Lie algebra ${\Scr S}=\bigoplus {\Scr S}_i$, where ${\Scr S}_i$ are the solvable algebras associated with each of the three factors  ${\Scr S}_i=({\bf E}_i,\,{\bf h}_i)$: $[{\bf h}_i,\,{\bf E}_j]=\delta_{ij}\,{\bf E}_i$.
 The coset representative $L$ is an element of the corresponding solvable group defined by the following exponentialization prescription:
\begin{equation}
L(\phi^s)=\exp(\phi^s\,t_s)=\prod_{i=1}^3 e^{a_i {\bf E}_i}e^{\varphi_i {\bf h}_i}\,.\label{L4stu}
\end{equation}
For the sake of completeness we give below the explicit matrix forms of the solvable Lie algebra generators in the ${\Scr R}_v$ representation (${\Scr R}_v[t_s]=(t_s{}^M{}_N)$):\footnote{With an abuse of notation we denote the abstract generator and its matrix representation in this symplectic basis by the same symbols, ${\bf h}_i,\,{\bf E}_i$: ${\bf h}_i\equiv {\Scr R}_v[{\bf h}_i],\,E_i\equiv {\Scr R}_v[{\bf E}_i]$.}
\begin{align}
\varphi_i {\bf h}_i&=-\frac{1}{2}{\rm diag}(\varphi _1+\varphi _2+\varphi _3,-\varphi _1+\varphi _2+\varphi _3,\varphi _1-\varphi _2+\varphi _3,\varphi
   _1+\varphi _2-\varphi _3,\nonumber\\
   &-\varphi _1-\varphi _2-\varphi _3,\varphi _1-\varphi _2-\varphi _3,-\varphi _1+\varphi
   _2-\varphi _3,-\varphi _1-\varphi _2+\varphi _3)\,,\nonumber\\
   a_i {\bf E}_i &=\left(\begin{matrix}{\bf A} & {\bf 0}\cr {\bf B} & -{\bf A}^T\end{matrix}\right)\,\,;\,\,\,
  {\bf A}=\left(
\begin{array}{llll}
 0 & 0 & 0 & 0 \\
 a_1 & 0 & 0 & 0 \\
 a_2 & 0 & 0 & 0 \\
 a_3 & 0 & 0 & 0
\end{array}
\right)\,\,,\,\,\,{\bf B}=\left(
\begin{array}{llll}
 0 & 0 & 0 & 0 \\
 0 & 0 & a_3 & a_2 \\
 0 & a_3 & 0 & a_1 \\
 0 & a_2 & a_1 & 0
\end{array}
\right)\,.
\end{align}
Exponentiating the above matrices according to (\ref{L4stu}), one derives the symplectic representation (\ref{hybrid}) of the coset representative $\mathbb{L}(\phi)={\Scr R}_v[L]=(\mathbb{L}^M{}_{{N}})$.\footnote{In this case $\mathcal{S}={\bf 1}$ in (\ref{hybrid}) and thus we do not underline the second index.} The reader can verify that
$$\Omega^M(\phi^s)=\mathbb{L}^M{}_{{N}}(\phi^s)\,\Omega^N(\phi^s=0)\,.$$
In this symplectic basis each ${\rm SL}(2,\mathbb{R})$-factor in $G$ is generated by the triple ${\bf h}_i,\,{\bf E}_i,\,{\bf E}_i^T$, satisfying the relations: \begin{equation}
[{\bf h}_i,\,{\bf E}_j]=\delta_{ij}\,{\bf E}_i\,\,;\,\,\,[{\bf h}_i,\,{\bf E}^T_j]=-\delta_{ij}\,{\bf E}^T_i\,\,;\,\,\,\,[{\bf E}_i,\,{\bf E}^T_j]=2\delta_{ij}\,{\bf h}_i\,.\label{hEalg}\end{equation}
 It is convenient to choose the following basis of generators for $G$:
\begin{equation}
t_a=t_{x,i}\,\,:\,\,\,t_{1,i}=-2{\bf h}_i\,\,;\,\,\,t_{2,i}={\bf E}_i-{\bf E}^T_i\,\,;\,\,\,\,t_{3,i}={\bf E}_i+{\bf E}^T_i\,,
\end{equation}
where  $x=1,2,3$ labels the generators within each $\mathfrak{sl}(2,\mathbb{R})$-factor, labeled by $i=1,2,3$.
The commutation relations of the isometry algebra read:
\begin{equation}
[t_{x,i},\,t_{y,j}]=-2\,\delta_{ij}\,\epsilon_{xy}{}^z\,t_{z,i}\,,
\end{equation}
where the $z$ index is raised by the invariant metric $\eta_{xy}={\rm diag}(+1,-1,+1)$.\par
 For the purpose of computing the scalar potential, it is convenient
 to evaluate the holomorphic Killing vectors
 $k_{x,i}$ which have the following form
\begin{equation}
k_{1,i}=-2\,z^i\,\partial_i\,\,;\,\,\,k_{2,i}=(1+(z^i)^2)\,\partial_i\,\,;\,\,\,k_{3,i}=(1-(z^i)^2)\,\partial_i\,,
\end{equation}
no summation over $i$.
These are in turn expressed in terms of
momentum maps ${\Scr P}_a$ (no summation over $i$):
\begin{align}
{\Scr P}_a &=-\overline{V}^M\,t_{a
M}{}^N\,\mathbb{C}_{NL}\,V^L\,,\nonumber\\
{\Scr P}_{1, i}&=-i\,\frac{z^i+\bar{z}^i}{z^i-\bar{z}^i}\,,\,\,{\Scr P}_{2,i}=i\,\frac{1+|z^i|^2}{z^i-\bar{z}^i}\,,\,\,{\Scr P}_{2,i}=i\,\frac{1-|z^i|^2}{z^i-\bar{z}^i}\,,
\end{align}
where we have used Eq. (\ref{PVtV}).
The reader can verify that Eqs. (\ref{kpal}) are satisfied.
Being ${\Scr R}_v={\bf (2,2,2)}$, by means of a symplectic transformation, the 8 basis elements for the ${\Scr R}_v$-representation can be changed to the following:
\begin{equation}
V^M=V^{(\alpha_1,\alpha_2,\alpha_3)}=(V^{(1,1,1)},V^{(1,1,2)},V^{(1,2,1)},V^{(1,2,2)},V^{(2,1,1)},V^{(2,1,2)},V^{(2,2,1)},
V^{(2,2,2)})\,,
\end{equation}
where $\alpha_i=1,2$ label the doublet representation of the $i^{th}$ ${\rm SL}(2,\mathbb{R})$-factor. In this basis the global symmetry generators $t_a$ of $\mathfrak{g}^{(SK)}=\mathfrak{sl}(2,\mathbb{R})^3$ read:
 \begin{align}
 (t_{x,1})_{\beta_1,\beta_2, \beta_3}{}^{\gamma_1,\gamma_2, \gamma_3}&=(s_{x_1})_{\beta_1}{}^{\gamma_1}\delta_{\beta_2}^{\gamma_2}\delta_{\beta_3}^{\gamma_3}\,,;\,\,
 (t_{x,2})_{\beta_1,\beta_2, \beta_3}{}^{\gamma_1,\gamma_2, \gamma_3}=(s_{x_2})_{\beta_2}{}^{\gamma_2}\delta_{\beta_1}^{\gamma_1}\delta_{\beta_3}^{\gamma_3}\,,;\nonumber\\
 (t_{x,3})_{\beta_1,\beta_2, \beta_3}{}^{\gamma_1,\gamma_2, \gamma_3}&=(s_{x_3})_{\beta_3}{}^{\gamma_3}\delta_{\beta_1}^{\gamma_1}\delta_{\beta_2}^{\gamma_2}\,,
 \end{align}
 where $s_x=(\sigma_1,\,i\sigma_2,\,\sigma_3)$.\par
 A distinctive feature of the STU model is its characteristic \emph{triality} symmetry under the exchange of the three scalars and thus of the three factors in $G^{(SK)}$.
 Some more details about the geometric structure of the STU model can be found in Appendix \ref{STUstruc}, where it is shown that this model is a characteristic truncation of the maximal supergravity.

\subsubsection{The Embedding Tensor.}
The most general gauge symmetry which can be introduced in the model is defined by generators $X_M$ of the form (\ref{N2gaugen}), where $t_m$, $m=1,2,3$, are ${\rm SO}(3)$-generators, satisfying the constraints (\ref{lc})-(\ref{qc3}).
Let us solve first these constraints in the components $\Theta_M{}^a$ of the embedding tensor, where $a=(x,i)$ and we write $M=(\alpha_1,\alpha_2,\alpha_3)$.
The embedding tensor $\Theta_M{}^a$ takes the following form:
\begin{equation}
\Theta_M{}^a=\{\Theta_{(\alpha_1,\alpha_2,\alpha_3)}{}^{x_1},\,\Theta_{(\alpha_1,\alpha_2,\alpha_3)}{}^{x_2},\,
\Theta_{(\alpha_1,\alpha_2,\alpha_3)}{}^{x_3}\}\,\,\in\,\,\, {\bf (2,2,2)}\times [{\bf (3,1,1)}+{\bf (1,3,1)}+{\bf (1,1,3)}]\,,
\end{equation}
where $x_i$ run over the adjoint (vector)-representations of the three $\mathfrak{sl}(2)$ algebras.
 Since:
\begin{equation}
{\bf (2,2,2)}\times [{\bf (3,1,1)}+{\bf (1,3,1)}+{\bf (1,1,3)}]=3\times
 {\bf (2,2,2)}+{\bf (4,2,2)}+{\bf (2,4,2)}+{\bf (2,2,4)}\,,
\end{equation}
each component of the embedding tensor can be split into its irreducible parts according to the above decomposition:
\begin{align}
\Theta_{(\alpha_1,\alpha_2,\alpha_3)}{}^{x_1}&=(s^{x_1})_{\alpha_1}{}^\beta\,\xi^{(1)}_{\beta\,\alpha_2\,\alpha_3}+
\Xi_{\alpha_1,\alpha_2,\alpha_3}{}^{x_1}\in {\bf (2,2,2)}+{\bf (4,2,2)} \,,\nonumber\\
\Theta_{(\alpha_1,\alpha_2,\alpha_3)}{}^{x_2}&=(s^{x_2})_{\alpha_2}{}^\beta\,\xi^{(2)}_{\alpha_1\,\beta\,\alpha_3}+
\Xi_{\alpha_1,\alpha_2,\alpha_3}{}^{x_2}\in {\bf (2,2,2)}+{\bf (2,4,2)}\,,\nonumber\\
\Theta_{(\alpha_1,\alpha_2,\alpha_3)}{}^{x_3}&=(s^{x_3})_{\alpha_3}{}^\beta\,\xi^{(3)}_{\alpha_1\,\alpha_2\,\beta}+
\Xi_{\alpha_1,\alpha_2,\alpha_3}{}^{x_3}\in {\bf (2,2,2)}+{\bf (2,2,4)}\,,\nonumber\\
\end{align}
The tensors $ \Xi_{\alpha_1,\alpha_2,\alpha_3}{}^{x_i}$ in the ${\bf (4,2,2)},\,{\bf (2,4,2)},\,{\bf (2,2,4)}$ representations are defined by the vanishing of the appropriate gamma-trace, namely:
\begin{equation}
 \Xi_{\beta,\alpha_2,\alpha_3}{}^{x_1} (s_{x_1})_{\alpha_1}{}^\beta=\Xi_{\alpha_1,\beta,\alpha_3}{}^{x_2} (s_{x_2})_{\alpha_2}{}^\beta=\Xi_{\alpha_1,\alpha_2,\beta}{}^{x_3} (s_{x_3})_{\alpha_3}{}^\beta=0\,.
 \label{purlo}
\end{equation}
 Let us now define embedded  gauge generators $X_{MN}{}^P$:
 \begin{align}
 X_{(\alpha_1,\alpha_2,\alpha_3),(\beta_1,\beta_2,\beta_3)}{}^{(\gamma_1,\gamma_2,\gamma_3)}&=\Theta_{(\alpha_1,\alpha_2,\alpha_3)}{}^{x_1}\,(s_{x_1})_{\beta_1}{}^{\gamma_1}\delta_{j_
 2}^{\gamma_2}\delta_{\beta_3}^{\gamma_3}+
 \Theta_{(\alpha_1,\alpha_2,\alpha_3)}{}^{x_2}\,(s_{x_2})_{\beta_2}{}^{\gamma_2}\delta_{\beta_1}^{\gamma_1}
 \delta_{\beta_3}^{\gamma_3}+\nonumber\\
 &+\Theta_{(\alpha_1,\alpha_2,\alpha_3)}{}^{x_3}\,(s_{x_3})_{\beta_3}{}^{\gamma_3}\delta_{\beta_2}^{\gamma_2}
 \delta_{\beta_1}^{\gamma_1}\,.
 \label{fischietto}
 \end{align}
As mentioned in Sect. \ref{cgaet2}, the linear constraint (\ref{lconstr1imp2}) implies:
 \begin{align}
& X_{(\alpha_1,\alpha_2,\alpha_3),(\beta_1,\beta_2,\beta_3)}{}^{(\alpha_1,\alpha_2,\alpha_3)}=0\,\,\Rightarrow\,\,\,\,\xi^{(1)}_{\alpha_1\,\alpha_2\,\alpha_3}+
 \xi^{(2)}_{\alpha_1\,\alpha_2\,\alpha_3}+\xi^{(3)}_{\alpha_1\,\alpha_2\,\alpha_3}=0\,,\nonumber\\
&X_{(\alpha_1,\alpha_2,\alpha_3),(\beta_1,\beta_2,\beta_3),(\gamma_1,\gamma_2,\gamma_3)}+X_{(\gamma_1,\gamma_2,\gamma_3),(\beta_1,\beta_2,\beta_3),(\alpha_1,\alpha_2,\alpha_3)}+
X_{(\beta_1,\beta_2,\beta_3),(\alpha_1,\alpha_2,\alpha_3),(\gamma_1,\gamma_2,\gamma_3)}=0\,\Rightarrow\nonumber\\&\Rightarrow\,\, \Xi_{\alpha_1,\alpha_2,\alpha_3}{}^{x_i}=0\,.
 \end{align}
This rules out the ${\bf (4,2,2)},\,{\bf (2,4,2)},\,{\bf (2,2,4)}$ representations leaving us only with three tensors in the ${\bf (2,2,2)}$ representation.
Explicitly the linearly constrained embedding tensor reads as follows:
 \begin{align}
\Theta_{(\alpha_1,\alpha_2,\alpha_3)}{}^{a}&=\{(s^{x_1})_{\alpha_1}{}^\beta\,\xi^{(1)}_{\beta\,\alpha_2\,\alpha_3},
\,(s^{x_2})_{\alpha_2}{}^\beta\,\xi^{(2)}_{\alpha_1\,\beta\,\alpha_3},\,
(s^{x_3})_{\alpha_3}{}^\beta\,\xi^{(3)}_{\alpha_1\,\alpha_2\,\beta}\}\,,
\end{align}
where the tensors in the ${\bf (2,2,2)}$ are further subject by the condition
\begin{equation}
\xi^{(1)}_{\alpha_1\,\alpha_2\,\alpha_3}+
 \xi^{(2)}_{\alpha_1\,\alpha_2\,\alpha_3}+\xi^{(3)}_{\alpha_1\,\alpha_2\,\alpha_3}=0
\end{equation}
The first of the quadratic conditions (\ref{qc3}) for $\Theta_M{}^a$ has the form:
\begin{equation}
 \epsilon^{\alpha_1 \beta_1}\epsilon^{\alpha_2 \beta_2}\epsilon^{\alpha_3 \beta_3}\,\Theta_{(\alpha_1,\alpha_2,\alpha_3)}{}^{a}\,\Theta_{(\beta_1,\beta_2,\beta_3)}{}^{b}=0\,,\label{qcnoFI}
\end{equation}
on top of which the closure conditions have to be imposed.
 By means of a MATHEMATICA computer code we were able to solve the quadratic constraints. We do not display them here, since, in Section \ref{noFIgauge}, we show how to classify the orbits into which such solutions are organized and it will be sufficient to consider only one representative for each orbit.

\subsubsection{{The Gaugings with no Fayet-Iliopoulos Terms}}
 \label{noFIgauge}
 We first consider the case of no Fayet-Iliopoulos terms, namely
 ($\Theta_M{}^m=0$). Since $\Theta_M{}^m={\Scr P}_M^m=0$, $m=1,2,3$, the scalar potential has no contribution from the gravitino shift-tensor and thus it is non-negative and reads:
 \begin{equation}
 {V}={V}(z,\bar{z},q)=g^2\,k_M^ik_N^{\bar{\jmath}}g_{i\bar{\jmath}}\overline{V}^M\,V^N\,.\label{VnoFI}
 \end{equation}
  We can use the global symmetry $G^{(SK)}$  of the
 theory to simplify our analysis. Indeed the field equations and
 Bianchi identities are invariant if we $G^{(SK)}$-transform the
 field and embedding tensors at the same time. This is in particular
 true for the scalar potential $V(\phi, \Theta)$, see Eq. (\ref{Vinvar}). Notice that we can have
other formal symmetries of the potential which are not in $G^{(SK)}$.
Consider for instance the symplectic transformation:
\begin{equation}
\mathcal{S}={\rm
diag}(1,\varepsilon_i,1,\varepsilon_i)\,,\label{Sep}
\end{equation}
where $\varepsilon_i=\pm 1$,
$\varepsilon_1\varepsilon_2\varepsilon_3=1$. These transformations
correspond to the isometries $z^i\rightarrow \varepsilon_i\,z^i$,
which however do not preserve the physical domain defined by the
lower-half plane for each complex coordinate, ${\rm Im}(z^i)<0$, in which $\mathcal{I}_{\Lambda\Sigma}<0$.
Therefore embedding tensors connected by such transformations are to
be regarded as physically inequivalent.
\par
 We have shown in the previous section that the embedding tensor, solution to the linear constraints, in the absence of Fayet-Iliopoulos terms, is parameterized by two independent tensors $\xi^{(2)},\,\xi^{(3)}$ in the
 ${\bf (2,2,2)}$ of $G^{(SK)}$.
 These are then subject to the quadratic constraints that restrict the $G^{(SK)}$-orbits of these two quantities.
 We can think of acting by means of $G^{(SK)}$ on $\xi^{(2)}$, so as
 to make it the simplest possible. By virtue of Eq. (\ref{Vinvar}) this will not
 change the physics of the gauged model (vacua, spectra,
 interactions), but just make the analysis simpler.
 \par
Let us recall that the $G^{(SK)}$-orbits of a single object, say
$\xi^{(2)M}$, in the
${\Scr R}_v = {\bf (2,2,2)}$ representation
are described by a quartic invariant $I_4(\xi^{(2)})$, see (\ref{qinvar}), defined as:
\begin{equation}
I_4(\xi^{(2)})=-\frac{2}{3}\,t_{a\,M
N}\,t^a{}_{PQ}\,\xi^{(2)M}\xi^{(2)N}\xi^{(2)P}\xi^{(2)Q}\,.
\end{equation}
For the classification of these orbits we can make contact with the literature on black hole solutions to ungauged extended supergravities and their classification with respect to the action of the global symmetry group $G$, see \cite{Ferrara:1997uz,Bellucci:2006xz}.
The orbits in the ${\bf (2,2,2)}$-representation are classified as follows\footnote{\label{finestruct} Strictly speaking, for all models in the second line of table 2, there is a further fine structure (see  \cite{Borsten:2011ai}) in some of the orbits classified above which  depends on other invariant quantities. We shall take care of this finer splitting by appropriately parametrizing our representatives so that different values of the parameters will correspond to the different sub-orbits.}:
\begin{itemize}
\item[i)] Regular, $I_4>0$, triality-invariant;
\item[ii)]Regular, $I_4>0$, non  triality-invariant;
\item[iii)]Regular, $I_4<0$;
\item[iv)]\emph{Light-like}, $I_4=0$, $\partial_M I_4\neq 0$;
\item[v)]\emph{Critical}, $I_4=0$, $\partial_M I_4= 0$, $t_A{}^{MN}\,\partial_M \partial_N\,I_4\neq 0$ ;
\item[vi)]\emph{Doubly critical}, $I_4=0$, $\partial_M I_4= 0$, $t_A{}^{MN}\,\partial_M \partial_N\,I_4=
0$ ,
\end{itemize}
where $\partial_M\equiv \partial/\partial \xi^{(2)M}$. The quadratic
constraints (\ref{qcnoFI}) restrict $\xi^{(2)}$ (and $\xi^{(3)}$)
to be either in the  \emph{critical} or in the
\emph{doubly-critical} orbit. Let us analyze the two cases
separately.
\paragraph{ $\xi^{(2)}$ Critical.}
The quadratic constraints imply $\xi^{(3)}=0$ and thus the embedding
tensor is parameterized by $\xi^{(1)}=-\xi^{(2)}$, namely the diagonal
of the first two $\mathrm{SL}(2,\mathbb{R})$ groups in $\mathrm{G}_{SK}$. We
can choose  a representative of the orbit in the form:\footnote{Other choices are mapped into this one by triality.}
\begin{equation}
\xi^{(2)}=g\,(0,1,c,0,0,0,0,0)\,.
\end{equation}
The gauge group is $G_g={\rm SL}(2,\mathbb{R})$ and the scalar potential reads:
\begin{equation}
{V}=
g^2\,e^{-\varphi_1-\varphi_2-\varphi_3}\,\left((a_1+c\,a_2)^2+(e^{\varphi_1}-c\,e^{\varphi_2})^2\right)\,.
\end{equation}
The truncation to the dilatons ($a_i \, = \, 0$) is a consistent one:
\begin{equation}
\left.\frac{\partial {V}}{\partial
a_i}\right\vert_{a_i=0}=0\,,
\end{equation}
and
\begin{equation}
\left.{V}\right\vert_{a_i=0}=g^2\,\left(e^{-\frac{1}{2}(-\varphi_1+\varphi_2+\varphi_3)}
-c\,e^{-\frac{1}{2}(\varphi_1-\varphi_2+\varphi_3)}\right)^2\,.
\end{equation}
The above potential has an extremum if $c>0$, for
$e^{\varphi_1}=c\,e^{\varphi_2}$, while it is runaway if $c<0$. In the former case the extrema are of Minkowski type and not unstable (being the scalar potential non-negative). The
sign of $c$ is changed by a transformation of the kind (\ref{Sep})
with $\varepsilon_1=-\varepsilon_2=-\varepsilon_3=1$. For the reason
outlined above, in passing from a negative to a positive $c$, the
critical point of the potential moves to the unphysical domain
(${\rm Im}(z^i) > 0$). In fact the sign of $c$ labels some of the sub-orbits mentioned in Footnote \ref{finestruct}. The gauging for $c=-1$ coincides with the one
considered  in \cite{Fre:2002pd}, in the absence of Fayet-Iliopoulos terms.\footnote{In that paper a different parametrization of the scalar manifold was used. It is called the \emph{Calabi-Vesentini parametrization} \cite{CV} of the manifolds listed in the second line of Table 2, and it refers to a symplectic frame in which no prepotential exists. Here, since we consider dyonic gaugings which allow for electric and magnetic components of the embedding tensor, we do not need to make any initial choice of the symplectic frame.}
\paragraph{ $\xi^{(2)}$ Doubly-Critical.}
We can choose  a representative of the orbit in the form:
\begin{equation}
\xi^{(2)}=g\,(1,0,0,0,0,0,0,0)\,.
\end{equation}
In this case $\xi^{(3)}$ is non-vanishing and reads:
\begin{equation}
\xi^{(3)}=g'\,(1,0,0,0,0,0,0,0)\,.
\end{equation}
The gauging is electric ($\Theta^\Lambda=0$) and the gauge
generators $X_\Lambda=(X_0,X_I)$, $I=1,2,3$, satisfy the following
commutation relations:
\begin{equation}
[X_0,\,X_I]=M_I{}^J\,X_J,\,\,\,\,M_I{}^J={\rm
diag}(-2\,(g+g'),2\,g,\,2\,g')\,,
\end{equation}
all other commutators being zero. This gauging describes a
Scherk-Schwarz reduction from $D=5$, see discussion in Sect. \ref{gaugE6}, in which the semisimple global
symmetry generator defining the reduction is the 2-parameter
combination $M_I{}^J$ of the $\mathfrak{so}(1,1)^2$ global symmetry
generators of the $D=5$ parent theory (in the maximal theory of Sect. \ref{gaugE6} this group was ${\rm E}_{6(6)}$).\par Just as in the corresponding gauging of the maximal theory, scalar potential is
axion-independent and reads:
\begin{equation}
{V}= (g^2+g
g'+g^{'2})\,e^{-\varphi_1-\varphi_2-\varphi_3}\,.
\end{equation}
It is runaway, as one would expect from the discussion in Footnote \ref{fotss}.
\subsubsection{ Adding ${\rm U}(1)$ Fayet-Iliopoulos Terms}
Let us now consider adding components of the embedding tensor along
one generator of the ${\rm SO}(3)$ global symmetry group:
$\theta_M=\Theta_M{}^{m=1}$, see Sect. \ref{FIsec}. The constraints on $\theta_M$ come from
(\ref{qc1})-(\ref{qc3}), and read:
\begin{equation}
\theta_M\,\mathbb{C}^{MP}\,X_{PN}{}^Q=0\,\,,\,\,\,X_{PN}{}^Q\,\theta_Q=0\,.
\end{equation}
while the constraints on $\Theta_M{}^a$ are just
the same as before and induce the same restrictions on the orbits of
$\xi^{(2)},\,\xi^{(3)}$.  Clearly if $X_{PN}{}^Q=0$, namely
$\Theta_M{}^a=0$, no special K\"ahler isometries are gauged and there are no
constraints on $\theta_M$. We shall consider this case
separately.\par The potential reads
\begin{align}
{V}&= {V}_{1}+ {V}_{2}\,,
\end{align}
where $V_1$ is given in (\ref{VnoFI}) was constructed, for the various orbits, in the previous Section, while $V_2$ has the form (\ref{VFI}) or, equivalently (\ref{VFI2}), in terms of the composite fields ${\Scr Z}$ and ${\Scr Z}_i$. Due to the similarity between $V_2$ and the effective black hole potential $V_{BH}$, see Eq. (\ref{VBHV}), the study of the extrema of the former parallels the corresponding analysis of the latter.
Let us now study the full scalar potential in the relevant cases.
\paragraph{ $\xi^{(2)}$ Critical (stable de Sitter vacuum).}
In this case, choosing
\begin{equation}
\xi^{(2)}=g\,(0,1,c,0,0,0,0,0)\,.
\end{equation}
we find for $\theta_M$ the following general solution to the
quadratic constraints:
\begin{equation}
\theta_M=(0,\frac{f_1}{c},\,f_1,\,0,\,0,\,f_2,\,\frac{f_2}{c},\,0)\,,
\end{equation}
where $f_1,\,f_2$ are constants.
\par The scalar potential reads:
\begin{equation}
\label{guliashi}
{V}=g^2\,e^{-\varphi_1-\varphi_2-\varphi_3}\,\left((a_1+c\,a_2)^2+(e^{\varphi_1}-c\,e^{\varphi_2})^2\right)-
\frac{e^{-\varphi_3}}{c}\,\left[(f_1+f_2\,a_3)^2+f_2^2\,e^{2\,\varphi_3}\right]\,.
\end{equation}
It has an extremum only for
$c<0$ and:
\begin{equation}
a_1=-c \,a_2,\,a_3=-\frac{f_1}{f_2}\,,\,\,\,\varphi_1=\varphi_2+\log(-c),\,\varphi_3=-\log\left(\left\vert\frac{2c
g}{f_2}\right\vert\right)\,,\,\,\,\,.
\end{equation}
The potential at the extremum is
\begin{equation}
{V}_0= 4\,|g\,f_2|>0\,,
\end{equation}
while the squared scalar mass matrix reads:
\begin{equation}
\left.(\partial_r\partial_s {V}\,g^{st})\right\vert_0={\rm
diag}(2,2,1,1,0,0)\times {V}_0\,.
\end{equation}
The gauge group is $G_g={\rm SL}(2,\mathbb{R})$ and, by Goldstone theorem, the two vectors gauging the non-compact
generators (which, as such, do not stabilize the vacuum) become massive. The two null eigenvalues in the scalar spectrum are therefore associated with the corresponding Goldstone bosons.
In this way we retrieve the stable de Sitter vacuum of \cite{Fre:2002pd} the two
parameters $f_1,\,f_2$ being related to the coupling constant of that model and to the de
Roo-Wagemans' angle \cite{deRoo:1985jh}. This angle parametrizes an ${\rm Sp}(4,\mathbb{R})$-transformation which is not in the global symmetry group $G$ of the model, just as the $\omega$-rotation in the maximal supergravity.
\paragraph{ $\xi^{(2)}$ Doubly-Critical.}
In this case the constraints on $\theta_M$ impose:
\begin{align}
(g+g')\theta_1&=0\,,\,g\,\theta_2=0\,,\,g'\,\theta_3=0\,,\,g\,\theta^0=g'\,\theta^0=0\,,\,g\,\theta^1=g'\,\theta^1=0\,,\,
g\,\theta^2=g'\,\theta^2=0\,,\nonumber\\
g\,\theta^3&=g'\,\theta^3=0\,.
\end{align}
Under these conditions, unless $g=g'=0$, which is the case we shall
consider next, the FI contribution to the scalar potential vanishes.
\paragraph{ Case $\Theta_M{}^a=0$: Pure Fayet-Iliopoulos gauging (anti-de Sitter vacuum).}
In this case, we can act on $\theta_M$ by means of $G^{(SK)}$ and
reduce it to the following normal form:
\begin{equation}
\theta_M=(0,f_1,f_2,f_3,f^0,0,0,0)\,.
\end{equation}
The scalar potential reads:
\begin{align}
{V}&=-\sum_{i=1}^3\,e^{-\varphi_i}\,\left(f_if^0\,(a_i^2+e^{2\varphi_i})+f_j
f_k\,\right)\,,
\end{align}
where $i\neq j\neq k\neq i$. Clearly if $f_i=0$, the potential vanishes and the model has a continuum of Minkowski vacua.
In general the truncation to the dilatons is consistent
and we find:
\begin{equation}
\label{doppiocritFI}
\left.{V}\right\vert_{a_i=0}=-\sum_{i=1}^3\,\left(f_if^0\,e^{\varphi_i}+f_j
f_k\,e^{-\varphi_i}\right)\,,
\end{equation}
which has an extremum, if all the $f_i$ are non-vanishing, for the following values of the dilatons
\begin{equation}
e^{2\varphi_i}=\frac{f_j f_k}{f_if^0}\,,
\end{equation}
and the potential at the extremum reads:
\begin{equation}
\mathcal{V}_0=-2\,\sqrt{f^0 f_1 f_2 f_3}\left({\rm sign}(f_1 f_2)+{\rm sign}(f_1 f_3)+{\rm sign}(f_2 f_3)\right)\,,
\end{equation}
This extremum exists only if $f^0 f_1 f_2 f_3>0$. This implies that
$\theta_M$ should  be either in the orbit $i)$ or in the orbit $ii)$.   Using the
analogy between $\theta_M$ and black hole charges, these two orbits
correspond to BPS and non-BPS with $I_4>0$ black holes. The extremum
condition for $V_{BH}$ in Eq. (\ref{VBHV}) fixes the scalar fields at the horizon
according to the attractor mechanism. Now the potential has an
additional term $-4\,|{\Scr Z}|^2$ which, however, for the orbits $i)$,
$ii)$, has the same extrema as $V_{BH}$ since its derivative with
respect to $z^i$ is $-4 \,{\Scr D}_i {\Scr Z}\,\overline{{\Scr Z}}$ which vanishes for the $i)$
orbit since at the extremum of $V_{BH}$ (BPS black hole horizon)
${\Scr D}_i {\Scr Z}=0$, and for the $ii)$ orbit since at the extremum of $V_{BH}$
(black hole horizon) ${\Scr Z}=0$.
\par We conclude that in the ``BPS'' orbit $i)$ the extremum corresponds to an AdS-vacuum where the scalar mass
spectrum reads as follows:
\begin{equation}
\left.(\partial_r\partial_s {V}\,g^{st})\right\vert_0={\rm
diag}\left(\frac{2}{3},\frac{2}{3},\frac{2}{3},\frac{2}{3},\frac{2}{3},\frac{2}{3}\right)\times
{V}_0<0\,,
\end{equation}
where $V_0=-6\,\sqrt{f^0 f_1 f_2 f_3}=-3/L^2<0$.
The vacuum is stable since $m_s^2 \,L^2=-2$ and the BF bound (\ref{BFBound}) is satisfied.\par
These models have provided a useful supergravity framework where to study black hole solutions in anti-de Sitter spacetime \cite{adsblackholes1,adsblackholes2,adsblackholes3,adsblackholes4,adsblackholes5}.\par
In the ``non-BPS'' orbit $ii)$ the potential has a de Sitter extremum which, however, is not stable, having tachyonic directions:
\begin{equation}\left.(\partial_r\partial_s {V}\,g^{st})\right\vert_0={\rm
diag}\left(-2,-2,2,2,2,2\right)\times
{V}_0\,,
\end{equation}
where $V_0=2\,\sqrt{f^0 f_1 f_2 f_3}>0$.

\section{The Other Extended $D=4$ Supergravities at a Glance}\label{glance}
Here we briefly review the main facts about the extended models in four dimensions with $\mathcal{N}=3,4,5,6$ and their gaugings, restricting ourselves to a group theoretical characterization of the embedding tensor. We omit $\mathcal{N}=7$ since it coincides with the maximal theory. For a general overview of ungauged supergravities see \cite{Andrianopoli:1996ve}.
\subsection{$\mathcal{N}=6$ Supergravity}
 The scalar fields are 30 and span the special K\"ahler manifold (see Table 1):
\begin{equation}
{\Scr M}_{{\rm scal}}=\frac{G}{H}=\frac{{\rm SO}^*(12)}{{\rm U}(6)}\,.\label{Msost6}
\end{equation}
The R-symmetry group is $H_R=H={\rm U}(6)$ and the global symmetry group is $G={\rm SO}^*(12)$ \cite{Cremmer:1978ds}. As mentioned in Sect. \ref{dcsl}, the model only describes the gravitational multiplet. The graviton, as usual, is a singlet with respect to both $G$ and $H$, while the electric and magnetic vector fields $A^M_\mu$ transform in the symplectic representation ${\Scr R}_v={\bf 32}_c$. The fermions comprise the six gravitinos $\psi_{A\mu}$ and $20+6$ dilatinos $\chi_{ABC},\,\chi_A$, $A,B,C,\dots=1\,\dots,\,6 $ (we take $\chi_A$ to have the same chirality as $\chi_{ABC}$). As explained in Sect. \ref{dcsl}, the composite field strengths $F^{\underline{\Lambda}}_{\mu\nu}$, which are functions of the scalar fields and the vectors, as well as the central and matter charges ${\Scr Z}^{\underline{\Lambda}}$, functions of the scalar fields and the quantized charges, transform covariantly  with respect to $H$ in a representation ${\bf R}_v$ of $H$. In particular the index $\underline{\Lambda}$ split into an antisymmetric couple $[AB]$ and a value $\bullet$ corresponding to an ${\rm SU}(6)$-singlet. Consequently the ${\Scr Z}_{\underline{\Lambda}}$ split into the central charges ${\Scr Z}_{AB}$ of the supersymmetry algebra and an extra matter charge ${\Scr Z}_\bullet$ which does not correspond to any matter multiplet. The $G$ and $H$-representations of the vectors, fermions and composite field strengths $F^{\underline{\Lambda}}_{\mu\nu}$ are summarized in Table \ref{N6content}.
\begin{table}[H]
\begin{center}
\begin{tabular}{|c||c|c|c|c|c|c|}
  \hline
    & $A^M_\mu$ &  $\psi_{A\mu}$& $\chi_{ABC}$& $\chi_{A}$ & $F^{AB}_{\mu\nu}$& $F^{\bullet}_{\mu\nu}$\\
    \hline  \hline
 ${\rm SO}^*(12)$ & ${\bf 32}_c$ & ${\bf 1}$ & ${\bf 1}$ & ${\bf 1}$ & ${\bf 1}$& ${\bf 1}$\\
  \hline
 ${\rm U}(6)$ & ${\bf 1}_0$ & ${\bf 6}_{+\frac{1}{2}}$ & ${\bf 20}_{+\frac{3}{2}}$ & ${\bf 6}_{-\frac{5}{2}}$ & $\overline{{\bf 15}}_{-1}$& ${\bf 1}_{+3}$\\
  \hline
\end{tabular}
  \caption{Relevant representations with respect to $G$ and $H$.}\label{N6content}
\end{center}
\end{table}
The representation of the embedding tensor is contained in product of ${\Scr R}_{v*}$ and the adjoint of the global symmetry group $G$ which decomposes as follows
\begin{equation}
{\bf 32}_c\times {\bf 66}\,\rightarrow \,{\bf 32}_c+{\bf 1728}+{\bf 352}_s\,.
\end{equation}
The linear constraint (\ref{linear2}) sets to zero all the representations in the above decomposition which occur in the three-fold symmetric product
\begin{equation}
X_{(MNP)}~\in~ ({\bf 32}_c\times {\bf 32}_c\times {\bf 32}_c)_{{\rm sym.}}
\,\rightarrow\;{\bf 32}_c+{\bf 1728}+ {\bf 4224}\,.
\end{equation}
Thus the representation singled out by (\ref{linear2}) is \cite{Andrianopoli:2008ea,Roest:2009sn}:
\begin{equation}
\Theta_M{}^\alpha\in {\Scr R}_\Theta={\bf 352}_s\,.
\end{equation}
As far as the quadratic constraints are concerned, they amount to setting to zero the representation ${\Scr R}_{\Theta\Theta}={\bf 66}+{\bf 2079}+{\bf 462}_s$ in the symmetric product of two ${\Scr R}_\Theta$ \cite{Roest:2009sn}.
The fermion-shift tensors enter as $O(g)$-terms in the supersymmetry transformation rules of the fermion fields:
\begin{equation}
\delta\psi_{A\mu}=\dots+i\,g\,\mathbb{S}_{AB}\gamma_\mu\epsilon^B\,,\,\,\,\delta \chi_{ABC}=\dots+g\,\mathbb{N}_{ABC}{}^D\epsilon_D\,\,,\,\,\,\,\chi_{A}=\dots+g\,\mathbb{N}_{A}{}^B\epsilon_B\,.
\end{equation}
Their $H$-representations are:
\begin{equation}
\mathbb{S}_{AB}\in {\bf 21}_{+1}\,\,,\,\,\,\mathbb{N}_{ABC}{}^D\in ({\bf 105+15})_{+1}\,\,,\,\,\,\mathbb{N}_{A}{}^D\in {\bf 35}_{-3}\,.
\end{equation}
These are the only representations which occur, together with their conjugate, in the branching of ${\Scr R}_{\Theta}={\bf 352}_s$ with respect to ${\rm U}(6)$.
\par
As mentioned in Sect. \ref{dcsl}, the ungauged $\mathcal{N}=6$ theory can be obtained as a consistent truncation of the maximal one, defined by branching the R-symmetry and ${\rm E}_{7(7)}$-representations in the latter with respect to the subgroups ${\rm U}(6)\times {\rm SU}(2)$ and ${\rm SO}^*(12)\times {\rm SU}(2)$ of ${\rm SU}(8)$ and ${\rm E}_{7(7)}$, respectively, and \emph{retaining only the ${\rm SU}(2)$-singlets}. In particular the
index labeling the fundamental of ${\rm SU}(8)$ splits into the index $A=1,\dots, 6$ and the remaining values $7,8$ labeling the ${\bf 2}$ of ${\rm SU}(2)$. The gravitini $\psi_{7\,\mu},\,\psi_{8\,\mu}$ are thus truncated out.
In this perspective we can identify $F^\bullet_{\mu\nu}$ with $F^{78}_{\mu\nu}$ and  $\chi_A=\chi_{A78}$.\par
The same relation between the $\mathcal{N}=6$ and $\mathcal{N}=8$ theories ceases to hold at the gauged level. To see this let us consider the branchings of the relevant ${\rm E}_{7(7)}$-representations with respect to the maximal subgroup ${\rm SO}^*(12)\times {\rm SU}(2)$:
\begin{align}
{\bf 56}&\rightarrow {\bf (32_c,\,1)}+{\bf (12,\,2)}\,,\nonumber\\
{\bf 133}&\rightarrow {\bf (66,\,1)}+{\bf (1,\,3)}+{\bf (32_s,\,2)}\,,\nonumber\\
{\bf 912}&\rightarrow {\bf (32_c,\,3)}+{\bf (352_s,\,1)}+{\bf (12+220,\,2)}\,.
\end{align}
The restriction to the ${\rm SU}(2)$-singlets singles out, in the branching of the ${\bf 912}$, the representation ${\bf (352_s,\,1)}$ identified above with ${\Scr R}_\Theta$. Moreover if we perform the analogous branching of the representation ${\Scr R}_{\Theta\Theta}={\bf 133}+{\bf 8645}$ of the quadratic constraint in the $\mathcal{N}=8$ theory, and restrict to the ${\rm SU}(2)$-singlets, we find, besides the ${\bf 66}+{\bf 2079}+{\bf 462}_s$, an extra representation  ${\bf 66}$. This is also contained in the symmetric product of two ${\bf 352}_s$ representations and thus will in general imply further constraints on the orbit of the embedding tensor. We conclude that a generic gauged $\mathcal{N}=6$ model is not a truncation of a maximal theory \cite{Roest:2009sn}.

The bosonic sector of the ungauged $\mathcal{N}=6$ theory is the same as that of a particular $\mathcal{N}=2$ supergravity coupled to 15 vector multiplets and no hypermultiplets, whose scalar manifold is (\ref{Msost6}). The two theories clearly differ in the fermionic sector and in the possible deformations. In the $\mathcal{N}=2$ model the global symmetry group is ${\rm SO}^*(12)\times {\rm SU}(2)$, ${\rm U}(2)$ is the R-symmetry while ${\rm SU}(6)$ is $H_{{\rm matt.}}$ acting on the 15 vector multiplets. The ${\Scr Z}_{AB}$ are now matter charges while the central charge of the supersymmetry algebra is  ${\Scr Z}_\bullet$. This model can be obtained from the ungauged maximal theory by truncating out, in the bosonic sector, all the doublet representations with respect to the ${\rm SU}(2)$ factor.
 FI terms $\Theta_M{}^x$, $x=1,2,3$, can be introduced, see Section \ref{FIsec}, besides the component of the embedding tensor in the ${\bf (352_s,\,1)}$, \cite{Andrianopoli:2008ea,Roest:2009sn} so that
\begin{equation}
{\Scr R}_\Theta={\bf (352_s,\,1)}+{\bf (32_c,\,3)}\,.
\end{equation}
Note that the FI terms correspond to gauging the global ${\rm SU}(2)$ symmetry. This induces minimal couplings of the vector fields to the fermions which are the only fields transforming non-trivially under this group.

Just as for the $\mathcal{N}=6$ case, not all these $\mathcal{N}=2$ gauged models can be obtained as truncations
of maximal supergravities. Indeed the extra ${\bf (66,\,1)}$ component in the quadratic constraints of the latter is not present in the $\mathcal{N}=2$ model.
\subsection{$\mathcal{N}=5$ Supergravity}
The scalar fields are 10 and span the special K\"ahler manifold (see Table 1):
\begin{equation}
{\Scr M}_{{\rm scal}}=\frac{G}{H}=\frac{{\rm SU}(1,5)}{{\rm U}(5)}\,.\label{Msc5}
\end{equation}
The duality representation of the global symmetry group $G={\rm SU}(1,5)$ is ${\Scr R}_v={\bf 20}$ \cite{Cremmer:1978ds}. The theory describes a single supergravity multiplet. The fermionic sector consists of the five gravitinos $\psi_{A\,\mu}$, $A=1,\dots, 5$, and $10+1$ dilatinos $\chi_{ABC},\,\chi$, where we define $\chi$ to have opposite chirality with respect to the other $\chi_{ABC}$. The composite field strengths are the $10$ $F^{AB}_{\mu\nu}$ with their complex conjugates (similarly the we only have central charges ${\Scr Z}^{AB},\,{\Scr Z}_{AB}$). The relevant $G$ and $H$-representations are summarized in the table
\begin{table}[H]
\begin{center}
\begin{tabular}{|c||c|c|c|c|c|}
  \hline
    & $A^M_\mu$ &  $\psi_{A\mu}$& $\chi_{ABC}$& $\chi$ & $F^{AB}_{\mu\nu}$\\
    \hline\hline
 ${\rm SU}(1,5)$ & ${\bf 20}$ & ${\bf 1}$ & ${\bf 1}$ & ${\bf 1}$ & ${\bf 1}$\\
  \hline
 ${\rm U}(5)$ & ${\bf 1}_0$ & ${\bf 5}_{+\frac{1}{2}}$ & $\overline{{\bf 10}}_{+\frac{3}{2}}$ & ${\bf 1}_{+\frac{5}{2}}$ & $\overline{{\bf 10}}_{-1}$\\
  \hline
\end{tabular}
  \caption{Relevant representations with respect to $G$ and $H$.}\label{N5content}
\end{center}
\end{table}
As mentioned in Sect. \ref{dcsl}, the ungauged $\mathcal{N}=5$ theory can be obtained as a consistent truncation of the maximal one, defined by branching the R-symmetry and ${\rm E}_{7(7)}$-representations with respect to the subgroups ${\rm U}(5)\times {\rm SU}(3)$ and ${\rm SU}(1,5)\times {\rm SU}(3)$ of ${\rm SU}(8)$ and ${\rm E}_{7(7)}$, respectively, and \emph{retaining only the ${\rm SU}(3)$-singlets}. In particular the eight gravitinos split into the five $\psi_{A\mu}$, in the $({\bf 5,1})_{+\frac{1}{2}}$ which are retained in the truncation and the three $\psi_{6\mu},\,\psi_{7\mu},\,\psi_{8\mu}$, in the $({\bf 1,3})_{-\frac{5}{6}}$, which are truncated out. The extra dilatino $\chi$ is identified with the charge conjugate of the spin-$1/2$ field $\chi_{678}$.\par
Along the same lines followed for the $\mathcal{N}>5$ models, we can derive the representation constraint (\ref{linear2})  as the condition that the embedding tensor be in the ${\rm SU}(1,5)$-representation:
\begin{equation}
{\Scr R}_\Theta={\bf 70}+{\bf 70}'\,,
\end{equation}
in the decomposition of the product ${\bf 20}\times {\bf 35}$. These are precisely the ${\rm SU}(3)$-singlet representations in the branching of the ${\bf 912}$, describing the embedding tensor of the maximal theory, with respect to ${\rm SU}(1,5)\times {\rm SU}(3)$.
They are described by tensors of the form $\theta_{mn,p},\,\theta^{mn,p}$, antisymmetric in the couple $mn$ and satisfying the condition $\theta_{[mn,p]}=0=\theta^{[mn,p]}$, $m,n,p=1,\dots,6$ labeling the fundamental representation of ${\rm SU}(1,5)$. The gauge generators in the fundamental of ${\rm SU}(1,5)$ have the form:
\begin{equation}
(X_{m_1 m_2 m_3})_m{}^n=\Theta_{m_1 m_2 m_3}{}^\alpha\,t_{\alpha\,m}{}^n=\theta_{[m_1 m_2\,|m}\delta_{m_3]}^n+\epsilon_{m_1 m_2 m_3 n_1 n_2 m}\,\theta^{n_1 n_2,n}\,,
\end{equation}
subject to the quadratic constraints (\ref{quadratic1}) and (\ref{quadratic2}).
\par
The $O(g)$-fermion shifts read:
\begin{equation}
\delta\psi_{A\mu}=\dots+i\,g\,\mathbb{S}_{AB}\gamma_\mu\epsilon^B\,,\,\,\,\delta \chi_{ABC}=\dots+g\,\mathbb{N}_{ABC}{}^D\epsilon_D\,\,,\,\,\,\,\chi=\dots+g\,\mathbb{N}_{A}\epsilon^A\,.
\end{equation}
where the $H$-representations of the shift-tensors are
\begin{equation}
\mathbb{S}_{AB}\in {\bf 15}_{+1}\,\,,\,\,\,\mathbb{N}_{ABC}{}^D\in ({\bf \overline{40}+10})_{+1}\,\,,\,\,\,\mathbb{N}_{A}\in {\bf 5}_{+3}\,.
\end{equation}
\subsection{$\mathcal{N}=4$ Supergravity}\label{n4sugrag}
$\mathcal{N}=4$ supergravity describes the gravity supermultiplet, consisting of the graviton, four gravitinos $\psi_{A\,\mu}$, $A=1,\dots,4$, six vectors with field strengths $F^{AB}_{\mu\nu}$, four dilations $\chi_{ABC}$ and a complex scalar field $S$, coupled to $n$ vector multiplets each consisting of a vector, four gauginos and six scalar fields.
The scalar manifold has the form (see Table 1):\footnote{The $\mathcal{N}=4$ ungauged supergravity with $n=6$ vector multiplets originates from the compactification of the gauge-invariant sector of the heterotic or Type I theories in ten dimensions, on a six-torus, see for instance \cite{Chamseddine:1980cp}. It coincides with the consistent truncation of the maximal theory in $D=4$ to the NS-NS fields, to be discussed in Sect. \ref{Tdualcomp}.}
\begin{equation}
{\Scr M}_{{\rm scal}}=\frac{G}{H}=\frac{{\rm SL}(2,\mathbb{R})}{{\rm SO}(2)}\times \frac{{\rm SO}(6,n)}{{\rm SO}(6)\times {\rm SO}(n)}\,.\label{Msc4}
\end{equation}
The symplectic duality representation of the global symmetry group $G={\rm SL}(2,\mathbb{R})\times {\rm SO}(6,n)$ is ${\Scr R}_v={\bf (2,n+6)}$. The isotropy group $H$ is the product of $H_R={\rm U}(4)$ and $H_{{\rm matt.}}={\rm SO}(n)$.  The relevant $G$ and $H$-representations are summarized in the Table \ref{N4content}.
\begin{table}[H]
\begin{center}
\begin{tabular}{|c||c|c|c|c|c|c|}
  \hline
    & $A^M_\mu$ &  $\psi_{A\mu}$& $\chi_{ABC}$& $\lambda_{I A}$ & $F^{AB}_{\mu\nu}$ & $F^{I}_{\mu\nu}$\\
    \hline \hline
 $G$ & ${\bf (2,n+6)}$ & ${\bf 1}$ & ${\bf 1}$ & ${\bf 1}$ & ${\bf 1}$& ${\bf 1}$\\
  \hline
 $H$ & ${\bf (1,1)}_0$ & ${\bf (4,1)}_{+\frac{1}{2}}$ & ${\bf (\bar{4},1)}_{+\frac{3}{2}}$ & ${\bf ({4},n)}_{-\frac{1}{2}}$ & ${\bf (6,1)}_{-1}$ & ${\bf (1,n)}_{+1}$\\
  \hline
\end{tabular}
  \caption{Relevant representations with respect to $G$ and $H$.}\label{N4content}
\end{center}
\end{table}
Gaugings of the four-dimensional half-maximal theory were studied in a large number of different contexts, see for instance \cite{Bergshoeff:1985ms,deRoo:1985jh,D'Auria:2002tc,D'Auria:2002th,deRoo:2003rm,Angelantonj:2003rq,Angelantonj:2003up,D'Auria:2003jk,deRoo:2006ms}. A systematic analysis of the most general gauging, using the embedding tensor formalism, was performed in \cite{Schon:2006kz}. The linear constraint defines for the embedding tensor the following representation:
\begin{equation}
{\Scr R}_\Theta={\bf (2,6+n)}+{\bf \left(2,\bigwedge{}^3(6+n)\right)}\,,
\end{equation}
described by tensors of the form:
\begin{equation}
\xi_{\tau\,{\tt M}}\,,\,\,\,f_{\tau\,{\tt MNP}}=f_{\tau\,[{\tt MNP}]}\,,
\end{equation}
where $\tau=1,2$ labels the doublet representation (the two values correspond to the positive and negative eigenvalues of the $\sigma^3$ Pauli generator) while the indices ${\tt M},{\tt N},{\tt P}=1,\dots 6+n$ label the fundamental of ${\rm SO}(6,n)$. These latter indices are lowered and raised using the ${\rm SO}(6,n)$-invariant metric $\eta_{{\tt MN}}$, with six ``$+1$'' and $n$ ``$-1$'' eigenvalues, and its inverse. The symplectic index $M$ is then identified with the couple $(\tau,\,{\tt M})$ and the symplectic invariant matrix is written as follows: $\mathbb{C}_{MN}=\epsilon_{\tau\sigma}\eta_{{\tt MN}}$.\par
The $O(g)$-fermion shifts read:
\begin{equation}
\delta\psi_{A\mu}=\dots+i\,g\,\mathbb{S}_{AB}\gamma_\mu\epsilon^B\,,\,\,\,\delta \chi_{ABC}=\dots+g\,\mathbb{N}_{ABC}{}^D\epsilon_D\,\,,\,\,\,\,\delta\lambda_{IA}=\dots+g\,\mathbb{N}_{IA}{}^B\epsilon_B\,.
\end{equation}
where the $H$-representations of the shift-tensors are
\begin{equation}
\mathbb{S}_{AB}\in {\bf (10,1)}_{+1}\,\,,\,\,\,\mathbb{N}_{ABC}{}^D\in {\bf (\overline{10}+{6},1)}_{+1}\,\,,\,\,\,\mathbb{N}_{IA}{}^B\in {\bf (15+1,n)}_{-1}\,.
\end{equation}
 The quadratic constraints were worked out in \cite{Schon:2006kz} to be:
\begin{align}
   \xi_\tau^{{\tt M}} \xi_{\sigma{{\tt M}}} &= 0 \, , \nonumber \\
   \xi^{{\tt P}}_{(\tau}  f_{\sigma){{\tt P}}{{\tt M}}{{\tt N}}} &= 0 \, , \nonumber \\
   3 f_{\tau{{\tt R}}[{{\tt M}}{{\tt N}}} {f_{\sigma{{\tt P}}{{\tt Q}}]}}^{{\tt R}} + 2 \xi_{(\tau[{{\tt M}}} f_{\sigma){{\tt N}}{{\tt P}}{{\tt Q}}]} &= 0 \; , \nonumber \\
   \epsilon^{\tau\sigma} \left( \xi_{\tau}^{{\tt P}} f_{\sigma{{\tt P}}{{\tt M}}{{\tt N}}} + \xi_{\tau{{\tt M}}} \xi_{\sigma{{\tt N}}} \right) &= 0 \, , \nonumber \\
   \epsilon^{\tau\sigma} \left( f_{\tau{{\tt M}}{{\tt N}}{{\tt R}}} {f_{\sigma{{\tt P}}{{\tt Q}}}}^{{\tt R}} - \xi^{{\tt R}}_\tau f_{\sigma{{\tt R}}[{{\tt M}}[{{\tt P}}} \eta_{{{\tt Q}}]{{\tt N}}]}
       - \xi_{\tau[{{\tt M}}} f_{{{\tt N}}][{{\tt P}}{{\tt Q}}]\sigma} + \xi_{\tau[{{\tt P}}} f_{{{\tt Q}}][{{\tt M}}{{\tt N}}]\sigma} \right) &= 0 \, .
   \label{N4qc}
\end{align}
In Sections \ref{n8fluxc} and \ref{Tdualcomp} gaugings are considered of two $\mathcal{N}=4$ models with $n=6$ vector multiplets.
Prior to gauging these theories could be obtained as consistent truncations of the maximal one and only differ in the symplectic frame.
One describes the common NS-NS sector of the Type II theories compactified on a six-torus, the other Type IIB theory on a $T^6/\mathbb{Z}_2$-orientifold.
In the former, see Sect. \ref{Tdualcomp}, the ${\rm SL}(2,\mathbb{R})$-factor in $G$ has a non-perturbative duality action and acts transitively on the four-dimensional dilaton $\phi_4$ and $\tilde{B}$, while the whole classical T-duality group ${\rm O}(6,6)$ has a diagonal duality action and coincides with $G_{el}$, so that the vector index $\Lambda$ can be identified with ${\tt M}$. The vector fields transform in the ${\bf 12}$ of ${\rm O}(6,6)$ and have the following identification: $A^\Lambda_\mu=A^{{\tt M}}_\mu=(G^{{\tt u}}_\mu,\,B_{{\tt u}\,\mu})$. The T-duality invariant set of generalized fluxes studied in the same section corresponds to the following choice of the embedding tensor:
\begin{equation}
\xi_{\tau\,{\tt M}}=0\,\,,\,\,\,f_{2\,{\tt MNP}}=0\,\,,\,\,\,\,-f_{1\,{\tt MN}}{}^{{\tt P}}=T_{{\tt MN}}{}^{{\tt P}}=\{H_{{\tt uvw}},\,T_{{\tt uv}}{}^{{\rm w}},\,Q_{{\tt u}}{}^{{\tt vw}},\,R^{{\tt uvw}}\}\,.\label{Ttotal2}
\end{equation} In this case the gauge generators read:
$$X_{{\tt M}}=-\frac{1}{2}\,T_{{\tt MNP}}\,t^{{\tt NP}}\,,$$
$t^{{\tt NP}}=-t^{{\tt PN}}$ being the ${\rm O}(6,6)$-generators, closing the commutation relations (\ref{tso66}). The gauging is electric, so that
the locality constraint is automatically satisfied. We are left with the closure constraints (\ref{quadratic2}), namely the requirement that the gauge algebra close inside $\mathfrak{so}(6,6)$, which boil down to (\ref{Xquadso66c}):
\begin{equation}
T_{{\tt R}[{\tt MN}}T_{{\tt PQ}]}{}^{{\tt R}}=0\,.
\end{equation}
This condition directly follows from (\ref{N4qc}) for this choice of the embedding tensor.\par
When studying the Type IIB flux-compactification on a $T^6/\mathbb{Z}_2$-orientifold, see end of Sect. \ref{Tdualcomp}, the ${\rm SL}(2,\mathbb{R})$ factor in $G$ is the Type IIB S-duality group ${\rm SL}(2,\mathbb{R})_{{\rm IIB}}$ and the symplectic section is characterized by an electric group which is:
\begin{equation}
G_{el}=\left[{\rm SL}(2,\mathbb{R})_{{\rm IIB}}\times {\rm GL}(6,\mathbb{R})\right]\ltimes e^{{\tt N}^{[{\bf (15',1)}_{+2}]}}\,,
\end{equation}
where ${\tt N}^{[{\bf (1,15')}_{+2}]}$ is spanned by the nilpotent Peccei-Quinn generators $t^{{\tt u_1 u_2 u_3 u_4}}$, associated with constant shifts on the axions $C_{{\tt u_1 u_2 u_3 u_4}}$, and having an off-diagonal duality action.
 The vector fields are $A^\Lambda_\mu=(C_{{\tt u}\mu},\,B_{{\tt u}\mu})$.\par
 The fluxes of the RR  and  NS-NS 3-form field strengths correspond to the following components of the embedding tensor:
\begin{equation}
f_{\tau\,{\bf uvw}}=(F_{{\tt u}{\tt v}{\tt w}},\,H_{{\tt u}{\tt v}{\tt w}})\,,
\end{equation}
all other components being zero. The quadratic constraints (\ref{N4qc}) impose no further conditions on these quantities. As mentioned in Sect. \ref{n8fluxc}, these models were considered in  \cite{Giddings:2001yu,Frey:2002hf,D'Auria:2002tc,D'Auria:2003jk} and feature Minkowski vacua with spontaneous supersymmetry breaking to $\mathcal{N}'=0,1,2,3$ at different scales, in which the dilaton field is fixed by the fluxes. This feature is strictly related to the fact that is this symplectic frame the group ${\rm SL}(2,\mathbb{R})_{{\rm IIB}}$, acting on $\phi$ and $\rho$, has a perturbative action. The same was true in the $\mathcal{N}=2$ models describing Type IIB flux-compactifications on $K3\times T^2/\mathbb{Z}_2$, see Sect. \ref{K3T2Z2}, where the axio-dilaton field $u$ was fixed in the vacuum. This model is not a truncation of a maximal gauged supergravity since the quadratic constraints in the latter, as explained in Sect. \ref{n8fluxc}, imply the extra condition (\ref{tad}), which can be interpreted as the requirement of tadpole cancelation in the absence of localized sources.
\subsection{$\mathcal{N}=3$ Supergravity}
$\mathcal{N}=3$ supergravity \cite{Castellani:1985ka} describes the gravity supermultiplet consisting of the graviton, three gravitinos $\psi_{A\,\mu}$, $A=1,\dots,3$, three vectors with field strengths $F^{AB}_{\mu\nu}$ and one dilatino $\chi_{ABC}=\chi\,\epsilon_{ABC}$, coupled to $n$ vector multiplets, each consisting of a vector, four gauginos $\lambda_{IA},\,\lambda_I$, and six scalar fields.
The scalar manifold has the form (see Table 1):
\begin{equation}
{\Scr M}_{{\rm scal}}=\frac{G}{H}= \frac{{\rm SU}(3,n)}{{\rm S}[{\rm U}(3)\times {\rm U}(n)]}\,.\label{Msc3}
\end{equation}
The symplectic duality representation of the global symmetry group $G={\rm SU}(3,n)$ is ${\Scr R}_v={\bf (3+n)}+\overline{{\bf (3+n)}}$. The isotropy group $H$ is locally isomorphic to the product of $H_R={\rm U}(3)$ and $H_{{\rm matt.}}={\rm SU}(n)$.
The relevant $G$ and $H$-representations are summarized in the Table \ref{N3content}.
\begin{table}[H]
\begin{center}
\begin{tabular}{|c||c|c|c|c|c|c|c|}
  \hline
    & $A^M_\mu$ &  $\psi_{A\mu}$& $\chi$& $\lambda_{I A}$ &$\lambda_{I}$ & $F^{AB}_{\mu\nu}$ & $F^{I}_{\mu\nu}$\\
    \hline \hline
 $G$ & ${\bf (3+n)}+\overline{{\bf (3+n)}}$ & ${\bf 1}$ & ${\bf 1}$ & ${\bf 1}$ & ${\bf 1}$& ${\bf 1}$& ${\bf 1}$\\
  \hline
 $H$ & ${\bf (1,1)}_0$ & ${\bf (3,1)}_{+\frac{1}{2}}$ & ${\bf (1,1)}_{+\frac{3}{2}}$ & ${\bf (3,n)}_{\frac{n+6}{2n}}$ &${\bf (1,n)}_{\frac{3(n+2)}{2n}}$ & ${\bf (3,1)}_{-1}$ & ${\bf (1,\bar{n})}_{-\frac{3}{n}}$\\
  \hline
\end{tabular}
  \caption{Relevant representations with respect to $G$ and $H$. The fermions $\lambda_{IA}$ and $\lambda_I$ have opposite chirality.}\label{N3content}
\end{center}
\end{table}
In our notations  the fundamental representation ${\bf 3+n}$ of ${\rm SU}(3,n)$ splits under ${\rm U}(3)\times {\rm SU}(n)$ as follows:
\begin{equation}
{\bf 3+n}\rightarrow {\bf (3,1)}_{-1}+{\bf (1,n)}_{+\frac{3}{n}}\,\,\Rightarrow\,\,\, v^{{\tt M}}=(v^{AB},\,v_I)\,,
\end{equation}
where $v^{{\tt M}}$ is a generic vector in the ${\bf 3+n}$, ${\tt M}=1,\dots, 3+n$. The representation $\underline{{\Scr R}}_v^c$ splits with respect to $H$ into ${\bf R}_v+ \underline{{\bf R}}_v$, where ${\bf R}_v={\bf (3,1)}_{-1}+{\bf (1,\bar{n})}_{-\frac{3}{n}}$, so that the composite field strengths $F_{\mu\nu}^{\underline{\Lambda}}$, as well as the central and matter charges ${\Scr Z}^{\underline{\Lambda}}$, split as follows:
\begin{equation}
F_{\mu\nu}^{\underline{\Lambda}}=(F_{\mu\nu}^{AB},\,F_{\mu\nu}^{I})\,\,,\,\,\,\,\,{\Scr Z}^{\underline{\Lambda}}=({\Scr Z}^{AB},\,{\Scr Z}^{I})\,.
\end{equation}
We see that in the $\underline{{\Scr R}}_v^c$ basis, ${\rm SU}(3,n)$ is not block-diagonal, as it should be since, in our conventions, its non-compact generators in this basis should be off-diagonal, see Eq. (\ref{kmatrixf}). \par
As far as gauging are concerned, the embedding tensor is restricted by the linear constraint to transform in representation described by an irreducible ${\rm SU}(3,n)$-tensor $\theta_{{\tt MN}}{}^{{\tt P}}=\theta_{[{\tt MN}]}{}^{{\tt P}}$ and its conjugate:
\begin{equation}
\Theta_M{}^\alpha\,:\,\,\,(\theta_{{\tt MN}}{}^{{\tt P}},\,\theta^{{\tt MN}}{}_{{\tt P}})\,,
\end{equation}
subject to the quadratic constraints (\ref{quadratic1}), (\ref{quadratic2}).
The $O(g)$-fermion shifts read:
\begin{equation}
\delta\psi_{A\mu}=\dots+i\,g\,\mathbb{S}_{AB}\gamma_\mu\epsilon^B\,,\,\,\,\delta \chi=\dots+g\,\mathbb{N}^D\epsilon_D\,\,,\,\,\,\,\delta\lambda_{IA}=\dots+g\,\mathbb{N}_{IA}{}^B\epsilon_B\,\,,\,\,\,\,
\delta\lambda_{I}=\dots+g\,\mathbb{N}_{IB}\epsilon^B\,.
\end{equation}
where the $H$-representations of the shift-tensors are
\begin{equation}
\mathbb{S}_{AB}\in {\bf (6,1)}_{+1}\,\,,\,\,\,\mathbb{N}^D\in {\bf (\overline{3},1)}_{+1}\,\,,\,\,\,\mathbb{N}_{IA}{}^B\in {\bf (8+1,n)}_{+\frac{3}{n}}
\,\,,\,\,\,\mathbb{N}_{IA}\in {\bf (3,n)}_{2+\frac{3}{n}}\,.
\end{equation}
Gaugings of semisimple groups inside the ${\rm SO}(3,n)$ electric subgroup of $G$ were recently studied in \cite{Karndumri:2016miq}, although a thorough analysis of the possible gaugings using the embedding tensor formalism is still missing.
\section{Gauged Supergravities from String/M-Theory Flux Compactifications}\label{gsfsmfc}
As mentioned in the introduction, \emph{traditional} compactifications of string/M-theory, i.e. compactifications on Ricci-flat manifolds in the absence of fluxes, typically yield ungauged lower-dimensional theories which are plagued by massless scalar fields (some related to the moduli on the internal manifold), which parametrize a continuum of degenerate Minkowski vacua.
This is the case of toroidal compactifications that were touched upon in Sect. \ref{N8MAB}, and which yield ungauged maximal supergravities in various dimensions. Similarly Calabi-Yau compactifications to four-dimensions of Type II theories, in the absence of internal fluxes, yield ungauged $\mathcal{N}=2$ theories in which the moduli associated with the shape and size of the internal manifold are not fixed  by any dynamics. As mentioned in the Introduction, switching on fluxes of RR or NS-NS field-strengths across cycles of the internal manifold induces, in the low-energy theory, a gauging, the corresponding embedding tensor being defined by the non-trivial background quantities. In the case of Calabi-Yau compactifications, the gauge generators typically involve quaternionic isometries. This flux-induced gauging implies the presence in the action of a scalar potential which (partially) lifts the original vacuum degeneracy. We refer the reader to the excellent, comprehensive reviews \cite{Grana:2005jc,Douglas:2006es,Blumenhagen:2006ci} for the topic of flux-compactifications in the presence of branes. In the following Sections we shall consider instances of flux-compactifications for the sake of illustrating a lower-dimensional, ``bottom-up'' kind of approach to the study of string dynamics on flux-backgrounds, which is based on gauged supergravities. In all the cases considered the low-energy effective theory is an extended supergravity. Without entering into the details of the compactifications, from general arguments we can infer the defining properties of the four-dimensional description:
field content, amount of supersymmetry, symplectic frame and the embedding tensor which defines the gauge group. This suffices, by supersymmetry, to uniquely determine the effective gauged supergravity description and with it the full non-linear interactions among the low-lying string modes which would be much harder to determine from a direct ``top-down'' approach.
\subsection{Toroidal Reductions}\label{toroidalc}
We start dealing with toroidal reductions, see for instance \cite{Scherk:1979zr,Cremmer:1997ct},
which are particularly simple and allow us to make contact with our previous discussion of Sect. \ref{N8MAB}.\par
Consider a theory in $D+n$ dimensions, the space-time being the product of a $D$-dimensional non-compact one $M_D$ and an internal torus $T^n$ of dimensions $n$. We split the $D+n$ coordinates as follows:
\begin{align}
(x^{\hat{\mu}})=(x^\mu,\,x^{\upalpha})\,\,,\,\,\,\,\,\mu=0,1,\dots,D-1\,\,,\,\,\,\,\upalpha=D,\dots, D+n-1\,.
\end{align}
Let us write the following ansatz for the higher-dimensional vierbein matrix $\mathbb{V}_{\hat{\mu}}{}^{\hat{a}}$ and its inverse $\mathbb{V}_{\hat{a}}{}^{\hat{\mu}}$:
\begin{align}
\mathbb{V}_{\hat{\mu}}{}^{\hat{a}}=\left(\begin{matrix}\Delta \, V_\mu{}^a & \alpha\,\upPhi_\upalpha{}^{\hat{\upalpha}}\,G^\upalpha_\mu\cr {\bf 0} & \upPhi_\upalpha{}^{\hat{\upalpha}}\end{matrix}\right)\,\,;\,\,\,\,\mathbb{V}_{\hat{a}}{}^{\hat{\mu}}=
\left(\begin{matrix}\Delta^{-1} \, V_a{}^\mu & -\alpha\,\Delta^{-1}\,V_a{}^\mu\,G^\upalpha_\mu\cr {\bf 0} & \upPhi_{\hat{\upalpha}}{}^{\upalpha}\end{matrix}\right)\,,
\end{align}
where $\hat{a}=0,\dots, D+n-1$, $a=0,\dots,D-1$ and $\hat{\upalpha}=D,\dots, D+n-1$ are the rigid indices on the whole space-time, its $D$-dimensional part and the torus, respectively (clearly $\upPhi_{\hat{\upalpha}}{}^{\upalpha}$ denotes the inverse matrix of $\upPhi_\upalpha{}^{\hat{\upalpha}}$). We have also set  $\alpha=\sqrt{16 \pi G_D}$, $G_D$ being the $D$-dimensional Newton's constant. The metrics on the whole $(D+n)$-dimensional space-time, on the torus and on $M_D$ read:
\begin{equation}
\hat{g}_{\hat{\mu}\hat{\nu}}=\sum_{\hat{a}}\mathbb{V}_{\hat{\mu}}{}^{\hat{a}}\mathbb{V}_{\hat{\nu}}{}^{\hat{b}}\eta_{\hat{a}\hat{b}}\,\,,\,\,\,\,
g_{\upalpha\upbeta}=-\sum_{\hat{\upalpha}=D}^{D+n-1}\upPhi_\upalpha{}^{\hat{\upalpha}}\upPhi_\upbeta{}^{\hat{\upalpha}}\,\,,\,
\,\,\,g_{\mu\nu}=\sum_{a=0}^{D-1}V_\mu{}^aV_\nu{}^b\eta_{ab}\,.
\end{equation}
The compactification on $T^n$ is effected by simply assuming the various components of the higher-dimensional fields not to depend on the internal coordinates $x^\upalpha$ (\emph{dimensional reduction}). The $(D+n)$-dimensional metric yields in the $D$-dimensional theory, the $D$-dimensional metric $g_{\mu\nu}$, $n$ Kaluza-Klein vectors $G^\upalpha_\mu$ and scalar fields describing the moduli of the internal torus $\upPhi_{\hat{\upalpha}}{}^{\upalpha}$, all functions of $x^\mu$ only.
Requiring the gravity term to be written in the Einstein frame fixes $\Delta$ as follows:
\begin{equation}
\Delta={\rm Vol}(T^n)^{-\frac{1}{D-2}}={\rm det}(\upPhi_\upalpha{}^{\hat{\upalpha}})^{-\frac{1}{D-2}}\,.
\end{equation}
The higher-dimensional theory was invariant under general  space-time reparametrization:
\begin{equation}
x^{\hat{\mu}}\rightarrow x^{\hat{\mu}}+\xi^{\hat{\mu}}(x^{\hat{\nu}})\,,\label{repinvdn}
\end{equation}
under which $\mathbb{V}_{\hat{\mu}}{}^{\hat{a}}$ and a $p$-form field of components $\hat{A}_{\hat{\mu}_1\dots\hat{\mu}_p}$ transform as follows:
\begin{align}
\delta_\xi \mathbb{V}_{\hat{\mu}}{}^{\hat{a}}&=\xi^{\hat{\nu}}\,\partial_{\hat{\nu}}\mathbb{V}_{\hat{\mu}}{}^{\hat{a}}+\partial_{\hat{\mu}}\xi^{\hat{\nu}}\,\mathbb{V}_{\hat{\nu}}{}^{\hat{a}}\,,\nonumber\\
\delta_\xi \hat{A}_{\hat{\mu}_1\dots\hat{\mu}_p}&=\xi^{\hat{\nu}}\,\partial_{\hat{\nu}}\hat{A}_{\hat{\mu}_1\dots\hat{\mu}_p}+
p\,\partial_{[\hat{\mu}_1}\xi^{\hat{\nu}}\,\hat{A}_{\hat{\nu}|\,\hat{\mu}_2\dots\hat{\mu}_q]}\,.\label{gentrass}
\end{align}
The local symmetries of the theory also include the gauge transformations associated with the $p$-form fields
\begin{equation}\hat{A}^{(p)}\rightarrow \hat{A}^{(p)}+d\hat{\Xi}^{(p-1)}+\dots\label{AXI}\end{equation}
where the ellipses refer to possible terms depending on the lower-order forms and their gauge transformation parameters.
Restricting the fields to $x^\mu$ only, we are left with the $D$-dimensional reparametrization-invariance $$x^\mu\rightarrow x^{{\mu}}+\xi^{{\mu}}(x^{\nu})\,,$$ and the invariance under  reparametrizations of the internal manifold depending on $x^\mu$: \begin{equation}x^\upalpha\rightarrow x^\upalpha+\xi^\upalpha(x^\mu)\,.\label{KKgau1}\end{equation} This latter amounts to a gauge transformation on the Kaluza-Klein vectors:
\begin{equation}
G^\upalpha_\mu\rightarrow G^\upalpha_\mu-\frac{1}{\alpha}\partial_\mu \xi^\upalpha\,.\label{KKgau2}
\end{equation}
There is a further invariance of the lower-dimensional theory which is remnant of the reparametrization invariance in $(D+n)$-dimensions. It corresponds to infinitesimal transformations (\ref{repinvdn}) of the form:
\begin{equation}
\delta x^\upalpha=\xi^\upalpha(x^\upbeta)=\xi_\upbeta{}^\upalpha\,x^\upbeta\,,\label{globgln}
\end{equation}
where $\xi_\upbeta{}^\upalpha$ is a constant $n\times n$ matrix. These parameters can be viewed as infinitesimal generators of global ${\rm GL}(n,\mathbb{R})$-transformations with respect to which all fields of the lower-dimensional theory fall into representations according to their internal index structure. For instance we have:
\begin{equation}
\delta G^\upalpha_\mu=-\xi_\upbeta{}^\upalpha\, G^\upbeta_\mu\,\,,\,\,\,\delta\upPhi_\upalpha{}^{\hat{\upalpha}}=
\xi_\upalpha{}^\upbeta\,\upPhi_\upbeta{}^{\hat{\upalpha}}\,.\label{APhitra}
\end{equation}
As for the other $D$-dimensional fields originating from antisymmetric tensors in higher-dimensions, they are defined so as to be invariant under (\ref{KKgau1}) and (\ref{KKgau2}).
This is effected by  expanding the higher-dimensional fields along the following basis of the cotangent space:
\begin{equation} dx^\mu,\,{\tt V}^\upalpha\equiv dx^\upalpha+\alpha\,G^\upalpha_\mu\,dx^\mu\,,\label{basisdxV}\end{equation}
where the one-forms ${\tt V}^\upalpha$ are clearly invariant under (\ref{KKgau1}) and (\ref{KKgau2}).
In this basis the $(D+n)$-dimensional metric reads:
\begin{equation}
ds^2_{D+n}=\hat{g}_{\hat{\mu}\hat{\nu}}\,dx^{\hat{\mu}} dx^{\hat{\nu}}= \Delta^2\,ds^2_D+g_{\upalpha\upbeta}\,{\tt V}^\upalpha{\tt V}^\upbeta\,,\label{dsdndsd}
\end{equation}
where $ds^2_D\equiv g_{\mu\nu}\,dx^\mu dx^\nu$. A $p$-form field $\hat{A}^{(p)}$ in $(D+n)$ dimensions is expanded as follows (unhatted components are fields in $D$-dimensions):
\begin{equation}
\hat{A}^{(p)}={A}^{(p)}+A^{(p-1)}_{\upalpha}\wedge {\tt V}^\upalpha+\frac{1}{2}\,A^{(p-2)}_{\upalpha\upbeta}\wedge {\tt V}^\upalpha\wedge {\tt V}^\upbeta+\dots\label{Apexpand}
\end{equation}
and so is the corresponding field strength:
\begin{equation}
\hat{F}^{(p+1)}={F}^{(p+1)}+F^{(p)}_{\upalpha}\wedge {\tt V}^\upalpha+\frac{1}{2}\,F^{(p-1)}_{\upalpha\upbeta}\wedge {\tt V}^\upalpha\wedge {\tt V}^\upbeta+\dots\label{Fpexpand}
\end{equation}
By construction the components $A^{(p-\ell)}_{\upalpha_1\dots\upalpha_{\ell}}$ are invariant under (\ref{KKgau1}) and (\ref{KKgau2}). The last of Eqs. (\ref{gentrass}) and the above  definitions imply that, under the ${\rm GL}(n,\mathbb{R})$-transformations (\ref{globgln}), the lower-dimensional fields $A^{(p-\ell)}_{\upalpha_1\dots\upalpha_{\ell}}$ transform in the $\bigwedge{}^\ell{\bf n}$ representation:
\begin{equation}
\delta A^{(p-\ell)}_{\upalpha_1\dots\upalpha_{\ell}}=\ell\,\xi_{[\upalpha_1}{}^\upbeta\,A^{(p-\ell)}_{\upbeta|\,\upalpha_2\dots\upalpha_{\ell}]}\,.\label{AVV}
\end{equation}
Since the matrix $\upPhi_\upalpha{}^{\hat{\upbeta}}$ transforms under ${\rm GL}(n,\mathbb{R})$ as in (\ref{APhitra}) but is defined modulo the action of an ${\rm SO}(n)$-transformation on its right index ${\hat{\upbeta}}$, it can be viewed as an element of the following coset manifold:
\begin{equation}
\upPhi_\upalpha{}^{\hat{\upbeta}}\in \frac{{\rm GL}(n,\mathbb{R})}{{\rm SO}(n)}\,,\label{slnson}
\end{equation}
which is the subspace of the full scalar manifold ${\Scr M}_{{\rm scal}}$ describing the moduli of the internal torus. \par
The action of the resulting $D$-dimensional theory has therefore a characteristic ${\rm GL}(n,\mathbb{R})$-global invariance. Note, however, that the ${\rm O}(1,1)$-factor of this global symmetry group, which acts as a rescaling of the internal volume, $g_{\upalpha\upbeta}\rightarrow e^{2\lambda}\,g_{\upalpha\upbeta}$, is  not just the result of a transformation of the form (\ref{globgln}) with $\xi_{\upalpha}{}^\upbeta=\lambda\,\delta_{\upalpha}^\upbeta$. Indeed invariance of the $D$-dimensional action requires combining the latter with a global scaling symmetry transformation  of the higher-dimensional theory \cite{Cremmer:1997ct}: $\hat{g}\rightarrow \Omega^2\,\hat{g}$. Under this combined action the $D\times D$ block $\hat{g}_{{\mu}{\nu}}$ of $\hat{g}_{\hat{\mu}\hat{\nu}}$ does transform, but in such a way that the  $D$-dimensional metric $g_{\mu\nu}$, as defined in (\ref{dsdndsd}), is left unaltered, and thus is ${\rm GL}(n,\mathbb{R})$-invariant.\par
Just as the fields, also the gauge parameters $\hat{\Xi}^{(q)}$ of the higher-dimensional theory can be expanded in the basis (\ref{basisdxV}):
\begin{equation}
\hat{\Xi}^{(q)}={\Xi}^{(q)}+{\Xi}^{(q-1)}_\upalpha\wedge {\tt V}^\upalpha+\frac{1}{2}\,{\Xi}^{(q-2)}_{\upalpha\upbeta}\wedge {\tt V}^\upalpha\wedge {\tt V}^\upbeta+\dots\label{XiVV}
\end{equation}
From (\ref{XiVV}), (\ref{AXI}) we derive the gauge transformations for the different components $A^{(p-\ell)}_{\upalpha_1\dots \upalpha_\ell}$.
%\begin{equation}
%\delta A^{(p-\ell)}_{\upalpha_1\dots \upalpha_\ell}=d{\Xi}^{(p-\ell-1)}_{\upalpha_1\dots \upalpha_\ell}+(-1)^p\,\alpha\,{\Xi}^{(p-\ell-2)}_{\upalpha_1\dots \upalpha_\ell \upalpha}\,F^\upalpha\,,
%\end{equation}
%where we have used $d{\tt V}^\upalpha=\alpha\,F^\upalpha$.
We can also choose for $\hat{\Xi}^{(p-1)}$ the following form:
\begin{equation}
\hat{\Xi}^{(p-1)}=\frac{1}{p!}\Sigma_{\upalpha \upalpha_1\dots \upalpha_{p-1}}\,x^\upalpha\,dx^{\upalpha_1}\wedge \dots \wedge dx^{\upalpha_{p-1}}\,,\label{globalsigma}
\end{equation}
where $\Sigma_{\upalpha \upalpha_1\dots \upalpha_{p-1}}$ is a constant tensor.
Computing $d\hat{\Xi}^{(p-1)}$ and expressing it in the basis (\ref{basisdxV}), we find:
\begin{align}
d\hat{\Xi}^{(p-1)}&=\frac{1}{p!}\Sigma_{\upalpha_1 \dots \upalpha_{p}}\,{\tt V}^{\upalpha_1}\wedge\dots \wedge {\tt V}^{\upalpha_p}-\frac{\alpha}{(p-1)!}\Sigma_{\upalpha\upalpha_1 \dots \upalpha_{p-1}}\,G^\upalpha\,\wedge {\tt V}^{\upalpha_1}\wedge\dots \wedge {\tt V}^{\upalpha_{p-1}}+\nonumber\\&+\frac{\alpha^2}{2(p-2)!}\,\Sigma_{\upalpha\upbeta\upalpha_1 \dots \upalpha_{p-2}}\,G^\upalpha\wedge G^\upbeta\wedge {\tt V}^{\upalpha_1}\wedge\dots \wedge {\tt V}^{\upalpha_{p-2}}+\dots
\end{align}
from which we can infer the corresponding transformations of the $D$-dimensional fields $A^{(p-\ell)}_{\upalpha_1\dots \upalpha_\ell}$ originating from $\hat{A}^{(p)}$.
In particular the components  $A_{\upalpha_1\dots \upalpha_p}$ in the expansion (\ref{AVV}) are axionic scalar fields transforming by the following constant shift:
\begin{equation}
\delta A_{\upalpha_1\dots \upalpha_p}=\Sigma_{\upalpha_1\dots \upalpha_p}\,.\label{deltaAsigma}
\end{equation}
The transformations induced by gauge parameters of the form (\ref{globalsigma}) are global symmetries of the $D$-dimensional theory. In particular (\ref{deltaAsigma}) is an isometry of the scalar manifold ${\Scr M}_{{\rm scal}}$. In the solvable parametrization of ${\Scr M}_{{\rm scal}}$, the scalar fields  $A_{\upalpha_1\dots \upalpha_p}$  parametrize nilpotent generators $t^{\upalpha_1\dots \upalpha_p}$ in ${\tt N}$, see Eq. (\ref{SCN}), and the corresponding constant shifts are generated by the same isometries.
The remaining generators in ${\tt N}$ belong to the solvable Lie algebra defining the solvable parametrization of the submanifold (\ref{slnson}). These are parametrized by the moduli $\gamma_\upalpha{}^\upbeta$ associated with the off-diagonal components of the internal metric, see Appendix \ref{appdualsugras}, and generate isometries acting on the same moduli.
Being ${\tt N}$ nilpotent, the scalars parametrizing it occur in the sigma-model action polynomially. This is not the case
of the dilatonic scalars parametrizing the  Cartan subalgebra ${\tt C}$ of the solvable Lie algebra ${\Scr S}$. These consist of the moduli associated with the radii of the internal torus and, in the string theory compactifications, of the dilaton.
Each generator in ${\tt N}$, and thus the corresponding axionic scalar field, has characteristic gradings with respect to the  ${\rm O}(1,1)$ global symmetries generated by elements of ${\tt C}$. The collection of these gradings defines the so-called \emph{weight} of the nilpotent generators in ${\tt N}$, relative to ${\tt C}$. Choosing a basis for ${\tt C}$, if $r$ is its dimension, a weight is a collection of $r$ numbers arranged in a vector $\overrightarrow{W}$, which can be lexiographically ordered. By construction, if ${\tt N}$ contains generators with weight $\overrightarrow{W}$, it does not contain generators with weight $-\overrightarrow{W}$. Toroidal compactifications of $D=11$ or of a $D=10$ supergravity yield lower-dimensional theories which are maximally supersymmetric and  whose  scalar manifolds are symmetric of the generic form ${\Scr M}_{{\rm scal}}=G/H$, $\mathfrak{g}$ being a \emph{split} (or maximally non-compact) Lie algebra, see paragraph at the end of Sect. \ref{ghsect}. In this case the solvable Lie algebra ${\Scr S}$ is a Borel subalgebra of $\mathfrak{g}$,  ${\tt C}$ is a non-compact Cartan subalgebra of $\mathfrak{g}$ and the weights $\overrightarrow{W}$ identifying the generators in ${\tt N}$ are the \emph{positive roots} of $\mathfrak{g}$ relative to ${\tt C}$. The negative roots define nilpotent generators in $\mathfrak{g}$ which \emph{are not} in the solvable Lie algebra ${\Scr S}$. Their interpretation in terms of symmetries of the higher dimensional theory is more obscure. An example of a symmetry generated by an element of ${\tt C}$ is the rescaling of the volume of the internal space: $g_{\upalpha\upbeta}\rightarrow e^{2\lambda}\,g_{\upalpha\upbeta}$. From inspection of the dimensionally reduced Lagrangian we can infer that the corresponding grading of the scalars $A_{\upalpha_1\dots \upalpha_p}$ is $k=p/2$:
\begin{equation}
g_{\upalpha\upbeta}\rightarrow e^{2\lambda}\,g_{\upalpha\upbeta}\,:\,\,\,A_{\upalpha_1\dots \upalpha_p}\rightarrow\,e^{2k\lambda}\,A_{\upalpha_1\dots \upalpha_p}\,=\,e^{p\lambda}\,A_{\upalpha_1\dots \upalpha_p}\,.
\end{equation}
In general all bosonic fields can be associated with weights of $\mathfrak{g}$: the axionic scalar fields through the corresponding nilpotent generator in ${\tt N}$, the vector fields as elements, together with their duals, of a symplectic vector acted on by the representation ${\Scr R}_v$, the background quantities (fluxes) as components of the embedding tensor  $\Theta$ in the representation ${\Scr R}_\Theta$. \footnote{See  Appendix \ref{appdualsugras} for a detailed discussion of this correspondence.} A generic field $\Phi$ whose field strength $F[\Phi]$ has the following index structure:
\begin{equation}
F[\Phi]_{\mu_1\dots\mu_q\,\upalpha_1\dots\upalpha_p}{}^{\upbeta_1\dots\upbeta_\ell}\,,\label{Phieldstr}
\end{equation}
with respect to the internal volume dilation $g_{\upalpha\upbeta}\rightarrow e^{2\lambda}\,g_{\upalpha\upbeta}$ has the following grading $k_\Phi$:
\begin{equation}
k_\Phi=\frac{1}{2}\left(p-\ell-\frac{n(q-1)}{D-2}\right)\,.\label{kappaphi}
\end{equation}
This means that the action of the corresponding ${\rm O}(1,1)$-transformation on it is: $\Phi\rightarrow \,e^{2k_\Phi \lambda}\,\Phi$.\par
With respect to this action the nilpotent subalgebra ${\tt N}$ of ${\Scr S}$, spanned by the axionic fields, has a graded structure:
\begin{equation}
{\tt N}=\bigoplus_k {\tt N}^{(k)}\,\,:\,\,\,\,e^{2h\lambda}\,{\tt N}^{(k)}\,e^{-2h\lambda}=e^{2k\lambda}\,{\tt N}^{(k)}\,,
\end{equation}
where only positive gradings $k$ contribute to the direct sum and define the ${\rm O}(1,1)$-grading of the scalar fields spanning the corresponding subspace, $h\in {\tt C}$ being the infinitesimal generator of the transformation. The axions describing the off-diagonal components of the internal metric parametrize nilpotent generators of ${\rm SL}(n,\mathbb{R})$, which commutes with this  ${\rm O}(1,1)$ group, and thus have vanishing grading. Indeed the corresponding field strengths have the index structure (\ref{Phieldstr}) with $q=1,\,p=1,\,\ell=1$.
\paragraph{Adding fluxes.}
In order to introduce a constant flux of the field strength $\hat{F}^{(p+1)}$ across a $(p+1)$-cycle of the internal torus, we shift the corresponding elementary field $\hat{A}^{(p)}$ by a background part $\hat{A}^{(p)}_{bg}$ which is independent of the $x^\mu$ and linear in the $x^\upalpha$:
\begin{equation}
\hat{A}^{(p)\prime}(x^{\hat{\mu}})=\hat{A}^{(p)}(x^\mu)+\hat{A}^{(p)}_{bg}(x^\upalpha)\,,\label{ApAbg}
\end{equation}
where
\begin{equation}
\hat{A}^{(p)}_{bg}\equiv \frac{1}{(p+1)!}\,g_{\upalpha\upalpha_1\dots \upalpha_p}\,x^\upalpha\,dx^{\upalpha_1}\wedge \dots \wedge dx^{\upalpha_{p}}\,,
\end{equation}
$g_{\upalpha\upalpha_1\dots \upalpha_p}$ being a constant tensor.
The $\hat{A}^{(p)}$ only depends on $x^\mu$ and describes the fluctuation of the field about the background, defined by $\hat{A}^{(p)}_{bg}$. It is expanded as in (\ref{Apexpand}). Similarly the field strength is shifted by a background term representing its flux:
\begin{equation} \hat{F}^{(p+1)\prime}(x^{\hat{\mu}})=\hat{F}^{(p+1)}(x^\mu)+\hat{F}^{(p+1)}_{bg}(x^\upalpha)\,,
\end{equation}
where $\hat{F}^{(p+1)}$ is expanded in the field strengths of the various components of $\hat{A}^{(p)}$, while
$\hat{F}^{(p+1)}_{bg}$ reads:
\begin{equation}
\hat{F}^{(p+1)}_{bg}=\frac{1}{(p+1)!}\,g_{\upalpha_1\dots \upalpha_{p+1}}\,dx^{\upalpha_1}\wedge \dots \wedge dx^{\upalpha_{p+1}}\,.
\end{equation}
The expansion of this component in the basis (\ref{basisdxV}), by writing $dx^\upalpha={\tt V}^\upalpha-\alpha\,G^\upalpha$, yields flux-dependent terms which modify the $D$-dimensional field strengths $F^{(p+1)},\,F^{(p)}_\upalpha,\dots$ in the expansion of $\hat{F}^{(p+1)}$. These redefinitions encode a gauge symmetry of the lower-dimensional theory induced by the form-flux. This local symmetry involves the nilpotent shift-isometries $t^{\upalpha_1\dots \upalpha_p}$, the gauge generators $X$ being expressed in terms of them through an embedding tensor defined by the flux-components $g_{\upalpha_1\dots \upalpha_{p+1}}$. We shall illustrate this in a simple example below. \par
We can also add \emph{geometric fluxes} $T_{\upbeta\upgamma}{}^\upalpha$ by ``deforming'' the geometry of the internal torus into that of a \emph{twisted torus}.
As mentioned in the Introduction, a {twisted torus} ${\Scr T}^n$ is a compact space which is locally described as an $n$-dimensional group manifold $G_T$. A compactification on ${\Scr T}^n$ proceeds along the same lines as that on $T^n$, replacing the closed 1-forms $dx^\upalpha$ on the latter by
 the left-invariant one-forms $\sigma^\upalpha$, $\upalpha=1,\dots, n$, on $G_T$, satisfying the Maurer-Cartan equation:
 \begin{equation}
 d\sigma^\upalpha=-\frac{1}{2}\,T_{\upbeta\upgamma}{}^\upalpha\,\sigma^\upbeta\wedge \sigma^\upgamma\,,\label{mctwist}
 \end{equation}
 where $T_{\upbeta\upgamma}{}^\upalpha$, also known as ``twist tensor'', are the structure constants satisfying the Jacobi identity $T_{[\upbeta\upgamma}{}^\upalpha\,T_{\upsigma]\upalpha}{}^\updelta=0$. Thus the forms ${\tt V}^\upalpha$ in the metric ansatz (\ref{dsdndsd}) are now defined as: ${\tt V}^\upalpha=\sigma^\upalpha+\alpha\,G^\upalpha$.
 The background internal metric is $ds^2_{n,bg}=-\sum_\upalpha \sigma^\upalpha\otimes \sigma^\upalpha$ and $\upPhi_\upalpha{}^{\hat{\upalpha}}$ describe $x^\mu$-dependent fluctuations about this metric.
   The one-forms $\sigma^\upalpha$ can be written as $\sigma^\upalpha=\sigma_\upbeta{}^\upalpha(x^\upgamma)\,dy^\upbeta$ and are obtained by acting on the $dy^\upbeta$, describing the torus cohomology, by means of the vielbein matrix $\sigma_\upbeta{}^\upalpha(x^\upgamma)$ on ${\Scr T}^n$, named ``twist matrix'', which the name of ``twisted torus'' for ${\Scr T}^n$ comes from.
The ansatz for the internal metric, in the basis $dy^\upalpha$, now depends on both the internal coordinates through the background metric and external coordinates through $\upPhi_\upalpha{}^{\hat{\upalpha}} (x^\mu)$ and reads:
\begin{equation}
g_{\upalpha\upbeta}(x^\mu,\,x^\updelta)=-\sigma_\upalpha{}^\upgamma(x^\updelta)\sigma_\updelta{}^\upsigma(x^\updelta) \upPhi_\upgamma{}^{\hat{\upgamma}}(x^\mu)\upPhi_\upsigma{}^{\hat{\upgamma}}(x^\mu)\,.
\end{equation}
Note that it is obtained from the corresponding quantity $-\upPhi_\upalpha{}^{\hat{\upgamma}}\upPhi_\upbeta{}^{\hat{\upgamma}}$ in the ordinary toroidal compactification by \emph{twisting} it through the twist matrix $\sigma_\upalpha{}^\upgamma(x^\updelta)$. An analogous twist, with respect to the toroidal reduction ansatz, occurs for all the fields with internal indices $\upalpha$.

 Let $X_\upalpha$ be the generators of $G_T$ dual to $\sigma^\upalpha$: $\sigma^\upalpha(X_\upbeta)=\delta^\upalpha_\upbeta$. They satisfy the commutation relations:
 \begin{equation}
 [X_\upalpha,\,X_\upbeta]=T_{\upalpha\upbeta}{}^\upgamma\,X_\upgamma\,.
 \end{equation}
  The internal diffeomorphisms $\delta x^\upalpha=\xi^\upalpha(x^\mu)$ which induced, in the toroidal reduction, a ${\rm U}(1)^n$ gauge symmetry in the lower-dimensional theory, are now replaced by the right-action, on a representative ${\bf g}$ of $G_T$, of a $x^\mu$-dependent element of the same group. The 1-forms ${\tt V}^\upalpha$ transform covariantly under such transformation provided $G^\upalpha_\mu$ transform as non-Abelian gauge vectors of $G_T$. To see this
  let us consider the right action of $G_T$ on itself by means of a transformation ${\bf g}_R^{-1}(x^\mu)\in G_T$: ${\bf g}\rightarrow {\bf g}\,{\bf g}_R^{-1}(x^\mu)$, for any ${\bf g}\in G_T$. The left-invariant one-forms $\sigma^\upalpha$ are defined by projecting ${\bf g}^{-1}d{\bf g}$ on the basis $X_\upalpha$ of generators of $G_T$ and transform as follows:
  \begin{equation}
  \sigma^\upalpha\,\rightarrow \, {\bf g}_R^{-1}{}_\upbeta{}^\upalpha\,\sigma^\upalpha+ ({\bf g}_R d {\bf g}_R^{-1})^\upalpha\,.
  \end{equation}
  The basis of 1-forms ${\tt V}^\upalpha\equiv \sigma^\upalpha+ \alpha\,G^\upalpha_\mu dx^\mu$ transforms covariantly under this group action:
  \begin{equation}
  {\tt V}^\upalpha\,\rightarrow \, {\bf g}_R^{-1}{}_\upbeta{}^\upalpha\, {\tt V}^\upbeta\,,
  \end{equation}
  provided $G^\upalpha_\mu$ transform as a $G_T$-gauge connection:
  \begin{equation}
  G^\upalpha_\mu\,\rightarrow\,{\bf g}_R^{-1}{}_\upbeta{}^\upalpha\, G^\upbeta-\frac{1}{\alpha}\,({\bf g}_R d {\bf g}_R^{-1})^\upalpha\,,
  \end{equation}
  which is the non-Abelian version of (\ref{KKgau2}).
  The expressions of the $D$-dimensional field strengths $F^{(p+1-\ell)}_{\upalpha_1\dots \upalpha_\ell}$ in (\ref{Fpexpand}) now contain $T$-dependent terms containing the Kaluza-Klein vectors $G^\upalpha_\mu$, which make them covariant under local $G_T$-transformations. These terms do not occur in the toroidal reduction since they arise when acting by the exterior derivative on the one-forms $\sigma^\upalpha$ inside ${\tt V}^\upalpha$ in (\ref{Apexpand}) and expressing $d\sigma^\upalpha$ in terms of $\sigma^\upbeta\wedge \sigma^\upgamma$ through (\ref{mctwist}).\par
  The resulting $D$-dimensional theory features a non-Abelian gauge symmetry group $G_T$ gauged by the Kaluza-Klein vectors $G^\upalpha$.  The quantity $T_{\beta\gamma}{}^\alpha$ can be also viewed as a \emph{torsion} on the original torus and is an example of geometric flux. The gauged isometries are the generators of a subgroup $G_T$ of the off-shell global symmetry group ${\rm GL}(n,\mathbb{R})$. Denoting by $t_\upalpha{}^\upbeta$ a basis of the Lie algebra of ${\rm GL}(n,\mathbb{R})$ the gauge generators $X_\upalpha$ of $G_T$ have the form
  \begin{equation}
  X_\upalpha=T_{\upalpha\upgamma}{}^\upbeta\,t_\upbeta{}^\upgamma=-\Theta_{\upalpha,\,\upgamma}{}^\upbeta\,t_\upbeta{}^\upgamma \,,
  \end{equation}
   and the $G_T$-covariant derivatives read: $\mathcal{D}_\mu=\partial_\mu-\alpha\,G^\upalpha_\mu\,X_\upalpha$.
   We see that also the geometric flux $T_{\upalpha\upgamma}{}^\upbeta$ enters the lower-dimensional theory as components of the embedding tensor.
   The induced scalar potential in the lower-dimensional theory has the following form \cite{Scherk:1979zr}:\footnote{Recall that $g_{\upalpha\upbeta}<0$.}
   \begin{equation}
   V=-\frac{1}{4\alpha^2}\,\delta^{-\frac{2}{D-2}}\left(2\,T_{\upalpha\upbeta}{}^\upgamma T_{\upgamma\updelta}{}^\upalpha\,g^{\upbeta\updelta}+
   T_{\upalpha\upbeta}{}^\upgamma T_{\upalpha'\upbeta'}{}^{\upgamma'}\,g_{\upgamma\upgamma'} g^{\upalpha\upalpha'}g^{\upbeta\upbeta'}\right)\,,\label{VTwist}
   \end{equation}
 where $\delta\equiv {\rm det}(\upPhi_\upalpha{}^{\hat{\upalpha}})$. This potential, if $G_T$ is semisimple,  is unbounded from below \cite{Scherk:1979zr}. Vacua can be found for specific choices of $T_{\upalpha\upbeta}{}^\upgamma$ corresponding to non-semisimple $G_T$. For example one can choose, as the only non-vanishing components of this tensor, $T_{D\upbeta}{}^\upgamma$, where $\upbeta,\,\upgamma=D+1,\dots D+n$, with the condition $T_{D\upbeta}{}^\upgamma=-T_{D\upgamma}{}^\upbeta$. The matrix $M=T_{D\upbeta}{}^\upgamma$ is a compact generator of the global symmetry group ${\rm SL}(n-1,\mathbb{R})$ of the $(D+1)$-dimensional theory obtained by ordinary toroidal dimensional reduction of the $(D+n)$-dimensional one. The $D$-dimensional theory then originates from a Scherk-Schwarz reduction of the $(D+1)$-dimensional one, see discussion in Section \ref{gaugE6}, with ``twist matrix'' $M$, and the resulting gauge group is an example of  ``flat group''. The difference between this compactification and the Cremmer-Scherk-
Schwarz (or generalized Scherk-Schwarz reductions) discussed, in the case of the maximal four-dimensional theory, in Section \ref{gaugE6}, is that in the latter case the matrix $M$ is chosen as a compact generator of the whole global symmetry group of the $(D+1)$-dimensional theory, while now $M$ is chosen only within the $\mathfrak{so}(n-1)$ algebra, and thus features less independent mass parameters, in number equal to the rank of ${\rm SO}(n-1)$. For instance a dimensional reduction of eleven-dimensional supergravity on this kind of twisted torus to $D=4$,  amounts to an ordinary toroidal reduction to $D+1=5$, followed by a Scherk-Schwarz one, with twist matrix $M=T_{D\upbeta}{}^\upgamma$, from five to four. In general this matrix can be any generator of the ${\rm SL}(6,\mathbb{R})$ symmetry of the five-dimensional theory, acting on the internal metric moduli. We have a flat gauge group in $D=4$, with Minkowski vacua, only if $M$ is chosen to be compact ($M=-M^T$), see Footnote \ref{fotss}, and thus a generator of ${\rm SO}(6)$. In this case $M$ would depend on three independent parameters $m_1,m_2,m_3$, corresponding to the rank of  ${\rm SO}(6)$, which provide the masses, and thus the scale of supersymmetry breaking, of the resulting gauged $D=4$ model. In the generalized Scherk-Schwarz reduction, we start from the maximal theory in $D+1=5$ dimensions, and choose as twist matrix $M$ a generic compact generator of the larger global symmetry group ${\rm E}_{6(6)}$. In this case, as an element of $\mathfrak{usp}(8)$, $M$ would depend on a larger number of independent parameters, equal to the rank of this algebra, which is 4.\par
   Form-fluxes can be introduced in the twisted-torus compactification by adding to the corresponding field strength in $(D+n)$-dimensions a background term of the form:
  \begin{equation}
\hat{F}^{(p+1)}_{bg}=\frac{1}{(p+1)!}\,g_{\upalpha_1\dots \upalpha_{p+1}}\,\sigma^{\upalpha_1}\wedge \dots \wedge \sigma^{\upalpha_{p+1}}\,.
\end{equation}
This induces in the $D$-dimensional theory a further gauge invariance associated with the local shift-symmetries (\ref{deltaAsigma}), the corresponding embedding tensor being identified with $g_{\upalpha_1\dots \upalpha_{p+1}}$.
The Bianchi identity on $\hat{F}^{(p+1)\prime}$ then implies closure of $\hat{F}^{(p+1)}_{bg}$ which in turn requires the following quadratic condition: $T_{[\upalpha_1\upalpha_2}{}^\upbeta g_{\upalpha_3\dots \upalpha_{p+2}]\upbeta}=0$. Once $g_{\upalpha_1\dots \upalpha_{p+1}}$ and  $T_{\upalpha\upgamma}{}^\upbeta$ are identified with components of the embedding tensor in the lower-dimensional supergravity, the above constraint, together with the Jacobi identity on $T_{\upalpha\upgamma}{}^\upbeta$, simply follow from (\ref{quadratic1}), (\ref{quadratic2}).\par
\paragraph{Dimensional reduction from $D=11$.}
Consider $D=11$ supergravity, whose bosonic sector consists in the metric $\hat{g}_{\hat{\mu}\hat{\nu}}$ and a 3-form field $\hat{A}^{(3)}=\frac{1}{3!}\,A_{\hat{\mu}\hat{\nu}\hat{\rho}}\,dx^{\hat{\mu}}\wedge dx^{\hat{\nu}}\wedge dx^{\hat{\rho}}$. Dimensional reduction on a seven-torus to $D=4$ dimensions yields the following bosonic fields:
\begin{align}
\hat{g}_{\hat{\mu}\hat{\nu}}&\,\,\rightarrow\,\,\,g_{\mu\nu},\,G^\upalpha_\mu,\,\upPhi_\upalpha{}^{\hat{\upalpha}}\,,\nonumber\\
\hat{A}^{(3)}&\,\,\rightarrow\,\,\,A_{\mu\nu\rho}\,,\,\,A_{\mu\nu\,\upalpha}\,,\,\,A_{\mu\,\upalpha\upbeta}\,,\,\,
A_{\upalpha\upbeta\upgamma}\,.
\end{align}
The tensor $A_{\mu\nu\rho}$ is non-dynamical, while $A_{\mu\nu\,\upalpha}$ can be dualized into seven scalar fields $\tilde{A}^\upalpha$ in the ${\bf 7}'_{+4}$ of ${\rm GL}(7,\mathbb{R})$ (for convenience we write for each representation the grading $4 k_\Phi/3$, $k_\Phi$ being given in (\ref{kappaphi})). $A_{\mu\,\upalpha\upbeta}$ are $21$ vectors in the ${\bf 21}_{-1}$ which, together with the Kaluza-Klein vectors $G^\upalpha_\mu$ in the ${\bf 7}'_{-3}$, form the 28 vectors of the resulting four-dimensional maximal supergravity. The components $A_{\upalpha\upbeta\upgamma}$ are 35 scalar fields in the ${\bf 35}_{+2}$ which, together with seven $\tilde{A}^\upalpha$ and the 28 $\upPhi_\upalpha{}^{\hat{\upalpha}}$ define the $70$ scalars of the maximal theory, consistently with the group theoretical analysis of Sect. \ref{N8MAB}. In particular the nilpotent shift-isometries $t^{\upalpha\upbeta\upgamma},\,t_\upalpha$ parametrized by $A_{\upalpha\upbeta\upgamma}$ and $\tilde{A}^\upalpha$, respectively, generate the spaces ${\tt N}^{{\tiny [{\bf 35}_{+2}]}}$ and ${\tt N}^{{\tiny [{\bf 7}_{+4}]}}$ within ${\Scr S}$, see Eq. (\ref{solvMth}).\par
 The 4-form flux $g_{\upalpha_1\upalpha_2\upalpha_3 \upalpha_{4}}$ belongs to the representation ${\bf 35}'_{+5}$ which is present in the brancing of ${\Scr R}_\Theta={\bf 912}$ with respect to ${\rm GL}(7,\mathbb{R})$, given in the last Table of Sect. \ref{N8MAB}. One can also consider the flux $g_{\mu\nu\rho\sigma}$, as in the Freund-Rubin solution, which corresponds to the flux $g_7$ of the dual 7-form field strength across the whole torus. The representation of the latter is ${\bf 1}_{+7}$ and occurs as well in the branching of the ${\bf 912}$. The geometric flux $T_{\upalpha\upgamma}{}^\upbeta$, on the other hand, is described by the representation ${\bf (140+7)}_{+3}$ in the ${\bf 912}$. The gauge group originating from the presence of $g_{\upalpha_1\upalpha_2\upalpha_3 \upalpha_{4}}$ and  $g_7$ on a twisted torus compactification of $D=11$ supergravity was studied in \cite{dft1,dft2,D'Auria:2005rv}. In particular $T_{\upalpha\upgamma}{}^\upbeta$ defines magnetic components of the embedding tensor which associates, in the gauge connection, the translational isometries $t_\upalpha$ on the scalar fields  $\tilde{A}^\upalpha$ with the magnetic vector fields $A^{\upalpha\upbeta}_\mu$: $A^{\upalpha\upbeta}_\mu\,T_{\upalpha\upbeta}{}^\upgamma\,t_\upgamma$.
 \paragraph{Dimensional reduction from $D=10$.}
 In ten dimensions maximal supergravity can either have the form of the non-chiral Type IIA theory, or of the chiral Type IIB, see Sect. \ref{N8MAB}. The two have a common sector describing the NS-NS string zero-modes and differ in the RR sector.
 The former is obtained by reducing eleven-dimensional supergravity on a circle of radius $R_{10}=e^{\sigma_{10}}$ (along the eleventh-dimension parametrized by $x^{10}$) and writing the eleven-dimensional metric in the following form
 \begin{equation}
 ds^2_{D=11}=e^{-\sigma_{10}}\,ds^2_{{\rm IIA,\,S}}-e^{2\,\sigma_{10}}\,({\tt V}^{10})^2\,,
\end{equation}
where ${\tt V}^{10}\equiv dx^{10}+\alpha\,\hat{C}^{(1)}$ and $ds^2_{{\rm IIA,\,S}}$ is the Type IIA space-time metric in the \emph{string frame} in which the kinetic terms of the NS-NS fields in the Lagrangian are multiplied by a characteristic factor $e^{-2\,\phi}$: ${\Scr L}_{{\rm NS-NS,\, S}}=e^{-2\,\phi}\,(-\frac{R}{2}+\dots)$. The Kaluza-Klein vector is the RR 1-form field of the ten-dimensional theory and the dilaton $\phi$ is related to the radial modulus $\sigma_{10}$ as follows: $\phi=\frac{3}{2}\,\sigma_{10}$. The ten-dimensional metric in the Einstein-frame is given by:
\begin{equation}
ds^2_{{\rm IIA,\,E}}=e^{-\frac{\phi}{2}}\,ds^2_{{\rm IIA,\,S}}\,.
\end{equation}
An analogous relation holds for the Type IIB metric in the two frames.\par
Consider now the compactification of a Type II (A or B) theory on a six-torus to four dimensions. We denote now by ${\tt u},\,{\tt v},\dots=4,\dots,9$ the indices labeling the internal coordinates and by $\hat{{\tt u}},\,\hat{{\tt v}},\dots$ their rigid counterpart.
The ansatz for the metric, in the string frame, reads:
\begin{equation}
s^2_{{\rm II,\,S}}=\Delta^2\,ds^2_4+g^{(S)}_{{\tt u}{\tt v}}\,{\tt V}^{{\tt u}}\,{\tt V}^{{\tt v}}\,,
\end{equation}
where $g^{(S)}_{{\tt u}{\tt v}}$ is the internal metric in the string frame and $\Delta={\rm det}(\upPhi^{(S)}{}_\upalpha{}^{\hat{\upalpha}})^{-\frac{1}{2}}\,e^{\phi}$.
 Let us expand the ten-dimensional Kalb-Ramond field $\hat{B}^{(2)}$ as in (\ref{Apexpand}):
 \begin{equation}
 \hat{B}^{(2)}=B^{(2)}+B^{(1)}_{{\tt u}}\wedge {\tt V}^{{\tt u}}+\frac{1}{2}\,B_{{\tt u}{\tt v}}\,{\tt V}^{{\tt u}}\wedge {\tt V}^{{\tt v}}\,,\label{Bpexp}
 \end{equation}
where $B^{(1)}_{{\tt u}}$ are six vector fields in the ${\bf 6}_{-1}$ representation of ${\rm GL}(6,\mathbb{R})$, and $B_{{\tt u}{\tt v}}$ are 15 axionic scalars in the ${\bf 15}_{+1}$ representation of the same group (the gradings now are $k_\Phi$ given in (\ref{kappaphi})). They are associated with shift isometries $t^{{\tt u}{\tt v}}$ of the scalar manifold spanning the nilpotent subspace ${\tt N}^{[{\tiny {\bf 15}_{+1}}]}$ of ${\Scr S}$. Under such transformations $B_{{\tt u}{\tt v}}$ transform as (\ref{deltaAsigma}):
\begin{equation}
\delta B_{{\tt u}{\tt v}}=\Sigma_{{\tt u}{\tt v}}\,.\label{Bisometry}
\end{equation}
This is just part of the transformations induced by the choice of the $\hat{B}^{(2)}$ gauge transformation parameter $\hat{\Xi}^{(1)}$ given in (\ref{globalsigma}): $d\hat{\Xi}^{(1)}=\frac{1}{2}\,\Sigma_{{\tt u}{\tt v}}\,dx^{{\tt u}}\wedge dx^{{\tt v}}$. We also have:
\begin{align}
\delta B^{(2)}&=\frac{\alpha^2}{2}\,\Sigma_{{\tt u}{\tt v}}\,G^{{\tt u}}\wedge G^{{\tt v}}\,,  \nonumber\\
\delta B^{(1)}_{{\tt u}}&=\alpha\,\Sigma_{{\tt u}{\tt v}}\, G^{{\tt v}}\,.\label{dualitygb}
\end{align}
Both $B^{(2)},\,B^{(1)}_{{\tt u}}$ are also associated with a local gauge invariance parametrized by $\Xi^{(1)}(x),\,\xi_{{\tt u}}(x)$. The second of Eqs. (\ref{dualitygb}) describes the global duality transformation on the NS-NS vector $B_{{\tt u}\,\mu}$ induced by the isometry (\ref{Bisometry}).
 The group ${\rm GL}(6,\mathbb{R})$, acting transitively on the internal metric moduli $g_{{\tt u}{\tt v}}$, together with the transformations generated by $\Sigma_{{\tt u}{\tt v}}\,t^{{\tt u}{\tt v}}$ in ${\tt N}^{[{\tiny {\bf 15}_{+1}}]}$ and extra symmetry transformations generated by $\tilde{\Sigma}^{{\tt u}{\tt v}}\,t_{{\tt u}{\tt v}}$,\footnote{The generators $t_{{\tt u}{\tt v}}$ have opposite weights with respect to $t^{{\tt u}{\tt v}}$, relative to the Cartan subalgebra of $\mathfrak{gl}(6,\mathbb{R})$.} close a $36+15+15=66$ dimensional group, which is the classical $T$-duality group ${\rm O}(6,6)$. This group acts transitively on $g_{{\tt u}{\tt v}},\,B_{{\tt u}{\tt v}}$ which span the symmetric manifold:
 \begin{equation}
 (g_{{\tt u}{\tt v}},\,B_{{\tt u}{\tt v}})\,\in\,\,\,\,\frac{{\rm O}(6,6)}{{\rm O}(6)\times {\rm O}(6)}\,.\label{o66man}
 \end{equation}
This space can also be written as follows:
  \begin{equation}
  \frac{{\rm O}(6,6)}{{\rm O}(6)\times {\rm O}(6)}\,\sim \,\frac{{\rm GL}(6,\mathbb{R})}{{\rm SO}(6)}\ltimes e^{{\tt N}^{[{\tiny {\bf 15}_{+1}}]}}\,,
  \end{equation}
  where $\sim$ stands, as usual, for an isometric mapping.
  This simply means that we can chose for it a coset representative of the form $L(g_{{\tt u}{\tt v}},\,B_{{\tt u}{\tt v}})=
  L(B_{{\tt u}{\tt v}})\,L(g_{{\tt u}{\tt v}})$, where the latter factor belongs to ${\rm GL}(6,\mathbb{R})/{\rm SO}(6)$ while the former is $L(B_{{\tt u}{\tt v}})=\exp(\frac{1}{2}\,B_{{\tt u}{\tt v}}\,t^{{\tt u}{\tt v}})\in e^{[{\tt N}^{{\tiny {\bf 15}_{+1}}}]}$. The action of ${\rm O}(6,6)$ on the moduli $g_{{\tt u}{\tt v}},\,B_{{\tt u}{\tt v}}$ is easily described by
  arranging these two matrices in a $12\times 12$  ${\rm O}(6,6)$-symmetric matrix $\mathcal{M}^{(g,B)}$ defined as follows:
  \begin{equation}
  \mathcal{M}^{(g,B)}[g,\,B]\equiv \left(\begin{matrix}g-B\,g^{-1}\,B & -B\,g^{-1}\cr g^{-1}\,B & g^{-1}\end{matrix}\right)\,,\label{Mgb}
  \end{equation}
  where $g\equiv (g_{{\tt u}{\tt v}}),\,B\equiv (B_{{\tt u}{\tt v}})$.
  We have \cite{Buscher:1987sk,Giveon:1994fu}:
  \begin{align}
  {\bf g}\in {\rm O}(6,6)&:\,(g,\,B)\rightarrow\,\,(g',\,B')\,\,;\,\,\,\,\,\mathcal{M}^{(g,B)}[g',\,B']={\bf g}^T\,\mathcal{M}^{(g,B)}[g,\,B]\,{\bf g}\,.\label{traM66}
  \end{align}
 Notice that $\mathcal{M}^{(g,B)}$ plays the role for the manifold (\ref{o66man}) that $\mathcal{M}$ in (\ref{M}) has for manifolds $G/H$ embedded in ${\rm Sp}(2n_v,\mathbb{R})/{\rm U}(n_v)$. The former is a pseudo-orthogonal matrix while the latter is a symplectic one and (\ref{traM66}) is analogous to (\ref{traM}). $T$-duality transformations along internal directions ($R_{{\tt u}_\ell}\rightarrow \alpha'/R_{{\tt u}_\ell}$) are implemented by acting on $\mathcal{M}^{(g,B)}$ by means of matrices of the form (\ref{outaut}).
 The vector fields $G^{{\tt v}}_\mu,\,B_{{\tt u}\,\mu}$ transform under ${\rm O}(6,6)$ in the fundamental ${\bf 12}$ representation.\par
  Let us now dimensionally reduce the field strength $\hat{H}^{(3)}\equiv d \hat{B}^{(2)}$ associated with the $B$-field.
  The expansion (\ref{Fpexpand}) and (\ref{Bpexp}) yield the following definitions of the four-dimensional field strengths:
  \begin{align}
  H^{(3)}&=dB^{(2)}-\alpha\,B^{(1)}_{{\tt u}}\wedge F^{{\tt u}}\,,\nonumber\\
    H^{(2)}_{{\tt u}}&=dB^{(1)}_{{\tt u}}-\alpha\,B_{{\tt u}{\tt v}}\wedge F^{{\tt v}}\,,\nonumber\\
     H^{(1)}_{{\tt u}{\tt v}}&= dB_{{\tt u}{\tt v}}\,.
  \end{align}
Let us now add a background flux $H_{{\tt u}{\tt v}{\tt w}}$ for $\hat{H}^{(3)}$ by adding to $\hat{B}^{(2)}$ a background contribution as in (\ref{ApAbg}). $\hat{H}^{(3)}$ gets shifted accordingly:
\begin{equation}
\hat{H}^{(3)}\rightarrow \hat{H}^{(3)}+\frac{1}{3!}\,H_{{\tt u}{\tt v}{\tt w}}\,dx^{{\tt u}}\wedge dx^{{\tt v}}\wedge dx^{{\tt w}}\,.
\end{equation}
The four-dimensional field strengths are then modified as follows:
  \begin{align}
  H^{(3)}&=dB^{(2)}-\alpha\,B^{(1)}_{{\tt u}}\wedge F^{{\tt u}}-\frac{\alpha^3}{3!}\,H_{{\tt u}{\tt v}{\tt w}}\,G^{{\tt u}}\wedge G^{{\tt v}}\wedge G^{{\tt w}}\,,\nonumber\\
    H^{(2)}_{{\tt u}}&=dB^{(1)}_{{\tt u}}-\alpha\,B_{{\tt u}{\tt v}}\wedge F^{{\tt v}}+\frac{\alpha^2}{2}\,H_{{\tt u}{\tt v}{\tt w}}\, G^{{\tt v}}\wedge G^{{\tt w}}\,\,,\nonumber\\
     H^{(1)}_{{\tt u}{\tt v}}&= dB_{{\tt u}{\tt v}}-\alpha\,H_{{\tt u}{\tt v}{\tt w}}\,G^{{\tt w}}\,,\nonumber\\
   F^{{\tt u}}&= dG^{{\tt u}}\,.  \label{Hdefs}
  \end{align}
The definitions of the field strengths encode the symmetries of the theory. In particular we see that, in the presence of the $H$-flux, $H^{(2)}_{{\tt u}}$ has become a non-Abelian field strength, while $ H^{(1)}_{{\tt u}{\tt v}}$ has the form of a covariant derivative. The expressions in (\ref{Hdefs}) are indeed invariant under the following local transformations:
\begin{align}
G^{{\tt u}}_\mu &\rightarrow\,\, G^{{\tt u}}_\mu+\partial_\mu\xi^{{\tt u}}\,,\nonumber\\
B_{{\tt u}{\tt v}} &\rightarrow\,\,B_{{\tt u}{\tt v}}+\alpha\,H_{{\tt u}{\tt v}{\tt w}}\,\xi^{{\tt w}}\,,\nonumber\\
B_{{\tt u}\,\mu}&\rightarrow\,\,B_{{\tt u}\,\mu}+\partial_\mu \xi_{{\tt u}}+\alpha^2\,H_{{\tt u}{\tt v}{\tt w}}\,G^{{\tt v}}_\mu\,\xi^{{\tt w}}\,,\nonumber\\
B^{(2)}&\rightarrow\,\,B^{(2)}+d\Xi^{(1)}+\alpha\,\xi_{{\tt u}}\,F^{{\tt u}}+\frac{\alpha^3}{2}\,H_{{\tt u}{\tt v}{\tt w}}\, \xi^{{\tt u}}\,G^{{\tt v}}\wedge G^{{\tt w}}\,.\label{thevariationsB}
\end{align}
From the structure of the vector field strengths we can infer the structure of the gauge algebra while from the covariant derivative of the scalar fields we infer the expression of the gauge generators in terms of isometries of the scalar manifold, through the embedding tensor. Here we are focusing on the NS-NS sector. The gauge generators have the form $X_{{\tt u}},\,X^{{\tt u}}$ and are gauged by $G^{{\tt u}}_\mu$ and $B_{{\tt u}\,\mu}$, respectively:
\begin{equation}
\Omega_{g\,\mu}=G^{{\tt u}}_\mu\,X_{{\tt u}}+B_{{\tt u}\,\mu}\,X^{{\tt u}}\,.
\end{equation}
From the second on (\ref{Hdefs}) we infer the commutation relation between the gauge generators:
\begin{equation}
[X_{{\tt u}},\,X_{{\tt v}}]=\alpha^2\,H_{{\tt u}{\tt v}{\tt w}}\,X^{{\tt w}}=-\alpha\,X_{{\tt u}{\tt v}{\tt w}}\,X^{{\tt w}}\,,
\end{equation}
all other commutators being zero. From the third of (\ref{Hdefs}), or from the second of (\ref{thevariationsB}), we can identify $X_{{\tt v}}$ with the following combination of isometries:
\begin{equation}
X_{{\tt u}}=\frac{1}{2}\,\Theta_{{\tt u},\,{\tt v}{\tt w}}\,t^{{\tt v}{\tt w}}=-\frac{1}{2}\,\alpha\,H_{{\tt u}{\tt v}{\tt w}}\,t^{{\tt v}{\tt w}}\,.
\end{equation}
Note that $X^{{\tt u}}$ are not identified with any isometries. They can be viewed as central charges with trivial action on the physical fields \cite{Angelantonj:2003rq}. The corresponding parameters only enter the gauge transformation of $B_{{\tt u}\,\mu}$. We have discussed this possible structure of the gauge algebra in Sect. \ref{backelectric}: $X^{{\tt u}}$ play the role of the generators $X_a$ in (\ref{gaualgcc}) when $h_{IJ}{}^a\neq 0$. If $\mathfrak{I}$ the normal Abelian subgroup of $G_g$ generated by $X^{{\tt u}}$, only $G_g/\mathfrak{I}$ is actually embedded in ${\rm O}(6,6)$:
 \begin{equation}
 {\rm Adj}(G_g)= {\rm Adj}(G_g/\mathfrak{I})\,\hookrightarrow\,\,{\rm Fund}({\rm O}(6,6))\,,
 \end{equation}
 where ${\rm Fund}({\rm O}(6,6))$ stands for the fundamental representation ${\bf 12}$ of ${\rm O}(6,6)$.
 This simple example illustrates how an internal flux induces a gauge symmetry in the lower-dimensional theory, the corresponding embedding tensor being identified with the flux itself.\par
Consider now a less simple example within the Type IIB theory, including the RR fields in the analysis. As mentioned in Sect. \ref{N8MAB}, this supergravity, at the classical level, features a global invariance defined by the group ${\rm SL}(2,\mathbb{R})_{{\rm IIB}}$. With respect to this symmetry, the two two-forms $\hat{B}^{(2)\,\sigma}=(\hat{C}^{(2)},\,\hat{B}^{(2)})$ transform in a doublet representation, labeled by  $\sigma=1,2$. The two scalars $\phi,\,\rho$ span the coset ${\rm SL}(2,\mathbb{R})_{{\rm IIB}}/{\rm SO}(2)$. The field strengths are defined as follows:
\begin{align}
\hat{\mathcal{F}}^{(3)\,\sigma}&= d\hat{B}^{(2)\,\sigma}\,\,;\,\,\,\,\hat{{F}}^{(3)\,\sigma}=L(\rho)^\sigma{}_\delta\, \hat{\mathcal{F}}^{(3)\,\delta}\,,\nonumber\\
\hat{{F}}^{(5)}&=d\hat{C}^{(4)}-\frac{1}{2}\,\epsilon_{\sigma\delta}\,\hat{B}^{(2)\,\sigma}\wedge d\hat{B}^{(2)\,\delta}={}^*\hat{{F}}^{(5)}\,,\label{Fsdefs}
\end{align}
$\hat{C}^{(4)}$ being an ${\rm SL}(2,\mathbb{R})_{{\rm IIB}}$-singlet and $$L(\rho)\equiv \left(\begin{matrix}1 & -\rho\cr 0& 1\end{matrix}\right)\,.$$
The dimensionally reduced bosonic fields arrange in representations of ${\rm SL}(2,\mathbb{R})_{{\rm IIB}}\times {\rm GL}(6,\mathbb{R})$ and are:
\begin{align}
g_{\hat{\mu}\hat{\nu}}&\rightarrow\,\,\,g_{\mu\nu},\,G^{{\tt u}}_\mu,\,\upPhi_{{\tt u}}{}^{\hat{\tt u}}\,,\nonumber\\
\hat{B}^{(2)\sigma}&\rightarrow\,\,\,B^\sigma_{\mu\nu},\,B^\sigma_{\mu\,{\tt u}},\,B^\sigma_{{\tt uv}}\,,\nonumber\\
\hat{C}^{(4)}&\rightarrow\,\,\,C_{\mu {\tt uvw}},\,C_{{\tt u}_1\dots {\tt u}_4}\,,\nonumber\\
\rho,\,\phi&\rightarrow\,\,\,\rho,\,\phi\,.
\end{align}
The scalar fields $B^{\sigma}_{{\tt u}{\tt v}}$ are associated with translational isometries $t_{\sigma}^{{\tt u}{\tt v}}$ in the ${\bf (2,15)}_{+1}$ of ${\rm SL}(2,\mathbb{R})_{{\rm IIB}}\times {\rm GL}(6,\mathbb{R})$. The 2-forms $B^\sigma_{\mu\nu}$
are dualized into scalars $\tilde{B}^\sigma=(\tilde{C},\,\tilde{B})$ in the ${\bf (2,1)}_{+3}$. Due to the self-duality of $\hat{F}^{(5)}$,  $C_{\mu\nu {\tt uv}}$ are not dynamically independent of the scalars $C_{{\tt u}_1\dots {\tt u}_4}$ in the ${\bf (1,15')}_{+2}$. The vectors $C_{\mu {\tt uvw}}$ belong to the ${\bf (1,20)}_{0}$ and consist of 10 electric and 10 magnetic fields. Finally the vectors $B^\sigma_{\mu\,{\tt u}}$ transform in the ${\bf (2,6)}_{-1}$.\par
Let us switch on NS-NS and RR 3-form fluxes by writing:
 \begin{equation}
 \hat{B}^{(2)\,\sigma}=B^{(2)\,\sigma}+B^{(1)\,\sigma}_{{\tt u}}\wedge {\tt V}^{{\tt u}}+\frac{1}{2}\,B^{\sigma}_{{\tt u}{\tt v}}\,{\tt V}^{{\tt u}}\wedge {\tt V}^{{\tt v}}+ \hat{B}^{(2)\,\sigma}_{bg}(x^{{\tt u}})\,,\label{BCpexp}
 \end{equation}
where:
\begin{equation}
d\hat{B}^{(2)\,\sigma}_{bg}=\frac{1}{3!}\,F^\sigma_{{\tt u}{\tt v}{\tt w}}\,dx^{{\tt u}}\wedge dx^{{\tt v}}\wedge dx^{{\tt w}}\,,
\end{equation}
$F^{\sigma=2}_{{\tt u}{\tt v}{\tt w}}$ being the $H$-flux considered above, while $F^{\sigma=1}_{{\tt u}{\tt v}{\tt w}}=F_{{\tt u}{\tt v}{\tt w}}$ is the RR flux. The resulting theory is a maximal $D=4$ supergravity containing forms of all order and featuring a flux-induced gauge group. In order to obtain a gauged model with a gauge group embedded inside ${\rm E}_{7(7)}$ we need to dualize all forms to lower-order ones. Having done this, the gauge connection has the following form:
\begin{equation}
\Omega_{g\,\mu}=G^{{\tt u}}_\mu\,X_{{\tt u}}+B^\sigma_{{\tt u}\,\mu}\,X_{\sigma}^{{\tt u}}+\frac{1}{3!}\,C_{{\tt uvw}\,\mu}\,X^{{\tt uvw}}\,,
\end{equation}
The gauge generators are expressed in terms of ${\rm E}_{7(7)}$ ones as follows:
\begin{equation}
X_{{\tt u}}\propto \alpha\,F^\sigma_{{\tt u}{\tt vw}}\,t_\sigma^{{\tt vw}}\,\,,\,\,\,\,
X^{{\tt u}}_\sigma\propto\epsilon_{\sigma\delta} F^\delta_{{\tt u}_1{\tt u}_2{\tt u}_3}\,t^{{\tt u}{\tt u}_1{\tt u}_2{\tt u}_3}\,\,,\,\,\,X^{{\tt uvw}}\propto\epsilon^{{\tt uvw}{\tt u}_1{\tt u}_2{\tt u}_3}\, F^\sigma_{{\tt u}_1{\tt u}_2{\tt u}_3}\,t_\sigma\,,
\end{equation}
where $t_\delta$ generate the shifts of $\tilde{B}^\delta$, in ${\tt N}^{[{\tiny {\bf (2,1)}_{+3}}]}$, and $t^{{\tt u}_1{\tt u}_2{\tt u}_3{\tt u}_4}$ of $C_{{\tt u}_1{\tt u}_2{\tt u}_3{\tt u}_4}$, in ${\tt N}^{[{\tiny {\bf (1,15')}_{+2}}]}$. We see that the embedding tensor is identified with the fluxes $F^\delta_{{\tt u}_1{\tt u}_2{\tt u}_3}$ and transforms in the ${\bf (2,20)}_{+3}$. The gauge algebra closes in $\mathfrak{e}_{7(7)}$ provided the quadratic constraints are satisfied. These can be readily worked out from (\ref{quadratic1}) to have the form
\begin{equation}
\epsilon^{{\tt v}_1{\tt v}_2{\tt v}_3{\tt u}_1{\tt u}_2{\tt u}_3}\epsilon_{\sigma\delta} F^\sigma_{{\tt v}_1{\tt v}_2{\tt v}_3}F^\delta_{{\tt u}_1{\tt u}_2{\tt u}_3}=0\,\,\Leftrightarrow\,\,\,\,\int_{T^6} \hat{F}^{(3)}_{bg}\wedge \hat{H}^{(3)}_{bg}=0\,.\label{qcondiF}
\end{equation}
As mentioned in Sect. \ref{n8fluxc}, this condition is nothing but that for the tadpole cancelation in the absence of localized sources. The gauge algebra reads:
\begin{equation}
[X_{{\tt u}},\,X_{{\tt v}}]\propto F^\sigma_{{\tt u}{\tt v}{\tt w}}\,X^{{\tt w}}_\sigma\,\,,\,\,\,\,[X_{{\tt u}},\,X^{{\tt w}}_\sigma]\propto \frac{1}{6}\,\epsilon_{\sigma\delta}\,F^\delta_{{\tt u}{\tt u}_1{\tt u}_2}\,X^{{\tt w}{\tt u}_1{\tt u}_2}\,,
\end{equation}
where we used the following relations among ${\rm E}_{7(7)}$-generators:
\begin{equation}
[t_\sigma^{{\tt u}_1{\tt u}_2},\,t_\delta^{{\tt u}_3{\tt u}_4}]=\epsilon_{\sigma\delta}\,t^{{\tt u}_1{\tt u}_2{\tt u}_3{\tt u}_4}\,\,;\,\,\,\,[t^{{\tt u}_1{\tt u}_2{\tt u}_3{\tt u}_4},\,t_\sigma^{{\tt u}_5{\tt u}_6}]=2\,\epsilon^{{\tt u}_1{\tt u}_2{\tt u}_3{\tt u}_4{\tt u}_5{\tt u}_6}\,t_\sigma\,.
\end{equation}
An alternative way to deduce condition (\ref{qcondiF}) is in the model which is directly obtained by dimensional reduction, prior to the dualizations. One finds the following gauge transformations for the fields originating from the 2-forms:
 \begin{align}
\delta G^{{\tt u}}_\mu &=\partial_\mu\xi^{{\tt u}}\,,\nonumber\\
\delta B^\sigma_{{\tt u}{\tt v}} &=\alpha\,F^\sigma_{{\tt u}{\tt v}{\tt w}}\,\xi^{{\tt w}}\,,\nonumber\\
\delta B^\sigma_{{\tt u}\,\mu}&= \partial_\mu \xi^\sigma_{{\tt u}}+\alpha^2\,F^\sigma_{{\tt u}{\tt v}{\tt w}}\,G^{{\tt v}}_\mu\,\xi^{{\tt w}}\,,\nonumber\\
\delta B^{(2)\,\sigma}&=d\Xi^{(1)\,\sigma}+\alpha\,\xi^\sigma_{{\tt u}}\,F^{{\tt u}}+\frac{\alpha^3}{2}\,F^\sigma_{{\tt u}{\tt v}{\tt w}}\, \xi^{{\tt u}}\,G^{{\tt v}}\wedge G^{{\tt w}}\,.\label{thevariationsB2}
\end{align}
It is straightforward to show that the field strengths of the axions $C_{{\tt u}_1\dots {\tt u}_4}$
  \begin{align}
      F^{(1)}_{{\tt u}_1\dots {\tt u}_4}&=dC_{{\tt u}_1\dots {\tt u}_4} -3\,\epsilon_{\sigma \delta}\,B^\sigma_{[{\tt u}_1{\tt u}_2} \mathcal{F}^{(1)\,\delta}_{{\tt u}_3{\tt u}_4]}-2\,\epsilon_{\sigma \delta}\,B^{(1)\,\sigma}_{[{\tt u}_1} F^{\delta}_{{\tt u}_2{\tt u}_3{\tt u}_4]}\,, \label{Fdefs}
  \end{align}
are invariant under the above transformations for a suitable transformation of $C_{{\tt u}_1\dots {\tt u}_4}$ provided  condition (\ref{qcondiF}) holds. Equivalently one can derive this  quadratic condition from the Bianchi identities on the field strengths.
\subsection{Duality Covariant Compactifications}\label{Tdualcomp}
As discussed in Sect. \ref{n8fluxc}, a minimal set of fluxes which contains $H_{{\tt uvw}}$ and is covariant
under $T$-duality must also contain the torsion $T_{{\tt uv}}{}^{{\rm w}}$ in the ${\bf (84+6)}_{+2}$, and the
non-geometric fluxes $Q_{{\tt u}}{}^{{\tt vw}},\,R^{{\tt uvw}}$ in the ${\bf (84'+6')}_{+1}$ and ${\bf 20}_{0}$, respectively. Consider, for the sake of simplicity, the NS-NS sector of the Type II theories. It consists of the dilaton $\phi$, the metric $\hat{g}_{{\tt uv}}$ and the B-field $\hat{B}^{(2)}$. Upon toroidal dimensional reduction to $D=4$, and dualization of all forms to lower-order ones, this sector is described by an $\mathcal{N}=4$ consistent truncation of the $\mathcal{N}=8$ theory.
The vector fields are:
\begin{align}
A^\Lambda_\mu=(G^{{\tt u}}_\mu,\,B_{{\tt u}\,\mu})\,,\nonumber
\end{align}
while the scalar fields consist in
\begin{equation}
\phi^s=(\phi_4,\,\tilde{B},\,g_{{\tt u}{\tt v}},\,B_{{\tt u}{\tt v}})\,,
\end{equation}
and span a scalar manifold of the form (\ref{subsl2so66}):
\begin{equation}
{\Scr M}_{{\rm scal}}^{\mathcal{N}=4}=\frac{{\rm SL}(2,\mathbb{R})}{{\rm SO}(2)}[\phi_4,\,\tilde{B}]\times \frac{{\rm SO}(6,\,6)}{{\rm SO}(6)\times {\rm SO}(6)}[g_{{\tt u}{\tt v}},\,B_{{\tt u}{\tt v}}]\subset\,\frac{{\rm E}_{7(7)}}{{\rm SU}(8)/\mathbb{Z}_2}\,,
\end{equation}
Note that $A^\Lambda_\mu$ transform in the ${\bf 12}$ of ${\rm SO}(6,\,6)$ while ${\rm SL}(2,\mathbb{R})$ has a non-perturbative duality action,  being ${\Scr R}_v={\bf (2,12)}$ with respect to ${\rm SL}(2,\mathbb{R})\times {\rm SO}(6,\,6)$. The presence of the minimal $T$-duality invariant set of fluxes $H_{{\tt uvw}},\,T_{{\tt uv}}{}^{{\rm w}},\,Q_{{\tt u}}{}^{{\tt vw}},\,R^{{\tt uvw}}$ induces a gauging in the $\mathcal{N}=4$ supergravity. The gauge algebra contains generators $X_{{\tt u}}$ gauged by the Kaluza-Klein vectors $G^{{\tt u}}_\mu$ as well as generators $X^{{\tt u}}$ gauged by  $B_{{\tt u}\mu}$.
These generators belong to $\mathfrak{so}(6,6)$ and are expressed in terms of $t_{{\tt u}}{}^{{\tt v}}$, generators of $\mathfrak{gl}(6,\mathbb{R})$, and the nilpotent generators $t^{{\tt u}{\tt v}}$,  $t_{{\tt u}{\tt v}}$ in the ${\bf 15}_{+1}$ and ${\bf 15}'_{-1}$, respectively, through an embedding tensor consisting of the $H,T,Q,R$-fluxes (we perform a suitable rescaling of the generators to get rid of the $\alpha$ factor and to simplify the commutation relations):
 \begin{align}
 X_{{\tt u}}&=-\frac{1}{2}\,Q_{{\tt u}}{}^{{\tt vw}}\,t_{{\tt vw}}+T_{{\tt uv}}{}^{{\tt w}}\,t_{{\tt w}}{}^{{\tt v}}-\frac{1}{2}\,H_{{\tt uvw}}\,t^{{\tt vw}}\,,\nonumber\\
 X^{{\tt u}}&=-\frac{1}{2}\,R^{{\tt uvw}}\,t_{{\tt vw}}-Q_{{\tt w}}{}^{{\tt uv}}t_{{\tt v}}{}^{{\tt w}}-\frac{1}{2}\,T_{{\tt vw}}{}^{{\tt u}}\,t^{{\tt vw}}\,.\label{XHTQR}
 \end{align}
 These generators can be grouped in an ${\rm O}(6,6)$-vector $X_\Lambda\equiv (X_{{\tt u}},\,X^{{\tt u}})$, so that Eqs. (\ref{XHTQR}) can be recast in the following compact form:
 \begin{equation}
 X_\Lambda=\frac{1}{2}\,X_{\Lambda\Sigma\Gamma}\,t^{\Sigma\Gamma}\,,\label{Xtso66}
 \end{equation}
 where $t^{\Sigma\Gamma}=-t^{\Gamma\Sigma}$ are the generators of ${\rm O}(6,6)$. These generators are chosen so as to satisfy the following commutation relations
 \begin{equation}
 [t^{\Lambda_1\Sigma_1},\,t^{\Lambda_2\Sigma_2}]=\eta^{\Sigma_1\Lambda_2}\,t^{\Lambda_1\Sigma_2}+\eta^{\Lambda_1\Sigma_2}\,t^{\Sigma_1\Lambda_2}-
 \eta^{\Lambda_1\Lambda_2}\,t^{\Sigma_1\Sigma_2}-\eta^{\Sigma_1\Sigma_2}\,t^{\Lambda_1\Lambda_2}\,,\label{tso66}
 \end{equation}
 where $\eta^{\Lambda\Sigma}$ is the ${\rm O}(6,6)$-invariant metric in the fundamental representation, whose non-vanishing entries are $\eta_{{\tt u}}{}^{{\tt v}}=\eta^{{\tt v}}{}_{{\tt u}}=\delta_{{\tt u}}^{{\tt v}}$. In the fundamental representation the generators have the following form: $(t^{\Lambda\Sigma})_\Gamma{}^\Delta=2\,\delta^{[\Lambda}_\Gamma\,\eta^{\Sigma]\Delta}$. \par
 Consistency of the corresponding gauging in the $\mathcal{N}=4$ model, see Sect.\ref{n4sugrag}, requires the following linear constraint on the components of the embedding tensor: $X_{\Lambda\Sigma\Gamma}=X_{[\Lambda\Sigma\Gamma]}$. The gauging is electric, involving only the vectors $G^{{\tt u}}_\mu,\,B_{{\tt u}\,\mu}$, and thus the locality constraint (\ref{quadratic1}) is satisfied. The only restriction comes from the closure constraint (\ref{quadratic1}), namely the condition that the gauge algebra closes inside $\mathfrak{so}(6,6)$:
 \begin{equation}
 [X_\Lambda,\,X_\Sigma]=-X_{\Lambda\Sigma}{}^\Gamma\,X_\Gamma=T_{\Lambda\Sigma}{}^\Gamma\,X_\Gamma\,,\label{closureXso66}
 \end{equation}
 where $X_{\Lambda\Sigma}{}^\Gamma=X_{\Lambda\Sigma\Pi}\eta^{\Pi\Gamma}$ and we have defined $T_{\Lambda\Sigma}{}^\Gamma=-X_{\Lambda\Sigma}{}^\Gamma$. Using Eqs. (\ref{Xtso66}) and (\ref{tso66}) the reader can easily verify that (\ref{closureXso66}) amounts to the following condition:
 \begin{equation}
 X_{[\Lambda\Sigma}{}^\Gamma\,X_{\Delta \Pi]\Gamma}=0\,.\label{Xquadso66c}
 \end{equation}
 In components Eq. (\ref{closureXso66}) can be written as follows \cite{Wecht:2007wu}:
\begin{align}
[X_{{\tt u}},\,X_{{\tt v}}]&=H_{{\tt uvw}}X^{{\tt w}}+T_{{\tt uv}}{}^{{\tt w}}\,X_{{\tt w}}\,,\nonumber\\
[X_{{\tt u}},\,X^{{\tt v}}]&=Q_{{\tt u}}{}^{{\tt vw}}X^{{\tt w}}-T_{{\tt uw}}{}^{{\tt v}}\,X^{{\tt w}}\,,\nonumber\\
[X^{{\tt u}},\,X^{{\tt v}}]&=R^{{\tt uvw}}X_{{\tt w}}+Q_{{\tt w}}{}^{{\tt uv}}\,X^{{\tt w}}\,.\label{Tdualgau}
\end{align}
Note that the gauge algebra and its structure are manifestly ${\rm O}(6,6)$-covariant.
We have thus grouped the $H,T,Q,R$-fluxes in a single ${\rm O}(6,6)$-covariant twist-tensor
\begin{equation}
T_{\Lambda\Sigma}{}^\Gamma=-X_{\Lambda\Sigma}{}^\Gamma=\{H_{{\tt uvw}},\,T_{{\tt uv}}{}^{{\rm w}},\,Q_{{\tt u}}{}^{{\tt vw}},\,R^{{\tt uvw}}\}\,,\label{Ttotal}
\end{equation}
so that \emph{the action of a T-duality on these fluxes is obtained by transforming this tensor through the corresponding ${\rm O}(6,6)$-transformation}.
The scalar potential has the form (we omit a positive proportionality factor)
   \begin{equation}
   V\propto\,-\left(2\,T_{\Lambda\Delta}{}^\Gamma T_{\Sigma\Gamma}{}^\Delta\,\mathcal{M}^{(g,B)\Lambda\Sigma}+
   T_{\Lambda\Sigma}{}^\Gamma T_{\Lambda'\Sigma'}{}^{\Gamma'}\,\mathcal{M}^{(g,B)}_{\Gamma\Gamma'} \mathcal{M}^{(g,B)\,\Lambda\Lambda'}\mathcal{M}^{(g,B)\,\Sigma\Sigma'}\right)\,,\label{VTwist2}
   \end{equation}
where the (negative definite) matrix $\mathcal{M}^{(g,B)}[g,\,B]_{\Lambda\Sigma}$ and its inverse $\mathcal{M}^{(g,B)}[g,\,B]^{\Lambda\Sigma}$ were defined in (\ref{Mgb}). Note that the potential in (\ref{VTwist2}) is a generalization of the one originating from a twisted torus reduction, given in (\ref{VTwist}), where $\mathcal{M}^{(g,B)}[g,\,B]$ plays the role of the internal metric and $T_{\Lambda\Sigma}{}^\Gamma$ of the torsion $T_{{\tt uv}}{}^{{\rm w}}$. Indeed if we choose as the only non-vanishing components of $T_{\Lambda\Sigma}{}^\Gamma$ the twist-tensor $T_{{\tt uv}}{}^{{\rm w}}$, the gauge algebra reduces to that of the group $G_T$ ($[X_{{\tt u}},\,X_{{\tt v}}]=T_{{\tt uv}}{}^{{\rm w}}\,X_{{\tt w}}$) and the above scalar potential to the one in (\ref{VTwist}). This formal analogy between the scalar potentials (\ref{VTwist}) and (\ref{VTwist2}) suggests the existence of some more general, T-duality covariant, geometric structure underlying this kind of compactifications. We shall elaborate on this issue below.

Certain compactifications in the presence of $H,T,Q,R$-fluxes are $T$-dual to known (geometric) string compactifications. Other combinations of these generalized fluxes, however, are not. The corresponding theories fit the definition of \emph{intrinsically (or truly) non-geometric} models given in the Introduction.
\par Compactifications in the presence of this $T$-duality invariant set of fluxes can be given a natural unified description in the context of \emph{generalized geometry} \cite{Hitchin:2004ut},\cite{Gualtieri:2007ng},\cite{Grana:2005jc}, where the tangent space $T$ to the internal manifold, parametrized by the vectors generating the diffeomorphisms, is doubled to include the cotangent one $T^*$ parametrized by the one-forms associated with gauge transformations of the $B$-field. One considers a single bundle $T\oplus T^*$ on $M_{int}$, on which generalized vielbein $E_\Lambda{}^{\underline{\Lambda}}$ and a corresponding generalized metric $\mathcal{M}^{(g,B)}[g,\,B]_{\Lambda\Sigma}=-E_\Lambda{}^{\underline{\Lambda}}E_\Sigma{}^{\underline{\Lambda}}$ of the form (\ref{Mgb}), are defined. They depend on both the metric and the $B$-field moduli. Different patches of the bundle are connected by transition functions in ${\rm O}(6,6)$. In this framework classical $T$-duality is \emph{geometrized}, that is it becomes part of the (generalized) geometry of the internal manifold as structure group.
The effective four-dimensional theory is obtained by defining a factorized ansatz for the generalized vielbein:
\begin{equation}
E_\Lambda{}^{\underline{\Gamma}}(x^{\hat{\mu}})=\mathring{E}_\Lambda{}^{\Sigma}(x^{{\tt u}})\hat{E}_\Sigma{}^{\underline{\Gamma}}(x^\mu)\,,\label{genSS}
\end{equation}
where $\mathring{E}_\Lambda{}^{\Sigma}(x^{{\tt u}})$ depends on the background moduli $\mathring{g}_{{{\tt uv}}},\,\mathring{B}_{{{\tt ub}}}$ and generalizes the twist matrix $\sigma_{{\tt u}}{}^{{\tt v}}$ encoding the geometry of the twisted torus, while $\hat{E}_\Sigma{}^{\underline{\Gamma}}(x^\mu)$ contains the $D=4$ scalars ${g}_{{{\tt uv}}},\,{B}_{{{\tt uv}}}$, fluctuations of the ten-dimensional fields about their background values. The $H,T,Q,R$-fluxes are encoded in the generalized  twist matrix
 $\mathring{E}_\Lambda{}^{\Sigma}(x^{{\tt u}})$ in the same way as $\sigma_{{\tt u}}{}^{{\tt v}}(x^{{\tt u}})$ contains the geometric flux $T_{{\tt uv}}{}^{{\tt w}}$, see Eq. (\ref{mctwist}).
Therefore, just as the geometric flux $T_{{\tt uv}}{}^{{\tt w}}$ is introduced by twisting the geometry of the torus $T^n$ into that of a twisted-torus ${\Scr T}^n$, in the context of generalized geometry, the generalized $H,T,Q,R$-fluxes are introduced by twisting the frame $E^\Lambda$ by means of a twist matrix $\mathring{E}$. These fluxes are part of a single ${\rm O}(6,6)$-covariant twist-tensor $T_{\Lambda\Sigma}{}^\Gamma$, see (\ref{Ttotal}).
\par
 A different approach to the definition of a unified description of $T$-dual flux-compactifications, is that of Double Field Theory (DFT) \cite{Siegel:1993xq,Siegel:1993th},\cite{Hull:2009mi,Hull:2009zb,Hohm:2010jy,Hohm:2010pp,Hull:2014mxa}, inspired by string field theory, see \cite{Aldazabal:2013sca} for a review. In this framework the space-time coordinates are doubled to include, besides the ordinary ones $x^{\hat{\mu}}$ dual to the momentum modes, new coordinates $x_{\hat{\mu}}$ dual to the winding modes. This produces the same doubling of the tangent space as in generalized geometry and the same generalized vielbein and metric $\mathcal{M}^{(g,B)}[g,\,B]_{\Lambda\Sigma}$ are introduced, this time however depending on the coordinates of the larger base manifold. Both approaches (generalized geometry and DFT) provide a $T$-duality covariant framework of dimensional reduction in which the classical $T$-duality group acts on the extended tangent bundle as a structure group. The gauge symmetry in four-dimensions is the remnant of the invariance of the DFT under generalized diffeomorphisms on the doubled coordinates which include ordinary space-time diffeomorphisms as well as gauge redefinitions of the $B$-field.
If a certain \emph{section condition} (or \emph{strong constraint}) on the coordinate dependence of the fields and gauge parameters is fulfilled, the whole construction is consistent: The generalized diffeomorphisms close and leave the DFT invariant.
If this condition is satisfied, moreover, locally a frame can be chosen in which the fields of the DFT only depend on the $x^{\hat{\mu}}$-coordinates. By choosing a suitable twist matrix of the form $\mathring{E}_\Lambda{}^{\Sigma}(x^{{\tt u}},\,x_{{\tt u}})$ a generalized Scherk-Schwarz ansatz can be devised which yields a class of gauged $\mathcal{N}=4$ four-dimensional supergravities, whose embedding tensor is described by the generalized twist tensor $T_{\Lambda\Sigma}{}^\Gamma$. The quadratic constraints of the resulting half-maximal theory follows from the closure of the generalized diffeomorphims. The strong constraint on the doubled-coordinate dependence in DFT, however, seems to pose further restrictions on the embedding tensor, which restricts the lower-dimensional $\mathcal{N}=4$ theory to models obtained as truncations of a gauged maximal ones. In order embed in DFT the most general half-maximal gauged model with embedding tensor $T_{\Lambda\Sigma}{}^\Gamma$, which comprises geometric and non-geometric fluxes, a relaxation of the strong constraint is thus called for. \par
Similar mathematical constructions have been devised in order to geometrize the whole classical global symmetry group $G={\rm E}_{11-D(11-D)}$ of $D$-dimensional maximal supergravity. This is done by extending the tangent space of the internal manifold in order to support the representation ${\Scr R}_v$ of $G$, which is thus  promoted to structure group.
This is the rough idea behind the \emph{extended generalized geometry} \cite{Hull:2007zu,Pacheco:2008ps}. A different construction is that of \emph{exceptional field theory} \cite{Hohm:2013pua,Hohm:2013uia,Hohm:2014qga}, in which the extension of the tangent space results from an enlargement of the whole space-time, now spanned by the $D$-dimensional space-time coordinates $x^\mu$ and by a new set of internal coordinates $\mathbb{Y}^M$ in the representation ${\Scr R}_{v}$ of $G$.
For example if $D=5$ $G={\rm E}_{6(6)}$, the vectors transform in the ${\Scr R}_{v}={\bf 27}'$ and thus there are 27 extra coordinates; if $D=4$, $G={\rm E}_{7(7)}$ and the extra coordinates are 56; if $D=3$ $G={\rm E}_{8(8)}$ and $\mathbb{Y}^M$ belong to the ${\bf 248}$, and so on (see Sect. \ref{vohd} for an overview of the maximal theories in various dimensions and in particular Table \ref{tabrepp} for the corresponding representations ${\Scr R}_{v}$ of the 1-forms).
 Just as the coordinates $x^\upalpha,\,x_\upalpha$ in DFT are dual to the momenta $p_\upalpha$ and winding modes $w^\upalpha$ along the corresponding internal directions, respectively, the generalized internal coordinates $\mathbb{Y}^M$ in exceptional field theory are naturally interpreted as dual to the 1-form charges $\Gamma^M$ of the $D$-dimensional theory \cite{Bossard:2015foa}(i.e. the four-dimensional electric and magnetic charges for $D=4$).\par
 Similarly to DFT, exceptional field theory is invariant under generalized diffeomorphisms on the extended set of coordinates which close provided \emph{section constraints} are satisfied. These conditions guarantee that locally all fields effectively only depend on the physical internal coordinates, which are $10-D$ ($x^{{\tt u}}$) if we are describing the Type II theories, or $11-D$ ($x^{\upalpha}$) if we are describing the eleven dimensional supergravity. Thus exceptional field theory provides a description of these higher-dimensional theories in which covariance with respect to the global symmetry group of a lower-dimensional maximal supergravity is manifest. This framework therefore unveils hidden symmetries of the eleven or ten-dimensional maximal theories, at the price of losing manifest Lorentz-invariance in the corresponding dimensions.\par
The section constraints, can be expressed as the following condition:
\begin{equation}
\left.\frac{\partial }{\partial \mathbb{Y}^M}\otimes \frac{\partial }{\partial \mathbb{Y}^N}\right\vert_{{\Scr R}'_2}=0\,,\label{scstr}
\end{equation}
where the derivatives on the left-hand-side either act on two different fields, or on a same one and ${\Scr R}'_2$ is the conjugate of the $G$-representation in which the 2-forms transform, see Table \ref{tabrepp}. Condition (\ref{scstr}) is the requirement that the product of the two gradients, in the representation ${\Scr R}_{v*}\otimes_s {\Scr R}_{v*}$, should have no component in the ${\Scr R}'_2$ representation. If $D=4$ we have ${\Scr R}'_2={\bf 133}$. In this case however there is an additional section constraint which involves the projection to the singlet representation through the matrix $\mathbb{C}$, so that the conditions  read:\footnote{Also in the formulation of the exceptional field theory with manifest $D=3$ global symmetry group ${\rm E}_{8(8)}$, the section constraints are stronger than (\ref{scstr}), since they involve the projections of $\partial_M\otimes \partial_N$ on the representations ${\bf 1}+{\bf 248}+{\bf 3875}$, while ${\Scr R}'_2$ is just ${\bf 1}+{\bf 3875}$.} \begin{equation}
t_\alpha{}^{MN}\partial_M\otimes \partial_N=0\,\,,\,\,\mathbb{C}^{MN}\partial_M\otimes \partial_N=0\,,\label{sconstr4}\end{equation} where $\partial_M$ stands for the partial derivative with respect to $\mathbb{Y}^M$. We can readily verify that eleven-dimensional supergravity, as well as the ten-dimensional Type II theories, satisfy the section constraints (\ref{sconstr4}). The fields of the former theory only depend on $x^\mu$ and on the seven coordinates $y^\upalpha$ of the internal torus. These latter can be identified with the components corresponding to ${\bf 7}'_{-3}$ in the branching of the ${\bf 56}$ representation of $\mathbb{Y}^M$ with respect to ${\rm GL}(7,\mathbb{R})$, see the first of Eqs. (\ref{eq:2M}). Restricting the dependence of the fields in the ${\rm E}_{7(7)}$-covariant exceptional field theory to $x^\mu$ and $y^\upalpha$ only, the section constraint is satisfied since the product of two gradients with respect to the internal coordinates would have a grading $+6$ which is absent in the branching of the ${\bf 133}$ representation, see the second of Eqs. (\ref{eq:2M}). The same holds for the projection to the singlet representation. A similar argument applies to the Type II theories. In Type IIA and IIB the six coordinates $x^{{\tt u}}$ of the torus are identified with ${\bf 6}'_{-2}$ in the branchings (\ref{eq:1IIA}) and (\ref{eq:1IIB}), respectively, of the ${\bf 56}$ with respect to ${\rm GL}(6,\mathbb{R})$. The product of two gradients with respect to such coordinates has grading $+4$ which is absent in the branchings (\ref{eq:2IIA}) and (\ref{eq:2IIB}). This implies that the restriction of this product to the ${\bf 133}$ is empty. Since the same is true for the projection to the singlet, the section constraint is satisfied.
\par
On the extended tangent space generalized internal vielbein and metric are defined. In the $D=4$ case the latter is nothing but the matrix $\mathcal{M}_{MN}$ of (\ref{M}), which depends on all the scalar fields  of the $D$-dimensional theory, now functions of the full set of $D+2 n_v=4+56=60$ coordinates. For generic $D$ the generalized internal metric $\mathcal{M}$ has the form $-\mathbb{L}\mathbb{L}^\dagger$, where $\mathbb{L}(\phi)$ is the coset representative in the representation ${\Scr R}_{v*}$.

 The effective $D$-dimensional gauged maximal theory, describing a consistent truncation of the ten or eleven-dimensional one on a given background, is obtained by writing a Scherk-Schwarz-like ansatz for the various fields, generalizing the one in (\ref{genSS}), in which the dependence on the internal coordinates of the background fields is encoded in a \emph{generalized twist matrix}, belonging to the group $G$, which, under certain conditions, factorizes out in the equations of motion, so as to yield the field equations of the $D$-dimensional theory in the $x^\mu$-dependent fluctuations.
Analogous constructions are defined in the context of extended generalized geometry where, however, only the tangent space is extended. In both cases the twist-tensor is the $G$-covariant tensor $X_{MN}{}^P$ of the gauged supergravity.\par
These frameworks have proven particularly useful in order to show that certain gauged $D$-dimensional maximal supergravities are consistent truncations of higher-dimensional ones on suitable backgrounds. However, not all gauged $D$-dimensional models could be embedded in this way within higher-dimensional ones.
The section constraint in exceptional field theory, for instance, restricts the possible gauged lower-dimensional supergravities which can be uplifted as consistent truncations of eleven or ten-dimensional ones. Examples of gauged  supergravities whose string or M-theory interpretation is as yet obscure are the ``$\omega$-rotated'' ${\rm SO}(p,q)$-model discussed in Sect. \ref{dyonicg}. To date, relaxing the section constraint so as to provide an ultra-violet completion of these models within superstring or M-theory is a challenging open problem. This would shed light on the CFTs dual to the broad class of new AdS vacua of these theories.

\paragraph{A different $\mathcal{N}=4$ truncation of the maximal theory.} As a final remark, let us point out that there is an $\mathcal{N}=4$ model which was mentioned in Sect. \ref{n8fluxc}, and which is different from the truncation considered at the beginning of this section. It originates from Type IIB theory on a $T^6/\mathbb{Z}_2$-orientifold \cite{Giddings:2001yu,Frey:2002hf,D'Auria:2002tc,D'Auria:2003jk}.
The vector fields are $A^\Lambda_\mu=(B^\sigma_{{\tt u}\mu})=(C_{{\tt u}\mu},\,B_{{\tt u}\mu})$ and transform in the ${\bf (2,6)}$ of ${\rm SL}(2,\mathbb{R})_{{\rm IIB}}\times {\rm SL}(6,\mathbb{R})$. The symplectic frame is different from the one of the model describing the NS-NS sector in that now the ${\rm SL}(2,\mathbb{R})_{{\rm IIB}}$-factor has a perturbative duality action and the electric global symmetry group $G_{el}$ contains ${\rm SL}(2,\mathbb{R})_{{\rm IIB}}\times {\rm GL}(6,\mathbb{R})$, corresponding to the $A$ and $D$-blocks in (\ref{ge}).
The scalar fields consist in $\phi,\,\rho$, parametrizing the first factor in (\ref{subsl2so66}), and  $g_{{\tt u}{\tt v}}$, $C_{{\tt u}_1{\tt u}_2{\tt u}_3{\tt u}_4}$, acted on by ${\rm SO}(6,\,6)$. The generators $t^{{\tt u}_1{\tt u}_2{\tt u}_3{\tt u}_4}$ of the Peccei-Quinn shift-symmetries on $C_{{\tt u}_1{\tt u}_2{\tt u}_3{\tt u}_4}$, which are still in $G_{el}$ and have an off-diagonal duality action, defining the $C$-block in (\ref{ge}).

%Let us focus only on the following field strengths/covariant derivatives
%  \begin{align}
%    \mathcal{F}^{(2)\,\sigma}_{{\tt u}}&=dB^{(1)\,\sigma}_{{\tt u}}-\alpha\,B^\sigma_{{\tt u}{\tt v}}\wedge F^{{\tt v}}+\frac{\alpha^2}{2}\,F^\sigma_{{\tt u}{\tt v}{\tt w}}\, G^{{\tt v}}\wedge G^{{\tt w}}\,\,,\nonumber\\
%       \mathcal{F}^{(1)\,\sigma}_{{\tt u}{\tt v}}&= dB^\sigma_{{\tt u}{\tt v}}-\alpha\,F^\sigma_{{\tt u}{\tt v}{\tt w}}\,G^{{\tt w}}\,,\nonumber\\
%      F^{(1)}_{{\tt u}_1\dots {\tt u}_4}&=dC_{{\tt u}_1\dots {\tt u}_4} -3\,\epsilon_{\sigma \delta}\,B^\sigma_{[{\tt u}_1{\tt u}_2} \mathcal{F}^{(1)\,\delta}_{{\tt u}_3{\tt u}_4]}-2\,\epsilon_{\sigma \delta}\,B^{(1)}\sigma_{[{\tt u}_1} F^{\delta}_{{\tt u}_2{\tt u}_3{\tt u}_4]}\,. \label{Fdefs}
%  \end{align}
%From which we can infer the structure of the gauge group induced by the fluxes. It contains generators $X_{{\tt u}},\,X^{{\tt u}}_\sigma$ gauged by $G^{{\tt u}}_\mu$ and $B^\sigma_{{\tt u}\,\mu}$, respectively.
%where
%\begin{align}
%X_{{\tt u}}=\alpha\,F^\sigma_{{\tt u}{\tt vw}}\,t_\sigma^{{\tt vw}}\,\,,\,\,\,\,
%X^{{\tt u}}_\sigma=2\,\epsilon_{\sigma\delta} F^\delta_{{\tt u}_1{\tt u}_2{\tt u}_3}\,t^{{\tt u}{\tt u}_1{\tt u}_2{\tt u}_3}\,.
%\end{align}
%These generators close an algebra

\subsection{Type IIB on $K3\times T_2/\mathbb{Z}_2$ with Fluxes and Branes}\label{K3T2Z2}

In this Section we review the gauged supergravity description of the Type IIB theory compactified on a
$K3\times T_2/\mathbb{Z}_2$-orientifold  in the presence of fluxes \cite{Tripathy:2002qw,Andrianopoli:2003jf} and space-filling $D3$ and $D7$-branes \cite{Angelantonj:2003zx}.\par
Let us define the ungauged four-dimensional theory originating from the compactification in the absence of fluxes. The manifold $K3$ is a compact K\"ahler manifold of complex dimension 2 and vanishing first Chern class (i.e. a Calabi-Yau 2-fold), see \cite{Aspinwall:1996mn} for an excellent review of its geometry.
 Let us use, only in the present section, for the space-time and internal coordinates the following notation:
 \begin{align}
 M_{D=10}[x^{\hat{\mu}}]&=\,M_{D=4}[x^\mu]\times K3[w^{{\tt s}},\,\bar{w}^{\bar{{\tt s}}}]\times T^2[x^p]\,,\nonumber\\
 \hat{\mu}&=0,\dots, 9\,\,;\,\,\,\,\,{\tt s},\,\bar{{\tt s}}=1,2\,\,;\,\,\,\,p=8,9\,,
 \end{align}
 where $w^{{\tt s}}$ denote the two complex $K3$-coordinates and the 2-torus is chosen along the directions $8$ and $9$.\par
 The compactification of Type IIB superstring on a $K3\times T_2/\mathbb{Z}_2$-orientifold is effected by truncating the zero-modes of the superstring theory on $K3\times T_2$ to the invariant-sector with respect to the action of the orientifold group $\mathbb{Z}_2=\{{\rm Id},\,\upOmega\,I_2\,(-)^{F_L}\}$, where $\upOmega$ is the worldsheet parity, $I_2$ is parity along the $T^2$ directions and $F_L$ is the fermion number in the string left-moving sector. The resulting theory is an ungauged $\mathcal{N}=2$ theory.
The $\mathbb{Z}_2$-even zero-modes described by the four-dimensional theory are:
\begin{align}
\underline{{\rm scalars}}&:\,\,\,\phi,\,\rho,\,g_{{\tt s}{\tt t}},\,g_{\overline{{\tt s}{\tt t}}},\,g_{{\tt s}\bar{{\tt t}}},\,g_{pq}\,,C_{{\tt s}\bar{{\tt t}}pq},\,C_{\overline{{\tt s}{\tt t}}pq},\,C_{{{\tt s}{\tt t}}pq},\,C_{{\tt s}_1{\tt s}_2\bar{{\tt t}}_1\bar{{\tt t}}_2}\,,\nonumber\\
\underline{{\rm vectors}}&:\,\,\,\{A^\Lambda_\mu\}\equiv\{B^\sigma_{p\,\mu}\}=\{C_{p\,\mu},\,B_{p\,\mu}\}\,,
\end{align}
where we have defined the composite index $\Lambda\equiv(\sigma,p)$, $\sigma=1,2$, $p=1,2$ and the internal metric is taken in the ten-dimensional Einstein frame. Let us count the number of the above fields and their supermultiplet arrangement.
The metric muduli of $T^2$, $g_{pq}$, are three and can be described as follows:
\begin{equation}
g_{pq}\,\,:\,\,\,\,\begin{cases}{\rm volume:}\,\,\,\,\,\,e^\varphi\equiv{\rm Vol}(T^2)\equiv \sqrt{{\rm det}(g_{pq})}\cr
\mbox{ complex structure modulus:}\,\,\,\,\,\,t=\frac{g_{89}-i\,e^\varphi}{g_{88}}\,, \end{cases}\label{tdef}
\end{equation}
where we have described the volume of the 2-torus by dilatonic scalar $\varphi$ (not to be confused with the ten-dimensional dilaton $\phi$).
Recalling our general discussion on Calabi-Yau manifolds, $g_{{\tt s}{\tt t}},\,g_{\overline{{\tt s}{\tt t}}}$ are the complex structure moduli of $K3$, while $g_{{\tt s}\bar{{\tt t}}}$ parametrize the possible choices of the K\"ahler form. The former are related to the periods of the holomorphic $(2,0)$ form $\boldsymbol{\Omega}$. The Hodge-numbers of $K3$ are $h_{0,0}=h_{2,2}=1$, $h_{1,0}=h_{0,1}=h_{2,1}=h_{1,2}=0$, $h_{2,0}=h_{0,2}=1$ and $h_{1,1}=20$. On $H^{2}(K3,\mathbb{Z})$  one defines the following symmetric inner product $(\boldsymbol{\alpha},\,\boldsymbol{\beta})\equiv \int_{K3} \boldsymbol{\alpha}\wedge\boldsymbol{\beta}=(\boldsymbol{\beta},\,\boldsymbol{\alpha})$. A basis $\boldsymbol{\omega}_I$, where $I,J=1,\dots,\,22$ (only in this Section), can be chosen so that  $(\boldsymbol{\omega}_I,\,\boldsymbol{\omega}_J)$ is diagonal of the form:
\begin{align}
\eta_{IJ}=(\boldsymbol{\omega}_I,\,\boldsymbol{\omega}_J)={\rm diag}(+1,+1,+1,-1,\dots,\,-1)\,,
\end{align}
with signature $(3,19)$: the positive and negative eigenvalues correspond to self-dual and anti-self-dual 2-forms, respectively. We can label the former by $x=1,2,3$ and the latter by $a=1,\dots 19$. By virtue of this inner product we can then write the following isomorphism:
\begin{equation}
H^{2}(K3,\mathbb{Z})\,\sim\,\,\Gamma^{3,\,19}\,,
\end{equation}
$\Gamma^{3,\,19}$ denoting a lattice with signature $(3,19)$.
Thus each two form in $H^{2}(K3,\mathbb{Z})$ can be represented by an integer vector in $\Gamma^{3,\,19}$.
The complex structure and the  K\"ahler moduli (except the volume) span a manifold of the form:
\begin{equation}
\mbox{complex structures}\,+\,\mbox{K\"ahler moduli (except the volume) }\,\,\in \,\,\,\,\frac{{\rm SO}(3,19)}{{\rm SO}(3)\times {\rm SO}(19)}\,,\label{so319}
\end{equation}
We can parametrize this space in terms of a $3\times 19$ matrix ${\bf e}=(e^x{}_a)$, $x=1,2,3$, $a=1,\dots,19$ by writing the coset representative in the form of the following $22\times 22$ matrix\footnote{The two indices of $e^x{}_a$ will, in the sequel, be raised and lowered by deltas, so their upper or lower position is not relevant.}
\begin{equation}
L({\bf e})=(L({\bf e})^I{}_J)\equiv \left(\begin{matrix}({\bf 1}+{\bf e}{\bf e}^T)^{\frac{1}{2}} & -{\bf e}\cr -{\bf e}^T & ({\bf 1}+{\bf e}^T{\bf e})^{\frac{1}{2}}\end{matrix}\right)\,.
\end{equation}
The reduction of the RR four-form $\hat{C}^{(4)}$ yields:
\begin{equation}
\hat{C}^{(4)}=C^I\,\boldsymbol{\omega}_I\wedge dx^8\wedge dx^9+C_{(K3)}\,\boldsymbol{\omega}_{K2}\,,
\end{equation}
where $\boldsymbol{\omega}_{K2}$ is the $(2,2)$-volume form of $K3$. The moduli $C^I$ correspond to $C_{{\tt s}\bar{{\tt t}}pq},\,C_{\overline{{\tt s}{\tt t}}pq},\,C_{{{\tt s}{\tt t}}pq}$, while $C_{(K3)}$ to $C_{{\tt s}_1{\tt s}_2\bar{{\tt t}}_1\bar{{\tt t}}_2}$.
Inspection of the sigma-model in four-dimensions shows that the 22 axions $C^I$ couple to the $K3$ metric moduli $e^x{}_a$ and to the volume of the torus ${\rm Vol}(T^2)$. These $80$ scalars belong to 20 hypermultiplets and span a quaternionic manifold of the form:
\begin{equation}
{\Scr M}_{QK}=\frac{{\rm SO}(4,20)}{{\rm SO}(4)\times {\rm SO}(20)}=\left[{\rm SO}(1,1)\times \frac{{\rm SO}(3,19)}{{\rm SO}(3)\times {\rm SO}(19)}\right]\ltimes e^{{\tt N}^{{\tiny [{\bf 22}_{+}]}}}\,,
\end{equation}
where ${\tt N}^{{\tiny [{\bf 22}_{+}]}}$ is an Abelian nilpotent space spanned by the 22 RR axions $C^I$, its generators being denoted by $t_I$, and ${\rm SO}(1,1)$ is parametrized by the dilatonic scalar $\varphi$ describing the volume of the 2-torus. The fields $C^I,\,{\rm Vol}(T^2)$ and $e^x{}_a$, consistently with them being hyper-scalars, do not couple the vectors. They are also not coupled to the dilaton $\phi$, the RR scalar $\rho$, the $K3$-volume, $C_{(K3)}$ and the complex structure of the torus. The latter scalars define three complex coordinates $u,s,t$ spanning, in the absence of branes, the special K\"ahler part of the scalar manifold:
\begin{align}
{\Scr M}_{SK}&=\left(\frac{{\rm SL}(2,\mathbb{R})}{{\rm SO}(2)}\right)_s\times \left(\frac{{\rm SL}(2,\mathbb{R})}{{\rm SO}(2)}\right)_t\times\left(\frac{{\rm SL}(2,\mathbb{R})}{{\rm SO}(2)}\right)_u\,,\nonumber\\
s&\equiv C_{(K3)}-i\,{\rm Vol}(K3)\,,\nonumber\\
u&\equiv \rho-i\,e^{-\phi}\,,
\end{align}
 $t$ being defined in (\ref{tdef}). The isometry group ${\rm SL}(2,\mathbb{R})_u$ of the last factor is the (classical) ten-dimensional Type IIB duality group ${\rm SL}(2,\mathbb{R})_{{\rm IIB}}$, acting on the axio-dilaton modulus $u$. The four vector fields $\{A^\Lambda_\mu\}\equiv\{B^\sigma_{p\,\mu}\}$ transform in the ${\bf (2,2)}$ of ${\rm SL}(2,\mathbb{R})_u\times {\rm SL}(2,\mathbb{R})_t$. This group has therefore a block-diagonal duality action of the vector fields. This is not the case for the remaining factor  ${\rm SL}(2,\mathbb{R})_s$ in $G^{(SK)}$, whose action on $B^\sigma_{p\,\mu}$ is non-perturbative. To show this we note that the vectors are coupled to the same power of ${\rm Vol}(K3)$, which is the dilatonic scalar defining the imaginary part of $s$:
 \begin{equation}
 -e\,e^{-\phi}\,{\rm Vol}(K3)\,\hat{g}^{pq}(\partial_{[\mu}B_{\nu]\,p})\partial^{[\mu}B^{\nu]}{}_{q}\,\,;\,\,\,-e\,e^{\phi}\,{\rm Vol}(K3)\,\hat{g}^{pq}(\partial_{[\mu}C_{\nu]\,p})\partial^{[\mu}C^{\nu]}{}_{q}\,,
 \end{equation}
 where $\hat{g}_{pq}$ is the torus metric normalized so that ${\rm det}(\hat{g}^{pq})=1$, and thus depending only on $t$. The above feature implies that all vectors have the \emph{same grading} with respect to the $K3$-volume dilation: ${\rm Vol}(K3)\rightarrow e^{4\lambda}\,{\rm Vol}(K3)$. This transformation is generated by the Cartan subalgebra of ${\rm SL}(2,\mathbb{R})_s$ and thus the effect of the non-Cartan generators of the same group is to transform electric field strengths into magnetic ones and vice-versa, i.e. ${\rm SL}(2,\mathbb{R})_s$ has a non-perturbative duality action. Indeed the real part of $s$ enters the matrix $\mathcal{R}_{\Lambda\Sigma}$ and couples to the vectors as follows:
 \begin{equation}
 \epsilon^{\mu\nu\rho\sigma}\,\epsilon^{pq}\,C_{(K3)}\,\partial_{\mu}B_{\nu\,p}\,\partial_{\rho}C_{\sigma\,q}\,,
 \end{equation}
 so that the corresponding shift symmetry $C_{(K3)}\rightarrow\,C_{(K3)}+\xi$ has indeed a non-perturbative duality action.\par
 Let us now consider stacks of $n_3$ space-filling $D3$-branes and $n_7$ space-filling $D7$-branes wrapped around $K3$, see \cite{Blumenhagen:2006ci} for an in-depth discussion of this microscopic setting. The
low--energy brane dynamics is described by a SYM theory on their
world volume. We shall consider the super Yang-Mills theories on the D3/D7
branes to be in the Coulomb phase (namely the branes to be
separated from each other), so that the gauge group and the
massless bosonic modes on the world volume theories are:
\begin{eqnarray}
\mbox{D3:}&&\mbox{gauge group = }\,{\rm
U}(1)^{n_3}\,\,;\,\,\,\mbox{bosonic 0--modes:
}\,A^r_\mu\,\,\,y^r=y^{8,r}+t\,y^{9,r}\,\,(r=1,\dots, n_3)\,,\nonumber\\
\mbox{D7:}&&\mbox{gauge group = }\,{\rm
U}(1)^{n_7}\,\,;\,\,\,\mbox{bosonic 0--modes:
}\,A^k_\mu\,\,\,x^k=x^{8,k}+t\,x^{9,k}\,\,(k=1,\dots,
n_7)\,,\nonumber
\end{eqnarray}
where $y^r$ and $x^k$ are complex scalars describing the position
of each D3, D7-brane along $T^2$ respectively.
Now the vector fields are $4+n_3+n_7$:
\begin{equation}
\{A^\Lambda_\mu\}=\{B^\sigma_{p\,\mu},\,A^k_\mu,\,A^r_\mu\}
\end{equation}
 The four-dimensional effective action
receives now the contributions from the actions $S_{Dp}$ on the D3 and D7 world volumes:\footnote{The tension $T_p$ and charge $\mu_p$ read: $T_p=\mu_p=(2\pi)^{-p}\,(\alpha')^{-(p+1)/2}$.}
\begin{equation}
S_{Dp}=-T_p\,\int_{\Sigma_{p+1}} d^{p+1}\sigma\,e^{-\phi}\,\sqrt{{\rm det}\left(\hat{g}^{(S)}+\hat{B}^{(2)}+2\pi\alpha' {F}\right)}+\mu_p\,\int \sum_\ell\,\hat{C}^{(p+1-\ell)}\wedge e^{(\hat{B}^{(2)}+2\pi\alpha' {F})}\,,
\end{equation}
where $F$ are the field strengths of the world volume vectors, $\hat{g}^{(S)}$ is the ten-dimensional metric in the string frame, and the argument of the determinant is pulled back on the world-volume $\Sigma_{p+1}$, where the two integrals are computed. Expanding the above action for $p=3$ and $p=7$ to lowest order in $\alpha'$ we can infer general information about the low-energy four-dimensional theory. For instance the kinetic term of the $Dp$ brane-vectors has the form:
\begin{equation}
-\frac{e}{4\,g_{YM}^2}\,F_{\mu\nu}\,F^{\mu\nu}\,\,;\,\,\,g_{YM}^{-2}=\begin{cases}T_3\, (2\pi\alpha')^2\,e^{-\phi} \,\,;\,\,\,p=3\cr T_7\, (2\pi\alpha')^2\,{\rm Vol}(K3)\,\,;\,\,\,p=7\,.\end{cases}
\end{equation}
As far as the generalized $\theta$-terms are concerned, for the two kinds of branes the Chern-Simons terms yield:
\begin{align}
p=3:&\,\,\,\,\mu_3\,\rho\,\int_{M_{D=4}} F\wedge F\nonumber\\
p=7:&\,\,\,\,\mu_7\,C_{(K3)}\,\int_{M_{D=4}} F\wedge F\,.
\end{align}
From the above terms we can infer, by means of the same arguments used earlier for the bulk fields $B^\sigma_{\mu\,p}$, that ${\rm SL}(2,\mathbb{R})_t$ has a trivial action on the world-volume vectors $A^k_\mu,\,A^r_\mu$,
${\rm SL}(2,\mathbb{R})_s$ has a trivial action on the $D3$-brane vectors and non-perturbative on the $D7$-brane ones, while ${\rm SL}(2,\mathbb{R})_u$ has a trivial action on the $D7$-brane vectors and non-perturbative on the $D3$-brane ones. From the kind of duality action (perturbative or non-perturbative) of the isometry group $G^{(SK)}$ we can infer the symplectic frame of the four-dimensional effective supergravity. The situation is summarized on Table \ref{tabpertnp}.
\begin{table}
\begin{center}
\begin{tabular}{|c|c|c|c|}
  \hline
    & $B^\sigma_{\mu\,p}$ & $A^r_\mu$ & $A^k_\mu$ \\
    \hline
 ${\rm SL}(2,\mathbb{R})_t$ & pert. & trivial & trivial \\\hline
  ${\rm SL}(2,\mathbb{R})_u$ & pert. & non-pert. & trivial \\\hline
  ${\rm SL}(2,\mathbb{R})_s$ & non-pert. & trivial & non-pert. \\
  \hline
\end{tabular}
  \caption{Duality action of ${\rm SL}(2,\mathbb{R})^3$-group.}\label{tabpertnp}
\end{center}
\end{table}
The spinors of the theory are
the gravitini $\psi^A_\mu$, the gaugini $\lambda^{i\,A}$
($i=1,\dots, n_v$) and the hyperini $\zeta^\alpha$.
Since in our model
\begin{equation}
H^{(QK)}={\rm SO}(4)\times {\rm SO}(20)={\rm SU}(2)\times H_{{\rm matt}}^{(QK)}\,,
\end{equation}
where $H_{{\rm matt}}^{(QK)}={\rm SU}(2)\times {\rm SO}(20)$, we can split the symplectic index $\alpha$ of this group as follows $\alpha=(A',{\tt A})$ where $A'=1,2$ is the doublet index of the ${\rm SU}(2)$ factor and ${\tt A}=1,\,\dots,\,20$ labels the hypermultiplets.\par The embedding of the fundamental representation of ${\rm SO}(3,19)$, acting on the $K3$ metric moduli, into that of ${\rm SO}(4,20)$, acting on all the hyperscalars, is defined by splitting the index of the latter as:  $m=0,1,2,3=(0,x)$, defining the positive-signature directions, and ${\tt A}=1,\dots,20=(a,20)$ defining the negative-signature ones, $x,a$ labeling the fundamental representation of ${\rm SO}(3,19)$. The structure of the hypermultiplets can be summarized as follows:
\begin{equation}
\left[\begin{matrix}\zeta^{A',a}\cr C^a,\,e^x{}_a\end{matrix}\right]\,,\,\,\left[\begin{matrix}\zeta^{A',20}\cr C^x,\,\varphi\end{matrix}\right]\,.
\end{equation}

\subsubsection{Special K\"ahler Geometry} In the presence of $D3$ and $D7$ branes, the spacial K\"ahler manifold is spanned by the $n=3+n_3+n_7$ complex scalars: $z^i=(s,t,u,x^k,y^r)$. In the special coordinate frame the geometry of this manifold is totally defined by a cubic holomorphic prepotential ${\Scr F}(z)$, see Eq. (\ref{prepotentialF}), of the following form \cite{Angelantonj:2003zx}:
 \begin{equation}
 \label{prepot}
 {\Scr F}(s,t,u,x^k,y^r)\,=\, stu-\frac{1}{2}\,s \,x^k
    x^k-\frac{1}{2}\,u\,
    y^r y^r\,,
\end{equation}
which correctly reproduces the scalar-scalar interactions in the sigma-model kinetic term.
The components of the holomorhic symplectic section $\Omega_{(sc)}^M(z)=(X_{(sc)}^\Lambda,\,F_{(sc)\,\Lambda})$ are then deduced from the prepotential in the special coordinate frame as explained in Sect. \ref{SK}: In a patch in which $X_{(sc)}^0=1$, $X_{(sc)}^\Lambda(z)=(1,z^i)$ and $F_{(sc)\,\Lambda}(z)$ is given by (\ref{prepotentialF}).
The
K\"ahler potential $\mathcal{K}$ is evaluated using (\ref{Komapp}), or by (\ref{Kprepot}), and reads:
\begin{eqnarray}
K &=& -\log [-8\,({\rm Im}(s)\,{\rm Im}(t){\rm
Im}(u)-\frac{1}{2}\,{\rm Im}(s)\,({\rm
Im}(x)^k\,)^2-\frac{1}{2}\,{\rm Im}(u)\,({\rm
Im}(y)^r\,)^2)] \,.
\end{eqnarray}
The geometry is of cubic type, see Eq. (\ref{cubicF}) and below. In the absence of D-branes ($n_3=n_7=0$), the special K\"ahler manifold would just be that of the STU model discussed in Sect. \ref{STUgeometry0}. If only one kind of branes is present, the manifold would be symmetric of the form:
\begin{align}
&\left(\frac{{\rm SL}(2,\mathbb{R})}{{\rm SO}(2)}\right)_s\times \frac{{\rm SO}(2,2+n_7)}{{\rm SO}(2)\times {\rm SO}(2+n_7)}\,\,\,,\,\,\,\,n_3=0\,,\nonumber\\
&\left(\frac{{\rm SL}(2,\mathbb{R})}{{\rm SO}(2)}\right)_u\times \frac{{\rm SO}(2,2+n_3)}{{\rm SO}(2)\times {\rm SO}(2+n_3)}\,\,\,,\,\,\,\,n_7=0\,.
\end{align}
If however both branes are present, the manifold is homogeneous non-symmetric of the kind $L(0,n_7,n_3)$ in the classification of \cite{deWit:1992wf}. The corresponding sigma-model metric was constructed, using the solvable Lie algebra parametrization, in \cite{D'Auria:2004cu}.\par
The non-minimal scalar-vector couplings are however not correctly reproduced in the special-coordinate symplectic frame, because the isometries do not have the right duality action summarized in Table \ref{tabpertnp}. The components
$X^\Lambda,\,F_\Sigma$ of the holomorphic symplectic section $\Omega^M(z)$ which
correctly describe our problem, are chosen by performing a
constant symplectic transformation on $\Omega_{(sc)}^M(z)$, so that they read:
\begin{eqnarray}
X^0 &=& \frac{1}{{\sqrt{2}}}\,(1 - t\,u +
\frac{(x^k)^2}{2})\,\,\,\,,\,\,\,\,\, X^1 = -\frac{t +
u}{{\sqrt{2}}}\,,
\nonumber \\
X^2 &=&  -\frac{1}{{\sqrt{2}}}\,({1 + t\,u -
\frac{(x^k)^2}{2}})\,\,\,\,,\,\,\,\,\,X^3 = \frac{t -
u}{{\sqrt{2}}}\,,
\nonumber \\
X^k &=& x^k\,\,\,\,,\,\,\,\,\,X^r = y^r\,,
\nonumber\\
F_0 &=& \frac{s\,\left( 2 - 2\,t\,u + (x^k)^2 \right) +
u\,(y^r)^2}{2\,{\sqrt{2}}}\,\,\,\,,\,\,\,\,\, F_1 =
  \frac{-2\,s\,\left( t + u \right)  +
  (y^r)^2}{2\,{\sqrt{2}}}\nonumber\\
F_2&=&
  \frac{s\,\left( 2 + 2\,t\,u - (x^k)^2 \right)  -
  u\, (y^r)^2}{2\,{\sqrt{2}}}\,\,\,\,,\,\,\,\,\,
 F_3 = \frac{2\,s\,\left( -t + u \right)  +
 (y^r)^2}{{2\,\sqrt{2}}}\nonumber\\
F_k &=& -
  s\,x^k
 \,\,\,\,,\,\,\,\,\,
F_r = -u\,y^r\, .
\end{eqnarray}
Notice that the scalar $s$ only appears in $F_\Lambda$ but not in $X^\Lambda$. Therefore we cannot write $F_\Lambda=F_\Lambda(X)$ and the new symplectic frame admits no prepotential function $F(X)$.
\subsubsection{Introducing Fluxes} Let us consider the effect of switching on fluxes of the
three--form field strengths across cycles of the internal
manifold. The only components of $\hat{F}_{bg}^{(3)\,\sigma}=d\hat{B}_{bg}^{(2)\,\sigma}$
which survive the orientifold projection are: $\hat{F}_{bg}^{(3)\,\sigma}=
F^{\sigma\, I}{}_{p}\, \boldsymbol{\omega}_I\wedge dx^p$. We can describe these
flux components in terms of four integer vectors $f_\Lambda{}^I$,
$\Lambda=0,\dots, 3$ :
\begin{eqnarray}
F^{\sigma\, I}{}_{p}&\equiv &F_\Lambda{}^I=\frac{4\,\pi^2}{R^3}\,
\alpha^\prime\, f_\Lambda{}^I\,\,\,;\,\,\,\,
f_\Lambda{}^I\,\in\,\Gamma^{3,19}\,,
\end{eqnarray}
where  $R$ is the linear size of the internal manifold and last
property follows from the flux quantization condition \cite{Grana:2005jc}
\begin{equation}
\frac{1}{(2\pi\,\sqrt{\alpha'})^{p-1}}\int\hat{F}^{(p)}_{bg}\,\in \,\mathbb{Z}\,.
\end{equation}
\par Just as in the simpler case of toroidal compactifications discussed in Sect. \ref{toroidalc}, the
presence of fluxes implies local invariance in the low--energy
supergravity. A way to see this is to consider the dimensional
reduction of the kinetic term for $C_{(4)}$:
\begin{eqnarray}
D=10&\rightarrow & D=4 \nonumber\\F_{(5)}\wedge
{}^*F_{(5)}&\longrightarrow & (\partial C^I-f_{\Lambda}{}^I\,
A^\Lambda_\mu)^2\,,
\end{eqnarray}
where de have redefined $ f_{\Lambda}{}^I$ by an appropriate rescaling and the four-form field strength is defined as in (\ref{Fsdefs}). Here and in the following we absorb the coupling constant in the definition of the embedding tensor. The Stueckelberg--like
kinetic terms for $C^I$ in four dimensions are clearly invariant
under the local translations $C^I\rightarrow C^I+f_\Lambda{}^I
\,\xi^\Lambda$, $\xi^\Lambda$ being four local parameters,
provided the bulk vectors are subject to the gauge transformation
$A^\Lambda_\mu\rightarrow A^\Lambda_\mu+\partial_\mu \xi^\Lambda$.
Thus from general arguments we expect that in the presence of the above fluxes, the low--energy supergravity should be
invariant under a four-dimensional Abelian gauge group ${G}_g$,
subgroup of $G^{(QK)}$ whose generators $X_\Lambda=f_\Lambda{}^I\, t_I$
are gauged by the bulk-vectors. The embedding tensor is electric and coincides with the fluxes $\Theta_\Lambda{}^I=f_\Lambda{}^I$. The ${\mathcal N}=2$ supergravity
originated from the flux compactification is obtained therefore
by gauging the subgroup ${G}_g$ of the isometry
group of ${\Scr M}_{QK}$.\par
We could also switch on magnetic fluxes of D7 vectors across 2-cycles of $K3$:
\begin{equation}
F^k_{bg}=F^{k\,I}\,\boldsymbol{\omega}_I\,,\,\,\,\,k=1,\dots, n_7\,.
\end{equation}
 These background quantities would induce, through the Chern-Simons term in the D7 world volume action,
minimal couplings of the vectors $A^k_\mu$ to the axions $C^I$ which amounts to gauging the corresponding translational isometries by $A^k_\mu$. In this case we would add the following entries to the embedding tensor:
\begin{equation}
f_{\Lambda=k+3}{}^I=F^{k\,I}\,,
\end{equation}
so that the D7-brane vectors are involved.\footnote{In principle one can also gauge the translational isometries $t_I$ by means of the D3-brane gauge vectors $A^r_\mu$. The interpretation of the corresponding embedding tensor $f_{\Lambda=3+n_7+r}{}^I=F^{r\,I}$, $r=1,\dots, n_3$, is less obvious.}
The embedding tensor $f_\Lambda{}^I$ is a collection of $n_v$ vectors, labeled by $\Lambda$, in $\mathbb{R}^{3,19}$, and can be conveniently split into positive and negative-norm vectors:
\begin{equation}
(f_\Lambda{}^I)=(f_\Lambda{}^x,\,h_\Lambda{}^a)\,,\,\, x=1,2,3\,,\,\,\,a=1,\dots, 19\,.
\end{equation}
The gauging procedure described in previous sections totally determines the four-dimensional effective action.
The Killing vectors $k^I_\Lambda$ and the momentum maps
${\Scr P}^x_\Lambda$ associated with the gauged isometries read:
\begin{align}
k^I_\Lambda&=&f_\Lambda{}^I\,\,;\,\,\,\,\,{\Scr
P}^x_\Lambda=e^{\varphi}\,L({\bf e})^{-1\,x}{}_I f_\Lambda{}^I=\,e^{\varphi}\,([(1+ee^t)^{\frac{1}{2}}]^x{}_y\,
f_\Lambda{}^y+e^x{}_a\, h_\Lambda{}^a),\,\,x=1,2,3\,.\label{kpk3}
\end{align}
In terms of these quantities the scalar potential can be evaluated using Eq. (\ref{potentialV}) and has the following form:
\begin{align}
V&=2\, e^{2\,\varphi}\,\left[(\delta_{xy}+2\,e_{xa}e_{ya})\,f_\Lambda{}^x\,f_\Sigma{}^y+
4\,f_{(\Lambda}{}^x\,e_{xa}[({\bf 1}+{\bf e}^T{\bf e})^{\frac{1}{2}}]_{ab}\,h_{\Sigma)}{}^b+
\right.\nonumber\\&
\left.+(\delta_{ab}+2\,e_{xa}e_{xb})\,h_\Lambda{}^a\,h_\Sigma{}^b\right]\,\bar{L}^\Lambda\,
L^\Sigma+\nonumber\\&
+e^{2\,\varphi}\,\left(U^{\Lambda\Sigma}-3\,\bar{L}^\Lambda\,
L^\Sigma\right)\,\left(f_\Lambda{}^x\,f_\Sigma{}^x+e_{xa}
e_{ya}\,f_\Lambda{}^x\,f_\Sigma{}^y+ 2\,[(1+{\bf e}\,{\bf
e}^T)^{\frac{1}{2}}]_{xy} e^x{}_a\,
f_{(\Lambda}{}^y\,h_{\Sigma)}^a+\right.\nonumber \\&\left. e^x{}_a
e^x{}_b\,h_\Lambda^a\,h_\Sigma^b\right)\,,\label{pot}
\end{align}
where the sum over repeated indices in understood.
 Once the potential is known then we can study the vacua of the
theory, that is bosonic backgrounds which extremize $V(\phi^s)$. If
we are interested in supersymmetric vacua we need to look for
bosonic backgrounds $(\phi^s_0)$ which admit a Killing spinor
$\epsilon$, namely directions in the supersymmetry parameter space
along which supersymmetry variations of the fermion fields vanish, see Eqs. (\ref{KSeqs}).\par
The Killing spinor equations have the following form:
\begin{align}
\delta_\epsilon\psi^A_\mu&=0\,\, \Leftrightarrow \,\,\,\delta_\epsilon \zeta^{A',\,20}=0\,\,\Leftrightarrow\,\,\,\,X^\Lambda\,{\Scr P}_\Lambda^x\,(\sigma^x)_A{}^C\epsilon_{CB}\epsilon^B=0\,,\label{detapsiKS}\\
\delta_\epsilon \zeta^{A',a}&=0\,\, \Rightarrow\,\,\,\, L({\bf e})^{-1\,a}{}_I\,f_\Lambda{}^I \,X^\Lambda=0\,,\label{detazetaaKS}\\
\delta_\epsilon \lambda^{\bar{\imath}\,A}&=0\,\, \Leftrightarrow \,\,\,g^{j\bar{\imath}}{\Scr D}_j X^\Lambda\,{\Scr P}_\Lambda^x\,(\sigma^x)_A{}^C\epsilon_{CB}\epsilon^B=0\,.\label{detalamKS}
\end{align}
To solve the second equation
we make the following position:
\begin{eqnarray}
e_x{}^a\, f^x_\Lambda&=&0=e_{x\,a}\,
h^a_\Lambda\,,\label{moduli}\\
h^a_\Lambda\, X^\Lambda&=&0\,.\label{tufix}
\end{eqnarray}
Conditions (\ref{moduli}) will fix  $K3$ complex structure moduli,
while Eq. (\ref{tufix}) will fix the $T^2$ complex structure $t$
and the axion/dilaton $u$. If (\ref{moduli}) hold, the momentum maps (\ref{kpk3}) acquire the simpler form: ${\Scr P}_\Lambda^x=e^\varphi\,f_\Lambda{}^x$.
 The Killing spinor equations (\ref{detapsiKS}),
$\delta_\epsilon \zeta^{A',\,20}=0$ and $\delta_\epsilon \psi_\mu^A=0$, on the other hand,
turn out to be equivalent for this gauging
and, together with (\ref{detalamKS}), will impose
restrictions on the fluxes.
\paragraph{${\mathcal N}=2$ vacua.}
These are bosonic backgrounds on which the Killing spinor equations must hold for any parameter $\epsilon^A$.
This implies that
\begin{equation}
X^\Lambda\,{\Scr P}_\Lambda^x={\Scr D}_iX^\Lambda\,{\Scr P}_\Lambda^x=0\,.
\end{equation}
Being ${\Scr P}_\Lambda^x$ real, the above conditions can be recast in the form:
\begin{equation}
\mathbb{L}_c^M{}_{\underline{N}}\,{\Scr P}_M^x=0\,,
\end{equation}
which in turn imply, since $\mathbb{L}_c$ is an invertible matrix, that in order to have ${\mathcal N}=2$ vacua, the momentum maps must vanish:
\begin{equation}
{\Scr P}_M^x=0\,\,\,\Leftrightarrow\,\,\,f_\Lambda{}^x=0\,,\label{f0n2}
\end{equation}
where we have used Eqs. (\ref{moduli}).
Eq. (\ref{f0n2}) can be restated as the requirement that no flux vector among
the  $f_\Lambda{}^I$ in $\Gamma^{3,19}$ have positive norm,
consistently with the results by Tripathy and Trivedi \cite{Tripathy:2002qw}.
Let us, for the sake of simplicity, choose as the only
non--vanishing components of the flux
\begin{eqnarray}
h_2{}^{a=1}&=&g_2\,\,\,;\,\,\,\,\,h_2{}^{a=2}=g_3\,.
\end{eqnarray}
Condition (\ref{tufix}) then implies:
\begin{eqnarray}
&&X^2=X^3=0\,\,\,\Leftrightarrow\,\,\,t=u \,,\,\,\,
1+t^2=\frac{(x^k)^2}{2}\,,
\end{eqnarray}
so that $t,u$ are fixed, while $s$ and the brane coordinates
$x^k,\,y^r$ remain moduli. Finally conditions (\ref{moduli}) imply
$e^x{}_{a=1,2}=0$. Since the two axions $C^{a=1,2}$ are Goldstone
bosons which provide mass to $A^2_\mu,\,A^3_\mu$, the whole two
hypermultiplets $a=1,2$ will not appear in the low--energy
effective theory. If the fluxes $F^{k\,I}$ are switched on, condition $X^\Lambda\,h_\Lambda{}^a=0$ will fix some of the $x^k$ scalars to zero.\par
This theory will be no--scale since the
potential at the minimum vanishes identically in the moduli.
\paragraph{${\mathcal N}=1,0$ vacua.} Let us look for ${\mathcal N}=1$ vacua by requiring
 the component
$\epsilon^2$ to be the Killing spinor. Upon implementation of
(\ref{moduli}), we obtain the following conditions:
\begin{eqnarray}
\begin{matrix}\delta_\epsilon \psi_\mu^A=0\cr \delta_\epsilon
\lambda^{i,A}=0\end{matrix}&\Rightarrow
&\begin{cases}(f_\Lambda{}^{x=1}+i\,f_\Lambda{}^{x=2})\, X^\Lambda=0\cr
(f_\Lambda{}^{x=1}+i\,f_\Lambda{}^{x=2})\, \partial_i
X^\Lambda=0\cr f_\Lambda{}^{x=3}=0\end{cases}\,.\label{psilam}
\end{eqnarray}
Condition $f_\Lambda{}^{x=3}=0$ in particular can be rephrased as
the statement that the flux should be defined by at most two
positive norm vectors in $\Gamma^{3,19}$, consistently with the
\emph{primitivity} condition on the complexified 3--form field
strength $\hat{G}^{(3)}\equiv \hat{F}^{(3)}-u\,\hat{H}^{(3)}$, as found in \cite{Tripathy:2002qw}.\par
Suppose, for the sake of simplicity, that the only non--vanishing
flux components are the following
\begin{eqnarray}
f_0{}^{m=1}&=&g_0\,\,\,;\,\,\,\,\,f_1{}^{m=2}=g_1\,\,\,;\,\,\,\,\,h_2{}^{a=1}=g_2\,\,\,;\,\,\,\,\,h_2{}^{a=2}=g_3\,,
\end{eqnarray}
then from the vanishing of the $D7$--brane gaugini variations  in
(\ref{psilam}) we have the condition $x^k=0$, namely that the D7
branes be stuck at the origin of $T^2$. Condition (\ref{tufix})
then implies:
\begin{eqnarray}
&&X^2=X^3=0\,\,\,\Leftrightarrow\,\,\,t=u=-i \,.
\end{eqnarray}
The four axions $C^{m=1,2},\,C^{a=1,2}$ are Goldstone bosons which
provide mass to all the bulk vectors. Finally conditions
(\ref{moduli}) will fix the 40 complex structure moduli of $K3$:
\begin{eqnarray}
e^x{}_{a=1,2}&=&0\,\,\,;\,\,\,\,\,e^{x=1,2}{}_{a>2}=0
\end{eqnarray}
leaving the 17 K\"ahler moduli $e^{x=3}{}_{a>2}$ unfixed. The
unfixed moduli will enter chiral multiplets in the effective
${\mathcal N}=1$ theory as the following complex scalars:
\begin{eqnarray}
s,\,y^r,\,C^{m=3}+i\,e^{\varphi},\,C^{a>2}+i\,e^{m=3}{}_{a>2}\,,
\end{eqnarray}
which span the scalar manifold:
\begin{eqnarray}
{\Scr M}_{scal}&=&\frac{{\rm U}(1,1+n_3)}{{\rm U}(1)\times {\rm
U}(1+n_3)}\times \frac{{\rm SO}(2,18)}{{\rm SO}(2)\times {\rm
SO}(18)}\,,
\end{eqnarray}
the former factor being parametrized by $s,\,y^r$. We have not
dealt with all conditions (\ref{psilam}) yet. In particular in the
effective ${\mathcal N}=1$ we can construct a superpotential using
${\mathcal N}=2$ quantities:
\begin{eqnarray}
W&=&[e^{-\phi}\,({\Scr P}^{x=1}_\Lambda+i\,{\Scr
P}^{x=2}_\Lambda)\, X^\Lambda]_{|\phi^s_0}\propto
g_0-g_1\,\,\,\mbox{moduli independent}\,.
\end{eqnarray}
On the other hand the expressions in (\ref{psilam})
$(f_\Lambda{}^{x=1}+i\,f_\Lambda{}^{x=2})\, X^\Lambda$ and
$(f_\Lambda{}^{x=1}+i\,f_\Lambda{}^{x=2})\, \partial_i X^\Lambda$
turn all out to be proportional to $g_0-g_1$. Therefore if $W=0$
we have ${\mathcal N}=1$ otherwise the vacuum will break all
supersymmetry. In both cases the potential at the minimum vanishes
identically in the moduli so that the effective supergravity is
no--scale.
\subsection{Mirror Covariant Gauging}\label{mcga}
As a last instance of gauged supergravity originating from flux compactifications we shall discuss in this section a
class of $\mathcal{N}=2$ gauged models which describe flux-compactifications of Type II theories on manifolds with ${\rm SU}(3)\times {\rm SU}(3)$-structure. In the context of Calabi-Yau compactifications the role of $T$-duality is played by mirror symmetry, defined in Sect. \ref{rtcy}. As pointed out in the same Section, the presence of fluxes (see \cite{Grana:2005jc,Douglas:2006es,Blumenhagen:2006ci} and references therein), through their back-reaction on the internal geometry, spoils the property of the internal manifold of being a Calabi-Yau. Requiring the four-dimensional effective theory to have $\mathcal{N}=2$ off-shell supersymmetries implies the internal manifold should have ${\rm SU}(3)$-structure. The effect of $T$-duality in toroidal compactifications, see Sect. \ref{Tdualcomp}, naturally extends to $H$-fluxes and geometric fluxes, relating them to the non-geometric $Q$ and $R$-fluxes. Similarly one may expect that also mirror symmetry can be consistently defined in the presence of fluxes and that a mirror-invariant set of form, geometric and non-geometric fluxes can be defined.
As for the toroidal reductions, a suitable framework where to study the problem is that of compactifications of Type II theories on manifolds with generalized geometry. More specifically, requiring the four-dimensional effective theory to be an  $\mathcal{N}=2$ supergravity, one has to consider generalized geometries with ${\rm SU}(3)\times {\rm SU}(3)$-structure, see  \cite{Grana:2006hr} and references therein.
Extending mirror symmetry to the presence of fluxes amounts to stating that Type IIA and Type IIB theories on mirror-dual flux-backgrounds are equivalent. This in turn implies that the corresponding low-energy  $\mathcal{N}=2$ gauged-supergravities must coincide. As usual the background quantities are identified with the embedding tensor defining the corresponding gauging. The general features of the four-dimensional supergravity were derived within a top-down approach, using generalized geometry, in \cite{Grana:2006hr}, building on previous related works, see \cite{Gurrieri:2002iw,Gurrieri:2002wz} and references therein. The four-dimensional theory was constructed as a gauged supergravity in \cite{D'Auria:2007ay}, building on \cite{D'Auria:2004tr,D'Auria:2004wd}. We shall review here this second result. \par
We have seen in Sect. \ref{rtcy} that Type II compactifications on Calabi-Yau manifolds yield ungauged $\mathcal{N}=2$ supergravities in which the scalar manifold contains two submanifolds of special K\"ahler type: one describing the complexified K\"ahler moduli $w^a$, $a=1,\dots, h_{1,1}$, the other the complex structure moduli $z^i$, $i=1,\dots, h_{2,1}$. We have denoted by ${\Scr M}_{SK}^{(1)}$ the former, parametrized by $w^a$, and by ${\Scr M}_{SK}^{(2)}$ the latter, parametrized by $z^i$. Depending on whether we are in Type IIA or Type IIB theory, one of them is the manifold ${\Scr M}_{SK}$ describing the scalars in the vector multiplets, the other is contained in the quaternionic K\"ahler manifold ${\Scr M}_{QK}$ and is spanned by hyper-scalars.
%In this section we will find it convenient to change this notation and denote by ${\Sr M}_{SK}^{(2)}$, of coordinates $z^i$, the manifold associated with the vector multiplet sector, and by ${\Sr M}_{SK}^{(1)}$, of coordinates $w^a$, the one contained inside ${\Scr M}_{QK}$. This is consistent with the previous conventions only for Type IIB.
In Sect. \ref{rtcy} the kind of quaternionic K\"ahler manifolds ${\Scr M}_{QK}$ occurring in these compactifications were characterized as being in the image of the c-map.\footnote{We are always working at string tree-level.} In Appendix \ref{BGM} their general structure is discussed and it is shown that they all feature a characteristic Heisenberg algebra of isometries ${\Scr H}$, see Eq. (\ref{MQKHeis}). This algebra contains the Peccei-Quinn translations on the axions originating from the RR fields: $C_\Lambda,\,C^\Lambda$ in Type IIA, $\rho,\,C^a,\,\tilde{C},\,C_a$ in Type IIB theory.\par
Just as for toroidal compactifications, we wish to define a minimal set of background quantities which is mirror invariant and contains the $H$-flux. We have seen that the $T$-dual of the $H$-flux on $T^6$ is a geometric flux which is introduced as a deformation of the geometry of the internal torus, by replacing the closed one-forms $dx^{{\tt u}}$ on $T^6$ by non-closed left-invariant one-forms $\sigma^{{\tt u}}$, locally describing the twisted torus ${\Scr T}^6$. Then the non-geometric $Q$ and $R$-fluxes, which complete the $T$-duality invariant picture, are arranged in a constant, $T$-duality-covariant tensor $T_{\Lambda\Sigma}{}^\Gamma$ describing the twist of a larger set of forms within generalized geometry, see discussion in Sect. \ref{Tdualcomp}.\par
A similar approach
has been followed for Calabi-Yau compactifications. Here the ``twisted Calabi-Yau manifold'' ${\Scr X}$ is defined by replacing the harmonic forms on the original manifold $\mathcal{X}$ by a new set of ``twisted forms''  $\boldsymbol{\alpha}_M,\,{\bf e}_{\mathcal{A}}$ which are neither of pure degree nor closed \cite{Grana:2006hr} and which satisfy the following relations:
\begin{align}
d\boldsymbol{\alpha}_M&=Q^{[{\Scr X}]}{}_M{}^{\mathcal{A}}\,{\bf e}_{\mathcal{A}}\,\,,\,\,\,\,d{\bf e}_{\mathcal{A}}=\tilde{Q}^{[{\Scr X}]}{}_{\mathcal{A}}{}^M\,\boldsymbol{\alpha}_M\,,\label{gendefX}
\end{align}
where, setting ${\bf Q}^{[{\Scr X}]}\equiv (Q^{[{\Scr X}]}{}_M{}^{\mathcal{A}})$, we have defined
\begin{equation}
\tilde{Q}^{[{\Scr X}]}{}_{\mathcal{A}}{}^M=\mathbb{C}_{\mathcal{A}\mathcal{B}}\mathbb{C}^{MN}\,Q^{[{\Scr X}]}{}_N{}^{\mathcal{B}}\,\,\Leftrightarrow\,\,\,\,\tilde{{\bf Q}}^{[{\Scr X}]}\equiv \mathbb{C}\,{\bf Q}^{[{\Scr X}]\,T}\,\mathbb{C}^T
\,.
\end{equation}
Equations (\ref{gendefX}) are consistent since in the generalized geometry picture $\boldsymbol{\alpha}_M,\,{\bf e}_{\mathcal{A}}$ are not of pure degree.
The constant matrix ${\bf Q}^{[{\Scr X}]}$, which has the following general structure,
 \begin{equation}
 {\bf Q}^{[{\Scr X}]}=\left(\begin{matrix}\tilde{b}_{\Lambda}{}^A & {b}_{\Lambda\,B}\cr
 \tilde{a}^{\Gamma\,A}& {a}^{\Gamma}{}_B\end{matrix}\right)\,\,,\,\,\,\,\tilde{{\bf Q}}^{[{\Scr X}]}=\left(\begin{matrix}{a}^{\Gamma}{}_B& -{b}_{\Lambda\,B}\cr
 -\tilde{a}^{\Gamma\,A}& \tilde{b}_{\Lambda}{}^A \end{matrix}\right)\,,
 \end{equation}
 plays the role of the twist tensor $T_{\Lambda\Sigma}{}^\Gamma$ in (\ref{Ttotal}) in that it encodes all the $H,T,Q$ and $R$-fluxes. It defines a deformation of the original manifold
compatible with the condition that ${\Scr X}$ have ${\rm SU}(3)\times {\rm SU}(3)$-structure \cite{Grana:2006hr}.
Integrability of (\ref{gendefX}) implies:
\begin{align}
d^2\boldsymbol{\alpha}_M&=0\,\,\Leftrightarrow\,\,\,\,{\bf Q}^{[{\Scr X}]}\mathbb{C}{\bf Q}^{[{\Scr X}]\,T}={\bf 0}\,,\label{int1Q}\\
d^2{\bf e}_{\mathcal{A}}&=0\,\,\Leftrightarrow\,\,\,\,{\bf Q}^{[{\Scr X}]\,T}\mathbb{C}{\bf Q}^{[{\Scr X}]}={\bf 0}\,.\label{int2Q}
\end{align}
Take for instance the simpler deformation in the Type IIA theory:
\begin{equation}
d{\alpha}_\Lambda=b_{\Lambda\,a}\,{\bf e}^a\,\,,\,\,\,\,d{\beta}^\Lambda=-a^\Lambda{}_a\,{\bf e}^a\,\,,\,\,\,\,d{\bf e}_a=a^\Lambda{}_a\,{\alpha}_\Lambda+b_{\Lambda\,a}\,{\beta}^\Lambda\,\,,\,\,\,\,d{\bf e}^a=0\,,\label{abbg1}
\end{equation}
with the following $H$-flux
\begin{equation}
\hat{H}^{(3)}_{bg}=d\hat{B}^{(2)}_{bg}=a^\Lambda{}_0\,\alpha_\Lambda+b_{\Lambda\,0}\,\beta^\Lambda\,.\label{abbg2}
\end{equation}
This corresponds to the case in which only the $H$ and geometric fluxes are present and the forms have pure degree.
Conditions $d\hat{H}^{(3)}_{bg}=0$ and $d^2\boldsymbol{\alpha}_M=0=d^2{\bf e}_a$ imply:
\begin{equation}
a^\Lambda{}_0\,b_{\Lambda\,a}-a^\Lambda{}_a\,b_{\Lambda\,0}=0\,\,;\,\,\,\,\,
a^\Lambda{}_b\,b_{\Lambda\,a}-a^\Lambda{}_a\,b_{\Lambda\,b}=0\,.
\end{equation}
The ten-dimensional form-fields expand as follows:
 \begin{align}
\hat{B}^{(2)}&=B^{(2)}+u^a\,{\bf e}_a+\hat{B}^{(2)}_{bg}\,,\,\,\,\hat{C}^{3}={C}^{(3)}+A^{(1)\,a}\wedge {\bf e}_a+\mathcal{Z}^M\,\boldsymbol{\alpha}_M\,\,,\,\,\,\,\hat{C}^{(1)}=A^0\,.
\end{align}
From the expansion of the ten-dimensional field strengths we can infer the gauge symmetries induced in the four-dimensional theory. In particular we find:
 \begin{align}
\hat{F}^{(4)}&\equiv d\hat{C}^{(3)}-\hat{C}^{(1)}\wedge \hat{H}^{(3)}=d{C}^{3}-A^0\wedge dB^{(2)}+(d\tilde{A}^a-u^a\,dA^0)\wedge {\bf e}_a+\nonumber\\
&+(d\zeta^\Lambda-A^A\,a^\Lambda{}_A)\wedge\,\alpha_\Lambda-
(d\tilde{\zeta}_\Lambda+A^A\,b_{\Lambda A})\wedge\,\beta^\Lambda+ (\zeta^\Lambda\,b_{\Lambda a}+\tilde{\zeta}_\Lambda\,a^\Lambda{}_A)\,{\bf e}^a\,,
\end{align}
where we have defined, $\tilde{A}^a\equiv A^a+u^a\,A^0$ and $A^A\equiv (A^0,\,\tilde{A}^a)$, the index $A$ being $A=(0,a)$. We see that the axions $\mathcal{Z}^M$ appear in covariant derivatives of Stueckelberg form:
\begin{equation}
\mathcal{D}\mathcal{Z}^M\equiv d\mathcal{Z}^M-\mathbb{C}^{MN}\,b_{N\,A}\,A^A\,,
\end{equation}
where $b_{M\,A}\equiv (b_{\Lambda A},\,a^\Lambda{}_A)$. The above derivatives are invariant under the following local transformations:
\begin{equation}
\delta A^A =d\uplambda^A\,\,,\,\,\,\,\delta \mathcal{Z}^M=\mathbb{C}^{MN}\,b_{N\,A}\uplambda^A\,.
\end{equation}
We see that the background quantities introduced in (\ref{abbg1}) and (\ref{abbg2}) induce in the low-energy effective theory the gauging of the translational isometries associated with $\mathcal{Z}^M$, which are part of the Heisenberg algebra ${\Scr H}$.\par
We could also introduce RR fluxes across cycles of the internal manifold. In the Type IIA and IIB theories theory these are expanded in the even and odd cohomology bases, respectively:
\begin{align}
\mbox{Type IIA RR fluxes:}&\nonumber\\
&\{\hat{F}^{(0)}_{bg},\,\hat{F}^{(2)}_{bg},\,\hat{F}^{(6)}_{bg},\,\hat{F}^{(4)}_{bg}\}\equiv \{F^{\mathcal{A}}\}\,,\nonumber\\
\mbox{Type IIB RR fluxes:}&\nonumber\\
&\{\hat{F}^{(3)}_{bg}\}\equiv \{F^{M}\}\,.\label{RRfluxcy}\\
\end{align}
The local symmetry induced by the general deformation (\ref{gendefX}) is defined by gauging a Abelian subalgebra of the
Heisenberg algebra ${\Scr H}$ in $\mathfrak{g}^{(QK)}$ \cite{D'Auria:2004tr,D'Auria:2004wd}. Consider the ungauged  $\mathcal{N}=2$ model originating from the Type IIB theory. The scalar fields in the vector multiplets are the complex structure moduli $z^i$ spanning ${\Scr M}_{SK}^{(2)}$ with a ${\rm Sp}(2h_{2,1}+2,\mathbb{R})$-symplectic bundle. The manifold ${\Scr M}_{SK}^{(1)}$ of the complexified K\"ahler moduli, with its  ${\rm Sp}(2h_{1,1}+2,\mathbb{R})$-symplectic bundle, is contained within ${\Scr M}_{QK}^{{\rm (IIB)}}$, which has the general form (\ref{MQKHeis}):
 \begin{equation}
 {\Scr M}_{QK}^{{\rm (IIB)}}\sim \left({\rm O}(1,1)\times {\Scr M}^{(1)}_{SK}\right)\ltimes e^{{\Scr H}}\,.\label{MQKHeisIIB}
 \end{equation}
The notation introduced in Appendix \ref{BGM} should be adjusted to the Type IIB theory by replacing the ${\rm Sp}(2h_{2,1}+2,\mathbb{R})$ indices $M,N,\dots$, by the ${\rm Sp}(2h_{1,1}+2,\mathbb{R})$ ones $\mathcal{A,B},\dots$. Moreover the scalar $\varphi$ is to be identified with the opposite of the four-dimensional dilaton $\phi_4$, see Eqs. (\ref{redsiiba}),  and $a$ is to be intended as proportional to the NS-NS field $\tilde{B}$. The generators of ${\Scr H}$ will then be denoted by $t_{\mathcal{A}},\,t_\bullet$ and satisfy the commutation relations:
 \begin{align}
[t_{\mathcal{A}},\,t_{\mathcal{B}}]=-2\,\mathbb{C}_{{\mathcal{AB}}}\,t_\bullet\,,\label{Heistruc}
 \end{align}
 all other commutators being zero.
The corresponding Killing vectors and momentum maps are given in Eqs. (\ref{KHeis}) and (\ref{PHeis}). The reader can verify that the Killing vectors satisfy the correct commutation relations:
 \begin{align}
[k_{\mathcal{A}},\,k_{\mathcal{B}}]=2\,\mathbb{C}_{\mathcal{AB}}\,k_\bullet\,.
 \end{align}
Let us define the gauge algebra as a generic Abelian subalgebra of  ${\Scr H}$.
Its generators will read:
\begin{equation}
X_M=\Theta_M{}^{\mathcal{A}}\,t_{\mathcal{A}}+\Theta_M{}^{\bullet}\,t_\bullet=\Theta_M{}^{\mathcal{A}}\,t_{\mathcal{A}}+c_M\,t_\bullet\,,
\end{equation}
where we have defined, for the sake of simplicity, $c_M\equiv \Theta_M{}^{\bullet}$.
The locality constraint implies:
\begin{equation}
\Theta_M{}^{\mathcal{A}}\Theta_N{}^{\mathcal{B}}\mathbb{C}^{MN}=0\,\,;\,\,\,
\Theta_M{}^{\mathcal{A}}\,c_N{}\mathbb{C}^{MN}=0\label{quadthc}
\end{equation}
The requirement that the gauge algebra be Abelian implies, in light of Eq. (\ref{Heistruc}), that:
\begin{equation}
\Theta_M{}^{\mathcal{A}}\Theta_N{}^{\mathcal{B}}\mathbb{C}_{\mathcal{AB}}=0\,.\label{quadthc2}
\end{equation}
Note that if we identify $\Theta_M{}^{\mathcal{A}}$ with $Q^{[{\Scr X}]}{}_M{}^{\mathcal{A}}$, the above equation and the first of Eqs. (\ref{quadthc}) coincide with (\ref{int2Q}) and (\ref{int1Q}), respectively, while the second of Eqs. (\ref{quadthc}) follow from the closure of $\hat{F}_{bg}$, $d\hat{F}_{bg}=0$, once we identify $c_M$ with the RR fluxes (\ref{RRfluxcy}). For this reason, if we also denote the $(2\,h_{2,1}+2)\times (2\,h_{1,1}+2)$ matrix $\boldsymbol{\Theta}\equiv (\Theta_M{}^{\mathcal{A}})$ by
$\boldsymbol{\Theta}^{[{\rm IIB;}\,{\Scr X}]}$, we can make the following identification:
\begin{equation}
\boldsymbol{\Theta}^{[{\rm IIB;}\,{\Scr X}]}={\bf Q}^{[{\Scr X}]}\,.
\end{equation}
Note that the NS-NS part $\boldsymbol{\Theta}^{[{\rm IIB;}\,{\Scr X}]}$ of the embedding tensor, coinciding with the twist tensor ${\bf Q}^{[{\Scr X}]}$, is an ${\rm Sp}(2\,h_{2,1}+2,\mathbb{R})\times {\rm Sp}(2\,h_{1,1}+2,\mathbb{R})$-tensor, namely is manifestly covariant with respect to the symplectic structures associated with the two special-K\"ahler manifolds.

The gauging is dyonic in that it involves both electric and magnetic vector fields. As discussed in Sect. \ref{sec:4}, consistency of the construction requires the introduction of rank-2 antisymmetric tensor fields $B_{\alpha\,\mu\nu}$ in the adjoint representation of $G$ (in this case of $G^{(QK)}$). Let us review the construction of the scalar potential. The building blocks are the gauge Killing vectors and tri-holomorphic momentum maps which read:
\begin{equation}
k_M\equiv \Theta_M{}^{\mathcal{A}}\,k_{\mathcal{A}}+c_M\,k_\bullet\,\,;\,\,\,\,\,{\Scr P}^x_M\equiv \Theta_M{}^{\mathcal{A}}\,{\Scr P}^x_{\mathcal{A}}+c_M\,{\Scr P}^x_\bullet\,.
\end{equation}
Using (\ref{KHeis}) and (\ref{PHeis}) we find
\begin{align}
k_M&=(c_M-\Theta_{M\mathcal{A}}\,\mathcal{Z}^{\mathcal{A}})\,\frac{\partial}{\partial a}+\Theta_{M}{}^\mathcal{A}\,\frac{\partial}{\partial \mathcal{Z}^{\mathcal{A}}}\,,\nonumber\\
{\Scr P}_M^1&=e^{\phi_4}\,\Theta_{M\mathcal{A}}\,(\overline{V}_1+{V}_1)^{\mathcal{A}}\,,\nonumber\\
{\Scr P}_M^2&=i\,e^{\phi_4}\,\Theta_{M\mathcal{A}}\,(\overline{V}_1-{V}_1)^{\mathcal{A}}\,,\nonumber\\
{\Scr P}_M^3&=\frac{e^{2\phi_4}}{2}\,(c_M-2\,\Theta_{M\mathcal{A}}\,\mathcal{Z}^{\mathcal{A}})\,,
\end{align}
where the subscripts $1$ and $2$ indicate quantities defined on ${\Scr M}_{SK}^{(1)}$ and ${\Scr M}_{SK}^{(2)}$, respectively.
The non-trivial Stueckelberg-like covariant derivatives, required by the gauging procedure, read:
\begin{align}
\mathcal{D}_\mu\mathcal{Z}^{\mathcal{A}}\equiv \partial_\mu\mathcal{Z}^{\mathcal{A}}-A^M_\mu\,\Theta_M{}^{\mathcal{A}}\,,\label{DZstuck}\\
\mathcal{D}_\mu a\equiv \partial_\mu a-A^M_\mu\,(c_M-\Theta_{M\mathcal{A}}\,\mathcal{Z}^{\mathcal{A}})\,.\label{Dastuck}
\end{align}
Under a gauge transformation, parametrized by $\uplambda^M(x)$, we find:
\begin{align}
\delta A^M_\mu &=\partial_\mu\uplambda^M\,\,,\,\,\,\,\delta\mathcal{Z}^{\mathcal{A}}=\uplambda^M\,\Theta_M{}^{\mathcal{A}}\,\,,\,\,\,
\delta  a=\uplambda^M\,(c_M-\Theta_{M\mathcal{A}}\,\mathcal{Z}^{\mathcal{A}})\,,\nonumber\\
\delta\mathcal{D}_\mu\mathcal{Z}^{\mathcal{A}}&=0\,\,,\,\,\,\,\delta \mathcal{D}_\mu a=-\uplambda^M\,\Theta_{M\mathcal{A}}\,\mathcal{D}_\mu\mathcal{Z}^{\mathcal{A}}\,.
\end{align}
The scalar potential has the general form (\ref{potentialV}):\footnote{The coupling constant has been absorbed in a redefinition of the embedding tensor.}
  \begin{align}
{V}&=4\,h_{uv}k_M^uk_N^v\,\overline{V}_2^M\,V_2^N+(U_2^{MN}-3\,V_2^M\overline{V}_2^N){\Scr P}^x_N{\Scr P}^x_M=\nonumber\\
& =4\,h_{uv}k_M^uk_N^v\,\overline{V}_2^M\,V_2^N+\left(-\frac{1}{2}\mathcal{M}_2^{MN}-4\,V_2^M\overline{V}_2^N\right){\Scr P}^x_N{\Scr P}^x_M \,,\label{potentialVmirr}
\end{align}
where we have used Eq. (\ref{UMN}). Using the general expression for $h_{uv}$ in (\ref{localc}) we find:
\begin{align}
h_{uv}\,k_M^u k_N^v\, \overline{V}_2^M\,V_2^N&=\frac{e^{4\phi_4}}{4}\,\vert {V}_2^T\,
({\bf c}-2\,\underline{\boldsymbol{\Theta}}\,\mathcal{Z})\vert^2
-\frac{e^{2\phi_4}}{2}\,\overline{V}_2^T\,
\underline{\boldsymbol{\Theta}}\,
\mathcal{M}_1^{-1}\,\underline{\boldsymbol{\Theta}}^T\,V_2\,,
\end{align}
where we have defined:
\begin{equation}
{\bf c}\equiv (c_M)\,\,,\,\,\,\,\underline{\boldsymbol{\Theta}}\equiv \boldsymbol{\Theta}\mathbb{C}=(\Theta_{M\mathcal{A}})\,.
\end{equation}
The potential reds:
\begin{align}
V&=-\frac{e^{4\phi_4}}{4}\,({\bf c}^T-2\,\mathcal{Z}^T\underline{\boldsymbol{\Theta}}^T)\,\mathcal{M}_2^{-1}\,({\bf c}-2\,\underline{\boldsymbol{\Theta}}\mathcal{Z})-\nonumber\\
&-2\,e^{2\phi_4}\,\overline{V}_1^T\underline{\boldsymbol{\Theta}}^T\mathcal{M}_2^{-1}\,\underline{\boldsymbol{\Theta}}{V}_1
-2\,e^{2\phi_4}\,\overline{V}_2^T\underline{\boldsymbol{\Theta}}\mathcal{M}_1^{-1}\,\underline{\boldsymbol{\Theta}}{V}_2-\nonumber\\
&-8\,e^{2\phi_4}\,(|\overline{V}_2^T\underline{\boldsymbol{\Theta}}V_1|^2+|{V}_2^T
\underline{\boldsymbol{\Theta}}V_1|^2)\,.\label{Vpotmir}
\end{align}
To understand how, on a vacuum solution, the degrees of freedom are distributed by the embedding tensor to the various fields, we make contact with the discussion of Sect. \ref{backelectric} and perform the rank factorization of $\Theta$:
\begin{equation}
\Theta_M{}^{\mathcal{A}}=\sum_{I=1}^{r-1}\,\xi_M{}^I\,\xi_I{}^{\mathcal{A}}\,\,;\,\,\,\,
c_M=\Theta_M{}^{\bullet}=\xi_M{}^{r} \xi_{r}{}^{\bullet}\,,
\end{equation}
where $r-1$ is the rank of the rectangular matrix $\boldsymbol{\Theta}$. The quantities $\xi_M{}^I$, $I=1,\dots,r$ define the matrix $E$ which rotates the embedding tensor to the electric frame, see Eq. (\ref{EEm1}). Note that, in virtue of Eqs. (\ref{quadthc2}), the rank $r-1$ cannot exceed either $h_{1,1}+1$ or $h_{2,1}+1$: $r-1\le h_{1,1}+1$, $r-1\le h_{2,1}+1$. In Sect. \ref{backelectric} we have shown that the tensor fields $B_{m\,\mu\nu}$, $m=(\mathcal{A},\,\bullet)$ effectively enter the Lagrangian only in the $r$-combinations:
\begin{equation}
B_{I\,\mu\nu}\equiv\,\xi_I{}^{m}\,B_{m\,\mu\nu}\,,
\end{equation}
and that the truly electric vector fields, which participate in the minimal coupling, are the $r$ combinations:
\begin{equation}
A^I_\mu=A^M_\mu\,\xi_M{}^I\,\,,\,\,\,I=1,\dots, r\,.
\end{equation}
Their magnetic duals $A_{I\mu}=\eta_{MI}\,A^M_\mu$ are eaten by the $B_{I\,\mu\nu}$ which become massive. Similarly to what we did for the vectors $\xi_M{}^I$, we can associate with $\xi_I{}^\mathcal{A}$, $I=1,\dots, r-1$, a set of constant vectors $\eta^{I\,\mathcal{A}}$ such that: $\xi_I{}^\mathcal{A}\eta^{J\,\mathcal{B}}\mathbb{C}_{\mathcal{AB}}=\delta_I{}^J$, all other symplectic products being zero. The RR axions $\mathcal{Z}^\mathcal{A}$ split as follows:
\begin{equation}
\mathcal{Z}^\mathcal{A}\,\,\longrightarrow\,\,\,\,\mathcal{Z}^I\equiv -\eta^{I\,\mathcal{A}}\,\mathbb{C}_{\mathcal{AB}}\mathcal{Z}^\mathcal{B}\,\,,\,\,\,\mathcal{Z}_I=
\xi_{I}{}^{\mathcal{A}}\,\mathbb{C}_{\mathcal{AB}}\mathcal{Z}^\mathcal{B}\,\,,\,\,\,I=1,\dots, r-1\,.
\end{equation}
The reader can verify that the $r-1$ scalars $\mathcal{Z}^I$ are eaten by the vectors $A^I_\mu$, $I=1,\dots, r-1$,
since, under a gauge transformation
\begin{equation}
\delta \mathcal{Z}^I=\uplambda^M\,\xi_M{}^I\,\,,\,\,\,\,\delta \mathcal{Z}_I=0\,,
\end{equation}
while only the gauge-invariant $\mathcal{Z}_I$ enter the scalar potential since:
\begin{equation}
\Theta_M{}^{\mathcal{A}}\mathbb{C}_{\mathcal{AB}}\,\mathcal{Z}^{\mathcal{B}}=\xi_M{}^I\,\mathcal{Z}_I\,.
\end{equation}
Extremizing the scalar potential $V$ with respect to these scalars we find the conditions:
\begin{equation}
\frac{\partial V}{ \partial Z_I}=0\,\,\Rightarrow\,\,\,\,\,c_M-2\,\xi_M{}^I\,\mathcal{Z}_I=0\,,\label{fixZc}
\end{equation}
which  fix  $\mathcal{Z}_I$ in terms of the fluxes:
 $$\mathcal{Z}_I=-\frac{1}{2}\,\eta_{M\,I}\mathbb{C}^{MN}\,c_N\,.$$
 After $\mathcal{Z}_I$ are integrated out, the terms in the first line of (\ref{Vpotmir}) vanish and the potential reads:
 \begin{align}
V_{eff}&\equiv \left.V\right\vert_{\frac{\partial V}{ \partial Z_I}=0}=-2\,e^{2\phi_4}\,\overline{V}_1^T\underline{\boldsymbol{\Theta}}^T\mathcal{M}_2^{-1}\,\underline{\boldsymbol{\Theta}}{V}_1
-2\,e^{2\phi_4}\,\overline{V}_2^T\underline{\boldsymbol{\Theta}}\mathcal{M}_1^{-1}\,\underline{\boldsymbol{\Theta}}{V}_2-\nonumber\\
&-8\,e^{2\phi_4}\,(|\overline{V}_2^T\underline{\boldsymbol{\Theta}}V_1|^2+|{V}_2^T
\underline{\boldsymbol{\Theta}}V_1|^2)\,.\label{Veffmir}
\end{align}
Of the RR scalars $\mathcal{Z}^{\mathcal{A}}$, $r-1$ become longitudinal modes of the vector fields, $r-1$ are fixed by (\ref{fixZc}). We are thus left with $2h_{1,1}+2-2(r-1)$ flat directions of the scalar potential.\par
As we shall show below, the effective potential in (\ref{Veffmir}) is manifestly invariant if we replace ${\Scr X}$  by its  ``mirror'' $\hat{{\Scr X}}$, to be characterized shortly.
It is interesting to notice that the scalar potential $V_{eff}$ in (\ref{Veffmir}) can be written in terms of a superpotential $W$, first given in \cite{Berglund:2005dm}. To show this let us introduce the complex scalar $S={\rm Re}(S)+\frac{i}{8}\,e^{-2\phi_4}$, which spans a K\"ahler manifold with K\"ahler potential an metric:
\begin{equation}
\mathcal{K}_S=-\log(2\,{\rm Im}(S))\,\,,\,\,\,\,g_{S\bar{S}}=\frac{1}{4\,{\rm Im}(S)^2}\,.
\end{equation}
We define the following superpotential:
\begin{equation}
W=e^{\frac{\mathcal{K}_S}{2}}\,\overline{V}_2^T\,\underline{\boldsymbol{\Theta}}\,V_1=2\,e^{\phi_4}\,\overline{V}_2^T\,\underline{\boldsymbol{\Theta}}\,V_1\,.\label{Wmirr}
\end{equation}
The reader can easily verify that:
\begin{equation}
{\Scr D}_SW\equiv \left(\partial_S +\frac{1}{2}\,\partial_S\mathcal{K}_S\right)\,W=\partial_S\mathcal{K}_S\,W\,.
\end{equation}
Next we use the property (\ref{UMN}) for both ${\Scr M}_{{SK}}^{(1)}$ and ${\Scr M}_{{SK}}^{(2)}$ to write:
\begin{align}
\mathcal{M}_1^{\mathcal{AB}}&=-2\, U_{1\,a}{}^{(\mathcal{A}}\,g_1^{a\bar{b}}\,\overline{U}_{1\,\bar{b}}{}^{\mathcal{B})}-2 \,\overline{V}_1^{(\mathcal{A}}V_1^{\mathcal{B})}\,,\nonumber\\
\mathcal{M}_2^{MN}&=-2\, U_{2\,i}{}^{(M}g_2^{i\bar{\jmath}}\,\overline{U}_{2\,\bar{\jmath}}{}^{N)}-2 \,\overline{V}_2^{(M}V_2^{N)}\,.\nonumber
\end{align}
Replacing the above expressions in (\ref{Veffmir}), after some algebra we arrive at the following form for the effective potential:
\begin{equation}
V_{eff}=g_2^{i\bar{\jmath}}{\Scr D}^{(2)}_iW{\Scr D}^{(2)}_{\bar{\jmath}}\overline{W}+g_1^{a\bar{b}}{\Scr D}^{(1)}_aW{\Scr D}^{(1)}_{\bar{b}}\overline{W}+
g^{S\bar{S}}|{\Scr D}_SW|^2-3\,|W|^2\,,\label{VeffWmir}
\end{equation}
where ${\Scr D}^{(1)},\,{\Scr D}^{(2)}$ are the covariant derivatives on ${\Scr M}_{{SK}}^{(1)}$ and ${\Scr M}_{{SK}}^{(2)}$, respectively. Eq. (\ref{VeffWmir}) is the expression of $V_{eff}$ as an $\mathcal{N}=1$ potential  in terms of the superpotential $W$ defined on the K\"ahler manifold:
\begin{equation}
{\Scr M}_{K}\equiv \left(\frac{{\rm SL}(2,\mathbb{R})}{{\rm SO}(2)}\right)_S\times {\Scr M}_{{SK}}^{(1)}\times {\Scr M}_{{SK}}^{(2)}\,.
\end{equation}
\par
Consider now Type IIA theory on ${\Scr X}$. The scalar fields in the vector multiplets are the complexified K\"ahler moduli $w^a$ spanning ${\Scr M}_{SK}^{(1)}$ with the ${\rm Sp}(2h_{1,1}+2,\mathbb{R})$-symplectic bundle while the compex structure moduli $z^i$ span the submanifold ${\Scr M}_{SK}^{(2)}$ of ${\Scr M}_{QK}^{{\rm (IIA)}}$. The electric and magnetic vector fields are $A^{\mathcal{A}}_\mu=(A^A_\mu,\,A_{A\,\mu})$ while the RR scalars in ${\Scr M}_{QK}^{{\rm (IIA)}}$ are $\mathcal{Z}^M=(\zeta^\Lambda,\,\tilde{\zeta}_\Lambda)$. The gauging of the Heisenberg algebra ${\Scr H}=\{t_{M},\,t_\bullet\}$ is effected by introducing an embedding tensor with components $\Theta^{[{\rm IIA;}\,{\Scr X}]}{}_{\mathcal{A}}{}^M,\,c_{\mathcal{A}}$, so that the
gauge generators read:
\begin{equation}
X_{\mathcal{A}}=\Theta^{[{\rm IIA;}\,{\Scr X}]}{}_{\mathcal{A}}{}^{M}\,t_{\mathcal{A}}+c_{\mathcal{A}}\,t_\bullet\,.
\end{equation}
The components $c_{\mathcal{A}}$ represent the RR fluxes while $\boldsymbol{\Theta}^{[{\rm IIA;}\,{\Scr X}]}=(\Theta^{[{\rm IIA;}\,{\Scr X}]}{}_{\mathcal{A}}{}^{M})$ is related to the NS-NS flux-matrix ${\bf Q}^{[{\Scr X}]}$ as follows:
\begin{equation}
\boldsymbol{\Theta}^{[{\rm IIA;}\,{\Scr X}]}=\mathbb{C}^T\,{\bf Q}^{[{\Scr X}]\,T}\,\mathbb{C}=\tilde{{\bf Q}}^{[{\Scr X}]}\,.
\end{equation}
The quadratic constraints on $\boldsymbol{\Theta}^{[{\rm IIA;}\,{\Scr X}]}$ coming from the locality condition and the requirement that the gauge algebra be Abelian yield, just as in the Type IIB case, the conditions (\ref{int1Q}), (\ref{int2Q}). \par The effect of mirror symmetry is to exchange even and odd cohomologies: if $\hat{\mathcal{X}}$ is the mirror of $\mathcal{X}$, we have $h_{1,1}(\mathcal{X})=h_{2,1}(\hat{\mathcal{X}})$ and $h_{2,1}(\mathcal{X})=h_{1,1}(\hat{\mathcal{X}})$. Let ${\Scr X}$ and $\hat{{\Scr X}}$ be the deformations of $\mathcal{X}$ and $\hat{\mathcal{X}}$, respectively. Consider Type IIB compactified on the former and Type IIA on the latter. The two rectangular matrices $\boldsymbol{\Theta}^{[{\rm IIB;}\,{\Scr X}]}$ and $\boldsymbol{\Theta}^{[{\rm IIA;}\,\hat{{\Scr X}}]}$ have the same dimensions, and so the two vectors of RR fluxes: ${\bf c}^{[{\rm IIB;}\,{\Scr X}]}=(c_M),\,{\bf c}^{[{\rm IIA;}\,\hat{{\Scr X}}]}=(c_{\mathcal{A}})$. If mirror symmetry is to hold in the presence of fluxes, the classical gauged $\mathcal{N}=2$ theories originating from the two compactifications should coincide. This is the case if and only if the corresponding embedding tensors are the same, namely:
\begin{equation}
\boldsymbol{\Theta}^{[{\rm IIB;}\,{\Scr X}]}=\boldsymbol{\Theta}^{[{\rm IIA;}\,\hat{{\Scr X}}]}\,\,;\,\,\,{\bf c}^{[{\rm IIB;}\,{\Scr X}]}={\bf c}^{[{\rm IIA;}\,\hat{{\Scr X}}]}\,.
\end{equation}
This implies the following relation between the twist-tensors defining  ${\Scr X}$ and $\hat{{\Scr X}}$:
\begin{equation}
{\bf Q}^{[{\Scr X}]}=\mathbb{C}^T\,{\bf Q}^{[{\hat{\Scr X}}]\,T}\,\mathbb{C}\,.
\end{equation}
Note that if we compactify a same Type II theory on ${\Scr X}$ and on $ \hat{{\Scr X}}$ the NS-NS components of the  embedding tensors
defining the resulting gauged supergravities are related as follows:
\begin{equation}
\boldsymbol{\Theta}^{[{\rm II;}\,{\Scr X}]}=\mathbb{C}^T\,\boldsymbol{\Theta}^{[{\rm II;}\,\hat{{\Scr X}}]\,T}\,\mathbb{C}\,\,\Leftrightarrow\,\,\,\,\,\underline{\boldsymbol{\Theta}}^{[{\rm II;}\,{\Scr X}]}=-\underline{\boldsymbol{\Theta}}^{[{\rm II;}\,\hat{{\Scr X}}]\,T}\,.\label{mirrcorrth}
\end{equation}
Since the mirror map ${\Scr X}\leftrightarrow \hat{{\Scr X}}$, implies interchanging ${\Scr M}_{SK}^{(1)}$ with ${\Scr M}_{SK}^{(2)}$, $h_{1,1}$ with $h_{2,1}$, and thus the subscripts ``1 '' and ``2 '' in the scalar potential $V$, once the axions $\mathcal{Z}_I$ have been integrated out, (\ref{mirrcorrth}) implies that $V_{eff}$ in (\ref{Veffmir}) is manifestly symmetric: it does not change if  we compactify a same Type II theory on ${\Scr X}$ or on $ \hat{{\Scr X}}$. The same is true for the corresponding superpotential $W$ defined in (\ref{Wmirr}).\par
Solutions to the models discussed in the present Section were studied in \cite{Ferrara:2008xz,Cassani:2009na}.
 The scalar potential $V$ in (\ref{Vpotmir}) has been derived in the context of DFT in \cite{Blumenhagen:2015lta}.\par
 The vacua of these models were studied, for instance, in  \cite{Micu:2007rd,Cassani:2009na}

\section{A View on Higher Dimensions.}\label{vohd} For good reviews of supergravities in diverse dimensions and their gaugings see \cite{Tanii:1998px} and \cite{Samtleben:2008pe}. As mentioned in point ii) of Sect. \ref{ageneraldiscussion}, there are equivalent formulations of ungauged supergravities in $D$-dimensions obtained from one another by dualizing certain $p$-forms $C_{(p)}$ (i.e.\ rank-$p$ antisymmetric tensor fields) into $(D-p-2)$-forms $C_{(D-p-2)}$ through a Hodge-duality relation between the corresponding field strengths:
\begin{equation}
  dC_{(p)}={}^*dC_{(D-p-2)}+\dots\,.
\end{equation}
Such formulations feature in general different global symmetry groups. This phenomenon is called \emph{Dualization of Dualities} and was studied in \cite{Cremmer:1997ct}. The scalar fields in these theories are still described by a non-linear sigma model and in $D\ge 6$ the scalar manifold is homogeneous symmetric. Just as in four dimensions the scalars are non-minimally coupled to the $p$-form fields (see below) and the global symmetry group $G$ is related to the isometry group of the scalar manifold, which is thus maximal in the formulation of the theory in which the scalar sector is maximal, that is in which all forms are dualized to lower order ones. This prescription, however, does not completely fix the ambiguity related to duality in even dimensions $D=2k$, when order-$k$ field strengths, corresponding to rank-$(k-1)$ antisymmetric tensor fields $C_{(k-1)}$, are present. Indeed after having dualized all forms to lower-order ones we can still dualize $(k-1)$-forms $C_{(k-1)}$ into $(k-1)$-forms $\tilde{C}_{(k-1)}$. This is the electric-magnetic duality of the four-dimensional theory, related to the vector fields, and also occurs for instance in six dimensions with the 2-forms and in eight dimensions with the 3-forms.
Duality transformations interchanging  $C_{(k-1)}$ with $\tilde{C}_{(k-1)}$, and thus the corresponding field equations with  Bianchi identities, are encoded in the group $G$, whose action on the scalar fields, just as in four dimensions, is combined with a linear action on the $k$-form field strengths ${F}_{(k)}$ and their duals $\tilde{F}_{(k)}$:
\begin{align}
{\bf g}\in G \;:\quad
\begin{cases}
{F}_{(k)}\rightarrow {F}_{(k)}'=A[{\bf g}]\,{F}_{(k)}+B[{\bf g}]\,\tilde{F}_{(k)}\,,\cr \tilde{F}_{(k)}\rightarrow \tilde{F}_{(k)}'=C[{\bf g}]\,{F}_{(k)}+D[{\bf g}]\,\tilde{F}_{(k)}\,.
\end{cases}
\end{align}
As long as the block $B[{\bf g}]$ is non-vanishing, this symmetry can only be on-shell since the Bianchi identity for the transformed ${F}_{(k)}'$, which guarantees that the transformed elementary field $C_{(k-1)}'$ be well defined, only holds if the field equations $d\tilde{F}_{(k)}=0$ for $C_{(k-1)}$ are satisfied \cite{Tanii:1998px}:
\begin{equation}
  d{F}_{(k)}'=A[{\bf g}]\,d{F}_{(k)}+B[{\bf g}]\,d\tilde{F}_{(k)}=B[{\bf g}]\,d\tilde{F}_{(k)}=0\,,
\end{equation}
The field strengths ${F}_{(k)}$ and $\tilde{F}_{(k)}$ transform in a linear representation ${\Scr R}$ of $G$ defined by the matrix:
\begin{equation}
  {\bf g}\in G\,\,\,\,\stackrel{{\Scr R}}{\longrightarrow }\,\,\,{\Scr R}[{\bf g}]=\left(\begin{matrix} A[{\bf g}] & B[{\bf g}]\cr C[{\bf g}] & D[{\bf g}] \end{matrix}\right)\,.\label{RgD}
\end{equation}
Just as in four dimensions, depending on which of the $C_{(k-1)}$ and $\tilde{C}_{(k-1)}$ are chosen to be described as elementary fields in the Lagrangian, the action will feature a different global symmetry $G_{el}$, though the global symmetry group $G$ of the field equations and Bianchi identities remains the same.
The constraints on ${\Scr R}$ derive from the non-minimal couplings of the scalar fields to the $(k-1)$-forms which are a direct generalization of those in four dimensions between the scalars and the vector fields, see (\ref{bosoniclagr})%
\footnote{
the Hodge dual ${}^*\omega$ of a generic $q$-form $\omega$ is defined as:
\begin{equation}
{}*\omega_{\mu_1\dots \mu_{D-q}}=\frac{e}{q!}\,\epsilon_{\mu_1\dots\mu_{D-q}\nu_1\dots \nu_q}\,\omega^{\nu_1\dots \nu_q}\,,
\end{equation}
where $\epsilon_{01\dots D-1}=1$. One can easily verify that ${}^{**}\omega=(-)^{q(D-q)}\,(-)^{D-1}\,\omega$
}:
\begin{equation}
\L_{\rm kin,\,C}=
-\frac{e\varepsilon}{2k!}\left(\I_{\Lambda\Sigma}(\phi)\,F^\Lambda_{\mu_1\dots \mu_k}\,F^{\Sigma\,\mu_1\dots \mu_k}+\R_{\Lambda\Sigma}(\phi)\,F^\Lambda_{\mu_1\dots \mu_k}\,{}^*F^{\Sigma\,\mu_1\dots \mu_k}\right)\,,\label{kinC}
\end{equation}
where $\mu=0,\dots, D-1$, $\Lambda,\Sigma=1,\dots, n_k$, being $n_k$ the number of $(k-1)$-forms $C_{(k-1)}$ and $\varepsilon\equiv (-)^{k-1}$. The matrices $\I_{\Lambda\Sigma}(\phi),\,\R_{\Lambda\Sigma}(\phi)$ satisfy the following properties:
 \begin{equation}
 \I_{\Lambda\Sigma}=\I_{\Sigma\Lambda}<0\,,\,\,\,\,\,\, \R_{\Lambda\Sigma}=-\varepsilon\,\R_{\Sigma\Lambda}\,.
 \end{equation}
Just as we did in four dimensions, see Eq.\ (\ref{GF}), we define dual field strengths:
\begin{equation}
  G_{\Lambda\,\mu_1\dots\,\mu_k}\equiv\varepsilon\,\epsilon_{\mu_1\dots\,\mu_k\nu_1\dots\nu_k}\,
 \frac{\delta \L}{\delta F^\Lambda_{\nu_1\dots \nu_k}}\,\,\Rightarrow\,\,\,\, G_\Lambda=-\I_{\Lambda\Sigma}\,{}^*F^\Sigma-\varepsilon\, \R_{\Lambda\Sigma}\,F^\Sigma\,,\label{defGk}
\end{equation}
where we have omitted the fermion terms, and define the vector of field strengths:
\begin{align}
\mathcal{G}= (\mathcal{G}^M)\equiv\left(\begin{matrix} F^\Lambda \cr  G_\Lambda\end{matrix}\right)\,.
\end{align}
The definition (\ref{defGk}) can be equivalently written in terms of the \emph{twisted self-duality condition} \cite{Cremmer:1997ct}:
\begin{equation}
 {}^*\mathcal{G}=-\mathbb{C}_\varepsilon\,\mathcal{M}(\phi)\,\mathcal{G}\,,\label{TSDCD}
\end{equation}
which generalizes (\ref{FCMF}), where
\begin{equation}
\mathbb{C}_\varepsilon\equiv (\mathbb{C}_\varepsilon^{MN})\equiv
\left(\begin{matrix}
\Zero & \Id   \cr
\varepsilon\,\Id & \Zero
\end{matrix}\right)
\,,\label{Ce}
\end{equation}
$\Id$, $\Zero$ being the $n_k\times n_k$ identity and zero-matrices, respectively, and
\begin{equation}
\mathcal{M}(\phi)= (\mathcal{M}(\phi)_{MN})\equiv
\left(\begin{matrix}(\I-\varepsilon\,\R\I^{-1}\R)_{\Lambda\Sigma} &
-(\R\I^{-1})_\Lambda{}^\Gamma\cr \varepsilon (\I^{-1}\R)^\Delta{}_\Sigma & \I^{-1\,
\Delta \Gamma}\end{matrix}\right)\,.\label{Me}
\end{equation}
The reader can easily verify that:
\begin{equation}
\mathcal{M}^T\,\mathbb{C}_\varepsilon\mathcal{M}=\mathbb{C}_\varepsilon\,.
\end{equation}
For $\varepsilon=-1$, which is the case of the vector fields in four dimensions, $\mathbb{C}_\varepsilon$ is the symplectic invariant matrix
and $\mathcal{M}$ is a symmetric, symplectic matrix, while for $\varepsilon=+1$, which is the case of 2-forms in six dimensions,
$\mathbb{C}_\varepsilon$ is the ${\rm O}(n_k,n_k)$-invariant matrix and $\mathcal{M}$ a symmetric element of ${\rm O}(n_k,n_k)$.
The Maxwell equations read:
\begin{equation}
d\mathcal{G}=0\,.\label{MDd}
\end{equation}
In order for (\ref{RgD}) to be a symmetry of eqs.\ (\ref{TSDCD}) and (\ref{MDd}) we must have:
\begin{equation}
\mathcal{M}({\bf g}\star  \phi)={\Scr R}[{\bf g}]^{-T}\mathcal{M}( \phi){\Scr R}[{\bf g}]^{-1}\,,
\end{equation}
and
\begin{equation}
{\Scr R}[{\bf g}]^{T}\mathbb{C}_\varepsilon {\Scr R}[{\bf g}]=\mathbb{C}_\varepsilon\,.
\end{equation}
This means that in $D=2k$ dimensions \cite{Gaillard:1981rj}:
\begin{align}
&\mbox{$k$ even:}\,\,\,\,\,\,\,\,\,\,\,\,\,\,{\Scr R}[G]\subset {\rm Sp}(2n_k,\mathbb{R})\,,\nonumber\\
&\mbox{$k$ odd:}\,\,\,\,\,\,\,\,\,\,\,\,\,\,{\Scr R}[G]\subset {\rm O}(n_k,n_k)\,.
\end{align}
In terms of the coset representative $\mathbb{L}(\phi)_{M}{}^{\underline{N}}$ in the representation ${\Scr R}$ (in which the right index refers to a basis where the transformations in $H$ act by means of orthogonal matrices, see the definition of hybrid coset representative given in Sect. \ref{gsg}), the matrix $\mathcal{M}_{MN}(\phi)$ reads:
\begin{equation}
\mathcal{M}_{MN}(\phi)=-\mathbb{L}(\phi)_{M}{}^{\underline{P}}\,\mathbb{L}(\phi)_{N}{}^{\underline{P}}\,.
\end{equation}
All other forms of rank $p\neq k-1$, which include the vector fields in $D>4$, will transform in linear representations of $G$. The corresponding kinetic Lagrangian only feature the first term of (\ref{kinC}), with no generalized theta-term ($\R=0$).
In this case the kinetic matrix $\mathcal{I}_{\Lambda\Sigma}$ has the form:
\begin{equation}
\mathcal{I}_{\Lambda\Sigma}(\phi)=-\mathbb{L}(\phi)_{\Lambda}{}^{\underline{\Gamma}}\,\mathbb{L}(\phi)_{\Sigma}{}^{\underline{\Gamma}}\,,
\end{equation}
where $\mathbb{L}(\phi)_{\Lambda}{}^{\underline{\Gamma}}$ is the coset representative in the $G$-representation ${\Scr R}$ of the $p$-form.
\par
\begin{figure}[H]
\begin{center}
\centerline{\includegraphics[width=0.7\textwidth]{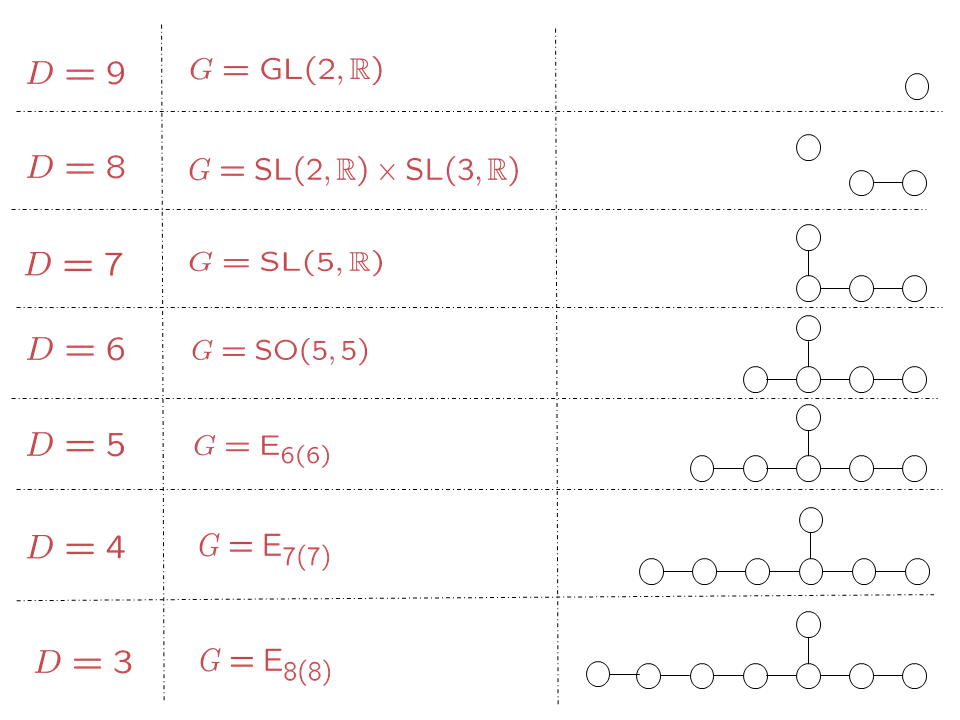}}
 \caption{\small The Dynkin diagrams associated with the global symmetry groups of the maximal supergravities in dimensions $3\le D\le 9$.}\label{figEs}
\end{center}
\end{figure}

If we compactify Type IIA/IIB or eleven-dimensional supergravity on a torus down to $D$-dimensions, we end up with an effective ungauged, maximal $D$-dimensional theory featuring form-fields of various order. Upon dualizing all form-fields to lower order ones, we end up with a formulation of the theory in which $G$ is maximal, and is described by the non-compact real form ${\rm E}_{11-D(11-D)}$ of the group ${\rm E}_{11-D}$. Here we use the symbol ${\rm E}_{11-D(11-D)}$ as a short-hand notation for the following groups:
\begin{align}
D&=9 \;:\quad\; G={\rm E}_{2(2)}\equiv {\rm GL}(2,\mathbb{R})\,,\nonumber\\
D&=8 \;:\quad\; G={\rm E}_{3(3)}\equiv {\rm SL}(2,\mathbb{R})\times {\rm SL}(3,\mathbb{R})\,,\nonumber\\
D&=7 \;:\quad\; G={\rm E}_{4(4)}\equiv {\rm SL}(5,\mathbb{R})\,,\nonumber\\
D&=6 \;:\quad\; G={\rm E}_{5(5)}\equiv {\rm SO}(5,5)\,,\nonumber\\
D&=5 \;:\quad\; G={\rm E}_{6(6)}\,,\nonumber\\
D&=4 \;:\quad\; G={\rm E}_{7(7)}\,,\nonumber\\
D&=3 \;:\quad\; G={\rm E}_{8(8)}\,.
\end{align}
Only for $D\le 5$, ${\rm E}_{11-D(11-D)}$ is a proper exceptional group. The Dynkin diagrams associated with the global symmetry groups in the various dimensions are related to one another as follows: The diagram of the $\mathfrak{e}_{11-D(11-D)}$ algebra in $D$-dimensions is obtained from that in $D-1$ dimensions by deleting the leftmost simple root, see Fig. \ref{figEs}. This defines simple inclusion relations among the various symmetry groups.
 The ungauged four-dimensional maximal supergravity was originally obtained from compactification of the eleven-dimensional one and dualization of all form-fields to lower order ones, in \cite{Cremmer:1978ds}, where the ${\rm E}_{7(7)}$ on-shell symmetry was found.\par
As mentioned earlier, in $D=10$ the Type IIA and IIB theories feature different global symmetry groups: $G_{{\rm IIA}}={\rm SO}(1,1)$ and $G_{IIB}= {\rm SL}(2,\mathbb{R})_{{\rm IIB}}$, respectively.
The latter encodes the conjectured S-duality symmetry of Type IIB string theory. In this theory $G_{{\rm IIB}}$ does not act as a duality group since the 5-form field strength is self-dual and is a $G_{{\rm IIB}}$-singlet.
\subsection{Tensor Hierarchy}\label{thi}
A $G$-covariant gauging \cite{deWit:2004nw,deWit:2005hv,Samtleben:2005bp,deWit:2008ta} is effected starting from the formulation of the ungauged theory in which $G$ is maximal and promoting a suitable global symmetry group of the Lagrangian $G_g\subset G$ to local symmetry. The choice of the gauge group is still completely encoded in a $G$-covariant embedding tensor $\Theta$:
\begin{equation}
\Theta\in {\Scr R}_{v*}\times \Adj(G)\,,\label{rtr}
\end{equation}
${\Scr R}_{v}$ being the representation of the vector fields, subject to a linear constraint dictated by supersymmetry which singles out in the above product a certain representation ${\Scr R}_\Theta$ for the embedding tensor, and a quadratic one expressing the $G_g$-invariance of $\Theta$. In general ${\Scr R}_{v}$ is the fundamental representation of $G$ and ${\Scr R}_{v*}$ its conjugate.

In Table \ref{tab:T-tensor-repr} we give, in the various $D$-dimensional maximal supergravities, the representations ${\Scr R}_\Theta$ of $\Theta$.\par
Just as in the duality covariant construction of the four-dimensional gaugings discussed above, one introduces all form-fields which are dual to the fields of the ungauged theory. All the form-fields will transform in representations of $G$ and dual forms of different order will belong to conjugate representations. In $D=2k$, in the presence of rank-$(k-1)$ antisymmetric tensors, this amounts to introducing the fields $\tilde{C}_{(k-1)\,\Lambda}$ dual to the elementary ones ${C}^\Lambda_{(k-1)}$, just as we did for the vector fields in four dimensions. Together they transform in the representation ${\Scr R}$ discussed above. By consistency each form-field is associated with its own gauge invariance. Only the fields of the original ungauged theory are described by kinetic terms, the extra fields enter in topological terms and in Stueckelberg-like combinations within the covariant field strengths. The latter, for a generic $p$-form field, can be schematically represented in the form (we suppress all indices)\footnote{This formula for $p=1$ should be modified by replacing $\mathcal{D}A^M$ by the correct definition of the non-Abelian field strengths $F^M_{\mu\nu}$: $\mathcal{D}A^M\rightarrow F^M=dA^M+X_{NP}{}^M\,A^N\wedge A^P$.}
\begin{equation}
F_{(p+1)}=\mathcal{D}C_{(p)}+Y_p[\Theta]\cdot C_{(p+1)}+\dots\,.
\end{equation}
where $Y_p[\Theta]$ is a constant \emph{intertwiner} tensor constructed out of $\Theta$ and of $G$-invariant tensors and $\cdot $ represents a contraction of indices in the appropriate representations, so that $Y_p[\Theta]\cdot C_{(p+1)}$ formally belongs to the same $G$-representation as $C_{(p)}$. On the other hand $Y_p[\Theta]$, for any $p$, should belong to the representation ${\Scr R}_\Theta$ of the embedding tensor. This allows to assign to each form-field to a unique representation ${\Scr R}_p$ of $G$, starting from the representations associated with the lowest order forms in the ungauged theory, see Table \ref{tabrepp}.
In particular the representation ${\Scr R}_2$ of the 2-forms is always contained in the symmetric product of two ${\Scr R}_1={\Scr R}_v$ and, in general, ${\Scr R}_p$ is contained in the product of ${\Scr R}_1$ and ${\Scr R}_{p-1}$. To understand the first property  let us recall that the $G$-tensor $X_{MN}{}^P$ naturally splits into the sum of its symmetric and antisymmetric components in the first two dindices:
\begin{equation}
X_{MN}{}^P=X_{(MN)}{}^P+X_{[MN]}{}^P\,.
\end{equation}
According to the closure constraint (\ref{quadratic2}), the former component vanishes when contracted with a gauge generator: $X_{(MN)}{}^P\,\Theta_P{}^\alpha$. The non-Abelian vector field strengths are then defined as in (\ref{FMdef}) and one can verify that, as we have shown in the four-dimensional case, their gauge variation contains a non-covariant piece proportional to $X_{(MN)}{}^P$. This is due to the failure of the generalized structure constants $X_{[MN]}{}^P$ to satisfy the Jacobi identity, see Eq. (\ref{nojacobi}).
Just as we did in $D=4$, we can dispose of these extra non-covariant terms by combining the field strengths $F^M_{\mu\nu}$ with the tensor fields $B_{{\tt I}\mu\nu}$, which, in $D$ dimensions will transform in a representation ${\Scr R}_2$, and defining:
\begin{equation}
\mathcal{H}^M_{\mu\nu}\equiv F^M_{\mu\nu}+Y_1^{M\,{\tt I}}\,B_{{\tt I}\mu\nu}\,.
\end{equation}
The non-covariant terms in $\delta F^M$ can be canceled by a corresponding transformation of the rank-2 antisymmetric tensors provided
the following property holds:
\begin{equation}
X_{(MN)}{}^P=-Y_1^{M\,{\tt I}}\,d_{{\tt I},\,MN}\,,
\end{equation}
where $d_{{\tt I},\,MN}$ is a $G$-invariant tensor and thus $Y_1^{M\,{\tt I}}$ belongs to the same $G$ representation as $X_{MN}{}^P$, i.e. as the embedding tensor. The above condition generalizes the linear constraint (\ref{linear22}) to a generic $D$-dimensional model. In the $D=4$ case ${\Scr R}_2={\bf 133}$, $Y_1^{M\,{\tt I}}=Z^{M\,\alpha}$ and $d_{{\tt I},\,MN}=t_{\alpha\,MN}$. The existence of $d_{{\tt I},\,MN}$ implies that ${\Scr R}_2$ has to be contained into the symmetric product of two ${\Scr R}_v$. In the five-dimensional maximal theory, $G={\rm E}_{6(6)}$, ${\Scr R}_v={\bf 27}'$ and ${\Scr R}_2={\bf 27}$, consistently with the general property that Hodge-dual fields transform in conjugate representations of $G$ (recall that in five dimensions vectors and 2-forms are Hodge dual to one another). The invariant quantity $d_{{\tt I},\,MN}$ is nothing but the characteristic ${\rm E}_{6(6)}$-invariant tensor $d_{MNP}$ defining the $D=5$ Chern-Simons term in the ungauged theory.
 \begin{table}[t]
\centering
\begin{tabular}{l l cccccc  }\hline
~&~&~&~&~&~&~& \\[-4mm]
~ &$~$& 1&2&3&4&5&6  \\   \hline
~&~&~&~\\[-4mm]
7   & ${\rm SL}(5)$ & ${\bf 10}'$  & ${\bf 5}$ & ${\bf 5}'$ &
${\bf 10}$ &  ${\bf 24}$ & ${\bf 15}'+{\bf 40}$  \\[1mm]
6  & ${\rm SO}(5,5)$ & ${\bf 16}_c$ & ${\bf 10}$ & ${\bf 16}_s$ &
  ${\bf 45}$ & ${\bf 144}_s$ &
$\!\!{\bf 10}\!+\! {\bf 126}_s\!+\! {\bf 320}\!\!$\\[.8mm]
5   & ${\rm E}_{6(6)}$ & ${\bf 27}'$ & ${\bf 27}$ & ${\bf 78}$
& ${\bf 351}$ & $\!\!{\bf 27}\!+\! {\bf 1728}\!\!$ &  \\[.5mm]
4   & ${\rm E}_{7(7)}$ & ${\bf 56}$ & ${\bf 133}$ & ${\bf 912}$ &
$\!\!\!{\bf 133}\!+\! {\bf 8645}\!\!\!$ &    \\[.5mm]
3   & ${\rm E}_{8(8)}$ & ${\bf 248}$ & ${\bf 1}\!+\! {\bf 3875}$ & ${\bf
  3875}\!+\! {\bf147250}$ & &
\\ \hline
\end{tabular}
%%%%%%%%%%%%%%%%%%%%%%%%%%%%%%%%%%%%%%%%%%%%%%%%%%%%%%%%%%%%%%%%%
\caption{\small
This Table is taken from \cite{deWit:2008ta} and shows the duality representations ${\Scr R}_p$ of
p-form fields for maximal supergravities in space-time dimensions $3\leq
D\leq 7$. Note that the representation ${\Scr R}_2$ of the 2-forms is always contained in the symmetric product of two ${\Scr R}_1={\Scr R}_v$. In general ${\Scr R}_p$ is contained in the product of ${\Scr R}_1$ and ${\Scr R}_{p-1}$.
}\label{tabrepp}
\end{table}
 The same assignment, for the various maximal theories, was derived using a different approach in \cite{Riccioni:2007au}.\par
The gauge variation of the $p$-form has the following schematic expression \cite{deWit:2008ta}:
\begin{equation}
\delta C_{(p)}=Y_p[\Theta]\cdot\Xi_{(p)}+\mathcal{D}\Xi_{(p-1)}+\dots\label{gfree}
\end{equation}
where the ellipses include $G_g$-transformations and, for $p>1$, non-covariant terms depending on $\delta A^M_\mu$, as last term on the right hand side of Eq. (\ref{Btra12}).
The embedding tensor defines through the tensors $Y_{p-1}[\Theta],\,Y_p[\Theta]$ a splitting of the $p$-forms into physical fields and unphysical components:
\begin{equation}
C_{(p)}^{{\rm (phys.)}}\equiv Y_{p-1}[\Theta]\cdot C_{(p)}\,\,,\,\,\,C_{(p)}=Y_p[\Theta]\cdot C_{(p)}^{{\rm (unphys.)}}+\mathring{C}_{(p)}\,,
\end{equation}
where $\mathring{C}_{(p)}$ only depends on ${C}_{(p)}^{{\rm (phys.)}}$.
Consistency requires the following orthogonality condition to be satisfied:
\begin{equation}
Y_{p-1}[\Theta]\cdot Y_p[\Theta]=0\,.\label{YYorth}
\end{equation}
It turns out that (\ref{YYorth}) is just a consequence of the quadratic constraint on $\Theta$ which expresses its gauge-invariance. From this property and from (\ref{gfree}) it follows that only the unphysical component of $C_{(p)}$ is affected by a gauge transformation of $C_{(p+1)}$, parametrized by $\Xi_{(p)}$.
Therefore the following mechanism is at work: $C_{(p)}^{{\rm (phys.)}} $ become massive by ``eating'' corresponding unphysical $(p-1)$-forms $C_{(p-1)}^{{\rm (unphys.)}} $, while $C_{(p)}^{{\rm (unphys.)}} $ are in turn gauged away and become degrees of freedom of massive $(p+1)$-forms.\footnote{For a rigorous treatment of this mechanism for the maximal theories in various dimensions we refer the reader to the original references cited at the beginning of this Section.} Just as in the four-dimensional model discussed in Sect. \ref{sec:4}, the embedding tensor defines the distribution of the physical degrees of freedom among the various fields by fixing the gauge freedom (\ref{gfree}) and solving the non-dynamical field equations. The linear and quadratic constraints guarantee the consistency of the construction. Therefore the formulation of a formally $G$-covariant gauging of a $D$-dimensional supergravity in terms of an embedding tensor requires the introduction of form-fields of all orders (up to the space-time dimension $D$).
The $G$-covariant selective couplings, discussed above, between forms of different order, determined by a single object $\Theta$, define the so-called \emph{tensor hierarchy} and was developed in the maximal theories, in \cite{deWit:2005hv,Samtleben:2005bp,deWit:2008ta} as a general $G$-covariant formulation of the gauging procedure in any dimension. It generalizes the duality-covariant gauging of four-dimensional supergravities, discussed in detail in Sect. \ref{sec:4}. In that case we had
\begin{equation}
Y_0[\Theta]^\alpha{}_M=\Theta_M{}^\alpha\,\,,\,\,\,Y_1[\Theta]^{M\,\alpha}=\frac{1}{2}\Theta^{M\,\alpha}\,,
\end{equation}
the reader can easily check that (\ref{YYorth}) is satisfied in virtue of the locality constraint (\ref{quadratic1}).
In principle we could have coupled our fields to three and four-forms, which are not dynamical.
Consistency requires $C_{(3)}$ to transform in the ${\bf 912}$ and $C_{(4)}$ in the ${\bf 133}+{\bf 8645}$, which is nothing but the representation ${\Scr R}_{\Theta\Theta}$ of the quadratic constraint $C_{MN}{}^\alpha=0$, see Sect. \ref{agtan}.
Let us just give the expression of $Y_2[\Theta]$ which intertwines between the representation ${\bf 133}$ of $B_{\alpha\,\mu\nu}$ and the ${\bf 912}$ of $C_{(3)}$:
\begin{equation}
Y_2[\Theta]_{\alpha,\,M}{}{}^\beta=-\Theta_M{}^\gamma\,{\rm f}_{\gamma\alpha}{}^\beta+t_{\alpha\,M}{}^N\,\Theta_N{}^\beta\,.
\end{equation}
The reader can easily verify that $Y_1[\Theta]\cdot Y_2[\Theta]=0$ is just the quadratic constraint (\ref{quadratic2}) on $\Theta$.\par
In a $D$-dimensional theory, forms of order $D-1$ and $D$ are not dynamical since their field strengths have no Hodge duals.
In general the representations of $C_{(D-1)}$ and $C_{(D)}$ are ${\Scr R}'_\Theta$ and ${\Scr R}'_{\Theta\Theta}$ conjugate to the representations of the embedding tensor and of the quadratic constraints, respectively. In the hierarchy picture we can treat the embedding tensor on an equal footing as all the other fields by coupling it with $C_{(D-1)}$ and $C_{(D)}$ through terms in the Lagrangian of the form \cite{deWit:2008ta}:
\begin{equation}
\Theta_M{}^\alpha\,\mathcal{D}C_{(D-1)}{}^M{}_\alpha+C_{MN}{}^\alpha\,C_{(D)}{}^{MN}{}_\alpha\,.
\end{equation}
The equation of motion of $C_{(D-1)}$  implies $\mathcal{D}\Theta=0$, while that of $C_{(D)}$ implements the quadratic constraints $C_{MN}{}^\alpha=0$, so that $\mathcal{D}\Theta=0$ is equivalent to stating that $\Theta$ is a constant tensor.
Therefore integrating out $C_{(D-1)}$ and $C_{(D)}$ we are left with $\Theta$ entering the theory as a constant spurionic object in the representation ${\Scr R}_\Theta$ and satisfying the quadratic constraints (\ref{quadratic2}) (or, equivalently, (\ref{quadratic1})).

In this formalism the maximal gauged supergravity in $D=5$ was constructed in \cite{deWit:2004nw}, generalizing previous works \cite{Gunaydin:1985cu,Andrianopoli:2000fi}; The general gauging of the six and seven -dimensional maximal theories were constructed in \cite{Bergshoeff:2007ef} and \cite{Samtleben:2005bp}, respectively, extending previous works \cite{Pernici:1984fe}; In $D=8$ the most general gaugings were constructed in \cite{Bergshoeff:2003ri},\cite{Andino:2016bwk}. We refer to these works for the details of the construction in the different cases.

\begin{table}[t]
\centering
\begin{tabular}{l l  l l  }\hline
~&~&~&~\\[-4mm]
$D$ &${\rm G}$& ${\rm H}$ & $\Theta$  \\   \hline
~&~&~&~\\[-4mm]
7   & ${\rm SL}(5)$ & ${\rm USp}(4)$  & ${\bf 10}\times {\bf
  24}= {\bf 10}+\underline{\bf 15}+  \underline{\overline{\bf 40}}+
{\bf 175}$
\\[1mm]
6  & ${\rm SO}(5,5)$ & ${\rm USp}(4) \times {\rm USp}(4)$ &
  ${\bf 16}\times{\bf 45} =
  {\bf 16}+ \underline{\bf 144} + {\bf 560}$ \\[.8mm]
5   & ${\rm E}_{6(6)}$ & ${\rm USp}(8)$ & ${\bf 27}\times{\bf
  78} =
  {\bf 27} + \underline{\bf 351} + \overline{\bf 1728}$  \\[.5mm]
4   & ${\rm E}_{7(7)}$ & ${\rm SU}(8)$  & ${\bf 56}\times{\bf 133} =
  {\bf 56} + \underline{\bf 912} + {\bf 6480}$   \\[.5mm]
3   & ${\rm E}_{8(8)}$ & ${\rm SO}(16)$ & ${\bf 248}\times{\bf 248} =
  \underline{\bf 1} + {\bf 248} + \underline{\bf 3875} +{\bf 27000}
  +  {\bf 30380}$
\\ \hline
\end{tabular}
\caption{\small
Decomposition of the embedding tensor $\Theta$ for maximal
supergravities in various space-time dimensions in terms of irreducible
${\rm G}$ representations
\cite{deWit:2002vt,deWit:2005hv}. Only the underlined representations
are allowed by supersymmetry. The R-symmetry group ${\rm H}$
is the maximal compact subgroup of ${\rm G}$.
}\label{tab:T-tensor-repr}
\end{table}
\section{Conclusions}
In this report we tried to give a comprehensive review of gauged extended supergravities. Special emphasis was given to the global symmetries of the ungauged models, which encode string/M-theory dualities and which survive as equivalences among theories after the gauging. The embedding tensor formalism allows to make this equivalence manifest and therefore it was central to our analysis.\par
Due to the vastness of the subject, in order to make the discussion as self-consistent as possible, choices had to be made as to which issues should be dealt with in detail. This unfortunately implied that a number of interesting topics had to be excluded from the review, or simply touched upon qualitatively. For instance we have focussed on four-dimensional theories giving just a brief overview of the higher-dimensional ones throughout the text, when dealing with compactifications, and in the final Section.\par A strong selection had also to be made on the applications of the gauging procedure to the description of string/M-theory dynamics of flux-backgrounds.
The topics of flux-compactifications and the physical effects of generalized fluxes, such as moduli stabilization, supersymmetry breaking, cosmological model building etc. definitely deserve a broader and more systematic treatment, which is however beyond the scope of the present report and which can be found in some excellent reviews, which are referred to throughout the manuscript.
\par
The frameworks of (extended) generalized geometry, DFT and exceptional field theory are particularly interesting since they  have recently allowed to make important progress in the direction of giving a higher-dimensional uplift to lower-dimensional gauged supergravity models. These approaches were only touched upon in Sect. \ref{Tdualcomp}. As mentioned in this Section,
an important open problem, is to broaden the class of lower-dimensional gauged supergravities which can be derived within these frameworks possibly through suitable Scherk-Schwarz ansaetze. There is constant progress in this direction \cite{Ciceri:2016dmd,Bosque:2016fpi,Cassani:2016ncu,Inverso:2016eet}. In exceptional field theory, for instance, results could be obtained by either deforming the theory or by relaxing, to a certain extent, the section constraint \cite{Ciceri:2016dmd}. Another perspective on the hidden gauge symmetries underlying eleven-dimensional or Type II theories is provided by the analysis of \cite{D'Auria:1982nx,Andrianopoli:2016osu}, of which it would be interesting to investigate possible connections, for instance, with DFT and exceptional field theory.   \par
One of the main purposes of this dissertation is to convey a general idea of the beautiful mathematical structure underlying  supergravity theories, and the central role of global and local symmetries in their construction. We hope to have achieved, at least in part, this objective.

\section{Acknowledgements}
I am deeply grateful to L. Andrianopoli, R. D'Auria and G. Inverso for a careful reading of the manuscript and useful comments. I am also very grateful to B. de Wit, S. Ferrara, T. Ortin, H. Samtleben and A. J. Di Scala for useful insights and stimulating discussions. This work was partially supported by \emph{Fondazione CRT} under the contract RF 2015.1084.
\appendix
\section{Appendices}
In these Appendices besides the definition of the notations and conventions used in this work, mathematical details related to topics dealt with in the main text are presented.
\subsection{Notations, Conventions and the Geometry of the Scalar Manifold}\label{nacv}
Let  $\mathcal{M}_4$ denote a curved  $D=4$ metric space-time, with Lorentzian metric tensor $g_{\mu\nu}(x)$, where $x=(x^\mu)$, $\mu=0,1,2,3$, describes a point in the space-time in a local coordinate system.
The vierbein basis is, as usual, defined by the condition:
\begin{equation}
g_{\mu\nu}(x)=V^a_\mu(x)\,V^b_\nu(x)\,\eta_{ab}\,,
\end{equation}
where $a,b,c,\dots=0,\dots, 3$ are the \emph{rigid indices} labeling the vierbein $V^a=V^a_\mu\,dx^\mu$ basis of $T^*\mathcal{M}_4$, or their duals $V_a=V_a^\mu\,\partial_\mu$ in $T\mathcal{M}_4$, where $V_a^\mu\,V^b_\mu=\delta_a^b$. The rigid metric is  $\eta={\rm diag}(+,-,-,-)$ (``mostly minus'' convention). The vierbein describe the local ``free-falling'' (or ``moving '') frame and are defined modulo a local Lorentz transformation $\boldsymbol{\Lambda}(x)=(\Lambda_b{}^a(x))$ to the right:
     \begin{equation}
V^a(x)\,\,\rightarrow\,\,\,\,V^b(x)\,\Lambda^{-1}{}_b{}^a(x)\,,\nonumber
\end{equation}
which leaves $\eta_{ab}$ invariant ($\Lambda_a{}^c \Lambda_b{}^d\,\eta_{cd}=\eta_{ab}$) and is the symmetry observed by the ``free-falling'' observer at $x=(x^\mu)$. Locally we can write $\boldsymbol{\Lambda}=\exp\left(\frac{1}{2}\theta^{ab}\,\mathcal{L}_{ab}\right)$, where the generators $\mathcal{L}_{ab}$ of the local Lorentz group close the following commutation relations:
\begin{equation}
[\mathcal{L}_{ab},\,\mathcal{L}_{cd}]=\eta_{bc}\,\mathcal{L}_{ad}+\eta_{ad}\,\mathcal{L}_{bc}
 -\eta_{bd}\,\mathcal{L}_{ac}-\eta_{ac}\,\mathcal{L}_{bd}\,.
\end{equation}
Let $\hat{\nabla}_\mu$ be a connection of \emph{metric type}, or \emph{metric-compatible}. This implies that, if $\hat{\Gamma}_{\mu\nu}^\sigma$ are the corresponding connection coefficients,
\begin{equation}
\hat{\nabla}_\mu g_{\nu\rho}=\partial_\mu g_{\nu\rho}-\hat{\Gamma}_{\mu\nu}^\sigma\,g_{\sigma\rho}-\hat{\Gamma}_{\mu\rho}^\sigma\,g_{\nu\sigma}=0\,.
\end{equation}
From this condition one finds:
\begin{equation}
\hat{\Gamma}^\sigma_{(\mu\nu)}=\frac{1}{2}\,g^{\sigma\gamma}\left(\partial_\mu g_{\gamma\nu}+\partial_\nu g_{\gamma\mu}-\partial_\gamma g_{\mu\nu}\right)+\frac{1}{2}\left(T_\mu{}^\sigma{}_\nu+T_\nu{}^\sigma{}_\mu\right)\,,
\end{equation}
where we have defined the \emph{torsion}:
\begin{equation}
T^\mu{}_{\nu\rho}\equiv \hat{\Gamma}^\mu_{\nu\rho}-\hat{\Gamma}^\mu_{\rho\nu}=2\,\hat{\Gamma}^\mu_{[\nu\rho]}\,.\label{Tmunudef}
\end{equation}
The coefficients $\hat{\Gamma}_{\mu\nu}^\sigma$ can then be computed as follows:
\begin{equation}
\hat{\Gamma}_{\mu\nu}^\sigma=\hat{\Gamma}_{[\mu\nu]}^\sigma+\hat{\Gamma}^\sigma_{(\mu\nu)}=\frac{1}{2}\,g^{\sigma\gamma}\left(\partial_\mu g_{\gamma\nu}+\partial_\nu g_{\gamma\mu}-\partial_\gamma g_{\mu\nu}\right)+K^\sigma{}_{\mu\nu}={\Gamma}_{\mu\nu}^\sigma+K^\sigma{}_{\mu\nu}\,,\label{GammaK}
\end{equation}
where:
\begin{equation}
K^\sigma{}_{\mu\nu}\equiv \frac{1}{2}\left(T_\mu{}^\sigma{}_\nu+T_\nu{}^\sigma{}_\mu+T^\sigma{}_{\mu\nu}\right)\,,
\end{equation}
 is called the \emph{contorsion}, and we have denoted by ${\Gamma}_{\mu\nu}^\sigma$ the torsionless Christoffel symbol:
 \begin{equation}
 {\Gamma}_{\mu\nu}^\sigma\equiv \frac{1}{2}\,g^{\sigma\gamma}\left(\partial_\mu g_{\gamma\nu}+\partial_\nu g_{\gamma\mu}-\partial_\gamma g_{\mu\nu}\right)={\Gamma}_{\nu\mu}^\sigma\,.
 \end{equation}
 It is known that in  supergravity the coupling of the gravitational field to its spin-$3/2$ superpartner, the gravitino, produces a torsion in the connection.\par
We give the expression of the Riemann curvature tensor:
\begin{equation}
R_{\mu\nu}{}^\sigma{}_\rho\equiv\partial_\mu \hat{\Gamma}_{\nu\rho}^\sigma-\partial_\nu \hat{\Gamma}_{\mu\rho}^\sigma-
\hat{\Gamma}_{\mu\rho}^\gamma\,\hat{\Gamma}_{\nu\gamma}^\sigma+\hat{\Gamma}_{\nu\rho}^\gamma\,\hat{\Gamma}_{\mu\gamma}^\sigma=-R_{\nu\mu}{}^\sigma{}_\rho=-R_{\mu\nu\,\rho}{}^\sigma\,,
\end{equation}
The definitions of the Ricci tensor and scalar are:
\begin{equation}
\mathcal{R}_{\mu\nu}\equiv R_{\mu\rho\nu}{}^\rho\,\,\,,\,\,\,\,\mathcal{R}=R_{\mu\nu}{}^{\mu\nu}\,.
\end{equation}
\paragraph{Spin connection.} The \emph{spin-connection} 1-form $\omega{}^a{}_b=\omega_{\mu}{}^a{}_b\,dx^\mu$ is
  defined by the following condition:
 \begin{equation}
 \partial_\mu V_\nu{}^a-\hat{\Gamma}_{\mu\nu}^\rho\,V_\rho{}^a+\omega_\mu{}^a{}_b\,V_{\nu}{}^b=0\,,\label{1vp}
\end{equation}
known as \emph{first vierbein postulate}. It is the connection associated with the invariance under local Lorentz transformations $\Lambda^a{}_b$ acting on the vierbein. Equation (\ref{1vp}) allows to express $\omega_{\mu}{}^a{}_b$ in terms of the connection coefficients $\hat{\Gamma}_{\mu\nu}^\rho$ and the vierbein. For vanishing torsion the whole space-time connection is of Levi-Civita type ($\hat{\Gamma}_{\mu\nu}^\rho={\Gamma}_{\mu\nu}^\rho$) and both $\Gamma_{\mu\nu}^\rho$ and the spin-connection only depend on the vierbein.\par
 Antisymmetrizing Eq. (\ref{1vp}) in $\mu\nu$ and using the definition (\ref{Tmunudef}),  we find:
\begin{equation}
T^a{}_{\mu\nu}\equiv V_\rho{}^a\,T^\rho{}_{\mu\nu}=2\,\hat{\Gamma}_{[\mu\nu]}^\rho\,V_\rho{}^a=\partial_{[\mu} V_{\nu]}{}^a+\omega_{[\mu}{}^a{}_b\,V_{\nu]}{}^b\,,
\end{equation}
where $T^a{}_{\mu\nu}$ is the torsion tensor with the upper rigid index.
We then define the \emph{torsion 2-form}:
\begin{equation}
T^a\equiv \frac{1}{2}\,T^a{}_{\mu\nu}\,dx^\mu\wedge dx^\nu=dV^a+\omega^a{}_b\wedge V^b\,.\label{cse1}
\end{equation}
Similarly one can compute the Riemann curvature tensor in the vierbein basis and find:
\begin{equation}
R_{\mu\nu}{}^a{}_b=2(\partial_{[\mu}\omega_{\nu]}{}^a{}_b+\omega_{[\mu}{}^a{}_c\,\omega_{\nu]}{}^c{}_b)\,,
\end{equation}
so that, defining the \emph{curvature 2-form} as follows:
\begin{equation}
R^a{}_b\equiv \frac{1}{2}\,R_{\mu\nu}{}^a{}_b\,dx^\mu\wedge dx^\nu\,,
\end{equation}
we have:
\begin{equation}
R^a{}_b=d\omega^a{}_b+\omega^a{}_c\wedge \omega^c{}_b\,.\label{cse2}
\end{equation}
It can be easily verified that, from the condition of  metric-compatibility of the connection $\hat{\nabla}_\mu g_{\nu\rho}=0$ and the definition of the vierbein matrices, the following property holds:
\begin{equation}
\omega^a{}_c\,\eta^{bc}=-\omega^b{}_c\,\eta^{ac}\,\,\Leftrightarrow\,\,\,\,\,\omega^a{}_b=-\omega_b{}^a\,,
\end{equation}
where the rigid indices $a,b,c,\dots$ are lowered and raised by $\eta_{ab}$ and its inverse matrix $\eta^{ab}$, respectively.
The following Bianchi identities hold:
\begin{align}
 dT^a+\omega^a{}_b\wedge T^b&=R^a{}_b\wedge V^b\,,\\
 dR^{ab}+\omega^a{}_c\wedge R^{cb}-\omega^b{}_c\wedge R^{ca}&=0\,.
\end{align}
Notice that, for a Levi-Civita connection, the torsion vanishes, so that $T^a=dT^a=0$. The Bianchi identity for $T^a$ then implies $R^a{}_b\wedge V^b=0$, that is:
\begin{equation}
R^a{}_b\wedge V^b=0\,\,\Leftrightarrow\,\,\,\,\,R_{[abc]}{}^d=0\,,
\end{equation}
where $R_{abc}{}^d\equiv V_a^\mu V_b^\nu R_{\mu\nu c}{}^d$.
If a field $\Phi(x)$ transforms in a representation ${\bf D}$ of the local Lorentz group we denote by $\nabla_\mu$ the covariant derivative only containing a space-time connection which consists in the Christoffel symbol $\Gamma_{\mu\nu}^\rho$ if $\Phi$ has curved indices $\mu,\,\nu,\dots$, and the spin connection:
\begin{equation}
\nabla\equiv d +\Gamma+\frac{1}{2}\,\omega_{ab}\,{\bf D}(\mathcal{L}^{ab})\,.
\end{equation}
Covariance of the above derivative with respect to a local Lorentz transformation $\Lambda_b{}^a$ requires the spin-connection to transform  as follows:
\begin{equation}
\omega^{ab}\,\,\rightarrow\,\,\,\omega^{\prime\, ab}=\Lambda^a{}_c\,\omega^{cd}\,\Lambda^{-1}{}_d{}^b+\Lambda^a{}_c d\Lambda^{-1\,cb}\,.\label{deltaomega1}
\end{equation}
The these conventions, the definition of the torsion tensor can then be written in the compact form: $T^a{}_{\mu\nu}=\nabla_{[\mu}V_{\nu]}{}^a$, where the Christoffel symbol $\Gamma_{\mu\nu}^\rho$ does not contribute to the right-hand -side being anti-symmetrized in its lower indices.
\paragraph{Spinor conventions.} Fermions on a curved space-time are described by fields transforming with respect to the local Lorentz group ${\rm SL}(2,\mathbb{C})$ in the spinor representation $\left(\frac{1}{2},0\right)\oplus\left(0,\frac{1}{2}\right)$, in which the generators $\mathcal{L}_{ab}$ act through the matrices
\begin{equation}
{\bf D}(\mathcal{L}_{ab})=\frac{\gamma_{ab}}{2}\,\,\,;\,\,\,\,\,\gamma_{ab}\equiv \frac{1}{2}\,[\gamma^a,\,\gamma^b]\,.\label{Lorentzspin}
\end{equation}
The constant $\gamma$-matrices $\gamma^a$ are defined by the condition $\{\gamma^a,\,\gamma^b\}=2\,\eta^{ab}$ and are chosen of the following form:
\begin{equation}
\gamma^a=\left(\begin{matrix}{\bf 0} & \sigma^a\cr \bar{\sigma}^a & {\bf 0}\end{matrix}\right)\,\,;\,\,\,\sigma^a=({\bf 1},\sigma^I)\,\,;\,\,\,\,\bar{\sigma}^a=({\bf 1},-\sigma^I)\,\,\,\,\,(I=1,2,3)\,,
\end{equation}
$\sigma^I$ being the three Pauli matrices:
\begin{equation}
\sigma^1=\left(\begin{matrix}0 & 1\cr 1 & 0\end{matrix}\right)\,\,;\,\,\,\,\sigma^2=\left(\begin{matrix}0 & -i\cr i & 0\end{matrix}\right)\,\,;\,\,\,\,\sigma^3=\left(\begin{matrix}1 & 0\cr 0 & -1\end{matrix}\right)\,.\label{Paulim}
\end{equation}
One can verify that:
\begin{equation}
\gamma^0\gamma^{a} \gamma^{0}=(\gamma^{a})^\dagger=\eta^{aa}\,\gamma^a\,.
\end{equation}
The metric-dependent matrices $\gamma^\mu(x)\equiv V_a{}^\mu(x)\,\gamma^a $, satisfy
\begin{equation}
\{\gamma^\mu,\,\gamma^\nu\}=2\,g^{\mu\nu}(x)\,.
\end{equation}
The constant matrix $\gamma^5$ is defined as:
\begin{equation}
\gamma^5\equiv\frac{i}{4!}\,\epsilon_{abcd}\gamma^a\gamma^b\gamma^c\gamma^d=\frac{i\,e}{4!}\,\epsilon_{\mu\nu\rho\sigma}\gamma^\mu\gamma^\nu\gamma^\rho\gamma^\sigma
=\left(\begin{matrix}-{\bf 1} & {\bf 0}\cr {\bf 0} & {\bf 1}\end{matrix}\right)\,,
\end{equation}
where ${e}\equiv {\rm det}(V_\mu{}^a)=\sqrt{|{\rm det}(g_{\mu\nu})|}$ and $\epsilon_{0123}=-\epsilon^{0123}=1$. Defining \begin{equation}
\gamma^{a_1\dots a_k}\equiv \gamma^{[a_1}\dots\gamma^{a_k]}\,,
\end{equation}
the following relations hold:
\begin{align}
\gamma^5\gamma_a&=-\frac{i}{3!}\,\epsilon_{abcd}\gamma^{bcd}\,\,;\,\,\,\gamma^5\gamma_{ab}=
-\frac{i}{2}\,\epsilon_{abcd}\gamma^{cd}\,,\label{gamma5mus0}\\
\gamma^5\gamma_{abc}&=i\,\epsilon_{abcd}\gamma^{d}\,\,;\,\,\,\gamma^5\gamma_{abcd}=i\,
\epsilon_{abcd}\,.\label{gamma5mus}
\end{align}
 We use for the complex conjugation of  Grassmann numbers the following convention: $(\xi_1\,\xi_2)^*=\xi_2^*\,\xi_1^*$.
 Dirac and complex conjugations on a spinor $\psi$ are defined, respectively,  as follows:
 \begin{equation}
 \bar{\psi}\equiv \psi^\dagger\gamma^0\,\,;\,\,\,\psi_c\equiv C\,\bar{\psi}^T\,,
 \end{equation}
 where the charge conjugation matrix $C$ is chosen to be $C=-i\gamma^2\,\gamma^0$ and satisfies the properties:
 \begin{align}
 C^{-1} \gamma^\mu\,C&=-(\gamma^\mu)^T\,\,;\,\,\,C=C^*=-C^T=-C^{-1}\,.
 \end{align}
The following relations hold:
\begin{align}
(C\gamma^{a_1\dots a_k})^T&=-(-1)^{\frac{k(k+1)}{2}}\,C\gamma^{a_1\dots a_k}\,,\nonumber\\
\bar{\chi}_c\gamma^{a_1\dots a_k}\lambda&=(-1)^{\frac{k(k+1)}{2}}\,\bar{\lambda}_c\gamma^{a_1\dots a_k}\chi\,,\nonumber\\
(\bar{\chi}_c\gamma^{a_1\dots a_k}\lambda)^*&=(-1)^k\,\bar{\chi}\gamma^{a_1\dots a_k}\lambda_c\,.\label{gprops}
\end{align}
The spinor representation can be reduced by imposing the  Majorana condition on spinors:
\begin{equation}
\psi=\psi_c=C\,\bar{\psi}^T\,.
\end{equation}
Of particular use is the basic Fierz identity:
\begin{align}
\lambda\,\bar{\chi}=-\frac{1}{4}\,(\bar{\chi}\lambda)-\frac{1}{4}\,(\bar{\chi}\gamma^5\lambda)\,\gamma^5-\frac{1}{4}\,(\bar{\chi}\gamma^\mu
\lambda)\,\gamma_\mu+\frac{1}{4}\,(\bar{\chi}\gamma^5\gamma^\mu\lambda)\,\gamma^5\gamma_\mu+\frac{1}{8}\,
(\bar{\chi}\gamma^{\mu\nu}\lambda)\,\gamma_{\mu\nu}\,.\label{Fierzbasic}
\end{align}
Applying the above identity to a  single spinor 1-form $\Psi=\Psi_\mu\,dx^\mu$, the only non-vanishing bilinears are $\bar{\Psi}\wedge \gamma^\mu\Psi$ and  $\bar{\Psi}\wedge \gamma^{\mu\nu}\Psi$, so that:
\begin{align}
\Psi \wedge\bar{\Psi}&=\frac{1}{4}\,(\bar{\Psi}\wedge \gamma^\mu
\Psi)\,\gamma_\mu-\frac{1}{8}\,
(\bar{\Psi}\wedge\gamma^{\mu\nu}\Psi)\,\gamma_{\mu\nu}\,,\label{Fierzpsi}
\end{align}
from which it follows:
\begin{equation}
\gamma^a\,\Psi\,\wedge\bar{\Psi}\,\wedge\gamma_a\Psi=0\,.\label{Fierz2}
\end{equation}
Below are other useful properties of the $\gamma$-matrices:
\begin{align}
\gamma_{\mu\nu}\gamma^\rho&=2\,\gamma_{[\mu} \delta_{\nu]}^\rho+\gamma_{\mu\nu}{}^\rho=2\,\gamma_{[\mu} \delta_{\nu]}^\rho+ie\,\epsilon_{\mu\nu}{}^{\rho\sigma}\gamma^5\gamma_{\sigma}\,,\nonumber\\
\gamma_{\mu\nu}\gamma^{\rho\sigma}&=\gamma_{\mu\nu}{}^{\rho\sigma}-4 \, \delta^{[\rho}_{[\mu}\gamma_{\nu]}{}^{\sigma]}-2\,\delta^{\rho\sigma}_{\mu\nu}\,,\nonumber\\
\gamma^{[\rho}\gamma_{\mu\nu}\gamma^{\sigma]}&=\gamma_{\mu\nu}{}^{\rho\sigma}+2\,\delta^{\rho\sigma}_{\mu\nu}=
2\,(\delta^{\rho\sigma}_{\mu\nu}+\frac{i e}{2}\,\epsilon_{\mu\nu}{}^{\rho\sigma}\,\gamma^5)\,,\nonumber\\
\gamma_\rho\gamma^{\mu_1\dots\mu_k}\gamma^\rho&=2(-1)^k(2-k)\,\gamma^{\mu_1\dots\mu_k}\,.\label{gamprop}
\end{align}
We define the self-dual and anti-self-dual  components of a rank-2, covariant antisymmetric tensor $F_{\mu\nu}$ as follows:
\begin{equation}
F_{\mu\nu}^{\pm}\equiv \frac{F_{\mu\nu}\pm i\,{}^*F_{\mu\nu}}{2}\,\,\,\,\,\,\Rightarrow\,\,\,\,{}^*F_{\mu\nu}^{\pm}=\mp i\,F_{\mu\nu}^{\pm}\,.
\end{equation}
From (\ref{gamma5mus0}) and the first of (\ref{gamprop}) we find:
\begin{equation}
F^+_{\mu\nu}\gamma^{\mu\nu}\epsilon_\bullet=F^-_{\mu\nu}\gamma^{\mu\nu}\epsilon^\bullet=0\,,
\end{equation}
and
\begin{align}
F^+_{\mu\nu}\gamma^{\mu\nu}\gamma_\rho\epsilon_\bullet=-4\,F^+_{\rho\nu}\gamma^\nu\epsilon_\bullet\,\,;\,\,F^+_{\mu\nu}\gamma^{\mu\nu}\gamma_\rho\epsilon^\bullet=0\,,\nonumber\\
F^-_{\mu\nu}\gamma^{\mu\nu}\gamma_\rho\epsilon^\bullet=-4\,F^-_{\rho\nu}\gamma^\nu\epsilon^\bullet\,\,;\,\,
F^-_{\mu\nu}\gamma^{\mu\nu}\gamma_\rho\epsilon_\bullet=0\,,
\end{align}
where, unless otherwise stated, we use the convention that lower and upper ${\rm SU}(\mathcal{N})$ indices ``$\bullet$'' are associated with positive and negative chiralities, respectively
\begin{equation}
\gamma^5\epsilon_\bullet=\epsilon_\bullet\,\,,\,\,\,\gamma^5\epsilon^\bullet=-\epsilon^\bullet\,.
\end{equation}
If $\epsilon_\bullet,\,\epsilon^\bullet$ are the chiral projections of a same Majorana spinor we have $\epsilon^\bullet=(\epsilon_\bullet)_c$.\par

\paragraph{The geometry of the scalar manifold.}
Let us here recall the relevant definitions and conventions about the scalar manifolds, referring the reader to standard textbooks \cite{Helgason,Nomizu} for a rigorous discussion of topic. As pointed out in Sect. \ref{ghsect} the scalar manifold $\Mscal$ in supergravity is a Riemannian, non-compact manifold. We shall always assume it to be simply connected. Let $n_s$ denote its real dimension, i.e. the number of real scalar fields of the theory, $s,t,r,\dots$ the curved indices (analogous to the space-time indices $\mu,\,\nu,\dots$) and  $\underline{s},\underline{r},\underline{t},\dots$ the rigid tangent-space indices (analogous to the space-time indices $a,\,b,\dots$). Let $\phi\equiv (\phi^s)$ be local coordinates. The positive definite metric tensor is denoted by ${\Scr G}_{st}(\phi)$. Just as for space-time, we introduce a set of vielbein 1-forms $\mathcal{P}^{\underline{s}}=\mathcal{P}_t{}^{\underline{s}}\,d\phi^t$ and a dual basis on the tangent space $K_{\underline{s}}=\mathcal{P}_{\underline{s}}{}^t\,\frac{\partial}{\partial\phi^t}$, where $\mathcal{P}_{\underline{s}}{}^t$ is the inverse of the $\mathcal{P}_t{}^{\underline{s}}$ matrix: $\mathcal{P}_t{}^{\underline{s}}\,\mathcal{P}_{\underline{s}}{}^r=\delta_t^r$.
 If $\eta_{\underline{s}\underline{t}}$ denotes the constant $H$-invariant matrix describing the metric in the basis $K_{\underline{s}}$, just as we did for space-time, we write:
\begin{equation}
{\Scr G}_{st}(\phi)=\mathcal{P}_s{}^{\underline{s}}(\phi)\mathcal{P}_t{}^{\underline{t}}(\phi)\eta_{\underline{s}\underline{t}}\,.
\end{equation}
 On $\Mscal$ we define a Levi-Civita (i.e. torsionless, metric-compatible) connection by the first vielbein postulate:
 \begin{equation}
 {\Scr D}_s\mathcal{P}_t{}^{\underline{r}}\equiv \partial_s \mathcal{P}_t{}^{\underline{r}}-\tilde{\Gamma}_{st}^r\,\mathcal{P}_r{}^{\underline{r}}+
 \mathcal{Q}_s{}^{\underline{r}}{}_{\underline{t}}\,\mathcal{P}_t{}^{\underline{t}}=0\,,\label{1vp2}
 \end{equation}
where  $\tilde{\Gamma}_{st}^r$ is the Christoffel symbol and $\mathcal{Q}_s{}^{\underline{r}}{}_{\underline{t}}$ the analogous to the  spin-connection on space-time and ${\Scr D}_s$ the corresponding covariant derivative.
Associated with this connection is the curvature rank-2 tensor:
\begin{equation}
R(\mathcal{Q})^{\underline{t}}{}_{\underline{s}}\equiv d\mathcal{Q}^{\underline{t}}{}_{\underline{s}}+\mathcal{Q}^{\underline{t}}{}_{\underline{r}}\wedge \mathcal{Q}^{\underline{r}}{}_{\underline{s}}=\frac{1}{2}\,R_{st}{}^{\underline{t}}{}_{\underline{s}}\,d\phi^s\wedge d\phi^t\,.
\end{equation}
 This connection defines the holonomy group $H$ and $R(\mathcal{Q})^{\underline{t}}{}_{\underline{s}}$ has values in the corresponding Lie algebra. Just as the Lorentz group in space-time, $H$ acts as local transformations on the rigid indices of $\mathcal{P}^{\underline{s}}$ and $K_{\underline{s}}$.
 The condition that the connection be metric implies:
\begin{equation}
\mathcal{Q}^{\underline{t}}{}_{\underline{r}}\,\eta^{\underline{r}\underline{s}}=-\mathcal{Q}^{\underline{s}}{}_{\underline{r}}\,
\eta^{\underline{r}\underline{t}}\,\,\Leftrightarrow\,\,\,\,\,\,\,\mathcal{Q}^{\underline{t}}{}_{\underline{r}}=
-\mathcal{Q}_{\underline{r}}{}^{\underline{t}}\,,\label{Qpropslc}
\end{equation}
where rigid indices, as usual, are raised by $\eta^{\underline{r}\underline{s}}$ and lowered by $\eta_{\underline{r}\underline{s}}$.\par
The metric may feature isometries, namely diffeomorphisms which leave the metric invariant:
 \begin{equation}
\phi^s\rightarrow \phi^{\prime s}(\phi)\,:\,\,\,\Gm_{s't'}(\phi'(\phi))\frac{\partial \phi^{\prime s^\prime}}{\partial \phi^s}\frac{\phi^{\prime t^\prime}}{\partial \phi^t}=\Gm_{st}(\phi)\,.\label{Gscalisom0}
 \end{equation}
 These close a group $G$.\par
A Riemannian manifold $\Mscal$ is \emph{homogenous} if its isometry group $G$ has a transitive action on it, namely any two points of the manifold are connected by an isometry transformation. The subgroup $H_P'$ of $G$ which leaves a point $P$ of homogeneous manifold $\Mscal$ invariant is the \emph{isotropy group} of $P$. The isotropy groups of any two points are isomorphic to a same group $H'$ and the homogeneous manifold can be described as a coset manifold $G/H'$. Indeed
if we fix a reference point $O$ in a homogeneous $\Mscal$, we can associate with every element ${\bf g}\in G$ a unique point $P\in \Mscal$ as follows
\begin{equation}
\forall {\bf g}\in G\,\,\rightarrow\,\,\,\,\exists P\in \Mscal\,\,:\,\,\,\,\,{\bf g}\cdot O=P\,,
\end{equation}
where $\cdot$ denotes the action of an element of $G$ on the manifold.
By definition of homogeneous manifolds the above mapping is onto. However it is not one-to-one since in general there are more elements of $G$ corresponding to a same point $P$. These are connected by the right action of an element of the isotropy group $H'=\{h\in G\vert\, h\cdot O=O\}$:
\begin{equation}
{\bf g}\cdot O=P\,\,,\,\,\,{\bf g}'\cdot O=P\,\,\Rightarrow\,\,\,\,\,{\bf g}^{-1}{\bf g}'\cdot O=O\,\,\Rightarrow\,\,\,\,\,{\bf g}'\in {\bf g}\,H'\,.
\end{equation}
Therefore there is a one-to-one correspondence between points of  $\Mscal$ end elements of the coset space $G/H'$, namely elements of $G$ modulo the right action of $H'$. The origin $O$ corresponds to the coset $\Id\,H'$, $\Id$ being the unit element of $G$.\par
Under general assumptions which hold for supergravity scalar manifolds, the connected component of the isotropy group $H'$ is contained in the connected component of the holonomy group $H$. For symmetric spaces the two components coincide. In our discussion, when dealing with the isotropy or the holonomy groups we shall always implicitly refer to the corresponding connected components since we are interested in the local properties of these groups, encoded in their Lie algebras. Therefore, by an abuse of notation, write $H'\subseteq H$ and $H'=H$ in the symmetric case.\par
Let us now illustrate how to compute the $H$-connection $\mathcal{Q}$ for a homogeneous, non-symmetric manifold $G/H'$. In this case, as pointed out in Sect. \ref{ghsect}, we cannot find a subspace $\mathfrak{K}$ of the isometry algebra $\mathfrak{g}$, complement to the isotropy algebra $\mathfrak{H}$, which closes by commutation on the latter, as in Eq. (\ref{KKH}). In general, if $\{K_{\underline{s}}\}$ is a basis of $\mathfrak{K}$, we have:
\begin{equation}
[K_{\underline{s}},\,K_{\underline{t}}]={\rm f}_{\underline{s}\underline{t}}{}^I\,{J_I}+{\rm f}_{\underline{s}\underline{t}}{}^{\underline{r}}\,K_{\underline{r}}\,,
\end{equation}
where, as usual, $J_I$ define a basis of the holonomy algebra which contains $\mathfrak{H}$.
For symmetric manifolds ${\rm f}_{\underline{s}\underline{t}}{}^{\underline{r}}=0$. Let us define the left-invariant one-form $\Omega$
\begin{equation}
\Omega=L^{-1}dL=\mathcal{P}+\mathcal{Q}'\,,
\end{equation}
where $\mathcal{P}$ and $\mathcal{Q}'$ are the components of $\Omega$ on $\mathfrak{K}$ and the isotropy algebra $\mathfrak{H}$, respectively. We can immediately see that $\mathcal{Q}'$ is not the $H$-connection of the manifold by writing the Maurer-Cartan equation $d\Omega+\Omega\wedge \Omega=0$ and projecting it on $\mathfrak{K}$:
\begin{equation}
d\mathcal{P}^{\underline{s}}+\mathcal{Q}^{\prime\,\underline{s}}{}_{\underline{t}}\wedge \mathcal{P}^{\underline{t}}+\frac{1}{2}\,
{\rm f}_{\underline{r}\underline{t}}{}^{\underline{s}}\,\mathcal{P}^{\underline{r}}\wedge \mathcal{P}^{\underline{t}}=0\,.\label{nsh1}
\end{equation}
In order to write it in the form
\begin{equation}
{\Scr D}\mathcal{P}^{\underline{s}}\equiv d\mathcal{P}^{\underline{s}}+\mathcal{Q}^{\underline{s}}{}_{\underline{t}}\wedge \mathcal{P}^{\underline{t}}=0\,,
\end{equation}
we need to define the connection 1-form $\mathcal{Q}^{\underline{s}}{}_{\underline{t}}$ as follows:
\begin{equation}
\mathcal{Q}^{\underline{s}}{}_{\underline{t}}=\mathcal{Q}^{\prime\underline{s}}{}_{\underline{t}}+\Delta \mathcal{Q}^{\underline{s}}{}_{\underline{t}}\,.\label{QQpDQ}
\end{equation}
If $\eta_{\underline{s}\underline{t}}$ is the constant $H'$-invariant metric computed in the non-coordinate basis $K_{\underline{s}}$, we require properties (\ref{Qpropslc}) to hold. From (\ref{nsh1}), writing $\Delta \mathcal{Q}^{\underline{s}}{}_{\underline{t}}=\Delta \mathcal{Q}_{\underline{r}}{}^{\underline{s}}{}_{\underline{t}}\,\mathcal{P}^{\underline{r}}$, we have:
\begin{equation}
\Delta \mathcal{Q}_{[\underline{r}}{}^{\underline{s}}{}_{\underline{t}]}=\frac{1}{2}\,
{\rm f}_{\underline{r}\underline{t}}{}^{\underline{s}}\,.
\end{equation}
Using the above property and the metric compatibility requirement (\ref{Qpropslc}) we can determine $\Delta \mathcal{Q}$:
\begin{equation}
\Delta \mathcal{Q}_{\underline{r}}{}^{\underline{s}}{}_{\underline{t}}=\frac{1}{2}\,\left({\rm f}_{\underline{r}\underline{t}}{}^{\underline{s}}+{\rm f}_{\underline{s}'\underline{t}}{}^{\underline{r}'}\eta_{\underline{r}'\underline{r}}\eta^{\underline{s}'\underline{s}}+
{\rm f}_{\underline{s}'\underline{r}}{}^{\underline{r}'}\eta_{\underline{r}'\underline{t}}\eta^{\underline{s}'\underline{s}}\right)\,.\label{DeltaQNomi}
\end{equation}
Equation (\ref{QQpDQ}) clearly shows that in the non-symmetric case the isotropy algebra $\mathfrak{H}$ is strictly contained in the holonomy one, which $\mathcal{Q}$ belongs to. Using $\mathcal{Q}$ we then define the curvature 2-form and the covariant derivatives occurring in the supergravity Lagrangian. Homogeneous special K\"ahler and quaternionic K\"ahler manifolds are characterized by restrictions on their respective holonomy groups, which imply conditions on $\eta_{\underline{s}\underline{t}}$.\par
We have seen in Sect. \ref{ghsect} that all homogeneous scalar manifolds $\Mscal$ in supergravity admit a global \emph{solvable parametrization}, which amounts to writing $\Mscal$ as globally isometric to a solvable Lie group manifold $G_S$ generated by a solvable Lie algebra ${\Scr S}\subset \mathfrak{g}$: $\Mscal\sim G_S=e^{{\Scr S}}$. This means that the two manifolds have the same metric properties provided we define on the tangent space to $G_S$ at its origin (the unit element of the group), which is isomorphic to ${\Scr S}$, a metric which, in a suitable basis, coincides with the metric $\eta_{\underline{r}\underline{s}}$ at the origin on $\Mscal$. Let $\{T_s\}$, $s=1,\dots, n_s$, be a basis of generators of ${\Scr S}$ which satisfy the following commutation relations:
\begin{equation}
[T_r,\,T_s]={\rm C}_{rs}{}^t\,T_t\,.\label{solstru}
\end{equation}
This subalgebra of $ \mathfrak{g} $ is not orthogonal to $\mathfrak{H}$ and, being it solvable, in a suitable basis of a matrix representation, all its generators can be put in an upper- (or lower-) triangular form. In particular we can project $T_r$ on the subspaces $\mathfrak{K},\,\mathfrak{H}$. Let us choose a basis $T_r$ so that their component on $\mathfrak{K}$ are precisely the basis elements $K_{\underline{s}}$ on which the metric of $\Mscal$ has the form $\eta_{\underline{r}\underline{s}}$:
\begin{equation}
T_s=K_{\underline{s}}+J_{\underline{s}}\,\,;\,\,\,\,K_{\underline{s}}\in \mathfrak{K}\,\,,\,\,\,J_{\underline{s}}\in \mathfrak{H}\,.\label{TsdecKJ}
\end{equation}
In symmetric spaces, for a suitable basis of the representation space, the above decomposition amounts to writing upper- (or lower-)
triangular matrices as sums of hermitian and anti-hermitian ones. In non-symmetric spaces some of the elements $K_{\underline{s}}$ may be nilpotent and thus coincide with the corresponding $T_s$, so that $J_{\underline{s}}=0$. We now introduce the following inner product on the tangent space to $G_S$ spanned by $\{T_s\}$: $(T_s,\,T_t)=\eta_{\underline{r}\underline{s}}$. This defines a metric on $G_S$ and we associate with it a Levi-Civita connection on the solvable group. In the solvable coordinate description,  the coset representative on $\Mscal$  is defined as an element of $G_S$:
\begin{equation}
L(\phi^s)=\exp(\phi^s\,T_s)\in G_S\,.
\end{equation}
Being a group element, if we define the left invariant 1-form $\Omega$, it will belong to ${\Scr S}$ and thus expand in $T_s$:
\begin{equation}
\Omega=L^{-1}dL=\mathcal{P}^{{s}}\,T_s=\mathcal{P}^{{s}}\,K_{\underline{s}}+
\mathcal{P}^{{s}}\,J_{\underline{s}}=\mathcal{P}+\mathcal{Q}\,,
\end{equation}
where we have used (\ref{TsdecKJ}).
We note that the components $\mathcal{P}^{{s}}$ of $\Omega$ along $T_s$ coincide with the vielbein 1-forms $\mathcal{P}^{\underline{s}}$ on $\Mscal$. They satisfy the Maurer-Cartan equations dual to (\ref{solstru}):
\begin{equation}
d\mathcal{P}^s+\frac{1}{2}{\rm C}_{rt}{}^s\,\mathcal{P}^r\wedge \mathcal{P}^t=0\,.
\end{equation}
The metric on $G_S$ at any point $\phi$ is obtained from that on its tangent space at the origin through the action of the left-translation by the element $L(\phi)$ of $G_S$, and  coincides with that on $\Mscal$:
\begin{equation}
ds^2=(\Omega,\,\Omega)=\mathcal{P}^s\,\mathcal{P}^t \eta_{\underline{r}\underline{s}}=\mathcal{P}^{\underline{s}}\,\mathcal{P}^{\underline{t}} \eta_{\underline{r}\underline{s}}\,.
\end{equation}
Being $\mathcal{P}^s=\mathcal{P}^{\underline{s}}$ we can define the Levi-Civita connection  associated with $\eta_{\underline{r}\underline{s}}$ by writing the above equation in the form
  \begin{equation}
 {\Scr D}\mathcal{P}^{\underline{r}}\equiv d \mathcal{P}^{\underline{r}}+
 \mathcal{Q}^{\underline{r}}{}_{\underline{t}}\,\wedge \mathcal{P}^{\underline{t}}=0\,,\label{1vp22}
 \end{equation}
  with $\mathcal{Q}^{\underline{r}}{}_{\underline{t}}$ satisfying the property (\ref{Qpropslc}). We find for this connection a formula analogous to (\ref{DeltaQNomi})
 \begin{equation}
\mathcal{Q}^{\underline{s}}{}_{\underline{t}}=\frac{1}{2}\,\left({\rm C}_{\underline{r}\underline{t}}{}^{\underline{s}}+{\rm C}_{\underline{s}'\underline{t}}{}^{\underline{r}'}\eta_{\underline{r}'\underline{r}}\eta^{\underline{s}'\underline{s}}+
{\rm C}_{\underline{s}'\underline{r}}{}^{\underline{r}'}\eta_{\underline{r}'\underline{t}}\eta^{\underline{s}'\underline{s}}\right)\,\mathcal{P}^{\underline{r}}\,,\label{Nomizu}
\end{equation}
 where ${\rm C}_{\underline{r}\underline{t}}{}^{\underline{s}}={\rm C}_{{r}{t}}{}^{{s}}$. This is the \emph{Nomizu connection} \cite{alek} in terms of which we define the constant components of the curvature tensor.

\paragraph{Conventions for the covariant derivatives.} In supergravity certain quantities, like the fermion-fields or the components of the coset-representative matrix (or the hybrid complex matrix $\mathbb{L}_c$), also transform with respect to the holonomy group $H$ of the scalar manifold, which is a local symmetry of the theory. We then generalize the definition of the covariant derivative and define:
\begin{equation}
{\Scr D}_\mu=\nabla_\mu+\mathcal{Q}_\mu\,,
\end{equation}
where $\mathcal{Q}_\mu\equiv \partial_\mu\phi^s\,\mathcal{Q}_s$ is the composite connection, which makes ${\Scr D}_\mu$ covariant with respect to local $H$-transformations. It acts in the $H$-representation of the field. For instance the covariant derivative of the gravitino field reads:
\begin{equation}
{\Scr D}_\mu\psi_{A\,\nu}=\nabla_\mu\psi_{A\,\nu}+
\mathcal{Q}_{\mu\,A}{}^B\,\psi_{B\,\nu}=\partial_\mu\psi_{A\,\nu}-\Gamma_{\mu\nu}^\rho\,\psi_{A\,\rho}+
\frac{1}{4}\,\omega_{\mu,\,ab}\gamma^{ab}\psi_{A\,\nu}+
\mathcal{Q}_{\mu\,A}{}^B\,\psi_{B\,\nu}\,.
\end{equation}
In the gauged theory, the derivative ${\Scr D}_\mu$ is further extended to a gauge-covariant one $\mathcal{ D}_\mu$:
\begin{equation}
\mathcal{ D}_\mu={\Scr D}_\mu-\Omega_{g\,\mu}\,,
\end{equation}
where $\Omega_{g\,\mu}$ is the gauge-connection, introduced in (\ref{gconnection}), acting in the representation of the field.
\subsection{Symplectic Representations}\label{spapp}
The group ${\rm Sp}(2n_v,\mathbb{R})$ consists of $2n_v\times 2n_v$ real matrices $S=(S^M{}_N)$ satisfying the
condition:
\begin{equation}
S^T\mathbb{C}S=\mathbb{C}\,,\label{scs}
\end{equation}
where $\mathbb{C}=(\mathbb{C}^{MN})=(\mathbb{C}_{MN})$ is given by (\ref{C}). Clearly if $S$ is symplectic, also $S^T$ is. If we write $S$ in block-form as
\begin{equation}
S=\left(\begin{matrix}
A & B\cr
C & D\end{matrix}\right)\,,
\end{equation}
the symplectic condition can be recast as follows:
\begin{equation}
A^T D-C^T B={\bf 1}\,\,,\,\,\,A^T C-C^T A={\bf 0}\,\,,\,\,\,\,B^TD-D^TB=0\,.\label{scfine}
\end{equation}
Being $S^T$ symplectic as well, we also have:
\begin{equation}
A D^T-B C^T={\bf 1}\,\,,\,\,\,A B^T-B A^T={\bf 0}\,\,,\,\,\,\,CD^T-DC^T=0\,.
\end{equation}
If the block $B$ is zero, the first equation implies: $A=D^{-T}$.\par
An infinitesimal generator ${\bf s} $ of ${\rm Sp}(2n_v,\mathbb{R})$ satisfies the condition (\ref{sympcondgen}), namely:
\begin{equation}
{\bf s}^T\mathbb{C}+\mathbb{C}{\bf s}={\bf 0}\,\,\Leftrightarrow\,\,\,\,\mathbb{C}{\bf s}=(\mathbb{C}{\bf s})^T\,.\label{sympcondgen2}
\end{equation}
If ${\bf s}$ has the following block-form:
\begin{equation}
{\bf s}=\left(\begin{matrix}
{\bf a} & {\bf b}\cr
{\bf c} & {\bf d}\end{matrix}\right)\,,
\end{equation}
Eq. (\ref{sympcondgen2}) can be written in the equivalent way:
\begin{equation}
{\bf b}={\bf b}^T\,\,,\,\,\,\,{\bf c}={\bf c}^T\,\,,\,\,\,\,{\bf a}=-{\bf d}^T\,.\label{sympcondgenblock}
\end{equation}
These generators close the algebra $\mathfrak{sp}(2n_v,\mathbb{R})$ of the symplectic group, which splits, according to the Cartan decomposition, into the subalgebra $\mathfrak{u}(n_v)$ of antisymmetric matrices, which generate  ${\rm U}(n_v)\subset {\rm Sp}(2n_v,\mathbb{R})$, and the subspace of symmetric matrices. By virtue of (\ref{sympcondgenblock}), a generator ${\bf u}$ in the former set has the general form:
 \begin{align}
{\bf u}&=\left(\begin{matrix}{\bf a} & {\bf b}\cr -{\bf b} & {\bf a}\end{matrix}\right)\,\,\,\,;\,\,\,\,
{\bf a}=-{\bf a}^T\,\,,\,\,{\bf b}^T={\bf b}\,,
\end{align}
while a generator in the latter space reads:
  \begin{align}
{\bf k}&=\left(\begin{matrix}{\bf a}' & {\bf b}'\cr {\bf b}' & -{\bf a}'\end{matrix}\right)\,\,\,\,;\,\,\,\,
{\bf a}'={\bf a}'^T\,\,,\,\,{\bf b}'^T={\bf b}'\,.\label{noncompsp}
\end{align}
Changing the basis to the complex one by means of the Cayley matrix
introduced in  (\ref{complexV}):
\begin{align}
\mathcal{A}&\equiv \frac{1}{\sqrt{2}}\,\left(\begin{matrix}{\bf 1} &
i\,{\bf 1}\cr {\bf 1} & -i\,{\bf 1} \end{matrix}\right)\,.
\end{align}
the generators of ${\rm U}(n_v)$ become block-diagonal, while the non-compact ones (\ref{noncompsp}) block-off-diagonal:
\begin{align}
\mathcal{A}\,{\bf u}\,\mathcal{A}^\dagger &=\left(\begin{matrix}{\bf a}-i\, {\bf b} & {\bf 0}\cr {\bf 0} & {\bf a}+i\, {\bf b}\end{matrix}\right)\,,\label{undiag}\\
\mathcal{A}\,{\bf k}\,\mathcal{A}^\dagger &=\left(\begin{matrix}{\bf 0}  & {\bf a}'+i\, {\bf b}'\cr  {\bf a}'-i\, {\bf b}'&{\bf 0} \end{matrix}\right)\,.\label{koffdiag}
\end{align}
Notice that ${\bf a}\pm i\, {\bf b}$, being ${\bf a}$ antisymmetric and ${\bf b}$ symmetric, represent a generator of ${\rm U}(n_v)$ in two  representations of which one is the complex conjugate of the other, according to the general decomposition:
\begin{equation}
{\bf 2n_v}\stackrel{{\rm U}(n_v)}\longrightarrow ({\bf n_v})_{+1}+\overline{({\bf n_v})}_{-1}\,.\label{2nvnvnv}
\end{equation}
Let us recall that we use for rising and lowering symplectic indices the North-West/South-East convention:
\begin{equation}
V^M=\mathbb{C}^{MN}\,V_N\,\,;\,\,\,\,V_M=V^N\,\mathbb{C}_{NM}\,.
\end{equation}
\paragraph{The hybrid complex coset representative  $\mathbb{L}_c(\phi)$.} In (\ref{hybridcomplex}) we defined the hybrid complex coset representative which describe the scalar fields in the vector multiplets. Its definition can be extended to the non-homogeneous special K\"ahler manifolds of $\mathcal{N}=2$ models. Let us summarize here its properties. We start writing  it in terms of the $n_v\times n_v$ blocks ${\bf f},\,{\bf h}$ defined in (\ref{hybridcomplex}), as follows
\begin{equation}
\mathbb{L}_c(\phi)=(\mathbb{L}_c^{N}{}_{\underline{M}})=\left(\begin{matrix}
{\bf f} & \bar{{\bf f}} \cr
{\bf h} &\bar{{\bf h}}
\end{matrix}\right)\,.\label{hybridcomplex2}
\end{equation}
Conditions (\ref{propsLc}) can be recast as follows:
\begin{align}
\mathbb{L}_c(\phi)^\dagger \mathbb{C}\,\mathbb{L}_c(\phi)&=\varpi\,\Leftrightarrow\,\,\,\begin{cases}{\bf f}^\dagger {\bf h}-{\bf h}^\dagger {\bf f}=-i\,{\bf 1} \cr {\bf f}^T{\bf h}={\bf h}^T{\bf f}
\end{cases}\,,\label{hfcon1}\\
\mathbb{L}_c(\phi)\, \varpi\,\mathbb{L}_c(\phi)^\dagger &=\mathbb{C}\,\Leftrightarrow\,\,\,\begin{cases}{\bf f}{\bf f}^\dagger =({\bf f}{\bf f}^\dagger)^T \cr {\bf h}{\bf h}^\dagger =({\bf h}{\bf h}^\dagger)^T\cr {\bf f}{\bf h}^\dagger-\bar{{\bf f}}{\bf h}^T=i\,{\bf 1}
\end{cases}\,.\label{hfcon2}
\end{align}
which are a compact rewriting of Eqs. (\ref{iduno})-(\ref{icinque}), (\ref{idsei}) and  (\ref{idsette}).\par
 From the first of (\ref{hfcon1}) we can derive the useful identity:
 \begin{equation}
 {\bf f}^{-1}=i\,({\bf f}^\dagger {\bf h} {\bf f}^{-1}-{\bf h}^\dagger)=i\,({\bf f}^\dagger {\Scr N}-{\bf h}^\dagger)=i\,{\bf f}^\dagger ({\Scr N}-\overline{{\Scr N}})=-2\,{\bf f}^\dagger \mathcal{I}\,,
 \end{equation}
 where we have used the definition (\ref{Nfh}) of ${\Scr N}=({\Scr N}_{\Lambda\Sigma})$ and $\mathcal{I}=(\mathcal{I}_{\Lambda\Sigma})={\rm Im}{\Scr N}$.\par
 The reader can verify that the above relations are equivalent to the following ones:
\begin{equation}
\mathbb{L}_c(\phi)^T \mathbb{C}\,\mathbb{L}_c(\phi)=i\mathbb{C}\,\,,\,\,\,\,\,\mathbb{L}_c(\phi) \mathbb{C}\,\mathbb{L}_c(\phi)^T=i\mathbb{C}\,.
\end{equation}
Using these properties, one can easily check that:
\begin{equation}
\mathbb{L}_c(\phi)^{-1}=i\,\left(\begin{matrix}
-{\bf h}^\dagger & {{\bf f}}^\dagger \cr
{\bf h}^T &-{{\bf f}}^T
\end{matrix}\right)\label{Lcm1}
\end{equation}
We have also used the matrix (\ref{otherL2}):
\begin{equation}
(\LL(\phi)_M{}^{\underline{N}})=\mathbb{L}_{c}^{-T}=i\,\left(\begin{matrix}
-\bar{{\bf h}} & {{\bf h}} \cr
\bar{{\bf f}} &-{{\bf f}}
\end{matrix}\right)\,,\label{otherL22}
\end{equation}
for the definition of the $\mathbb{T}$-tensor. In terms of $(\LL(\phi)_M{}^{\underline{N}})$ the matrix $\mathcal{M}_{MN}$ reads:
\begin{equation}
\mathcal{M}_{MN}=-\LL(\phi)_M{}^{\underline{N}}(\LL(\phi)_N{}^{\underline{N}})^*\,,
\end{equation}
summation over $\underline{N}$ being understood.
\subsection{$X$-Identities}\label{apXid}
Let us summarize there the constraints on the embedding tensor and the identities on the $X_{MN}{}^P$-tensor which follow from them.
The linear constraint (\ref{lconstr}), (\ref{lconstrnew}) is:
\begin{align}
X_{(MNP)}=-X_{(MN}{}^P\mathbb{C}_{P)Q}=0\,&\Leftrightarrow\,\,\,\,X_{(MN)}{}^P=-\frac{1}{2}\,\mathbb{C}^{PL}\,X_{LM}{}^Q\mathbb{C}_{QN}=
-\frac{1}{2}\,\mathbb{C}^{PL}\,\Theta_L{}^\alpha\,t_{\alpha\,MN}\,\Leftrightarrow\nonumber\\
&\Leftrightarrow\begin{cases}
2\,X_{(\Lambda\Sigma)}{}^\Gamma=X^\Gamma{}_{\Lambda\Sigma}\,,\cr
2\,X^{(\Lambda\Sigma)}{}_\Gamma=X_\Gamma{}^{\Lambda\Sigma}\,,\cr X_{(\Lambda\Sigma\Gamma)}=0\,.
\end{cases}
\end{align}
The quadratic constraints (\ref{quadratic1}), (\ref{quadratic2}) read:
  \begin{align}
   &\mathbb{C}^{MN}\Theta_M{}^\alpha\Theta_N{}^\beta=0\,\,;\,\,\,\,\,
   [X_M,\,X_N]=-X_{MN}{}^P\,X_P\,.\nonumber
   \end{align}
   The last identity implies, symmetrizing in $MN$, that:
\begin{equation}
X_{(MN)}{}^P\,X_P=0\,,\label{XsymP}
\end{equation}
which is Eq. (\ref{quad2n}) in the text.
 For those isometries which have a non-trivial duality action, and thus are described by the tensor $X_{MN}{}^P$, we can write (\ref{quadratic2}) in the form
(\ref{quadrdual}):
\begin{equation}
X_{MP}{}^R X_{NR}{}^Q-X_{NP}{}^R X_{MR}{}^Q+X_{MN}{}^R X_{RP}{}^Q=0\,.\label{Xid1}
\end{equation}
If we antisymmetrize both sides of (\ref{Xid1}) in $[MNP]$ we find:
\begin{equation}
2\,X_{[MP}{}^R X_{N]R}{}^Q=-X_{[MN}{}^R X_{R|P]}{}^Q=X_{[MN}{}^R X_{P]R}{}^Q-2\,\left[X_{MN}{}^R X_{(R|P)}{}^Q\right]_{[MNP]}\,,\label{Xid1.5}
\end{equation}
where the subscript $[MNP]$ stands for the antisymmetrization in the three indices. From the above identity we then find:
\begin{equation}
X_{[MN}{}^R X_{P]R}{}^Q=\frac{2}{3}\,\left[X_{MN}{}^R X_{(RP)}{}^Q\right]_{[MNP]}\,,\label{Xid2}
\end{equation}
Next let us write
\begin{align}
\left[X_{MP}{}^R X_{[NR]}{}^Q\right]_{[MNP]}&=\frac{1}{2}\left[X_{MP}{}^R X_{NR}{}^Q-X_{MP}{}^R X_{RN}{}^Q\right]_{[MNP]}=\nonumber\\
&=\frac{1}{2}\left[X_{MP}{}^R X_{NR}{}^Q-2\,X_{MP}{}^R X_{NR}{}^Q\right]_{[MNP]}=-\frac{1}{2}\left[X_{MP}{}^R X_{NR}{}^Q\right]_{[MNP]}=\nonumber\\&=
\frac{1}{3}\left[X_{MN}{}^R X_{(RP)}{}^Q\right]_{[MNP]}\,,\label{Xid22}
\end{align}
where we have used (\ref{Xid1.5}) and (\ref{Xid2}). From the above identity we deduce (\ref{nojacobi}):
\begin{equation}
X_{[MP]}{}^R X_{[NR]}{}^Q-X_{[NP]}{}^R X_{[MR]}{}^Q+X_{[NM]}{}^R X_{[PR]}{}^Q=-(X_{NM}{}^R\,X_{(PR)}{}^Q)_{[MNP]}\,,\nonumber
\end{equation}
which is the failure of the \emph{generalized structure constants} $X_{[MN]}{}^P$ to satisfy the Jacobi identities. Such obstruction is proportional to the tensor $X_{(PR)}{}^Q$ and thus, by virtue of (\ref{XsymP}), vanishes if both sides are contracted with the gauge generators $X_Q$. This ensures, as explained in the text, that the gauge field-strength $\mathcal{F}_{\mu\nu}\equiv F^M_{\mu\nu}\,X_M$ satisfy the Bianchi identities, namely that the gauge connection $A_\mu^M\,X_M$ be well defined.\par
From (\ref{Xid1}), antisymmetrizing in the indices $PN$, we also derive:
\begin{equation}
X_{MN}{}^R\,X_{[PR]}{}^Q-X_{MP}{}^R\,X_{[NR]}{}^Q=-X_{[NP]}{}^R\,X_{MR}{}^Q\,.\label{Xid3}
\end{equation}
From (\ref{Xid2}) and (\ref{Xid1}) the following identity follows:
\begin{equation}
\left(X_{(PQ)}{}^R\,X_{MN}{}^Q\right)_{[MNP]}+\left(X_{(MP)}{}^Q \,X_{NQ}{}^R-X_{[NP]}{}^Q\,X_{(MQ)}{}^R\right)_{[PN]}=0\,.
\end{equation}
Finally, by using (\ref{Xid3}), (\ref{Xid1.5}) and (\ref{Xid2}) we derive the following cubic identity:
\begin{equation}
\left(X_{RN}{}^M\,X_{(SP)}{}^N\,X_{LQ}{}^P\right)_{[LQRS]}=\frac{3}{4}\,X_{(PN)}{}^M\,X_{[RS}{}^N\,X_{LQ]}{}^P\,.
\end{equation}
The above identity is needed, for instance, in order to derive the Bianchi identity (\ref{Bid2n}).
\subsection{Projectors on the Embedding Matrix: a General Discussion}
\label{app-A}
%%%%%%%%%%%%%%%%%%%%%%%%%%%%%%%%%%%%%%%%%%%%%%%%%%%%%%%%%%%%%
In this appendix, taken form \cite{deWit:2002vt},  we explicitly construct the projectors onto the
irreducible representations in the tensor product of the fundamental
with the adjoint representation of an arbitrary simple group G. These
projectors for ${\rm G}={\rm E}_{7(7)}$, and ${\rm G}={\rm E}_{6(6)}$ have been used in
the main text to correctly identify the embedding matrices in $d=4$
and $d=5$, respectively.

Let us assume that the product of a fundamental representation ${\bf
D(\Lambda)}$ times the adjoint decomposes in the direct sum of ${\bf
D(\Lambda)}$ plus two other representations, ${\bf D_1}$ and~${\bf
D_2}$,
\begin{eqnarray}
{\bf D(\Lambda)}\times { Adj(G)}\rightarrow
{\bf D(\Lambda)}+{\bf D_1}+{\bf D_2}  \;.
\label{gendec}
\end{eqnarray}
As far as the lowest-dimensional fundamental representation in
concerned, the above branching rule holds true for any simple group
with the exception of ${\rm E}_8$ for which the fundamental coincides
with the adjoint representation.  The branching also holds for
orthogonal groups when the fundamental representation is replaced by
the spinor representation .  Denote $d_\Lambda={\rm dim}({\bf
D(\Lambda)})$, $d={\rm dim}({\rm G})$, and $\{t^\alpha\}$
($\alpha=1,\dots , d$) the generators of G in the ${\bf D(\Lambda)}$
representation. Furthermore, let $C_\theta,\,C_\Lambda$ be the
Casimirs of the adjoint and fundamental representations,
respectively. We define the invariant matrix $\eta^{\alpha\beta}={\rm
Tr}(t^\alpha t^\beta)$ and use it to rise and lower the adjoint
indices; it is related to the Cartan-Killing metric
$\kappa^{\alpha\beta}$ by
\begin{eqnarray}
\kappa^{\alpha\beta}&=&
\frac{d}{C_\Lambda d_\Lambda}\, \eta^{\alpha\beta}\;.
\end{eqnarray}
Using the definition of the Casimir operator,
$C_\Lambda\,{\bf 1}_{d_\Lambda} =\kappa_{\alpha\beta} t^\alpha
t^\beta$, we have the following relation
\begin{eqnarray}
{\rm f}_{\alpha\beta}{}^\gamma\, {\rm f}^{\alpha\beta}{}_\sigma
&=&-\frac{d}{d_\Lambda}\, C_{\rm r}\,\delta_{\sigma}^\gamma
\;,\qquad
\mbox{with} \quad
C_{\rm r}=\frac{C_\theta}{C_\Lambda}=
\frac{d_\Lambda}{d}\frac{g^{\vee}}{\tilde{I}_\Lambda}
\;,
\end{eqnarray}
where $g^{\vee}$ is the dual Coxeter number and $\tilde{I}_\Lambda$
is the Dynkin index of the fundamental representation.  In the simply
laced case there is a useful formula:
\begin{eqnarray}
C_{\rm r}&=&\frac{d_\Lambda}{d}\left(
\frac{d}{r}-1\right)\frac{1}{\tilde{I}_\Lambda}
\;,
\end{eqnarray}
with $r$ the rank of G.

Denote the projectors on the representations in (\ref{gendec}) by
$\mathbb{P}_{{\bf D(\Lambda)}},\,\mathbb{P}_{{\bf D_1} },\,
\mathbb{P}_{{\bf D_2} }$ which sum to the identity on ${\bf
D(\Lambda)}\times { Adj(G)}$.  These three projectors can be
expressed in terms of three independent objects, namely:
\begin{eqnarray}
\mathbb{P}_{{\bf D(\Lambda)M}}{}^{\alpha N} {}_\beta
&=&\frac{d_\Lambda}{d}\, (t^\alpha t_\beta)_M{}^N\,,\nonumber\\
\mathbb{P}_{{\bf D_1}M }{}^{\alpha N} {}_\beta &=& a_1 \,\delta^\alpha
{}_{\!\beta}\,\delta_M{}^N+a_2 \, (t_\beta t^\alpha)_M{}^N+a_3\,
(t^\alpha
t_\beta)_M{}^N\,,\nonumber\\
\mathbb{P}_{{\bf D_2}M }{}^{\alpha N} {}_\beta&=&
(1-a_1)\,\delta^\alpha{}_{\!\beta}\, \delta_M{}^N-a_2\, (t_\beta
t^\alpha)_M{}^N-({d_\Lambda/d}+a_3)\,(t^\alpha t_\beta)_M{}^N
\;,
\label{projectors}
\end{eqnarray}
with constants $a_1, a_2, a_3$. Making use of the fact that only three
representations appear in the decomposition (\ref{gendec}), these
coefficients may be determined by computing the contractions of
various products of the projectors (\ref{projectors}). This yield
\begin{eqnarray}
a_1 &=&\frac{d_\Lambda\left(4+(C_{\rm r}-4)
d)\right)+\Delta\left((C_{\rm r}-2)
d-2)\right)}{\left(10+d(C_{\rm r}-8)+d^2 (C_{\rm r}-2)\right)
d_\Lambda}\,, \nonumber\\
a_2 &=&-\frac{2\left(4+(C_{\rm r}-4)
d)\right)\left((d-1)d_\Lambda-2\Delta\right)}{\left(10+d(C_{\rm
r}-8)+d^2 (C_{\rm r}-2)\right) C_{\rm r} d }\,, \nonumber\\
a_3 &=&\frac{-d_\Lambda\left(4+(C_{\rm r}-4)
d)\right)\left(2+(C_{\rm r}-2)d\right)+\Delta\left(16(d-1)-10 (d-1)
C_{\rm r}+C_{\rm r}^2 d \right)}{\left(10+d(C_{\rm r}-8)+d^2 (C_{\rm
r}-2)\right) C_{\rm r} d } \;,
\nonumber
\end{eqnarray}
with $\Delta=\dim({\bf D_1})$. Moreover, $\Delta$ is determined to be
\begin{eqnarray}
\Delta &=& \frac{d_\Lambda}{2}\left[d-1+\frac{\sqrt{C_{\rm
r}}\left(10+d(C_{\rm r}-8)+d^2 (C_{\rm r}-2)\right) }{\sqrt{256
(d-1)+C_{\rm r} (100+4 d (5 C_{\rm r}-38)+(C_{\rm r}-2)^2
d^2)}}\right] \;.{\quad{~}}
\label{Delta}
\end{eqnarray}
In table \ref{projs} the relevant data are collected for all simple
Lie algebras except ${\rm E}_{8}$ (for which the relevant projectors
have been computed in \cite{Koepsell:1999uj}).
%%%%%%%%%%%%%%%%%%%%%%%%%%%%%%%%%%%%%%%%%%%%%%%%%%%%%%%%%%%%%%
%%%%%%%%%%%%%%%%%%%%%%%%%%%%%%%%%%%%%%%%%%%%%%%%%%%%%%%%%%%%%%
\begin{table}
\begin{center}
\begin{tabular}{l c c l l r r r}\hline
G & $g^{\vee}$ & $d_\Lambda$ & $\tilde{I}_\Lambda $ & $\Delta$       &
$a_1$   & $a_2$ & $a_3$ \\
\hline
${\rm A_r}$ & $r+1$ & $r+1$ &$\frac{1}{2}$    &
$\frac{1}{2}(r-1)(r+1)(r+2)$ &  $\frac{1}{2}$ &  $-\frac{1}{2}$&
$-\frac{1}{2 r}$  \\
${\rm B_r}$   & $2r-1$ & $2r+1$& 1    & $\frac{1}{3}r (4 r^2-1)$  &
$\frac{1}{3}$ & $-\frac{2}{3}$ & $0$ \\
${\rm B_r}$   & $2r-1$ & $2^{r}$& $2^{r-3}$    & $2^{r+1}\,r$  &
$\frac{2}{2r-1}$ & $-2^{r-1}\frac{1}{2r-1}$ & $2^{r-1}\,\frac{2r-7}{4
r^2-1}$ \\
 ${\rm C_r}$   & $r+1$ & $2r$& $\frac{1}{2}$    & $\frac{8}{3}r
(r^2-1)$  & $\frac{2}{3}$ & $-\frac{2}{3}$ & $-\frac{2}{1+2r}$ \\
${\rm D_r}$   & $2r-2$ & $2r$ & 1    & $\frac{2}{3}r (2r^2-3 r +1)$  &
$\frac{1}{3}$ &  $-\frac{2}{3}$ & $0$ \\
${\rm D_r}$   & $2r-2$ & $2^{r-1}$ &  $2^{r-4}$   & $2^{r-1}\,(2r-1)$
& $\frac{1}{r-1}$ &  $-2^{r-3}\frac{1}{r-1}$ &
$2^{r-3}\frac{(r-4)}{r\,(r-1)}$ \\
 ${\rm G_2}$   & 4 & 7 & 1    & $27$  & $\frac{3}{7}$ &
$-\frac{6}{7}$ & $-\frac{3}{14}$ \\
 ${\rm F_4}$   & 9 & 26 & 3    & $273$  & $\frac{1}{4}$ &
$-\frac{3}{2}$ & $\frac{1}{4}$ \\
${\rm E}_{6}$ & 12 & 27 & 3 & 351  &  $\frac{1}{5}$  & $-\frac{6}{5}$
& $\frac{3}{10}$ \\
${\rm E}_{7}$ & 18 & 56 &6       & 912   & $\frac{1}{7}$ &
$-\frac{12}{7}$ &   $\frac{4}{7}$  \\
\hline
\end{tabular}
\end{center}
\caption{\small
Coefficients for the projector $\mathbb{P}_{{\bf D_1}}$ for the
various algebras. }
\label{projs}
\end{table}

\subsection{Comparison with the Notations of {\tt 0705.2101}}\label{apcomparison}
In this review we adopt, for the supergravity fields, notations which are different from those used in the literature of maximal supergravity (see for instance \cite{deWit:2007mt}), in order to make contact with the literature of gauged $\N<8$,in particular $\N=2$, theories \cite{Andrianopoli:1996cm}. Denoting \emph{only here} by a hat the quantities in \cite{deWit:2007mt}, the correspondence between the two notations is:
\begin{align}
&\hat{\gamma}^\mu =i\gamma^\mu\,;\quad\hat{\gamma}_\mu =-i\gamma_\mu\,;\quad
\hat{\gamma}_5 =\gamma_5\,,\nonumber\\
\hat{\epsilon}_{\mu\nu\rho\sigma}=i\,{\epsilon}_{\mu\nu\rho\sigma}\,,\nonumber\\
&\hat{\epsilon}_i=\frac{1}{\sqrt{2}}\,\epsilon^A\,;\quad
\hat{\epsilon}^i=\frac{1}{\sqrt{2}}\,\epsilon_A\quad
(i=A)\,,\nonumber\\
&\hat{\psi}_{i\mu} =\sqrt{2}\,\psi^A_\mu\,;\quad
\hat{\psi}^i_{\mu} =\sqrt{2}\,\psi_{A\,\mu}\quad
(i=A)\,,\nonumber\\
&\hat{\chi}_{ijk}=\chi^{ABC}\,;\quad
\hat{\chi}^{ijk}=\chi_{ABC}\quad
([ijk]=[ABC])\,,\nonumber\\
&\hat{A}_{ij}=(\hat{A}_{ij})^*=A^{AB}\,;\quad
\hat{A}_i{}^{jkl}=(\hat{A}^i{}_{jkl})^*=A^A{}_{BCD}\quad
(i=A,\,j=B,\,k=C,\,l=D)\,,\nonumber\\
&\mathcal{V}^{\Lambda\,ij}=-\frac{i}{\sqrt{2}}\,f^\Lambda{}_{AB}\,;\quad
\mathcal{V}_{\Lambda}{}^{ij}=
\frac{i}{\sqrt{2}}\,h_{\Lambda AB}\,;\quad
(i=A,\,j=B)\,,\nonumber
\end{align}
where in the last line the $28\times 28$ blocks of $\mathcal{V}_M{}^{\underline{N}}$ have been put in correspondence with those of $\LL^M{}_{\underline{N}}$, the factor $\sqrt{2}$ originates from a different convention with the contraction of antisymmetric couples of ${\rm SU}(8)$-indices:\; $$\hat{V}_{ij}\hat{V}^{ij}=\frac{1}{2}\,V^{AB}\,V_{AB}\,.$$
We also have the following correspondences
\begin{align}
\hat{F}^{-\,ij}_{\mu\nu}&=\frac{i}{\sqrt{2}}\,F^{-}_{AB\,\mu\nu}\,\,,\,\,\,\,
\hat{H}^{-\,ij}_{\mu\nu}=\frac{i}{\sqrt{2}}\,H^{-}_{AB\,\mu\nu}\quad(i=A,\,j=B)\,,\nonumber\\
\hat{\mathcal{P}}_s^{ijkl}&=-\frac{1}{2}\,\mathcal{P}_{s\,ABCD}\quad (i=A,\,j=B,\,k=C,\,l=D)\,,\nonumber\\
\hat{\mathcal{Q}}_{ij}{}^{kl}&=\frac{1}{2}\,\mathcal{Q}^{AB}{}_{CD}\,\,\,\,;\,\,\,\hat{\mathcal{Q}}_{i}{}^{j}=2\,\mathcal{Q}^{A}{}_{B}\quad (i=A,\,j=B,\,k=C,\,l=D)\,.\nonumber\\
(\hat{T}_{ij})_{kl}{}^{pq}&=\frac{1}{4\sqrt{2}}\,(\mathbb{T}^{AB})^{CD}{}_{EF}\quad (i=A,\,j=B,\,k=C,\,l=D,\,p=E,\,q=F)\,,\nonumber\\(\hat{T}_{ij})^{klpq}&=
\frac{1}{4\sqrt{2}}\,(\mathbb{T}^{AB})_{CDEF}\quad (i=A,\,j=B,\,k=C,\,l=D,\,p=E,\,q=F)\,,\nonumber
\end{align}
analogous relations holding for the components of $\hat{T}^{ij}$.
\subsection{Fluxes and ${\rm E}_{7(7)}$ Weights}\label{appdualsugras}
In this appendix, which is taken form \cite{Trigiante:2007ki}, we illustrate is some detail how to associate the
known fluxes, in ten-dimensional Type II superstring compactified
on a six-torus, with components of the embedding tensor which
defines the corresponding four-dimensional gauged supergravity,
namely with elements of the ${\bf 912}$ representation of ${\rm
E}_{7(7)}$. An element of a Lie group $G$  representation  is
characterized by its transformation property under the action of
the maximal torus of $G$, generated by its Cartan subalgebra
(CSA). This property is encoded in the \emph{weights} of the
representation. Let $G$ be the global symmetry group of an
extended supergravity theory. Its maximal torus has a diagonal
action on the electric field strengths and their duals, and
therefore it is a symmetry of the ungauged Lagrangian, namely it
is contained in $G_e$. If fluxes are to be assigned, by
identification with components of the embedding tensor, to
$G$-representations, so as to restore on shell global $G$
invariance, they should couple in the action to the dilatonic
fields parametrizing the CSA of $G$ according to their weights. In
the maximal theory the CSA of ${\rm E}_{7(7)}$ is seven
dimensional and is parametrized, from the Type II superstring
point of view, by the six radial moduli of the internal torus
$R_{{\tt u}}=e^{\sigma_{{\tt u}}}$ (${{\tt u}}=4,\dots,9$) and by the ten-dimensional
dilaton $\phi$. The bosonic zero-modes of Type II superstring
theory consist in the ten-dimensional dilaton $\phi$, the metric
$\mathbb{V}_{\hat{\mu}}{}^{\hat{a}}$
($\hat{\mu},\,\hat{\nu}=0,\dots,9$ and
$\hat{a},\,\hat{b}=0,\dots,9$ are the curved and rigid ten-dimensional indices respectively), a NS-NS 2--form $\hat{B}^{(2)}$, odd
R--R forms, $\hat{C}^{(1)},\,\hat{C}^{(3)}$, in Type IIA theory and even R--R
forms, $\hat{C}^{(0)},\,\hat{C}^{(2)},\,\hat{C}^{(4)}$, in Type IIB. The ansatz for
the metric in the string frame reads
\begin{eqnarray}
\mathbb{V}_\mu{}^r=
e^{\phi_4}\,V_\mu{}^r\,\,\,;\,\,\,\,\mathbb{V}^{\hat{{\tt u}}}=\upPhi_{{\tt u}}{}^{\hat{{\tt u}}}\,(dx^{{\tt u}}+A^{{\tt u}}_\mu\,dx^\mu)\,,
\end{eqnarray}
where ${\tt u,v}=4,\dots, 9$ and $\hat{{\tt u}},\hat{{\tt v}}=4,\dots, 9$ are the
curved and rigid indices on the six torus respectively,
$V_\mu{}^r$ is the four-dimensional metric in the four-dimensional
Einstein frame, $A^{{\tt u}}_\mu$ are the six Kaluza Klein vectors,
$\phi_4=\phi-\frac{1}{2}\,\sum_{{\tt u}}\sigma_{{\tt u}}$ is the four-dimensional
dilaton and $\upPhi_{{\tt u}}{}^{\hat{{\tt v}}}$ are the metric moduli of the
internal torus, which can be identified with the coset
representative of ${\rm GL}(6,\mathbb{R})/{\rm SO}(6)$. By
suitably fixing the ${\rm SO}(6)$ symmetry we can adopt the
solvable Lie algebra representation of the manifold ${\rm
GL}(6,\mathbb{R})/{\rm SO}(6)$ \cite{Cremmer:1978ds,Andrianopoli:1996bq,Cremmer:1997ct} and write
$\upPhi_{{\tt u}}{}^{\hat{{\tt v}}}$ in the form
\begin{eqnarray}
\upPhi_{{\tt u}}{}^{\hat{{\tt v}}}&\equiv &
U\,e^{\sum_{{\tt u}=4}^9\sigma_{{\tt u}}\,H_{\epsilon_{{\tt u}}}}\,,\nonumber\\
U&=&\prod_{{\tt u<v}}e^{\gamma_{{\tt u}}{}^{{\tt v}}\,E_{{\tt u}}{}^{{\tt v}}}\,\,\,\,\,(\mbox{no
summation})\,,\label{Uu}
\end{eqnarray}
where $\epsilon_{{\tt u}}$ is an orthonormal basis of vectors, $E_{{\tt u}}{}^{{\tt v}}$
are the ${\rm SL}(6,\mathbb{R})$ shift generators corresponding to
the positive root $\epsilon_{{\tt u}}-\epsilon_{{\tt v}}$ and $\gamma_{{\tt u}}{}^{{\tt v}}$ are
the moduli parametrizing the off-diagonal components of the
internal metric. The internal metric will read
$g_{{\tt uv}}=-\sum_{\hat{{\tt w}}}\upPhi_{{\tt u}}{}^{\hat{{\tt w}}}\upPhi_{{\tt v}}{}^{\hat{{\tt w}}}$. Let
us define a representative of the maximal torus of ${\rm
E}_{7(7)}$ to have the form $\exp(H_{\vec{h}})$, where $\vec{h}$
is defined as
\begin{eqnarray}
\vec{h}(\sigma,\phi)&=&\sum_{u=4}^9\sigma_{{\tt u}}\,\epsilon_{{\tt u}}-\sqrt{2}\,\phi_4\,\epsilon_{10}=
\sum_{u=4}^9\hat{\sigma}_{{\tt u}}\,(\epsilon_{{\tt u}}+\frac{1}{\sqrt{2}}\,\epsilon_{10})-\frac{1}{2}\,\,\phi\,a\,,\nonumber\\
a&=&-\frac{1}{2}\,\sum_{u=4}^9\epsilon_{{\tt u}}+\frac{1}{\sqrt{2}}\,\epsilon_{10}\,,\label{hvect}
\end{eqnarray}
where $\epsilon_I=(\epsilon_{{\tt u}},\epsilon_{10})$ is an orthonormal
basis of seven dimensional vectors and
$\hat{\sigma}_{{\tt u}}=\sigma_{{\tt u}}-\phi/4$ are the radial moduli in the ten-dimensional Einstein frame. The four-dimensional Lagrangian,
resulting from the dualization of the 2-forms to scalar fields,
will contain the following terms
\begin{eqnarray}
e^{-1}{\Scr L}_{scal}&=&
\frac{1}{2}\partial_\mu\vec{h}\cdot\partial^\mu\vec{h}+\frac{1}{4}\,\sum_{u,v}\,e^{-2\,\alpha^{B}_{{\tt uv}}\cdot
\vec{h}}\,H_{\mu {\tt uv}}\,H^{\mu}{}_{{\tt uv}}+\frac{1}{2}\,\sum_{u<v}\,e^{-2\,\alpha_{{\tt u}}{}^{{\tt v}}\cdot
\vec{h}}\,{\tt P}_{\mu u}{}^{{\tt v}}\,{\tt P}^\mu{}_{{\tt u}}{}^{{\tt v}}+\nonumber\\
&&+\sum_k\frac{1}{2k!}\,\sum_{{{\tt u}}_1,\dots,{{\tt u}}_k}\,e^{-2\,\alpha^C_{{{\tt u}}_1\dots
{{\tt u}}_k}\cdot \vec{h}}\,F_{\mu {{\tt u}}_1\dots {{\tt u}}_k} F^\mu{}_{{{\tt u}}_1\dots {{\tt u}}_k}+\frac{1}{2}\,e^{-2\,\alpha^{B}\cdot
\vec{h}}\,\partial_\mu \tilde{B}\,\partial^\mu \tilde{B}\dots\,,\label{lscal}\nonumber\\&&\\
e^{-1}{\Scr L}_{vec}&=&-\frac{1}{4}\,\sum_{u}e^{-2\,W^{{\tt u}}\cdot \vec{h}
}\,F^{{\tt u}}_{\mu\nu}\,F^{u\,\mu\nu}-\frac{1}{4}\,\sum_{u}e^{-2\,W_{{\tt u}}^B\cdot
\vec{h}
}\,H_{\mu\nu\,u}H^{\mu\nu}{}_{{\tt u}}-\nonumber\\&&-
\sum_k\frac{1}{4(k-1)!}\,\sum_{{{\tt u}}_1,\dots,{{\tt u}}_{k-1}}\,e^{-2\,W^C_{{{\tt u}}_1\dots
{{\tt u}}_{k-1}}\cdot \vec{h}}\,F_{\mu\nu\,{{\tt u}}_1\dots
{{\tt u}}_{k-1}}F^{\mu\nu}{}_{{{\tt u}}_1\dots
{{\tt u}}_{k-1}}+\dots\,.\nonumber\\&&\label{lvec}
\end{eqnarray}
where ${\tt P}_{\mu\,u}{}^{{\tt v}}\equiv (U^{-1}\partial_\mu U)_{{\tt u}}{}^{{\tt v}}$ and the internal indices of the scalar and vector fields are
``dressed'' with the matrix $U$ in Eq. (\ref{Uu}), which only depends on the moduli $\gamma_{{\tt u}}{}^{{\tt v}}$ associated with the  off-diagonal entries of the metric:
\begin{equation}
F_{\mu_1\dots \mu_p\,{{\tt u}}_1\dots
{{\tt u}}_{\ell}}\equiv (U^{-1})_{{{\tt u}}_1}{}^{{{\tt v}}_1}\dots (U^{-1})_{{{\tt u}}_{\ell}}{}^{{{\tt v}}_{\ell}}(p\,\partial_{[\mu_1}A_{\mu_2\dots\mu_p]\,{\tt {{\tt v}}_1\dots
{{\tt v}}_{\ell}}}+\dots)\,,
\end{equation}
and an analogous definition holds for the components of the $H$-tensor.
The 2-form $B_{\mu\nu}$ has been dualized to the axion $\tilde{B}$ while
in the Type IIA theory the tensors $C_{\mu\nu u}$ were dualized to
$\tilde{C}^{{\tt u}}\equiv \epsilon^{{\tt u u_1\dots u_5} }\,C_{{\tt u_1\dots u_5}}/5!$ and in the Type IIB
theory $C_{\mu\nu}$ was dualized to the scalar $\tilde{C}=C_{4\dots 9}$. The
range of values of $k$ in the summations in (\ref{lscal}) is: $k=1,3,5$ in Type IIA and $k=0,2,4,6$ in Type
IIB. In
(\ref{lvec}), on the other hand, $k$ run over the following values:  $k=1,3$ in Type IIA and $k=0,2$ in Type
IIB.  The seven dimensional vectors $\alpha$ and $W$ in the
exponential factors of (\ref{lscal}) and (\ref{lvec}) have the
form
\begin{align}
\alpha_{{\tt u}}{}^{{\tt v}}&=\epsilon_{{\tt u}}-\epsilon_{{\tt v}}\,\,;\,\,\,\alpha^B_{{\tt uv}}=\epsilon_{{\tt u}}+\epsilon_{{\tt v}}\,\,;\,\,\,\alpha^B=\sqrt{2}\,\epsilon_{10}\,\,;\,\,\,
\alpha^C_{u_1\dots u_k}=a+\epsilon_{{\tt u}_1}+\dots
+\epsilon_{{\tt u}_k}\,,\label{alpha}\\
&(k=1,3,5,\,\,\mbox{IIA}\,\,,\,\,\,\,\,k=0,2,4,6,\,\,\mbox{IIB})\,,\nonumber\\
W^{{\tt u}}&=-\epsilon_{{\tt u}}-\frac{1}{\sqrt{2}}\,\epsilon_{10}\,\,;\,\,\,W^B_{{\tt u}}=\epsilon_{{\tt u}}-\frac{1}{\sqrt{2}}\,\epsilon_{10}\,\,;\,\,\,W^C_{{\tt u_1\dots
u_{k-1}}}=a+\epsilon_{{\tt u}_1}+\dots
+\epsilon_{{\tt u}_{k-1}}-\frac{1}{\sqrt{2}}\,\epsilon_{10}\,,\label{ww}\\
&(k=1,3,\,\,\mbox{IIA}\,\,,\,\,\,\,\,k=0,2,\,\,\mbox{IIB})\,,\nonumber
\end{align}
If we define the simple roots of the $\mathfrak{e}_{7(7)}$ algebra, see Figure \ref{fige70},
to be of the form\footnote{Recall that the root $\alpha_7$ is the ${\rm Spin}(6,6)$-spinorial weight ${\bf W_{32}}$, whose chirality depends on whether we are in the Type IIA or Type IIB description.}
\begin{align}
\alpha_{{\tt u}-3}&=&\epsilon_{{\tt u}}-\epsilon_{{\tt u}+1}\,\,\,({\tt u}=4,\dots,
8)\,;\,\,\,\,\alpha_6=\epsilon_8+\epsilon_9\,;\,\,\,\alpha_7=\begin{cases}{\bf W_{32_s}}=a&
\mbox{Type IIB}\cr {\bf W_{32_c}}=a+\epsilon_9 & \mbox{Type IIA}
\end{cases}\,,\label{iiaiib}
\end{align}
 the vectors in (\ref{alpha}) are the $\mathfrak{e}_{7(7)}$ positive
roots while those in (\ref{ww}), together with their opposite $-W$
(corresponding to the magnetic vector fields), are the weights of
the ${\bf 56}$ representation (in the Type IIB description, the
weights $W^C_{{{\tt u_1u_2u_3}}}$ are $20$ and correspond to the vectors
$C_{\mu {{\tt u_1u_2u_3}}}$ originating from the 4-form; by virtue of the
property of the 5-form field strength of being self dual, these 20
weights already include 10 weights corresponding to electric
vector fields and their opposite associated with the magnetic
duals).
We may follow a similar strategy in order to associate
fluxes with $\mathfrak{e}_{7(7)}$ weights, namely read off the
weight from the dilaton dependence of the term in the action of
the form $(\mbox{flux})^2$:
\begin{eqnarray}
e^{-2W^T_{{\tt uv}}{}^{{\tt w}}\cdot
\vec{h}}\,(T_{{\tt uv}}{}^{{\tt w}})^2\,\,,\,\,\,\,\,e^{-2W^H_{{\tt uvw}}\cdot
\vec{h}}\,(H^{(3)}_{{\tt uvw}})^2\,\,,\,\,\,\,\,e^{-2\,W^F_{{\tt u_1\dots u_{k+1}}}\cdot
\vec{h}}\,(F^{(k+1)}_{{\tt u_1\dots u_{k+1}}})^2\,,\label{lflux}
\end{eqnarray}
where the term containing $(T_{{\tt uv}}{}^{{\tt w}})^2$ is part of the
Scherk-Schwarz potential \cite{Scherk:1979zr}. The values of $k$ in the RR fluxes are: $k=-1,1,3,5$ in Type IIA theory, corresponding to the
internal components of the forms
$F^{(0)},\,F^{(2)},\,F^{(4)},\,F^{(6)}$, and $k=0,2,4$ in Type IIB
theory, corresponding to the field strengths
$F^{(1)},\,F^{(3)},\,F^{(5)}$. We may also consider RR fluxes
with four space-time indices which do not explicitly break Lorentz
invariance. By performing the dimensional reduction, we find the
general field-strength--weight correspondence:
\begin{eqnarray}
H^{(3)}_{{\tt u_1u_2u_3}}&\leftrightarrow &
W^H_{{\tt u_1u_2u_3}}=\epsilon_{{\tt u}_1}+\epsilon_{{\tt u}_2}+\epsilon_{{\tt u}_3}+\frac{1}{\sqrt{2}}\,\epsilon_{10}\,,\nonumber\\
T_{{\tt u_1u_2}}{}^{{\tt u_3}}&\leftrightarrow &
W^T_{{\tt u_1u_2}}{}^{{\tt u_3}}=\epsilon_{{\tt u}_1}+\epsilon_{{\tt u}_2}-\epsilon_{{\tt u}_3}+\frac{1}{\sqrt{2}}\,\epsilon_{10}\,,\nonumber\\
F^{(k+1)}_{\mu_1\dots\mu_\ell {\tt u_1\dots u_{s}}}&\leftrightarrow &
W^F_{\mu_1\dots\mu_\ell {\tt u_1\dots
u_s}}=-\frac{1}{2}\,\sum_{{\tt u}}\epsilon_{{\tt u}}+\epsilon_{{\tt u_1}}+\dots+\epsilon_{{\tt u_s}}+
\frac{2-\ell}{\sqrt{2}}\,\epsilon_{10}\,\,\,\,\,(\ell+s=k+1)\,.\nonumber\\&&\label{weightflux}
\end{eqnarray}
In the M-theory reduction on a torus the ${\rm O}(1,1)$ factor in
$G_e={\rm GL}(7,\mathbb{R})$ is generated by the Cartan operator
$H_\lambda$ where
\begin{eqnarray}
\lambda&=&\sum_{{\tt u}}\epsilon_{{\tt u}}+2\sqrt{2}\,\epsilon_{10}\,,
\end{eqnarray}
and the ${\rm O}(1,1)$-grading associated with the field strengths
in (\ref{weightflux}) are simply computed as the scalar product of
$\lambda$ with the corresponding weight $W$: $\lambda\cdot W$.
From the embedding of the ${\rm SL}(6,\mathbb{R})$ group,
corresponding to the six torus in the compactification of Type II
theories, inside ${\rm E}_{7(7)}$, we may deduce the ${\rm
SL}(6,\mathbb{R})$-representation  of each of the weights in
(\ref{weightflux}) and identify it, together with the relevant
${\rm O}(1,1)$ gradings, with representations in the branching of
the embedding tensor representation ${\bf 912}$. The embedding of
${\rm SL}(6,\mathbb{R})$ inside ${\rm E}_{7(7)}$ is defined by
identifying its simple roots with $\alpha_1\dots \alpha_5$.\par
\paragraph{Dualities.} Let us now consider the effect of dualities.
T-dualities $T^{(u_1,\dots, u_k)}$ along the internal directions $u_1,\dots, u_k$, are implemented by the following ${\rm O}(6,6)$ matrices:
\begin{equation}
T^{(u_1,\dots, u_k)}=\left(%
\begin{matrix}
\mathbf{1}_6-\mathbf{D}_{(k)} & \mathbf{D}_{(k)}\cr {\bf D}_{(k)} & \mathbf{1}_6-
\mathbf{D}_{(k)}
\end{matrix}
\right)\,,  \label{outaut}
\end{equation}
where each block is an $6\times 6$ matrix and the only non-vanishing entries of $\mathbf{D}_{(k)}$ are $\mathbf{D}_{(k)\,{\tt u}_\ell-3,\, {\tt u}_\ell-3}=1$, $\ell=1,\dots, k$. Such transformation, for odd $k$, belongs to the $\mathrm{O}(6)$ subgroup
of $\mathrm{O}(6,6)$ (it has negative determinant), and is an outer automorphism of the $D_6$, while for even $k$ it is part of the Weyl group of the same algebra \cite{Lu:1996ge}.
It has the effect of
changing the sign to $\epsilon_{{\tt u}_\ell}$, $\ell=1,\dots, k$, or, equivalently, to
their coefficients in $\vec{h}$:
\begin{equation}
\epsilon_{{\tt u}_\ell}\rightarrow -\epsilon_{{\tt u}_\ell}\,\,\,;\,\,\,\ell=1,\dots,
k\,.
\end{equation}
To see this let us consider the effect of $T^{({\tt u_1,\dots, u_k})}$ on the dilatonic
part of the coset representative ${\bf L}_6$ of $\mathrm{O}(6,6)/[\mathrm{O}(6)\times
\mathrm{O}(6)]$, which has the following form:
\begin{equation}
{\bf L}_6(\sigma_{{\tt u}})=\left(%
\begin{matrix}
(e^{\sigma_{{\tt u}}}\delta_{{\tt u}-3}{}^{{\tt v}-3}) & \mathbf{0}\cr {\bf 0} & (e^{-\sigma_{{\tt u}}}\delta_{{\tt u}-3}{}^{{\tt v}-3})%
\end{matrix}%
\right)\,\,,\,\,\,\,{\tt u,v}=4,\dots, 9\,.
\end{equation}
We see that:
\begin{equation}
(T^{({\tt u_1,\dots, u_k})})^{-1}{\bf L}_6(\sigma_{{\tt u}})T^{({\tt u_1,\dots, u_k})}={\bf L}_6(\sigma_{{\tt u}}^{\prime })\,,
\end{equation}
where $\sigma_{{\tt u}_\ell}^{\prime }=-\sigma_{{\tt u}_\ell},\,\sigma_{{\tt u}\neq {\tt u}_\ell}^{\prime
}=\sigma_{{\tt u}\neq {\tt u}_\ell}$, $\ell=1,\dots, k$, and thus $T^{({\tt u_1,\dots, u_k})}$ amounts to a $T$-duality along the internal directions $x^{{\tt u_\ell}}$ \cite{Lu:1996ge,Bertolini:1999uz}:
\begin{equation}
R^{\prime }_{{\tt u}_\ell}=e^{{\sigma}^{\prime }_{{\tt u}_\ell}}=e^{-{\sigma}
_{i_\ell}}=\frac{1}{R_{{\tt u}_\ell}}\,\,;\,\,\,\phi^{\prime }=\phi-
\sum_{\ell=1}^k{\sigma}_{{\tt u}_\ell}\,.
\end{equation}
 We can verify that under the effect of $T^{({\tt w})}$ the
weight of $H_{{\tt uvw}}$ is mapped into the weight of $T_{{\tt uv}}{}^{{\tt w}}$ and
moreover subsequent actions of $T^{({\tt v})}$ and $T^{({\tt u})}$ allow to
define the weights $W^Q{}_{{\tt u}}{}^{{\tt vw}}$ and $W^{R\,{\tt uvw}}$, associated
with the non-geometric fluxes $Q_{{\tt u}}{}^{vw}$ and $R^{{\tt uvw}}$, respectively
\begin{eqnarray}
W^H_{{\tt uvw}}&\stackrel{T^{({\tt w})}}{\longrightarrow }&
W^T_{{\tt uv}}{}^{{\tt w}}\stackrel{T^{({\tt v})}}{\longrightarrow
}W^Q{}_{{\tt u}}{}^{{\tt vw}}\stackrel{T^{({\tt u})}}{\longrightarrow
}\,\,W^{R\,{\tt uvw}}\,\nonumber\\
W^Q{}_{{\tt u}}{}^{{\tt vw}}&=&\epsilon_{{\tt u}}-\epsilon_{{\tt v}}-\epsilon_{{\tt w}}+\frac{1}{\sqrt{2}}\,\epsilon_{10}\,\,\,;\,\,\,\,\,
W^{R\,{\tt uvw}}=-\epsilon_{{\tt u}}-\epsilon_{{\tt v}}-\epsilon_{{\tt w}}+\frac{1}{\sqrt{2}}\,\epsilon_{10}\,.
\end{eqnarray}
Other duality transformations are implemented as  Weyl
transformations $\sigma_\alpha$ \cite{Lu:1996ge} corresponding to $\mathfrak{e}_{7(7)}$-roots
$\alpha$ which are not $\mathfrak{so}(6,6)$ roots and whose action on a weight $W$ is defined as follows
\begin{eqnarray}
W&\longrightarrow &\sigma_\alpha (W)=W-2\,\left(\frac{W\cdot
\alpha}{\alpha\cdot \alpha}\right)\,\alpha\,.
\end{eqnarray}
The effect of these
transformations on the dilatonic scalars in $\vec{h}$ can be inferred from the invariance of the scalar product
$\vec{h}\cdot \overline{W}$ in the exponents appearing in the various kinetic terms and potential terms: $\vec{h}\cdot \overline{W}=\sigma_\alpha (W)\cdot
\sigma_\alpha(\vec{h})$.
As a consequence of this the field which is described by the weight $W$ (which, for
a scalar field, is a positive root) in the original theory,
corresponds to the new weight $\sigma_\alpha (W)$ in the dual one, which features a new set of dilatonic scalars
$\sigma_{{\tt u}}^\prime,\,\phi^\prime$, entering the dilatonic vector
$\sigma_\alpha(\vec{h})$. The relation between
$\sigma_{{\tt u}}^\prime,\,\phi^\prime$ and $\sigma_{{\tt u}},\,\phi$ can be
deduced by the following condition
\begin{eqnarray}
\vec{h}(\sigma^\prime,\phi^\prime)&\equiv
&\sigma_\alpha(\vec{h}(\sigma,\phi))\,\,\,\Rightarrow\,\,\,\,\sigma^\prime=\sigma^\prime(\sigma,\phi)\,,\,\,\,\phi^\prime=\phi^\prime(\phi)\,.\label{htrans}
\end{eqnarray}
The non perturbative $S$--duality is implemented as a Weyl
transformation with respect to the vector $\alpha=a$ in Eq.
(\ref{hvect}), which is an $\mathfrak{e}_{7(7)}$ root in Type IIB
theory, but not in Type IIA theory, see equation (\ref{iiaiib}).
This represents the known fact that $S$--duality is a symmetry of
Type IIB theory (it corresponds to an ${\rm E}_{7(7)}$
transformation) but not of Type IIA theory (it maps Type IIA
superstring into M-theory). If we compute its action on the
dilatonic scalars, using (\ref{htrans}), we find
\begin{eqnarray}
\hat{\sigma}_{{\tt u}}^\prime &=
&\hat{\sigma}_{{\tt u}}\,\,\,;\,\,\,\,\phi^\prime=-\phi\,,
\end{eqnarray}
where $\hat{\sigma}_{{\tt u}}$ are the radial moduli in the ten-dimensional Einstein frame. One can verify, using the weight
representation in Eq. (\ref{weightflux}), that in Type IIB theory
\begin{eqnarray}
\sigma_a(W^H_{{\tt u_1 u_2 u_3}})&=&W^F_{{\tt u_1 u_2 u_3}}\,,
\end{eqnarray}
which is the known $S$--duality correspondence between the NS-NS
and the RR 3--form fluxes $H^{(3)}_{{\tt u_1 u_2 u_3}}$, $F^{(3)}_{{\tt u_1 u_2 u_3}}$. One can also verify that the torsion $T_{{\tt uv}}{}^{{\tt w}}$ is
inert under $S$--duality, while the action of $S$--duality on the
non--geometric fluxes gives rise to more general fluxes which we
can identify with components of the embedding tensor, knowing the
corresponding weights.\par
Weyl transformations associated with positive roots $\alpha$ different from the $\mathfrak{so}(6,6)$-ones or from $a$ in the Type IIB picture, define proper U-duality transformations.
\subsection{The Scalar Potential for the ${\rm SO}(8)_\omega$-Model in the ${\rm G}_2$-Invariant Sector}\label{Aso8om}
 The scalar potential in the $G_2$-invariant sector has the form \cite{Dall'Agata:2012bb}
\begin{equation}
	\label{G2potential}
	V(\vec \phi) = A(\vec \phi) - \cos (2 \omega) f(\phi_1,\phi_2) - \sin (2 \omega) f(\phi_2,\phi_1),
\end{equation}
having denoted by $\vec{\phi}=(\phi_1,\,\phi_2)$, $x \equiv e^{|\vec \phi|}$ and
\begin{eqnarray}
	A(\vec \phi) &=&\textstyle \frac{(1+ x^4)^3}{64 |\vec \phi|^4 \,x^{14}}\left[4(1+ x^4)^2(1-5 x^4+ x^8)(\phi_1^4+\phi_2^4) \right. \nonumber\\[1mm]
	&+& \left.\phi_1^2 \phi_2^2 (1+ 4x^4-106 x^8+4x^{12}+x^{16})\right],
\end{eqnarray}
which is an even function of $\phi_1$ and $\phi_2$ and symmetric in their exchange. The function $f$ is defined as
\begin{equation}
	\begin{array}{rl}	
	f(\phi_1,\phi_2) &= \frac{(-1+ x^4)^5 \,\phi_1^3}{64 |\vec \phi|^7 \,x^{14}}\left[4(1+5 x^4 + x^8)\phi_1^4 + \right.\\[2mm]
	&+ \left. 7 (1+ 6 x^4 + x^8) \phi_1^2 \phi_2^2 + 7 (1+x^4)^2 \phi_2^4\right],
	\end{array}
\end{equation}
The scalar potential is manifestly invariant under the following three discrete transformations:
\begin{equation}
	\label{symmetriespot}
	\left\{\begin{array}{l}
	\omega \leftrightarrow - \omega \\[2mm]
	\phi_2 \leftrightarrow - \phi_2
 	\end{array}\right.\,, \ \
	\left\{\begin{array}{l}
	\omega \leftrightarrow \omega + \frac{\pi}{2} \\[2mm]
	\vec\phi \leftrightarrow - \vec\phi
 	\end{array}\right.\,, \ \
	\left\{\begin{array}{l}
	\omega \leftrightarrow \omega-\frac{\pi}{4} \\[2mm]
	\phi_1 \to \phi_2 \\[2mm]
	\phi_2 \to -\phi_1
 	\end{array}\right.,
\end{equation}
the former being parity, the other two being in ${\rm E}_{7(7)}$. In particular the middle one is implemented by the symplectic matrix ${\bf S}=\mathbb{C}$, the latter by
\begin{equation}{\bf S}_T=\frac{1}{\sqrt{2}}\,\left(\begin{matrix}\boldsymbol{\Gamma} & \boldsymbol{\Gamma}\cr -\boldsymbol{\Gamma} & \boldsymbol{\Gamma} \end{matrix}\right)\,,\end{equation}
 $\boldsymbol{\Gamma}$ being the triality transformation defined in Sect. \ref{symfram}.
\subsection{The c-map}\label{BGM}
This Appendix is taken form \cite{Andrianopoli:2016eub}. Here we recall the formal steps to define the c-map of an  $\mathcal{N}=2$ supergravity Lagrangian describing a number of vector multiplets.\par
 Let us start from an $\mathcal{N}=2$ supergravity model of $n$ vector multiplets in four dimensions \cite{Cecotti:1988qn} whose bosonic Lagrangian has the following general form:
\begin{equation}
 {e}^{-1}{\Scr L}_4=-\frac{ {R}}{2}+g_{i\bar{\jmath}}\,\partial_\mu z^i\partial^\mu \bar{z}^{\bar{\jmath}}+\frac{1}{4}\, {F}^{\Lambda}_{ {\mu} {\nu}} \mathcal{I}_{\Lambda\Sigma}\, {F}^{\Sigma\, {\mu} {\nu}}
+\frac{1}{4}\, {F}^{\Lambda}_{ {\mu} {\nu}} \mathcal{R}_{\Lambda\Sigma}\,{}^* {F}^{\Sigma\, {\mu} {\nu}}\,,
\end{equation}
where $ {\mu,\nu}=0,1,2,3$ and $g_{i\bar{\jmath}}$ is the metric on the special K\"ahler manifold ${\Scr M}_{SK}$ spanned by $z^i$.\par
We can perform a dimensional reduction, along an internal circle, to three dimensions on a background with metric:
\begin{equation}
ds^2=e^{-2\varphi }\,g_{\hat \mu\hat \nu}\,dx^{\hat \mu} dx^{\hat \nu}-e^{2\varphi }\,(dx^3+\hat{A}^{(3)})^2\,,
\end{equation}
where  ${\hat \mu},{\hat \nu}=0,1,2$ and $g_{{\hat \mu}{\hat \nu}}=g_{{\hat \mu}{\hat \nu}}(x^{\hat \rho})$, $\hat{A}^{(3)}=\hat{A}^{(3)}_{\hat \mu}(x^{\hat \nu}) dx^{\hat \mu}$ are the $D=3$ metric and Kaluza-Klein vector (only in this Appendix we use hatted symbols to denote lower-dimensional quantities).
The vectors in $D=4$ reduce to three dimensional ones as follows:
\begin{equation}
A^\Lambda=\hat{A}^\Lambda_{\hat \mu}(x^{\hat \nu})\,dx^{\hat \mu}+\zeta^\Lambda(x^{\hat \nu})\,V^3\,\,,\,\,\,\,V^3=dx^3+\hat{A}^{(3)}\,.\label{43vec}
\end{equation}
where $V^3$ is proportional to the vielbein in the isometry direction and we have set $A^\Lambda_3= \zeta^\Lambda$.
The corresponding field strengths read:
\begin{equation}
F^\Lambda=\hat{F}^\Lambda+F^\Lambda_3\,V^3\,,
\quad \mbox{where }
\hat{F}^\Lambda=d\hat{A}^\Lambda+\zeta^\Lambda\,\hat{F}^{(3)}\,\,,\,\,F^\Lambda_3\equiv d\zeta^\Lambda \,\,,\,\,\hat{F}^{(3)}\equiv d\hat{A}^{(3)}\,.
\end{equation}
Next we consider the $D=3$ Lagrangian which is given by the four-dimensional one written in terms of three dimensional fields, plus a Chern-Simons term inducing the dualization of the $D=3$ vector fields $\hat{A}^{(3)}$, $\hat{A}^{\Lambda }$ to scalar degrees of freedom $a$, $\tilde\zeta_\Lambda$:
\begin{eqnarray}
{\hat e}^{-1}{\Scr L}_3&=&-\frac{\hat{R}}{2}+\partial_{\hat \mu} \varphi \partial^{\hat \mu} \varphi -\frac{e^{4\varphi }}{8} {\hat F}^{(3)}_{{\hat \mu}{\hat \nu}}\,{\hat F}^{(3)\,{\hat \mu}{\hat \nu}}+g_{i\bar{\jmath}}\,\partial_{\hat \mu} z^i\partial^{\hat \mu} \bar{z}^{\bar{\jmath}}+\nonumber\\
&&+\frac{e^{2\varphi }}{4}\,{{\hat F}}^{\Lambda}_{{{\hat \mu}}{{\hat \nu}}} \mathcal{I}_{\Lambda\Sigma}\,{{\hat F}}^{\Sigma\,{{\hat \mu}}{{\hat \nu}}}-\frac{e^{-2\varphi }}{2}\,\partial_{\hat \mu} \zeta^{\Lambda}\mathcal{I}_{\Lambda\Sigma}\,\partial^{\hat \mu} \zeta^{\Sigma}-\frac{1}{2\,{\hat e}}\,\epsilon^{{\hat \mu}{\hat \nu}{\hat \rho}}\,{\hat F}^\Lambda_{{\hat \mu}{\hat \nu}}\mathcal{R}_{\Lambda\Sigma}\partial_{\hat \rho} \zeta^\Sigma+\nonumber\\
&&+{\hat e}^{-1}{\Scr L}_{CS}\,,
\end{eqnarray}
where ${\hat e}\equiv \sqrt{{\rm det}(g_{\hat{\mu}\hat{\nu}})}$,
$
{\Scr L}_{CS}=\frac{1}{2}\,\epsilon^{{\hat \mu}{\hat \nu}{\hat \rho}}\,{\hat F}^\Lambda_{{\hat \mu}{\hat \nu}}\,\partial_{\hat \rho} \tilde{\zeta}_\Lambda-\frac{1}{4}\epsilon^{{\hat \mu}{\hat \nu}{\hat \rho}}\,{\hat F}^{(3)}_{{\hat \mu}{\hat \nu}}\,\omega_{\hat \rho}\,,
$
and we have defined $\epsilon_{{\hat \mu}{\hat \nu}{\hat \rho}}=\epsilon_{{\hat \mu}{\hat \nu}{\hat \rho} 3}$, so that $\epsilon_{012}=1$. The vector $\omega_{\hat \mu}$ is given in terms of scalar degrees of freedom and reads:
\begin{equation}
\omega_{\hat \mu}\equiv \partial_{\hat \mu} {a}+ \zeta^\Lambda\,\partial_{\hat \mu} \tilde{\zeta}_\Lambda-\partial_{\hat \mu} \zeta^\Lambda\,\tilde{\zeta}_\Lambda\,.
\end{equation}
Integrating out ${\hat F}^\Lambda_{{\hat \mu}{\hat \nu}}$ and ${\hat F}^{(3)}_{{\hat \mu}{\hat \nu}}$ we find the following equations:
\begin{eqnarray}
{\hat F}^{\Lambda\,{\hat \mu}{\hat \nu}}&=&\frac{e^{-2\varphi }}{{\hat e}}\epsilon^{{\hat \mu}{\hat \nu}{\hat \rho}}\,\mathcal{I}^{-1\,\Lambda\Sigma}(\mathcal{R}_{\Sigma\Gamma}\,\partial_{\hat \rho} \zeta^\Gamma-\partial_{\hat \rho} \tilde{\zeta}_\Sigma)\,,\nonumber\\
{\hat F}^{(3)\,{\hat \mu}{\hat \nu}}&=&-\frac{e^{-4\varphi }}{{\hat e}}\,\epsilon^{{\hat \mu}{\hat \nu}{\hat \rho}}\,\omega_{\hat \rho}\,.
\end{eqnarray}
Replacing the above solutions in ${\Scr L}_3$ we find the final expression of the three dimensional Lagrangian fully written in terms of scalar degrees of freedom and exhibiting manifest ${\rm Sp}(2n+2,\mathbb{R})$-structure \cite{Breitenlohner:1987dg}:
\begin{align}
{\hat e}^{-1}{\Scr L}_3&=-\frac{{{\hat R}}}{2}+\partial_{\hat \mu} \varphi \partial^{\hat \mu} \varphi +\frac{e^{-4\varphi }}{4} \omega_{\hat \mu}\,\omega^{\hat \mu}+g_{i\bar{\jmath}}\,\partial_{\hat \mu} z^i\partial^{\hat \mu} \bar{z}^{\bar{\jmath}}-\frac{e^{-2\varphi }}{2}\,\partial_{\hat \mu} \zeta^T \mathcal{I}\,\partial^{\hat \mu} \zeta-\nonumber\\
&-\frac{e^{-2\varphi }}{2}\,\left(\partial_{\hat \mu} \tilde{\zeta}^T-\partial_{\hat \mu}\zeta^T\mathcal{R}\right)\mathcal{I}^{-1}\left(\partial^{\hat \mu} \tilde{\zeta}-\mathcal{R}\partial^{\hat \mu}\zeta\right)=\nonumber\\
&=
-\frac{{{\hat R}}}{2}+\partial_{\hat \mu} \varphi \partial^{\hat \mu} \varphi +\frac{e^{-4\varphi }}{4} \omega_{\hat \mu}\,\omega^{\hat \mu}+g_{i\bar{\jmath}}\,\partial_{\hat \mu} z^i\partial^{\hat \mu} \bar{z}^{\bar{\jmath}}-\frac{e^{-2\varphi }}{2}\,\partial_{\hat \mu}\mathcal{Z}^M\mathcal{M}_{MN}\partial^{\hat \mu}\mathcal{Z}^N\,,\label{localc}
\end{align}
being  $\mathcal{Z}^M=\{\zeta^\Lambda,\,\tilde{\zeta}_\Lambda\}$ and $\mathcal{M}$ is the matrix defined in (\ref{M}).
Eq. (\ref{localc}) is the bosonic Lagrangian of $n+1$ hypermultiplets (containing the scalars $\{\zeta^\Lambda,\,\tilde{\zeta}_\Lambda$, $ z^i, \bar z^{\bar\jmath}, \varphi , a\}$) coupled to gravity in a $D=3,\,\mathcal{N}=4$ supergravity theory. The target space ${\Scr M}_{QK}$ of the sigma model is of quaternionic K\"ahler type \cite{Ferrara:1989ik}. One of the multiplets (corresponding to $\varphi , a$ and $\zeta^0,\,\tilde{\zeta}_0$) is the \emph{universal hypermultiplet} containing the degrees of freedom of the supergravity multiplet in $D=4$. Its four scalars span the characteristic submanifold ${\rm SU}(1,2)/{\rm U}(2)$ of ${\Scr M}_{QK}$.\par  The fields $(\mathcal{Z}^M,a)$ are acted on by the isometries
\begin{equation}
\delta\mathcal{Z}^M=\upalpha^M\,,\quad\quad \delta a =\upbeta -\upalpha^M \mathbb{C}_{MN} \mathcal{Z}^N\,,
\end{equation}
$\upalpha^M,\,\upbeta$ being constant parameters, which close a characteristic Heisenberg algebra ${\Scr H}$ \cite{Ferrara:1989ik}.

 However, since hypermultiplets couple in the same way both to $D=3$ and to $D=4$ supergravity,  it can be promoted to a $D=4$ Lagrangian  describing the coupling of $n+1$ hypermultiplets to $D=4$ supergravity, by just extending the range of indices to $0,\dots,3$.\par The scalar manifold ${\Scr M}_{QK}$ can be written as isometric to the following space:
 \begin{equation}
 {\Scr M}_{QK}\sim \left({\rm O}(1,1)\times {\Scr M}_{SK}\right)\ltimes e^{{\Scr H}}\,,\label{MQKHeis}
 \end{equation}
 where ${\rm O}(1,1)$ is parametrized by $\varphi $ and its generator will be denoted by $t_0$.
 ${\Scr M}_{QK}$ is homogeneous (symmetric) if and only if ${\Scr M}_{SK}$ is homogeneous (symmetric). In this case we can define the coset representative $L_{QK}$ of ${\Scr M}_{QK}$ in terms of that of ${\Scr M}_{SK}$ ($L_{SK}$) as follows:
 \begin{equation}
 L_{QK}=e^{a t_\bullet}\,e^{\mathcal{Z}^M\,t_M}\,L_{SK}\,e^{2\,\varphi \,t_0}\,,
 \end{equation}
 where we have introduced the generators $t_M,\,t_\bullet$ of the Heisenberg algebra which, together with $t_0$, satisfy the following commutation relations:
 \begin{align}
 [t_0,\,t_M]=\frac{1}{2}\,t_M\,,\,\,[t_0,\,t_\bullet]=t_\bullet\,,\,\,[t_M,\,t_N]=-2\,\mathbb{C}_{MN}\,t_\bullet\,.
 \end{align}
 The isometry generators $t_a$ of ${\Scr M}_{SK}$ commute with $t_0,\,t_\bullet$ while they have the following commutation relation with $t_M$:
 \begin{equation}
 [t_a,\,t_M]=-t_{a\,M}{}^N\,t_N\,.
 \end{equation}
The above construction shows that the $(2n+4)$-dimensional algebra of isometries $\mathfrak{o}(1,1)\oplus {\Scr H}$, generated by $\{t_0,\,t_M,\,t_\bullet\}$, is a characteristic of all the quaternionic manifolds in the image of the c-map.\par
In this appendix we have discussed the dimensional reduction to $D=3$ of a generic $\mathcal{N}=2$ supergravity model describing $n$ vector multiplets. This yields the image through the c-map of the special K\"ahler scalar manifold in four dimensions. When applied to the four-dimensional models originating from Calabi-Yau reduction
of Type IIA theory, this formal procedure defines the correspondence between ${\Scr M}_{SK}^{(1)}$ spanned by the complexified K\"ahler moduli $w^a$ in the $n=h_{1,1}$ vector multiplets and its c-map image, which we have denoted by  ${\Scr M}_{QK}^{({\rm IIB})}$ (the hyper-scalars in four dimensions span the same manifold in the three dimensional theory). Similarly in the Type IIB picture the manifold ${\Scr M}_{SK}^{(2)}$ spanned by the complex structure moduli is mapped in the quaternionic space that we have denoted by ${\Scr M}_{QK}^{({\rm IIA})}$.
Therefore both ${\Scr M}_{QK}^{({\rm IIA})}$ and ${\Scr M}_{QK}^{({\rm IIB})}$ describing the hyper-scalars in the
Type IIA and IIB theories on a same Calabi-Yau, have the metric structure illustrated above, with $\varphi$ replaced by $-\phi_4$, $a$ by $\tilde{B}$ in both theories and $z^i$ by $w^a$ and $\mathcal{Z}^M$ by $\mathcal{Z}^{\mathcal{A}}$ in the Type IIB one only, see Sect. \ref{rtcy}.
\paragraph{Geometry of the quaternionic manifold.}
Following \cite{Ferrara:1989ik}, let us rewrite the quaternionic metric in the following form
\begin{equation}
ds^{2}=h_{uv} dq^u\,dq^v=u\bar{u}+\bar{e}^{I}e_{I}+v\bar{v}+E_{I}\overline{E}^{I}\,,
\end{equation}
where the quantities in the right-hand-side are defined as follows:
\begin{align}
u  &  =ie^{-\varphi  }V^{T}%
%TCIMACRO{\\varphi {2102} }%
%BeginExpansion
\mathbb{C}
%EndExpansion
d\mathcal{Z,}\\
v  &  =d\varphi  +\frac{i}{2}e^{-2\varphi  }\chi\,,\text{ \ \ \ \ \ }%
\chi=da+\mathcal{Z}%
%TCIMACRO{\\varphi {2102} }%
%BeginExpansion
\mathbb{C}
%EndExpansion
d\mathcal{Z,}\\
E_{I}  &  =-ie^{-\varphi  }e^{-1}{}_I{}^{\bar{\imath}}\overline{U}_{\bar{\imath}}^{T}%
%TCIMACRO{\\varphi {2102} }%
%BeginExpansion
\mathbb{C}
%EndExpansion
d\mathcal{Z}\,,\\
e_{I}  &  =e_{i\,I}\,dz^{i},
\end{align}
where $e_{i\,I},\,e_{\bar{\imath}}{}^I$ are the complex vielbein on ${\Scr M}_{SK}$ and $V=\left(
V^{M }\right)  ,$  is the covariantly holomorphic symplectic section on the same manifold.

Let us define the $SU(2)$-connection $\omega^{x}$. This quantity is encoded in
the following $2\times 2$ matrix \textbf{p }:%
\begin{equation}
\mathbf{p=}\left(
\begin{array}
[c]{cc}%
\frac{1}{4}(v-\bar{v})+\frac{i}{2}\mathcal{Q} & -u\\
\bar{u} & -\frac{1}{4}(v-\bar{v})-\frac{i}{2}\mathcal{Q}
\end{array}
\right)  =\omega^{x}\left(  -i\frac{\sigma^{x}}{2}\right)  ,
\end{equation}
where $\mathcal{Q}$ is the K\"{a}hler connection defined in(\ref{connec}).
The components read%
\begin{align}
\omega^{x}  &  =i{\rm Tr}\left(  \sigma^{x}\mathbf{p}\right)  ,\nonumber\\
\omega^{1}  &  =i(u-\bar{u})=e^{-\varphi  }(\bar{V} +V )^{T}%
%TCIMACRO{\\varphi {2102} }%
%BeginExpansion
\mathbb{C}
%EndExpansion
d\mathcal{Z,}\nonumber\\
\omega^{2}  &  =u+\bar{u}=ie^{-\varphi  }(V -\bar{V} )^{T}%
%TCIMACRO{\\varphi {2102} }%
%BeginExpansion
\mathbb{C}
%EndExpansion
d\mathcal{Z,}\nonumber\\
\omega^{3}  &  =-\frac{i}{2}(\bar{v}-v-2i\mathcal{Q})=-\operatorname{Im}v-\mathcal{Q}=-\frac
{1}{2}e^{-2\varphi  }\chi-\mathcal{Q}.
\end{align}
The ${\rm SU}(2)$-curvature reads:%
\begin{align}
d\mathbf{p+p\wedge p}&=R^{x}\left(  -i\frac{\sigma^{x}}%
{2}\right)\,\,;\,\,\,\,\,
R^{x} =d\omega^{x}+\frac{1}{2}\epsilon^{xyz}\omega^{y}\mathbf{\wedge
}\omega^{z}=- K^{x}.
\end{align}
where $K^{x}$ is the hyper-K\"{a}hler 2-form defined in (\ref{Kxdef}). The ${\rm Sp}(2n+2,\mathbb{R})$-curvature is given in \cite{Ferrara:1989ik}.\par
The vielbein 1-forms $\mathcal{U}^{A\alpha}$, which are needed for writing the hyperini fermion shift tensor read:
\[
\mathcal{U}^{1\alpha}=\frac{1}{\sqrt{2}}\left(
\begin{array}
[c]{c}%
\bar{u}\\
\bar{e}^{I}\\
-v\\
-E_{I}%
\end{array}
\right)  ;\ \ \ \ \ \ \mathcal{U}^{2\alpha}=\frac{1}{\sqrt{2}}\left(
\begin{array}
[c]{c}%
\bar{v}\\
\bar{E}^{I}\\
u\\
e_{I}%
\end{array}
\right)  \text{ \ \ },
\]
The reader can verify that the above expressions satisfy the properties of $\mathcal{U}^{A\alpha}$ given in Sect. \ref{QMans}.\par
The Killing vectors $k_{\alpha}=\left(  k_{M },k_{\bullet
}\right)  $ associated with the Heinsenberg isometries read%
\begin{equation}
k_{M }=\frac{\partial}{\partial\mathcal{Z}^{M }}%
+\mathcal{Z}^{N}%
%TCIMACRO{\\varphi {2102} }%
%BeginExpansion
\mathbb{C}
%EndExpansion
_{NM}\frac{\partial}{\partial a};\text{ \ \ \ \ }k_{\bullet}%
=\frac{\partial}{\partial a}.\label{KHeis}
\end{equation}
One can verify that
\begin{equation}
\left[  k_{M },k_{N}\right]  =2%
%TCIMACRO{\\varphi {2102} }%
%BeginExpansion
\mathbb{C}
%EndExpansion
_{ MN }k_{\bullet},
\end{equation}
all others commutators vanish.  The momentum maps read%
\begin{equation}
\mathcal{P}_{\alpha}^{x}=\left\{  \mathcal{P}_{M }^{x},\mathcal{P}%
_{\bullet}^{x}\right\}  =-k_{\alpha}^{u}\omega_{u}^{x},
\end{equation}%
\begin{align}
\mathcal{P}_{M }^{x}  &  =(e^{-\varphi  }%
%TCIMACRO{\\varphi {2102} }%
%BeginExpansion
\mathbb{C}
%EndExpansion
_{ MN }(V +\bar{V} )^{N},ie^{-\varphi  }%
%TCIMACRO{\\varphi {2102} }%
%BeginExpansion
\mathbb{C}
%EndExpansion
_{ MN }(V -\bar{V} )^{N},-e^{-2\varphi  }%
%TCIMACRO{\\varphi {2102} }%
%BeginExpansion
\mathbb{C}
%EndExpansion
_{ MN }\mathcal{Z}^{N}),\nonumber\\
\mathcal{P}_{\bullet\text{ }}^{x}  &  =\mathcal{P}_{\bullet\text{ }}^{3}%
=\frac{e^{-2\varphi  }}{2}.\label{PHeis}
\end{align}
\subsection{More on STU Geometry}\label{STUstruc}
The STU model is a consistent truncation of the maximal theory: Its bosonic sector is defined by truncating the scalar and vector fields to the largest subset such that the only remaining, non-vanishing components of the composite vector fields strengths $F^{AB}_{\mu\nu}$ are:
\begin{equation}
F^{AB}_{\mu\nu}\,\stackrel{{\rm truncation}}{\longrightarrow} \,\,F^{12}_{\mu\nu},\,F^{34}_{\mu\nu},\,F^{56}_{\mu\nu},\,F^{78}_{\mu\nu}\,.\label{STUtruncn8}
\end{equation}
Similarly the corresponding central charges of the maximal theory ${\Scr Z}_{AB}$ truncate as follows: ${\Scr Z}_{AB}\,\rightarrow\,\,{\Scr Z}_{12},\,{\Scr Z}_{34},\,{\Scr Z}_{56},\,{\Scr Z}_{78}$. One of these four components coincides with the $\mathcal{N}=2$ central charge ${\Scr Z}$ of the resulting truncation,\footnote{Which of the four coincides with ${\Scr Z}$ depends on how the $\mathcal{N}=2$ residual supersymmetry is embedded in the original $\mathcal{N}=8$. For a study of the different embeddings of the STU model inside the maximal one, see for instance \cite{Bertolini:1999uz}} while the other three coincide with the matter charges $\overline{{\Scr Z}}_I$, $I=1,2,3$. Note that, by means of a ${\rm SU}(8)$ transformation, the $\mathcal{N}=8$ central charge matrix can always be skew-diagonalized, so that the only non-vanishing entries are just $Z_1={\Scr Z}_{12},\,Z_2={\Scr Z}_{34},\,Z_3={\Scr Z}_{56},\,Z_4={\Scr Z}_{78}$:
 \begin{equation}
 {\Scr Z}_{AB}\,\stackrel{{\rm SU}(8)}{\longrightarrow} \,\,\left(\begin{matrix}Z_1\,\epsilon & {\bf 0}& {\bf 0}& {\bf 0}\cr
{\bf 0}  &Z_2\,\epsilon & {\bf 0}& {\bf 0}\cr {\bf 0}&{\bf 0}  &Z_3\,\epsilon &  {\bf 0}\cr {\bf 0}&{\bf 0}&{\bf 0}  &Z_4\,\epsilon \end{matrix}\right)\,\,;\,\,\,\,\epsilon\equiv \left(\begin{matrix}0 & 1\cr -1 & 0\end{matrix}\right)\,.
 \end{equation}
 In doing this ${\rm SU}(8)$ is broken to ${\rm U}(1)^3 \times {\rm SU}(2)^4$. The ${\rm SU}(2)^4$ factor is the \emph{stabilizer} of the skew-diagonal form, namely the largest subgroup of ${\rm SU}(8)$ which leaves it unaltered, while ${\rm U}(1)^3$ is its \emph{normalizer}, which acts on the skew-eigenvalues by means of phase factors, leaving their overall phase invariant \cite{Andrianopoli:1997wi}. This latter group is the isotropy group of ${\Scr M}_{{\rm scal}}^{(STU)}$. Similarly the isometry group ${\rm SL}(2,\mathbb{R})^3$ of ${\Scr M}_{{\rm scal}}^{(STU)}$ can be characterized as the largest subgroup of ${\rm E}_{7(7)}$ which, acting simultaneously on the scalar and vector fields, does not switch on components of the composite field strengths $F^{AB}_{\mu\nu}$ other than the four in (\ref{STUtruncn8}). These consequently transform by a compensator in the normalizer group ${\rm U}(1)^3$. \par
The feature that the central and matter charges of the STU truncation of maximal supergravity coincide with the skew-eigenvalues of the central charge matrix ${\Scr Z}_{AB}$ implies that the most general single-center black hole solution to the former encodes all the duality-invariant properties of the generic solution of the same kind in the latter model. Such properties are described by ${\rm SU}(8)$-invariant functions of  ${\Scr Z}_{AB}$, which are five and are in one to one correspondence with the moduli of the four skew-eigenvalues $Z_\ell$ and their overall phase. For this reason the STU model has played an important role in the study of black hole solutions in maximal supergravity. \footnote{See \cite{Andrianopoli:2006ub} and references therein. For more recent works on the STU model in relation to the study of black hole solutions see, for instance, \cite{Bergshoeff:2008be,Bossard:2011kz} and \cite{Chow:2013tia}. } The STU model truncation of the maximal theory has also been used in order to study vacua of the ${\rm SO}(8)$-gauged model, see for instance \cite{Duff:1999rk} and references therein. In this case the $\mathcal{N}=2$ model features a ${\rm U}(1)^4$ gauge symmetry, corresponding to the Cartan subalgebra of ${\rm SO}(8)$.\par
Let us now give some more mathematical details related to the special geometry of the scalar manifold.
The symplectic coset representative $\mathbb{L}(\phi^s)^M{}_N$ in Sect. \ref{STUgeometry0} is constructed as follows. One starts from the hybrid matrix $\mathbb{L}_c$ defined in (\ref{LcN2}) and computes the real symplectic matrix $\mathbb{L}(\phi^s)^M{}_{\underline{N}}$ (note the underlined second index) using the Cayley matrix:
\begin{equation}
\mathbb{L}(\phi^s)^M{}_{\underline{N}}=\mathbb{L}_c(\phi^s)^M{}_{\underline{P}}\mathcal{A}^{\underline{P}}{}_{\underline{N}}=
\sqrt{2}\,\left({\rm Re}(V),\,{\rm
Re}(U_I),\,-{\rm Im}(V),{\rm Im}(U_I)\right)\,.
\end{equation}
Note that this matrix is not yet the coset representative $\mathbb{L}(\phi^s)^M{}_N$ in (\ref{L4stu}) since it is not the identity at the origin $\mathcal{O}$ defined by $\phi^s=0$ (i.e. $\varphi_i=0,\,a_i=0$). We thus define:
\begin{equation}
\mathbb{L}(\phi^s)^M{}_P\equiv \mathbb{L}(\phi^s)^M{}_{\underline{N}}\,\mathbb{L}(\mathcal{O})^{-1\, \underline{N}}{}_P\,.
\end{equation}
This matrix satisfies the property:
 \begin{equation}
 V^M(\phi^r)= \mathbb{L}(\phi^r)^M{}_N\,V^N(\phi^r\equiv  0)\,.
 \end{equation}
and is the coset representative in the solvable gauge. To show this we compute the generators:
\begin{equation}
{\bf h}_i=\left.\frac{\partial \mathbb{L}}{\partial
\varphi_i}\right\vert_{\phi^r\equiv 0}\,\,;\,\,\,{\bf
E}_i=\left.\frac{\partial \mathbb{L}}{\partial
a_i}\right\vert_{\phi^r\equiv 0}\,,
\end{equation}
which can be verified to close the solvable Lie algebra ${\Scr S}$, Borel subalgebra of $\mathfrak{g}^{(SK)}$.\par
The matrix $\mathcal{M}_{MN}$ is readily computed using the general formula
\begin{align}
\mathcal{M}_{MN}&=-\sum_{P=1}^8 \mathbb{L}(\phi^r)_M{}^P\mathbb{L}(\phi^r)_N{}^P\,,
\end{align}
and the matrices $\mathcal{I}_{\Lambda\Sigma}$ and $\mathcal{R}_{\Lambda\Sigma}$ read:
{\small \begin{align}
\mathcal{I}_{\Lambda\Sigma}&=\left(
\begin{array}{llll}
 -\frac{b_1 b_2 a_3^2}{b_3}-\frac{\left(a_2^2 b_1^2+\left(a_1^2+b_1^2\right) b_2^2\right) b_3}{b_1 b_2}
   & \frac{a_1 b_2 b_3}{b_1} & \frac{a_2 b_1 b_3}{b_2} & \frac{a_3 b_1 b_2}{b_3} \\
 \frac{a_1 b_2 b_3}{b_1} & -\frac{b_2 b_3}{b_1} & 0 & 0 \\
 \frac{a_2 b_1 b_3}{b_2} & 0 & -\frac{b_1 b_3}{b_2} & 0 \\
 \frac{a_3 b_1 b_2}{b_3} & 0 & 0 & -\frac{b_1 b_2}{b_3}
\end{array}
\right)\,,\nonumber\\
\mathcal{R}_{\Lambda\Sigma}&=\left(
\begin{array}{llll}
 2 a_1 a_2 a_3 & -a_2 a_3 & -a_1 a_3 & -a_1 a_2 \\
 -a_2 a_3 & 0 & a_3 & a_2 \\
 -a_1 a_3 & a_3 & 0 & a_1 \\
 -a_1 a_2 & a_2 & a_1 & 0
\end{array}
\right)\,,
\end{align}}
where $b_i\equiv e^{\varphi_i}$.
\bibliographystyle{unsrtnat}
\bibliography{gaugedsugra}
 \end{document}